\newif\ifprintfig
\printfigtrue
\documentclass{emulateapj}
\usepackage{graphicx}
\usepackage{lscape}
\newcommand\degree{{^\circ}}

\newcommand\Mpc{{\rm\,Mpc}}

\newcommand\Gyr{{\rm\,Gyr}}

\newcommand\kmsec{{\rm\,km\,s^{-1}}}
\newcommand\kms{\kmsec}

\newcommand\surfb{{\rm\,mag\,arcsec^{-2}}}

\newcommand\clock{\count0=\time \divide\count0 by 60
     \count1=\count0 \multiply\count1 by -60 \advance\count1 by \time
     \number\count0:\ifnum\count1<10{0\number\count1}\else\number\count1\fi}
\shortauthors{Dalcanton et al.}
\shorttitle{ANGST}

\begin{document}
%\draft

\title{The ACS Nearby Galaxy Survey Treasury}

\author{Julianne J. Dalcanton\altaffilmark{1},
Benjamin F. Williams\altaffilmark{1}, 
Anil C. Seth\altaffilmark{2},  
Andrew Dolphin\altaffilmark{3}, 
Jon Holtzman\altaffilmark{4}, 
Keith Rosema\altaffilmark{1},
Evan D. Skillman\altaffilmark{5}, 
Andrew Cole\altaffilmark{6},
L\'eo Girardi\altaffilmark{7}, 
Stephanie M. Gogarten\altaffilmark{1}, 
Igor D. Karachentsev\altaffilmark{8}, 
Knut Olsen\altaffilmark{9},
Daniel Weisz\altaffilmark{5}, 
Charlotte Christensen\altaffilmark{1}, 
Ken Freeman\altaffilmark{11}
Karoline Gilbert\altaffilmark{1}, 
Carme Gallart\altaffilmark{12},
Jason Harris\altaffilmark{13}, 
Paul Hodge\altaffilmark{1},
Roelof S. de Jong\altaffilmark{10},
%Adrienne Stilp, 
%Sarah Loebman, 
Valentina Karachentseva\altaffilmark{14}, 
Mario Mateo\altaffilmark{15}, 
Peter B. Stetson\altaffilmark{16}, 
Maritza Tavarez\altaffilmark{17}, 
Dennis Zaritsky\altaffilmark{13}, 
%Ricardo Covarrubias, 
%Andrew Becker\altaffilmark{1}, 
Fabio Governato\altaffilmark{1}, 
Thomas Quinn\altaffilmark{1}
}

\altaffiltext{1}{Department of Astronomy, Box 351580, University of Washington, Seattle, WA 98195; jd@astro.washington.edu; ben@astro.washington.edu; krosema@astro.washington.edu; stephanie@astro.washington.edu; christensen@astro.washington.edu; becker@astro.washington.edu; fabio@astro.washington.edu; trq@astro.washington.edu}
\altaffiltext{2}{CfA Fellow, Harvard-Smithsonian Center for Astrophysics, 60 Garden Street, Cambridge, MA 02138; aseth@cfa.harvard.edu}
\altaffiltext{3}{Raytheon, 1151 E. Hermans Road, Tucson, AZ 85756; adolphin@raytheon.com}
\altaffiltext{4}{Department of Astronomy, New Mexico State University, Box
30001, 1320 Frenger St., Las Cruces, NM 88003; holtz@nmsu.edu}
\altaffiltext{5}{Department of Astronomy, University of Minnesota, 116 Church St. SE, Minneapolis, MN 55455; dweisz@astro.umn.edu; skillman@astro.umn.edu}
\altaffiltext{6}{School of Mathematics and Physics, University of Tasmania, Hobart, Tasmania, Australia; andrew.cole@utas.edu.au}
\altaffiltext{7}{Osservatorio Astronomico di Padova -- INAF, Vicolo
dell'Osservatorio 5, I-35122 Padova, Italy; leo.girardi@oapd.inaf.it}
\altaffiltext{8}{Special Astrophysical Observatory, Russian Academy of Sciences, Nizhnji Arkhyz, Karachai-Circessia Republic 369167, Russia; ikar@luna.sao.ru}
\altaffiltext{9}{NOAO, National Optical Astronomy Observatory 950 N. Cherry Ave., Tucson, AZ 85719; kolsen@noao.edu}
\altaffiltext{10}{Space Telescope Science Institute, 3700 San Martin Dr., Baltimore, MD 21218; dejong@stsci.edu}
\altaffiltext{11}{Mount Stromlo Observatory, Research School of Astronomy and
Astrophysics, Mount Stromlo Observatory, The Australian National University, ACT 0200 Australia; kcf@mso.anu.edu.au}
\altaffiltext{12}{Instituto de Astrofísica de Canarias, Vía Láctea, s/n, 38200 La Laguna, Tenerife, SPAIN; carme@iac.es}
\altaffiltext{13}{Steward Observatory, University of Arizona, 933 North Cherry Avenue, Tucson, AZ 85721; jharris@as.arizona.edu; dennis@fishingholes.as.arizona.edu}
\altaffiltext{14}{Astronomical Observatory of Kiev University, Observatorna 3, 254053, Kiev, Ukraine; vkarach@observ.univ.kiev.ua}
\altaffiltext{15}{Department of Astronomy, University of Michigan, 830 Denninson Building, Ann Arbor, MI 48109-1090; mmateo@umich.edu}
\altaffiltext{16}{Dominion Astrophysical Observatory, Herzberg Institute of Astrophysics,National Research Council, 5071 West Saanich Road, Victoria, BC V9E 2E7, Canada; Peter.Stetson@nrc-cnrc.gc.ca}
\altaffiltext{17}{Forest Ridge School of the Sacred Heart, 4800 139th Ave SE, Bellevue, WA 98006; martavbrown@yahoo.com}
  
\begin{abstract}

  The ACS Nearby Galaxy Survey Treasury (ANGST) is a systematic survey
  to establish a legacy of uniform multi-color photometry of resolved
  stars for a volume-limited sample of nearby galaxies ($D<4\Mpc$).
  The survey volume encompasses 69 galaxies in diverse environments,
  including close pairs, small \& large groups, filaments, and truly
  isolated regions.  The galaxies include a nearly complete range of
  morphological types spanning a factor of $\sim\!10^4$ in luminosity
  and star formation rate.  The survey data consists of images taken
  with the Advanced Camera for Surveys (ACS) on the {\emph{Hubble
      Space Telescope}} (HST), supplemented with archival data and new
  Wide Field Planetary Camera (WFPC2) imaging taken after the failure
  of ACS.  Survey images include wide field tilings covering the full
  radial extent of each galaxy, and single deep pointings in uncrowded
  regions of the most massive galaxies in the volume.  The new wide
  field imaging in ANGST reaches median 50\% completenesses of
  $m_{F475W}\!=\!28.0$~mag, $m_{F606W}\!=\!27.3$~mag, and
  $m_{F814W}\!=\!27.3$~mag, several magnitudes below the tip of the
  red giant branch (TRGB).  The deep fields reach magnitudes
  sufficient to fully resolve the structure in the red clump (RC).
  The resulting photometric catalogs are publicly accessible and
  contain over 34 million photometric measurements of $>$14 million
  stars.  In this paper we present the details of the sample
  selection, imaging, data reduction, and the resulting photometric
  catalogs, along with an analysis of the photometric uncertainties
  (systematic and random), for both the ACS and WFPC2 imaging.  We
  also present uniformly derived relative distances measured from the
  apparent magnitude of the TRGB.
\end{abstract}

\vfill
\clearpage

%----------------------------------------------
\section{Introduction}  \label{introsec}

The study of nearby galaxies has been revolutionized by the
{\emph{Hubble Space Telescope}} (HST).  The high spatial resolutions
of WFPC2 and ACS reveal individual stars and parsec-scale structures,
permitting studies of stellar populations, star formation histories,
and stellar clusters for galaxies out to several megaparsecs.
However, despite the large number of HST projects on these topics,
past observations have been piecemeal and lack a unifying, coherent
observational strategy in spite of the considerable overlap in the
core scientific goals of many of the projects. Within a single galaxy,
or from galaxy to galaxy, the locations of the HST exposures have been
chaotic (having been chosen independently and for different purposes),
and the filters and depths of the exposures have been highly variable.
While past observations have provided dramatic insights into the star
formation histories of individual systems, the resulting archive
complicates any uniform comparative study of galaxies in the Local
Universe and dramatically reduces the scientific legacy of this
data set.

The ACS Nearby Galaxy Survey Treasury (ANGST) program aims to rectify
this situation by creating a uniform, multi-color archive of
observations of resolved stellar populations within a volume-limited
sample of nearby galaxies.  The survey provides complete and unbiased
sampling of the local universe, thereby maximizing the legacy impact
of the resulting data set, and enabling meaningful comparisons among
galaxies in the sample and with cosmological simulations.  Within this
volume, ANGST adds more than a hundred orbits of new high-quality
observations, and provides uniform reduction and photometry of both
the new and archival observations.  The resulting survey now
offers superb targets for future multi-wavelength surveys, including
the VLA-ANGST survey \citep{ott08} and the Spitzer Local Volume Legacy
Survey \citep[LVL;][]{kennicutt07}, by allowing one to tie the
multi-wavelength observations to the local star formation history
revealed by ANGST.

In this paper we describe the survey design of ANGST, including the
sample selection (\S\ref{samplesec}), the observing strategy for new
observations (tiling patterns, filter choices, exposure times, etc.)
using both ACS (\S\ref{ACSobservingsec}) and WFPC2
(\S\ref{WFPC2observingsec}), and the archival data employed by the
survey (\S\ref{archivalsec}).  We then present photometry for the
survey galaxies for both ACS (\S\ref{ACSphotsec}) and WFPC2
(\S\ref{WFPC2photsec}), tests of the photometric reliability
(\S\ref{phottestsec}), astrometry (\S\ref{astrometrysec}), and the
resulting data products included in this data release
(\S\ref{dataproductsec}).  In \S\ref{cmdsec} we plot color-magnitude
diagrams (CMDs) for all of the ANGST galaxies, and in \S\ref{TRGBsec}
we measure colors and magnitudes for the tip of the red giant branch
(TRGB), from which accurate relative distances are derived.  

%----------------------------------------------
\section{The Sample}  \label{samplesec}

\subsection{Sample Selection}  \label{sampleselectsec}

We drew the initial ANGST sample from the \citet{karachentsev04}
Catalog of Neighboring Galaxies (CNG), updated with revised distances
provided by I.\ Karachentsev.  We restricted the catalog to galaxies
beyond the zero velocity surface of the Local Group
\citep{vandenbergh00}, due to the efficacy of ground-based
observations within 2$\Mpc$ and the large number of existing HST
observations \citep[e.g.,][]{holtzman06}. We further restricted the
sample to $|b|\!>\!20\degree$, to avoid sample incompleteness at low
Galactic latitudes.

The choice of a maximum distance for the sample required balancing our
scientific goals against constructing an observationally efficient
program.  At large distances, a wider variety of galaxy environments
can be sampled, at the expense of larger photometric errors due to
increased crowding and longer exposure times.  We adopted an initial
outer radius cut of $3.5\Mpc$, within which deep CMDs could be derived
with only modest exposure times.  However, the Local Volume contains
mostly field galaxies until reaching the massive M81 group at
$\sim\!3.6\Mpc$ and the Cen A group at $\sim\!3.7\Mpc$.
Scientifically, the case for including at least one of these groups is
strong.  Without them, the limited range of environments sampled by a
$D\lesssim3.5\Mpc$ sphere would preclude studies of correlations
between star formation history, galaxy morphology, and local
environment.  Of the two groups, the M81 group was judged to be the
preferred target due to its high galactic latitude, low foreground
extinction, and highly complete membership information.  Galaxies in
the M81 group were drawn from \citet{karachentsev02}, but do not
include the newest candidate members reported in \citet{chiboucas08}.
We also included a second high-density environment centered on the
NGC~253 clump ($D\approx3.9\Mpc$) in the Sculptor filament
\citep{karachentsev03}, further increasing the range of environments
probed.  The extensions into the M81 group and NGC~253 clump of the
Sculptor group also improves coverage of luminous galaxies that are
poorly represented in the $D\!<\!3.5\Mpc$ volume.

The resulting sample of 69 galaxies is given in
Table~\ref{sampletable}, along with the distances adopted during
sample selection.  Notable changes from the published version of the
CNG include larger distances for UGC~8638, E059-01, and KKH60, which
took them out of the sample, revised closer distances for NGC~4163 and
DDO~183 which brought them into the sample, and elimination of HIJASS,
which has no detectable stars.  Distances for NGC~247, NGC~55,
DDO~187, UGC~8833, HS117, and KKH37 were also revised according to new
distances in \cite{karachentsev05}.  Other data compiled in
Table~\ref{sampletable} includes absolute total magnitudes in $B$,
morphological T-types, angular diameters ($D_{25}$ for large galaxies,
$D_{26.5}$ for some dwarfs), and HI line widths ($W_{50}$); all these
quantities are listed as originally compiled in the CNG, and details
can be found in \citet{karachentsev04}.  We also include apparent
total $K$-band magnitudes from the literature when available; these
are included for completeness only, and no attempt has been made to
bring these to a common aperture with the $B$-band magnitudes from
the CNG.  Table~\ref{sampletable} also indicates the original
{\emph{planned}} observational strategy for the sample galaxies; as we
discuss below, not all observations were carried out as planned, due
to the failure of ACS.

%-----------
\subsection{Properties of the Final Sample}  \label{samppropsec}

The volume-limited sample defined above contains a rich assortment of
galaxies.  The range of distances, luminosities (in $B$ and $K$), and
morphological types of the sample galaxies can be seen in
Figure~\ref{magdistfig}.  Galaxy absolute magnitudes span from
brighter than $M_B=-20$ (M81 and NGC~253, the dominant galaxies in the
M81 group and the Sculptor filament), down to fainter than $M_B=-9$,
comparable to the Carina dwarf spheroidal in the Local Group.
$K$-band total magnitudes were adopted from \citet{jarrett03} or
\citet{vaduvescu05} when available, or inferred from $B$ band
magnitudes assuming $B-K\sim2.86$, based on the estimates in
\citet{mannucci01} for dwarf irregular spectral types.

As for any volume-limited sample, the distribution of luminosities is
strongly weighted towards dwarf systems.  Roughly 90\% of the galaxies
in the ANGST volume are fainter than the Large Magellanic Cloud (LMC),
and 80\% are fainter than the Small Magellanic Cloud.  Integrating the
luminosities of the galaxies, 99\% of the $B$-band luminosity is
contained in galaxies brighter than $M_B=-13.7$ (33\% by number).  In
the $K$-band, which presumably is a better tracer of the stellar mass,
99\% of the luminosity is contained within only 17\% of the galaxies
($M_K<-17.5$).  The large number of low luminosity systems is also
reflected in the distribution of morphological types.  Only 17\% of
the galaxies have morphological types characteristic of spirals ($1\le
T \le 9$), while 58\% are classified as dwarf irregulars and 25\% as
dwarf ellipticals.  In spite of the large population of dwarf
ellipticals, there are no massive early types in the sample.
NGC~404 is classified as an S0, but has relatively low luminosity and
an extended gas disk \citep{delrio04}.  The earliest massive spiral in
the sample is M81, with a morphological type of Sab.

The sample galaxies reside in diverse environments.  There are at
least 4 distinct groups with a range of richnesses -- the dwarf
dominated NGC~3109 group, two clumps in the Sculptor filament (one at
NGC~55/NGC~300, and one at NGC~253/NGC~247), and the rich M81 group.
Several of \citet{tully06}'s ``dwarf groups'' are also included in the
ANGST survey volume (14$+$12, 14$+$13 14$+$07,14$+$08; the first two
are the NGC~3109 and NGC~55/NGC~300 groups mentioned above).  Group
membership is also included in Table~\ref{sampletable}.  Some of these
groups can be seen in Figure~\ref{3Dfig}, where we show the
3-dimensional distribution of the survey galaxies, using updated
distances from \S\ref{TRGBsec} below\footnote{These updated distances
  agree with the distances in Table~\ref{sampletable} to 10\% in
  almost all cases, as discussed in \S\ref{TRGBsec}.}.

%---------------------------------------------- 
\section{ACS Observing Strategy}  \label{ACSobservingsec}

When designing an observing strategy for the ANGST sample, we balanced
the limited number of orbits (295, down from an initial request of
555) against the goal of simultaneously recovering the star formation
history (SFH) of the volume and establishing a general purpose imaging
archive.  We aimed to maximize uniformity, depth, and versatility,
while making efficient use of the allocated orbits and the data
already in the archive.

As part of this strategy, we chose to allocate a larger fraction of
the orbits to the galaxies with the most stars, which contained either
99\% of the stars, or 99\% of the recently formed stars.  These
galaxies fall to the right of the vertical lines in the right and left
hand panels of Figure~\ref{magdistfig}.

The full radial extent of all galaxies was imaged in at least 2
filters.  For dwarfs, these wide field tiles could be acquired in a
single pointing.  Larger, angularly extended galaxies were each imaged
with a radial strip of overlapping ACS tiles extending from the
galaxy's center to its outskirts.  In addition to the wide fields,
deep fields with high completeness in the red clump were planned for
the 12 galaxies that dominate the $K$-band luminosity of the ANGST
volume; this depth provides strong constraints on the the ancient star
formation history \citep[e.g.,][]{dolphin02b}.  Another 16 galaxies within
$\sim\!2.5\Mpc$ would reach a comparable depth from their wide field
tilings alone.  In addition to the two filters in the standard wide
field tilings, the 23 galaxies which dominate the recent star
formation density (as assessed in the $B$-band) would be imaged in
three filters, to permit extinction corrections and multi-wavelength
source identification.  Fourth, archival imaging of comparable depth
to the new observations would be used when possible.

We now discuss the details of the wide-field tilings, the deep fields,
the choice of filters, and the exposure times.  In
\S\ref{WFPC2observingsec}, we discuss how this strategy was modified
after ACS failed during our program's execution.

\subsection{Wide Field Tiling}  \label{widesec}

The wide field tilings were designed for efficient multi-filter
coverage of each galaxy's radial extent.  Thanks to ACS's large
field-of-view (FOV), dwarf galaxies could be imaged with a single
pointing.  For smaller dwarf galaxies, the galaxy center was aligned
with the center of the WFC1 chip to avoid the chip gap's occluding the
center of the galaxy.  For the larger dwarf galaxies DDO~44 and
DDO~82, the center was placed slightly above the chip gap.  For
galaxies whose radial extents were larger than could be imaged in a
single pointing, we adopted a set of radial tiles extending from the
center of the galaxy out to the position of the deep field, along the
major axis in whichever direction required the smallest number of
tiles.  To allow flexible telescope scheduling, the tiles were allowed
to be at any multiple of a $90^\degree$ rotation from the major axis,
with a $\pm5^\degree$ leeway.  Adjacent tiles were overlapped by
$22\arcsec$ to allow complete coverage throughout the permitted range
of telescope roll angles.  All tiles were dithered to fill the chip
gap and to remove cosmic rays and hot pixels.  In the ANGST target
naming scheme, tiles are numbered from the outermost tile inwards.
The resulting field locations are shown superimposed on images from
the Digitized Sky Survey in
Figures~\ref{overlayfig1}-\ref{overlayfig4}, for both the ANGST
observations and the archival observations described below
(Table~\ref{obstable}~\&~\ref{archivetable}).

Among the galaxies eligible for a full radial strip, we did not image
M81 or M82.  The former had complete tiling through programs GO-10250
($F814W$ only) and GO-10584 ($F435W$, $F606W$, and some $F814W$ in
outer fields). M82 was tiled by STScI through program DD-10776
\citep{mutchler07}.  Only the M81 photometry from GO-10584 is
presented here; the $F814W$ tiling in GO-10250 was not aligned in
either pointing or rotation with the bluer observations in GO-10584,
and thus requires capabilities not included in
the current data processing pipeline.

\subsection{Deep Field Pointing}  \label{deepsec}

A single deep pointing was originally planned for each of the 12
galaxies which dominate the $K$-band luminosity (and presumably the
stellar mass) of the local universe.  The deep field exposure times
were chosen to provide high completeness in the red clump region of
the CMD, as described below.  However, deep exposures are subject to
significant stellar crowding, due to the increasing number of stars at
fainter magnitudes in the CMD.  When stellar fields become too
crowded, longer exposure times no longer decrease the photometric
errors or increase the number of detected stars.  Instead, the
photometric uncertainties are dominated by systematic errors produced
by crowded, blended point-spread functions (PSFs).  To avoid this
situation, the deep fields needed to be placed in regions of the
galaxies where photometric errors would not be dominated by crowding.

When placing the deep fields, we used the simulations of
\citet{olsen03} to calculate the surface brightness below which
photometric errors would be less than $0.1$ mag in the red clump.  This
limiting surface brightness depends on distance, the underlying
stellar population, and the pixel scale and PSF of the camera.  We
found typical ACS limiting surface brightness of $\mu_V\sim22.2-24.6\surfb$
for galaxies at $D=1.3-4\Mpc$.  These limits yielded of order 100K
stars per ACS FOV at our typical exposure time, which was consistent
with our previous experience with an ACS snapshot survey
\citep{seth05}.  The resulting limiting surface brightnesses were used
to identify appropriate field locations for each of the target
galaxies, using a combination of 2MASS, SDSS, and deep Malin
({\tt{http://www.aao.gov.au/images/}}) images to estimate the local
surface brightness along each galaxy's major axis.  The fields were
allowed to have any orientation, and were contiguous with the
outermost wide field tile.

\subsection{Filter Choice}  \label{filtersec}

Imaging was carried out in three filters for the galaxies that dominate
the recent star formation in the local volume (i.e., to the right of
the line in the left panel of Figure~\ref{magdistfig}), and two filters
for all others.  For the galaxies with 3 filter coverage, we used
$F475W+F606W+F814W$, which maximized the combination of wavelength
coverage and throughput.  The three filters are useful for identifying
X-ray counterparts, HII region nebulosity, and extinction (when
combined with future UV or NIR imaging).  Although the $F435W$ filter
allows for a larger wavelength baseline and disjoint wavelength
coverage with $F606W$, its throughput is much less than that of
$F475W$.

For the dwarf galaxies with 2 filter coverage, we used a $F475W+F814W$
filter combination, instead of the more commonly used $F606W+F814W$.
Although $F475W$ does not reach as far down the CMD as $F606W$ for a
given exposure time, it provides greater temperature sensitivity due
to the longer wavelength baseline of the $F475W-F814W$ color
combination.  For regions above the red clump, more scientific
information can be extracted from better temperature sensitivity than
from the slight gain in depth possible with $F606W+F814W$.  This
choice allowed us to better separate main sequence stars from the blue
helium burning sequence, and to derive stronger constraints on the
metallicity distribution of red giant branch (RGB) stars.  This effect
can be seen in Figures~\ref{cmdfig1}-\ref{cmdfig14}, when comparing
CMDs in $F475W+F814W$ and $F606W+F814W$ for galaxies with 3-filter
observations (such as DDO~190).  Given the very low extinctions
expected in low metallicity systems, a third filter was not deemed
necessary for the faintest dwarf galaxies.  For many of these, some
$F606W$ imaging is already available in the archive, largely from the
SNAP-9771 and SNAP-10210 programs.

For the deep fields, the scientific demand of constraining ancient
star formation requires the highest possible completeness in the red
clump.  Thus, the majority of time invested in deep fields was in the
more traditional $F606W+F814W$ color combination, which maximizes the
depth along the RGB at the expense of lower color sensitivity.  A
single orbit of $F475W$ was also included for continuity with the
wide-field observing strategy, and to allow the possibility for
extinction mapping in the future.

\subsection{Exposure Times}  \label{exptimesec}

Exposure times were chosen to achieve two separate goals.  For the
wide fields, the goal was efficient, multi-color imaging of the upper
regions of the CMD, allowing good constraints on the occupation of the
main sequence, the luminosity function of the blue and red helium
burning sequences, the color distribution of the RGB, and the
population of AGB stars.  For the deep field, the goal was high
completeness and photometric accuracy in the red clump.  We discuss
the details of the wide field and deep field observations in
\S\ref{wideexpsec} and \S\ref{deepexpsec} below.  A listing of the new
observations taken for this program can be found in
Table~\ref{obstable}.

\subsubsection{Wide Fields}  \label{wideexpsec}

The wide field observing strategy was shaped by the need to get up to
3 filters at each tile position.  In each filter we need at least a 2-
or 3-point dither pattern to reject cosmic rays and to cover the chip
gap.  Due to data volume constraints, two orbits are required to get
at least 2 images in each of the three filters.  For crowded areas, we
used the minimum 2 orbits for the wide-field tilings, while in the
outermost wide fields, where crowding was not a limiting factor on the
photometry, we used 3 orbits, 1 orbit per filter.  For dwarf galaxies,
we devoted one orbit to each of the 2 filters.  Total exposure times
can be found in Table~\ref{obstable}.  The typical photometric depths
(S/N=5) were 28.4 in $F475W$ and $F606W$, and
27.5 in $F814W$ for a single orbit.

\subsubsection{Deep Fields}  \label{deepexpsec}

The goal of the ANGST deep fields is to obtain an accurate census on
the number, magnitude and color of stars in the red clump.  These
stars place a strong constraint on the ancient SFH enabling the
possibility of breaking the age-metallicity degeneracy present along
the upper RGB.  We requested deep fields for 12 galaxies in the ANGST
volume with $M_K < -17.5$.  These 12 galaxies contain 99\% of the
K-band luminosity within our survey volume, and thus have dominated the
past total SFH.  The significant time investment required to obtain CMDs
reaching below the red clump meant that these exposures were limited
to a single field in each galaxy.

Exposure times were chosen to obtain S/N$\gtrsim$10 in both $F606W$
and $F814W$ for stars in the red clump.  In practice, we achieved this
by using the ACS Imaging Exposure Time Calculator to estimate the time
necessary to reach S/N=3.5 for a G0III star normalized to $M_V=+1.5$
(for $F606W$) and $M_I=+0.7$ (for $F814W$), a magnitude below the
theoretical red clump for a [Fe/H]=--1.3, 10~Gyr population in the
Padova isochrones ({\tt{http://pleiadi.pd.astro.it/}}).  To calculate
the appropriate red clump magnitude for each galaxy, reddenings and
extinctions were adopted from \citet{schlegel98}, and distance
moduli were chosen by carefully evaluating data from the literature
\citep{karachentsev02,karachentsev03,rekola05,mouhcine05,sakai01,
  karachentsev06, freedman94,sakai99, tikhonov03,maiz02,drozdovsky02,
  gieren04,gieren05,rizzi06, minniti99,mendez02,aparicio00} and from
our own TRGB measurements using archival data.  Exposure times were
turned into orbit estimates using the appropriate overheads and
available visibility times depending on the declination of the source.
For the two deep exposures of M81 and NGC~2976 that were obtained with
ACS before its failure, a single long exposure ($\sim$2700 sec) was
taken in each orbit.  Each visit contained an orbit in each filter
both to maximize our baseline for variable stars and to minimize the
risk of obtaining incomplete filter coverage in the event of
spacecraft failure.  A short 100s exposure was taken to permit
photometry of the brighter stars saturated in the full orbit
exposures, and a full orbit of $F475W$ data was included for
consistency with the wide fields and to enable the possibility of
internal reddening estimations.

\subsection{Parallels}  \label{parallelsec}

WFPC2 observations were taken in parallel with the ACS observations in
Table~\ref{obstable}.  These observations were divided evenly between
$F606W$ and $F814W$, and are $6\arcmin$ away from the center of the ACS
FOV.  Photometry of these fields will be reduced with the WFPC2
pipeline described below in \S\ref{WFPC2photsec}, but is not included
in this initial data release.

%----------------------------------------------
\section{WFPC2 Observing Strategy}  \label{WFPC2observingsec}

ACS observations for our program began in early September 2006.
Unfortunately, the wide field camera on ACS failed in late January
2007, $\sim$5 months into the execution of our program.  As a result,
we lost 147 orbits on the massive galaxies with deep fields
($M_K<-17.5$; 71\% lost), and lost 44 orbits on the fainter galaxies
(50\% lost), for a total of 191 orbits lost from the
original allocation of 295 orbits.  Of the 195 orbits that were to be
devoted to the deep fields, we received 41 orbits (79\% lost),
primarily for NGC~2976 and M81.  Given the uncertainties in the
upcoming HST servicing mission, we decided to continue the program
with WFPC2.

Following an appeal, the Telescope Time Review Board restored 116 of
the 191 lost orbits to execute deep single-pointing observations for
the nearest luminous galaxies and the very closest dwarfs (NGC~55,
NGC~4214, NGC~404, NGC~2403, NGC~3109, Sex~B, and IC~5152).  Time for
wide field observations was not granted.  For the majority of these
galaxies, sufficient data exist in the archive for tying the large
radius deep fields to the SFH of the galaxy as a whole, although with
a lack of complete radial coverage and uniformity.  However, NGC~55
and NGC~3109 did not have adequate radial coverage due to their large
angular extents.  Through a Director's Discretionary request
(DD-11307), we were awarded an additional 25 orbits to execute radial
tilings for these two remaining galaxies (5 pointings per galaxy, with
2 orbits per tile for NGC~3109 and 3 orbits per tile for NGC~55).

\subsection{WFPC2 Deep Fields}  \label{wfpc2deepsec}

Transferring the ANGST deep field observing strategy to WFPC2 required
a number of modifications.  The first significant change was in field
placement.  The wide-field chips of WFPC2 are undersampled compared to
ACS, leading to larger photometric errors due to crowding at
comparable surface brightnesses and exposure times.  We therefore had
to move the deep fields to even lower surface brightnesses (and thus
larger radii) than the original ACS deep field locations.  Using the
\citet{olsen03} simulations, we recalculated the surface brightness
limit at which our observations would become crowding limited.  These
revised limits were $\sim$1.5 mag fainter than for ACS.  These changes
required shifting the fields typically another $\sim$1.4 disk scale
lengths further out, increasing the risk that the WFPC2 FOV would fall
beyond any significant disk truncation, if present.  This appears to
have happened for IC~5152, but did not affect any of the other
observations.

The second adaptation was to accept slightly less photometric depth.
WFPC2's throughput is substantially worse than ACS's, and thus
matching the depth of the ACS deep fields would require a prohibitive
number of orbits.  However, our experience with the ACS deep fields
for M81 and NGC~2976 suggested that we could reach our scientific
goals with slightly shallower data, and thus we revised our target
depth to a SNR of 3 at 1.5 magnitudes below the middle of the red
clump.  The final change to the program was to eliminate the $F475W$
observations, where WFPC2's sensitivity is particularly poor.

When allocating orbits, we maximized the photometric accuracy in the
red clump (where $F606W-F814W\approx0.75$) by allocating twice as many
orbits to $F814W$ than to $F606W$.  A random-walk dither pattern was
adopted and full-orbit exposures were used; the number ($>$5) of
exposures made cosmic-ray rejection straightforward without the need
to CR-split the exposures during the orbit, allowing us to obtain the
maximum depth possible with each orbit.

The resulting images typically had between 5000 and 15000 stars per
WFPC2 chip. We checked our photometry on a chip-by-chip basis to
identify potential problems or offsets due to the well-known WF4 bias
anomaly.  Images of WF4 showed no obvious problems with the bias, nor
was the photometry noticeably worse, suggesting that the anomaly had
been properly addressed by STScI's WFPC2 data reduction pipeline
and/or that the chip was performing well at the time the observations
were performed.  We therefore are including WF4 data in the released
photometric catalogs.  These catalogs include a flag identifying the
chip that a star fell on in the reference image, allowing the user to
filter out WF4 data, if needed.

\subsection{WFPC2 Wide Field Tilings}  \label{wfpc2widesec}

For the WFPC2 wide field tilings of NGC~55 and NGC~3109, we aimed to
match the depth (in absolute magnitude) of the wide radial tiles in
the more distant systems of the ANGST survey, assuring that the WFPC2
tiles were at least as deep as the shallowest wide field tiles in the
survey.  This depth corresponds to a signal-to-noise of 5 and 50\%
completeness at $M_{F814W}=-0.5$ for the colors of the RGB.  At the
distances of NGC~3109 (1.3\Mpc; $m_{lim,F814W}=25.1$) and NGC~55
(2.1\Mpc; $m_{lim,F814W}=25.8$), we could reach this depth and
completeness in 2 orbits for NGC~3109 and 3 orbits for NGC~55,
including overheads for CR-SPLITs and guide-star acquisition, based on
comparable 2-orbit observations for Sextans~A \citep{dohm97}
and WFPC2 parallel data from the main ANGST ACS observations.

To produce a radial strip, we adopted a ``Groth strip'' tiling
strategy of interleaved chips, with an orientation set to maximize
schedulability for each target.  Unlike the original ANGST ACS
program, we did not tile all the way out to the deep fields, which had
to be moved even further out to cope with WFPC2's lower resolution.
We instead stopped the radial tiling where we are sure that we have
imaged most of the recent star formation.  To conserve orbits, tiles
were placed on whichever side of the galaxy presented the smallest
distance to the edge of the star-forming disk.

%----------------------------------------------
\section{Archival Data}  \label{archivalsec}

The original ANGST survey strategy was designed to take advantage of
archival data whenever it matched or surpassed the quality of the
proposed observations, in comparable filters.  The failure of ACS
during execution of the ANGST program further increased our reliance
on archival data.  In Table~\ref{archivetable} we summarize the
archival data sets that are incorporated into the ANGST data release,
along with papers that have published CMDs from these data
independently.  We have excluded data sets that have only one filter
at a single position, or that have severe offsets or misalignments
among multiple filters.  Photometry in the latter cases is
significantly compromised by the distortion of the ACS WFC, and cannot
be readily produced by the ANGST pipeline.  In future data releases,
we will incorporate such data as needed, most notably for the $F814W$
tilings of M81 by GO-10250.

%----------------------------------------------
\section{ACS Photometry}  \label{ACSphotsec}

Photometry was carried out on bias subtracted, flat-fielded
{\tt{*\_flt}} images (or {\tt{*\_crj}} images when available) produced
by the STScI ACS pipeline OPUS versions 2006\_5 through 2008\_1, which
used CALACS version 4.6.1.\ through 4.6.4.  For {\tt{*\_crj}} images,
the value of the readnoise reported in the CALACS header did not
reflect that the final image contains two co-added readouts.
In these cases, we multiplied the read noise listed in the header by
$\sqrt{2}$, so that it properly accounted for the combined read noise
of the co-added CR-SPLIT observations.  Failure to make this
correction would have produced systematic errors in the reported
photometric errors.

To measure stellar photometry, we used the software package
DOLPHOT\footnote{http://purcell.as.arizona.edu/dolphot}
\citep{dolphin02} including the ACS module.  This package is optimized
for measuring photometry of stars on dithered ACS images, where the
position angle of the multiple exposures are the same, and the shifts
between exposures are small ($\lesssim$30$''$).  To align images,
DOLPHOT makes a fast initial pass through the data to find bright
stars common to all of the frames, using approximate shifts supplied
by the user.  The final shifts between the images are then determined
based on these stars.  By this method, our exposures were able to be
aligned to $\sim$0.01$''$ precision.  The precision is slightly worse
($\sim$0.015$''$) for fields with small numbers of stars, and slightly
better ($\sim$0.005$''$) for more crowded fields.  When aligning
images, we currently do not incorporate time-dependent corrections to
the geometric distortion.  While improved distortion corrections would
help improve the astrometric solution for the frames, it has only a
second-order effect on our photometry, since the photometric accuracy
depends primarily on multiple images being aligned correctly relative
to each other, rather than relative to an undistorted frame.  As we
currently are only analyzing stacks of images with small positional
shifts, taken close together in time, temporal drifts in the geometric
distortion are not a limiting factor in our photometry.  They will,
however, be considered in future releases.

Although DOLPHOT operates on non-drizzled images, we also combined all
images into a single drizzled image using the {\tt multidrizzle} task
within PyRAF \citep{koekemoer02}, which allowed us to flag the cosmic
rays in the individual images.  Once the cosmic rays were identified,
the photometry was measured for all of the objects using the
individual, uncombined frames.

To calculate the flux of each star, DOLPHOT initially adopts the point
spread function (PSF) calculated by Tiny Tim \citep{krist95}, and
scales the PSF in flux to minimize residuals throughout the image
stack.  The shape and width of the Tiny Tim PSF has been shown to
match to the shape of the true PSF well throughout both ACS chips,
based on the extensive analysis presented by \citet{jee07}.  There are
slight deviations close to the bottom of Chip 1, but our tests in
\S\ref{phottestsec} find that these do not lead to noticeable
systematics in the magnitude errors.  DOLPHOT makes additional minor
adjustments to the Tiny Tim PSF by using the brightest and most
isolated stars to correct for PSF changes due to 
temperature variations of the telescope during orbit.  These
adjustments typically affect the photometry at the 0.01 magnitude
level.

DOLPHOT also uses the most isolated stars in the field to determine
aperture corrections to the PSF magnitudes, which accounts for any
systematic differences between the model and true PSF.  These
corrections were generally $\lesssim$0.05 mag for a given exposure.
DOLPHOT then applies the aperture corrections for each exposure,
corrects for the charge transfer efficiency of the ACS detector using
the coefficients given in the ACS-ISR 2004-006\footnote{Preliminary
  versions of revised ACS charge transfer efficiency (CTE) corrections
  have recently been released on STScI's web site, but were announced
  after all the data had been processed for this release.  These
  corrections will be used in subsequent releases, and updated on the
  data release website.}, combines the results from the individual
exposures, and converts the measured count rates to the VEGAMAG system
(a system where Vega is defined to have mag=0 in all filters)
using the \citet{sirianni05} zero points for each filter.  Note that
the \citet{sirianni05} zero points have since been updated in ACS-ISR
2007-02 to reflect the change in sensitivity due to the increase in
ACS's operating temperature in July 2006 after the switch to Side 2
electronics; these changes are of order $\sim0.015$ mag, and have been
propagated into the relevant photometric catalogs covered in this
release.  The resulting zero points match those in Table~5 of ACS-ISR
2007-02.  We have not yet propagated zero point changes due to
improvements in the calibration of the system throughput (ACS-ISR
2007-06); these changes seem to of the order of less than 0.01 in the
filters covered in this data release, but still have significant
unresolved uncertainties in the red wavelength regimes that dominate
much of the ANGST catalogs.

DOLPHOT makes use of all exposures of a field when measuring stellar
properties.  This technique results in a single raw photometry output
file for each field that contains measured properties of all objects
detected in the field, including the position, object type (point
source, extended, elongated, or indeterminate), combined magnitude,
magnitude error, signal-to-noise, sharpness, roundness, $\chi^2$ fit to
the PSF, crowding, and error flag (chip edge or saturated) in all
filters.  The output catalog contains these measurements for each star
in each individual exposure as well, providing the opportunity for
variability studies.

We spent significant time investigating optimal values for the dozens
of parameters that can be adjusted in DOLPHOT to maximize the quality
of the photometry measured from the data.  Three of the parameters
which had the strongest influence on our resulting photometry were the
{\tt{Force1}} parameter, the aperture radius ({\tt{RAper}}), and the
sky fitting parameter ({\tt{FitSky}}).

The {\tt{Force1}} parameter forces all sources detected to be fitted
as stars, assuming that separate culling will be performed on the
output file to discard non-point sources.  For our crowded fields,
this option was optimal, but required that special care be taken in
fitting the sky.  We found that with {\tt{FitSky}} set to 1 (fit the sky
in an annulus around each star), our photometry was much more heavily
affected by crowding, resulting in large crowding errors for nearly
all of the stars in our wide field data.  With {\tt{FitSky}}$=2$ (fit
the sky inside the PSF radius but outside the photometry aperture), we
needed a small aperture radius (4 pixels), and found systematic errors
($\sim$0.02 mag) in the recovered magnitudes of artificial stars added
to the data, using bright stars whose random photometric uncertainties
did not overwhelm the systematic error.  We found that setting
{\tt{FitSky}}$=3$ (fit the sky within the photometry aperture as a
2-parameter PSF fit) allowed a larger photometry radius (10 pixels)
with smaller aperture correction, and provided photometry with the
smallest crowding errors and no significant systematic errors.  Note
that in all these methods DOLPHOT subtracts the flux of neighboring
stars before measuring the sky and stellar flux; differences in the
operation of {\tt{FitSky}} therefore change the way residuals
propagate, but not the total flux of nearby stars.

We provide full photometry output for each field along with culled
catalogs containing the highest quality photometry.  We culled the raw
DOLPHOT output in two ways, releasing both a complete and a
conservative (but very high quality) catalog for each field.  
The complete catalog contains all sources that were not flagged by
DOLPHOT as extended, elongated, extremely sharp, highly saturated,
significantly cut off by the edge of the chip, or not detected at
high signal-to-noise (4.0 or higher in at least 2 filters).

In addition to the complete catalog, we also provide a more
conservative catalog of stellar photometry which has been culled to
remove highly uncertain photometry.  These catalogs have been filtered
to only allow objects with low sharpness ($({\tt{sharp}}_1 +
{\tt{sharp}}_2)^2 \le 0.075$ and crowding ($({\tt{crowd}}_1
+ {\tt{crowd}}_2) \le 0.1$) in both filters.  The sharpness parameter cut
removes extended objects such as background galaxies missed by the
earlier cuts.  The crowding parameter gives the difference between the
magnitude of a star measured before and after subtracting the
neighboring stars in the image.  When this value is large, it suggests
that the star's photometry was significantly affected by crowding
effects, and we therefore exclude it from our most conservative
catalogs.  Quality cuts based on the $\chi^2$ values were also
considered, but they were rejected when a correlation was found
between $\chi^2$ and the local background.

We found that these final cuts produce CMDs with well-defined features
in the uncrowded field, while retaining most of the stars in high
surface brightness regions.  However, the cuts in the more
conservative catalog may remove stars from certain interesting
regions, like stellar clusters.  We advise anyone interested in
studying clusters or identifying stellar counterparts for specific
sources to check the effects of the different parameter cuts.  In
\citet{gogarten08}, we found that relaxing the crowding parameter cuts
to ${\tt{crowd}}_{1} + {\tt{crowd}}_{2} \le 0.6$ recovered a number of
stars in clusters without dramatically compromising the quality of the
photometry.

Our final catalogs include stars that may contain some saturated
pixels, as long as the saturation was not so bad that the PSF could
not be reliably fitted.  Saturation limits our wide field photometry
to magnitudes fainter than $\sim$18 and our deep field photometry to
magnitudes fainter than $\sim$20.  In Table~\ref{phottable} we give
the level of 50\% photometric completeness for each observation, as
determined from initial artificial star tests.  These completeness
limits are for the field as a whole, but can be expected to vary
spatially within a field due to spatial variations in the degree of
crowding.

While these cuts do an excellent job of restricting the catalogs to
stellar sources, we have noted occasional limits to star-galaxy
separation near the photometric limits of the data, and spurious
``stellar'' sources in the diffraction spikes of extremely bright
stars.  If these issues are of critical importance for a particular
scientific project, we recommend additional culling using information
from galaxy-specific photometry packages such as SExtractor
\citep{bertin96} to mask out possible sources of contamination.  

We also note that star-galaxy separation is frequently impossible for
sources near the photometric limit, even in high-resolution HST data.
Some fraction of the faintest sources in the photometric catalogs are
therefore likely to be unresolved background galaxies.  We do not
think that these sources are a significant issue for most analyses,
however, since they represent a negligible fraction of the sources in
the main body of most galaxies.  To quantify this, we can use the
WFPC2 field for IC5152.  This field has a completeness limit of 26.45
magnitudes in $F814W$ and contains 325 objects in the cleaned
{\tt{*.gst}} catalog (described below), over an area of 5.65 square
arcminutes.  The field unfortunately fell beyond the radius where
IC5152's disk truncates, and thus the majority of the 325 objects are
likely to be either foreground MW stars or unresolved background
galaxies.  We can then take 57.5 stars per square arcminute to be the
upper limit for the contamination in observations of this depth.

We can scale the IC5152 data to other completeness limits, using the
observed galaxy number counts given in Figure~3b of
\citet{windhorst08}.  Over the range of depths for the ANGST data, the
slope in the galaxy number counts scales as $\log_{10}(N_1/N_2)
\approx 0.32(m_{lim,1}-m_{lim,2})$. We have applied this scaling
relation to the data in Table~\ref{phottable} to calculate the upper
limit on the fraction of sources that could potentially be
contaminants in each field.  After IC5152 itself (which has 100\%
contamination by definition), the next highest contamination fraction
is 35\% for an outer halo field of M81 (NGC3031-HALO-2), which lies
well beyond the main body of the galaxy.  All other contamination
fractions are less than 23\%, and 90\% have maximum contamination
fractions of less than 10\%.  Given that most ANGST targets take up
less than one third of the total chip area, the contribution of
unresolved galaxies to the CMD is likely to be less than 3\% within
the galaxy radius in almost all cases.

%----------------------------------------------
\section{WFPC2 Photometry}  \label{WFPC2photsec}

After the failure of ACS and the transfer of our program to WFPC2, we
adopted the WFPC2 pipeline previously used by \citet{holtzman06} for
their archival study of Local Group dwarfs.  We briefly summarize the
key features of the pipeline here, but refer the interested reader to
the more extensive documentation in \citet{holtzman06}.

The \citet{holtzman06} pipeline operates on images processed with the
standard STScI baseline processing.  Photometry is performed using
HSTphot \citep{dolphin00b}, a predecessor of DOLPHOT that is optimized
for WFPC2 images.  HSTphot shares DOLPHOT's basic strategy of using
Tiny Tim PSFs supplemented with image-based aperture corrections to
derive photometry from unstacked images that have not been distortion
corrected or drizzled.  HSTphot adopts the photometric calibration
given in \citet{holtzman95}, updated with improved calibrations from
{\tt{http://purcell.as.arizona.edu/wfpc2\_calib/}}.  Note that the
WFPC2 photometric system is defined such that Vega
has a magnitude in each WFPC2 filter corresponding to Vega's magnitude
in the nearest UBVRI filter; the different definition of the zeropoints in
the WFPC2 and ACS photometric systems leads to offsets of 
0.02-0.04 mag between the calibrated magnitudes of the two instruments
(see \S\ref{wfpc2acscompsec} below).

The only significant difference from the \citet{holtzman06} pipeline
is that the current version of HSTphot uses the latest (July 2008) CTE
corrections derived by A.\ Dolphin
({\tt{http://purcell.as.arizona.edu/wfpc2\_calib/}}).  Compared to the
\citet{dolphin02} prescription, the new CTE calibration no longer
assumes that background and stellar brightness factors are
independent, leading to somewhat fainter WFPC2 magnitudes than
previous calibrations, and many fewer systemic offsets in the
residuals.  Using a typical ANGST wide field observation as a
baseline, the switch to the new CTE correction changes the CTE
correction for a $V=22$ magnitude star from 0.052 mag to 0.115 mag in
$F606W$ and from 0.067 mag to 0.123 mag in $F814W$, for a background
sky level of $\sim\!100$ cts/pixel in both filters.  For a fainter
$V=28$ magnitude star, the CTE changes from 0.328 mag to 0.214 mag in
$F606W$ and from 0.438 mag to 0.238 mag in $F814W$.  The much lower
level of residuals in the new CTE calibration suggests accuracy in the
bright end of $0.01-0.02$ magnitudes.  At the fainter end, it is much
more difficult to assess any systematic offsets, as they are much
smaller than the photometric uncertainties.

After processing by the \citet{holtzman06} pipeline, we integrate the
photometry into the database shared by the main ACS pipeline.  Slight
differences in the WFPC2 and ACS keywords used the released data
tables are described below in \S\ref{dataproductsec}.

%----------------------------------------------
\section{Photometric Tests} \label{phottestsec}

The photometric pipeline produces catalogs of multi-filter photometry
and estimates of the photometric uncertainty for each measurement.
These uncertainties include Poisson flux errors, uncertainties in the
sky determination, and uncertainties in the subtraction of neighboring
objects.  They do not include systematic errors due to spatial and
temporal variation of the point spread function (i.e.,
\citet{jee07,rhodes06}, ACS ISR 07-12, ACS-ISR 06-01), in the absolute
calibration of the photometric system, and in the accuracy of the
adopted CTE corrections (which were in flux at the time that this data
was released).  To assess the degree of systematic errors, and the
accuracy of the reported uncertainties, we have performed a series of
consistency checks to measure shifts in the photometry of individual
stars measured multiple times, in different portions of the field of
view, and for different instruments.  All tests use the conservative
photometric catalogs, to allow the greatest sensitivity to systematic
errors.

\subsection{Repeated ACS Measurements} 

We first analyze the magnitude difference between stars measured in
two individual single-orbit $F814W$ ACS exposures from the M81 deep
field. The exposures were taken during a single visit, which minimizes
any temporal changes in the PSF.  The exposures also had only modest
(0.1--0.2$\arcsec$) dithers between them, allowing us to minimize
systematic errors in modeling the spatial variations in the PSF.
Crowding errors should likewise be minimal, given the low stellar
density within the field.  This test case therefore offers the ``best
case scenario'' for agreement between repeated measurements, and sets
a lower bound to our expected error distribution in less than optimal
cases.

To measure magnitude differences closer to the ``worst case scenario''
for ACS, we also analyze repeated measurements of stars that fall in
the overlap region in the wide field tiling of NGC~300, between WIDE1
and WIDE2.  These stars lie in the most highly distorted regions of
the ACS chip, and have close to the maximum possible offset in their
locations on the chips between the two images, making them a highly
sensitive test of the uncertainties produced by errors in the
point-spread function that DOLPHOT adopts from Tiny Tim.  The images
were also taken two days apart in separate visits, making them
somewhat sensitive to temporal changes in the PSF as well.  However,
as there was little change in the {\tt{Y}}-position of the stars, this
comparison has no sensitivity to systematic errors produced by CTE.

Finally, we measure the magnitude differences between the measured and
the ``true'' magnitudes of artificial stars, added to the same overlap
region analyzed above for NGC~300.  In this case, the stars are
recovered with a PSF that is identical to that used for generating the
artificial stars.  This case therefore minimizes effects due to PSF
uncertainty.  However, it remains sensitive to errors due to Poisson
variations in the flux and the sky background, and due to
contamination from nearby stars.  Note that the error distributions
are expected to be different from the previous two tests, which probed
magnitude differences between repeated measurements of stars with
identical crowding and sky backgrounds, not the magnitude differences
from truth.

In Figure~\ref{magdifffig}, we plot the cumulative distribution of
magnitude differences between repeated measurements of individual
stars, scaled by the quadrature sum of the uncertainties reported in
the individual measurements (i.e., $\Delta m/\sigma_m$, where $\Delta m
\equiv m_1 - m_2$ and $\sigma_m^2 = \sigma_{m1}^2 + \sigma_{m2}^2$ and
$m_1$ and $m_2$ refer to the measurements of a single star in two
different images).  The distributions are generated for all stars in a
limited magnitude range, in steps of 1 magnitude, with fainter bins plotted
with darker lines.  The distributions for brighter stars have
insufficient numbers of stars to be reliable, and are not plotted.  In
red, we plot the distribution of scaled magnitude differences that
would be expected if the magnitude differences were distributed as a
Gaussian with width $\sigma_m$.  The left panel contains only stars
in the overlap region ($\sim\!600$ pixels wide), and the right panel
contains stars for the whole frame.

The distributions of magnitude differences between repeated
measurements show a number of features.  First, even in the worst case
scenario of large positional shifts, the magnitude differences are
essentially unbiased.  The median magnitude difference is less than
5\% of the reported uncertainty in all cases, such that repeated
measurements of given isochrone features will converge on the same
magnitude, even when observed with different parts of chip, or with
multiple exposures.  There is a slight tendency, however, for the bias
to be somewhat larger when large positional shifts are present,
particularly for the brighter stars.  This indicates that there are
indeed small systematic errors in the assumed PSF that are more
noticeable when the wings of the PSF are well-exposed.  However, these
biases will be swamped by the intrinsic random and crowding errors, as
well as Poisson uncertainties, and thus can safely be neglected in
almost all practical applications.

The second feature of the distributions is their tendency to be wider
than a Gaussian whose width is set by the reported uncertainties.  The
true distributions are broader and more flat-topped than expected.
This leads to larger numbers of stars at a given magnitude difference
than one would predict for a perfect Gaussian error distribution.
This difference is most pronounced for the brightest stars.  However,
even the largest shifts do not produce measurable tails beyond
$5\sigma_m$, so while the shape of the error distribution differs
from a Gaussian, we do not detect more than 1-2 stars with $\Delta m >
5\sigma_m$ in our analysis regions.

We can get clues to the origin of the increased width by noting that
the discrepancy from a Gaussian is larger for brighter stars, for
which the distribution becomes closer to a uniform ``top hat''.  We
believe that a significant fraction of this broadening is actually due
to the limited precision of the errors and magnitudes reported by
DOLPHOT.  Catalog values of $\Delta m$ and $\sigma_m$ are quantized at
the 0.001 mag level, rather than being true continuous variables.
This quantization has the largest impact on the distribution of
$\Delta m/\sigma_m$ when errors and magnitude differences are close to
the level of quantization, as they are for the brightest stars.  Not
until the faintest magnitude bins do the errors approach the
distribution expected for a continuous variable.

In Figure~\ref{magdifffakefig} we show the distribution of magnitude
differences between the true and the measured magnitudes
($\Delta m\equiv m_{true}-m_{measured}$, and
$\sigma_m\equiv\sigma_{m,measured}$) of artificial stars added to and
recovered from the images, for a series of magnitude bins.  These
distributions are quite different than the distributions for repeated
measurements, as would be expected.  First, the distributions are
highly skewed, producing a large tail towards measured magnitudes that
are brighter than the true magnitude.  This skewing results when
artificial stars land on or near a star that would otherwise be
undetected.  The flux from the previously undetected star adds to the
artificial star, biasing the measured flux upwards.  Such a bias would
not be apparent in a repeated measurement, as both measurements would
share the same bias.  The skewing is most severe for the faintest
stars, where the additional flux from undetected stars produces the
largest fractional change in the detected flux.  Moreover, the sample
of faint recovered stars will be biased towards stars with heavily
contaminated fluxes, given that such stars are preferentially
detected; this effect is reduced for brighter stars, which are
detectable whether or not an undetected companion falls within the
PSF.

The second feature apparent in the comparison between true and
measured magnitudes is that the distributions are systematically
broader than a Gaussian with a standard deviation equal to the
magnitude uncertainty reported for the measured star.  This deviation
is not surprising, given that the uncertainties are not due solely to
Poisson counting statistics, and are thus unlikely to have
distributions that approach a perfect Gaussian.

\subsection{WFPC2-ACS Comparison}  \label{wfpc2acscompsec}

Due to the failure of ACS, we are releasing photometry both from WFPC2
and ACS.  Differences between these two photometric systems are expected
due to different instrumental responses, CTE corrections, and
absolute photometric calibrations between the two photometric systems.
We have made an initial assessment of the degree of possible systematic
offsets using observations of the dwarf elliptical DDO~44, which was
observed with WFPC2 (GO-8137) in January of 2001, and with ACS as part
of ANGST in September of 2006.  Both data sets were processed with the
respective WFPC2 and ACS pipelines described above.  Stars were
automatically matched between the two catalogs using the closest
positional match in right ascension and declination, after solving for
shifts and rotation between the two fields.  We consider only pairs
of stars that agreed in magnitude to within $10\sigma_m$, where
$\sigma_m$ is the magnitude error from the quadrature sum of the error
in each pair of stars; this procedure produced good matches for
$\gtrsim90$\% of the overlapping stars, though there are clearly
occasional spurious matches as well.  The resulting matched catalog
was restricted further to include only stars above the approximate
completeness limit of each data set ($m_{F814W}$ brighter than 26.0
and 26.1 for the WFPC2 and ACS data sets, respectively).  Comparisons
were made in the $F814W$ filter, which is the only overlapping filter
between the two sets of observations.  

Before comparing $m_{F814W,WFPC2}$ to $m_{F814W,ACS}$, we need to account
for the different zeropoint definitions in the ACS and WFPC2 photometric
systems. Both systems are relative to Vega, but the ACS system defines
Vega to have 0 magnitude in all ACS filters, while the WFPC2 system
defines Vega to have a magnitude corresponding to Vega's magnitude in
the nearest UBVRI filter. For F814W, Vega has $m_{F814W,WFPC2}=0.035$.
As a result, 0.035 mag must be added to the ACS F814W photometry to
compare the results on the same system
\footnote{Note that the \citet{bedin05} ACS calibration also
assumes that $m_{F814W,WFPC2}=0.0$ (giving $m_{zp,F814W,ACS}=25.492$),
so comparisons with this alternate calibration also require adding an
$+$0.035 offset; this correction was not made during the WFPC2-ACS
comparison in \citet{saviane08}, and thus their apparent agreement of
$m_{F814W,WFPC2}-m_{F814W,ACS}=0.003\pm0.005$ actually implies that
$m_{F814W,WFPC2}-m_{F814W,ACS}=0.038\pm0.005$ over the magnitude and
color range of $26<m_{F814W,ACS}<27.5$ and
$0<m_{F606W,ACS}-m_{F814W,ACS}<2$.}.
The systems will also differ for stars of a different color than
Vega to the extent that the system response of the ACS F814W 
filter$+$camera$+$detector system differs from that of WFPC2.

We examined the resulting magnitude differences as a function of
magnitude, color, and {\tt{Y}}-position on the chip.  At almost all
magnitude levels, the systematic errors are dominated by the random errors in
the photometry (which themselves are dominated by Poisson counting
variations and residual flux from crowding).  However, we do detect
residual systematic errors at the few percent level, which vary
steadily with {\tt{Y}} on either instrument, indicating low level
problems with the adopted CTE corrections in both WFPC2 and ACS.
Updated CTE corrections for ACS are in progress at STScI, but these
corrections were not ready in time for reducing the data for this
release.  Given that these corrections are typically swamped by random
errors and are smaller than current uncertainties in the stellar
isochrones that are used to interpret the CMDs, we decided to release
the data as is.  Subsequent releases will include
the new CTE corrections as they become available.  We also detected
possible signs of a color-dependence in the magnitude differences
between the WFPC2 and ACS $F814W$ VEGAMAGS, which appear to be
larger than expected based on synthetic filter curves.  A definitive
diagnosis of this dependence must wait until the improved CTE
corrections for ACS are implemented, but should the effect persist,
then there may be an additional few percent uncertainty in the
instrumental response of either WFPC2 or ACS or both.

%----------------------------------------------
\section{Astrometry}  \label{astrometrysec}

Astrometry for the photometric catalogs was initially taken from the
FITS headers of the original HST images, which have astrometry that is
accurate to $1-2\arcsec$.  Recently, the Hubble Legacy Archive (HLA)
improved on the default astrometry using the Guide Star Catalog (GSC),
the Sloan Digital Sky Survey (SDSS), and the Two Micron All Sky Survey
(2MASS).  The revised astrometric solutions have a typical RMS of
$0.1-0.3\arcsec$ in most cases.  There are many cases in our data set
where the RMS is much larger, due to using faint (or non-existent)
sources in the GSC, or cosmic rays in the image during matching.  In
these cases, new astrometric positions will have to be derived by
hand.  Because this process is almost always limited by the lack of
astrometric standards within nearby galaxies, in many cases
$1-2\arcsec$ uncertainties remain, and will have to be dealt with in
future releases by using a system of secondary astrometric standards
defined in wide-field ground-based imaging.

Relative photometry within a given field is usually accurate to a
fraction of a pixel, and the absolute position is good to a few pixels
in most cases.  However, in applications requiring sub-arcsecond
accuracy of the absolute astrometric position (e.g., such as slit
masks or comparisons with multi-wavelength data), users should
consider making an independent astrometric solution.  We will continue
to release improved astrometric solutions as they become available,
including time-dependent geometric distortion corrections as well.

%----------------------------------------------
\section{Data Products}  \label{dataproductsec}

Binary FITS tables of photometry for the ANGST sample have been
released through the Multimission Archive at STScI (MAST:
{\tt{http://archive.stsci.edu/prepds/angst/}}), and can also be
accessed interactively through the project website
({\tt{http://www.nearbygalaxies.org}}).  File names and field names
were taken from the image headers and are of the format
{\tt{PROPOSID-TARGNAME}}, where {\tt{PROPOSID}} is the value of the
header keyword ``PROPOSID'' and {\tt{TARGNAME}} is the value of the
header keyword ``TARGNAME.''  The naming conventions and column names
for the files are summarized below, and are contained in the headers
of the fits files themselves.

{\tt{*.param}}: DOLPHOT parameter files: These files provide the
 parameters used by DOLPHOT when measuring the photometry, and are
 useful for interpreting the columns in the raw photometry files.
 These files are currently only available on the project website.

 {\tt{*.phot}}: Raw photometry files: These large ASCII files contain
 the raw output from DOLPHOT.  Descriptions of the columns can be
 found in the DOLPHOT manual
 ({\tt{http://purcell.as.arizona.edu/dolphot/}}).  The listing of
 individual columns can be found on the project web site.

 {\tt{*.st.fits}}: Star files: these files contain the photometry of
 all objects classified as stars (object type $<=$ 2) with S/N $>$ 4
 and data flag $<$ 8.  Compared to the {\tt{*.gst}} files described
 below, these files will contain more objects and have higher
 completeness in crowded regions, at the expense of producing less
 well defined CMDs with more potential contamination from background
 galaxies.  Columns are {\tt{X}}, {\tt{Y}}, {\tt{RA}}, {\tt{DEC}},
 {\tt{MAG1\_ACS}} (or {\tt{MAG1\_WFPC2}} in WFPC2 files),
 {\tt{MAG1\_STD}}, {\tt{MAG1\_ERR}}, {\tt{CHI1}}, {\tt{SHARP1}},
 {\tt{CROWD1}}, {\tt{SNR1}}, {\tt{FLAG1}} (or {\tt{CHIP}} in WFPC2
 files), {\tt{MAG2\_ACS}} (or {\tt{MAG2\_WFPC2}} in WFPC2 files),
 {\tt{MAG2\_STD}}, {\tt{MAG2\_ERR}}, {\tt{CHI2}}, {\tt{SHARP2}},
 {\tt{CROWD2}}, {\tt{SNR2}}, {\tt{FLAG2}} ({or \tt{FLAG}} in WFPC2
 files).  These values are defined as follows.  {\tt{X}} and {\tt{Y}}
 positions are relative to positions on the drizzled reference image.
 The {\tt{MAG1}} and {\tt{MAG2}} values refer to the filters given in
 the file name and in the FITS header.  {\tt{ACS}} magnitudes are
 VEGAMAGs, which are calibrated by setting the zeropoint of each
 filter so that the magnitude of Vega is 0.0 \citep{sirianni05}.
 {\tt{WFPC2}} magnitudes are VEGAMAGs, which are calibrated by setting
 the zeropoint of each filter so that the magnitude of Vega is 0.035
 in most ANGST filters \citep{holtzman95}.  {\tt{STD}} magnitudes have
 been converted from VEGAMAGs to standard Johnson-Cousins magnitudes
 for the nearest Johnson-Cousins filter (B, V, or I) using the
 transformation equations of Sirianni et al. (2005; ACS) and Holtzman
 et al. (1995; WFPC2).  The DOLPHOT {\tt{CHI}} value indicates the
 goodness of the PSF fit, with values of $<$1.5-2.5 being reasonable
 for uncrowded well-exposed stars, and values of up to 4-5 being
 expected for either blended but unresolved stars, or for stars in
 crowded regions.  The {\tt{SHARP}} parameter indicates the deviation
 from a perfect PSF profile, with positive values indicating profiles
 that are too sharp (such as cosmic rays), and negative values indicating
 profiles that are too broad (such as unresolved blends, clusters, or
 background galaxies).  The {\tt{SNR}} value gives the signal-to-noise
 with which the star was detected. The {\tt{CROWD}} parameter is in
 magnitudes, and indicates how much brighter the star would have been
 if flux from nearby stars had not been subtracted.  {\tt{FLAG1}} and
 {\tt{FLAG2}} are the DOLPHOT quality flags for each filter as
 described in the manual.  {\tt{FLAG}} is the \citet{holtzman06}
 quality flag used in the Local Group Stellar Populations Archive.
 Further details can be found in the DOLPHOT manual.

{\tt{*.gst.fits}}: ``Good'' star files: These files contain the stars
that pass the conservative ANGST quality cuts for sharpness and
crowding ({\tt{sharp}}$_{1}$ + {\tt{sharp}}$_{2}$)$^2 \leq
$ 0.075 and {\tt{crowd}}$_{1}$ + {\tt{crowd}}$_{2} \leq $
0.1), in addition to the S/N and flag criteria.  Columns and header
information are the same as for the {\tt{*.st.fits}} files.

The field names, number of stars, and 50\% completeness limits can
be found in Table~\ref{phottable}.

The MAST archive for ANGST also includes copies of the reference
images to which all $X-Y$ positions are tied.  For ACS, the reference
image is a single dithered image in the deepest filter.  For WFPC2,
there are four reference images, one for each chip on the camera.

%----------------------------------------------
\section{Color-Magnitude Diagrams}  \label{cmdsec}

In Figures~\ref{cmdfig1}-\ref{cmdfig14} we present color magnitude
diagrams (CMDs) for all of the fields listed in
Tables~\ref{obstable}~\&~\ref{archivetable}.  The plotted photometry
is drawn from the high-quality ({\tt{*.gst.fits}}) catalogs.  As
described above, these quality cuts produce the most well-defined
features in the CMD, at the expense of completeness in high-crowding
regions (such as the densest stellar clusters). In regions of high
stellar density on the CMD, data are plotted as contoured Hess
diagrams, with contours drawn at levels of
1,1.5,2,2.5,3,4,6,8,12,16,20 $\times$10$^4$ stars mag$^{-2}$.
Characteristic photometric uncertainties are shown with error bars on
the left side of the CMDs in Figures~\ref{cmdfig1}-\ref{cmdfig14}.

The ANGST CMDs show a richness of detail, thanks to their depth, high
photometric accuracy, and large number of stars.  As a guide to
interpreting the many features visible in these CMDs, in
Figure~\ref{simcmdfig} we plot simulated CMDs that show the locations
of different stellar populations, as a function of age (right) and
metallicity (left).  The plots show the CMDs expected for a constant
star formation rate color-coded by age (left) and for an early burst
of star formation color-coded by metallicity (right), assuming
photometric uncertainties typical for our data at the inner (left
panel) and the outer (right) distances of the ANGST target galaxies.

As has been discussed extensively elsewhere \citep[e.g.,][and
references therein]{gallart05}, the simulated CMDs show how young
stellar populations are found primarily in the upper left of the CMD,
older stellar populations are found at lower luminosities and redder
colors along the red giant branch (RGB) and asymptotic giant branch
(AGB), and metal rich stars are found at redder colors for older
stellar populations.  We do not plot the metallicity dependence of
younger stars, since the color of the main sequence has essentially no
metal dependence for the filters used in this data release.

Also overlayed on Figure~\ref{simcmdfig} are tracks indicating the
typical locations of young main sequence stars, of blue and red core
helium burning stars (BHeB and RHeB), and asymptotic giant branch
stars.  Among these features, the blue and red core helium burning
sequences are the least widely known, since they are only visible when
the CMD is well populated.  We note that the HeB sequences can produce
potentially confusing features in the CMD.  In particular, the upper
end of the blue core helium burning sequence can be easily mistaken for
a ``double'' main sequence.  Additional vertical sequences sometimes
appear where BHeB stars pass through the instability strip, leading to
a nearly vertical spread in magnitude for variable stars observed only
at a single epoch.

%----------------------------------------------
\section{Magnitude of the Tip of the Red Giant Branch}  \label{TRGBsec}

In Table~\ref{trgbtable} we list the $F814W$ magnitude of the tip of
the red giant branch (TRGB) for each galaxy.  TRGB magnitudes were
determined using the edge-detection filter described in
\citet{mendez02} applied to a Gaussian-smoothed luminosity function as
in \citet{sakai96} and \citet{seth05}.  Although more sophisticated
techniques exist \citep[e.g.,][]{makarov06,frayn03}, the TRGB in our
sample is typically well-populated and falls well above the photometric
limit of the data, making our use of the widely used and calibrated
edge-detection technique adequate for an initial distance measurement.

The reported $F814W$ TRGB magnitude $m_{TRGB}$ and the associated
uncertainty were determined by running 500-750 Monte Carlo trials with
bootstrap resampling of the stars. In each trial, additional Gaussian
random errors are added to the stars, scaled to the magnitude of each
star's photometric error.  Each trial returned the magnitude
corresponding to the peak of the edge-detection response filter within
a 1 magnitude interval around the likely TRGB.  We generated a
histogram of the returned magnitudes, and fit the peak at $m_{TRGB}$
in the histogram with a Gaussian.  We take the mean and width of the
Gaussian to be the magnitude of the TRGB and its uncertainty.
Although the Monte Carlo process artificially increases the
photometric error (during randomization of magnitudes) and potentially
biases $m_{TRGB}$ by scattering stars preferentially above the tip, in
practice the effect of the added noise is negligible, since the
photometric uncertainties are extremely small at $m_{TRGB}$ in almost
every case.  Furthermore, we have verified visually that the method
above converges on a consistent part of the luminosity function, and
thus preserves the accuracy of the relative distances.

In some Monte Carlo trials, there are additional peaks in the
edge-detection response function that clearly do not correspond to the
TRGB (see Figure~\ref{TRGBbadfig}).  These spurious peaks are most
prevalent when there are a smaller number of stars, or an old
population of AGB stars with a well-defined peak luminosity.  In these
cases, we initialized the Gaussian fit with a mean chosen to be
centered on the peak corresponding most closely to the true TRGB.
Examples of the luminosity function, edge-detection response,
histogram of Monte Carlo TRGB magnitudes, and the CMD of the analyzed
stars are presented in Figures~\ref{TRGBfig}~\&~\ref{TRGBbadfig}.

Uncertainties in the measured value of $m_{TRGB}$ can include random
errors (due to small numbers of stars and to photometric errors) and
systematic errors (due to contamination from stars on the red helium
burning sequence, to uncertainties in the MW foreground extinction,
and to the unknown internal extinction).  We attempted to reduce our
systematic errors by considering only stars likely to be RGB or AGB
stars.  To do so, we selected stars that fell within model RGB
isochrones from \citet{girardi08} in the appropriate HST filter set,
extrapolated up into the region populated by AGB stars.  This process
was automated by first shifting the stars in magnitude and color based
on the estimated foreground extinction from
\citet{schlegel98}\footnote{The one exception is M82, for which the
  \citet{schlegel98} value is clearly contaminated by point source
  emission from M82 itself, leading to an erroneously high foreground
  extinction ($A_B=0.685$).  Instead, we took $A_B=0.25$, based upon
  regions immediately adjacent to M82.}, assuming $R_V=3.1$, 
$A_{F435W}/A_V=1.30$, $A_{F475W}/A_V=1.15$, $A_{F555W}/A_V=1.00$,
$A_{F606W}/A_V=0.87$, and $A_{F814W}/A_V=0.57$, based upon
\citet{girardi08} for typical temperatures of RGB stars.  We used
triangular interpolation of the isochrones to generate a regular grid
of metallicities as a function of color and magnitude for a uniform,
intermediate age population ($4\Gyr$). We then interpolated the
observed stars onto this grid to assign metallicities to each star,
and rejected stars with unphysical metallicities.  This process is
equivalent to assigning each star to a particular RGB isochrone, and
rejecting stars that are inconsistent with all plausible isochrones.
For the remaining stars, we used a robust bi-weight to find the peak
and width of the distribution of the logarithms of the inferred
metallicities.  We selected all stars whose metallicities fell within
$1.5\sigma$ of the peak in $\log{([Fe/H])}$.  We further excluded
stars with metallicities outside of the range 0.0002 and 0.006 (for
$4\Gyr$ isochrones); within this metallicity range, the $F814W$
magnitude of the TRGB varies by less than $\pm0.05$ mag, but outside it,
the TRGB becomes steadily fainter by several tenths of a magnitude,
blurring the TRGB discontinuity and introducing systematic errors when
converting $m_{TRGB}$ to distance.  We expect little dependence of
the TRGB absolute magnitude $M_{TRGB}$ on age or metallicity; the predicted
absolute magnitude of the TRGB depends primarily on the color of the RGB, and
more weakly upon the particular age$+$metallicity combination that generated
a particular RGB isochrone.  The final isochrone fitting procedure
cleanly isolated the bulk of RGB and AGB stars, while significantly
reducing contamination from non-RGB features
(Figures~\ref{TRGBfig}~\&~\ref{TRGBbadfig})\footnote{Note that
  although we derived the ``metallicity'' of each star, we do not
  treat these as actual measurements of the metallicity due to the
  likely presence of mixed stellar ages on the RGB; instead, we only
  use the inferred metallicity as a label for the RGB isochrone on which a star
  lies.  Likewise, the mean color that we report for the TRGB stars
  includes only those stars that made the various metallicity cuts,
  and does not reflect the color of the RGB as a whole.}.

We further reduced systematic biases due to internal extinction and
photometric errors by restricting our analysis to stars in regions of
low crowding within an individual field, when sufficient numbers of
stars were available ($>$30,000).  We chose a density threshold such
that at least 25\% of the area and 50\% of the stars were included in
the analysis.  This cut eliminated stars in the most crowded regions
with the highest internal extinction, while still preserving large
numbers of stars.  For galaxies with multiple pointings, the TRGB was
measured in whichever fields had the least crowding and lowest
probability of high internal or differential reddening, while still
having large numbers of stars.  When multiple clean fields were
available, we analyzed both, to compare our internal systematics and
to constrain the variation in internal extinction.  The resulting TRGB
magnitudes were frequently several tenths of a magnitude brighter than
those measured within the main body of a galaxy.  Beyond field
placement, however, we make no further attempt to correct for internal
extinction, although extinctions of several tenths of a magnitude are
certainly possible in the outer regions of massive galaxies (e.g.,
Holwerda et al 2008).

To transfer the measured TRGB magnitudes into initial distance
estimates, we used the measured mean color within 0.2$m$ of the TRGB
to pick a \citet{girardi08} isochrone with similar colors, from which
we then find the absolute magnitude $M_{TRGB}$ of the TRGB.  However,
due to the uncertain state of the ACS CTE correction, there are likely
to be systematic uncertainties present in the data that limit the
accuracy of the inferred distance to a few percent.  The likely
systematic uncertainties in the adopted value of $M_{TRGB}$ are even
larger, with different theoretical models and empirical calibrations
differing by as much as $0.2$ mag \citep[e.g., see Figure~8
of][]{gallart05}.  The reported distances are thus best used as
relative distances, rather than absolute ones.  The uncertainties
listed in Table~\ref{trgbtable} do not include these systematic
errors, and include only the Poisson uncertainties captured by
bootstrap resampling.  Thus, while the Gaussian fitting procedure
described above (and seen in the upper left panels of
Figures~\ref{TRGBfig}~\&~\ref{TRGBbadfig}) frequently reports formal
distance errors of a few percent, the true uncertainties are
undoubtedly larger.

The resulting data in Table~\ref{trgbtable} includes: the number of
stars within 1 magnitude of the TRGB ($N_{stars}$); the adopted
foreground extinction $A_V$; the mean color (within 0.2 mag of the TRGB)
of the stars used to measure the TRGB, for the particular filter
combination used; the predicted absolute magnitude of the TRGB at that
color, based on isochrones from \citet{girardi08}; the apparent
magnitude of the TRGB in $F814W$, uncorrected for extinction
($m_{TRGB}(raw)$); the extinction corrected TRGB magnitude
($m_{TRGB}$) and its uncertainty; the resulting extinction corrected
distance modulus $(m-M)_0$; and the inferred distance $D$ in Mpc.  The
resulting spatial configuration of galaxies is shown in
Figure~\ref{3Dfig}.

Figure~\ref{TRGBdifffig} plots the differences between the new
distance moduli and those inferred from distances in
Table~\ref{sampletable}, which had been used for initial sample
selection, as a function of increasing distance (left) and luminosity
(right).  The revised distances agree well with previously published
values.  The median change in distance modulus is only $-0.02$ mag for
the entire sample, indicating that there is little systematic
deviation between our adopted TRGB scale and those used in the
literature.  There is however, a modest tendency for past distances to
be systematically overestimated for the most massive galaxies.  If we
split the sample into galaxies that are brighter or fainter than
$M_B=-17$, the median offset is only -0.012 for the faint galaxies,
but increases to $-0.074$ for the more luminous galaxies.  We believe
that that offset is most likely due to past TRGB determinations using
stars closer to the galaxies' centers, where the extinction of dust is
larger, leading the TRGB magnitude to appear fainter.  In contrast,
our measurements use the outskirts of galaxies, where the internal
extinction is small, producing a brighter TRGB.  On the other hand, many
of the published distance estimates have made corrections for internal
extinction, unlike those we present in Table~\ref{trgbtable}

Because the most luminous galaxies tend to be found in the M81 group,
the left panel shows a hint of correlation between distance and the
offset in the distance modulus.  However, the Spearman rank
correlation coefficient is much smaller when using distance instead of
luminosity ($-0.09$ vs 0.30), indicating that the variation in
luminosity is the principal driver of the trend.

The dispersion about the mean difference is $0.05$ mag, comparable to
the precision of distances in Table~\ref{sampletable} and to their
published uncertainties.  There is no difference in the dispersion for
brighter and fainter galaxies.  Multiple observations of the same
galaxy (connected with a solid line in Figure~\ref{TRGBdifffig}) show
differences of typically less than $0.1$ mag (i.e.\ 10\% in distance).
Because these measurements were made in different regions of these
galaxies, some of this variation is likely to be due to differences in
internal extinction within the galaxy.  In such cases, the outermost
distance measurement is likely to be the least affected by dust,
although the reduced numbers of stars in such fields leads to larger
Poisson uncertainties.  We also found no systematic offsets between
distances determined with ACS and those measured with WFPC2 data.

We see no evidence that the revised distances would have changed our
initial sample selection.  The largest change in distance is for
UGCA~292, which is nearly 40\% further away than previously estimated
from its brightest stars \citep{makarova98}.  The distance to
MCG9-20-131 also appeared to decrease significantly; however, there is
some ambiguity as to whether or not the apparent TRGB is due to RGB or
AGB stars, or potentially even red supergiants.  Thus, the distance
may have a systematic offset, even though the formal error on the
magnitude of the tip is relatively small.  With the new distances,
NGC~247 and NGC~253 are much closer to each other.  The morphology of
the other groups remain essentially unchanged.

%----------------------------------------------
\section{Conclusions}   \label{conclusionsec}

The ACS Nearby Galaxy Survey Treasury is the now the largest
repository of uniform stellar photometry for nearby galaxies.  The
resulting catalogs contain millions of measurements that can be used
for studies of ancient and recent star formation histories
\citep{williams08, weisz08} and comparisons with multi-wavelength data
\citep{gogarten08,ott08}.  The raw images are a resource for searches
for stellar clusters, HII region nebulosity, and background light
sources.  

%----------------------------------------------
\acknowledgements 

We are happy to acknowledge the consistently professional and helpful
assistance from the staff at STSCI, including Alison Vick, Marco
Sirianni, Howard Bond, and Neill Reid.  We also are pleased to thank
Jay Anderson, Brent Tully, Abi Saha, Stan Vlcek, Pat Taylor, Sarah
Garner, and Richard Coffey for assistance at various times during the
project.  We also thank the referee for constructive comments.  JJD
acknowledges partial support from the Wyckoff Faculty Fellowship
during this work, and the hospitality of the MPIA and Carnegie
Observatories during some of the writing of this paper.  LG
acknowledges financial support from contract ASI-INAF I/016/07/0.  IK
and VK were partially supported by RFFI grant 07--02--00005 and grant
DFG-RFBR 06--02--04017.

This work is based on observations made with the NASA/ESA Hubble Space
Telescope, obtained from the data archive at the Space Telescope
Science Institute. Support for this work was provided by NASA through
grant number GO-10915 from the Space Telescope Science Institute,
which is operated by AURA, Inc., under NASA contract NAS 5-26555.
This research has made use of the NASA/IPAC Infrared Science Archive
and the NASA/IPAC Extragalactic Database (NED), which are both
operated by the Jet Propulsion Laboratory, California Institute of
Technology, under contract with the National Aeronautics and Space
Administration.  This research has made extensive use of NASA's
Astrophysics Data System Bibliographic Services.

{\it Facilities:} \facility{HST (ACS), HST (WFPC2)}

%----------------------------------------------

%\bibliographystyle{apj}  \bibliography{references}

%--------------
\clearpage
\LongTables
\begin{landscape}
%
% started with final.dat from last year (final_05.dat).
% updated distances for N247, N55, DDO187, U8833, HS117, KKH37 (using table in Karachentsev & Kashibadze 05)
% eliminated UGC 8638 and E059-01, which moved too far away
% added N4163 and DDO 183, which moved closer.  (don't have orbital info yet)
% removed KKH60 and HIJASS on Igor's recommendation (4/10 email)
%
%  Why isn't I342 in the sample?!?!?!  NGC has D=3.28.  Low b?  Local Group?
%
%  K band fluxes from 2MASS LGA and from Vaduvescu et al 2005 (dwarfs)
%
% name  dist     ra		dec             diam      M_B?    m_K    T     V_c      nwide deep 3filt arch
%         
\begin{deluxetable}{llccccccccccccc}
%\rotate
\setlength{\tabcolsep}{0.02in}
\tablewidth{557pt}
\tabletypesize{\scriptsize}
\tablecaption{ANGST Sample Galaxies \& Planned Observations}
\tablehead{
    \colhead{Galaxy} &
    \colhead{Alt.} &
    \colhead{Dist.} &
    \colhead{RA} &
    \colhead{Dec} &
    \colhead{Diam.} &
    \colhead{$M_B$} &
    \colhead{$m_K$} &
    \colhead{T} &
    \colhead{$W_{50}$} &
    \colhead{Group} &
    \colhead{\#} &
    \colhead{Deep?} &
    \colhead{3 Filt?} &
    \colhead{Arch?} \\
    \colhead{} &
    \colhead{Names} &
    \colhead{(Mpc)} &
    \colhead{(J2000)} &
    \colhead{(J2000)} &
    \colhead{($\prime$)} &
    \colhead{} &
    \colhead{} &
    \colhead{} &
    \colhead{$\kms$} &
    \colhead{} &
    \colhead{Fields} &
    \colhead{} &
    \colhead{} &
    \colhead{} 
}
\startdata                                                              
Antlia      &  & 1.3 &   10:04:04.0 &  -27:19:55  &  2.0  &   -9.38 &       &   10 &  22  &  14+12 &  1  & Y &   & 10210  \\
SexA	    & DDO75 & 1.3 &   10:11:00.8 &  -04:41:34  &  5.5  &  -13.71 &       &   10 &  63  &  14+12 &     &   &   & 7496\tablenotemark{a}\\
N3109	    & DDO236 & 1.3 &   10:03:07.2 &  -26:09:36  &  17.0 &  -15.18 & 9.3   &   9  &  116 &  14+12 &  3  & Y & Y &   \\
SexB	    & U5373 & 1.4 &   10:00:00.1 &  05:19:56   &  5.1  &  -13.88 &       &   10 &  38  &  14+12 &  1  & Y & Y &   \\
KKR25	    &  & 1.9 &   16:13:47.6 &  54:22:16   &  1.1  &   -9.94 &       &   10 &    & M31(p)?&  1  & Y &   & 8601\tablenotemark{a}  \\
KK230	    & KKR3 & 1.9 &   14:07:10.7 &  35:03:37   &  0.6  &   -8.49 &       &   10 &  21  & M31(p)?&  1  & Y &   & 9771  \\
E410-005    &  & 1.9 &   00:15:31.4 &  -32:10:48  &  1.3  &  -11.49 &       &   -1 &      &  14+13 &     &   &   & 10503  \\
E294-010    &  & 1.9 &   00:26:33.3 &  -41:51:20  &  1.1  &  -10.86 &       &   -3 &      &  14+13 &     &   &   & 10503\\
\enddata
\tablecomments{Distances, $M_B$, $W_{50}$, and T-type taken from CNG;
values for $m_K$ are total $K$ magnitudes from either the 2MASS Large
Galaxy Atlas \citep{jarrett03} or from \citet{vaduvescu05} for dwarfs;
Group membership from \citet{karachentsev05} or \citet{tully06}; \# of
ANGST Pointings includes any planned deep fields; Deep column
indicates planned observations that would reach high completeness in
red clump; Note that not all planned observations were executed, due
to ACS failure -- actual observations are given in
Table~\ref{obstable}; 3 Filter column indicates observations made in
$F475W$+$F606W$+$F814W$, rather than default $F475W$+$F606W$ for
dwarfs; Archival lists Proposal ID of archival data to be used, with
entries in parentheses indicating that archival data will be
supplemented with new observations; HI detection for KKR25 in the CNG
was not confirmed in later GMRT observations \citep{begum05}; (a)
WFPC2 Archival data.}
   \label{sampletable}
\end{deluxetable}

\clearpage
\end{landscape}
%------------------

%--------------
\clearpage
\LongTables
%\begin{landscape}
%
\begin{deluxetable}{llccccccc}
%\rotate
%\setlength{\tabcolsep}{0.02in}
\tabletypesize{\scriptsize}
\tablecaption{ANGST Observations}
\tablehead{
    \colhead{Field Name} &
    \colhead{RA} &
    \colhead{Dec} &
    \colhead{V3 PA} &
    \colhead{Aperture} &
    \colhead{Date} &
    \colhead{Instr.} &
    \colhead{Filter} &
    \colhead{Exptime} \\
    \colhead{} &
    \colhead{(J2000)} &
    \colhead{(J2000)} &
    \colhead{($\deg$)} &
    \colhead{} &
    \colhead{Range} &
    \colhead{} &
    \colhead{} &
    \colhead{(s)}
}
\startdata                                                              
NGC3109-WIDE1 &	10 02 41.8  &	-26 08 58   &	95.001  &	WFALL-FIX &	2007-11-02 &	WFPC2 &	F606W &	2700  \\
              &	            &	            &	        &	          &	           &	      &	F814W &	3900  \\
%NGC3109-WIDE1 &	10 02 41.81 &	-26 08 58.1 &	95.001  &	WFALL-FIX &	2007-11-02 &	WFPC2 &	F606W &	900  \\
%NGC3109-WIDE1 &	10 02 41.86 &	-26 08 58.1 &	95.001  &	WFALL-FIX &	2007-11-02 &	WFPC2 &	F606W &	900  \\
%NGC3109-WIDE1 &	10 02 41.92 &	-26 08 58.2 &	95.001  &	WFALL-FIX &	2007-11-02 &	WFPC2 &	F606W &	900  \\
%NGC3109-WIDE1 &	10 02 41.81 &	-26 08 58.1 &	95.001  &	WFALL-FIX &	2007-11-02 &	WFPC2 &	F814W &	1300  \\
%NGC3109-WIDE1 &	10 02 41.86 &	-26 08 58.1 &	95.001  &	WFALL-FIX &	2007-11-02 &	WFPC2 &	F814W &	1300  \\
%NGC3109-WIDE1 &	10 02 41.92 &	-26 08 58.2 &	95.001  &	WFALL-FIX &	2007-11-02 &	WFPC2 &	F814W &	1300  \\
%
NGC3109-WIDE2 &	10 02 49.9  &	-26 09 07   &	95.001  &	WFALL-FIX &	2007-11-08 &	WFPC2 &	F606W &	2700 \\
              &	            &	            &	        &	          &	           &	      &	F814W &	3900  \\
%NGC3109-WIDE2 &	10 02 49.81 &	-26 09 07.1 &	95.001  &	WFALL-FIX &	2007-11-08 &	WFPC2 &	F606W &	900  \\
%NGC3109-WIDE2 &	10 02 49.86 &	-26 09 07.1 &	95.001  &	WFALL-FIX &	2007-11-08 &	WFPC2 &	F606W &	900  \\
%NGC3109-WIDE2 &	10 02 49.92 &	-26 09 07.2 &	95.001  &	WFALL-FIX &	2007-11-08 &	WFPC2 &	F606W &	900  \\
%NGC3109-WIDE2 &	10 02 49.81 &	-26 09 07.1 &	95.001  &	WFALL-FIX &	2007-11-08 &	WFPC2 &	F814W &	1300  \\
%NGC3109-WIDE2 &	10 02 49.81 &	-26 09 07.1 &	95.001  &	WFALL-FIX &	2007-11-08 &	WFPC2 &	F814W &	1300  \\
%NGC3109-WIDE2 &	10 02 49.92 &	-26 09 07.2 &	95.001  &	WFALL-FIX &	2007-11-08 &	WFPC2 &	F814W &	1300  \\
%
NGC3109-WIDE3 &	10 02 57.81 &	-26 09 16.4 &	95.001  &	WFALL-FIX &	2007-11-06 &	WFPC2 &	F606W &	2400  \\
              &	            &	            &	        &	          &	           &	      &	F814W &	2400  \\
\enddata
\tablecomments{All data is from GO-10915, except NGC~55 and NGC~3109
  wide fields, which are from DD-11307.  Entries with reduced
  precision in the listings for RA and Dec indicate a representative
  center for dithered observations.  Multiple dates for a given field
  indicate the start times for the earliest and the latest
  observations of that field.  Note that many fine guidance sensor
  problems occured during observations for GO-10915, and that not all
  data in the archive under this PID is useable; only high quality
  observations are included in this table.}
\label{obstable}
\end{deluxetable}

\clearpage
%\end{landscape}
%------------------

%--------------
\clearpage
\LongTables
\begin{landscape}
\begin{deluxetable}{lllcccccccccc}
%\rotate
\setlength{\tabcolsep}{0.02in}
\tablewidth{557pt}
\tabletypesize{\scriptsize}
\tablecaption{Archival Observations}
\tablehead{
    \colhead{Catalog} &
    \colhead{Target} &
    \colhead{PID} &
    \colhead{RA} &
    \colhead{Dec} &
    \colhead{Ang.\ Sep.} &
    \colhead{V3 PA} &
    \colhead{Aperture} &
    \colhead{Date} &
    \colhead{Instr.} &
    \colhead{Filter} &
    \colhead{Exptime} &
    \colhead{Literature} \\
%-----------
    \colhead{Name} &
    \colhead{Name} &
    \colhead{} &
    \colhead{(J2000)} &
    \colhead{(J2000)} &
    \colhead{($^\prime$)} &
    \colhead{($\deg$)} &
    \colhead{} &
    \colhead{Range} &
    \colhead{} &
    \colhead{} &
    \colhead{(s)} &
    \colhead{CMD}
}
\startdata                                                              
%----------------------------------------------------------------------------
Antlia & ANTLIA&	10210&	10 04 03.77&	-27 19 47.2&	0.14&	247.479&	WFC1&	2005-04-02&	ACS&	F814W&	1174 & 1 \\
 &       &	     &	           &	           &	     &	       &	    &	          &	   &	F606W&	985 \\
%----------------------------------------------------------------------------
SexA & DDO75     &      7496&  10 10 56.9 &    -04 41 30  &   0.983 & 287.891&     WF3-FIX&    1999-04-05 &  WFPC2& F555W & 19200 & 2 \\
 &           &	     &	           &	           &	     &	      &	           &	1999-04-08 &	   & F814W & 38400 \\
%----------------------------------------------------------------------------
KKR25 & KKR25 & 8601 & 16 13 47.47 &    +54 21 27.7&    0.83&  358.523& WF3-FIX &2001-05-28 & WFPC2&    F606W&  600 & 9 \\
 &      &	    &	           &	           &	&	      &	   &	          &	   &	F814W&	600 \\
%----------------------------------------------------------------------------
KK230 & KK230&	9771&	14 07 10.70&	+35 03 37.0&	0.00&	18.337&	WFC&	2004-04-06&	ACS&	F606W&	1200 & 3 \\
 &      &	    &	           &	           &	&	      &	   &	          &	   &	F814W&	900 \\
\enddata
\tablecomments{Entries with reduced precision in the listings for RA
  and Dec indicate a representative center for dithered observations.
  Multiple dates for a given field indicate the start times for the
  earliest and the latest observations of that field.  Angular
  separation refers to separation between the aperture position and
  the catalog coordinates in Table~\ref{sampletable}. Field names
  indicate the titles given in the archive.  Observations lacking
  multiple filters at nearly the same position and alignment are not
  included. [a] Observations for N4190 were
  proprietary at the time this paper was submitted; their photometry
  will be added to the data release when it becomes public.
  References for published CMDs from the associated data are given in
  the last column: 1 $=$ \citet{tully06}; 2 $=$ \citet{dohm02b}; 3 $=$
  \citet{karachentsev06}; 4 $=$ \citet{weisz08}; 5 $=$
  \citet{seth05a}; 6 $=$ \citet{butler04}; 7 $=$ \citet{rizzi06}; 8
  $=$ \citet{caldwell98}; 9 $=$ \citet{karachentsev01}; 10 $=$
  \citet{karachentsev02b}; 11 $=$ \citet{karachentsev02c}; 12 $=$
  \citet{dolphin01}; 13 $=$ \citet{izotov02}; }
\label{archivetable}
\end{deluxetable}

\clearpage
\end{landscape}
%------------------

%--------------
\clearpage
\LongTables
%\begin{landscape}
%
\begin{deluxetable}{lclccccc}
%\rotate
%\setlength{\tabcolsep}{0.02in}
\tabletypesize{\scriptsize}
\tablecaption{Photometry}
\tablehead{
   \colhead{Catalog} &
    \colhead{Proposal} &
    \colhead{Target} &
   \colhead{Instrument} &
    \colhead{Filter} &
    \colhead{Exposure} &
    \colhead{$N_{stars}$} &
    \colhead{50\%}  \\
%-----------
    \colhead{Name} &
    \colhead{ID} &
    \colhead{name} &
    \colhead{} &
    \colhead{} &
   \colhead{Time ($s$)} &
    \colhead{} &
   \colhead{Completeness}\\
%-----------
    \colhead{} &
    \colhead{} &
    \colhead{} &
    \colhead{} &
    \colhead{} &
   \colhead{($s$)} &
    \colhead{} &
   \colhead{(mag)}
}
\startdata
%------------------------
Antlia/P29194 & 10210 & ANTLIA & ACS & F606W &    985 & 19207 &  28.03\\
Antlia/P29194 & 10210 & ANTLIA & ACS & F814W &   1174 & 19207 &  27.30\\
SexA/DDO75 & 7496 & DDO75 & WFPC2 & F555W &  19200 & 33251 &  27.34\\
SexA/DDO75 & 7496 & DDO75 & WFPC2 & F814W &  38400 & 33251 &  26.50\\
N3109 & 10915 & NGC3109-DEEP & WFPC2 & F606W &   2700 & 13329 &  26.76\\
N3109 & 10915 & NGC3109-DEEP & WFPC2 & F814W &   3900 & 13329 &  26.00\\
N3109 & 11307 & NGC3109-WIDE1 & WFPC2 & F606W &   2700 & 21868 &  26.61\\
N3109 & 11307 & NGC3109-WIDE1 & WFPC2 & F814W &   3900 & 21868 &  25.65\\
\enddata
\tablecomments{Exposure times may differ from those in Table~\ref{archivetable} when individual fields were unusable.}
\label{phottable}
\end{deluxetable}

\clearpage
%\end{landscape}
%------------------

%--------------
\clearpage
\LongTables
\begin{landscape}
\begin{deluxetable}{llcrccccccc}
%\rotate
\setlength{\tabcolsep}{0.02in}
\tabletypesize{\scriptsize}
\tablecaption{TRGB Measurements}
\tablehead{
    \colhead{Catalog} &
    \colhead{Target} &
    \colhead{Filters} &
    \colhead{$N_{stars}$} &
    \colhead{$A_V$} &
    \colhead{Mean} &
    \colhead{$M_{TRGB}$} &
    \colhead{$m_{TRGB}$} &
    \colhead{$m_{TRGB}$} &
    \colhead{$(m-M)_0$} &
    \colhead{$D$} \\
    \colhead{Name} &
    \colhead{Name} &
    \colhead{} &
    \colhead{} &
    \colhead{} &
    \colhead{Color} &
    \colhead{($F814W$)} &
    \colhead{(raw)} &
    \colhead{} &
    \colhead{} &
    \colhead{($\Mpc$)}
}
\startdata
   Antlia &                ANTLIA & F606W,F814W &    169 & 0.243 & 1.05 & -4.04 & 21.642 & $21.503\pm0.011$ & 25.546 & $1.29\pm0.015$ \\
     SexA &                 DDO75 & F555W,F814W &    276 & 0.139 & 1.34 & -3.95 & 21.835 & $21.756\pm0.033$ & 25.705 & $1.38\pm0.050$ \\
    N3109 &         NGC3109-WIDE1 & F606W,F814W &    355 & 0.201 & 1.04 & -4.00 & 21.608 & $21.493\pm0.018$ & 25.498 & $1.26\pm0.025$ \\
    N3109 &          NGC3109-DEEP & F606W,F814W &    263 & 0.201 & 1.01 & -3.98 & 21.687 & $21.572\pm0.014$ & 25.551 & $1.29\pm0.019$ \\
     SexB &             SEXB-DEEP & F606W,F814W &    754 & 0.095 & 0.98 & -3.95 & 21.824 & $21.770\pm0.024$ & 25.720 & $1.39\pm0.036$ \\
    KKR25 &                 KKR25 & F606W,F814W &     80 & 0.027 & 1.00 & -3.96 & 22.479 & $22.464\pm0.022$ & 26.428 & $1.93\pm0.047$ \\
    KK230 &                 KK230 & F606W,F814W &     47 & 0.043 & 1.00 & -4.00 & 22.492 & $22.467\pm0.026$ & 26.472 & $1.97\pm0.055$ \\
 E410-005 &            ESO410-005 & F606W,F814W &    194 & 0.042 & 1.06 & -4.05 & 22.493 & $22.469\pm0.042$ & 26.514 & $2.01\pm0.090$ \\
\enddata
\tablecomments{Color-dependent absolute magnitudes for the TRGB are
  taken from \citet{girardi08} isochrones. Mean colors are for the
  stars used to measure the TRGB, which are not necessarily all RGB
  stars, and include only stars within 0.2$^m$ of the TRGB.  $A_V$
  values are as reported by IRSA for coordinates in Table 1, with the
  exception of M82.  Extinction corrections from $A_V$ to the observed
  filters are adopted from \citet{girardi08}, as described in the
  text.  $m_{TRGB}$ was measured in the least crowded region of each
  galaxy. Measured distances for A0952$+$69, BK3N, and Holmberg IX are
  dominated by outer M81 stars, rather than RGB stars associated with
  the named galaxies. Listed uncertainties are dominated by
  photometric uncertainties and by stochasticity in the number of
  stars near the tip; systematic uncertainties (due to uncertainties
  in the assumed TRGB absolute magnitudes and extinction) are likely
  to be much larger, but are not included in the listed uncertainties.
  [a] The distance to MCG9-20-1 is ambiguous, as it was not clear if
  the observed tip was for the RGB or AGB; the true distance modulus
  may potentially be significantly fainter. }
\label{trgbtable}
\end{deluxetable}

\clearpage
\end{landscape}
%------------------

%--------------------------------------------------------------------

\ifprintfig

\begin{figure}[p]
\centerline{
\includegraphics[width=3.25in]{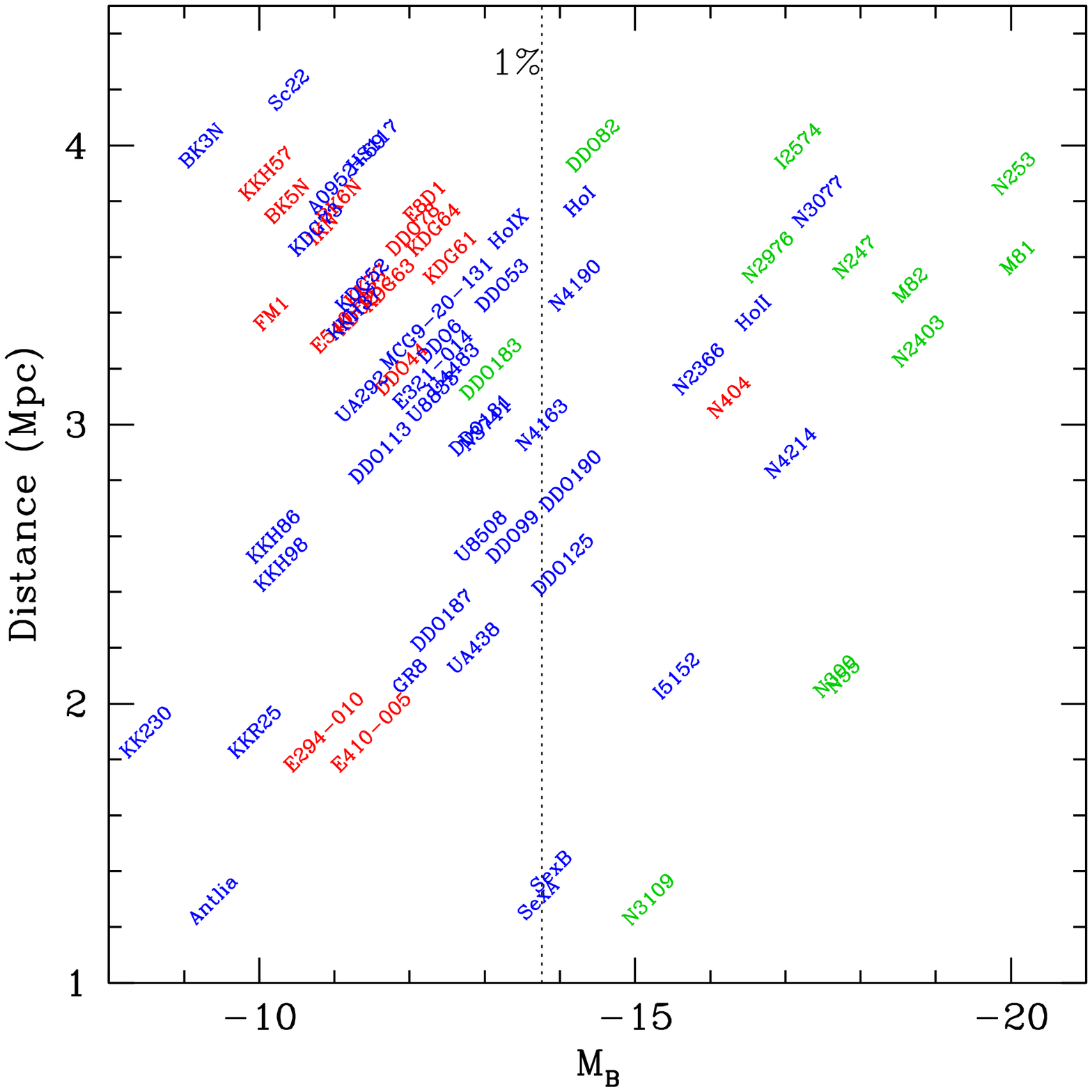}
\includegraphics[width=3.25in]{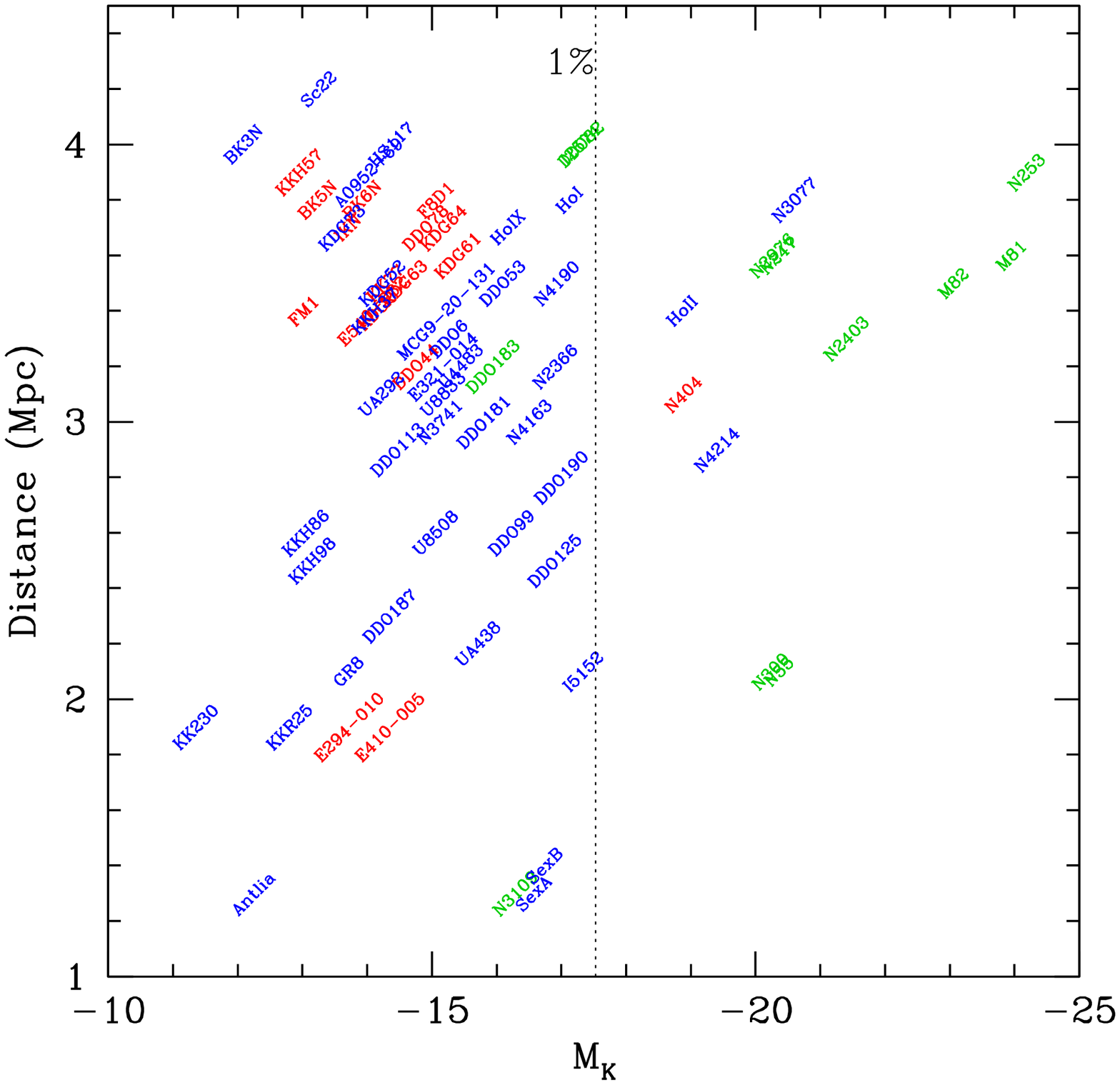}
}
\caption{The distribution of the ANGST sample galaxies in distance and
absolute magnitude (left: $B$-band; right: $K$-band).  Points are
color-coded by morphological type (red: $T\le0$; green: $1\le T \le
9$, blue: T=10).  The majority of early type galaxies are dwarf
ellipticals in the dense M81 group. The galaxies to the left of the
vertical line contain less than 1\% of the integrated $B$- or $K$-band
luminosity in the survey volume.  $K$-band absolute magnitudes have
been estimated for some of the low luminosity galaxies
\citet{mannucci01}. \label{magdistfig}}
\end{figure}
\vfill
\clearpage

%-------------------
\begin{figure}[p]
\centerline{
\includegraphics[height=6in]{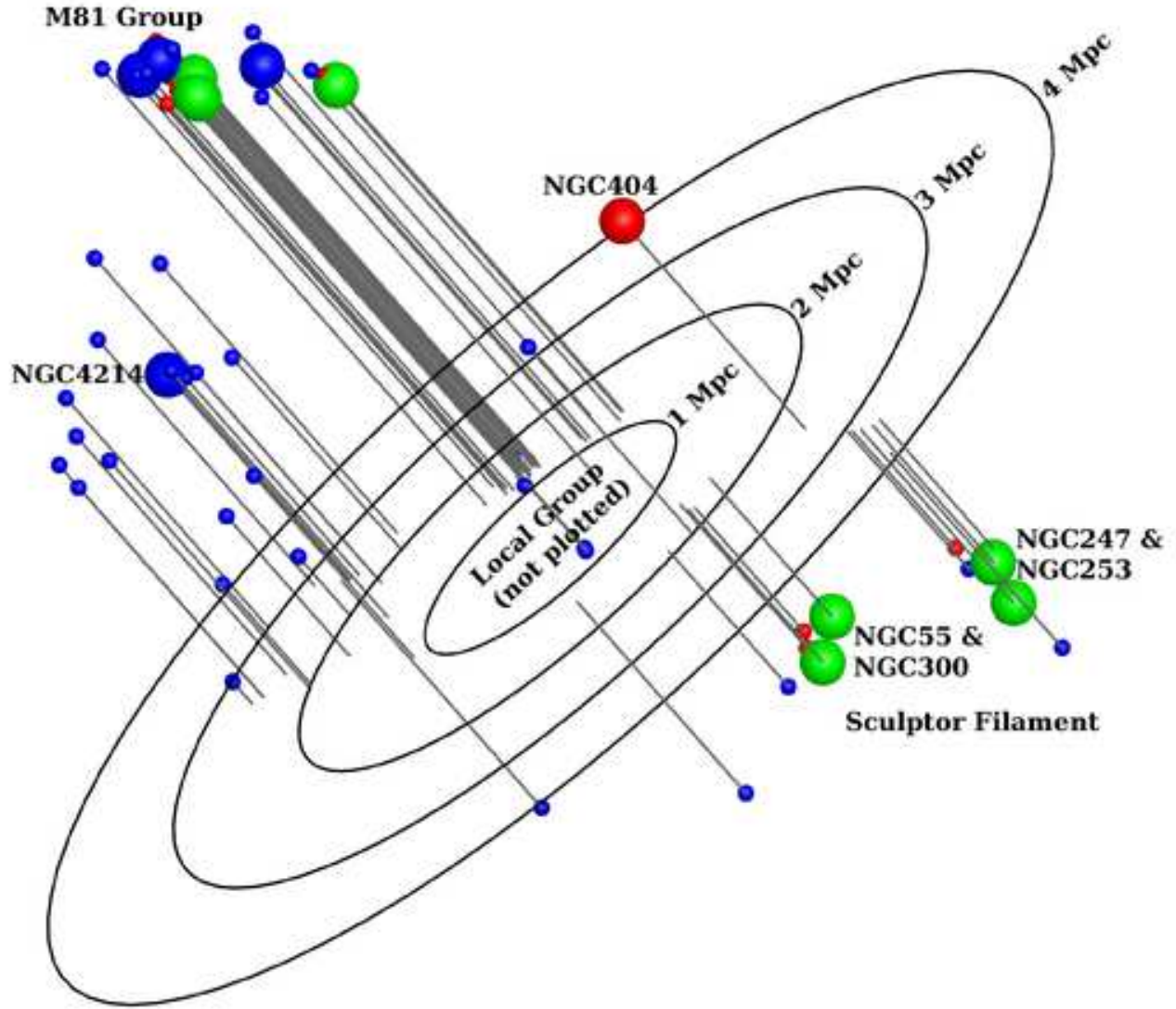}
}
\caption{The 3-dimensional space distribution of the ANGST sample
galaxies.  Galaxies are color-coded by morphological type (red:
$T\le0$; green: $1\le T \le 9$, blue: T=10), as in
Figure~\ref{magdistfig}.  Larger symbols indicate galaxies brighter than $M_B=-16.0$.  
The large clump of galaxies in the upper
left is the rich M81 group.  The two clumps under the plane along the
right hand axis are the closer NGC~55/NGC~300 clump and the more
distant NGC~247/NGC~253 subclump along the Sculptor filament.  The bright galaxy
at the center of the dwarf cloud on the left is NGC~4214, and the bright isolated 
early-type galaxy to the right is NGC~404.  The circles
are drawn at intervals of $1\Mpc$, along the equatorial plane.  Distances are taken
from Table~\ref{trgbtable}.  Note that not all galaxies within $4\Mpc$ are plotted, due to ANGST's $|b|\!>\!20\degree$ selection criteria.
\label{3Dfig}}
\end{figure}
\vfill
\clearpage

%-----
%\input{newoverlays.tex}
%------------------
\begin{figure}[p]
\centerline{
\includegraphics[width=1.562in]{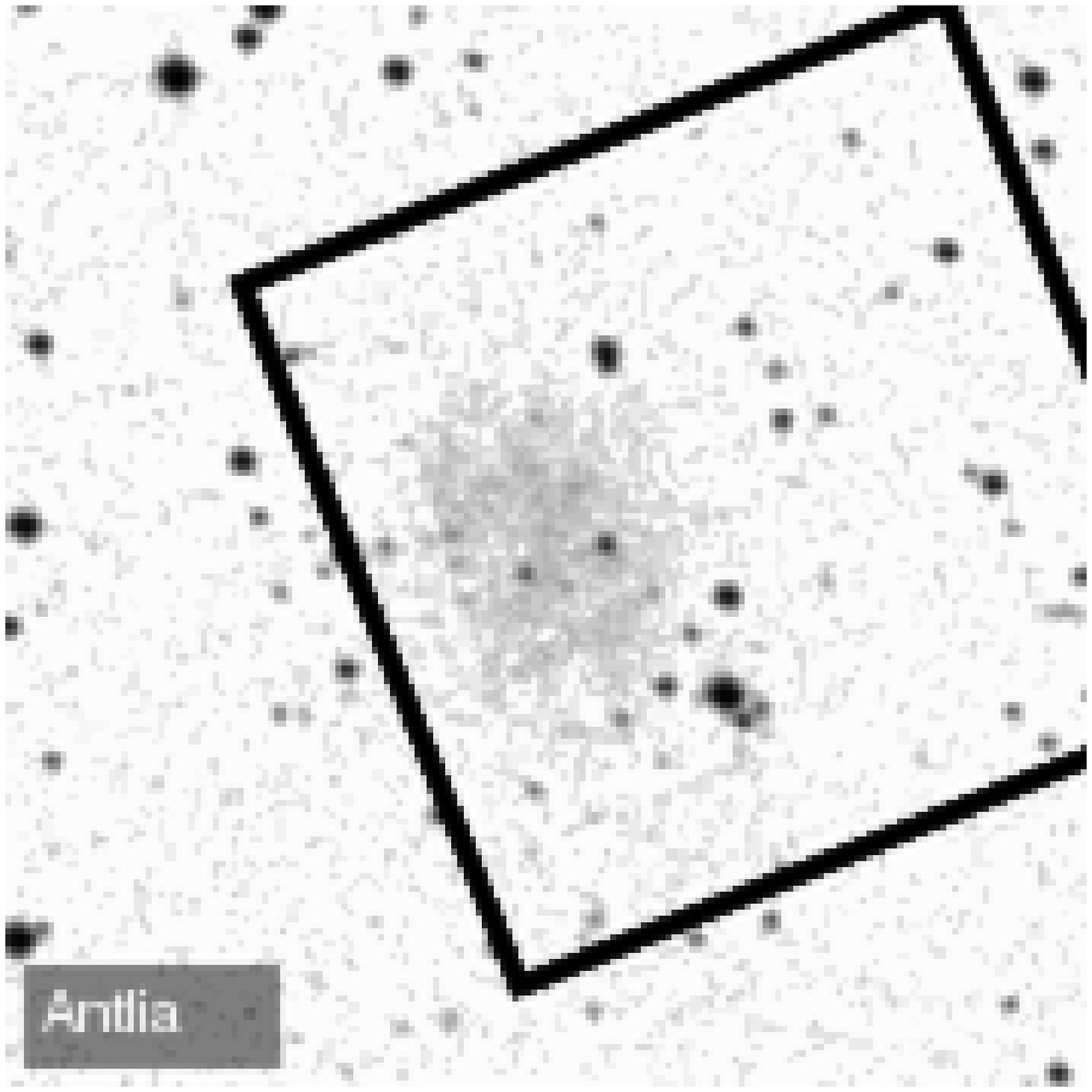}
\includegraphics[width=1.562in]{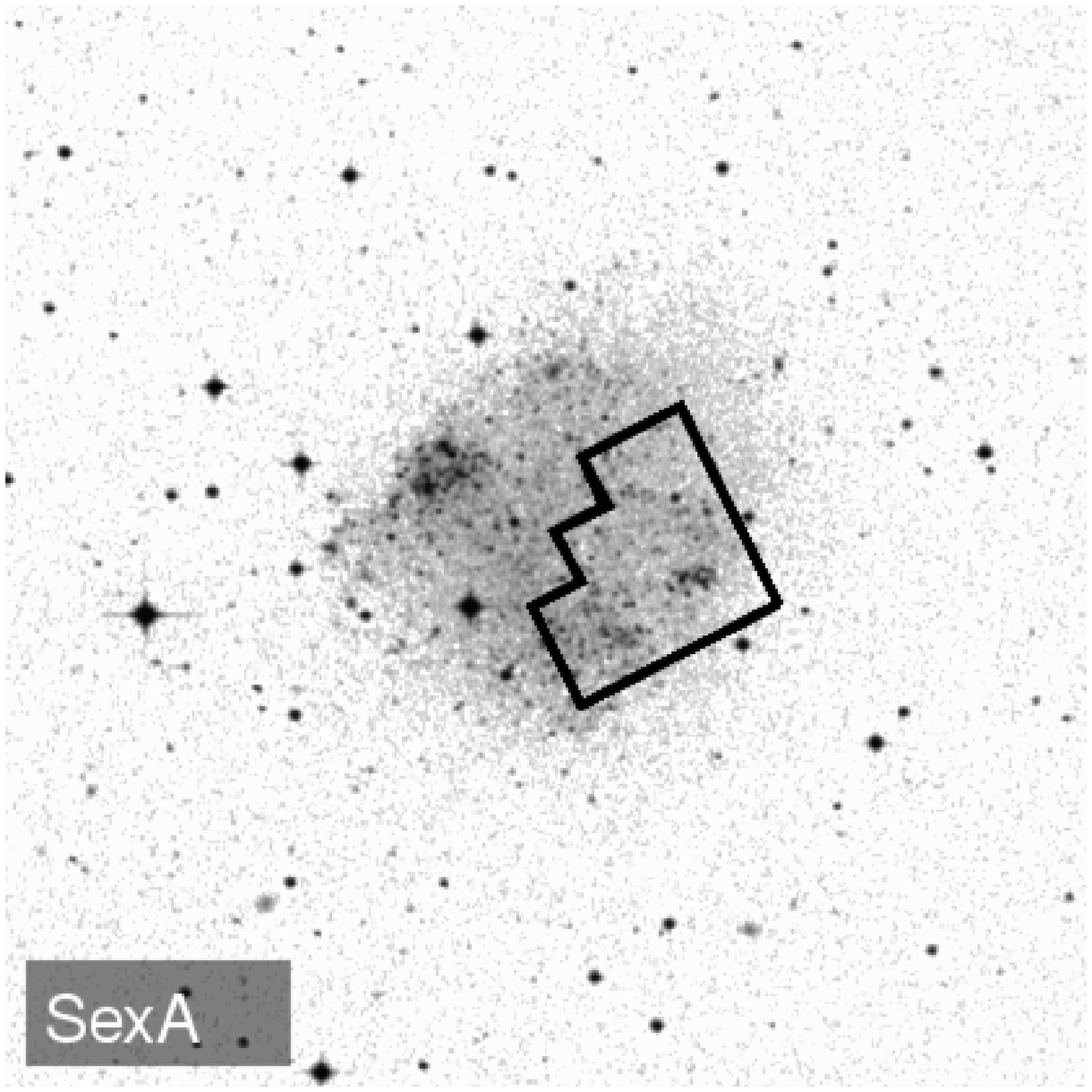}
\includegraphics[width=1.562in]{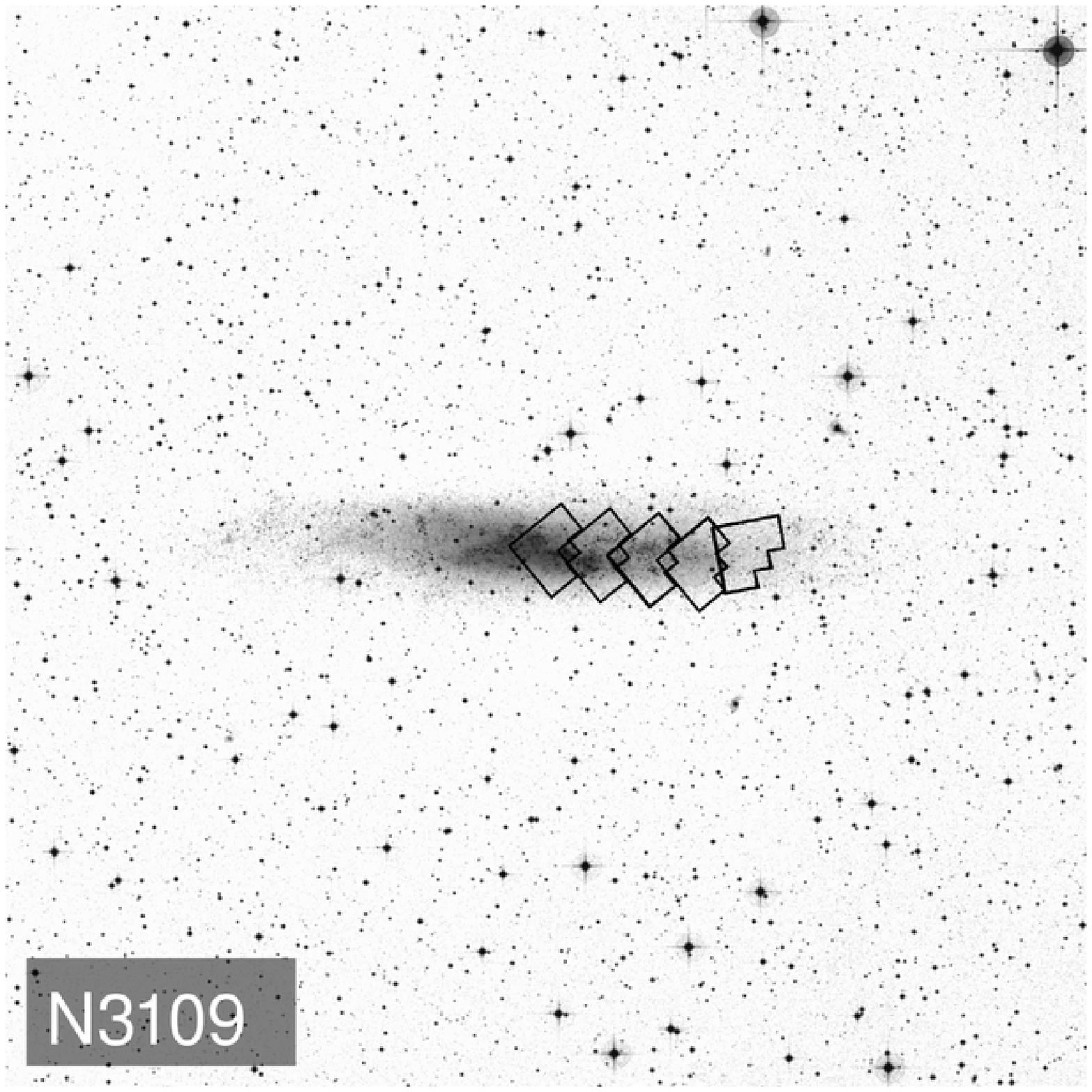}
\includegraphics[width=1.562in]{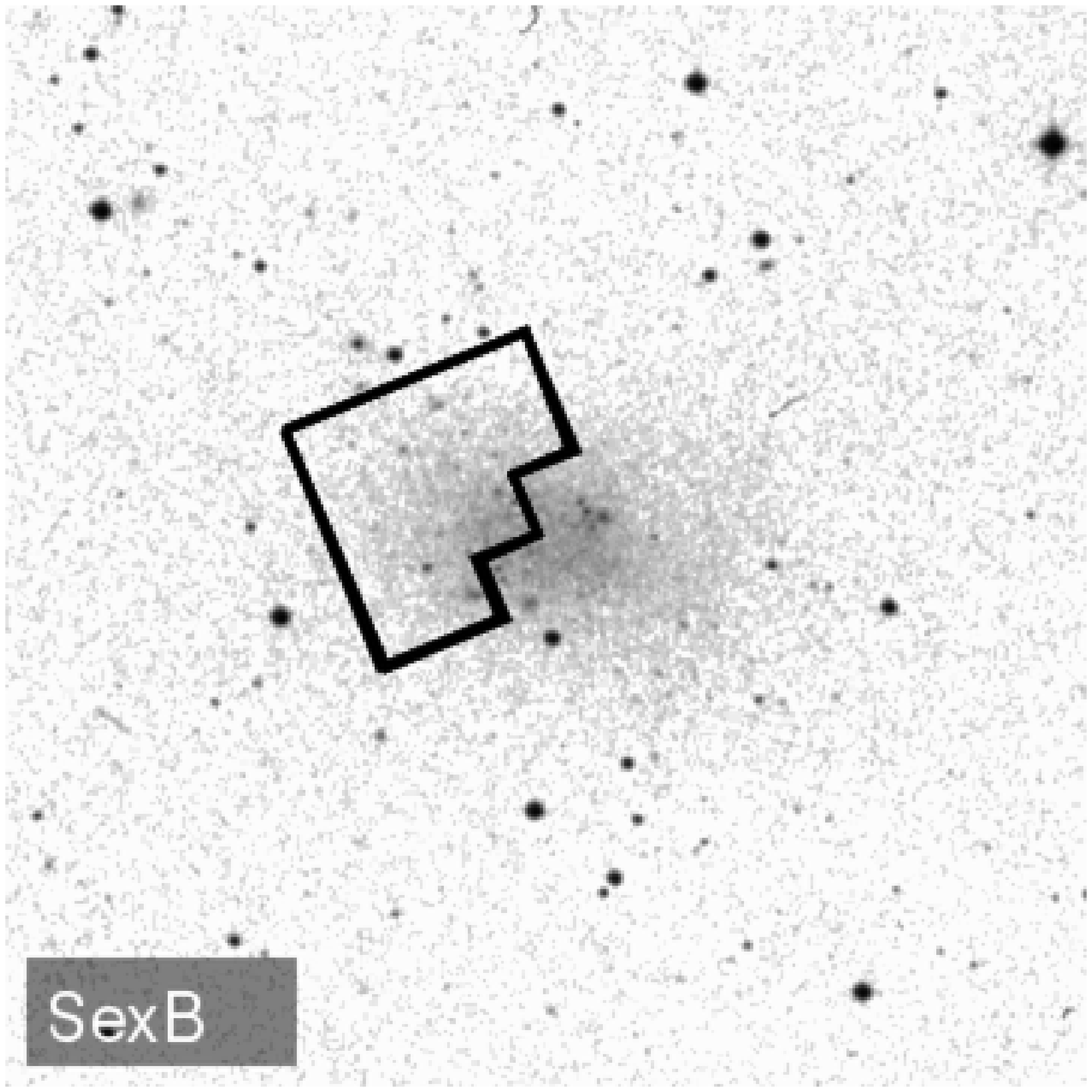}
}
\centerline{
\includegraphics[width=1.562in]{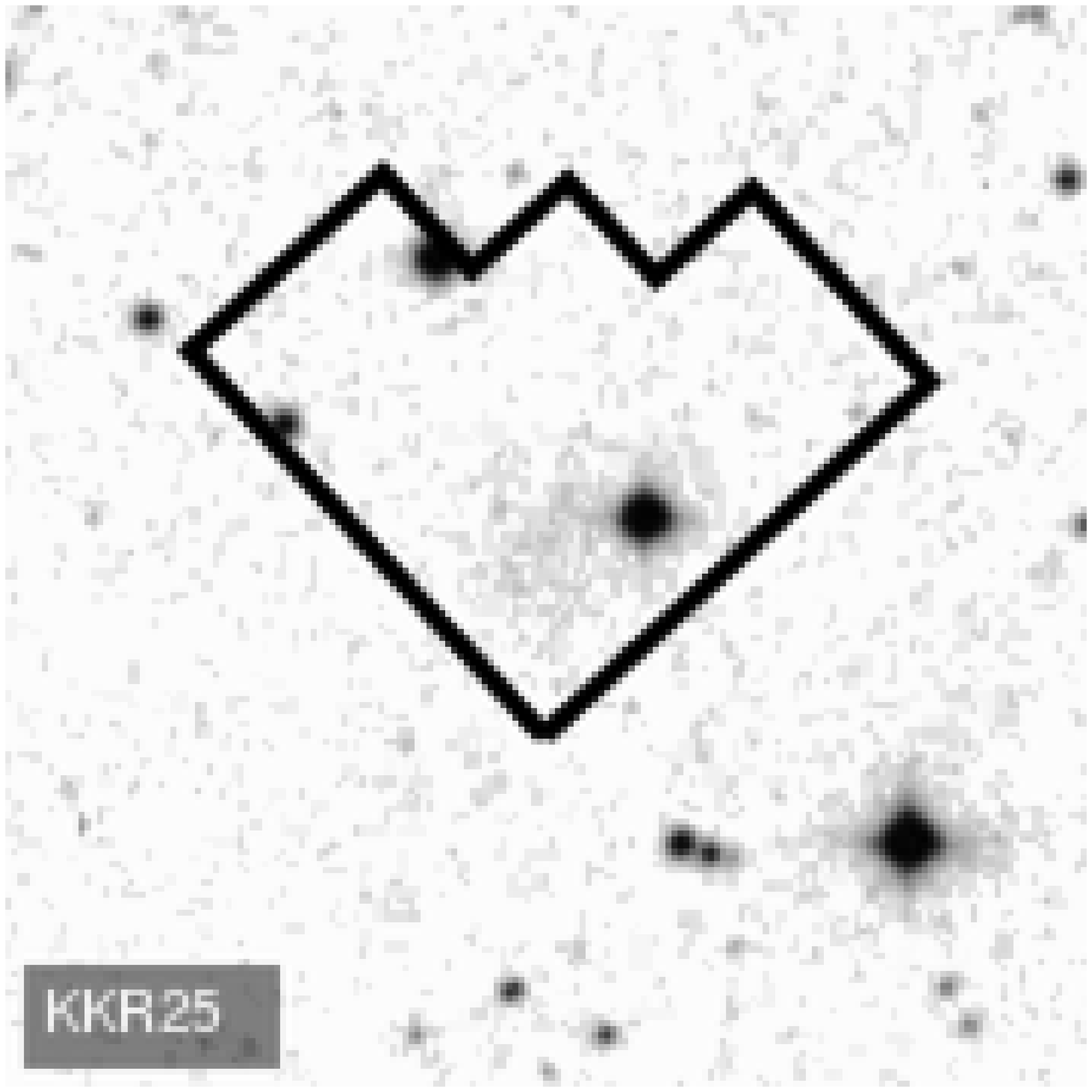}
\includegraphics[width=1.562in]{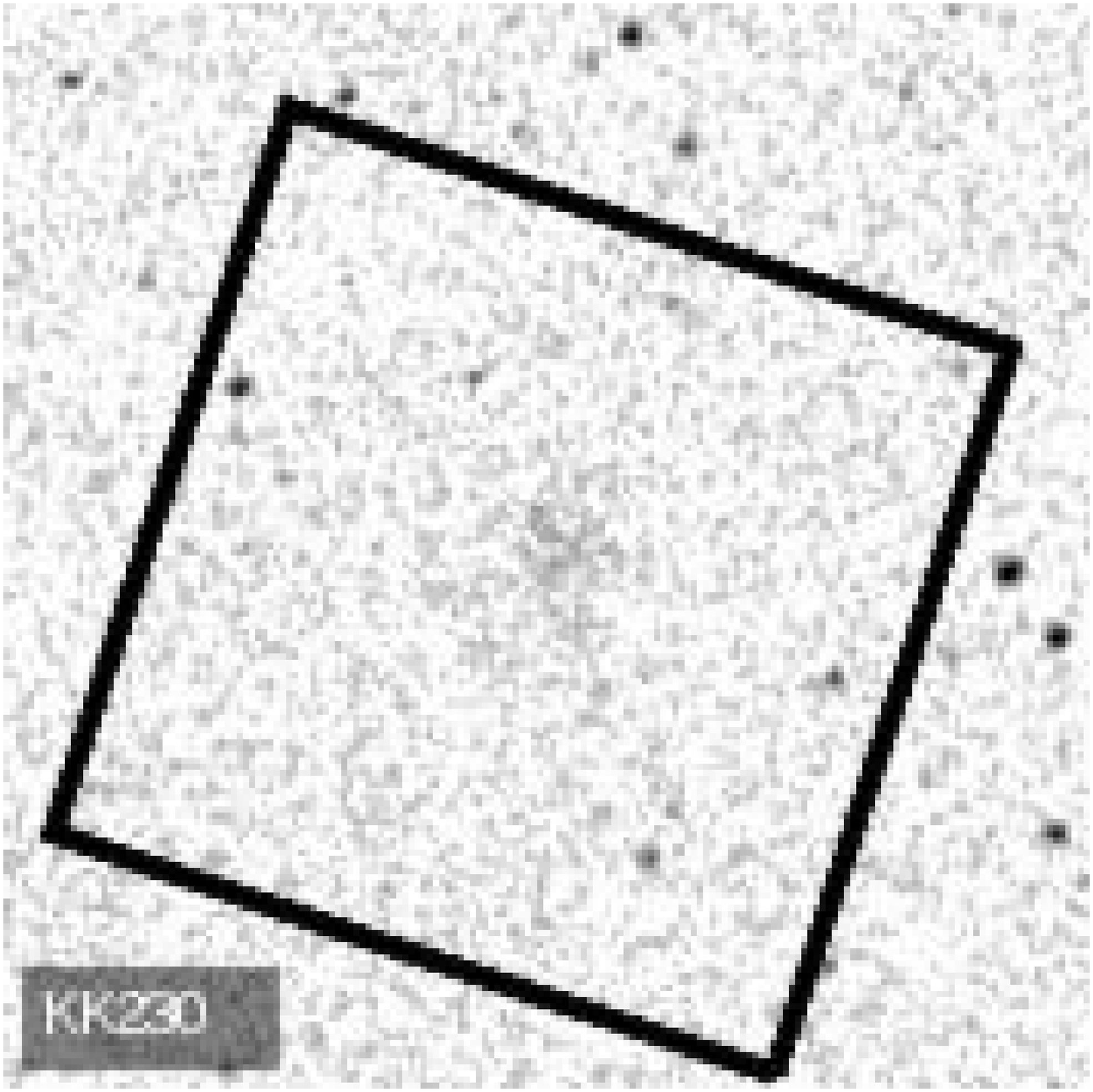}
\includegraphics[width=1.562in]{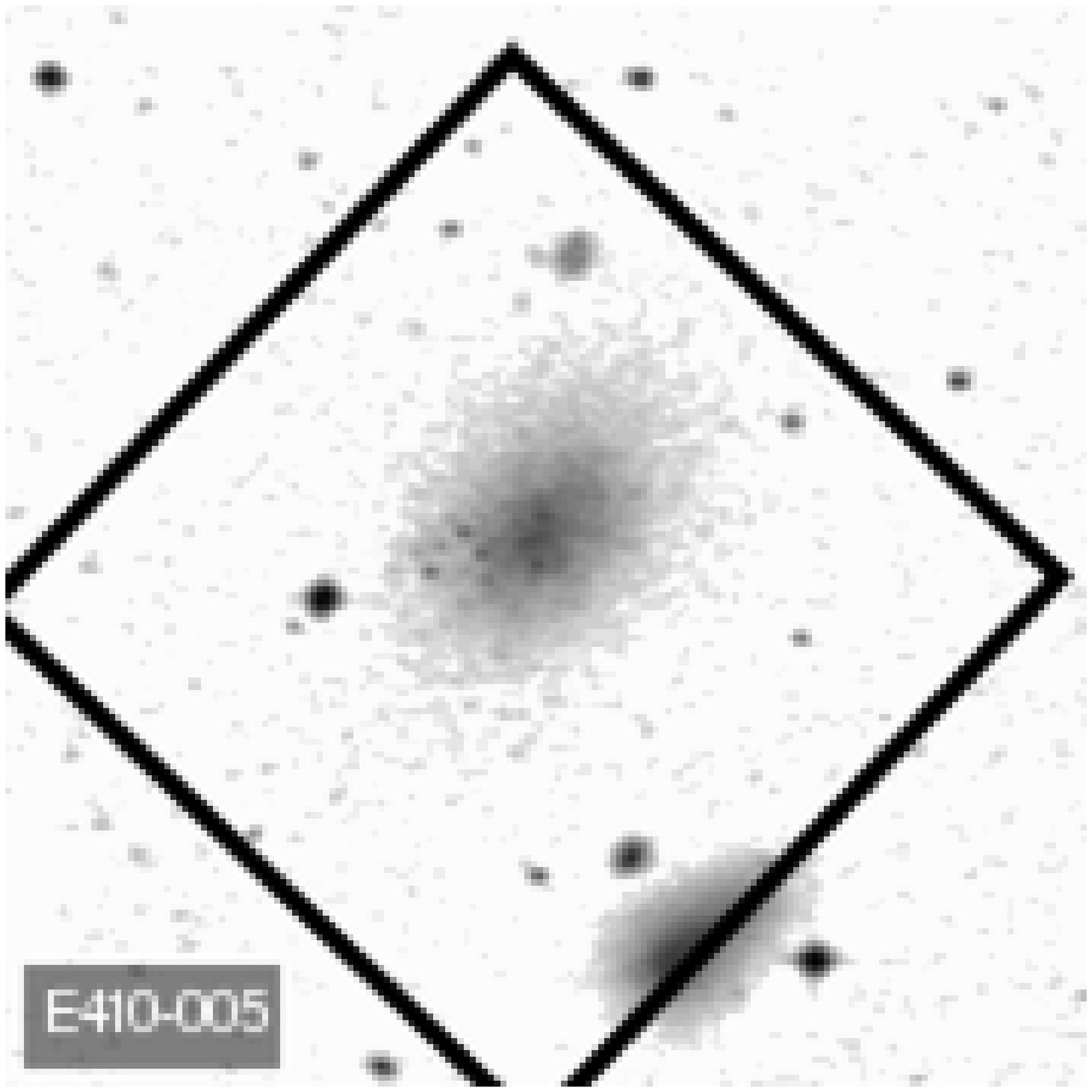}
\includegraphics[width=1.562in]{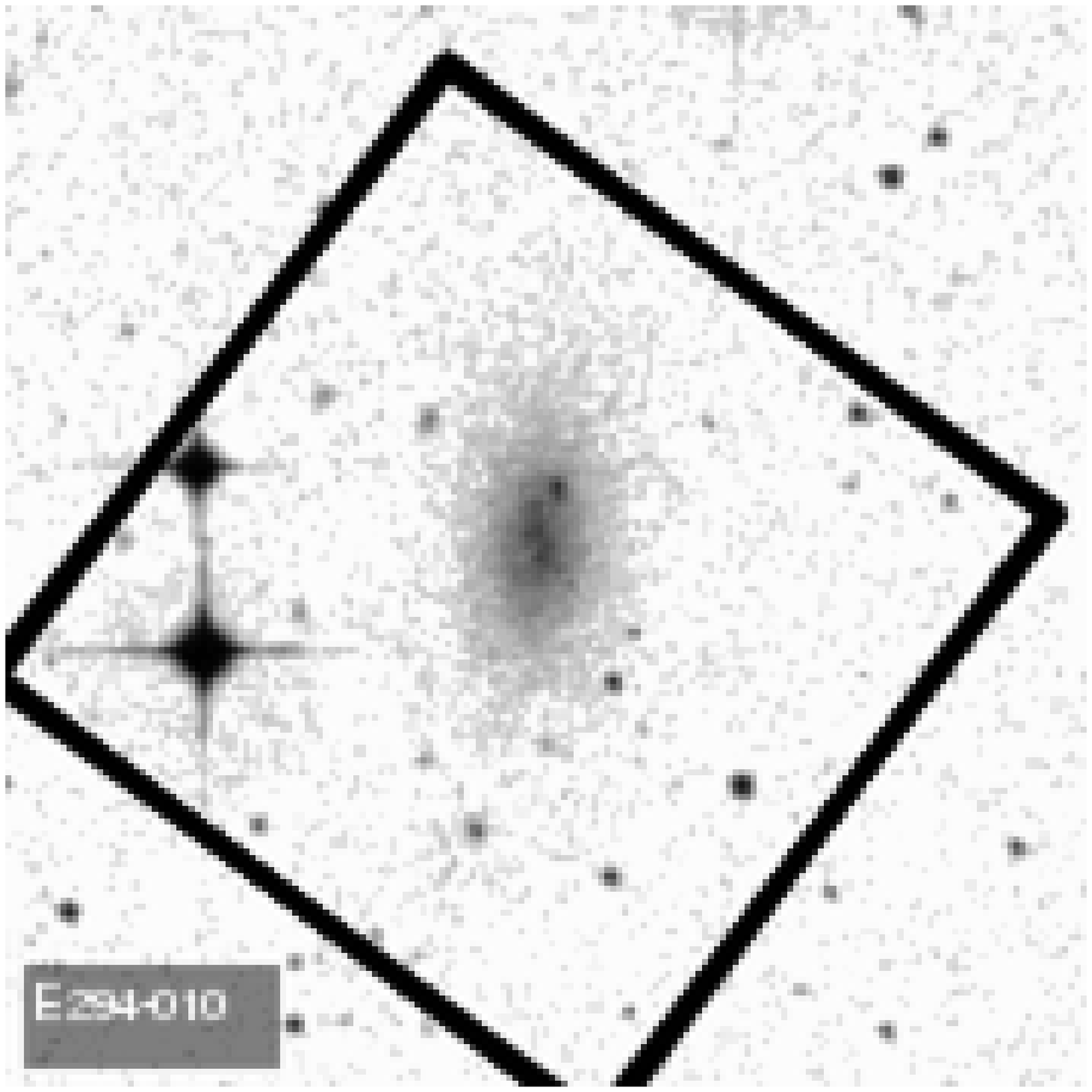}
}
\centerline{
\includegraphics[width=1.562in]{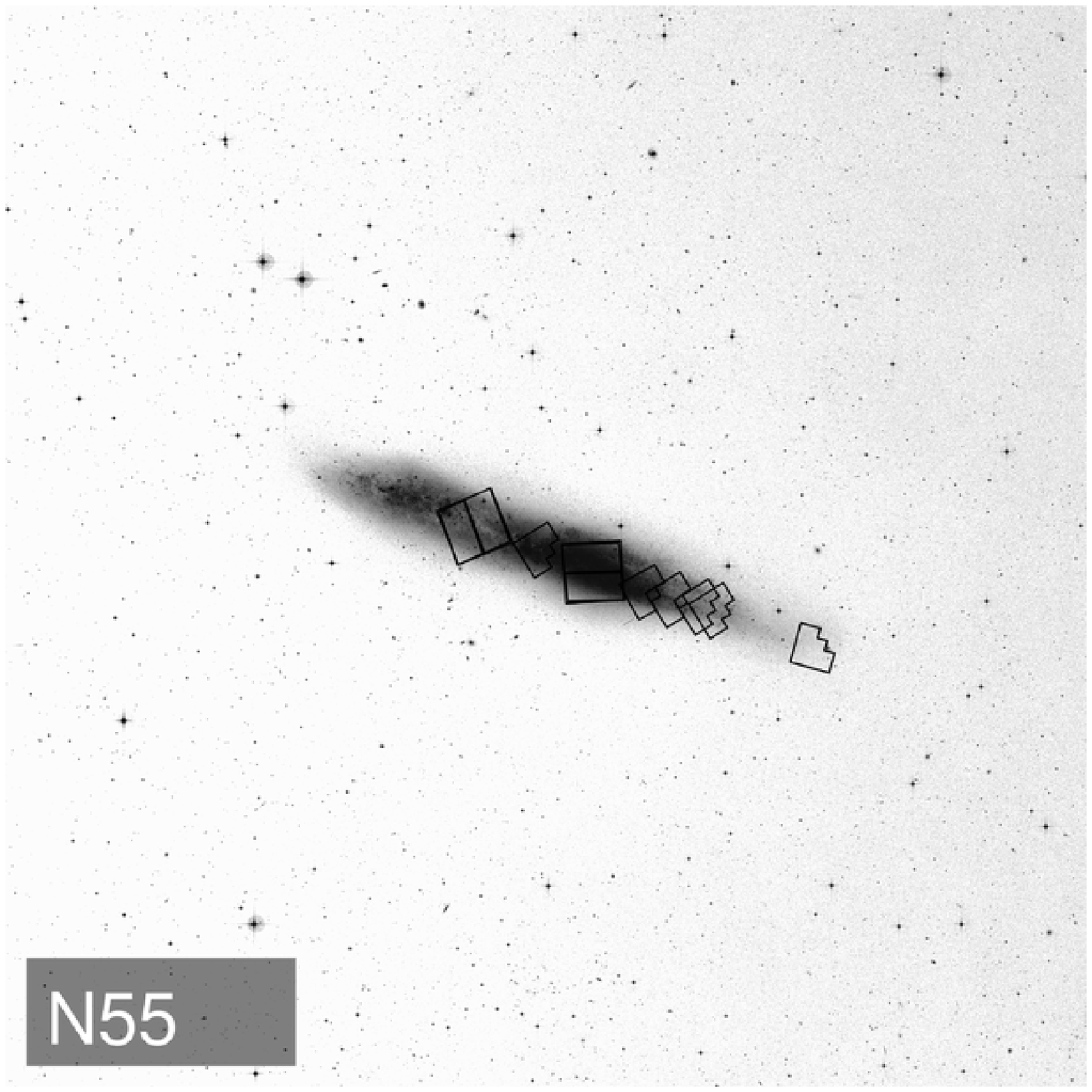}
\includegraphics[width=1.562in]{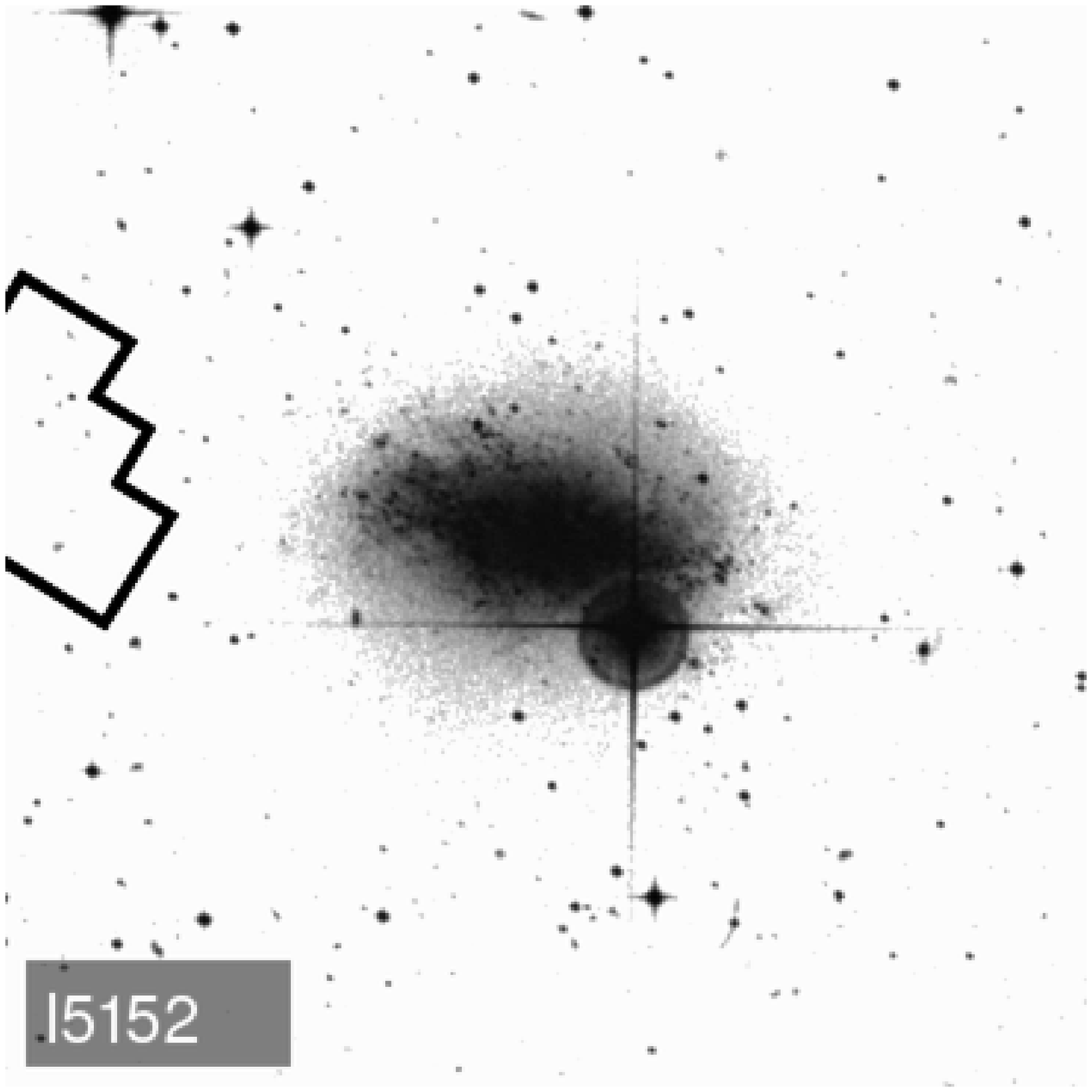}
\includegraphics[width=1.562in]{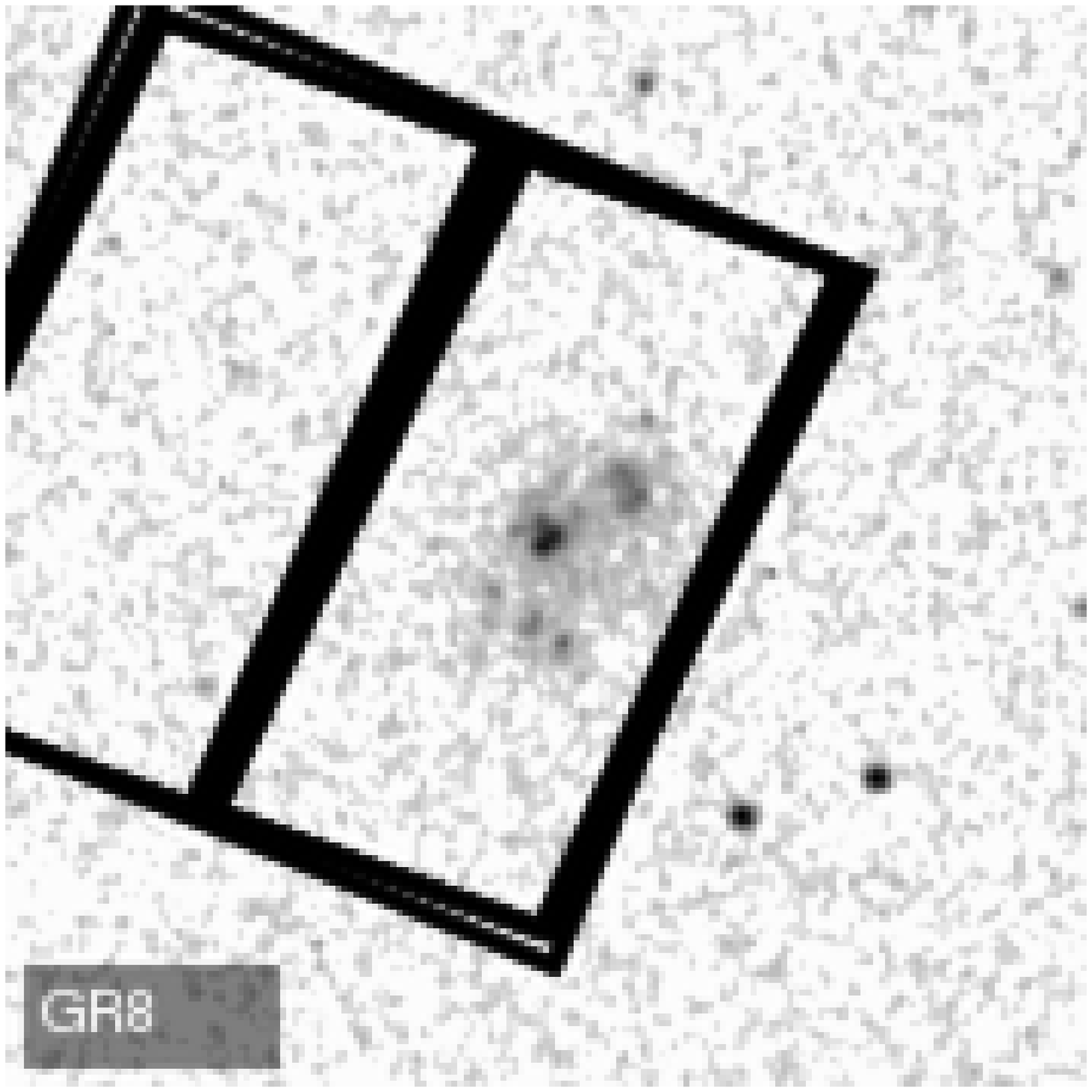}
\includegraphics[width=1.562in]{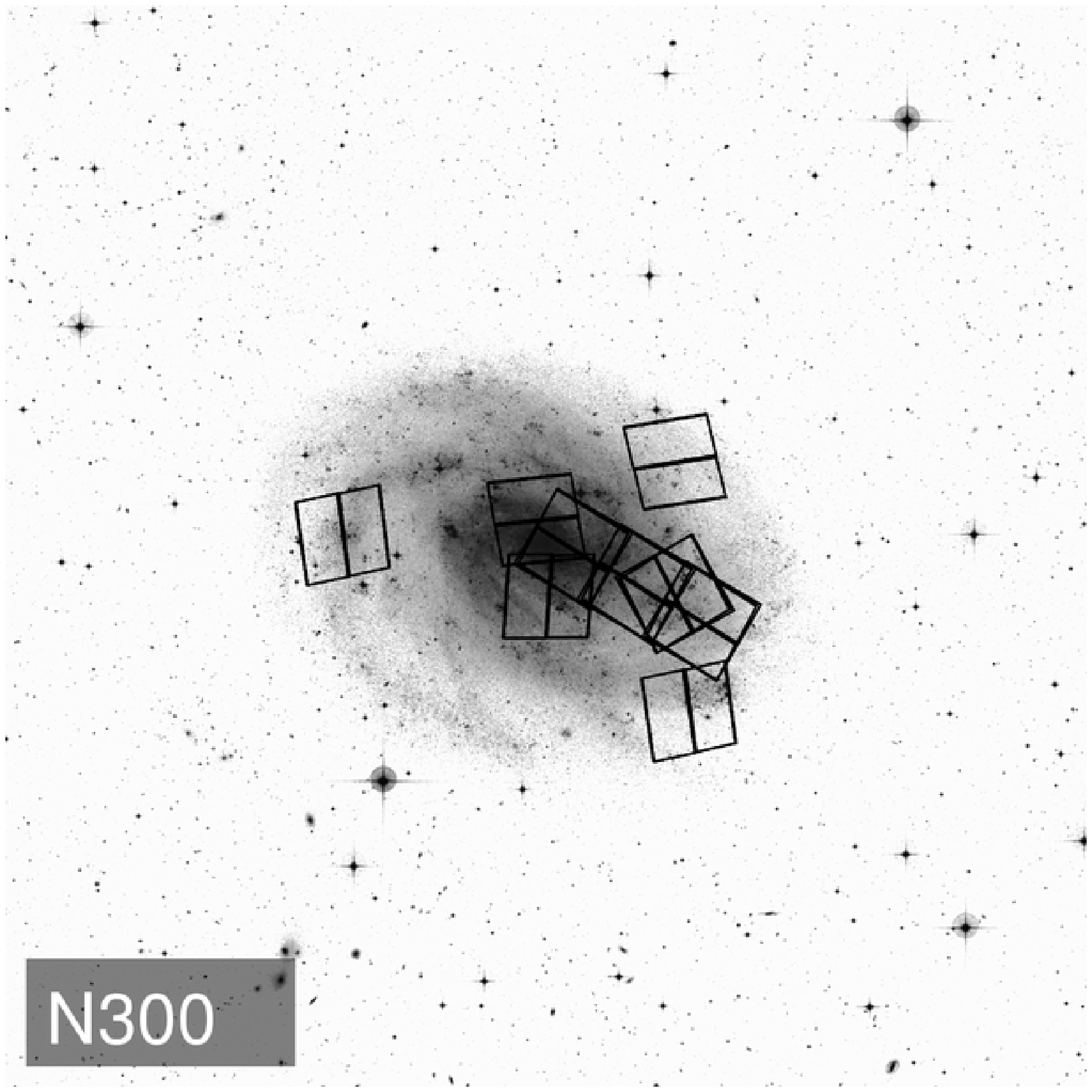}
}
\centerline{
\includegraphics[width=1.562in]{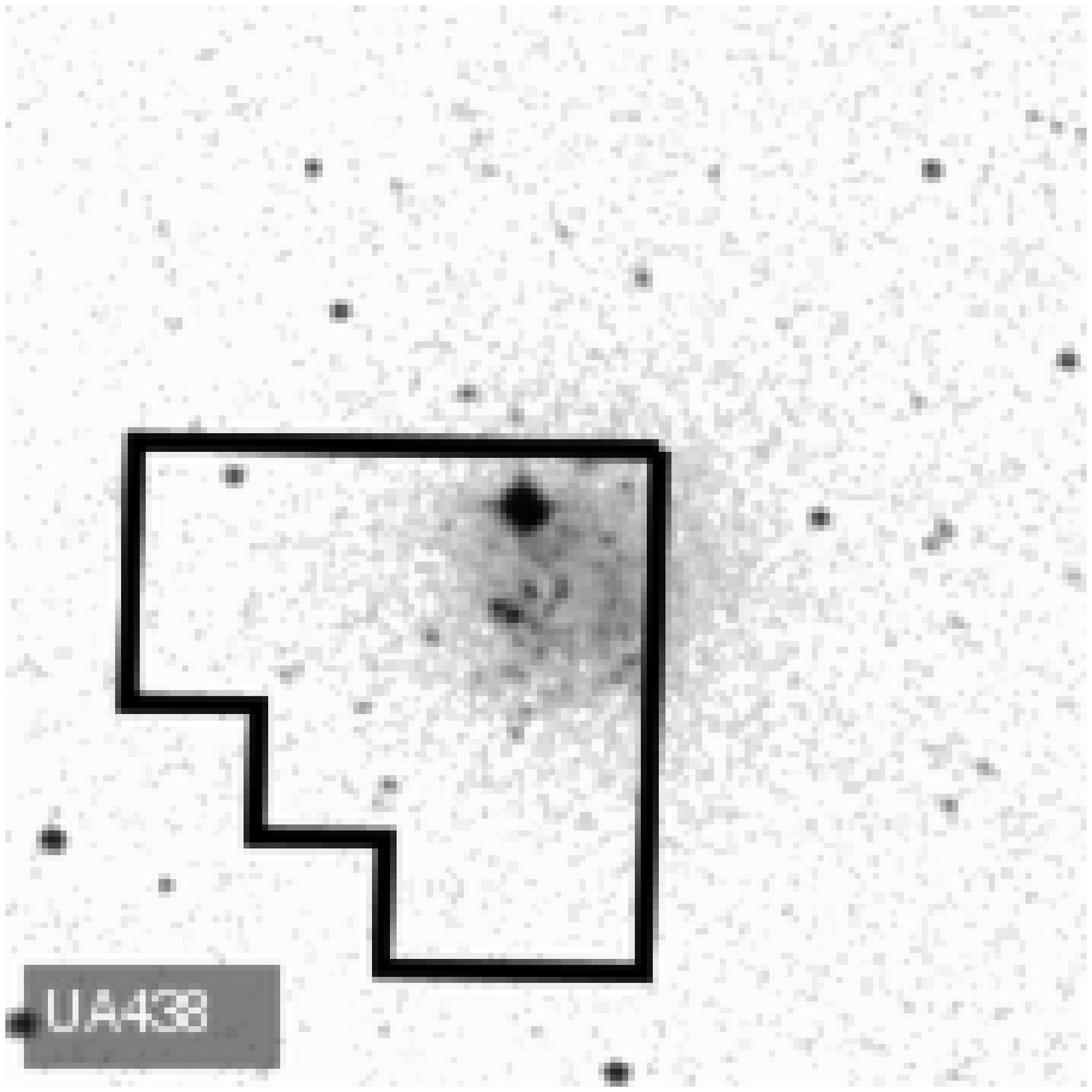}
\includegraphics[width=1.562in]{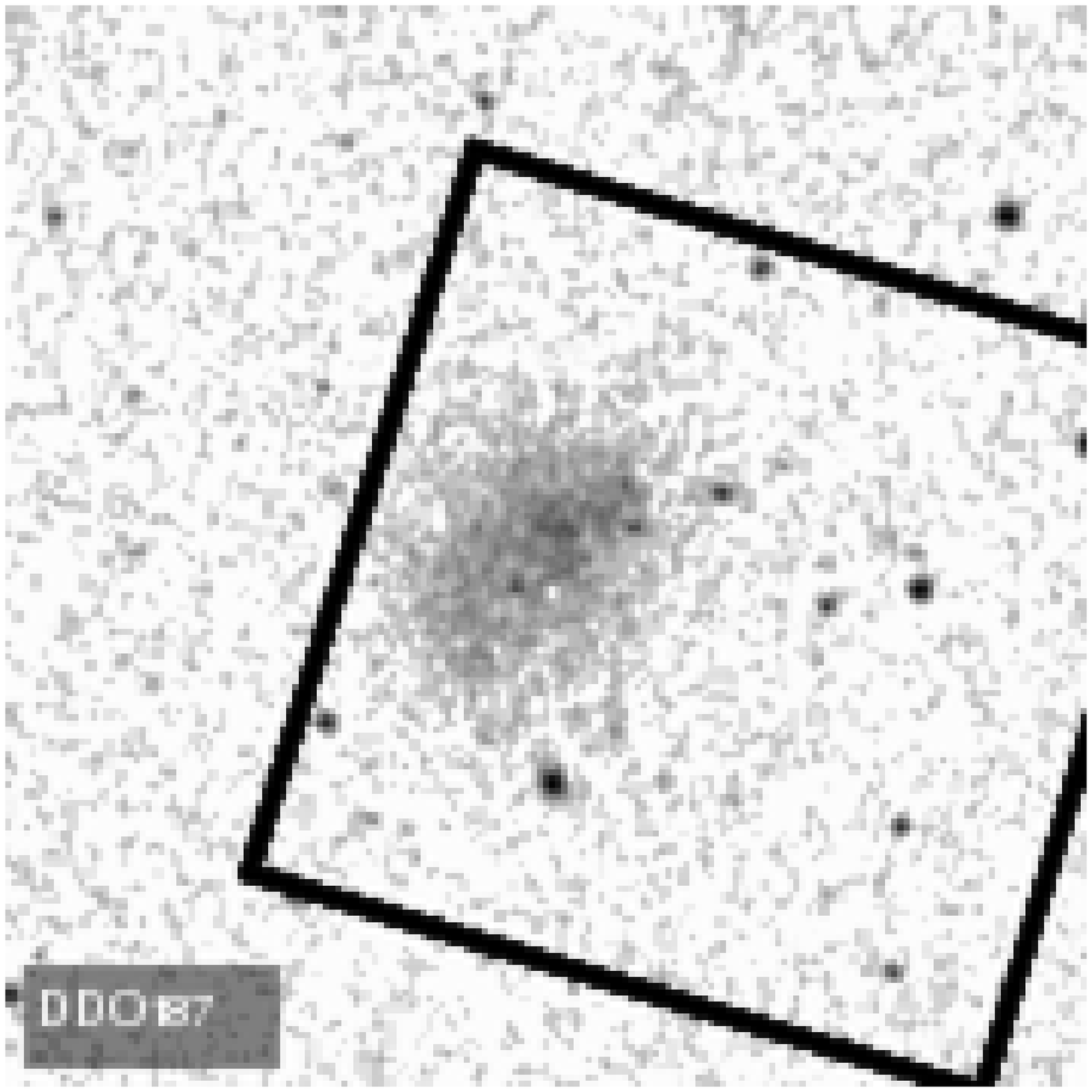}
\includegraphics[width=1.562in]{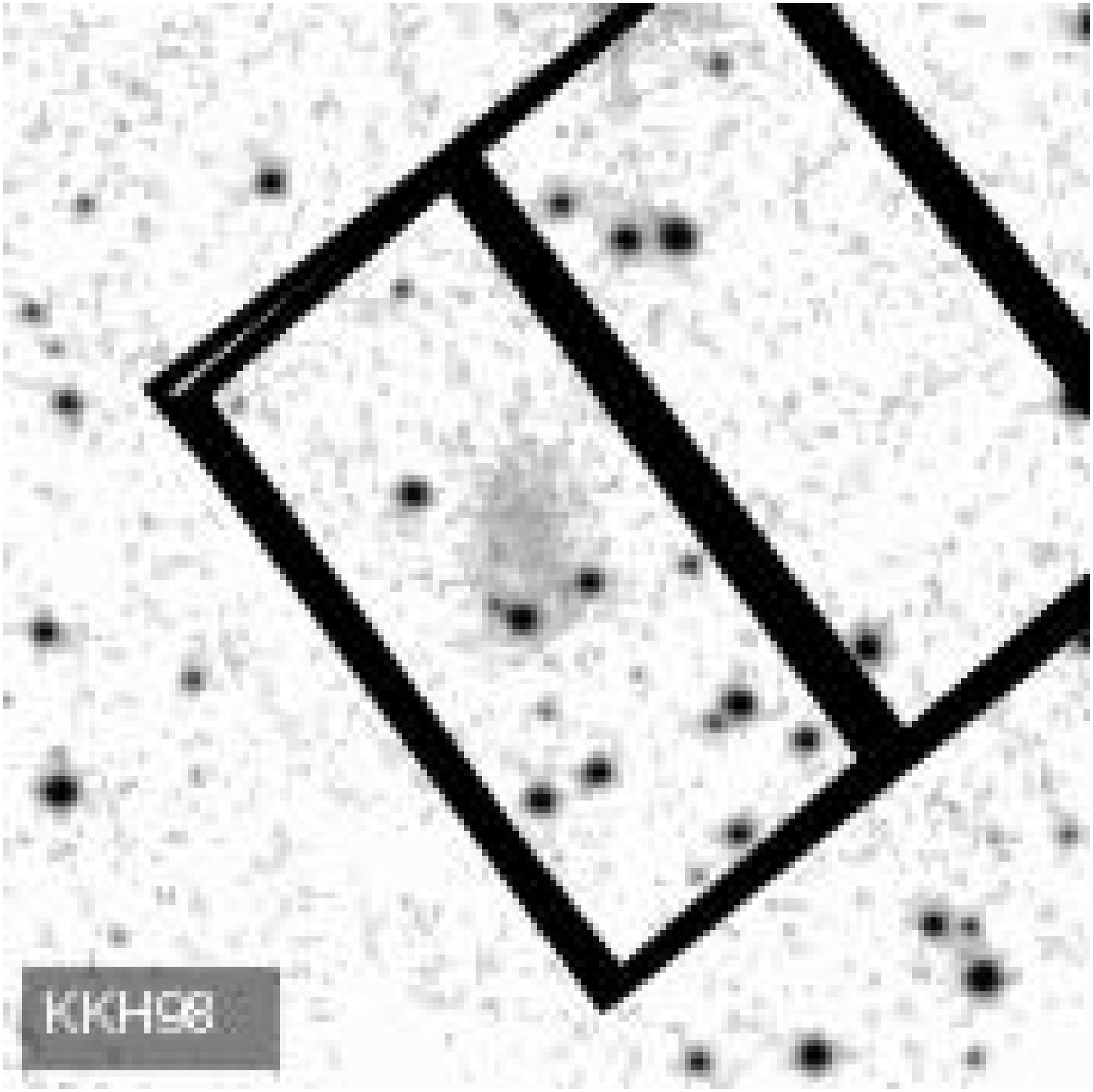}
\includegraphics[width=1.562in]{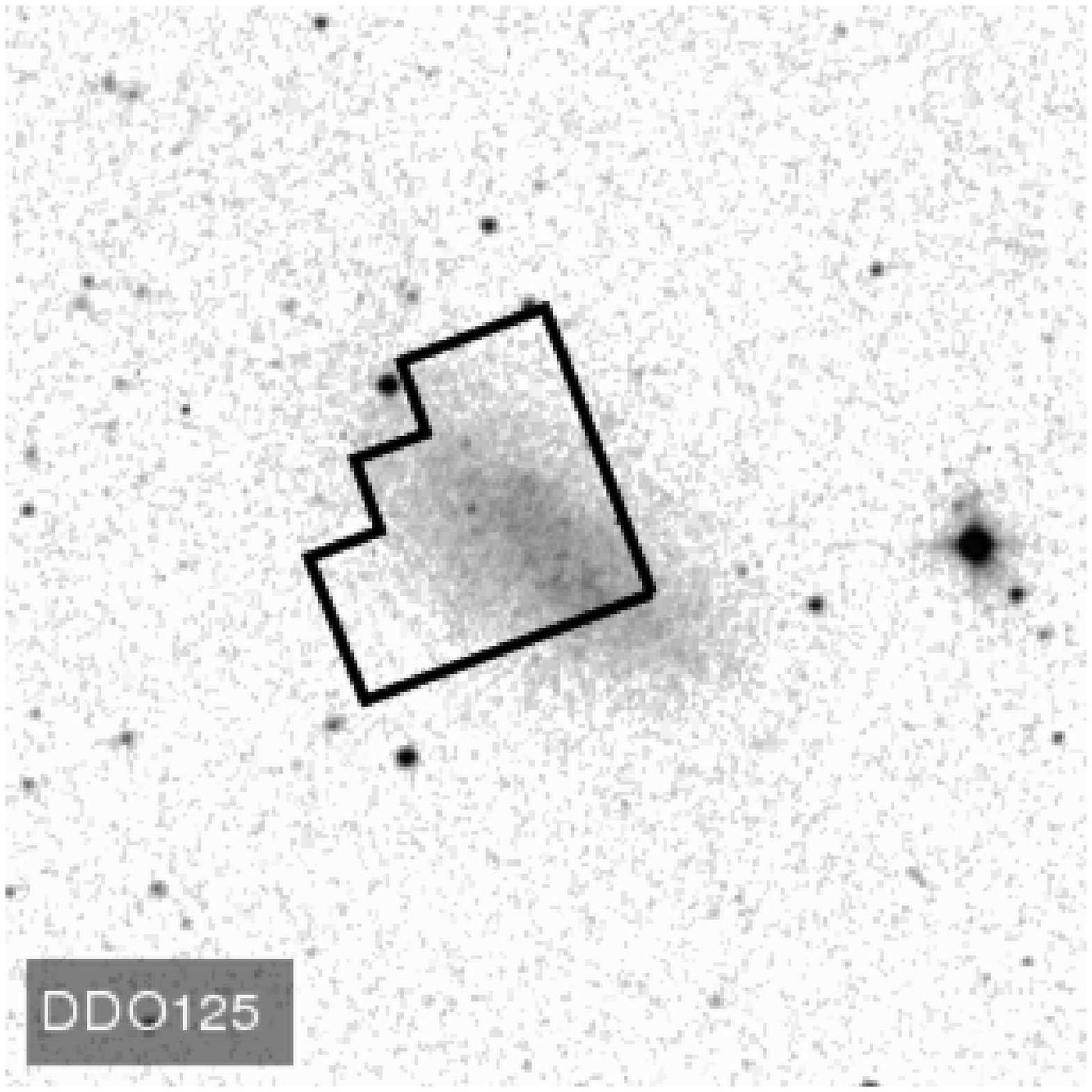}
}
\centerline{
\includegraphics[width=1.562in]{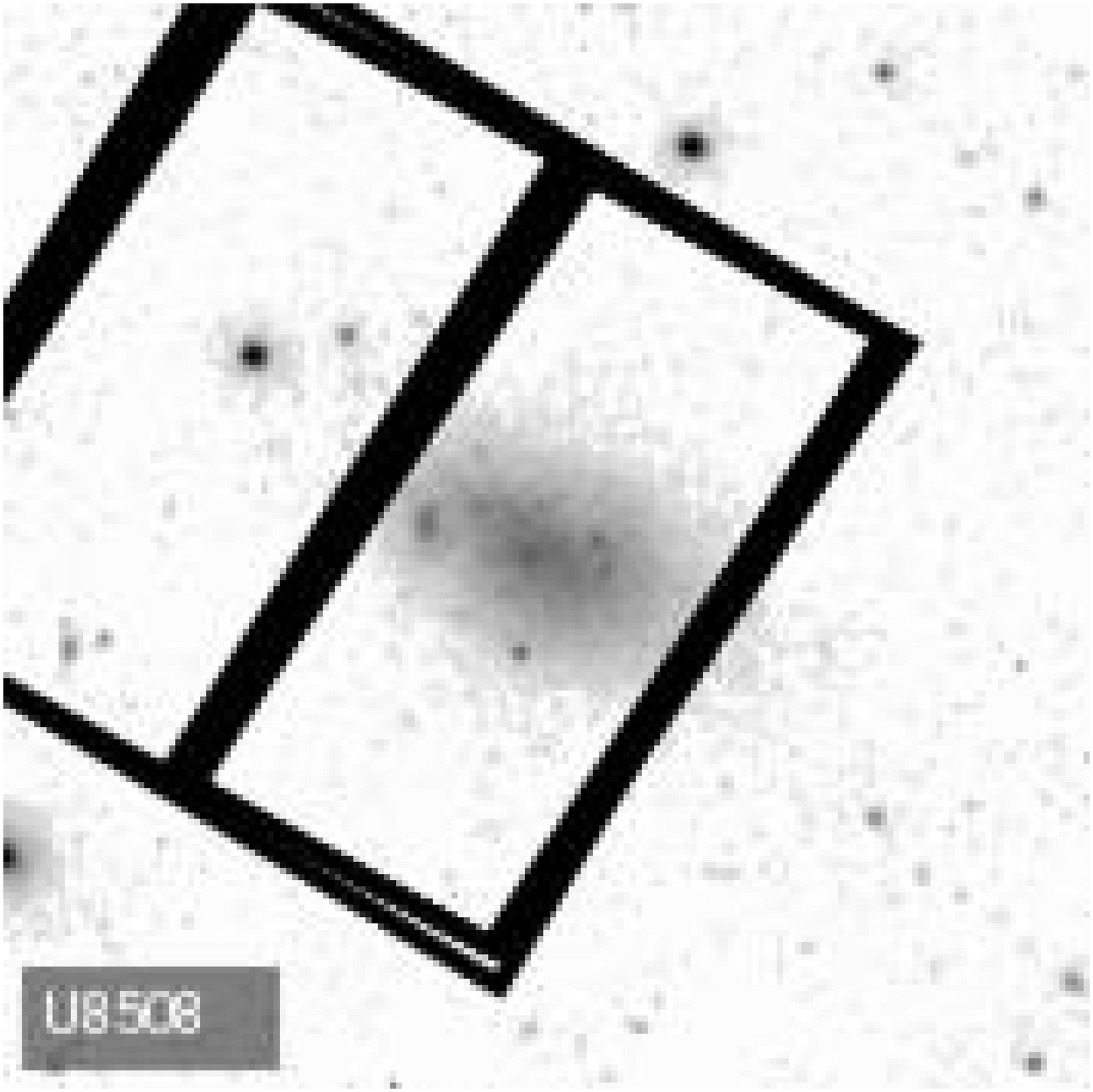}
\includegraphics[width=1.562in]{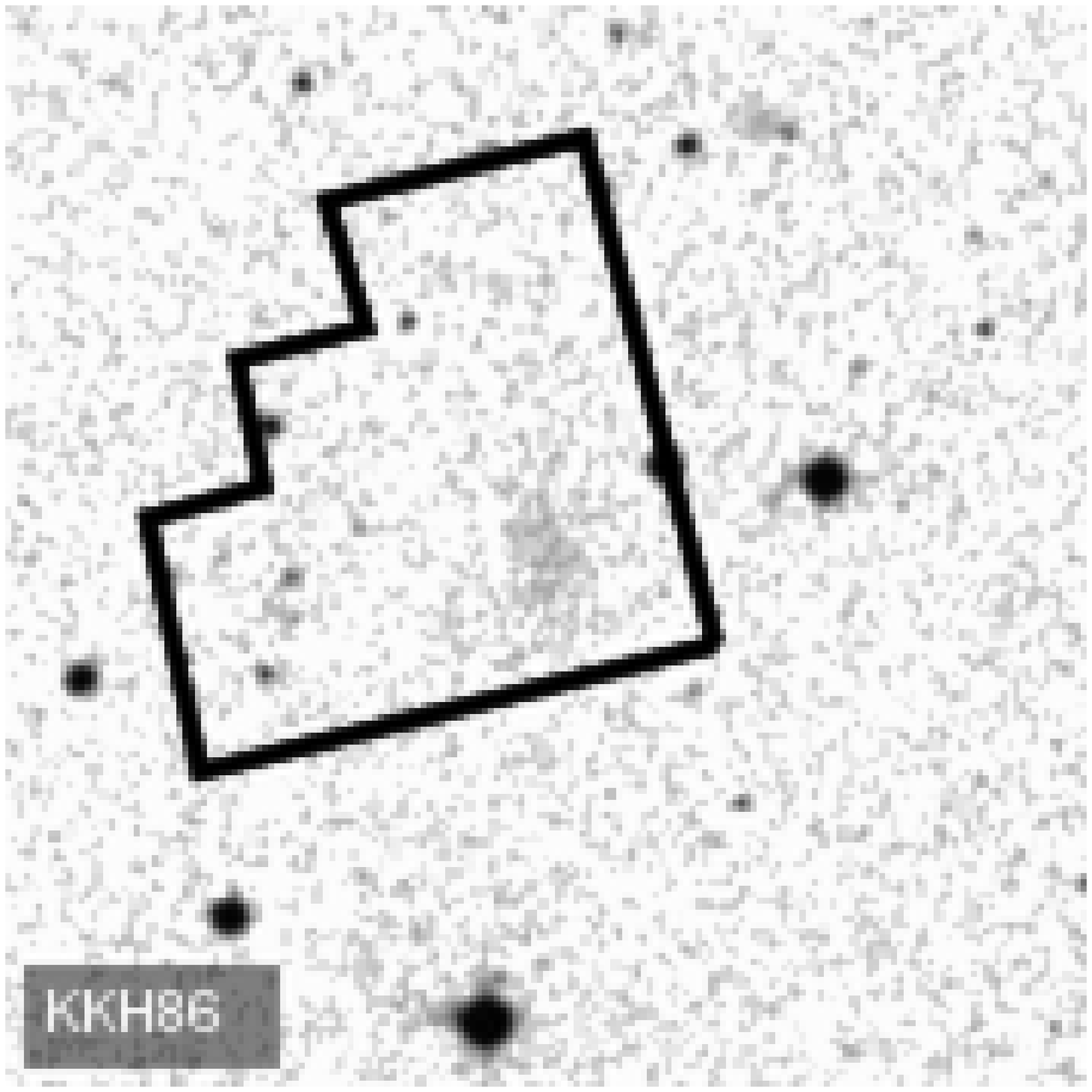}
\includegraphics[width=1.562in]{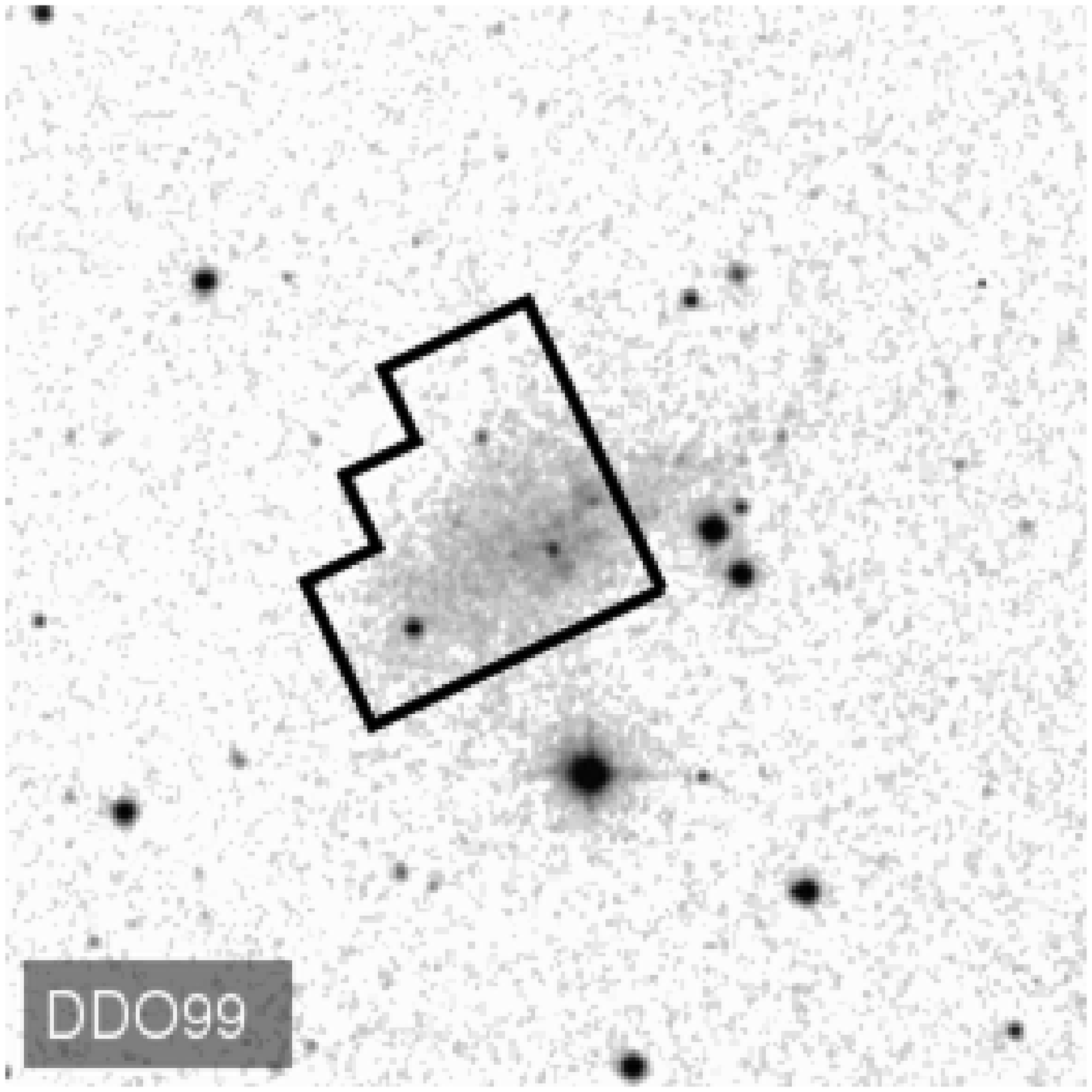}
\includegraphics[width=1.562in]{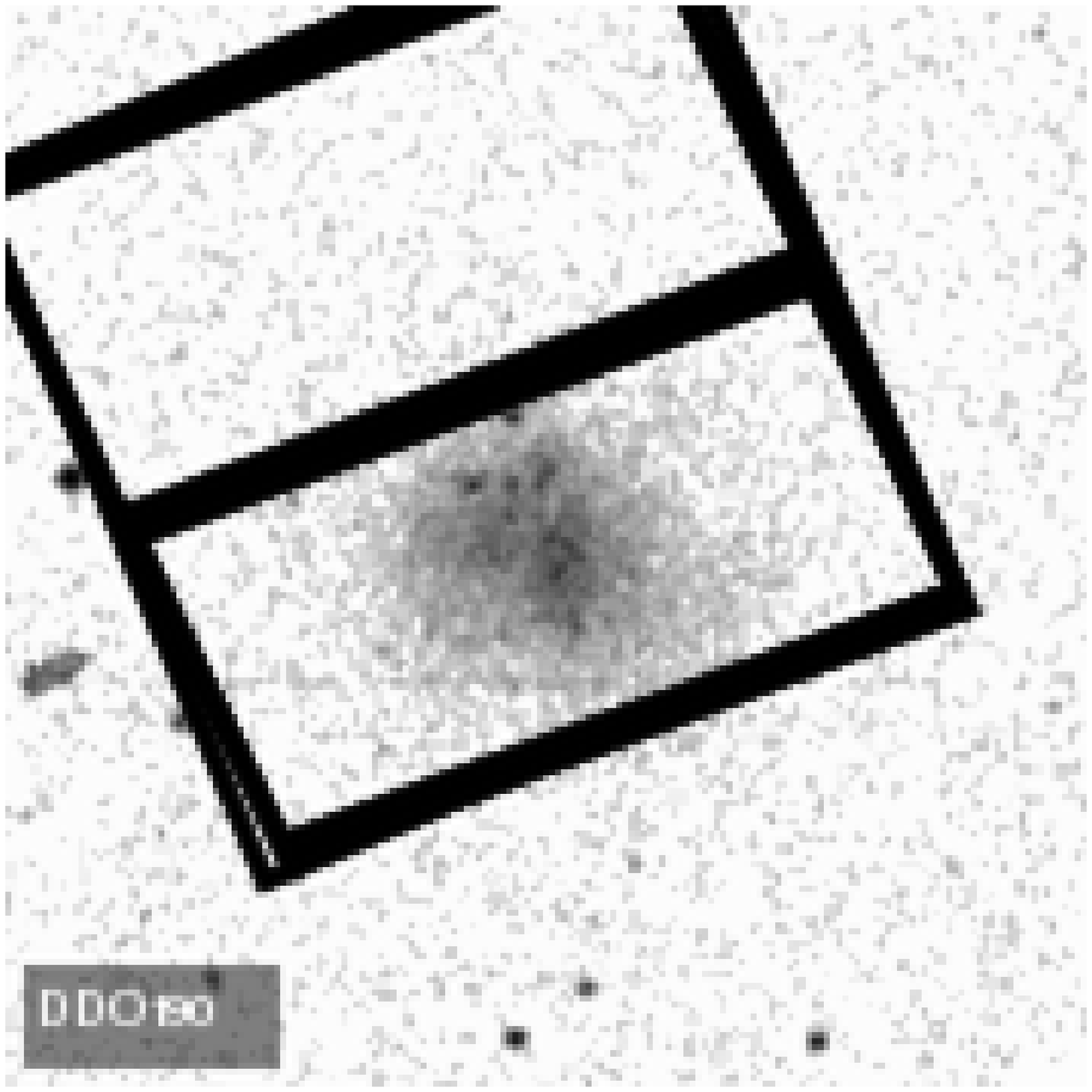}
}
\caption{
Field positions of images included in
(Table~\ref{obstable}~\&~\ref{archivetable}).
Figures are ordered from the upper left to the bottom right.
(a) Antlia; (b) SexA; (c) N3109; (d) SexB; (e) KKR25; (f) KK230; (g) E410-005; (h) E294-010; (i) N55; (j) I5152; (k) GR8; (l) N300; (m) UA438; (n) DDO187; (o) KKH98; (p) DDO125; (q) U8508; (r) KKH86; (s) DDO99; (t) DDO190; 
    \label{overlayfig1}}
\end{figure}
\vfill
\clearpage
 
%-------------------
\begin{figure}[p]
\centerline{
\includegraphics[width=1.562in]{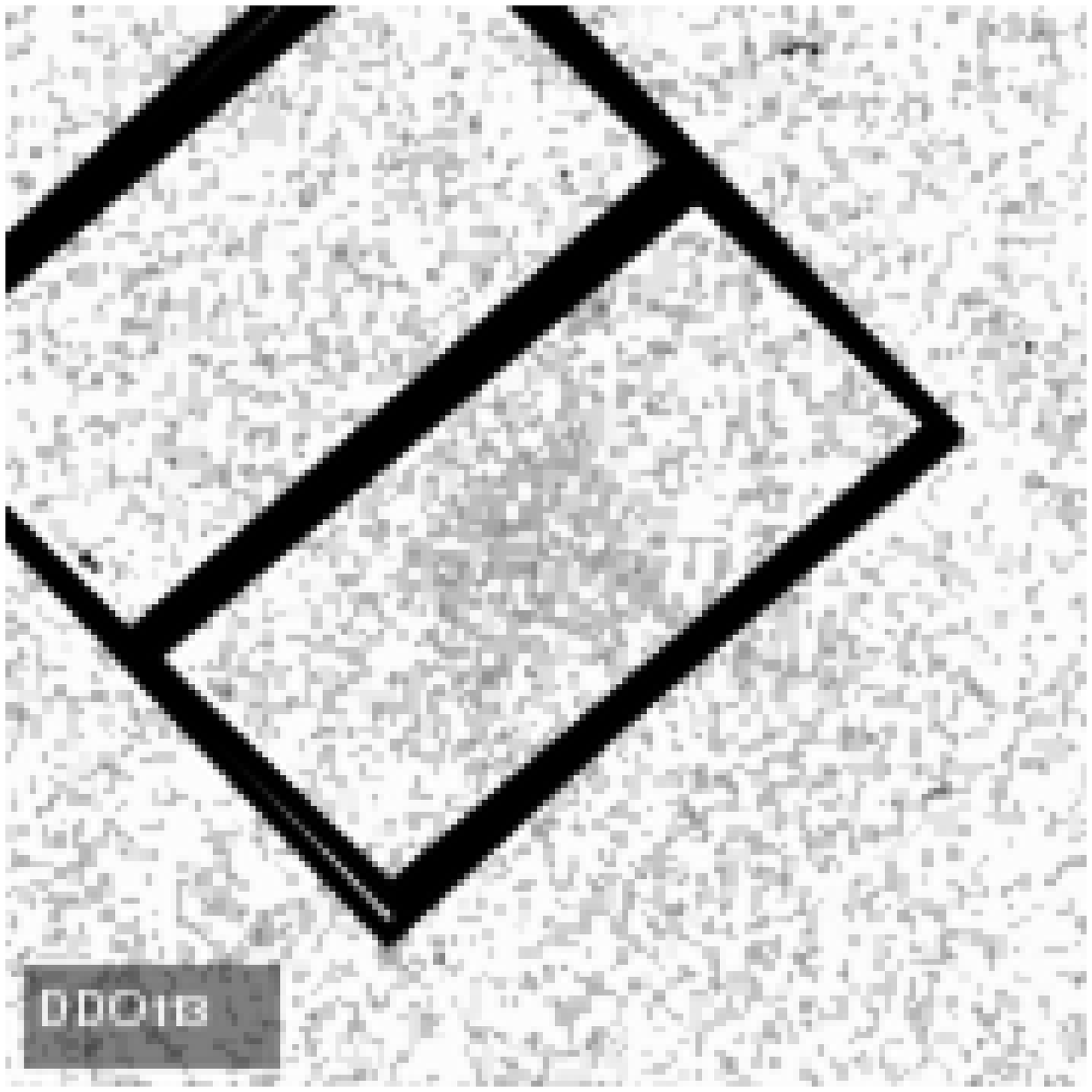}
\includegraphics[width=1.562in]{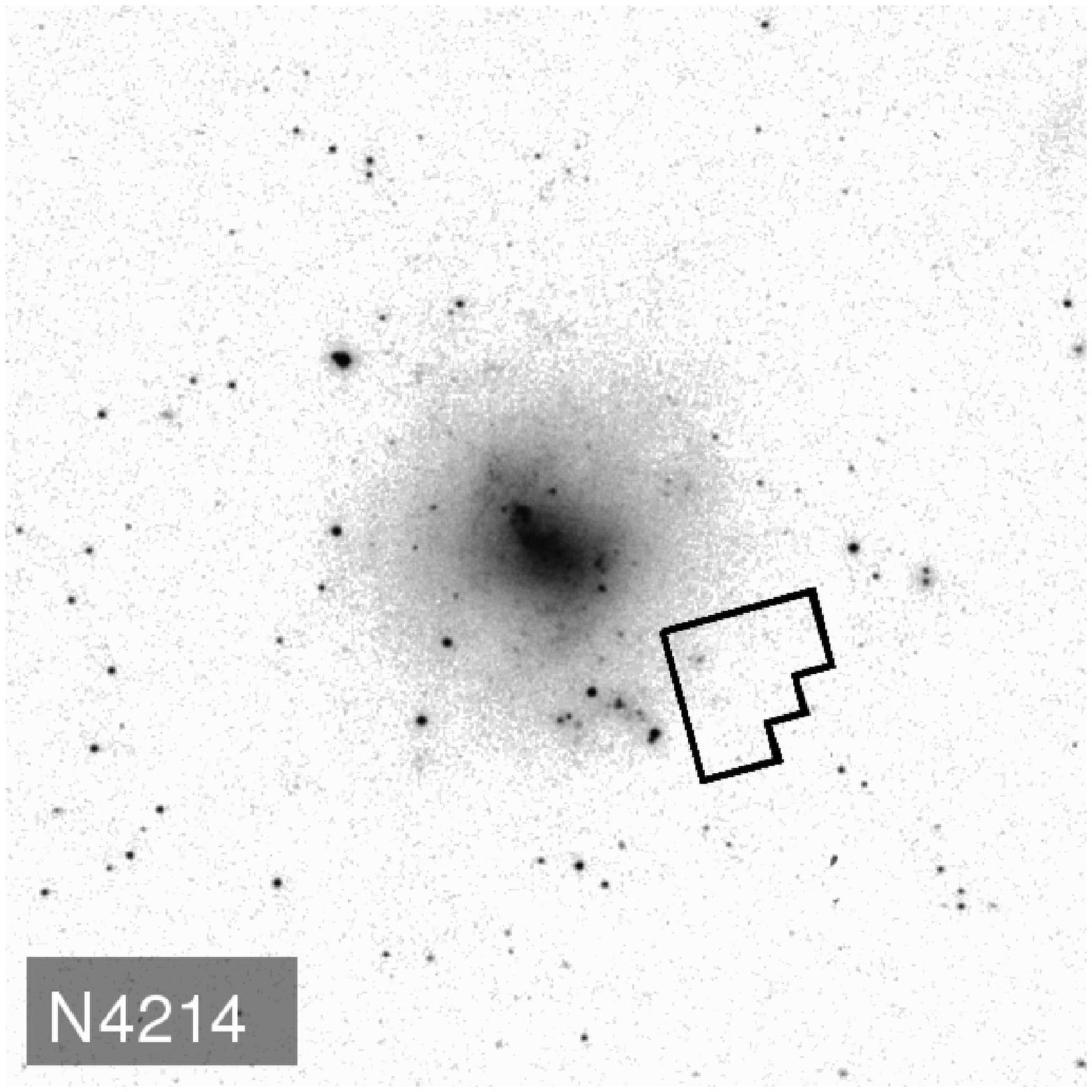}
\includegraphics[width=1.562in]{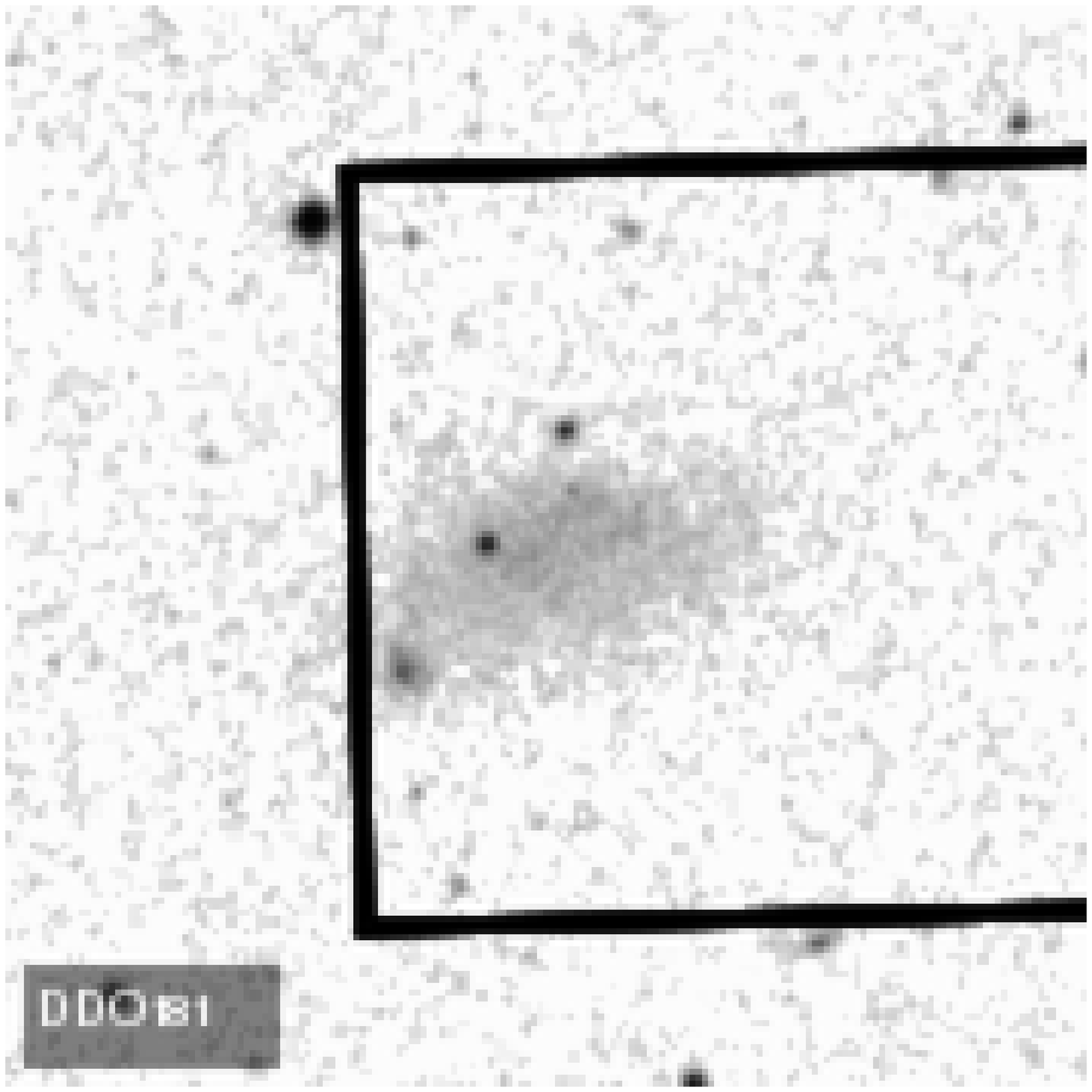}
\includegraphics[width=1.562in]{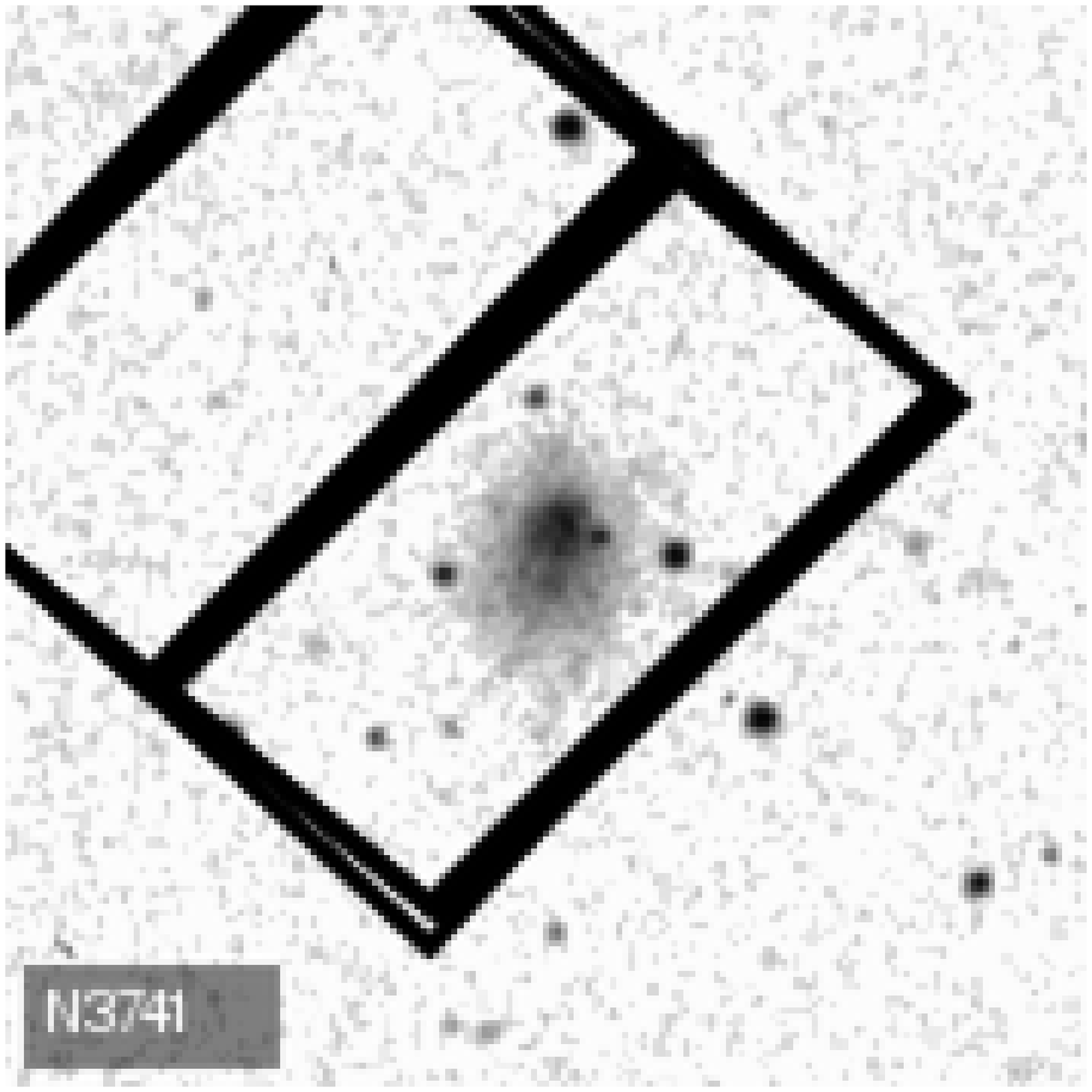}
}
\centerline{
\includegraphics[width=1.562in]{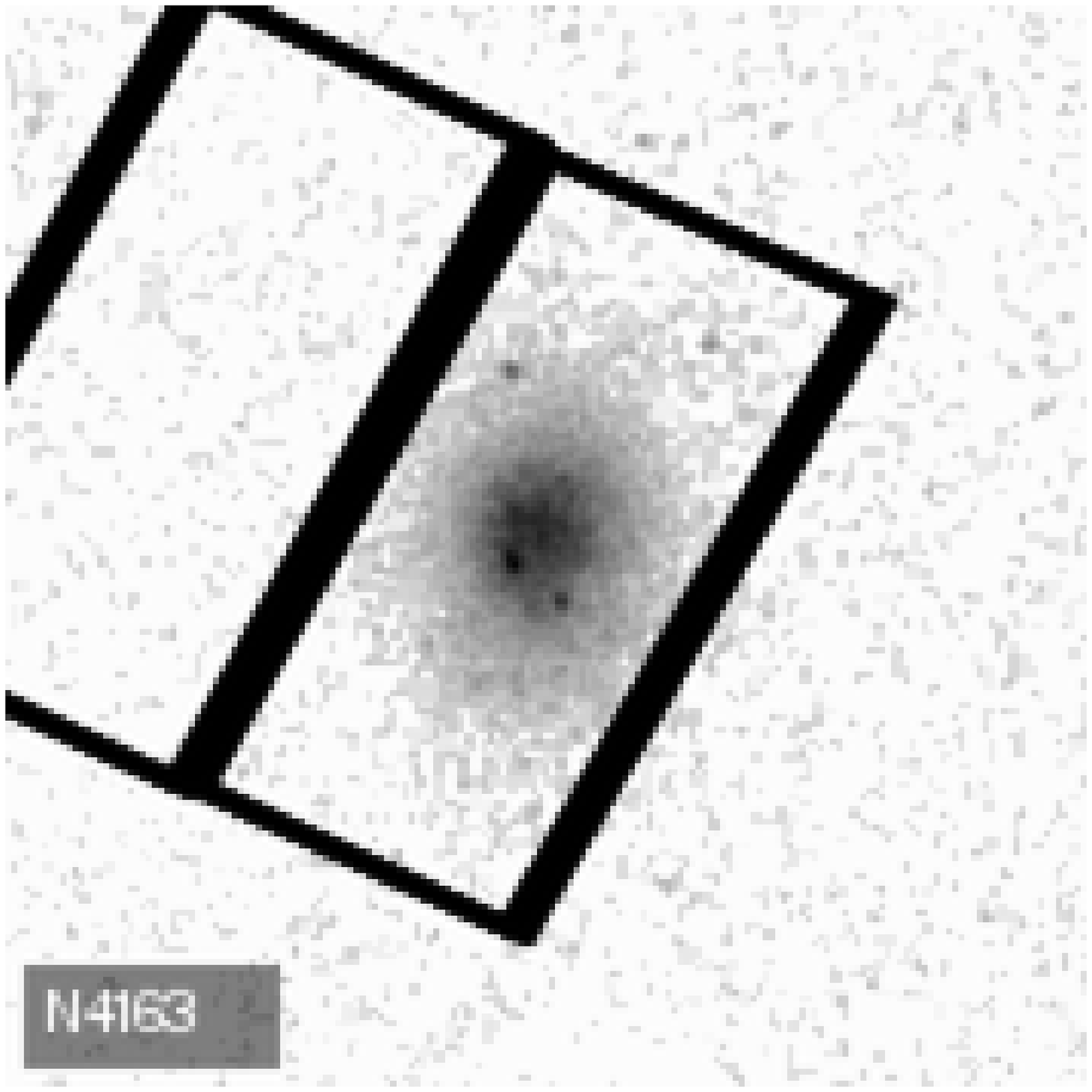}
\includegraphics[width=1.562in]{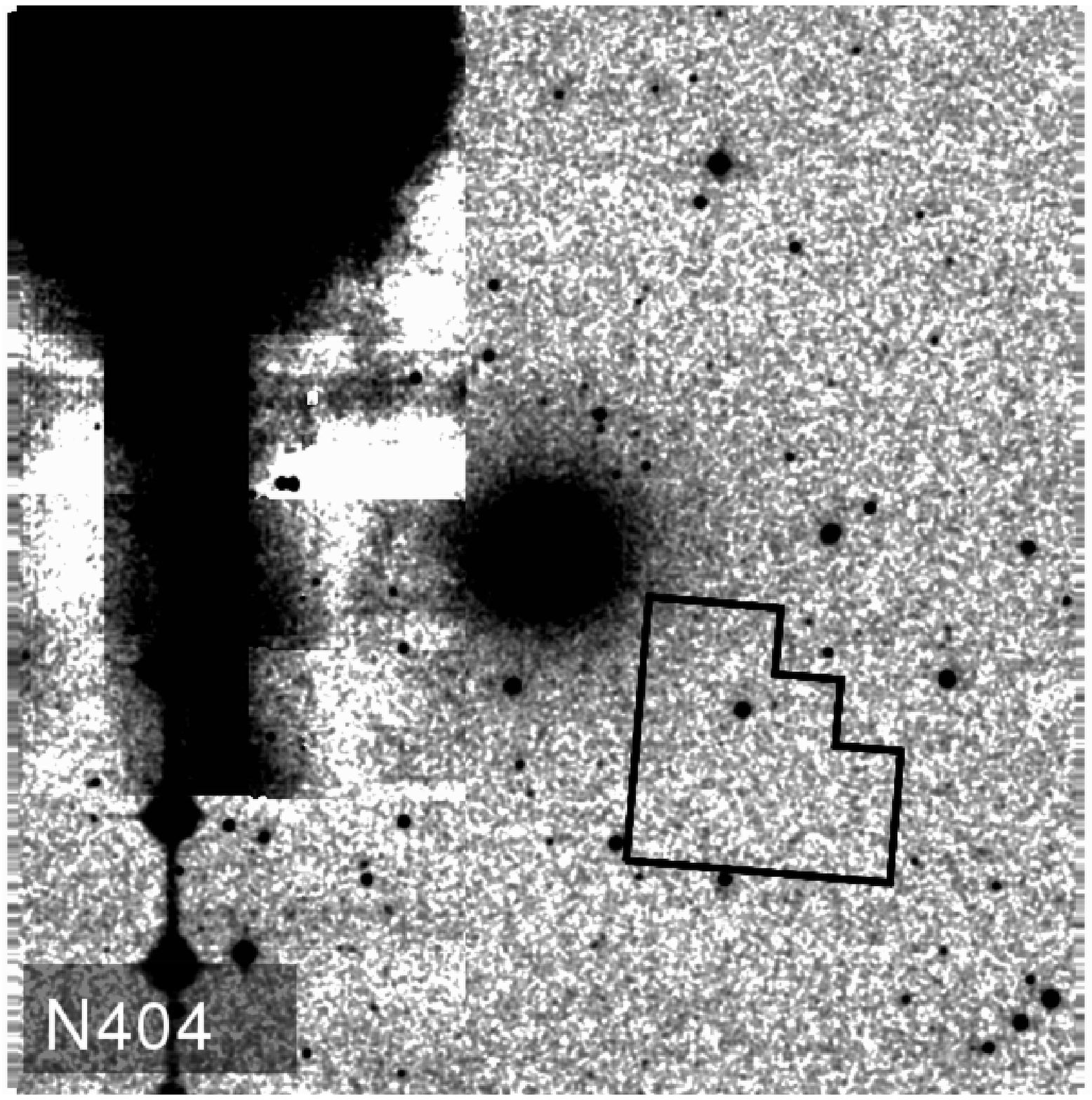}
\includegraphics[width=1.562in]{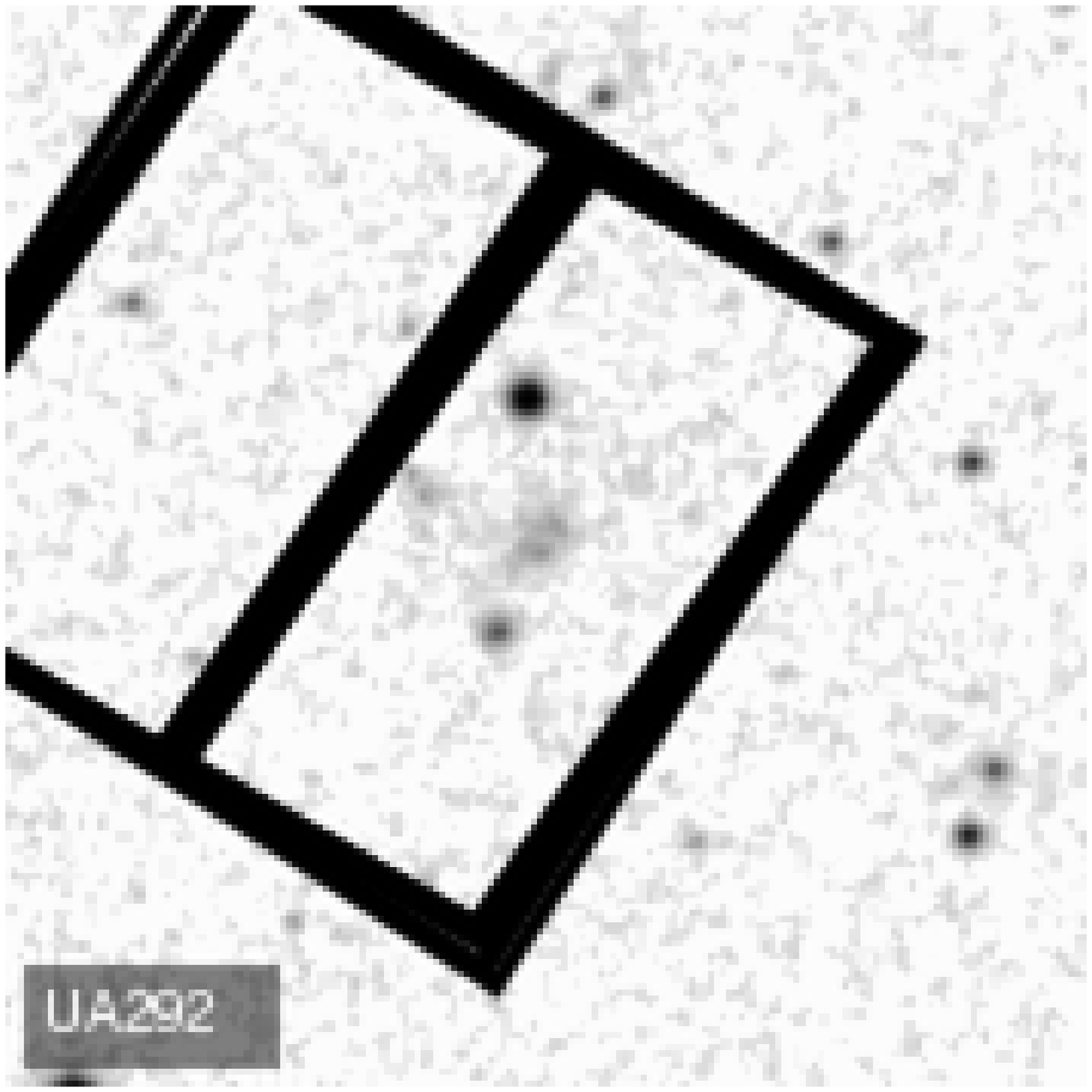}
\includegraphics[width=1.562in]{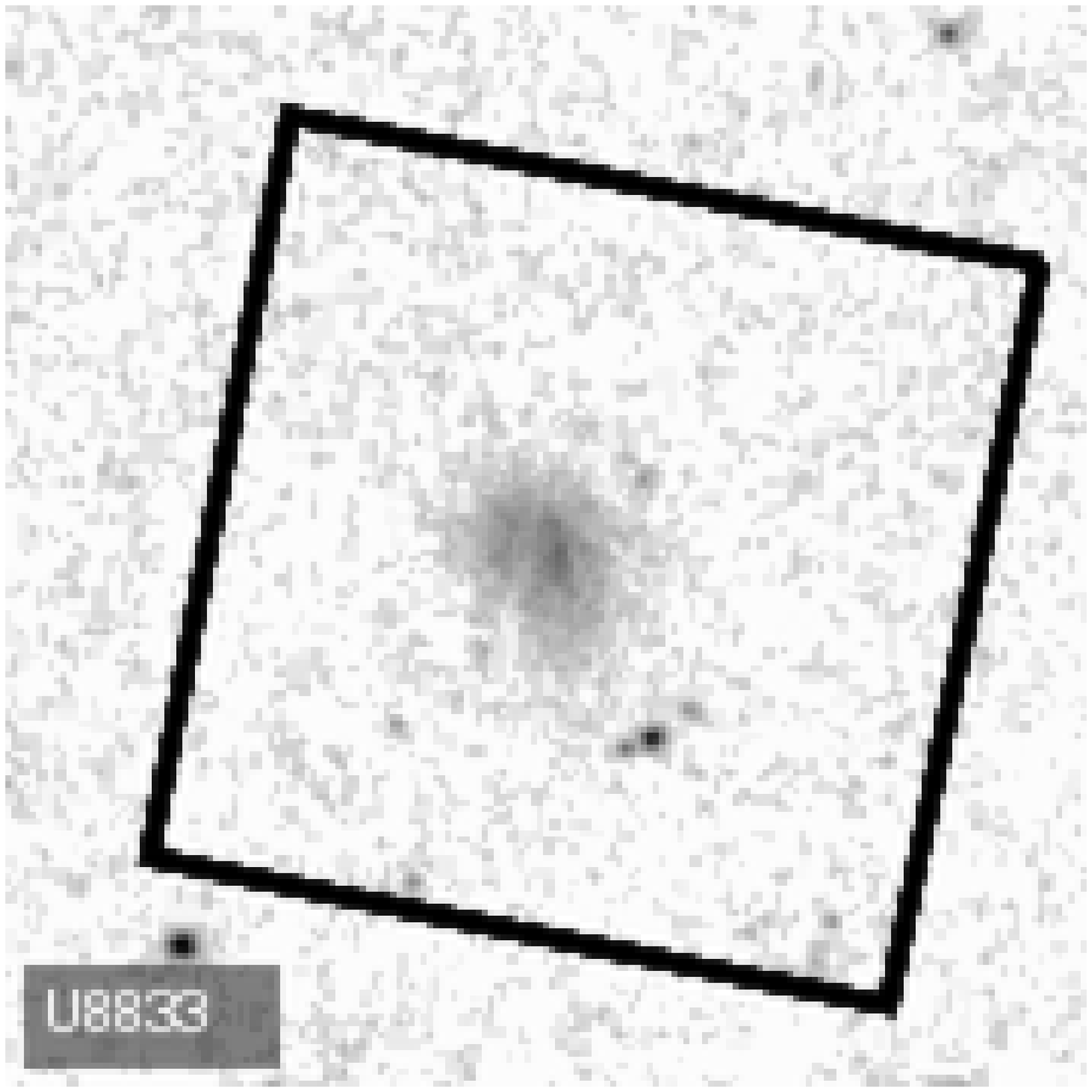}
}
\centerline{
\includegraphics[width=1.562in]{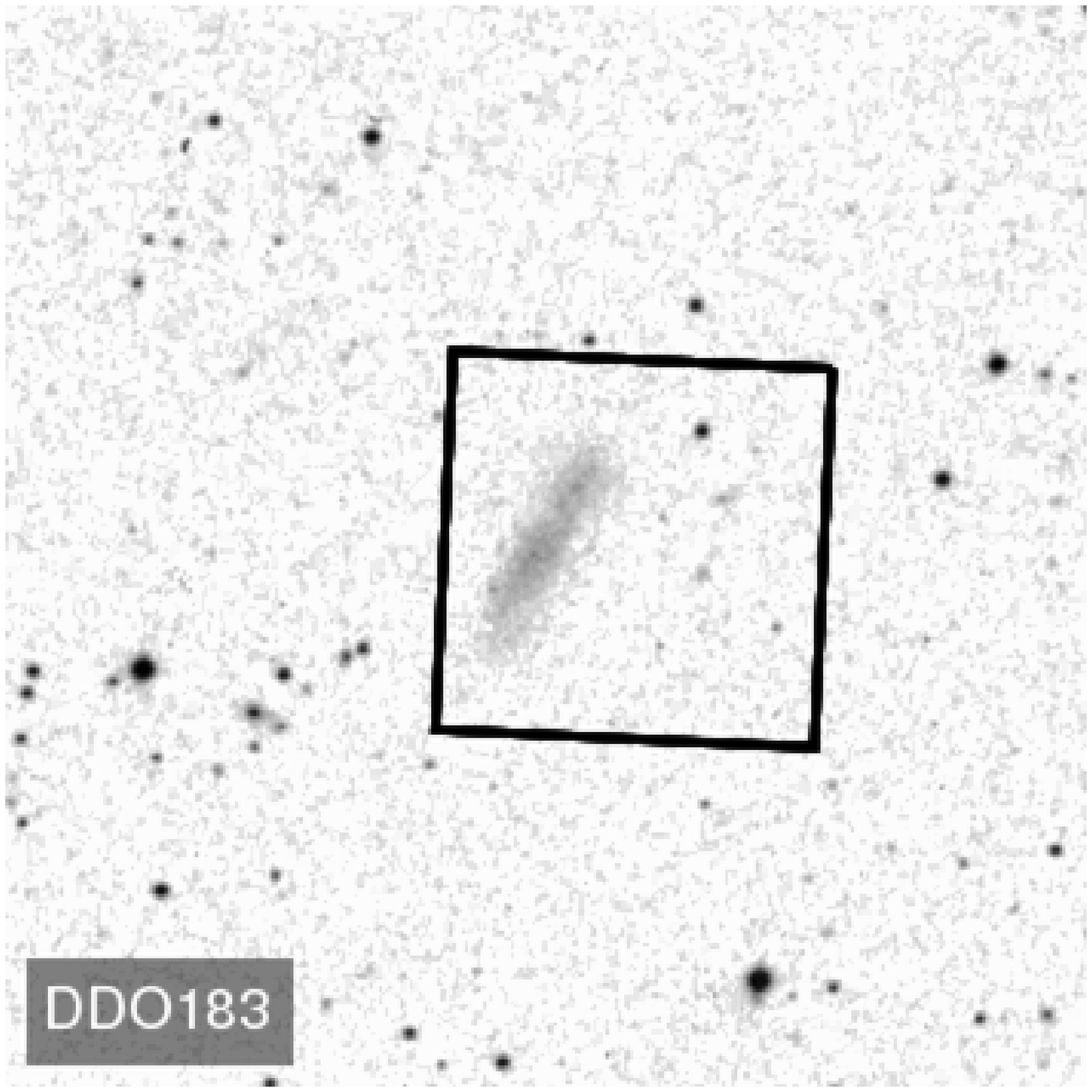}
\includegraphics[width=1.562in]{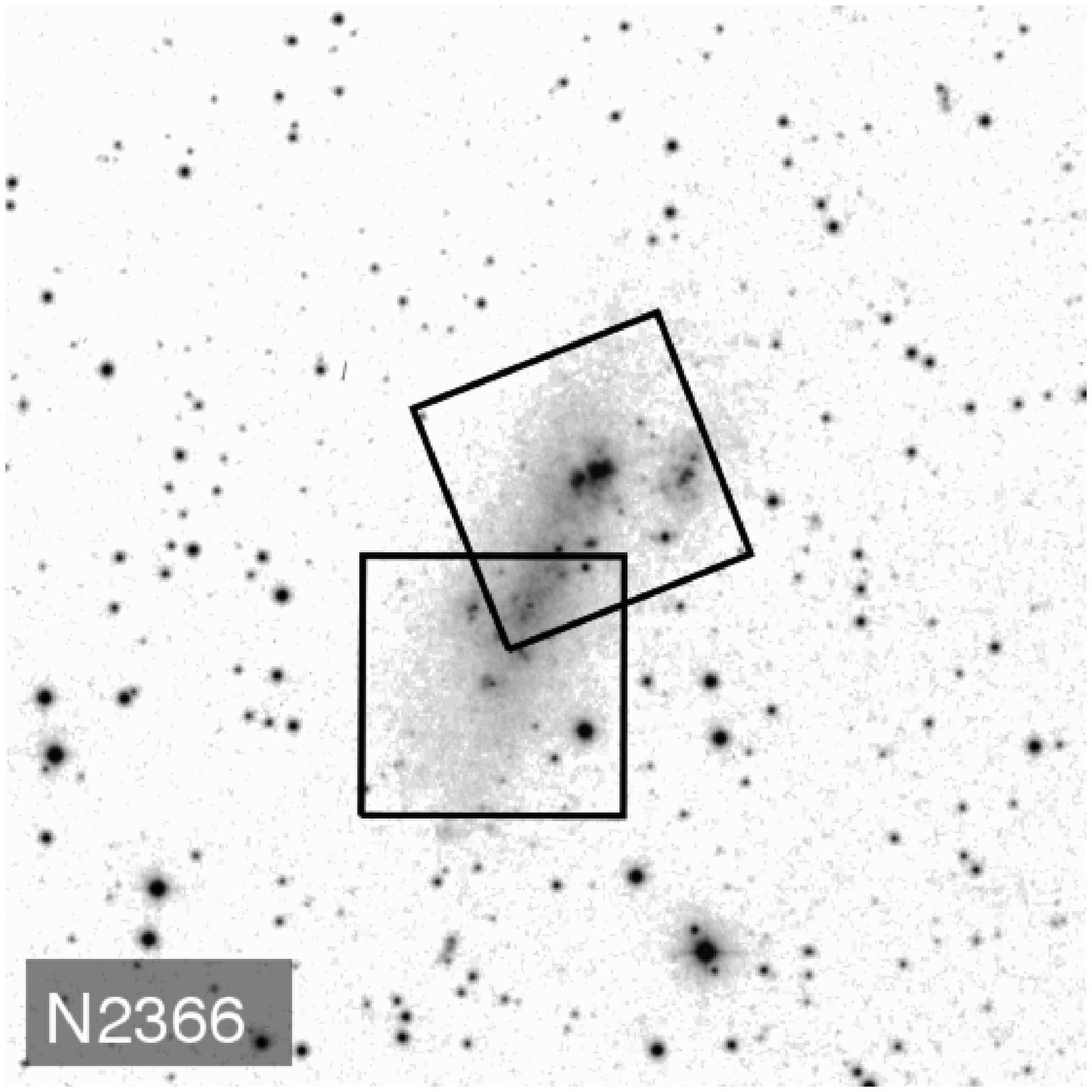}
\includegraphics[width=1.562in]{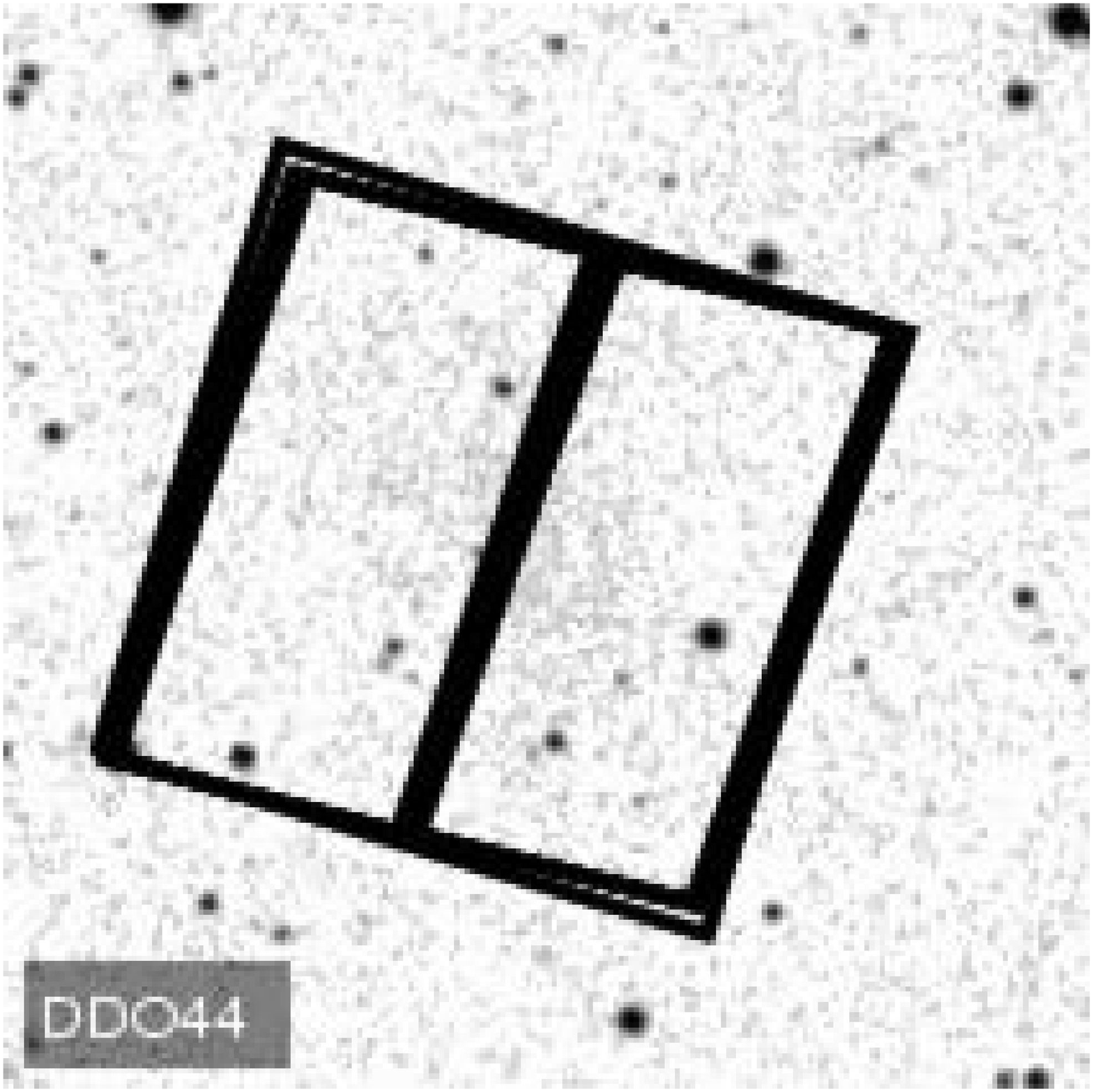}
\includegraphics[width=1.562in]{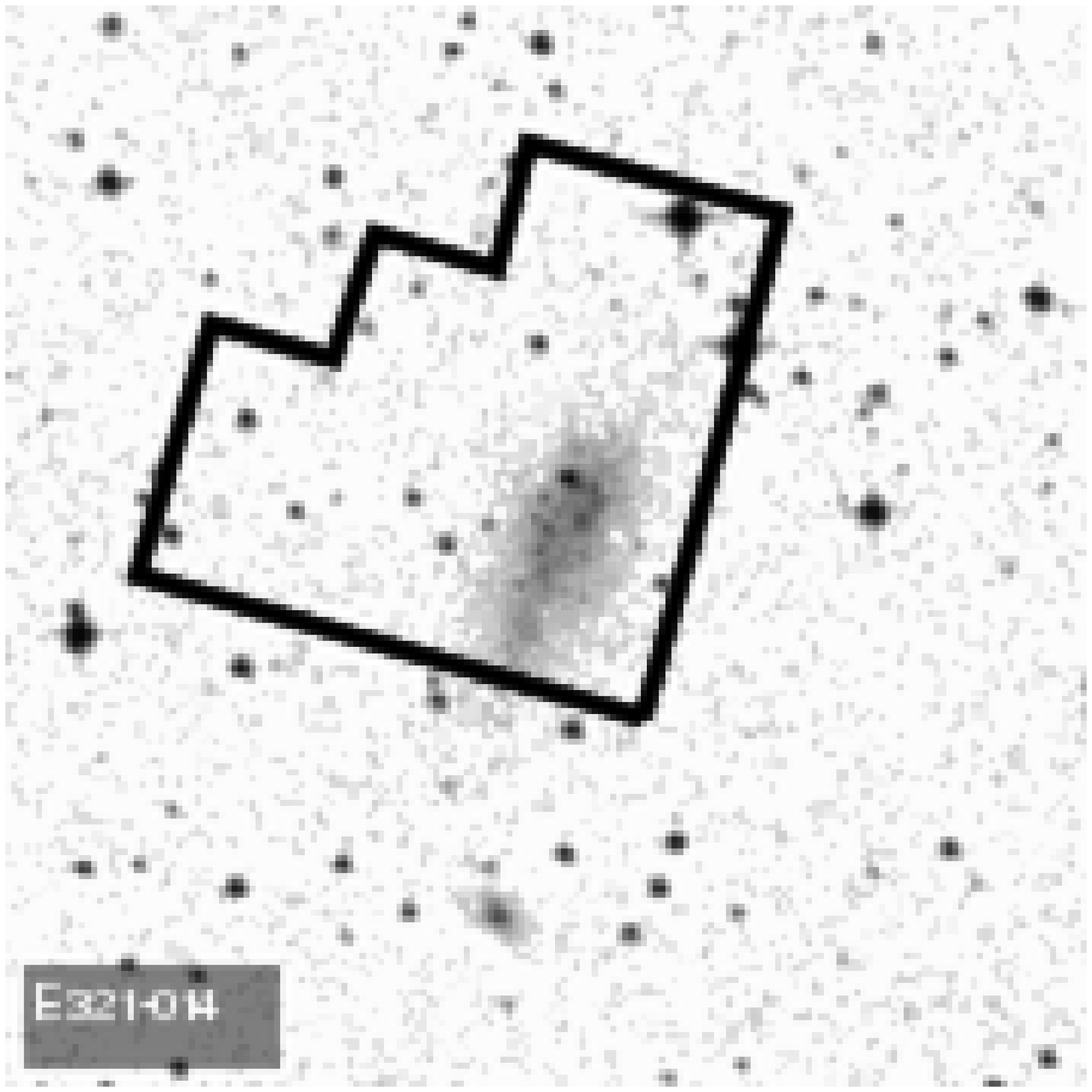}
}
\centerline{
\includegraphics[width=1.562in]{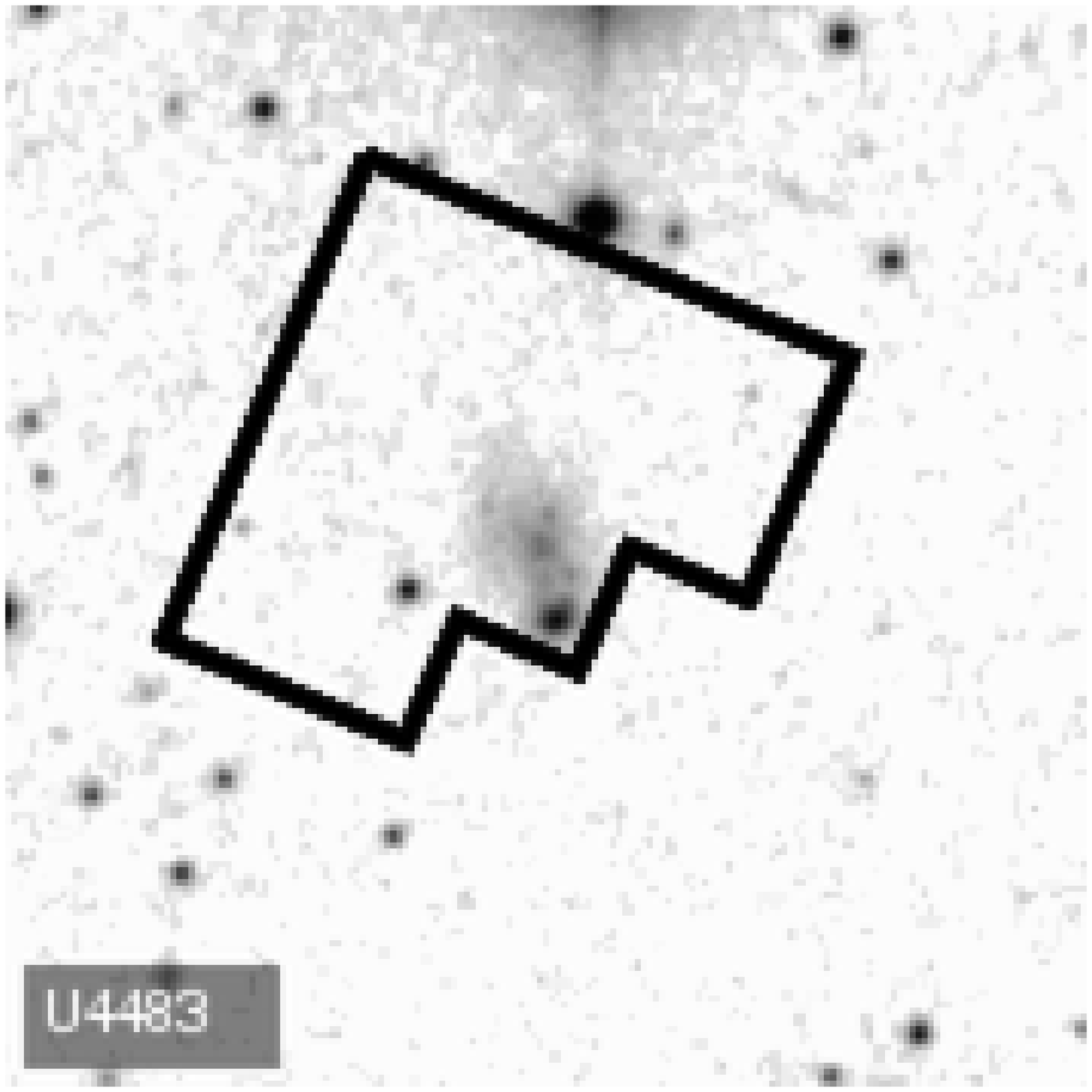}
\includegraphics[width=1.562in]{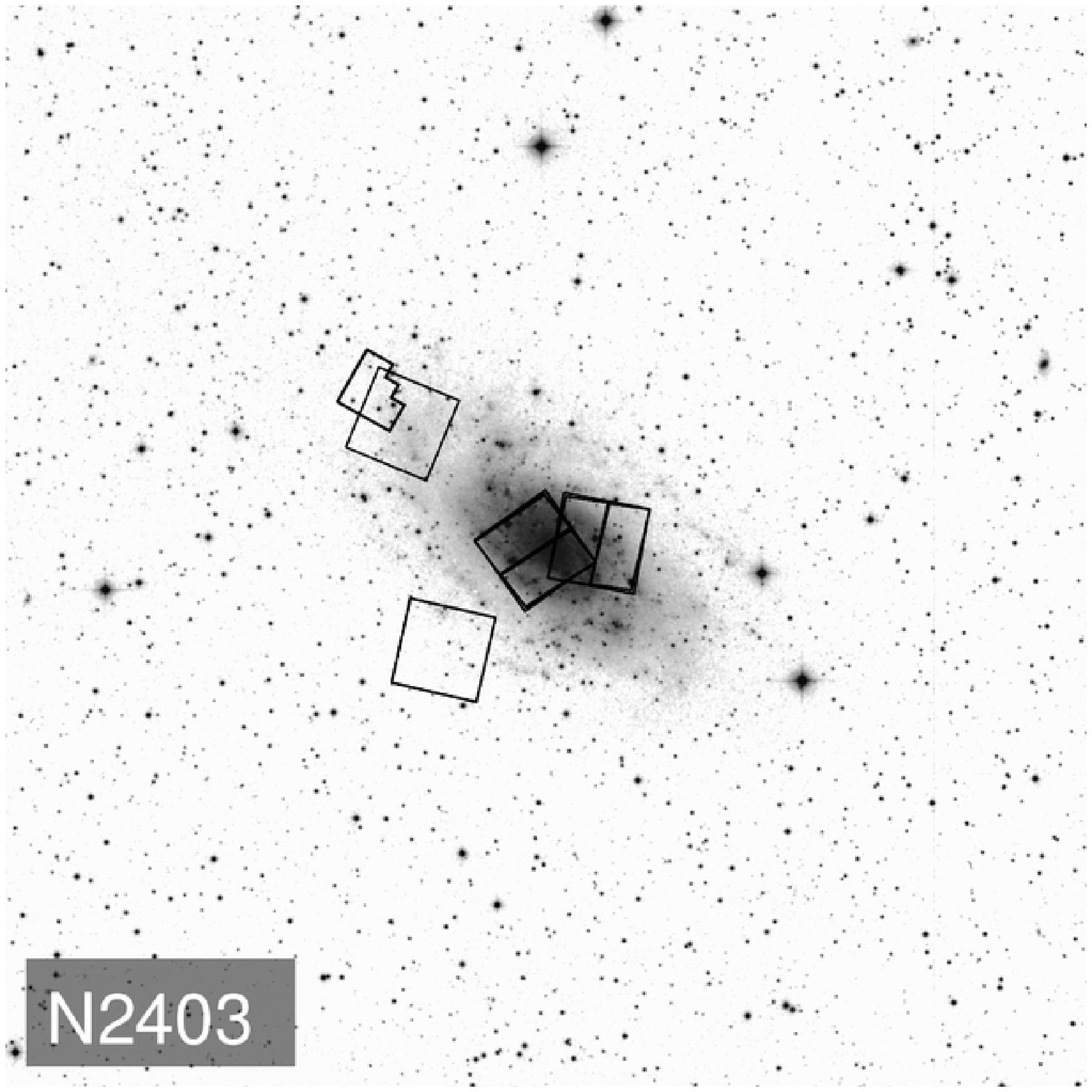}
\includegraphics[width=1.562in]{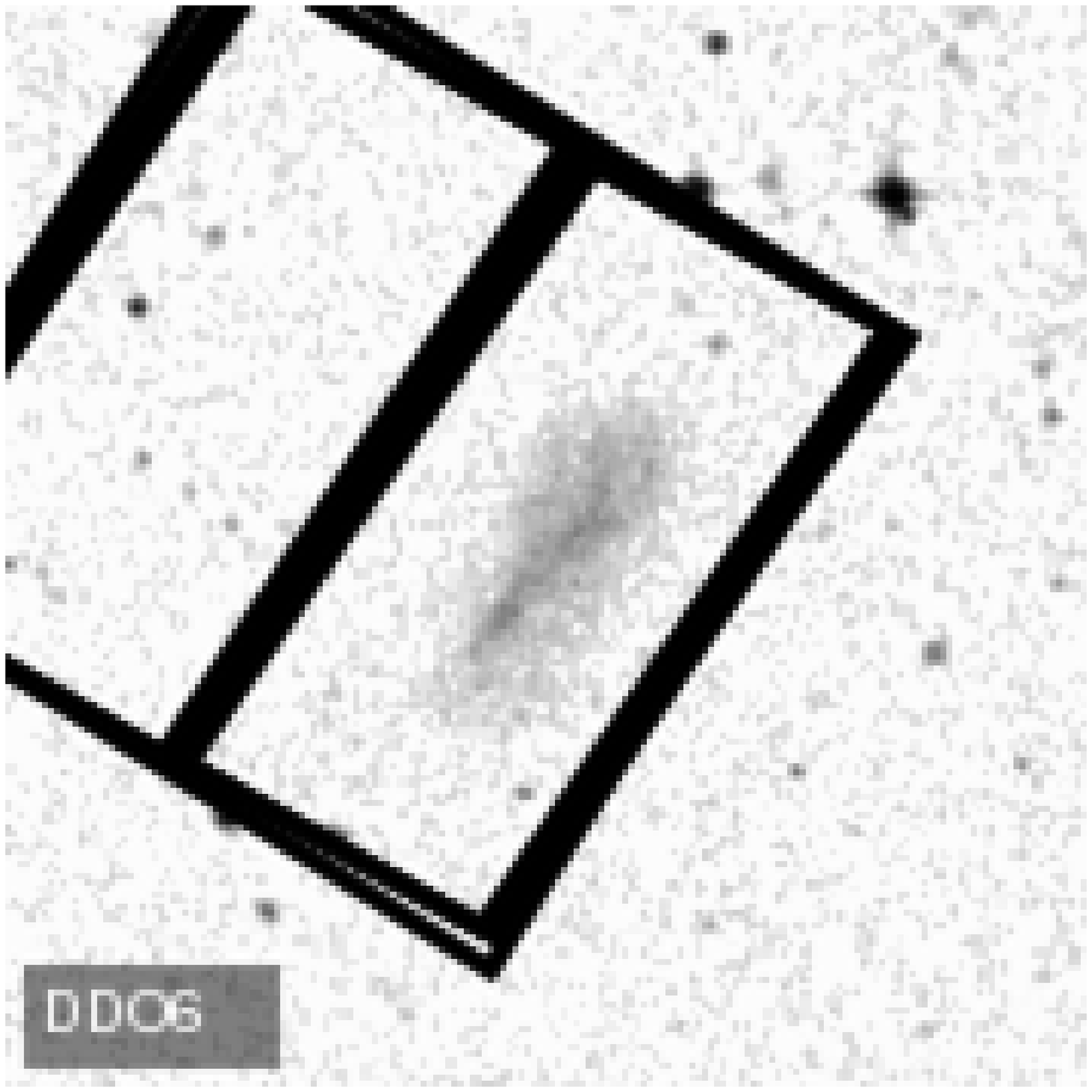}
\includegraphics[width=1.562in]{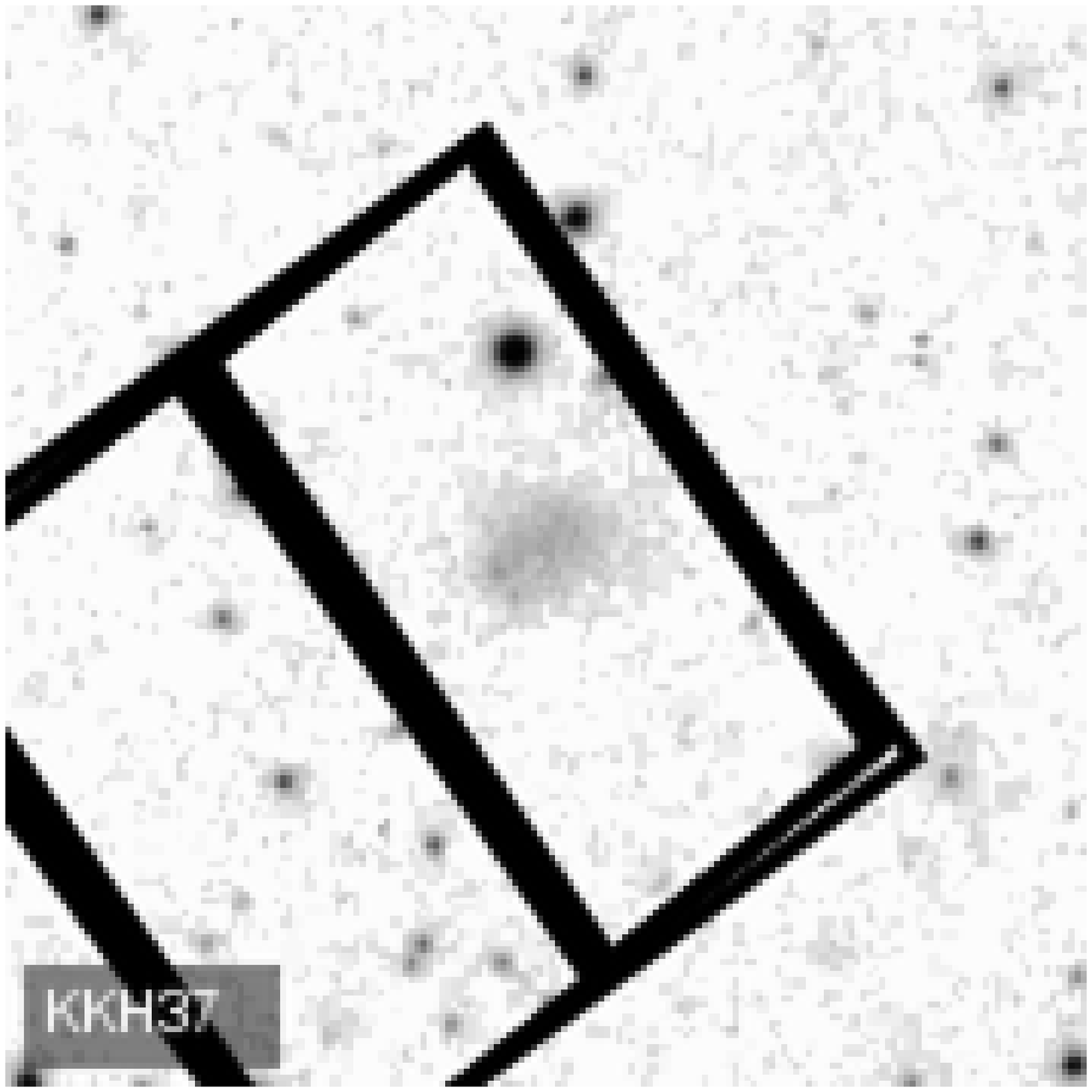}
}
\centerline{
\includegraphics[width=1.562in]{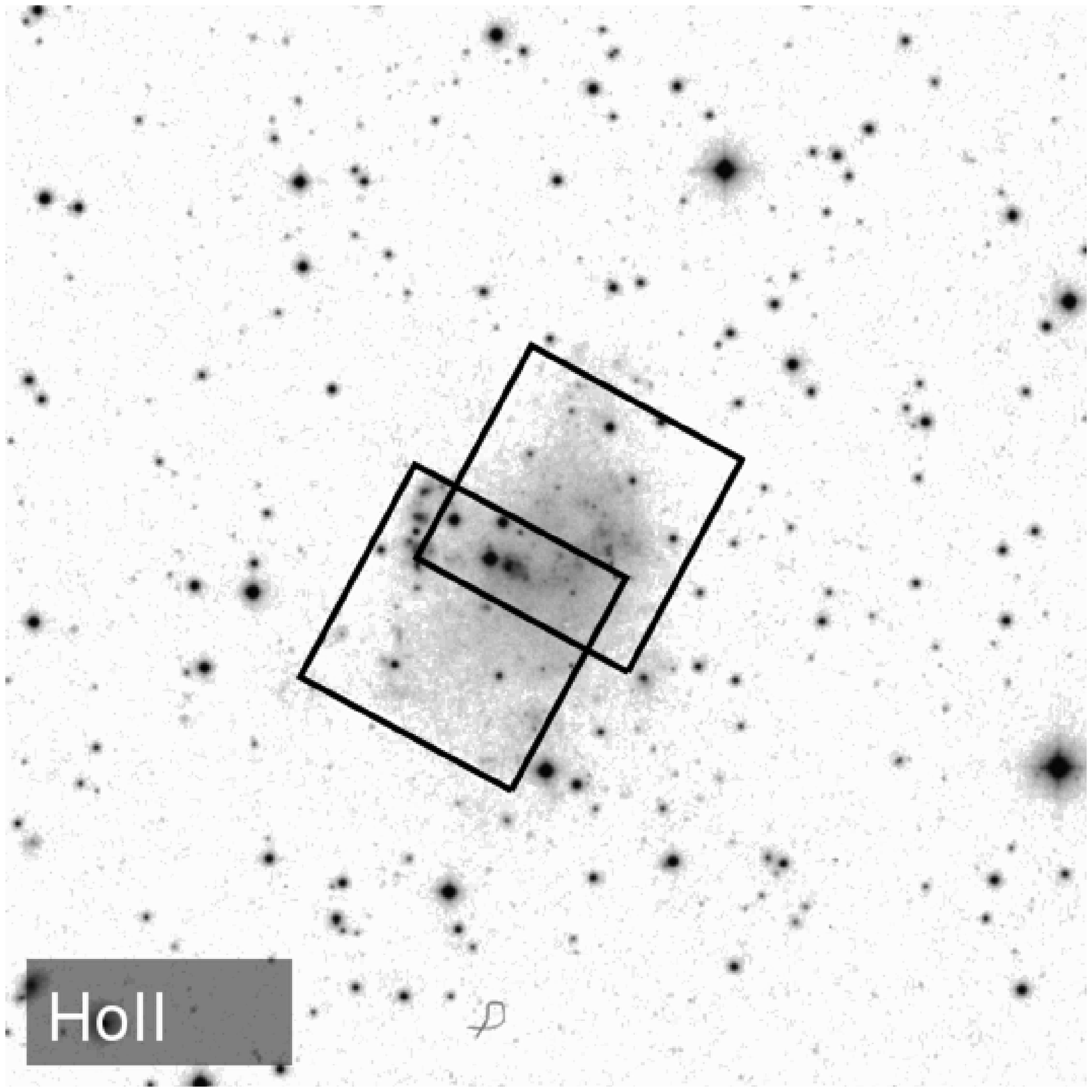}
\includegraphics[width=1.562in]{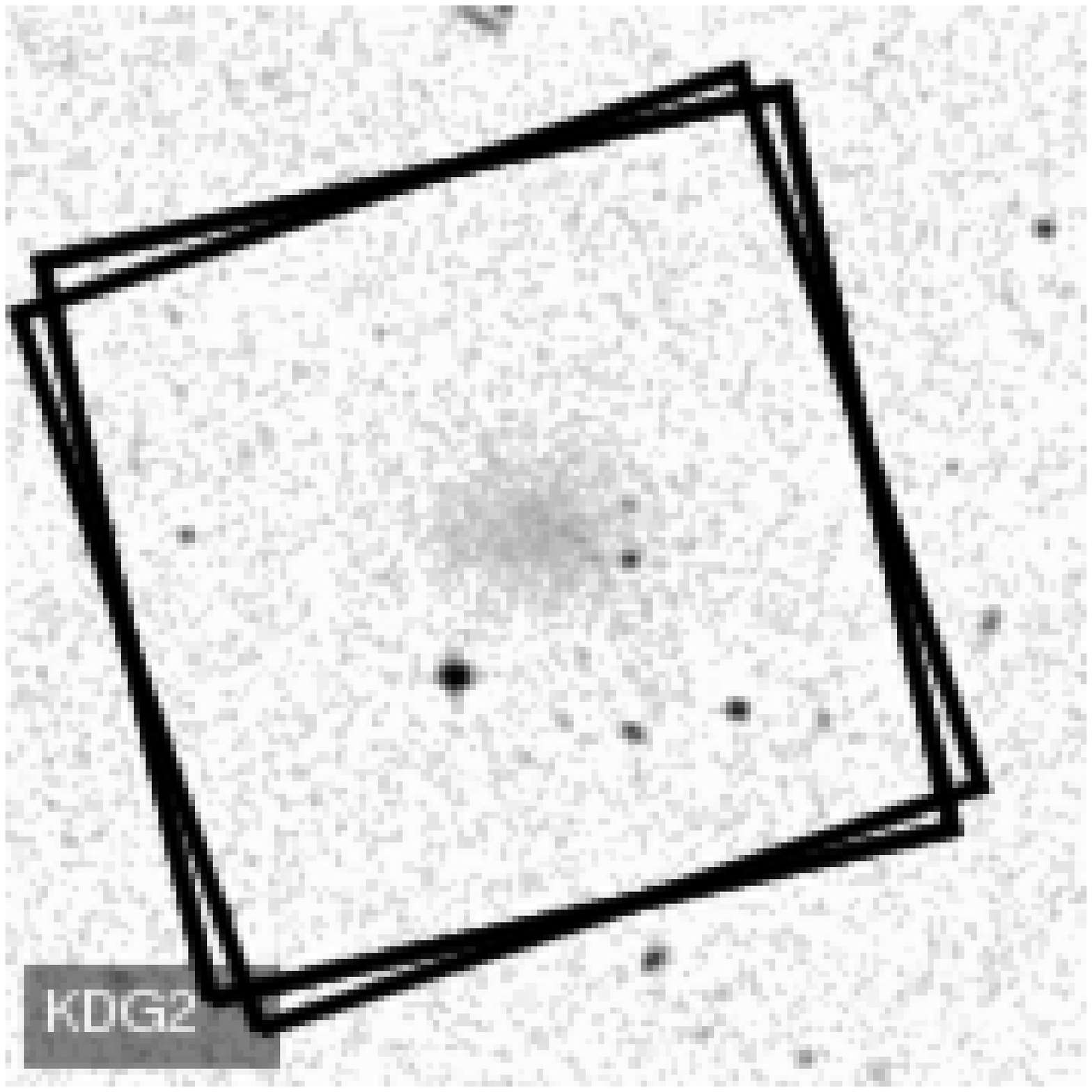}
\includegraphics[width=1.562in]{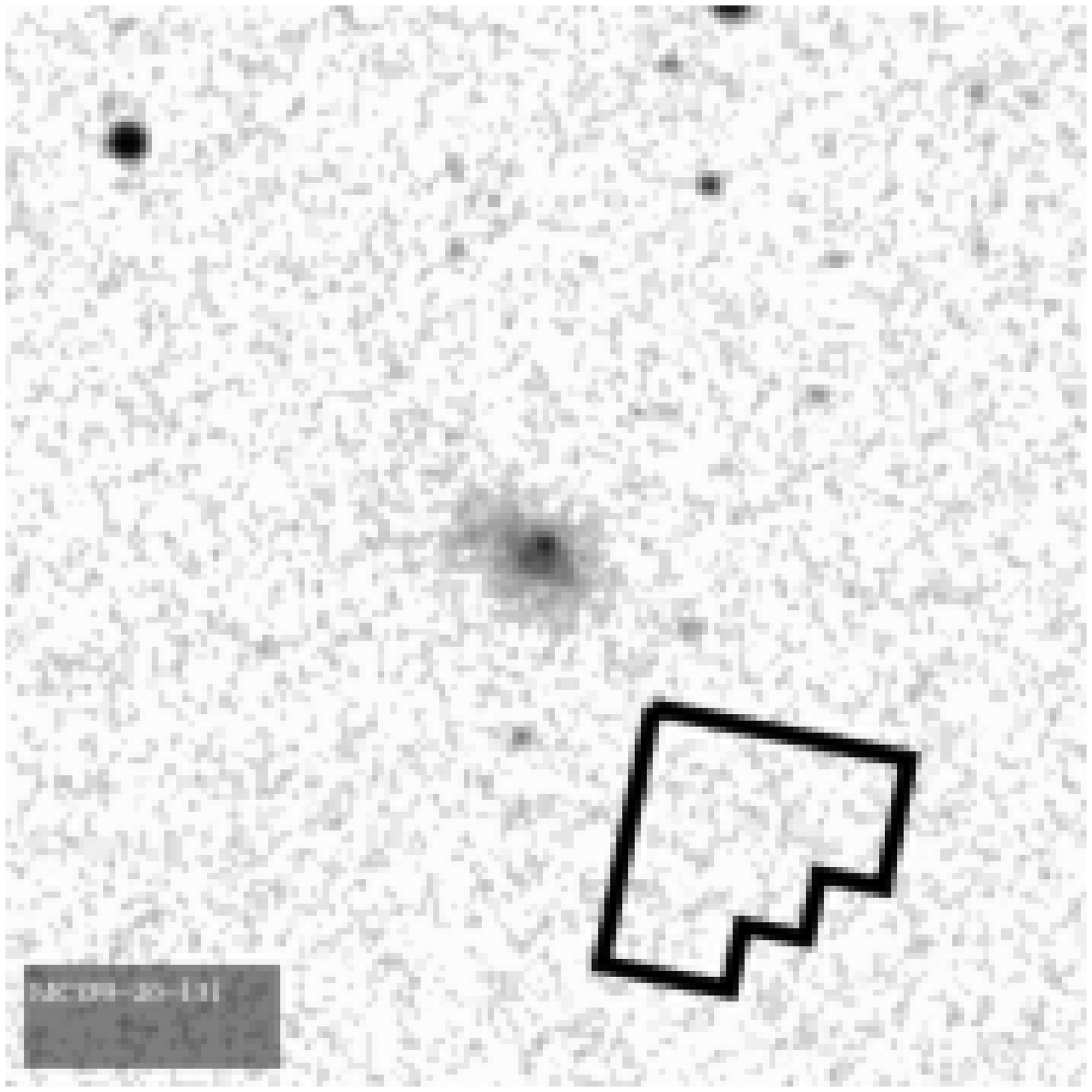}
\includegraphics[width=1.562in]{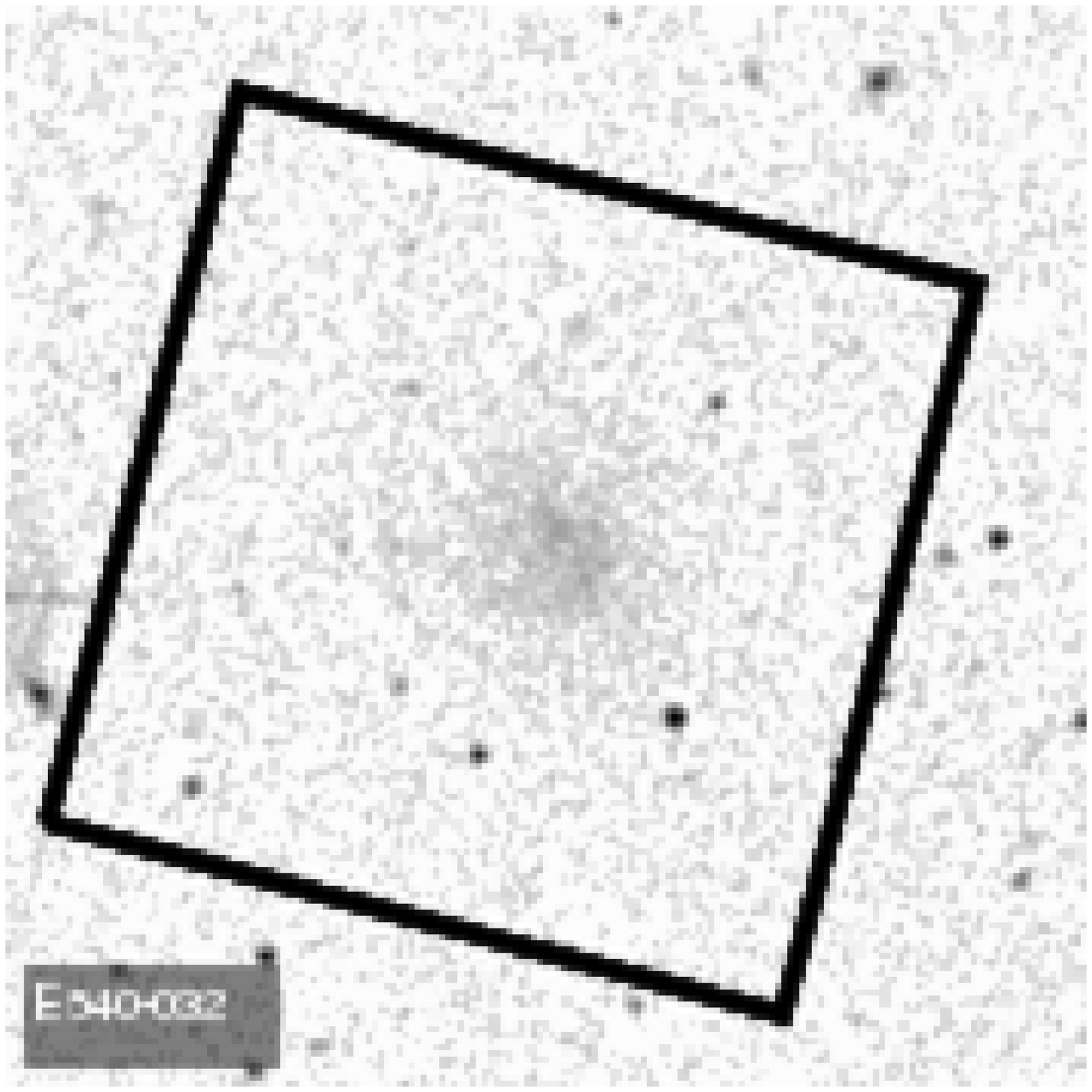}
}
\caption{
Field positions of images included in
Table~\ref{obstable}~\&~\ref{archivetable},
as described in Figure~\ref{overlayfig1}.
Figures are ordered from the upper left to the bottom right.
(a) DDO113; (b) N4214; (c) DDO181; (d) N3741; (e) N4163; (f) N404; (g) UA292; (h) U8833; (i) DDO183; (j) N2366; (k) DDO44; (l) E321-014; (m) U4483; (n) N2403; (o) DDO6; (p) KKH37; (q) HoII; (r) KDG2; (s) MCG9-20-131; (t) E540-032; 
    \label{overlayfig2}}
\end{figure}
\vfill
\clearpage
 
%-------------------
\begin{figure}[p]
\centerline{
\includegraphics[width=1.562in]{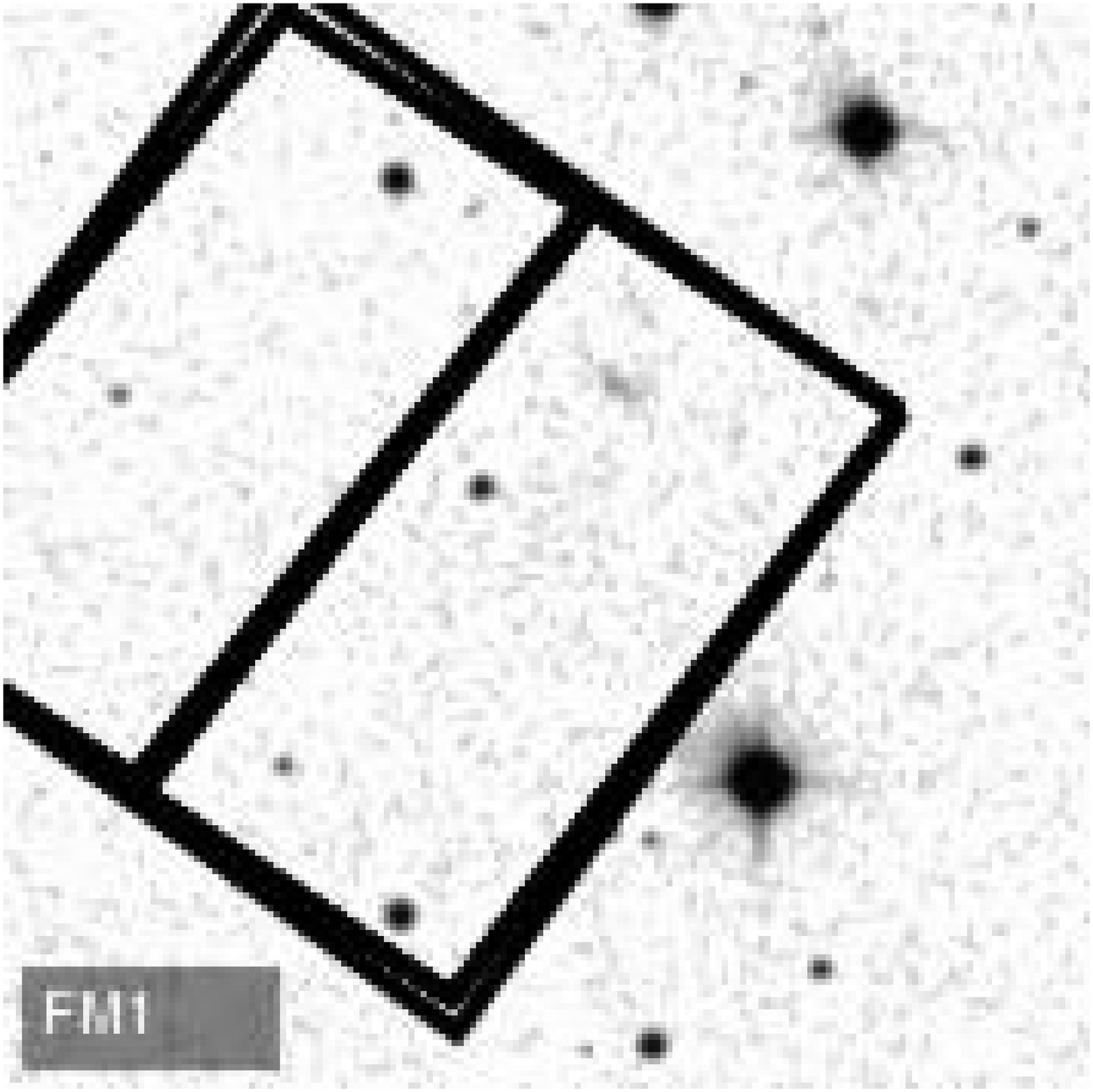}
\includegraphics[width=1.562in]{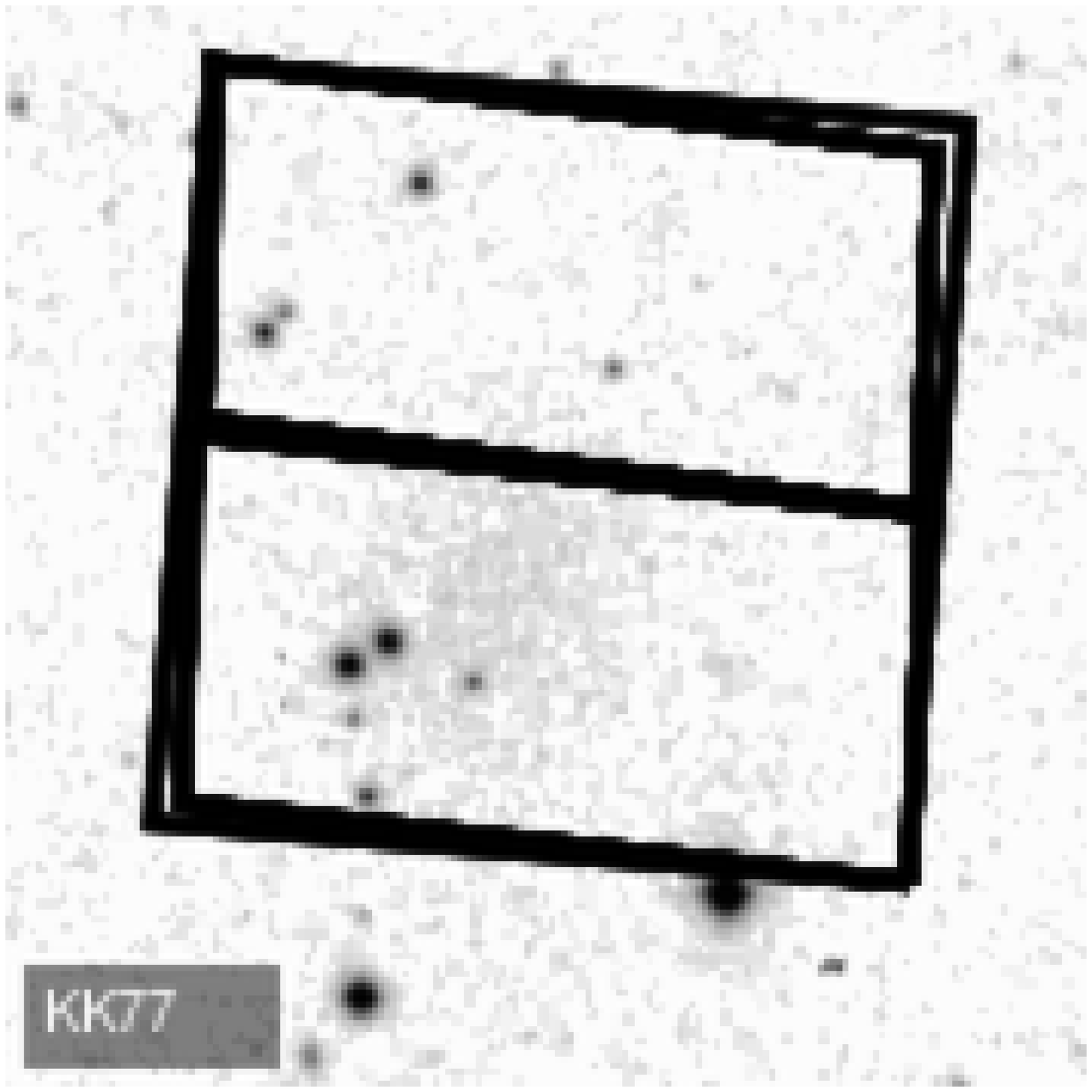}
\includegraphics[width=1.562in]{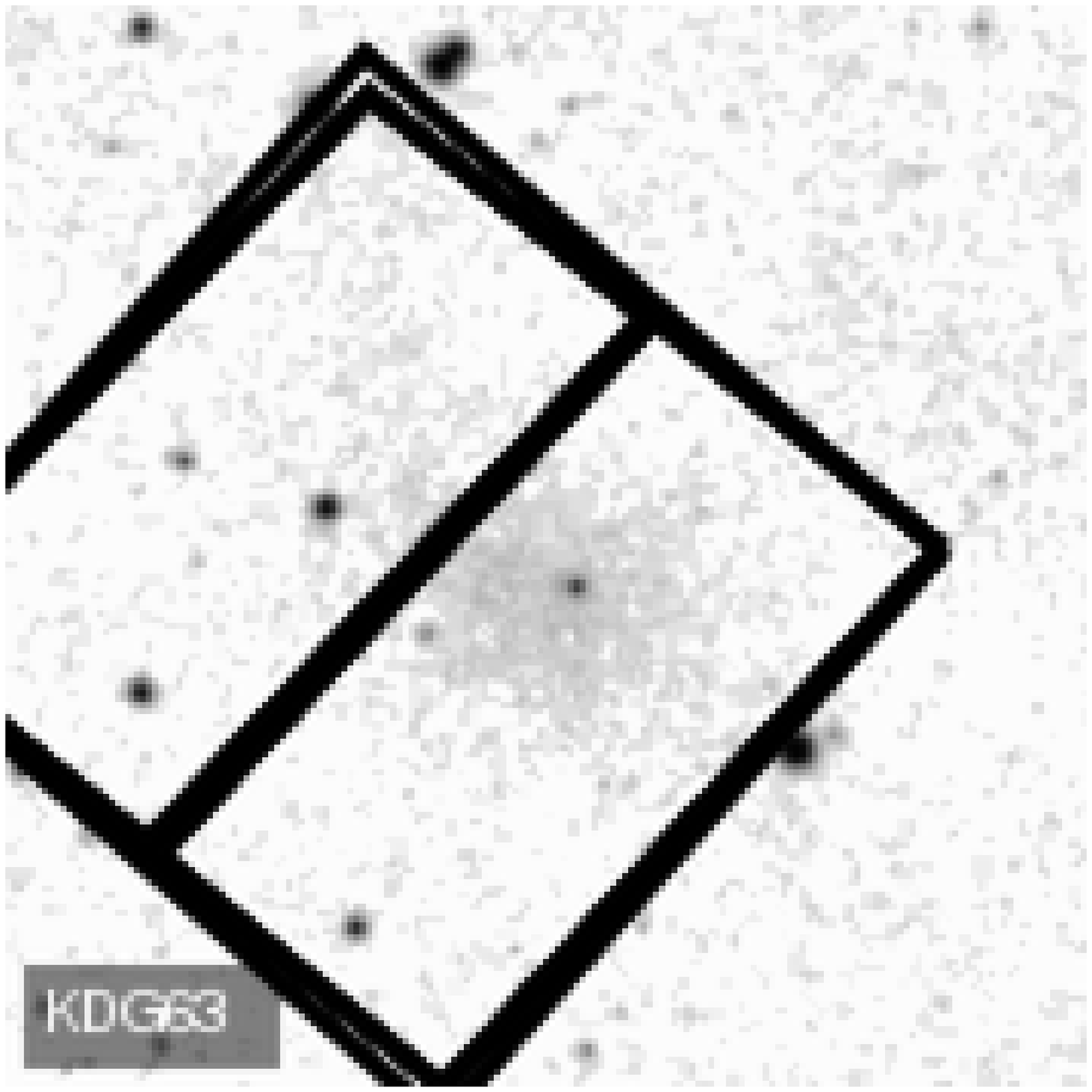}
\includegraphics[width=1.562in]{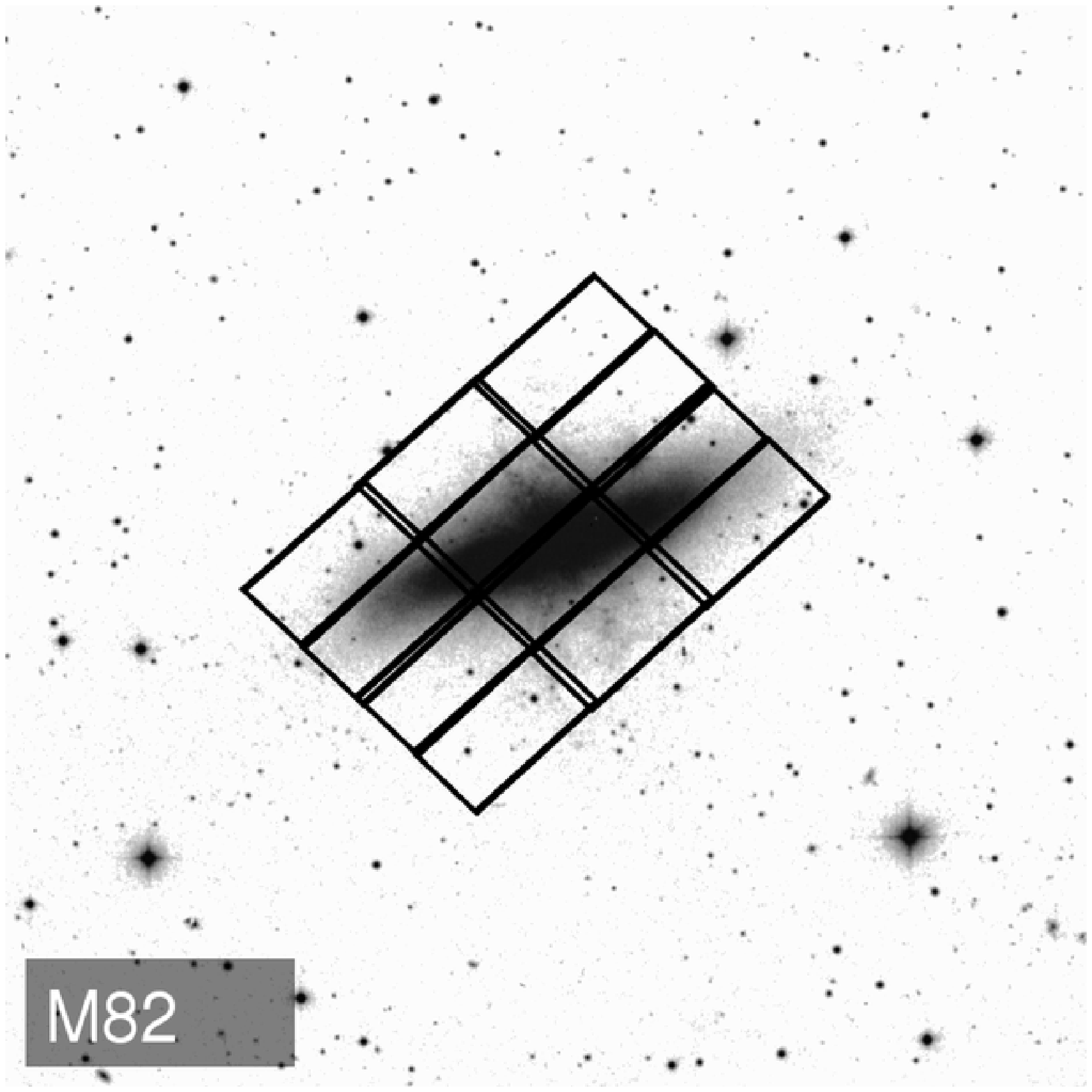}
}
\centerline{
\includegraphics[width=1.562in]{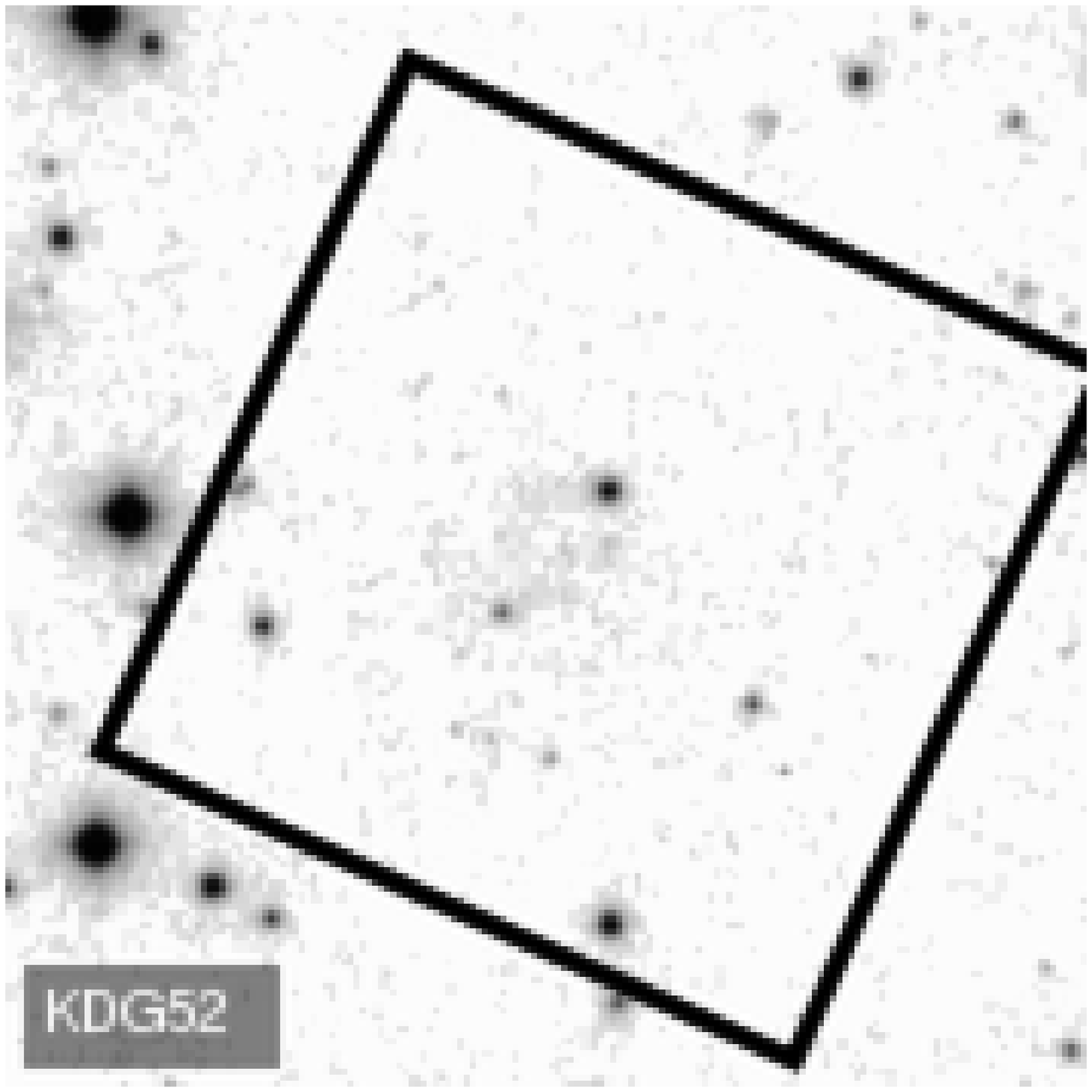}
\includegraphics[width=1.562in]{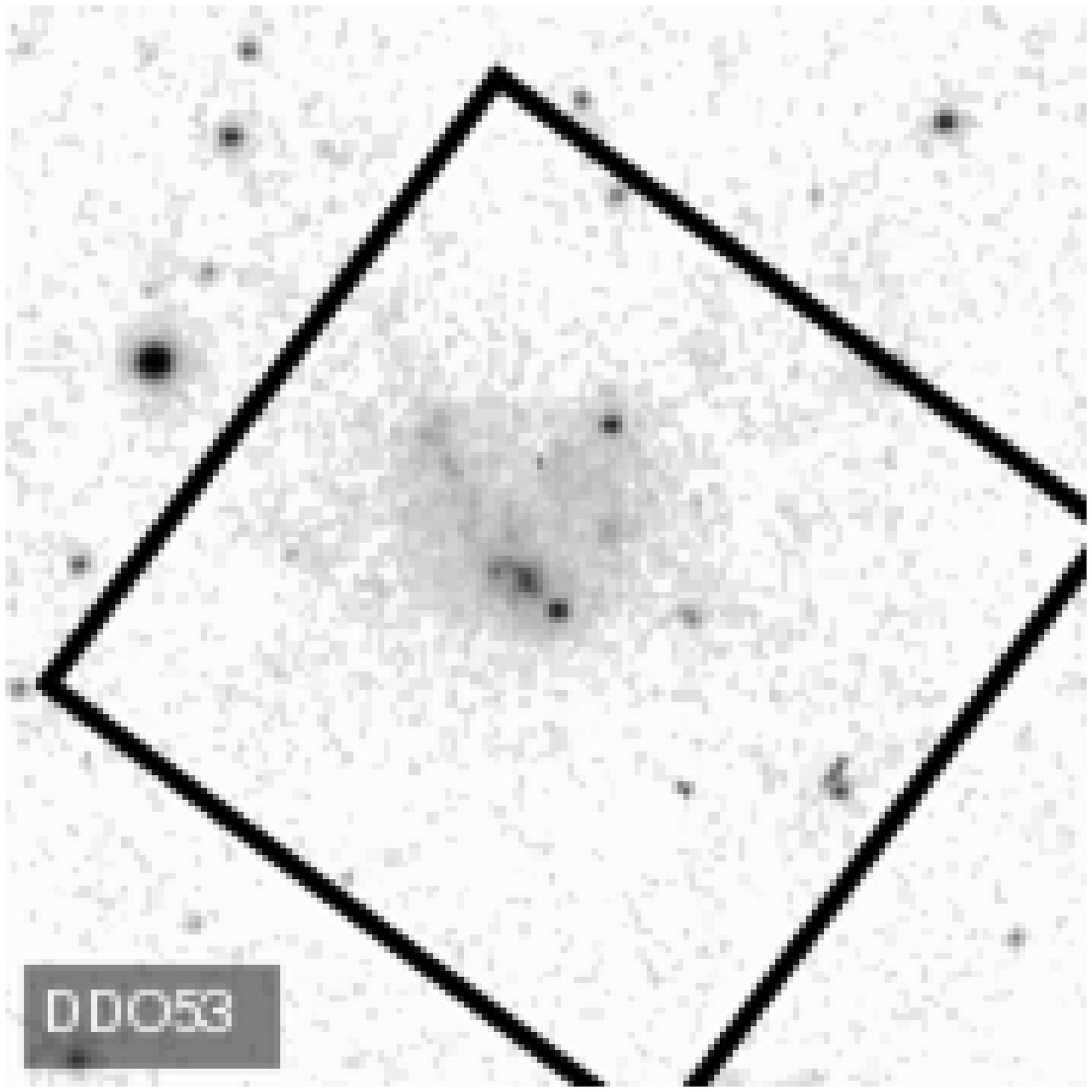}
\includegraphics[width=1.562in]{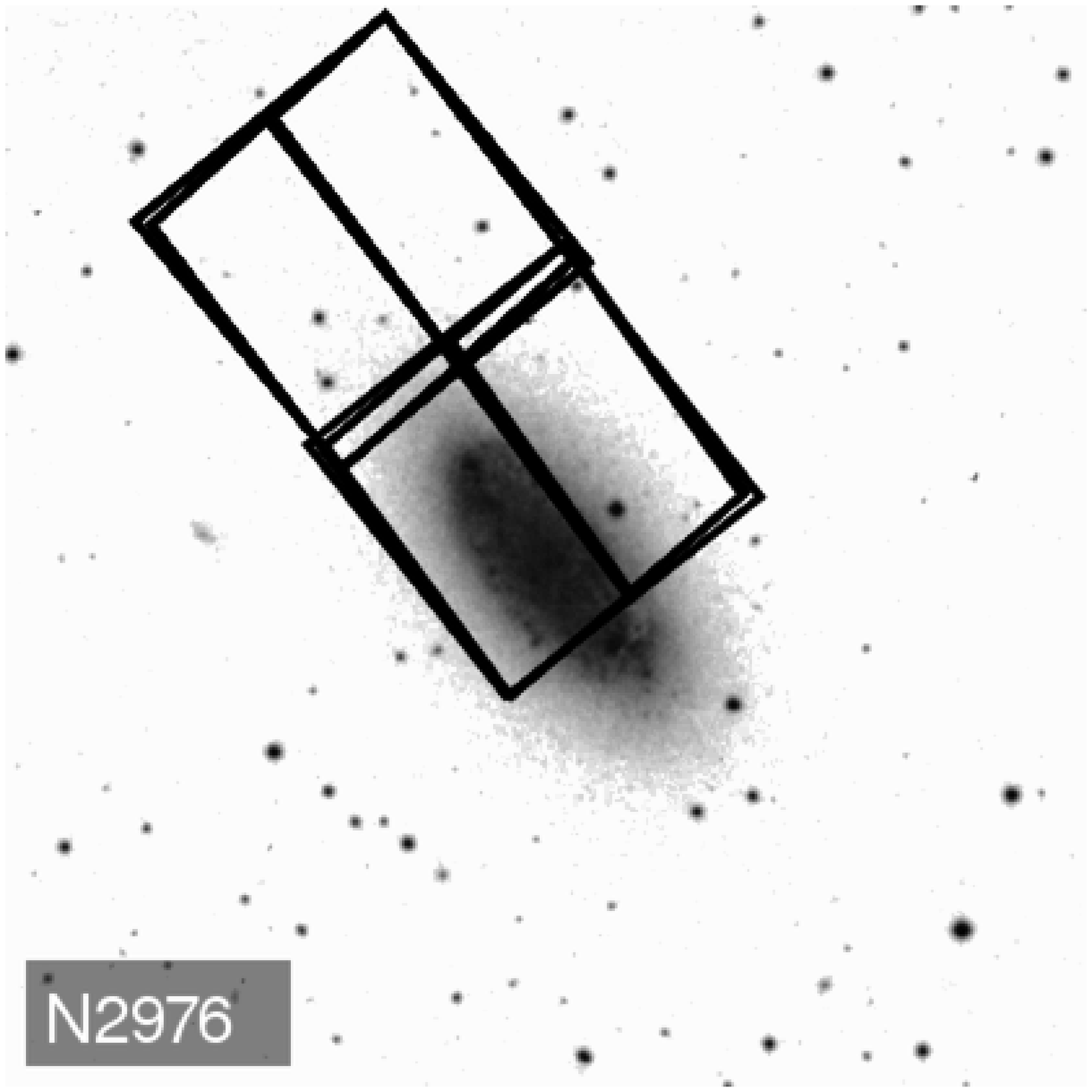}
\includegraphics[width=1.562in]{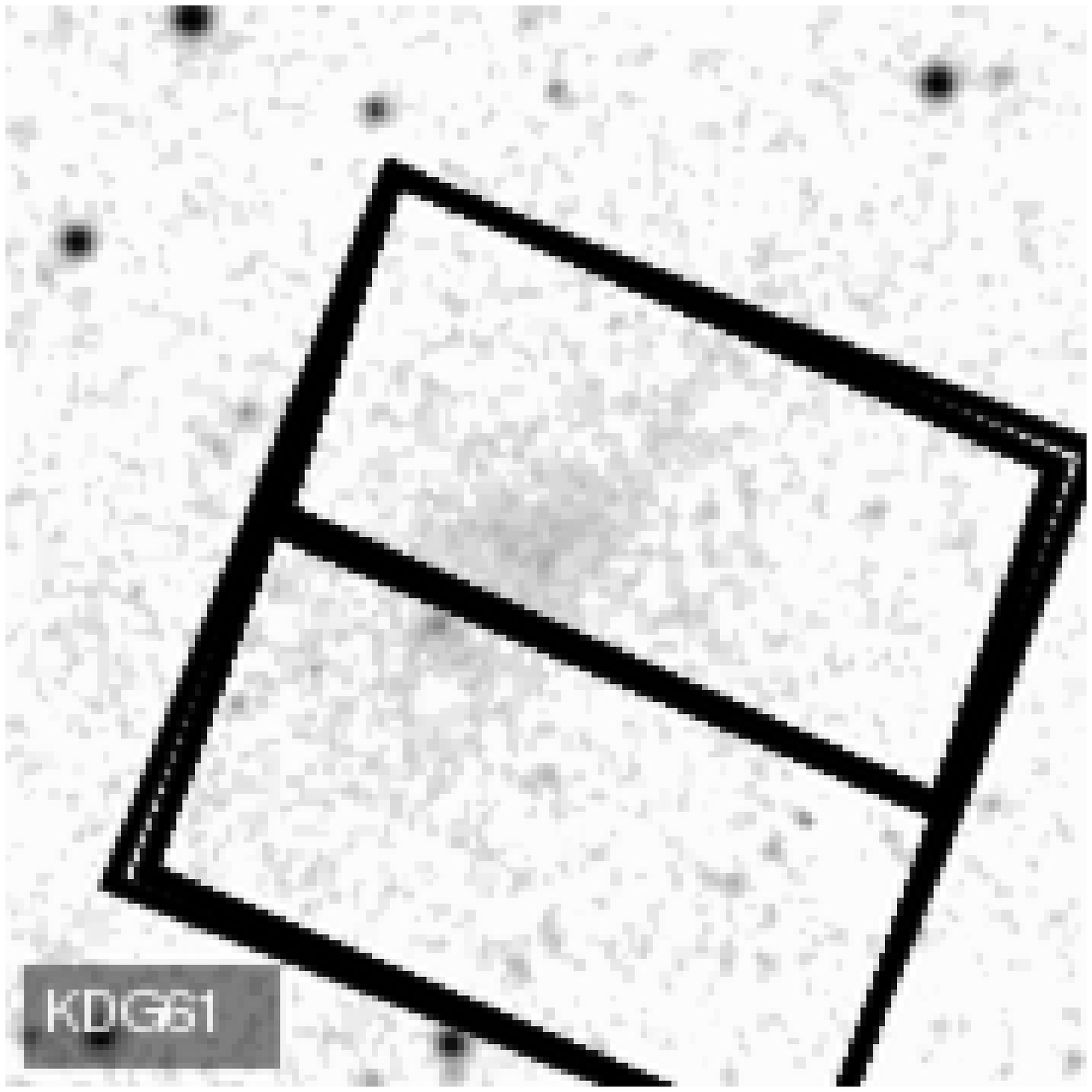}
}
\centerline{
\includegraphics[width=1.562in]{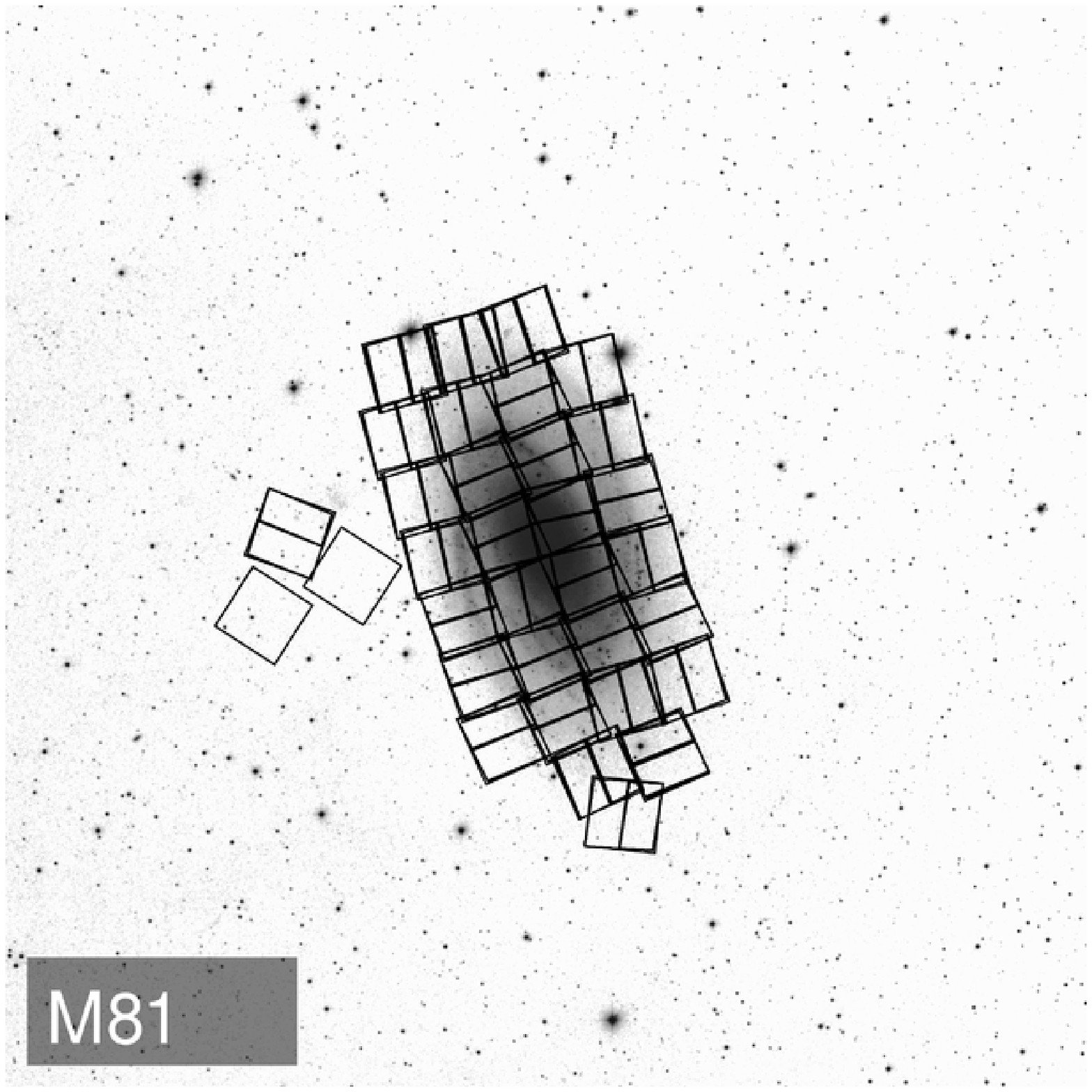}
\includegraphics[width=1.562in]{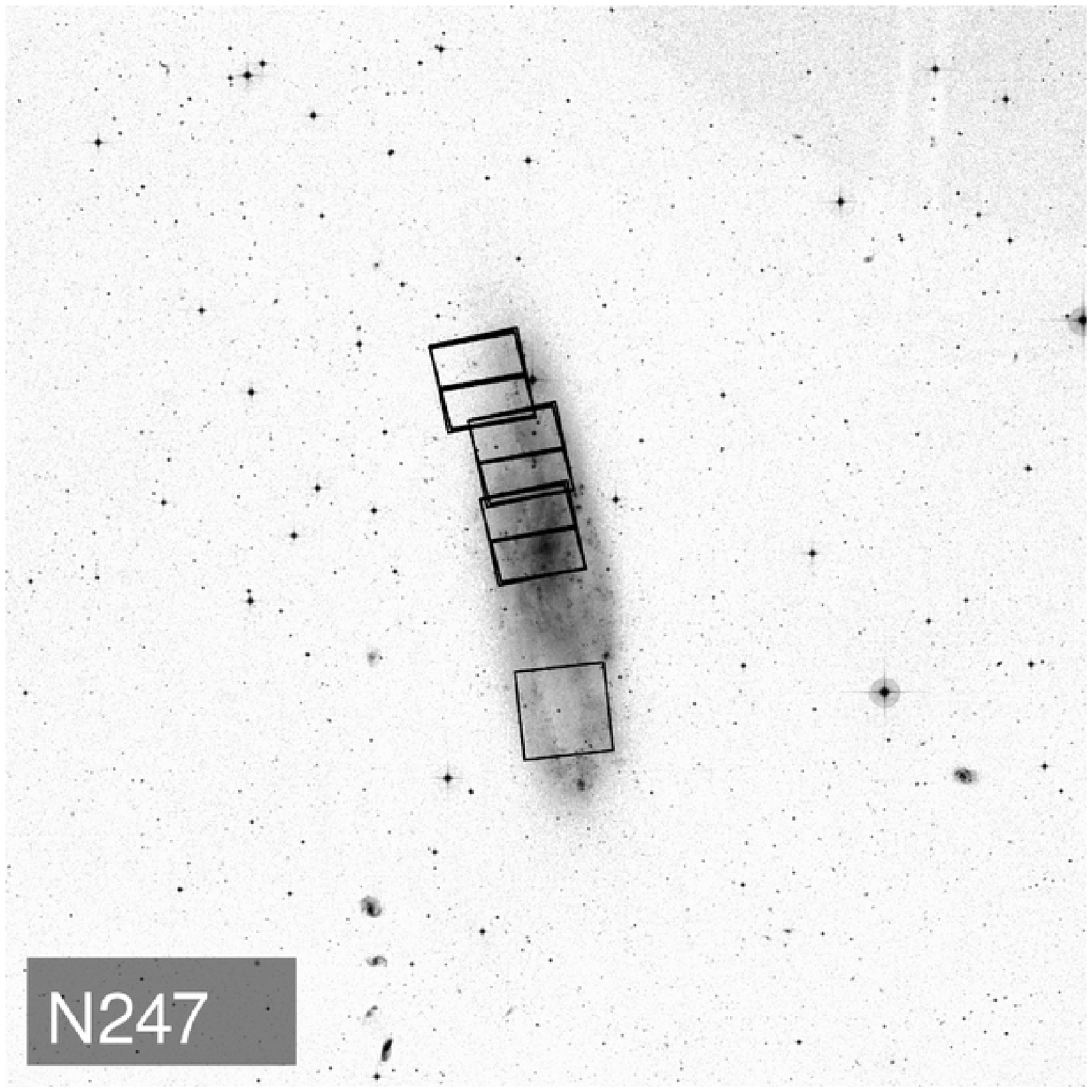}
\includegraphics[width=1.562in]{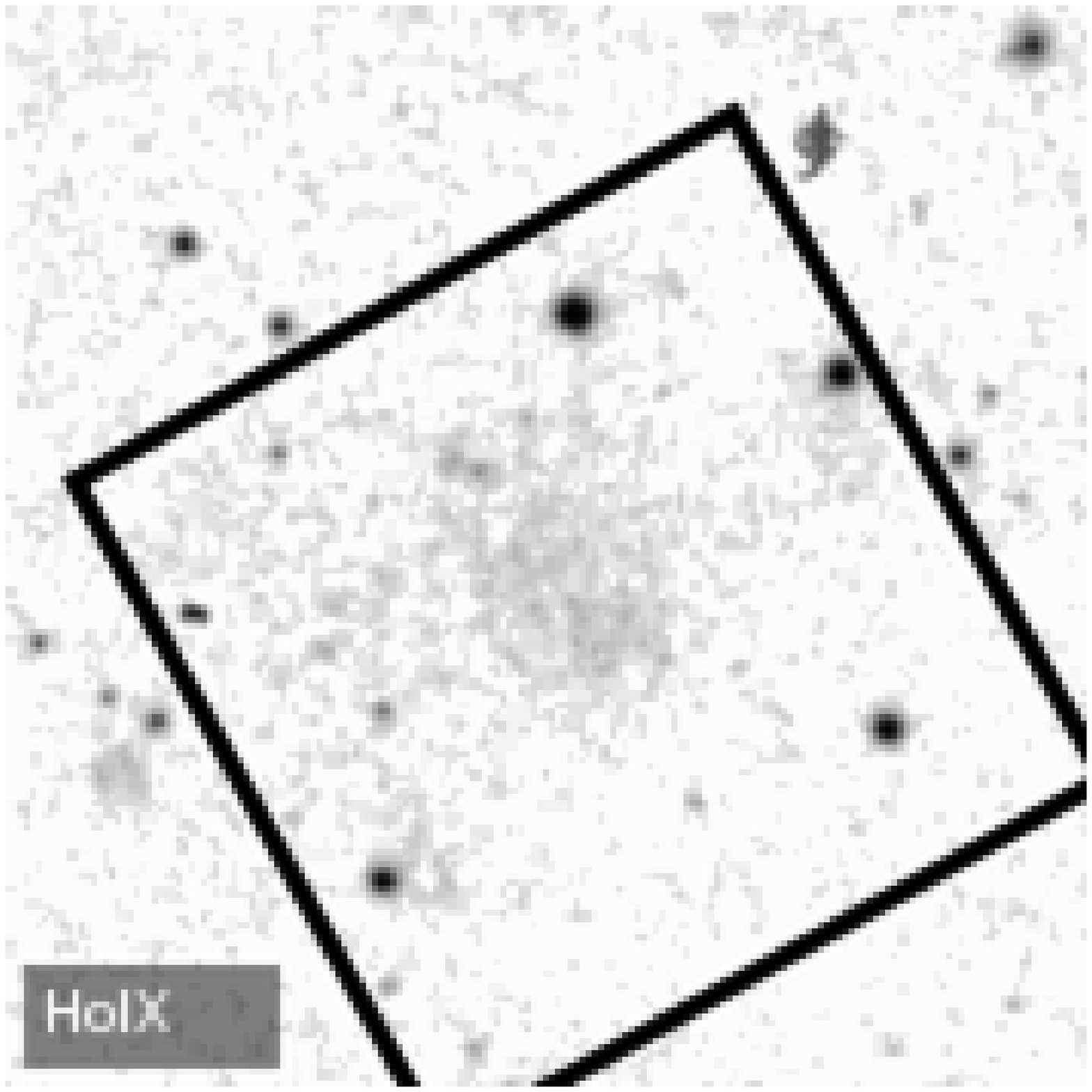}
\includegraphics[width=1.562in]{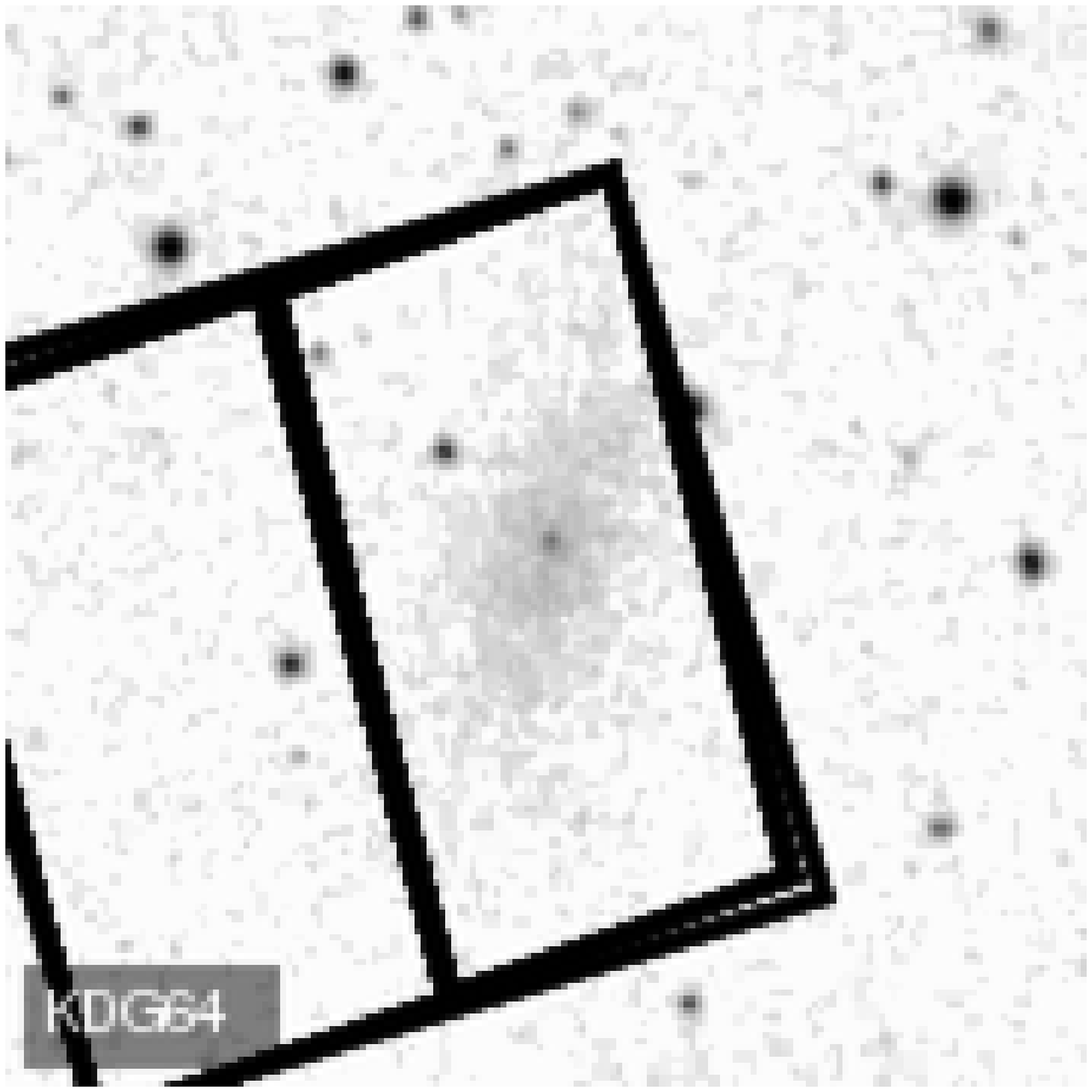}
}
\centerline{
\includegraphics[width=1.562in]{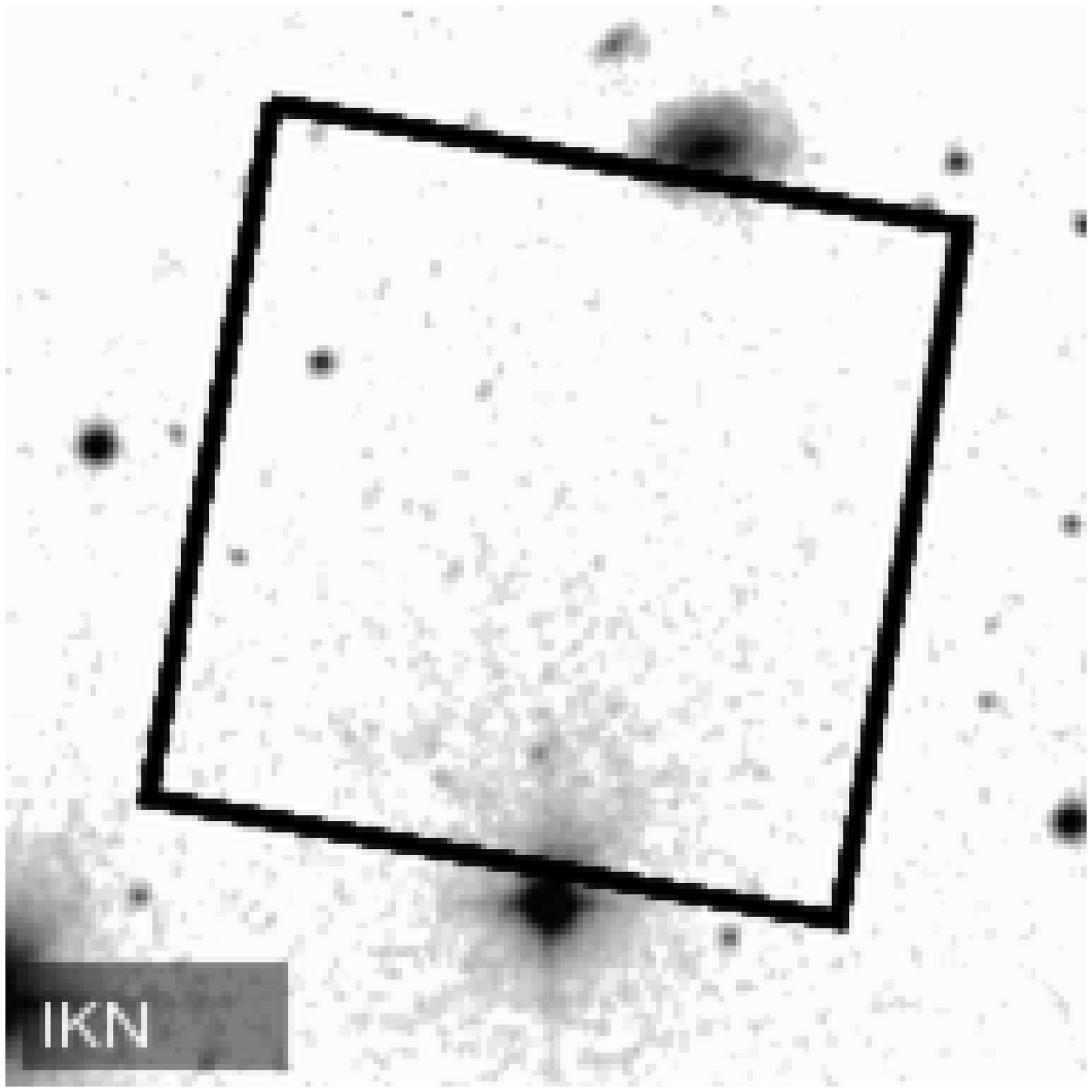}
\includegraphics[width=1.562in]{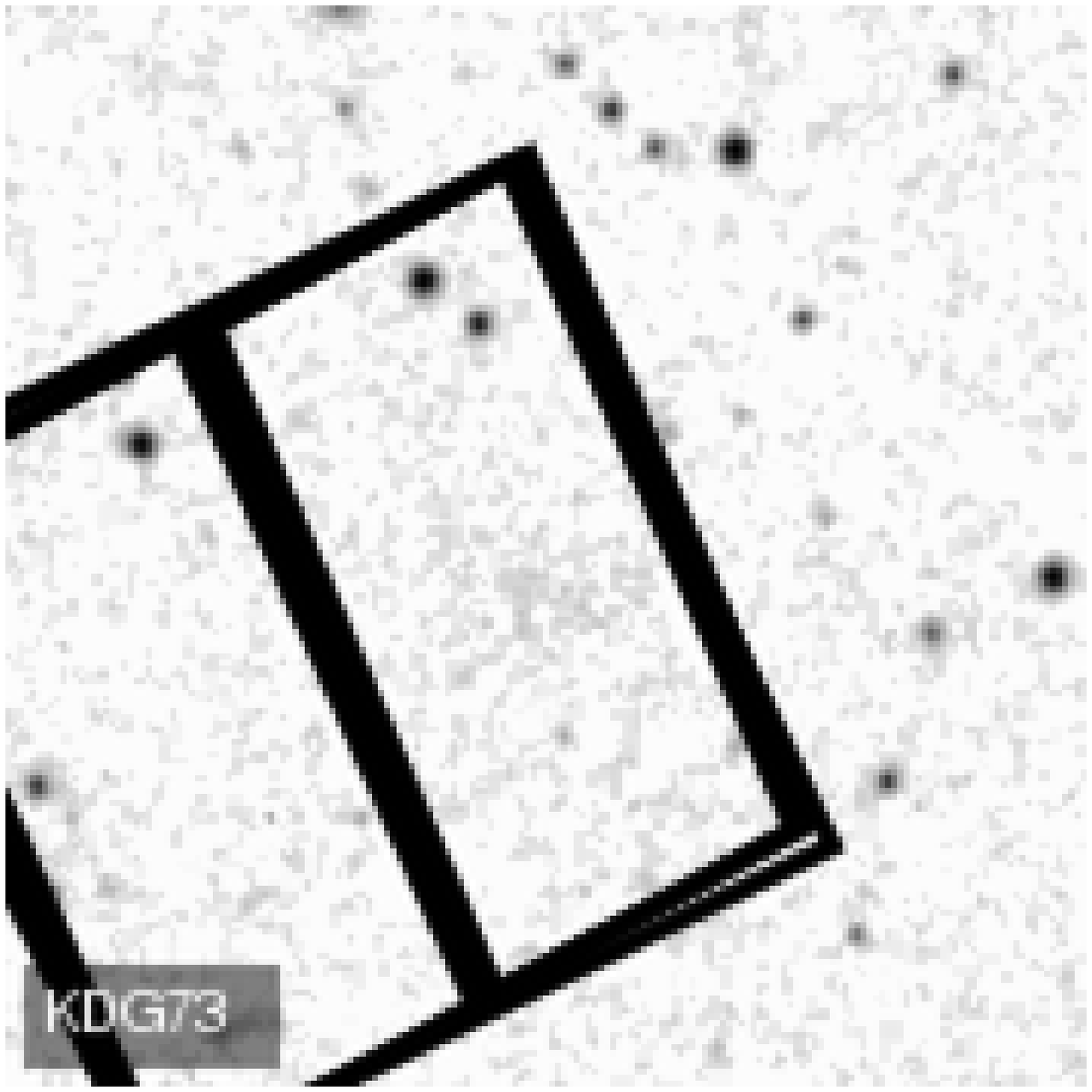}
\includegraphics[width=1.562in]{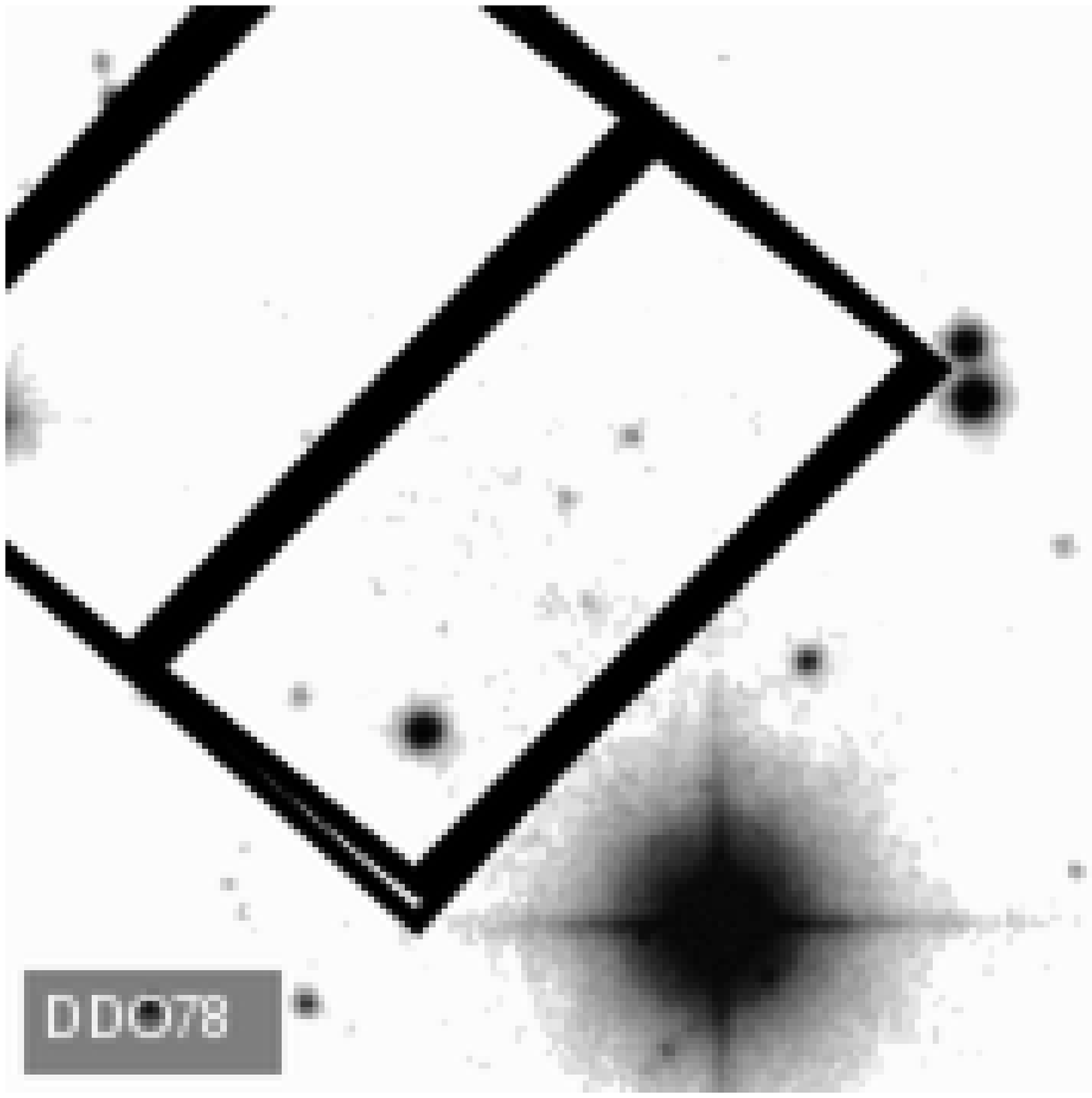}
\includegraphics[width=1.562in]{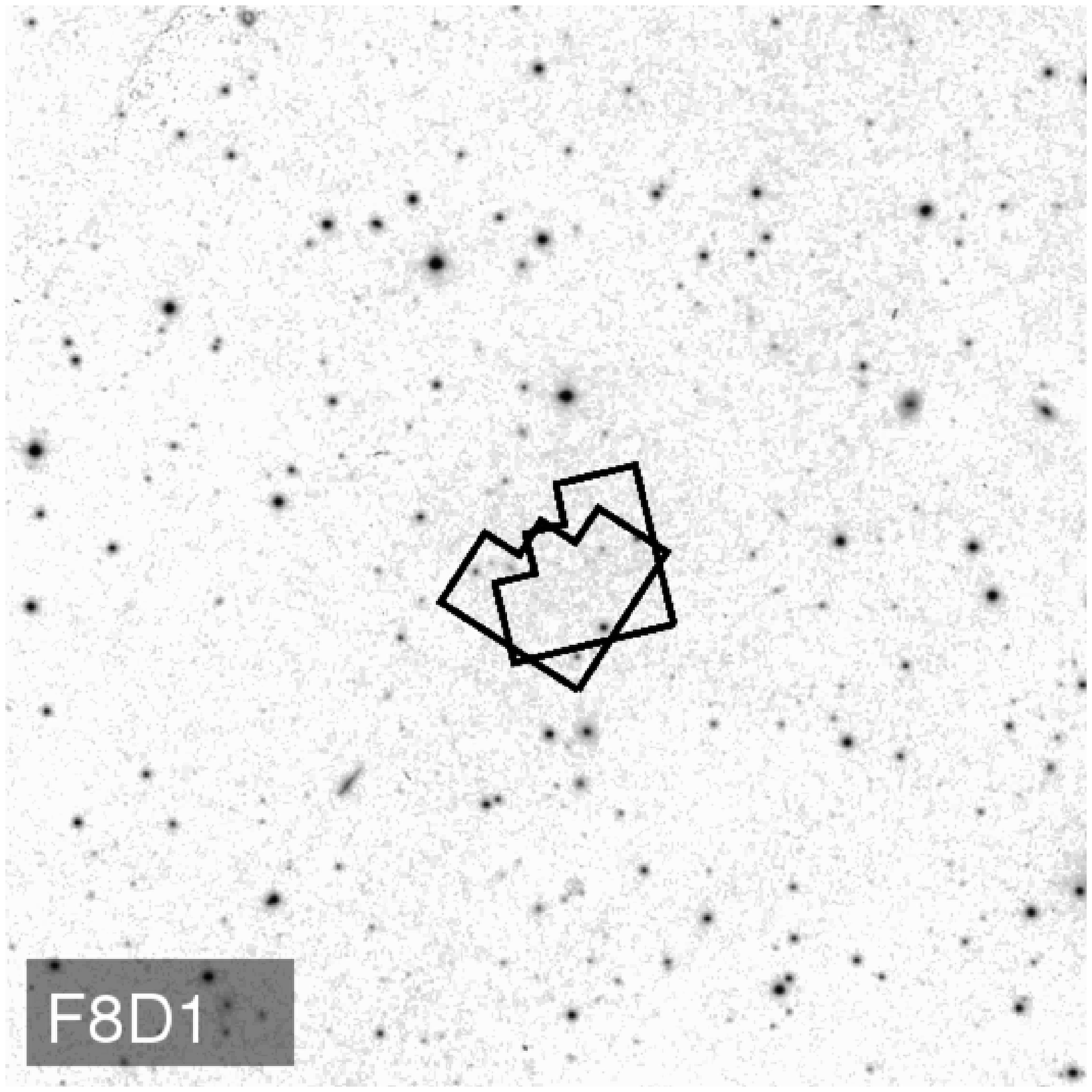}
}
\centerline{
\includegraphics[width=1.562in]{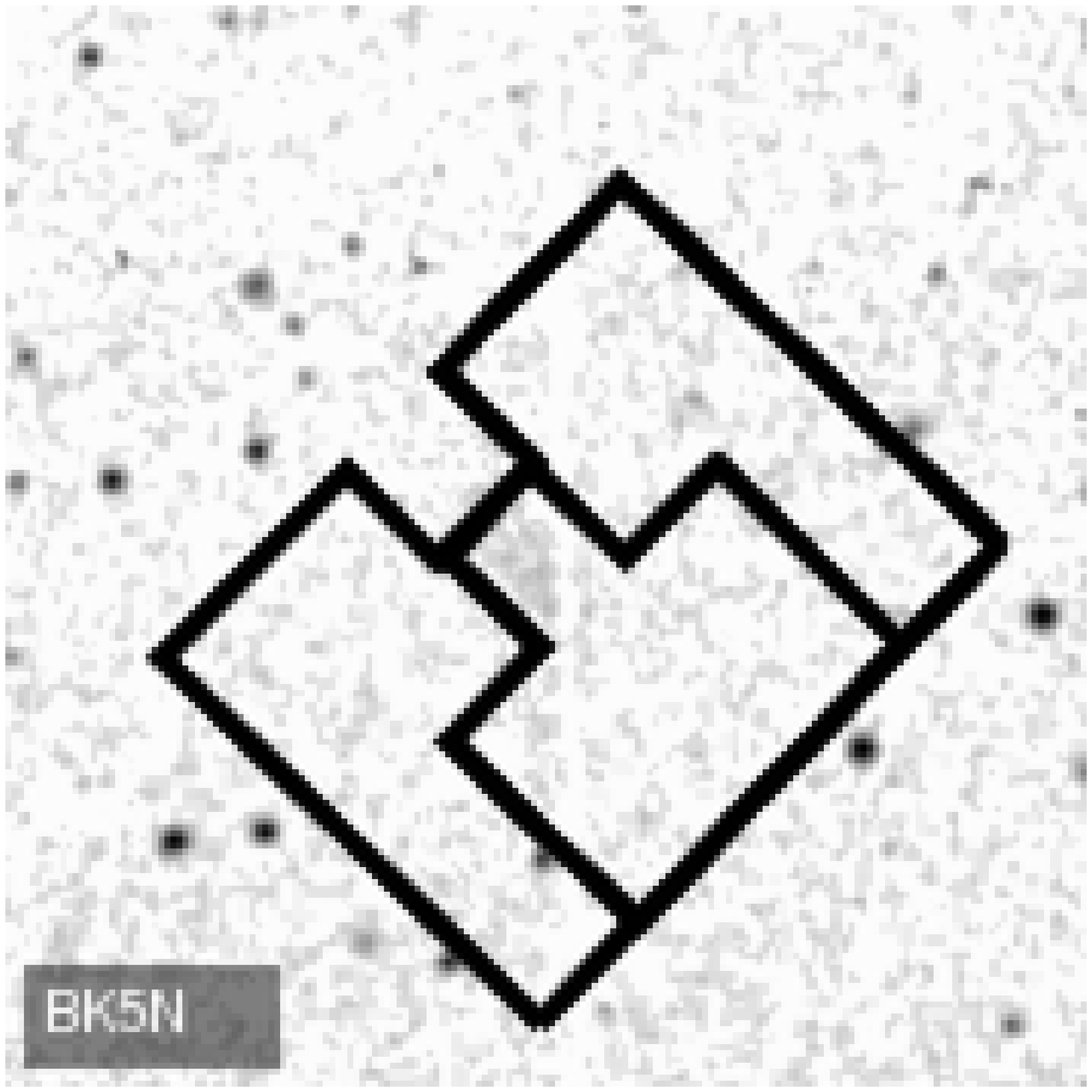}
\includegraphics[width=1.562in]{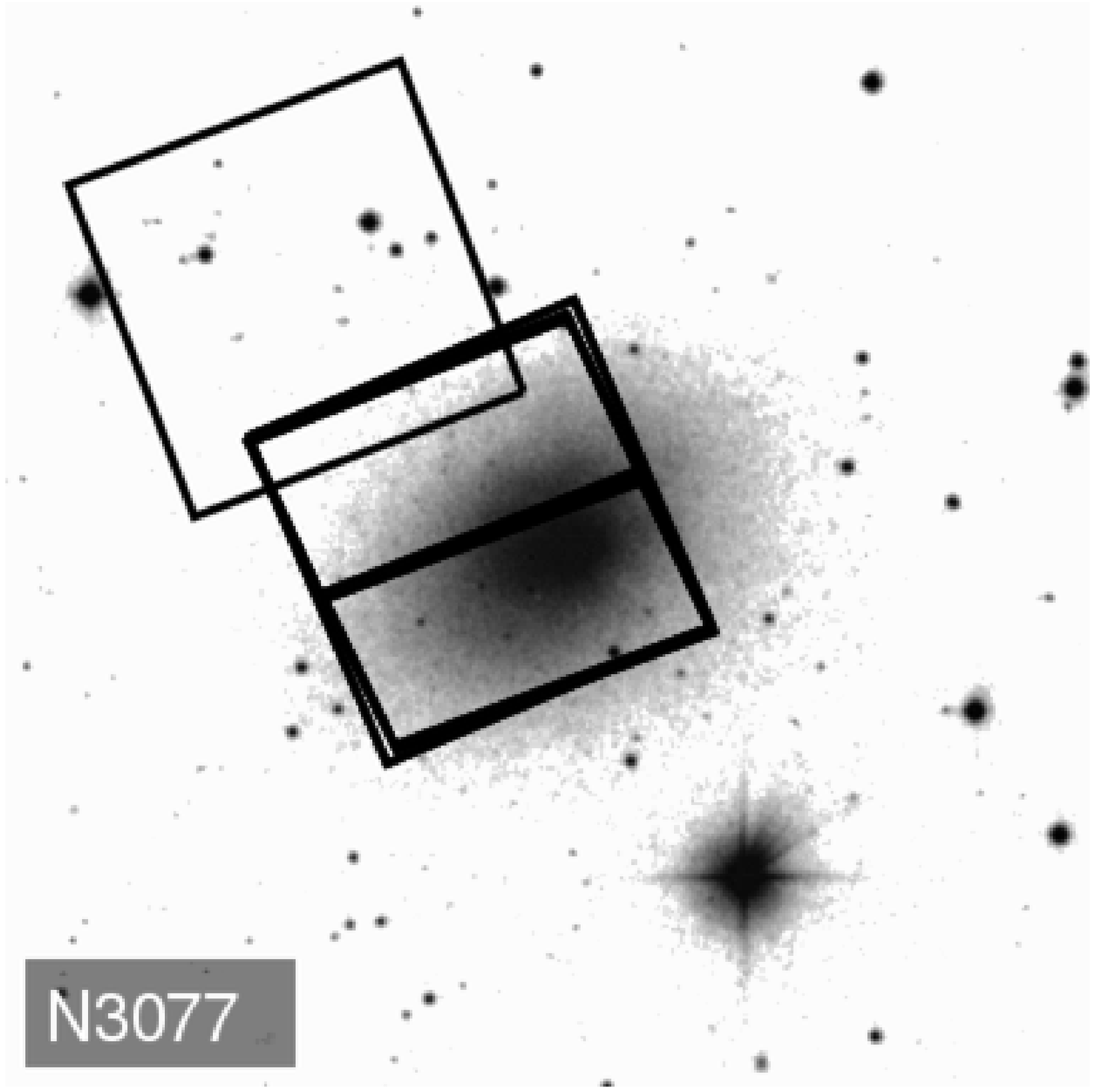}
\includegraphics[width=1.562in]{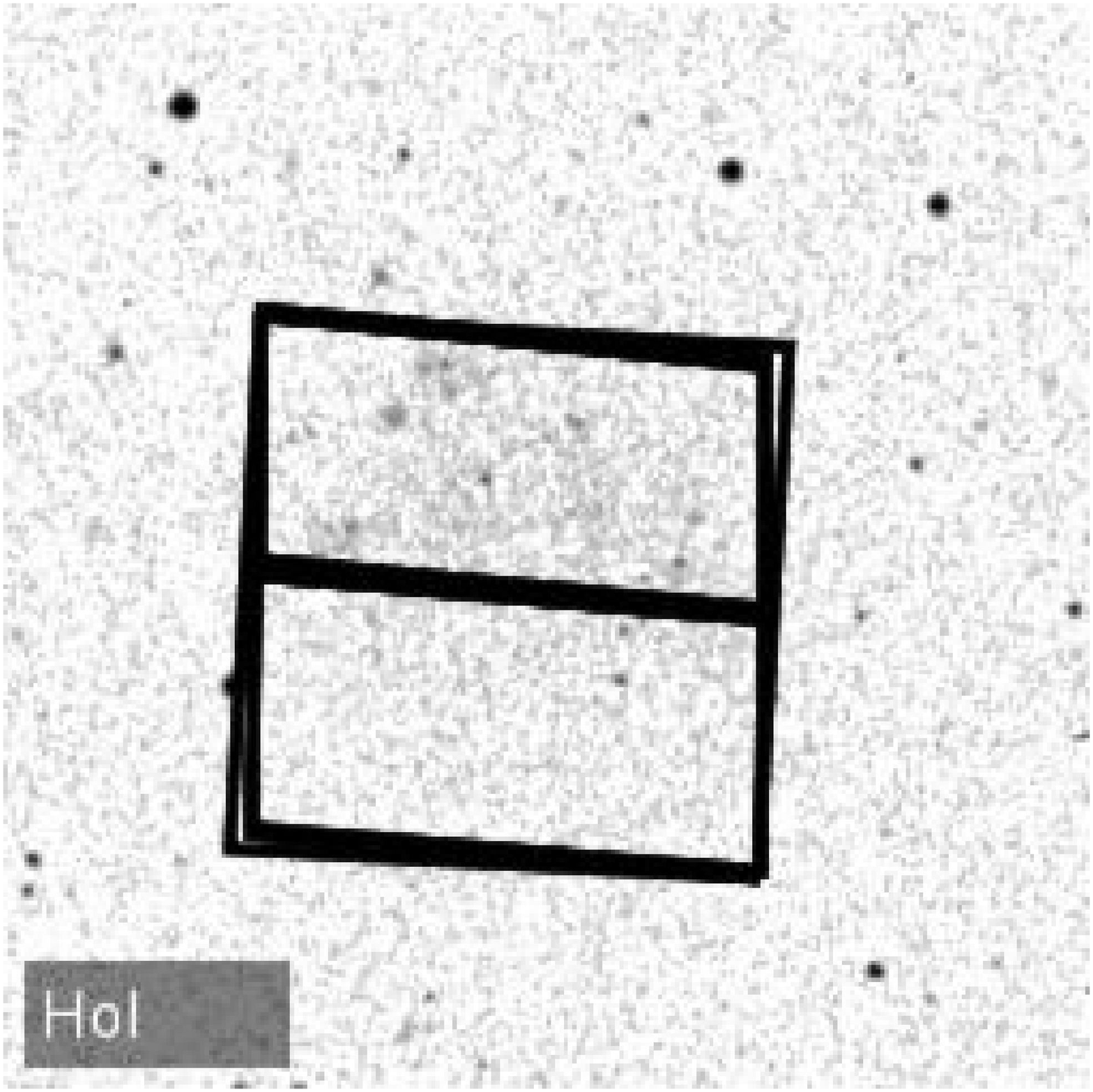}
\includegraphics[width=1.562in]{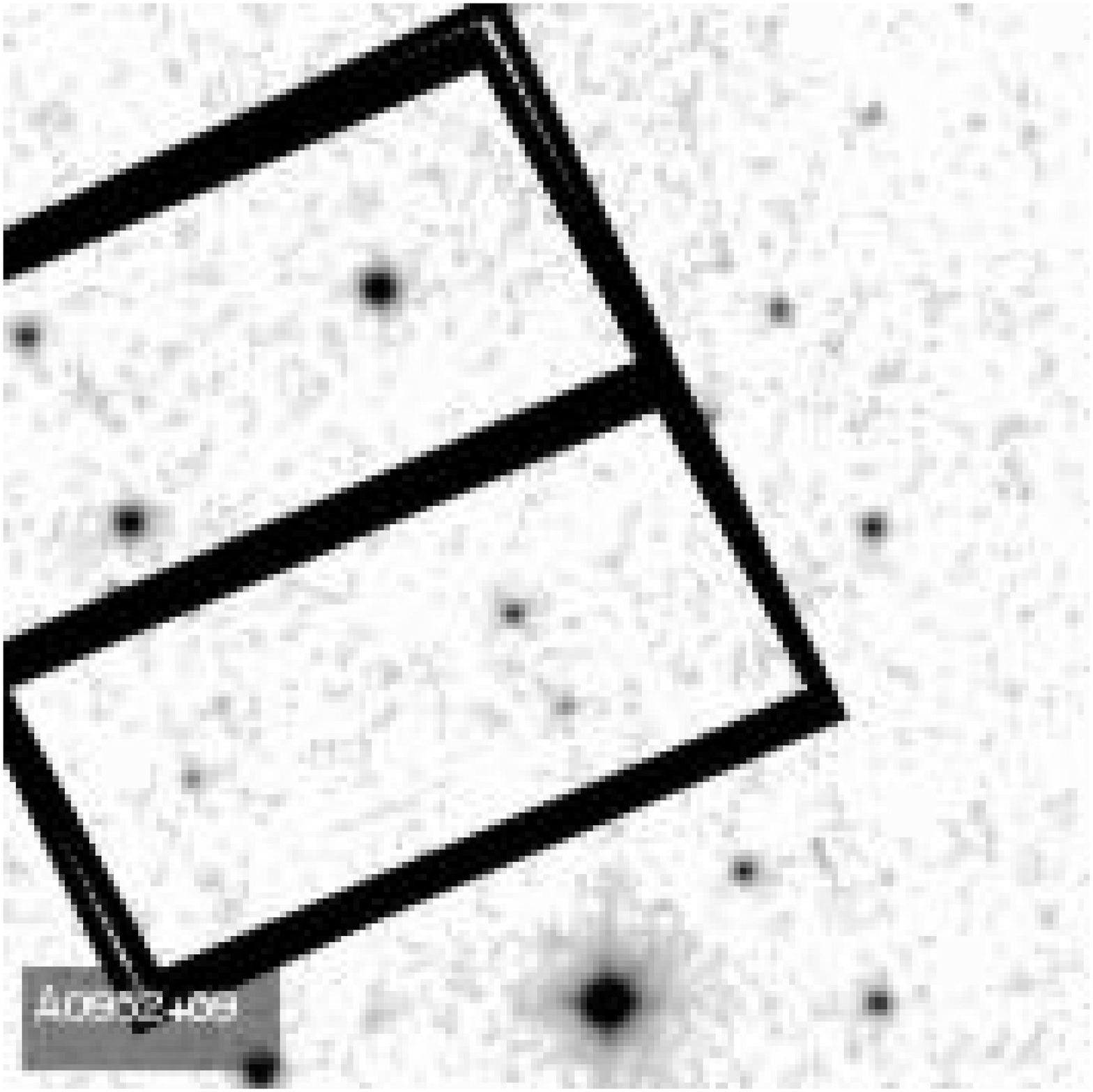}
}
\caption{
Field positions of images included in
Table~\ref{obstable}~\&~\ref{archivetable},
as described in Figure~\ref{overlayfig1}.
Figures are ordered from the upper left to the bottom right.
(a) FM1; (b) KK77; (c) KDG63; (d) M82; (e) KDG52; (f) DDO53; (g) N2976; (h) KDG61; (i) M81; (j) N247; (k) HoIX; (l) KDG64; (m) IKN; (n) KDG73; (o) DDO78; (p) F8D1; (q) BK5N; (r) N3077; (s) HoI; (t) A0952+69; 
    \label{overlayfig3}}
\end{figure}
\vfill
\clearpage
 
%-------------------
\begin{figure}[p]
\centerline{
\includegraphics[width=1.562in]{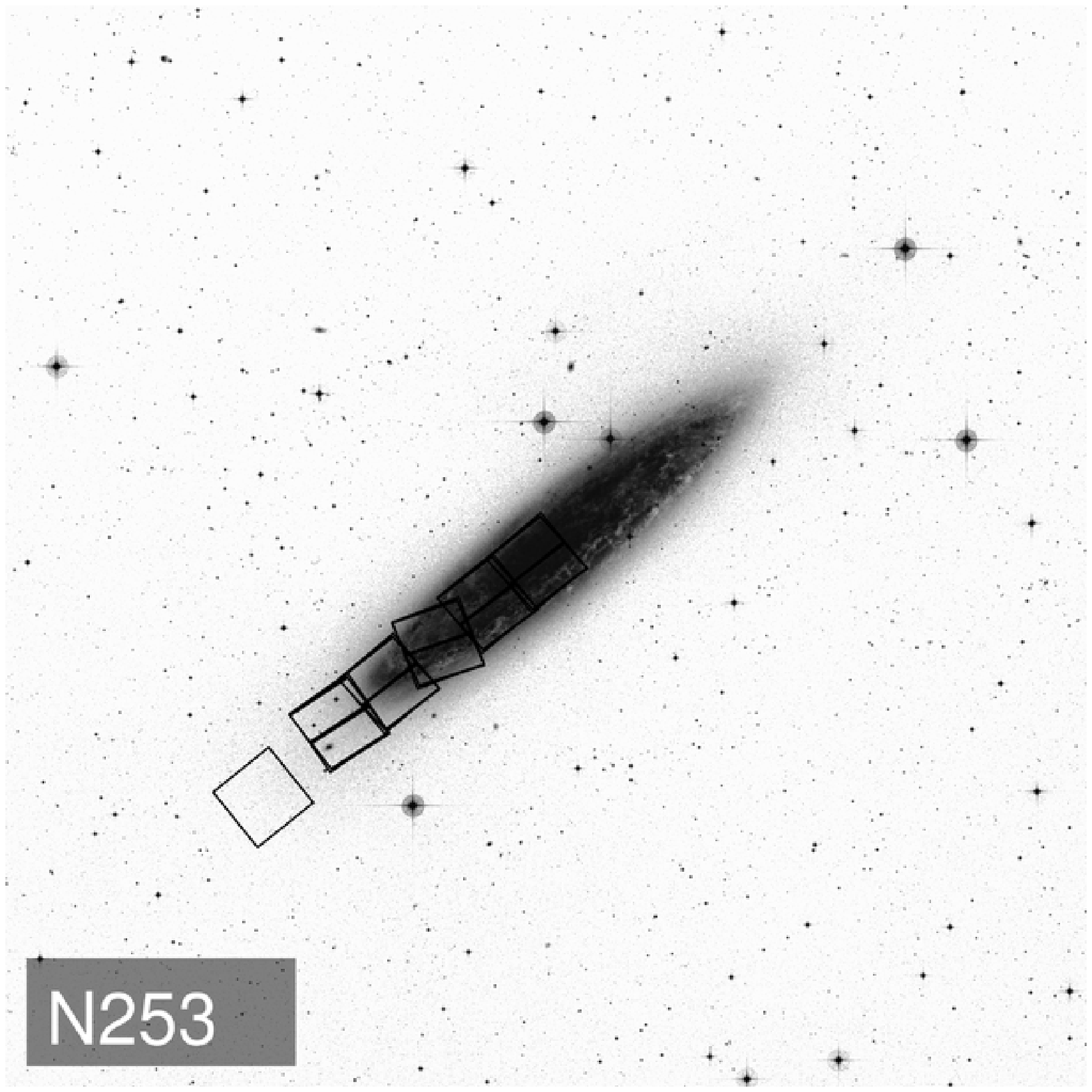}
\includegraphics[width=1.562in]{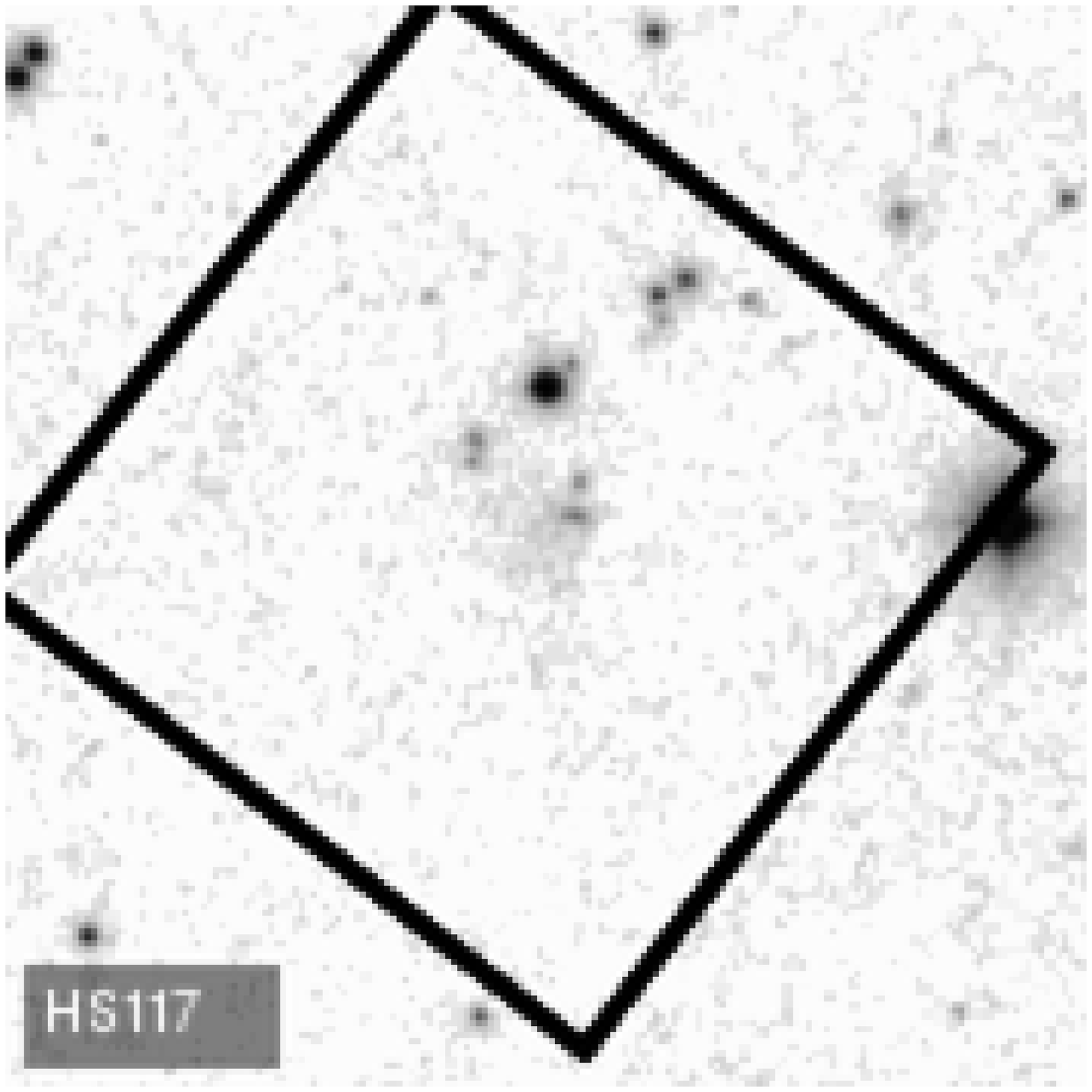}
\includegraphics[width=1.562in]{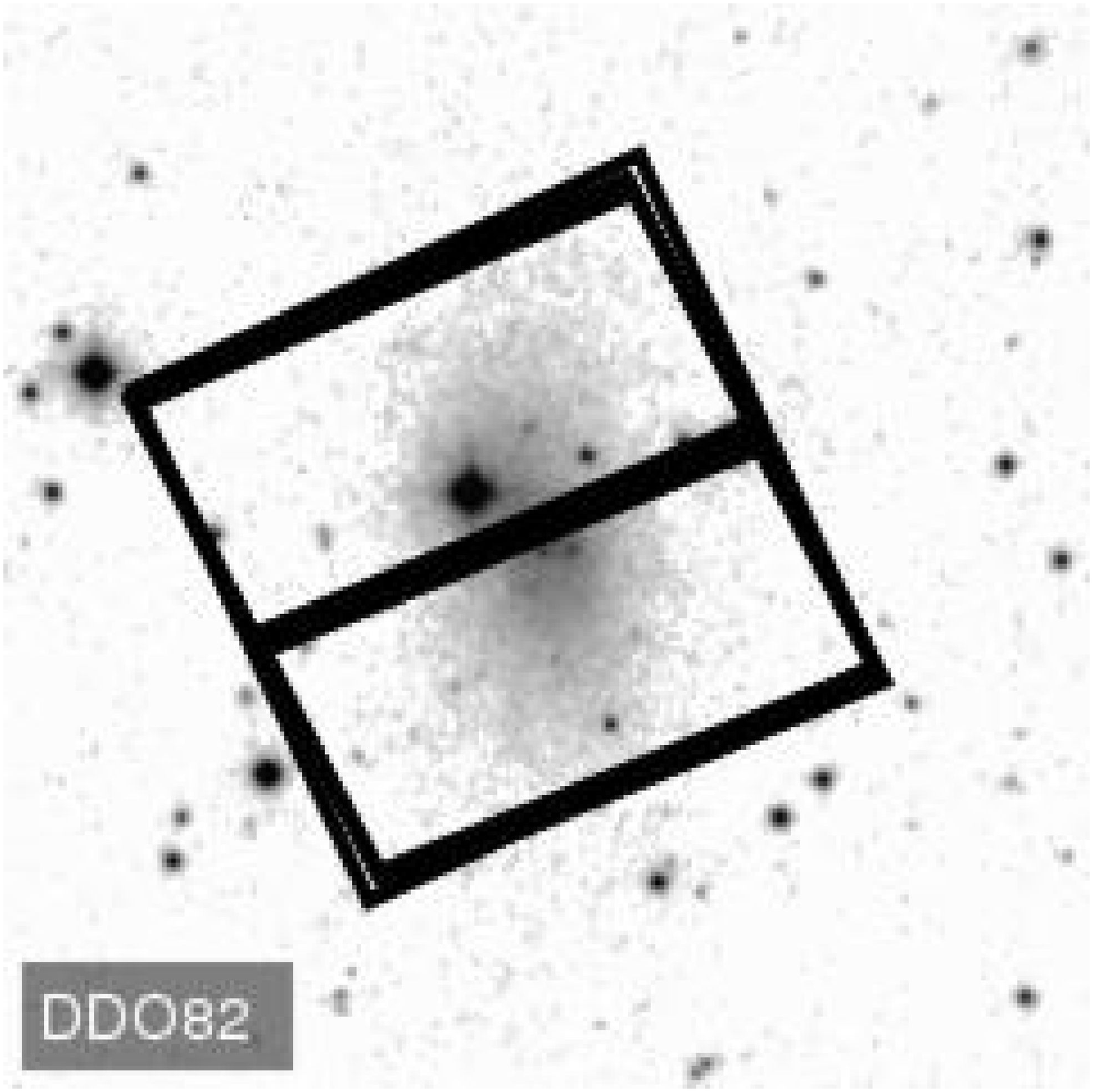}
\includegraphics[width=1.562in]{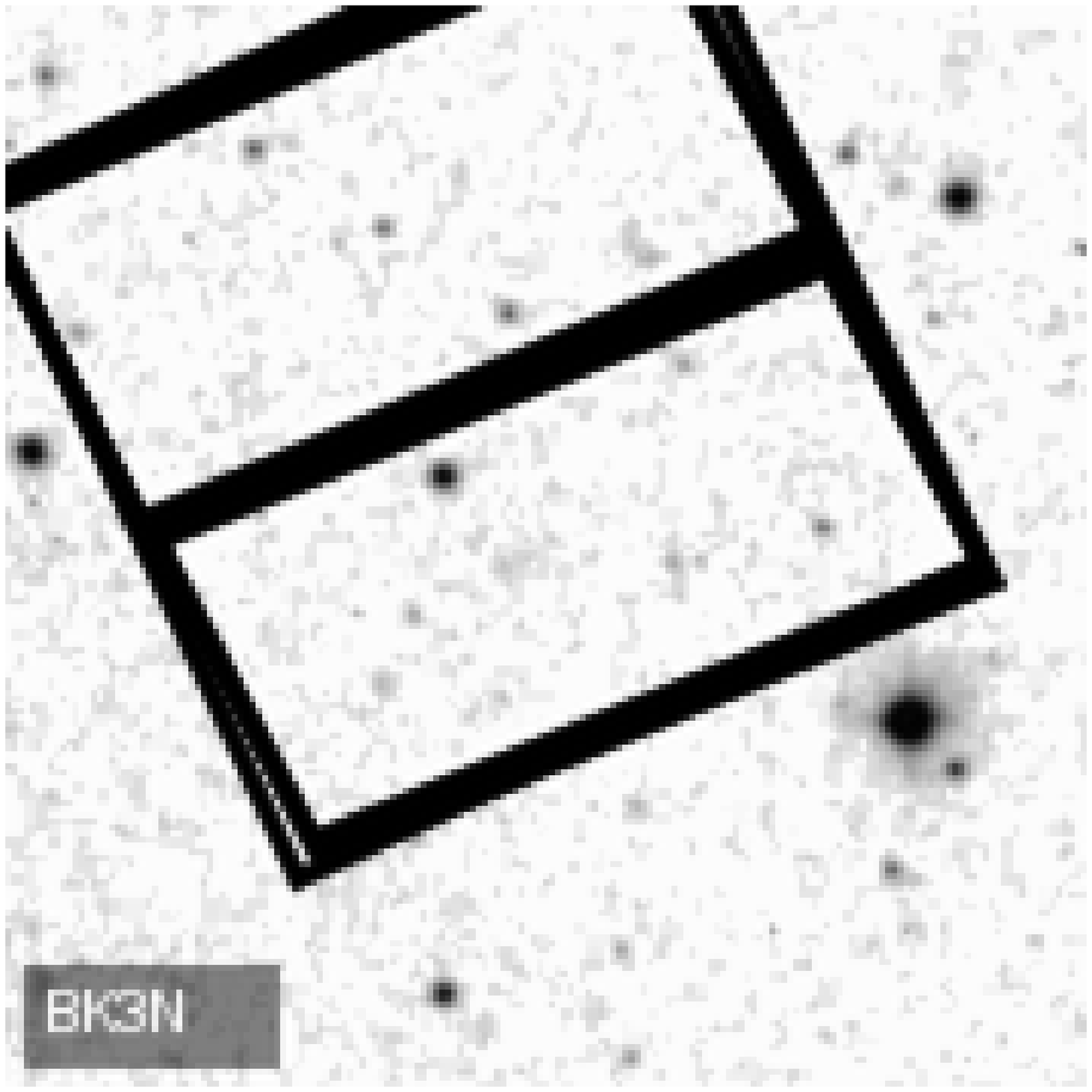}
}
\centerline{
\includegraphics[width=1.562in]{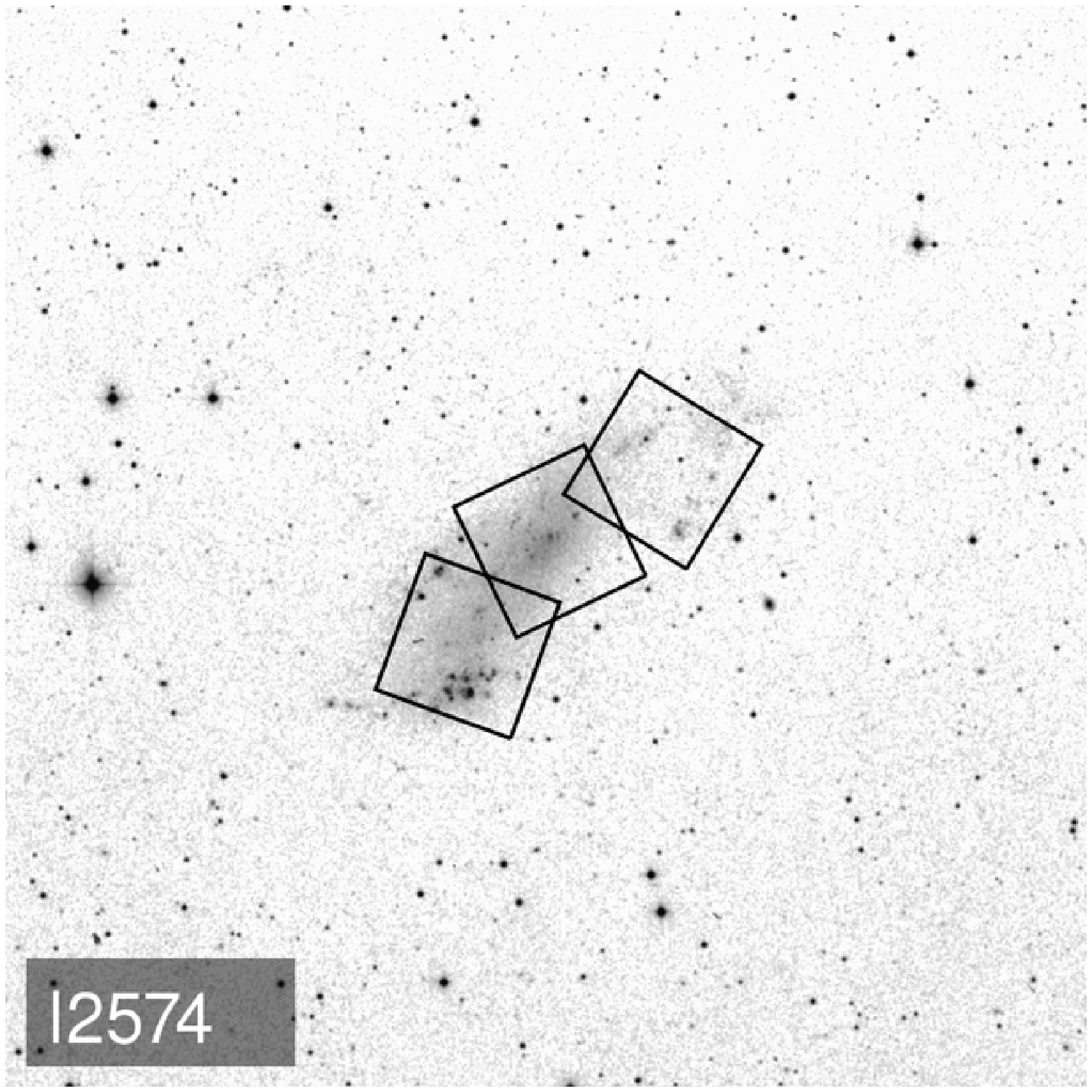}
\includegraphics[width=1.562in]{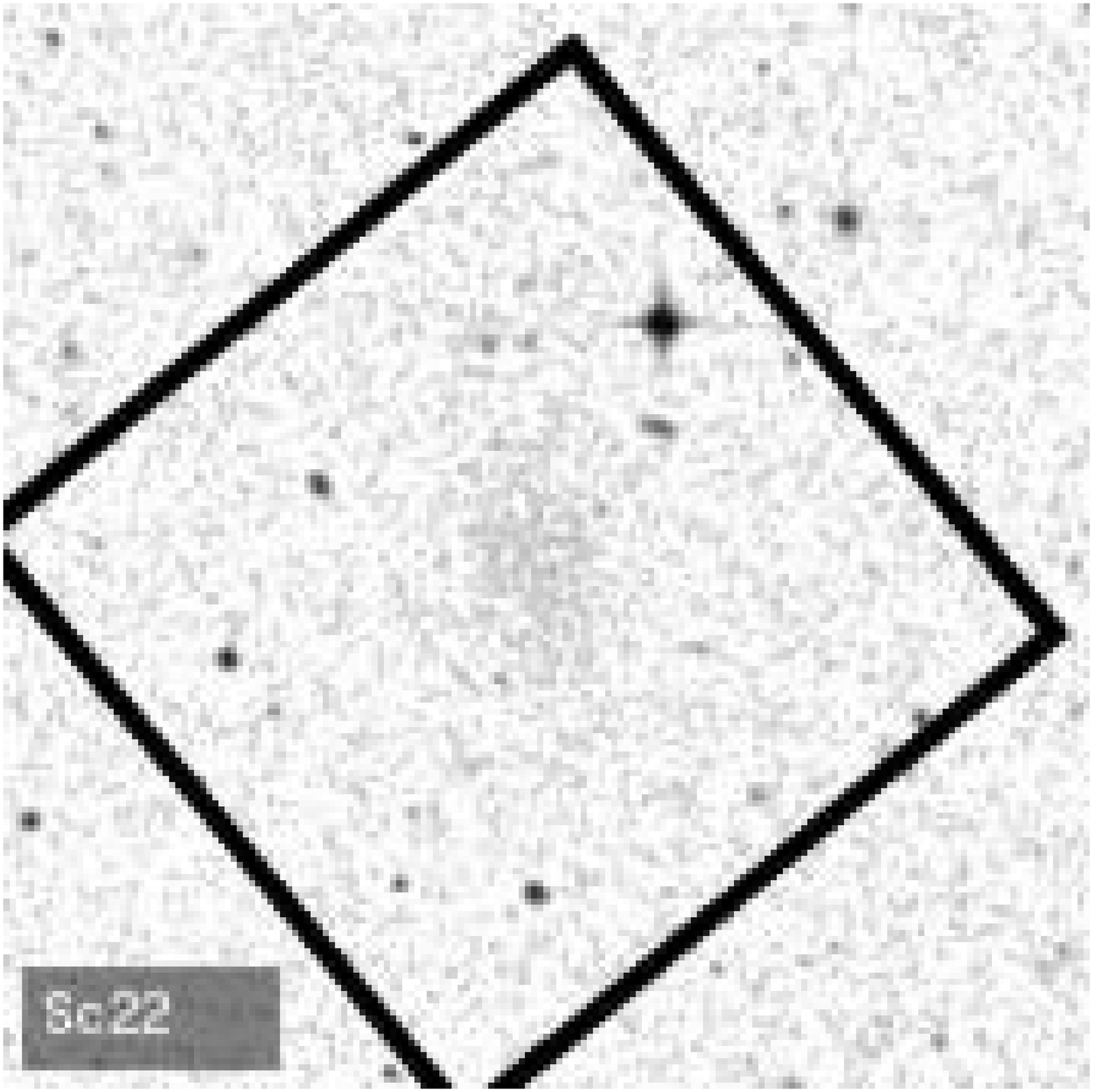}
}
\caption{
Field positions of images included in
Table~\ref{obstable}~\&~\ref{archivetable},
as described in Figure~\ref{overlayfig1}.
Figures are ordered from the upper left to the bottom right.
(a) N253; (b) HS117; (c) DDO82; (d) BK3N; (e) I2574; (f) Sc22; 
    \label{overlayfig4}}
\end{figure}
\vfill
%\clearpage

%-------------------
\begin{figure}[p]
\centerline{
\includegraphics[width=3.5in]{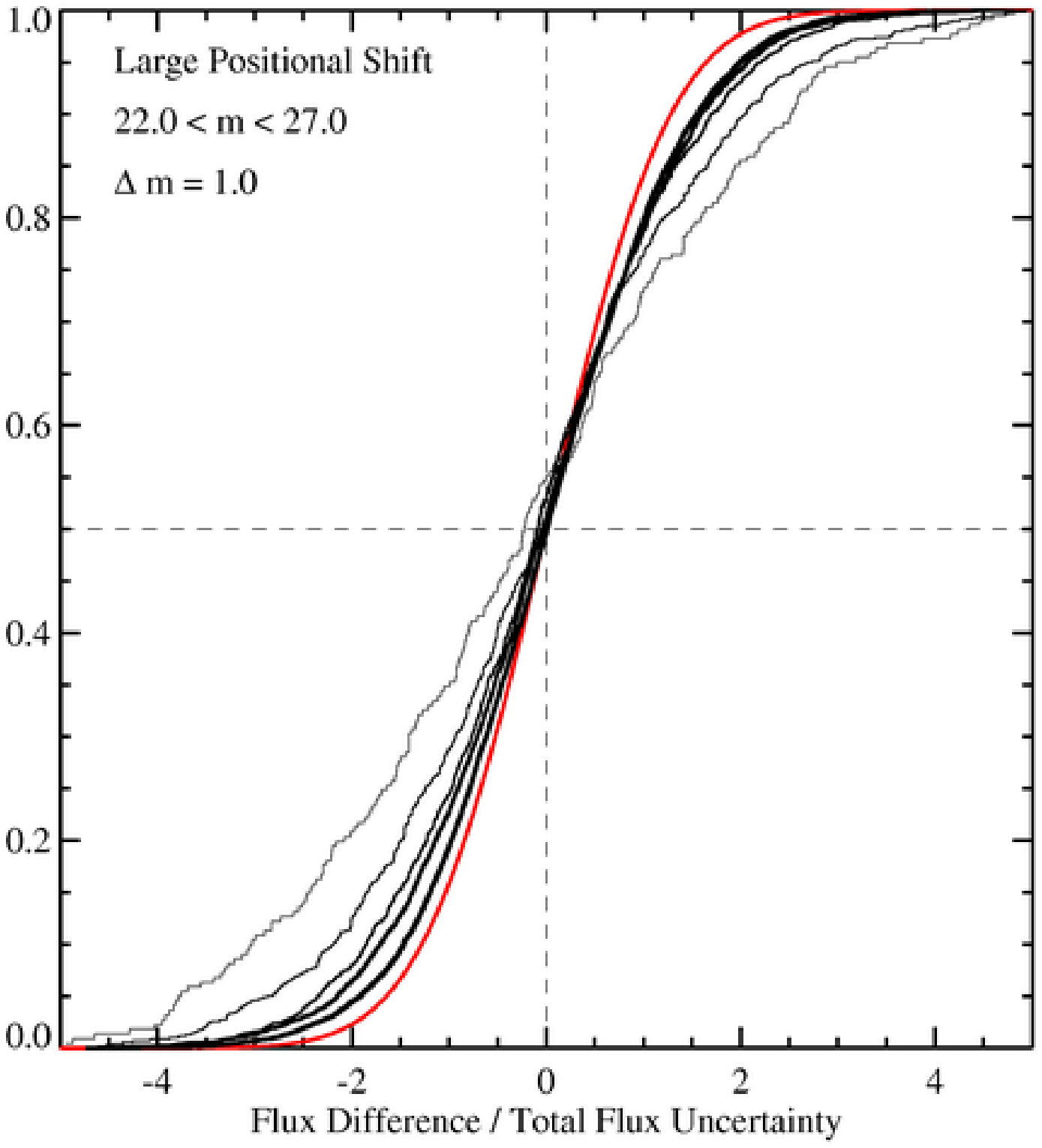}
\includegraphics[width=3.5in]{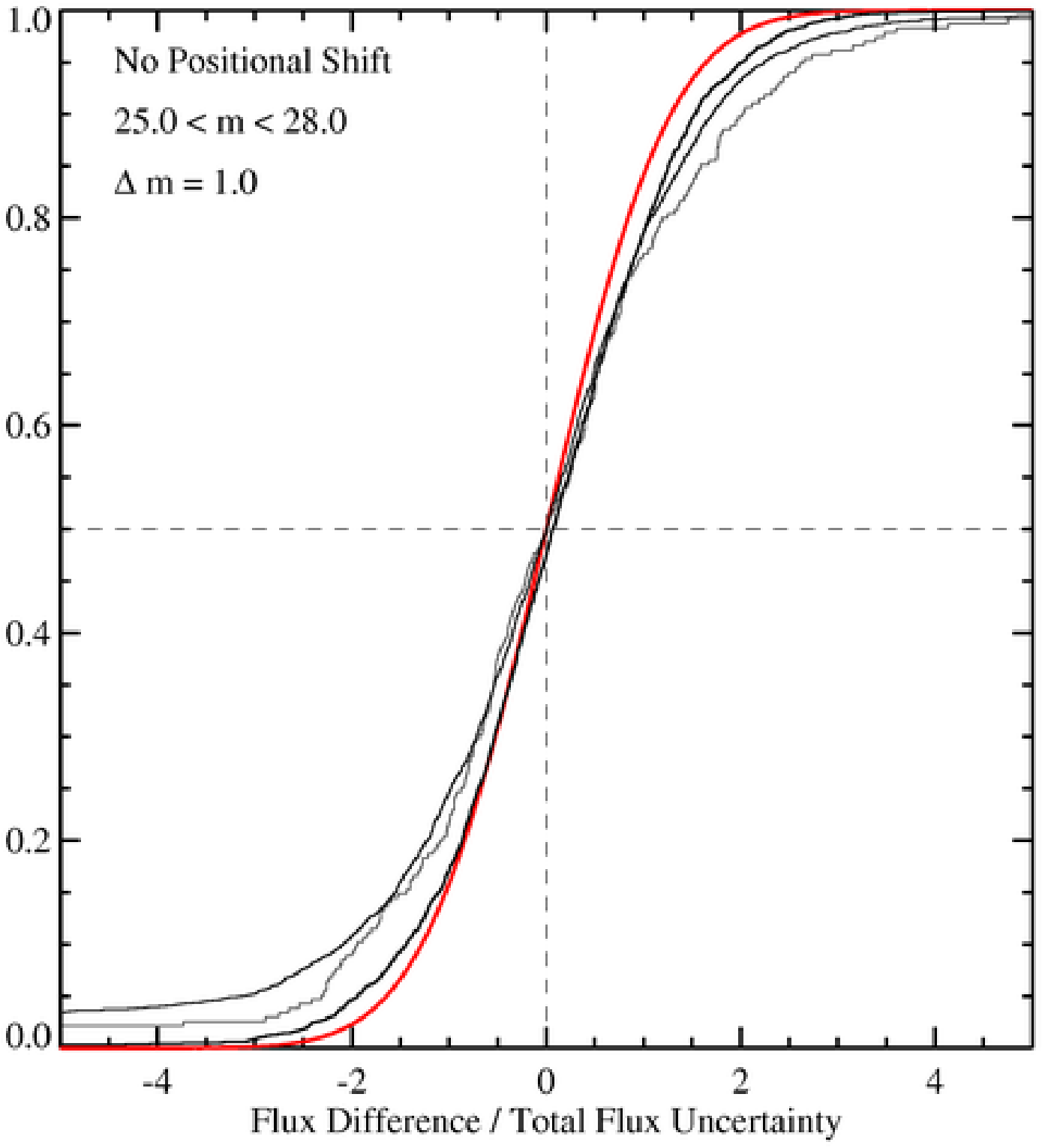}
}
\caption{The cumulative distribution of $F814W$ magnitude differences
between stars measured in two widely separated but overlapping ACS
exposures between NGC~300's WIDE1 and WIDE2 (left), and between stars
measured in successive 1-orbit ACS exposures at the same pointing in
M81's deep field (right).  Magnitude differences are scaled by the
reported magnitude error for each star, added in quadrature for the
case of repeat measurements.  The distributions are calculated in bins
of 1 magnitude, with the heaviest line indicating the faintest
bin.  The red curve indicates the expectation for a perfect Gaussian
error distribution.  In both cases, the distribution of errors is only
slightly broader than a Gaussian, and systematic errors are swamped by
photometric uncertainties.
	 \label{magdifffig}}
\end{figure}
\vfill
\clearpage

%\begin{figure}[p]
%\centerline{
%%\includegraphics[width=6.5in]{Figures/mag_vs_magerr.ps}
%\includegraphics[width=6.5in]{f4.eps}
%}
%%
%\caption{The $F814W$ magnitude errors reported for NGC~300 WIDE1, as a
%function of magnitude.  Vertical lines indicate the boundaries of the
%bins used in generating the magnitude difference distributions in
%Figure~\ref{magdifffig}.  The magnitude errors are quantized at the
%0.001 level, leading to the broader distributions seen in
%Figure~\ref{magdifffig} at bright magnitudes.  The red points indicate
%stars that have been flagged as having a small number of bad pixels
%(either 1 central pixel or 2-4 adjacent pixels).  The distinct
%sequence at higher errors are stars that were not well-measured in all
%of the dithered images, due to falling on a bad pixel or cosmic ray in
%some of the images.  Stars in the chip gap or near and edge have been
%clipped.
%	 \label{magvsmagerrfig}}
%%
%\end{figure}
%\vfill
%\clearpage

%-------------------
\begin{figure}[p]
\centerline{
\includegraphics[height=4.2in]{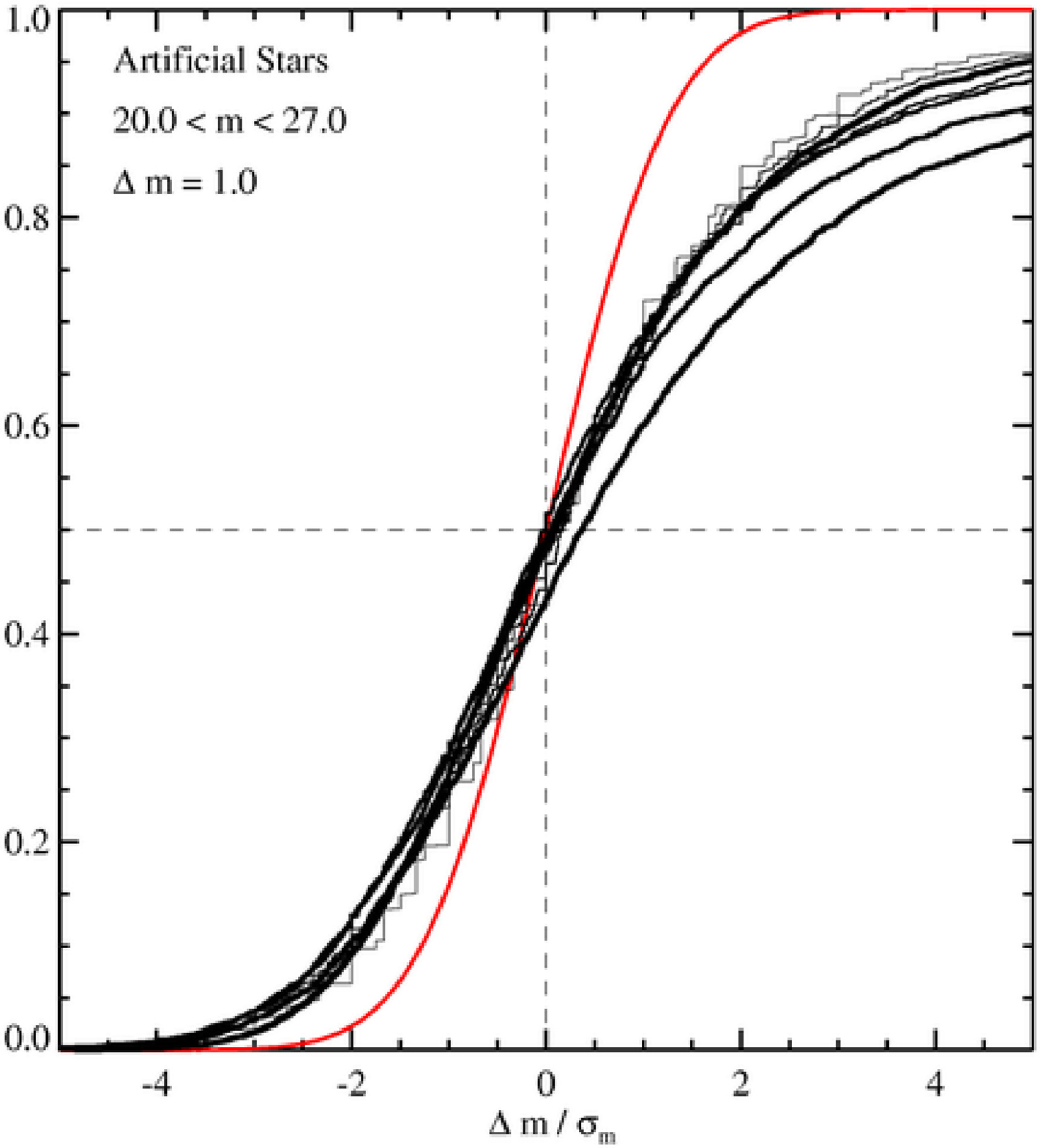}
}
\centerline{
\includegraphics[height=2in]{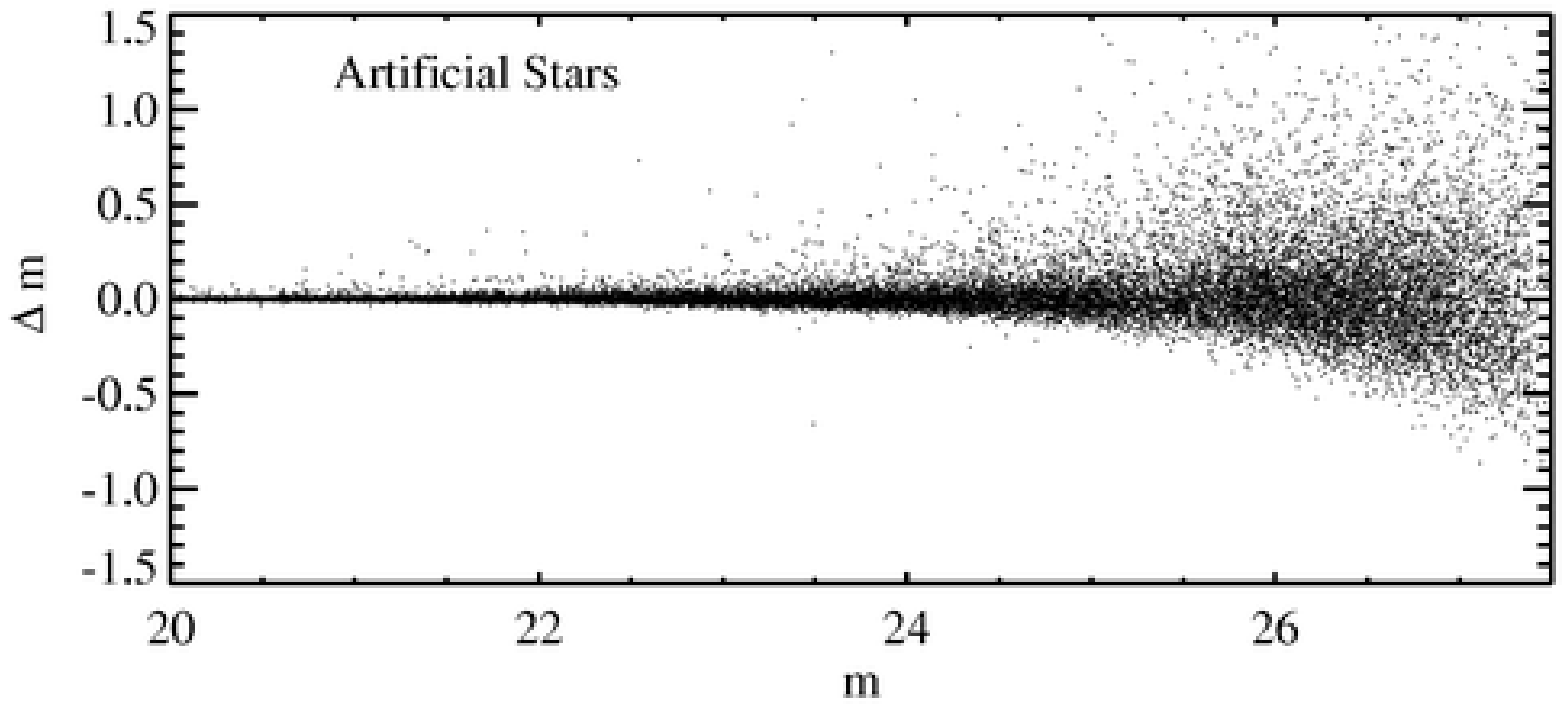}
}
\caption{The cumulative distribution of $F814W$ magnitude differences
between the true and recovered magnitudes in artificial stars from the
same overlap region from NGC~300's WIDE1.  The top panel shows the
cumulative distribution of magnitude differences, scaled by the
reported magnitude error for each star, added in quadrature for the
case of repeat measurements.  The distributions are calculated in bins
of 1 magnitude wide, with the heaviest line indicating the faintest
bin.  The red curve indicates the expectation for a perfect Gaussian
error distribution.  The bottom panel shows the the measured magnitude
differences as a function of magnitude.  Recovered magnitudes tend to
be somewhat brighter than true magnitudes, due to blending with
fainter unresolved sources.
	 \label{magdifffakefig}}
\end{figure}
\vfill
\clearpage

%-------------------

%\input{cmdfigs.tex}
%-------------------
%-------------------
\begin{figure}[p]
\centerline{
\includegraphics[width=1.625in]{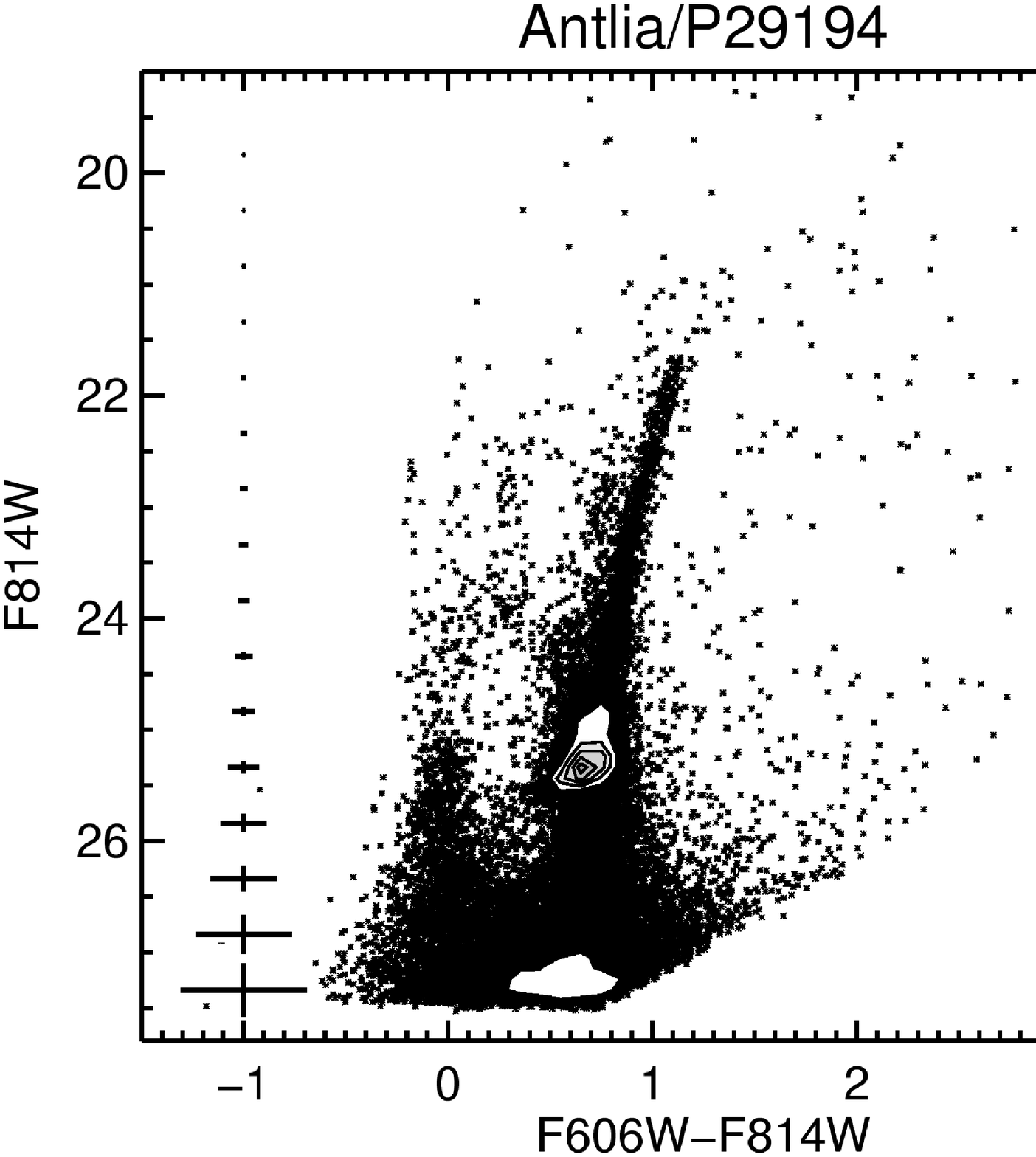}
\includegraphics[width=1.625in]{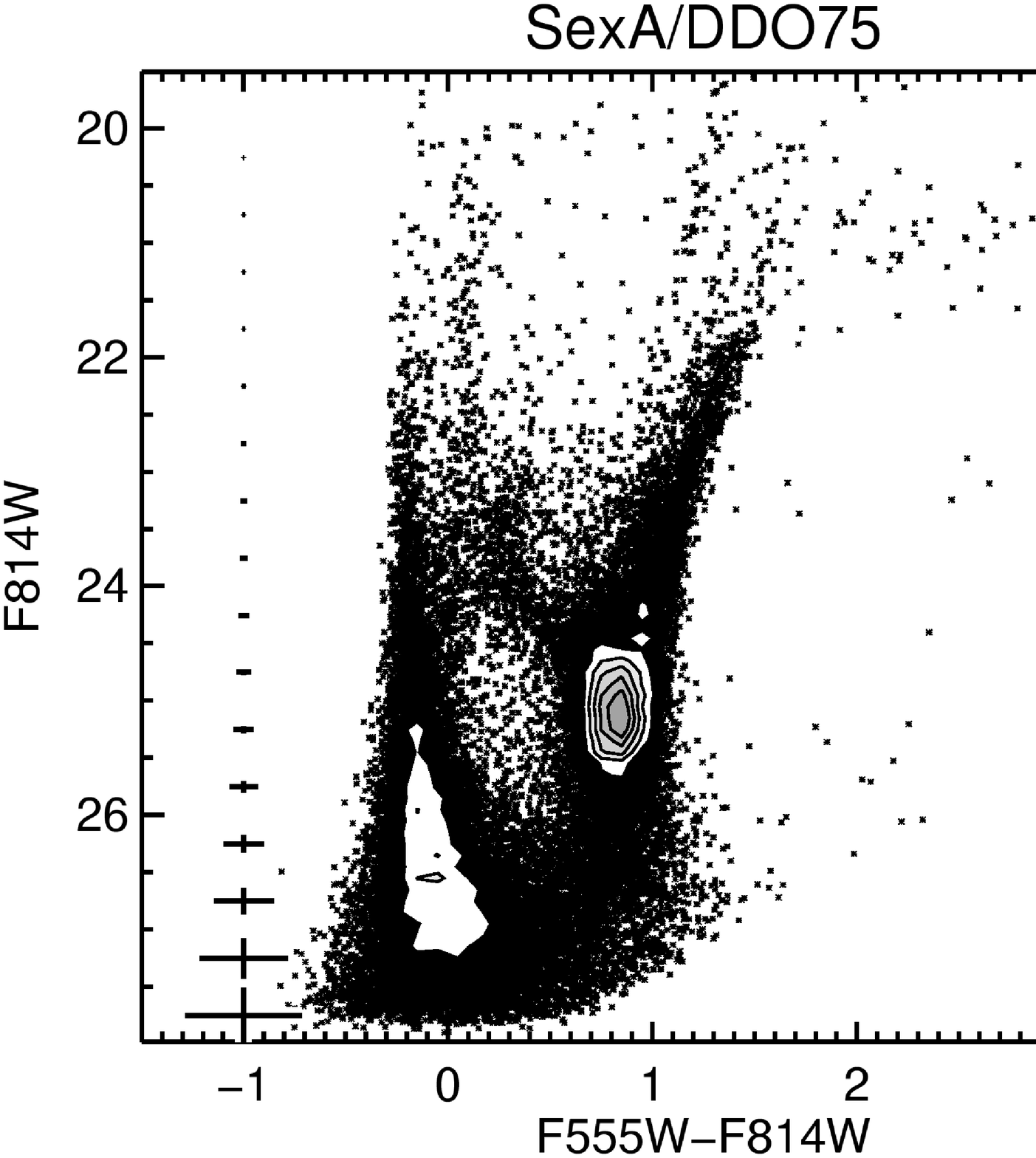}
\includegraphics[width=1.625in]{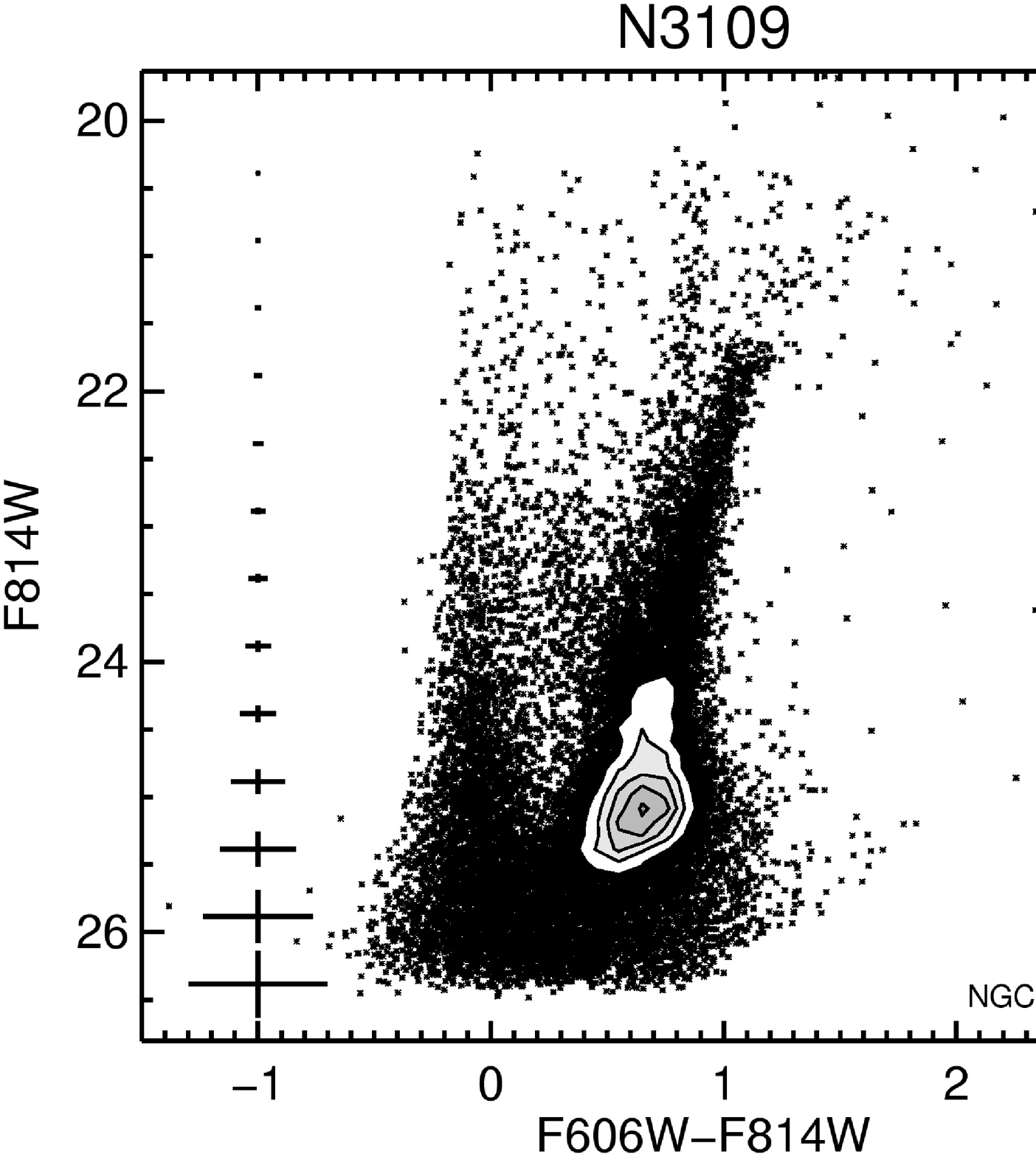}
\includegraphics[width=1.625in]{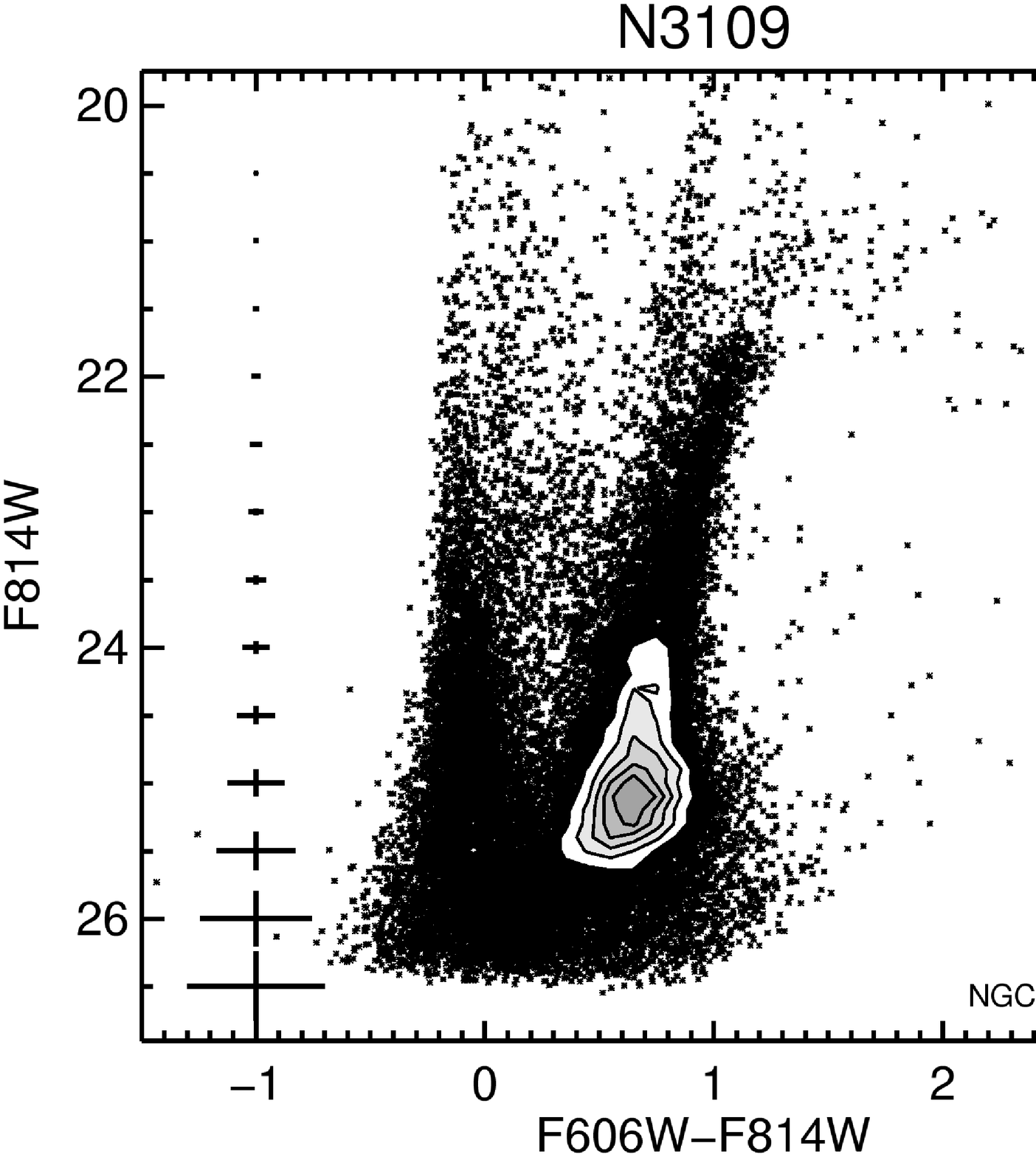}
}
\centerline{
\includegraphics[width=1.625in]{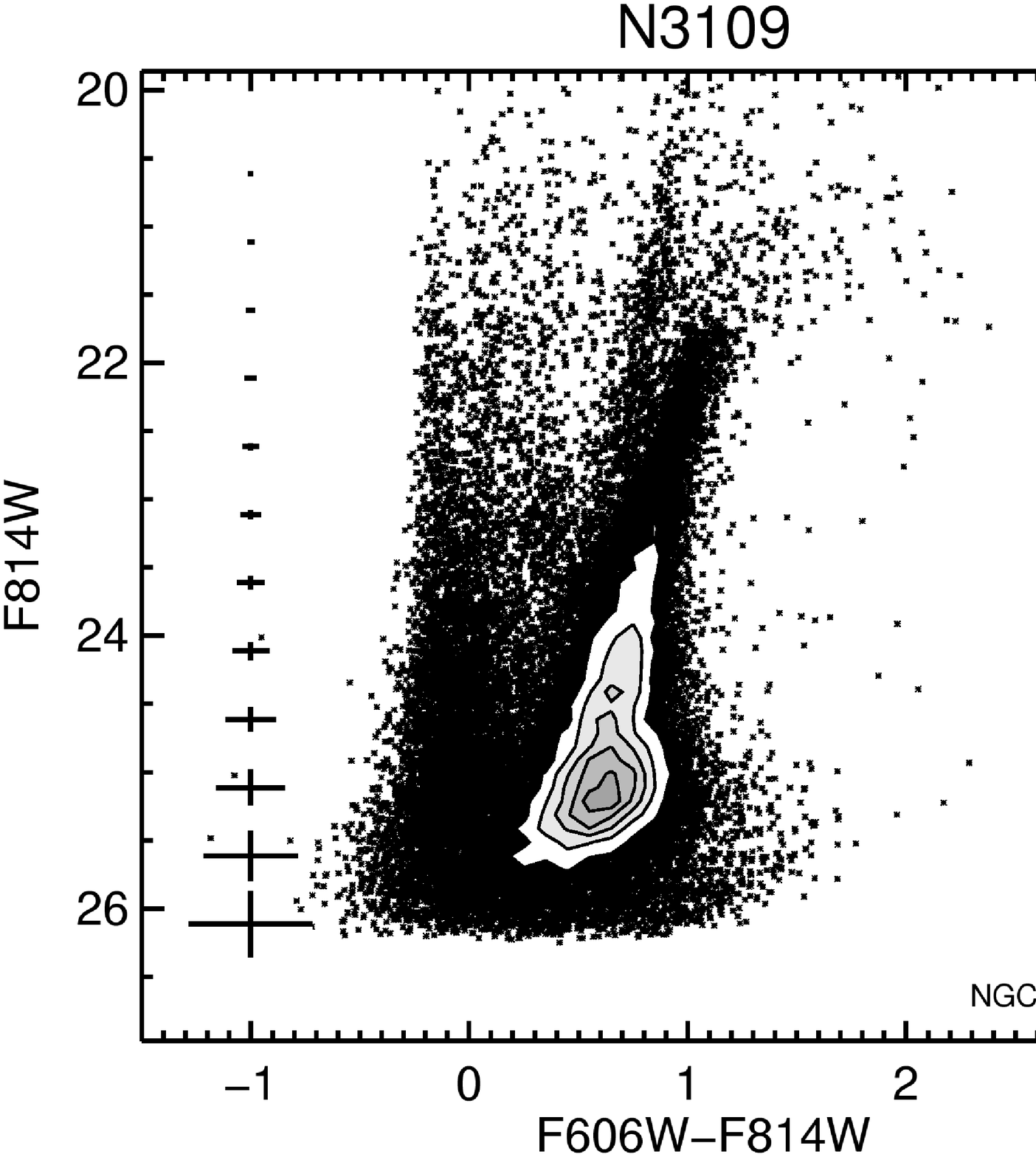}
\includegraphics[width=1.625in]{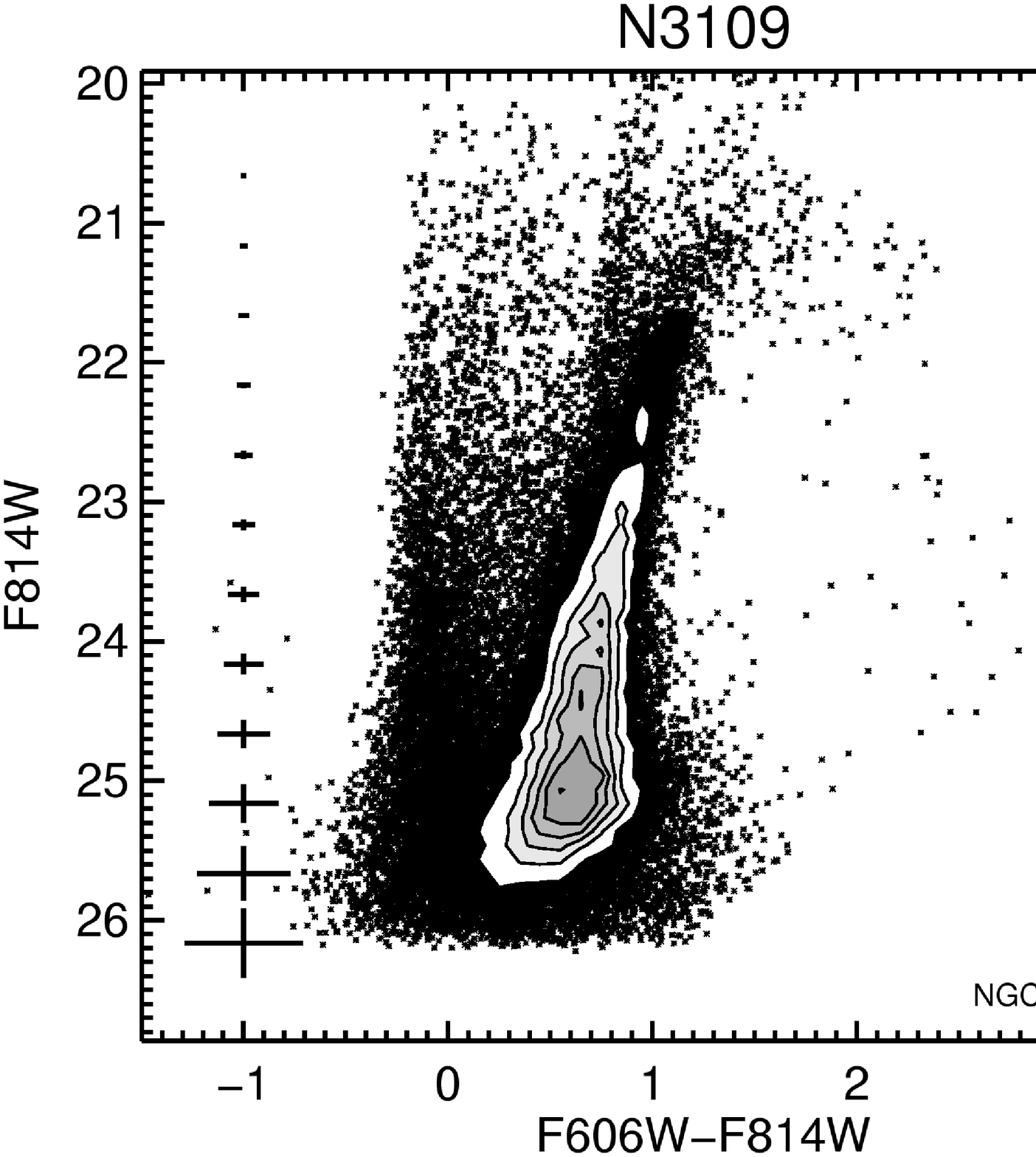}
\includegraphics[width=1.625in]{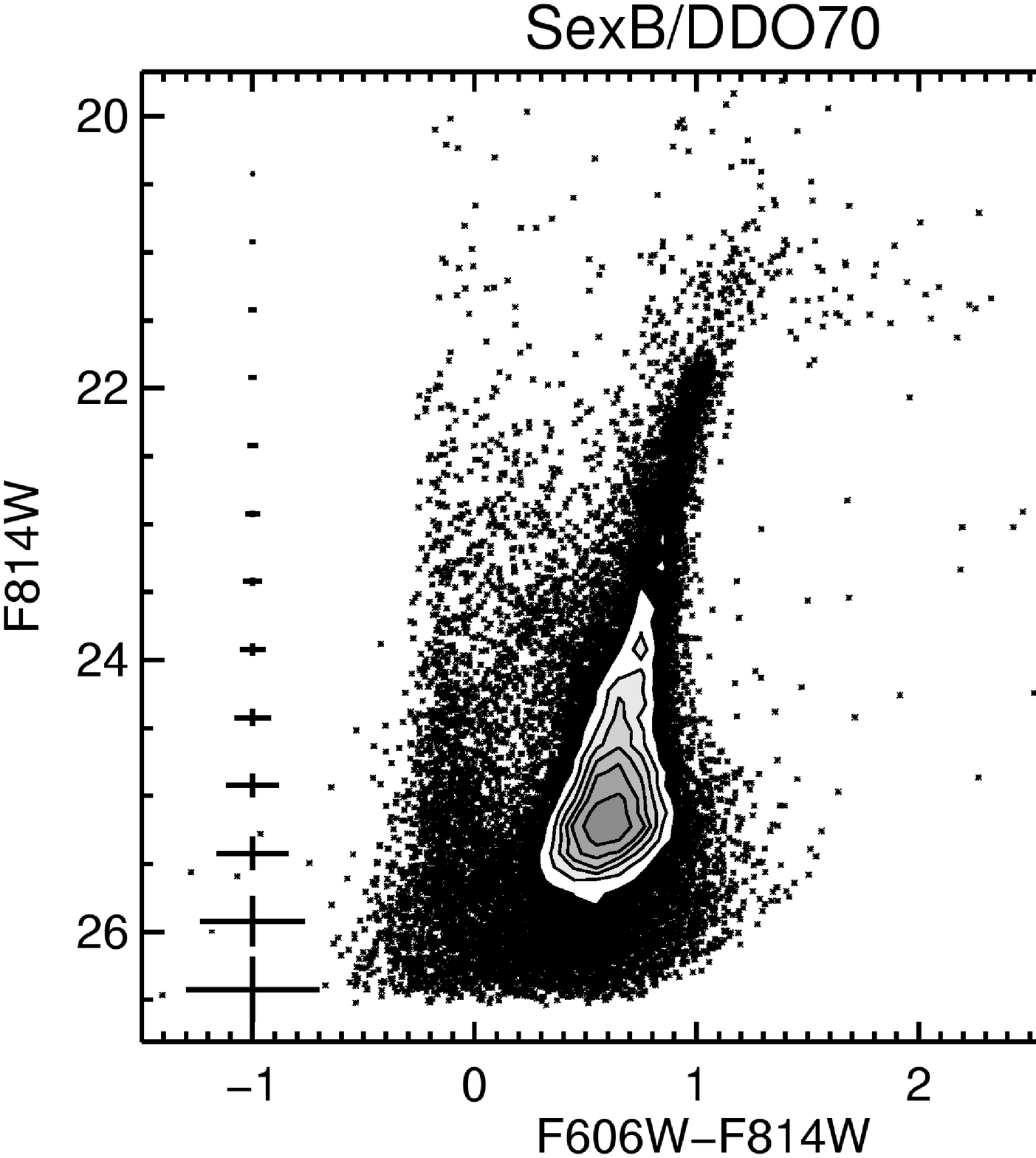}
\includegraphics[width=1.625in]{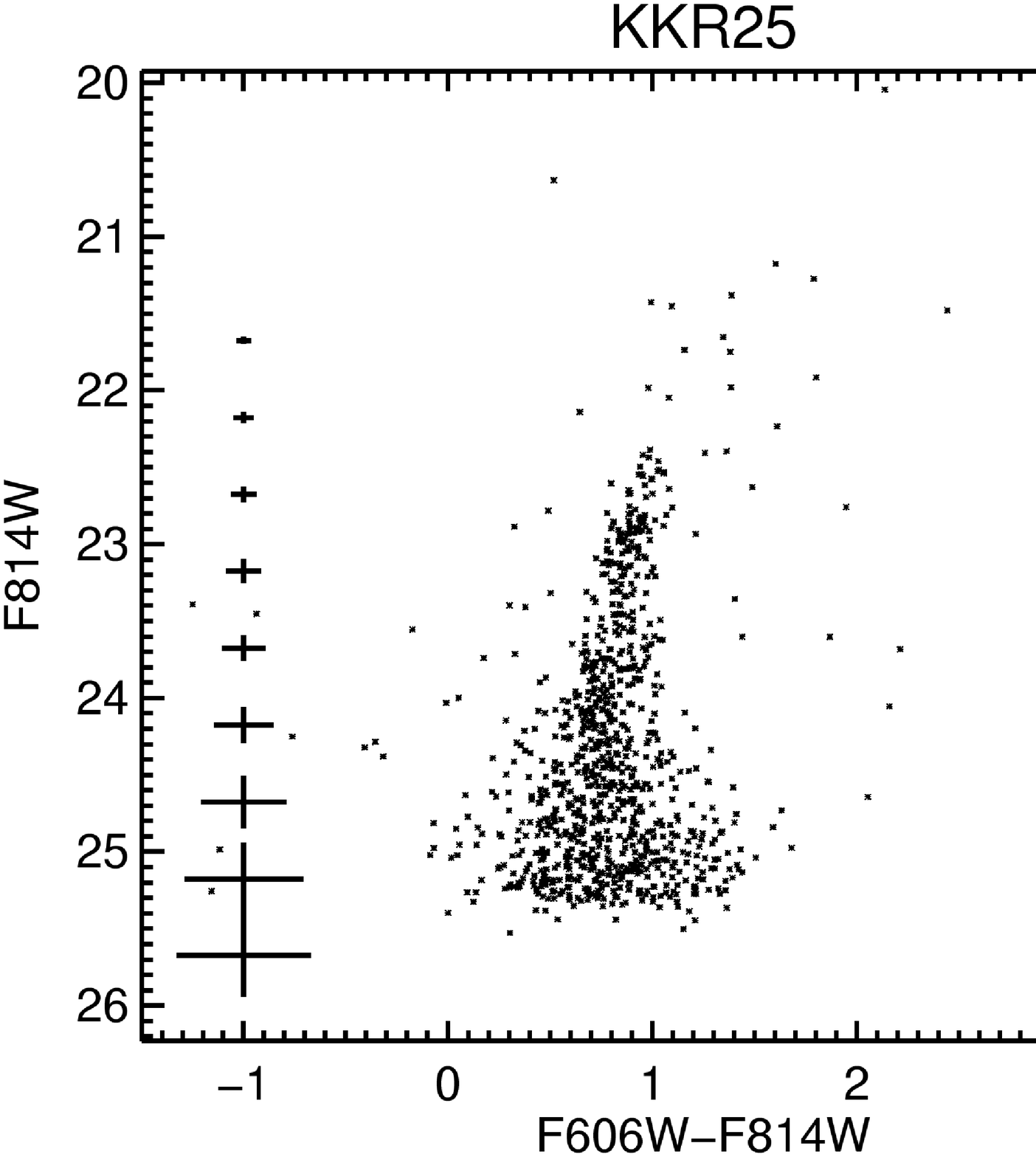}
}
\centerline{
\includegraphics[width=1.625in]{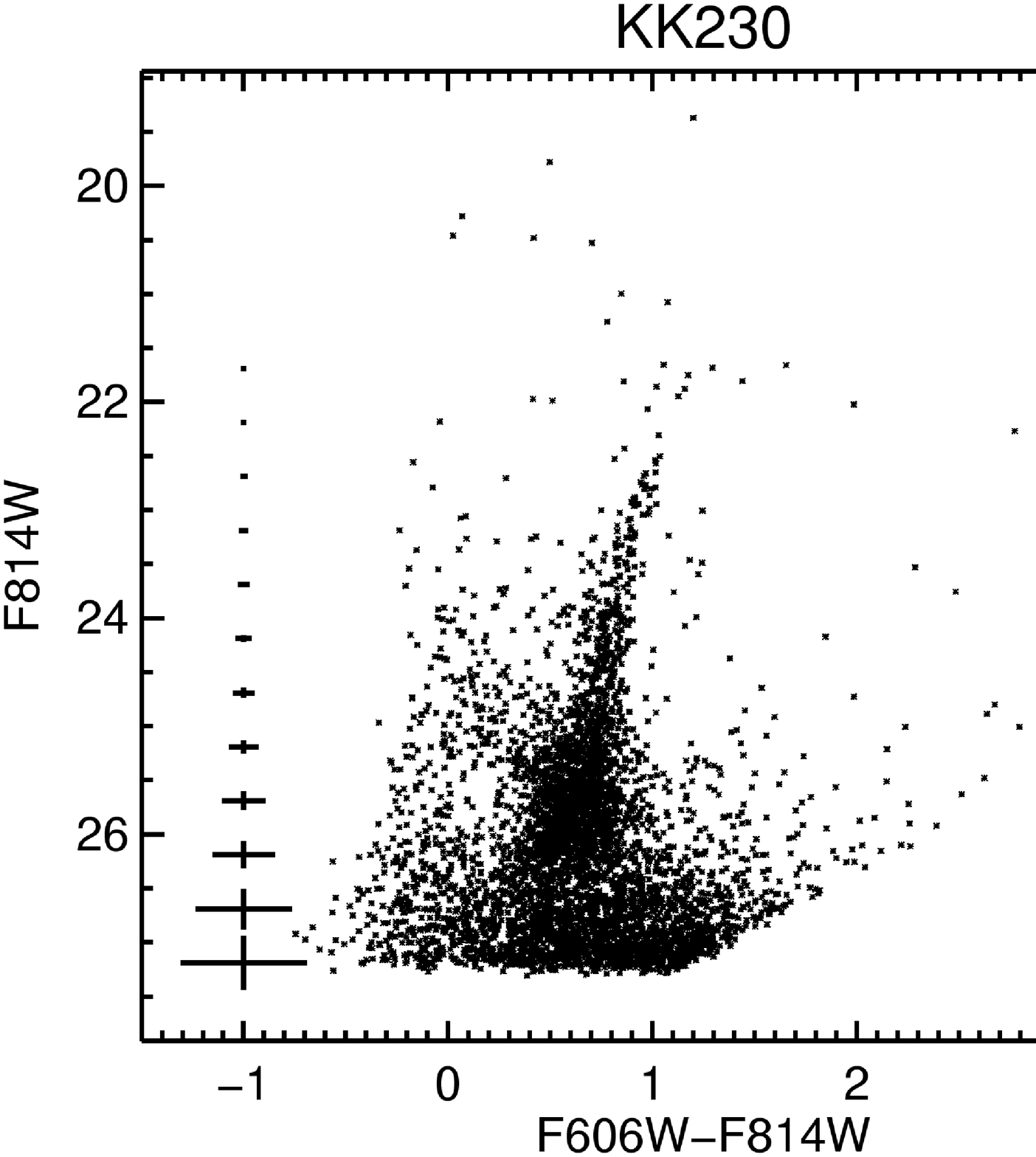}
\includegraphics[width=1.625in]{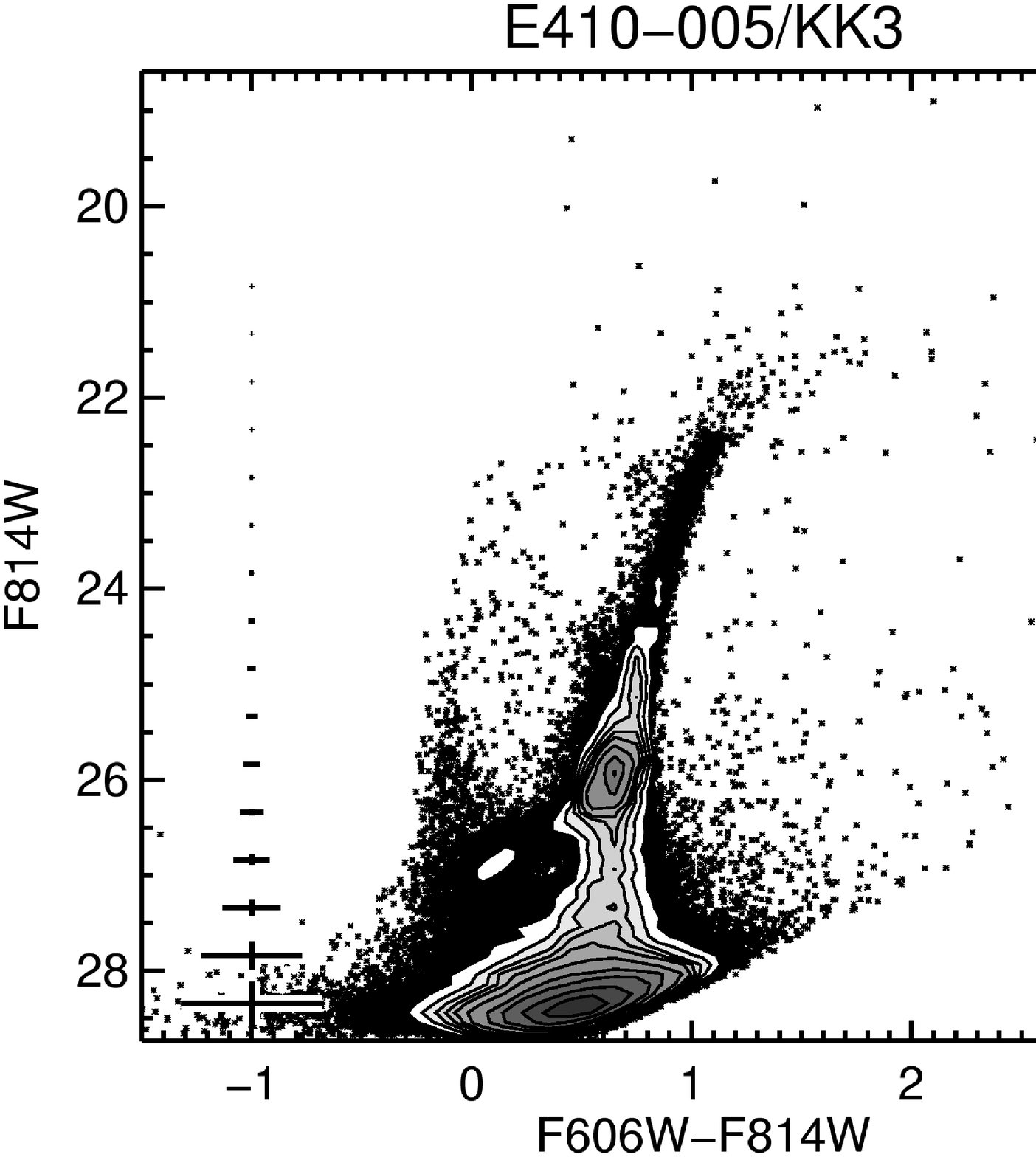}
\includegraphics[width=1.625in]{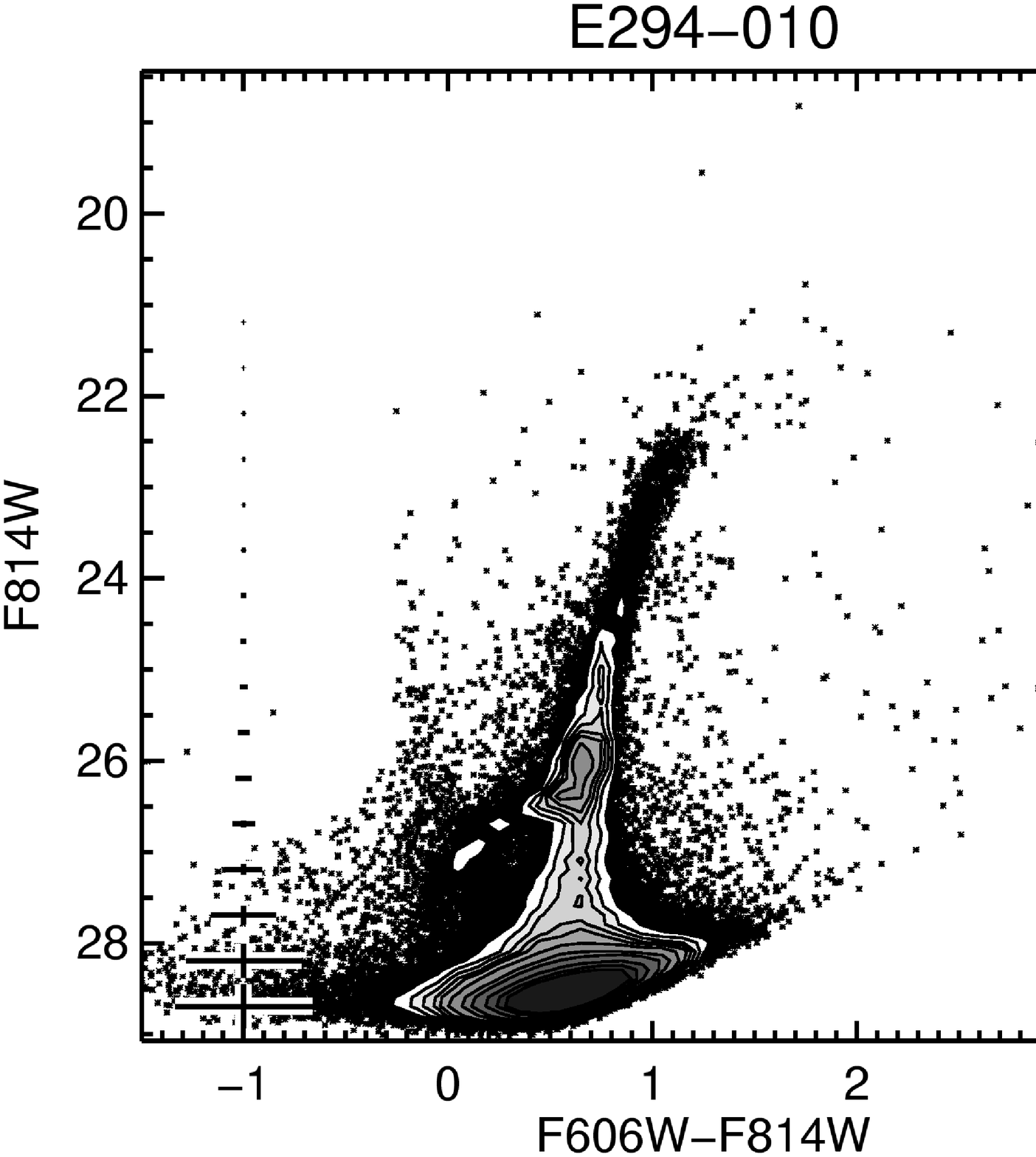}
\includegraphics[width=1.625in]{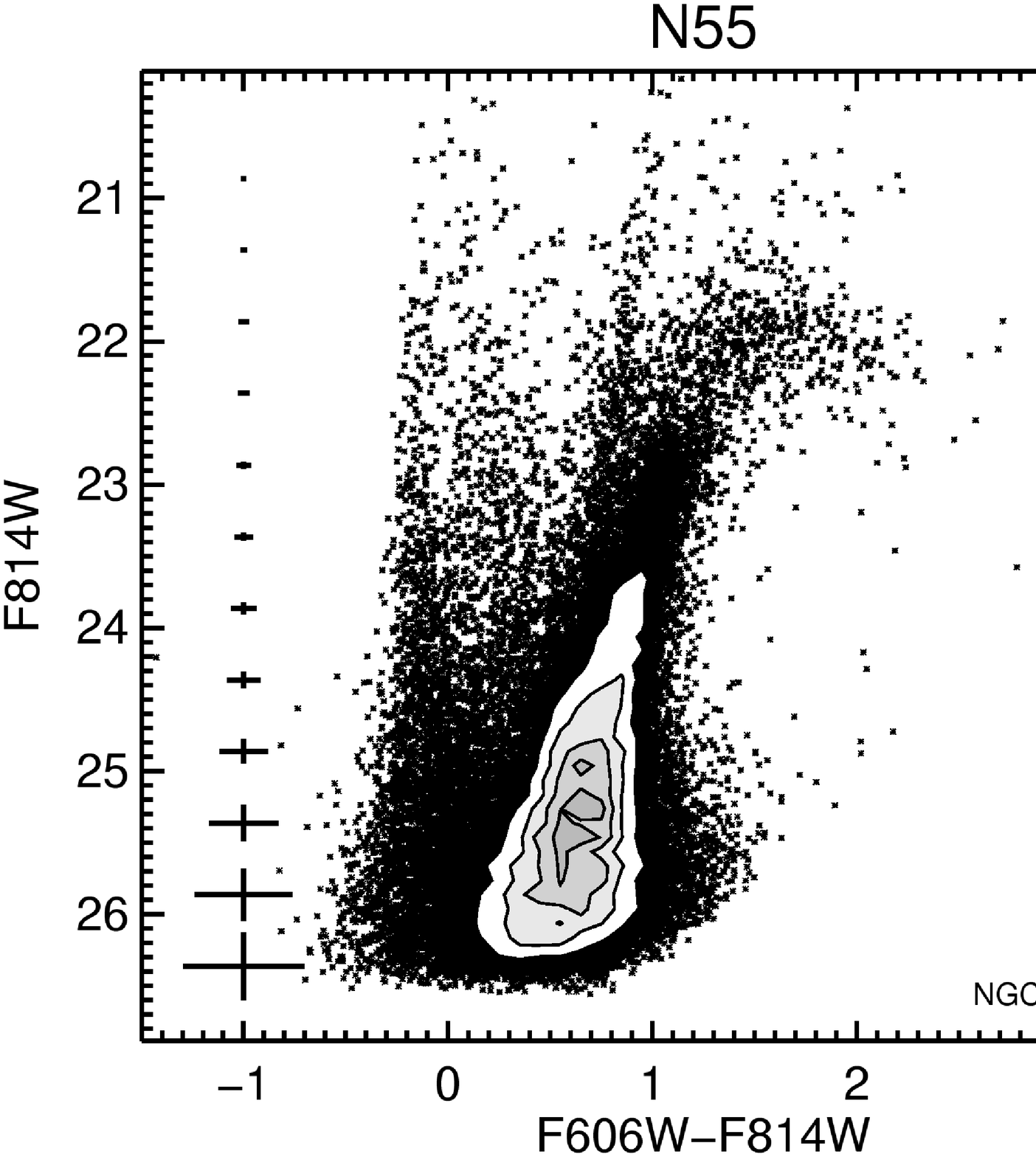}
}
\centerline{
\includegraphics[width=1.625in]{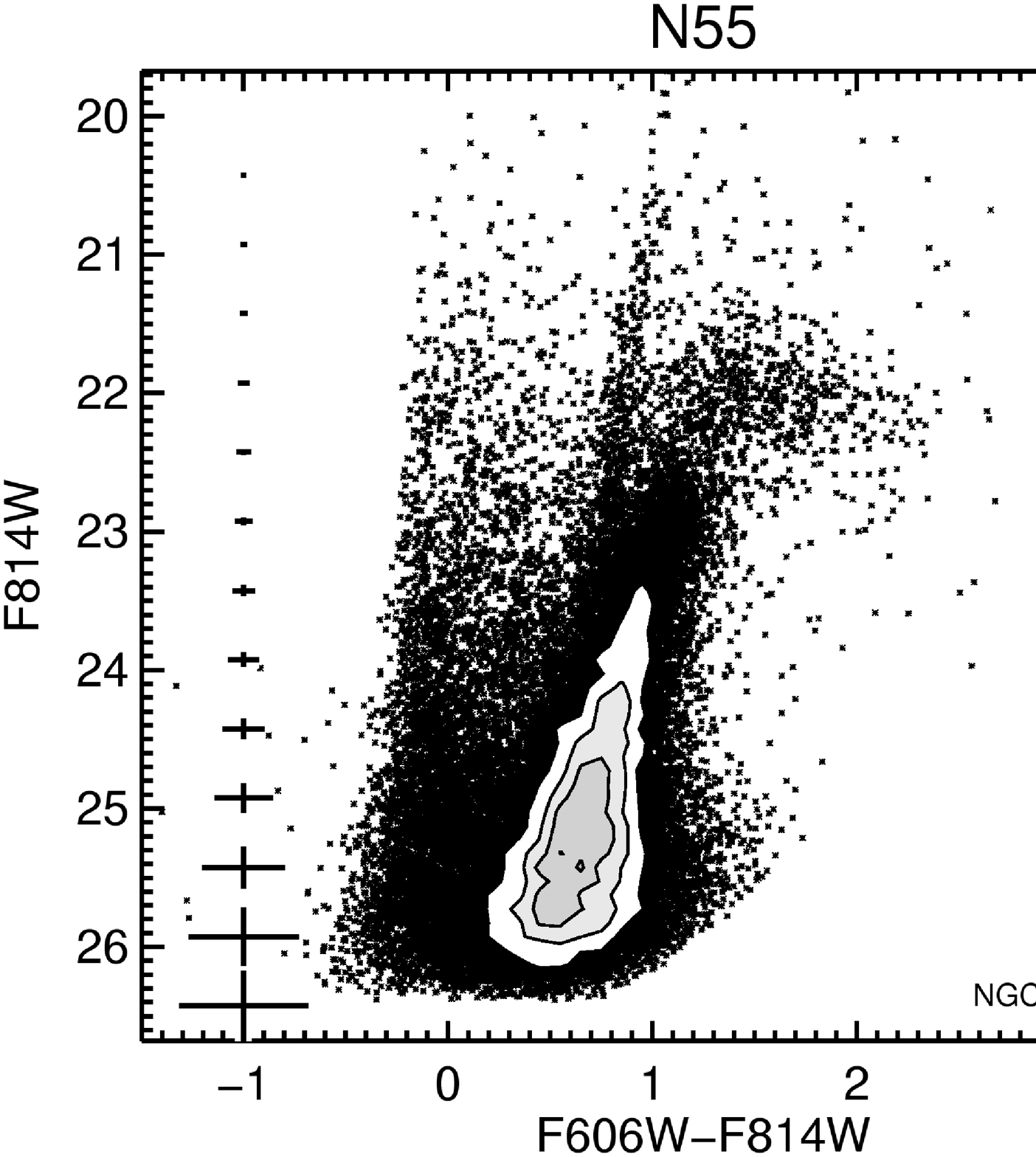}
\includegraphics[width=1.625in]{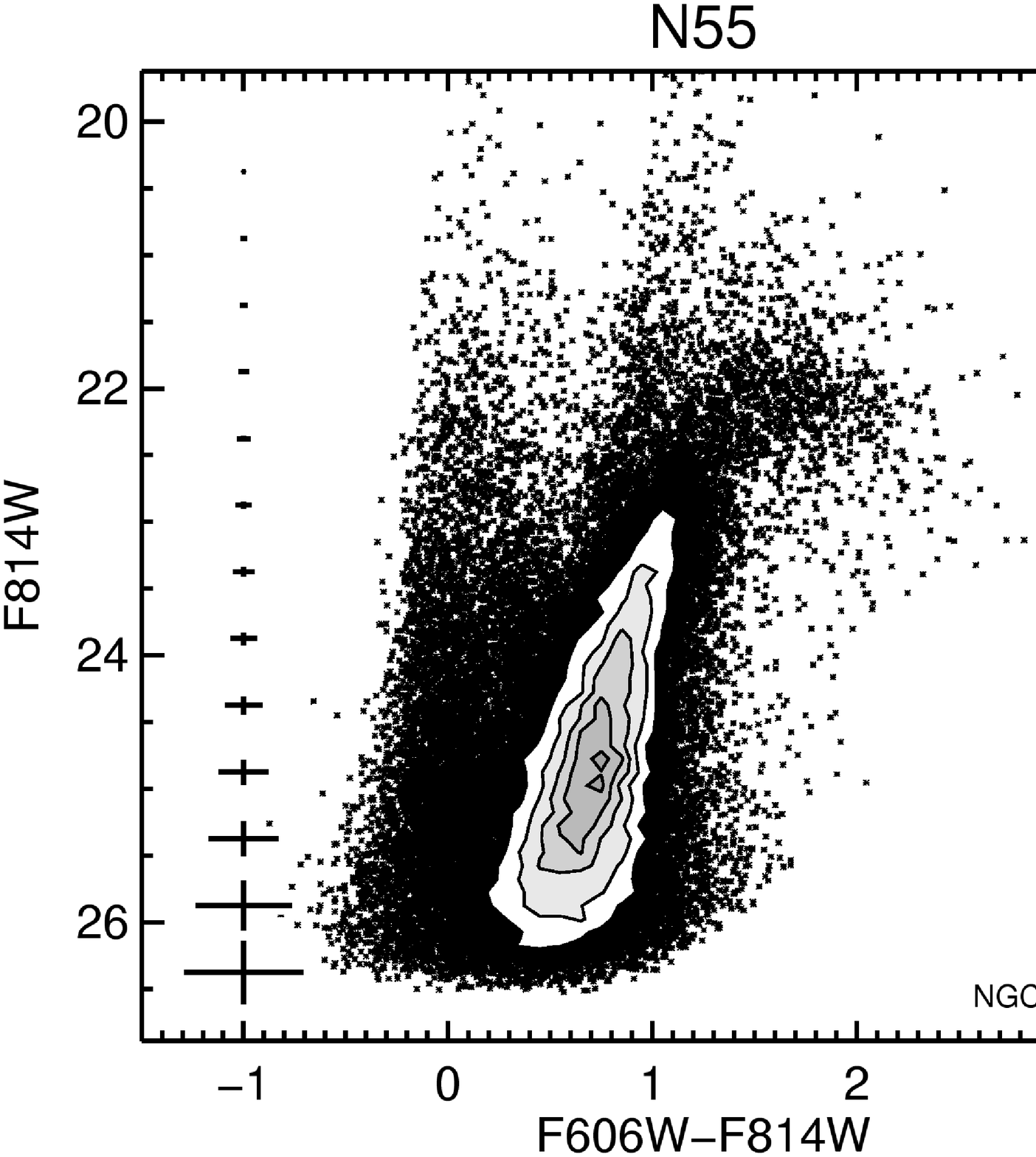}
\includegraphics[width=1.625in]{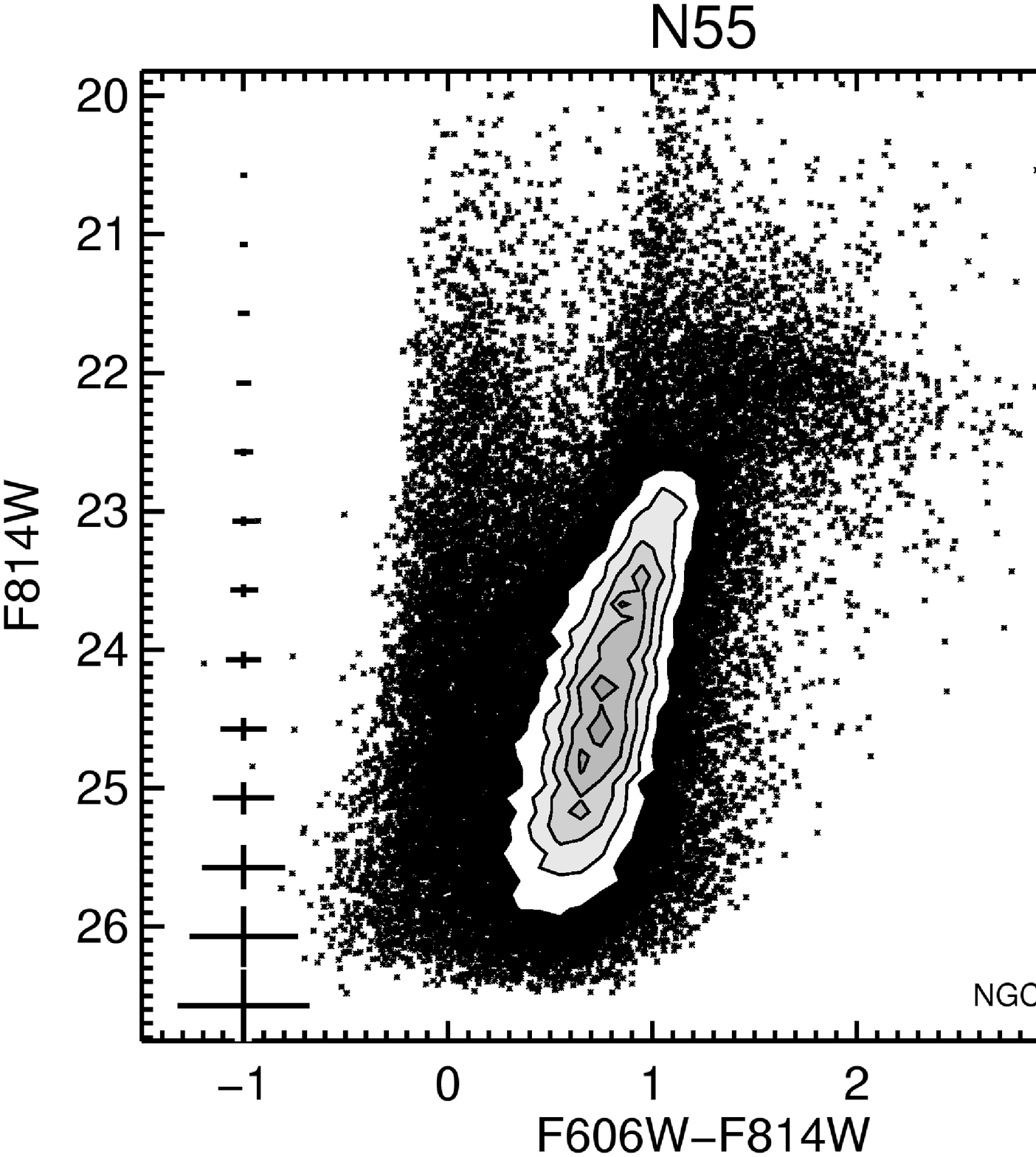}
\includegraphics[width=1.625in]{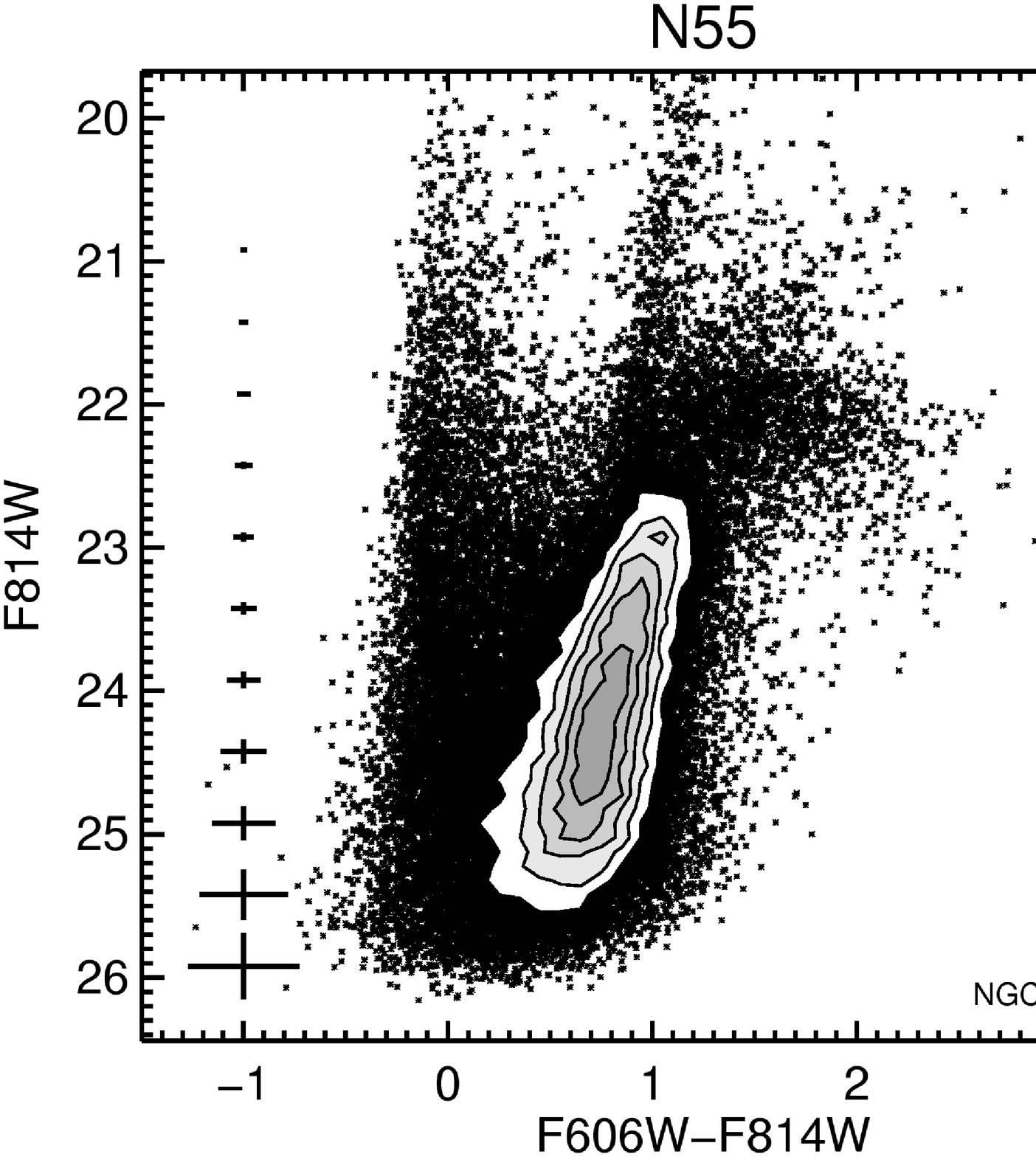}
}
\caption{
CMDs of galaxies in the ANGST data release
(Table~\ref{obstable}~\&~\ref{archivetable}), using photometry from
the conservative {\tt{*\_gst}} catalogs.  Stars are plotted as individual points in
regions of the CMD with few stars, and are plotted as a Hess diagram
otherwise.  The lower right of each plot shows the HST proposal ID and target
name, and an arrow indicating the
direction of the reddening vector. The ANGST/CNG Catalog
name is given at the top of each plot.  Error bars on the left indicate typical
photometric errors in each magnitude bin, but do not include
systematic errors derived from artificial star tests.
Galaxies are ordered as in Table~\ref{sampletable}.
Some fields have multiple CMDs, showing all possible filter combinations
on the color axis (e.g., $F475W-F606W$, $F475W-F814W$, $F606W-F814W$).
Figures are ordered from the upper left to the bottom right.
(a) Antlia; (b) SexA; (c) N3109; (d) N3109; (e) N3109; (f) N3109; (g) SexB; (h) KKR25; (i) KK230; (j) E410-005; (k) E294-010; (l) N55; (m) N55; (n) N55; (o) N55; (p) N55; 
    \label{cmdfig1}}
\end{figure}
\vfill
\clearpage
 
%-------------------
\begin{figure}[p]
\centerline{
\includegraphics[width=1.625in]{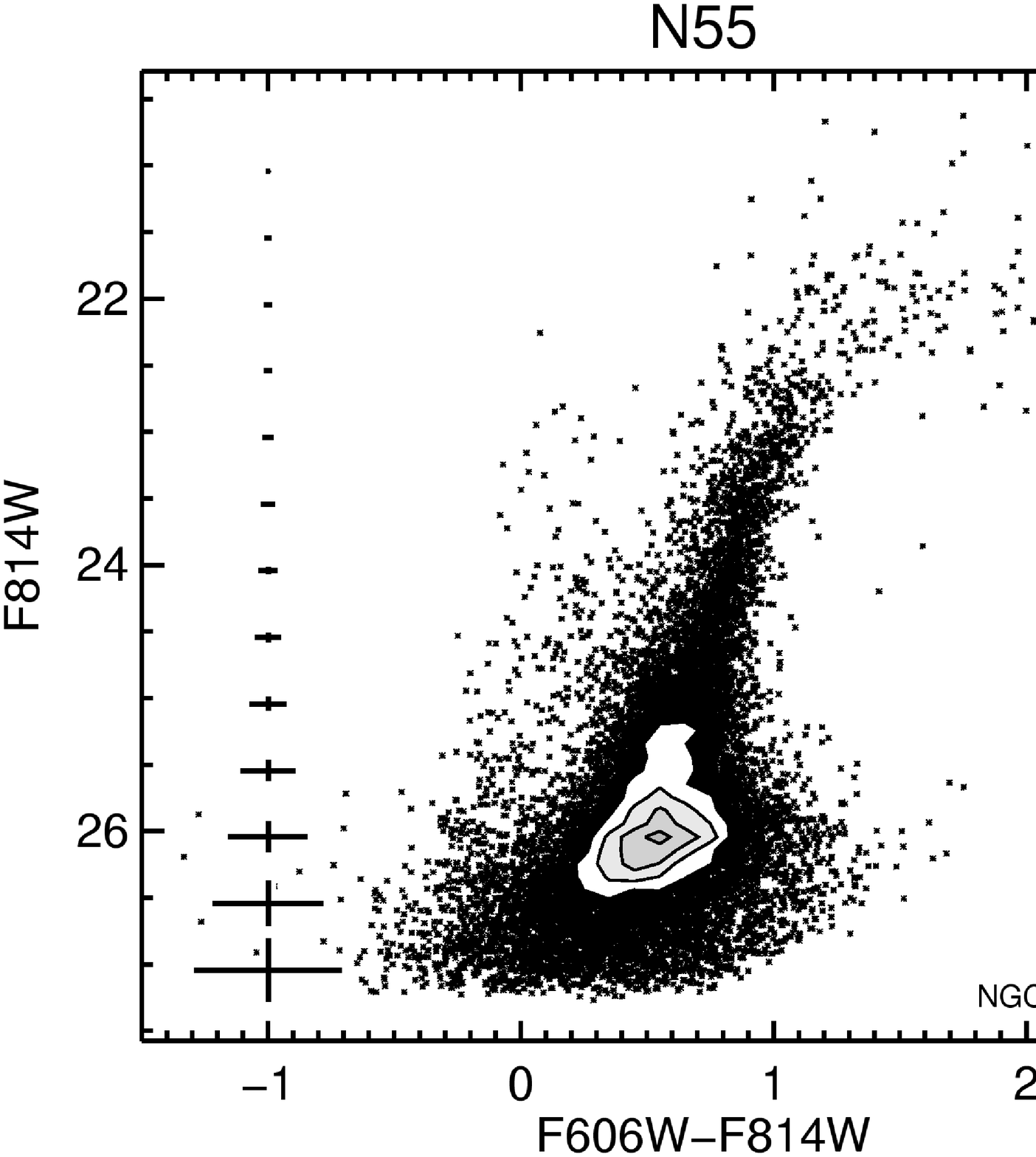}
\includegraphics[width=1.625in]{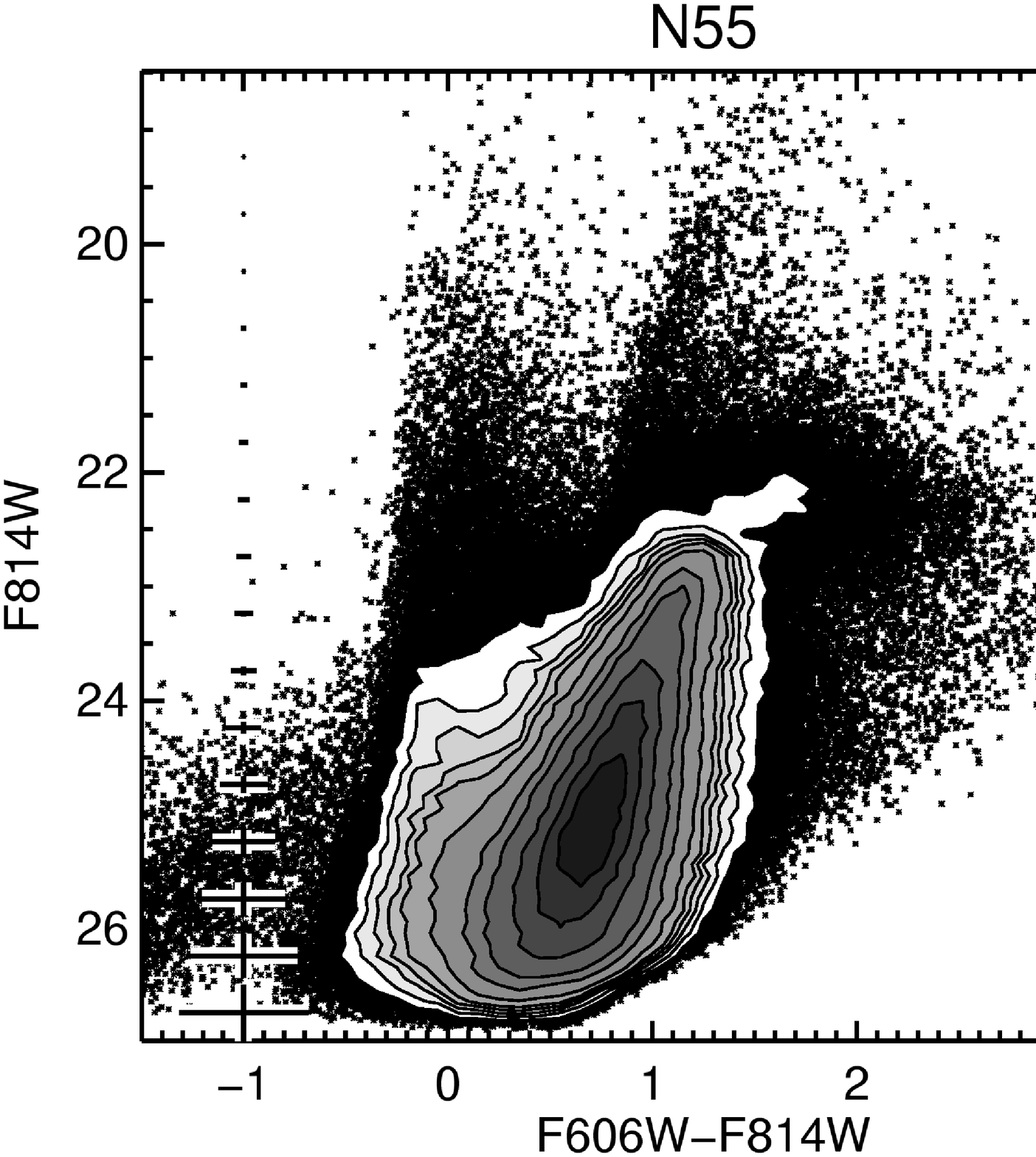}
\includegraphics[width=1.625in]{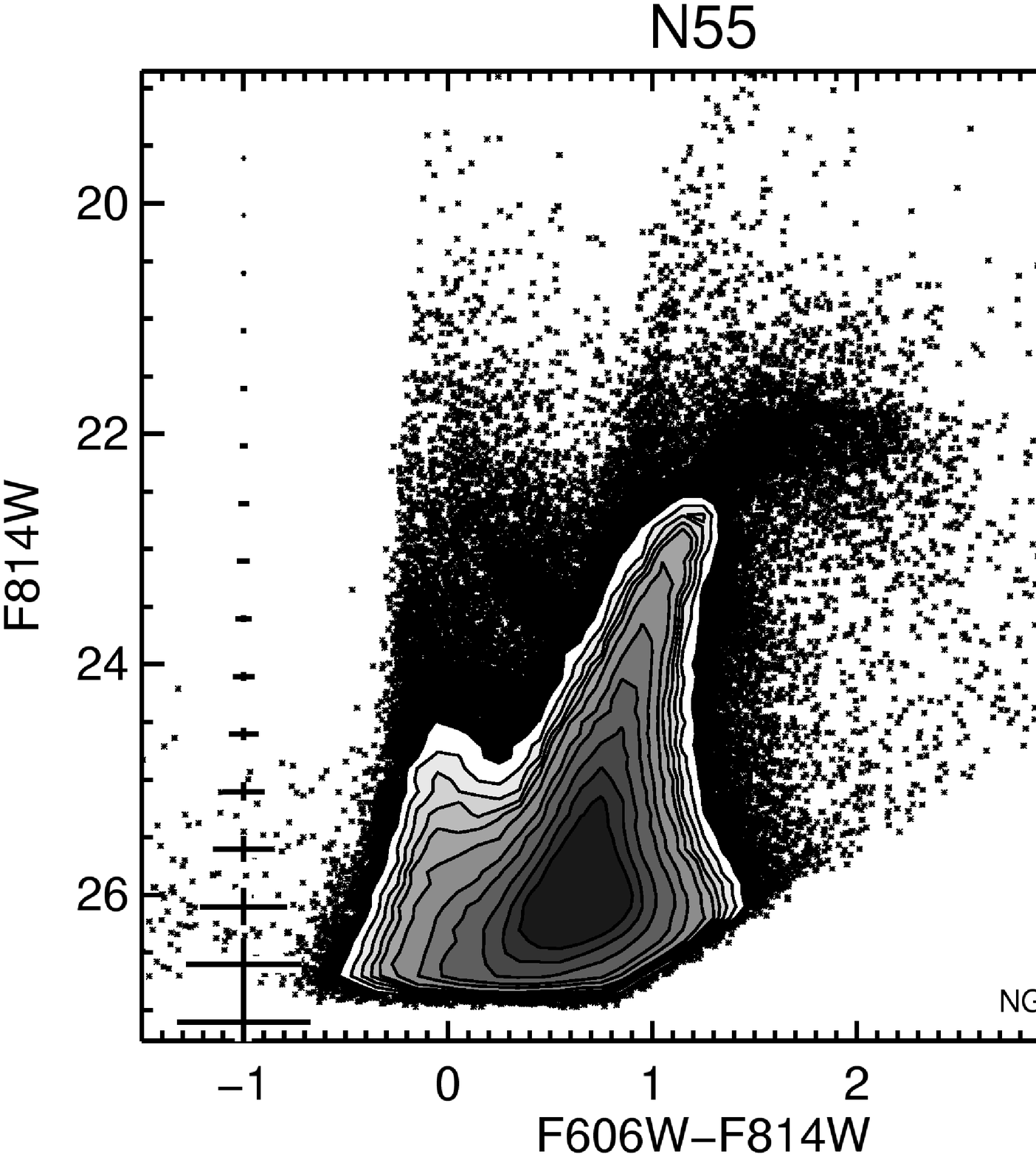}
\includegraphics[width=1.625in]{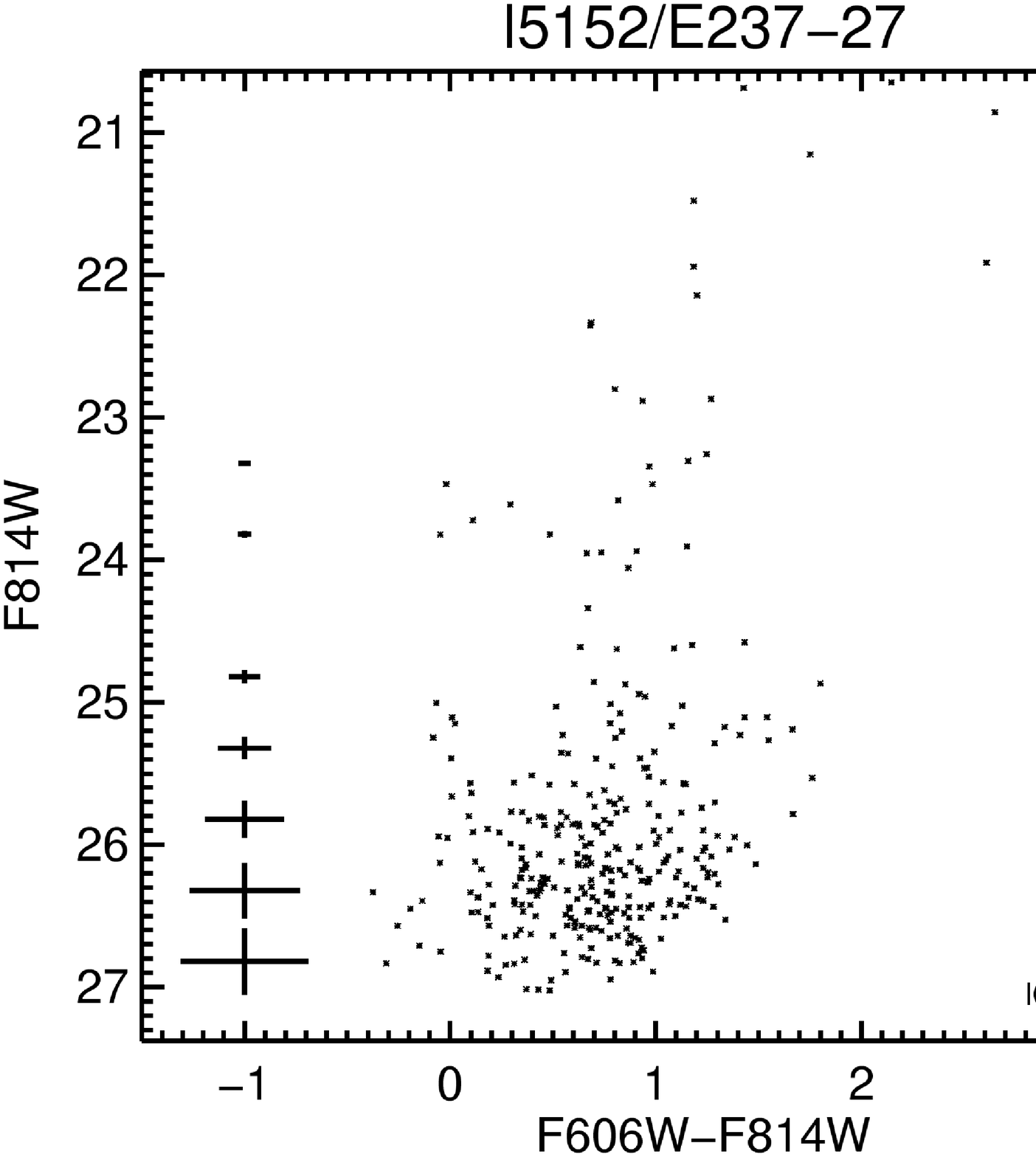}
}
\centerline{
\includegraphics[width=1.625in]{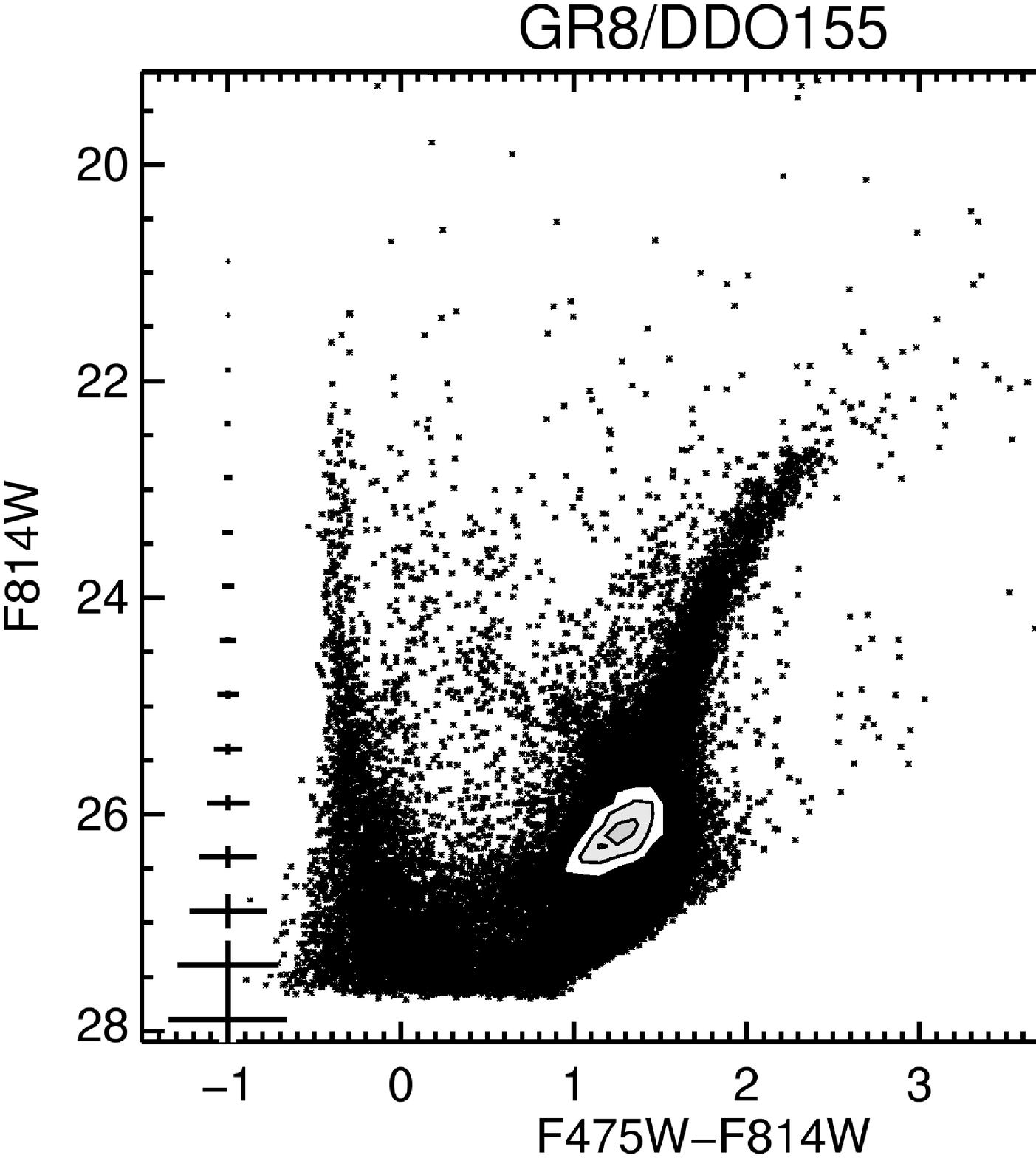}
\includegraphics[width=1.625in]{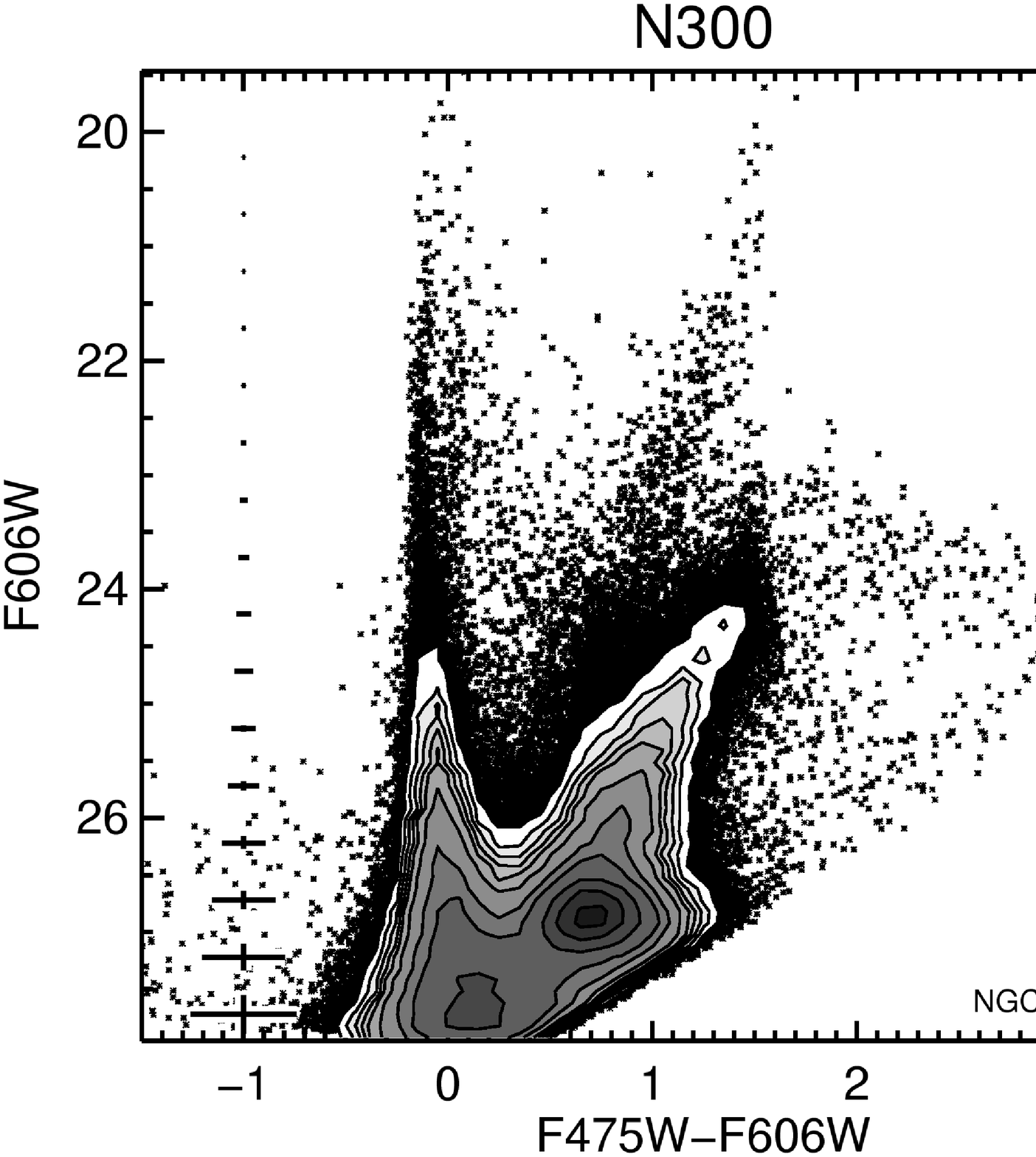}
\includegraphics[width=1.625in]{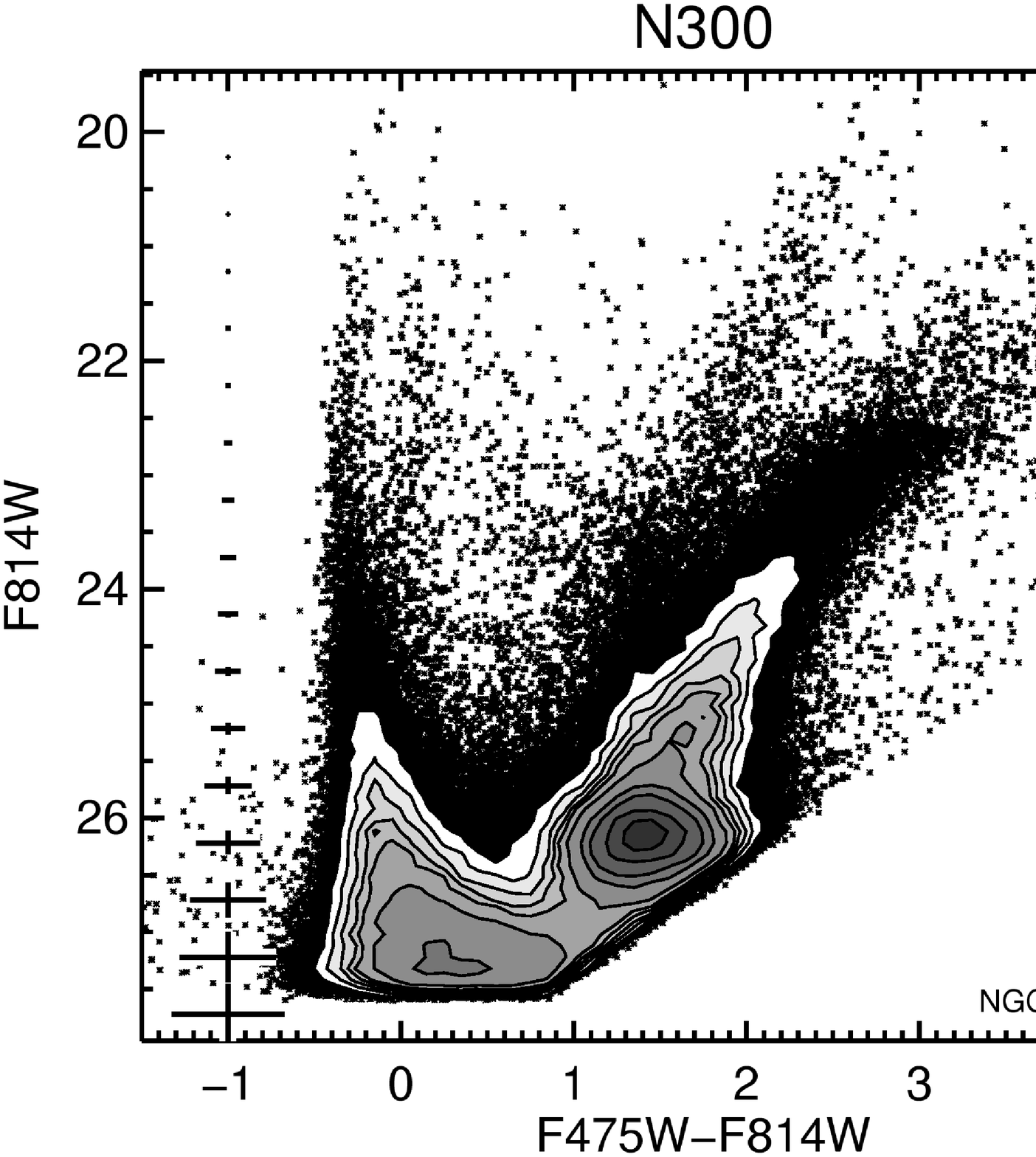}
\includegraphics[width=1.625in]{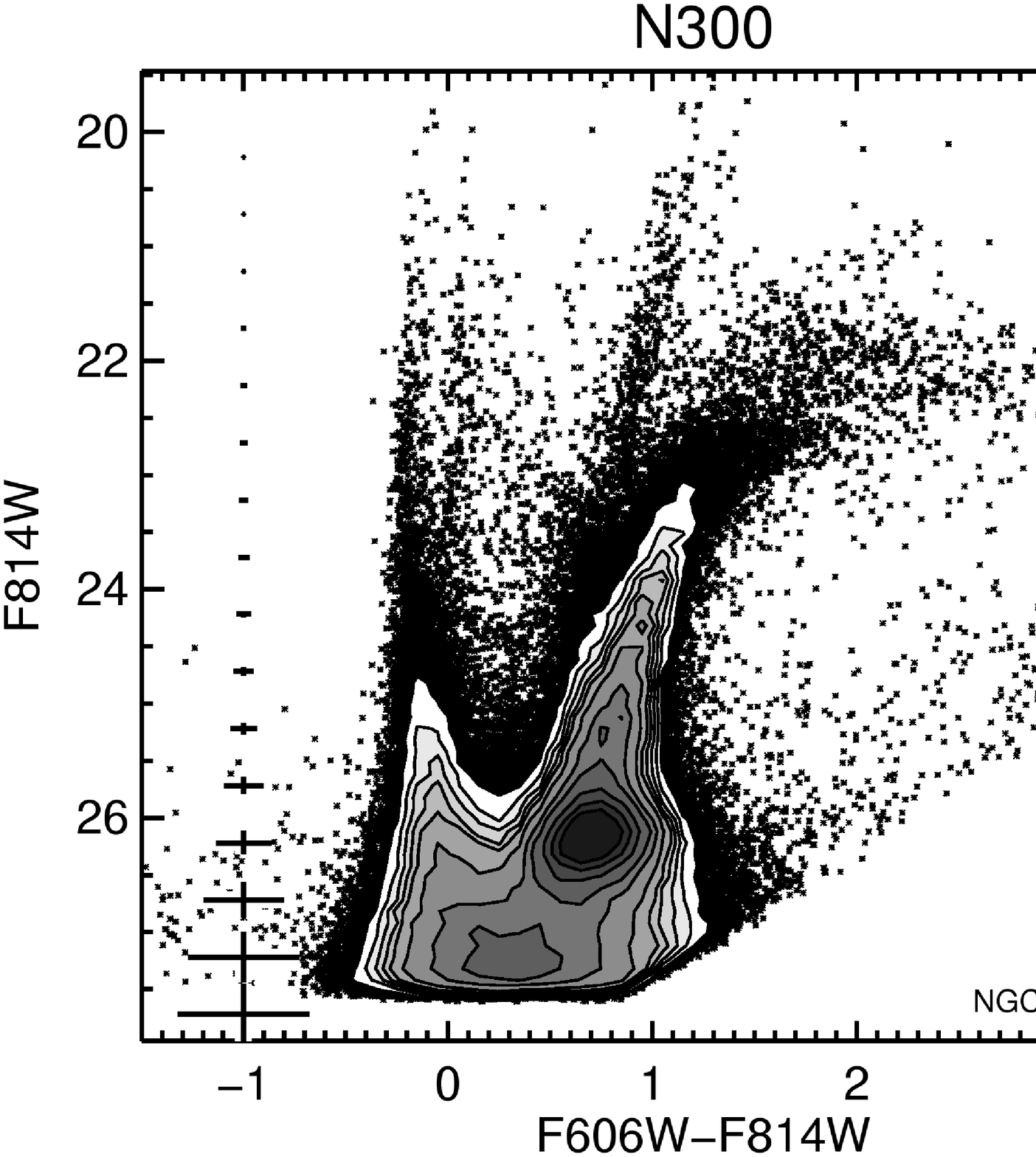}
}
\centerline{
\includegraphics[width=1.625in]{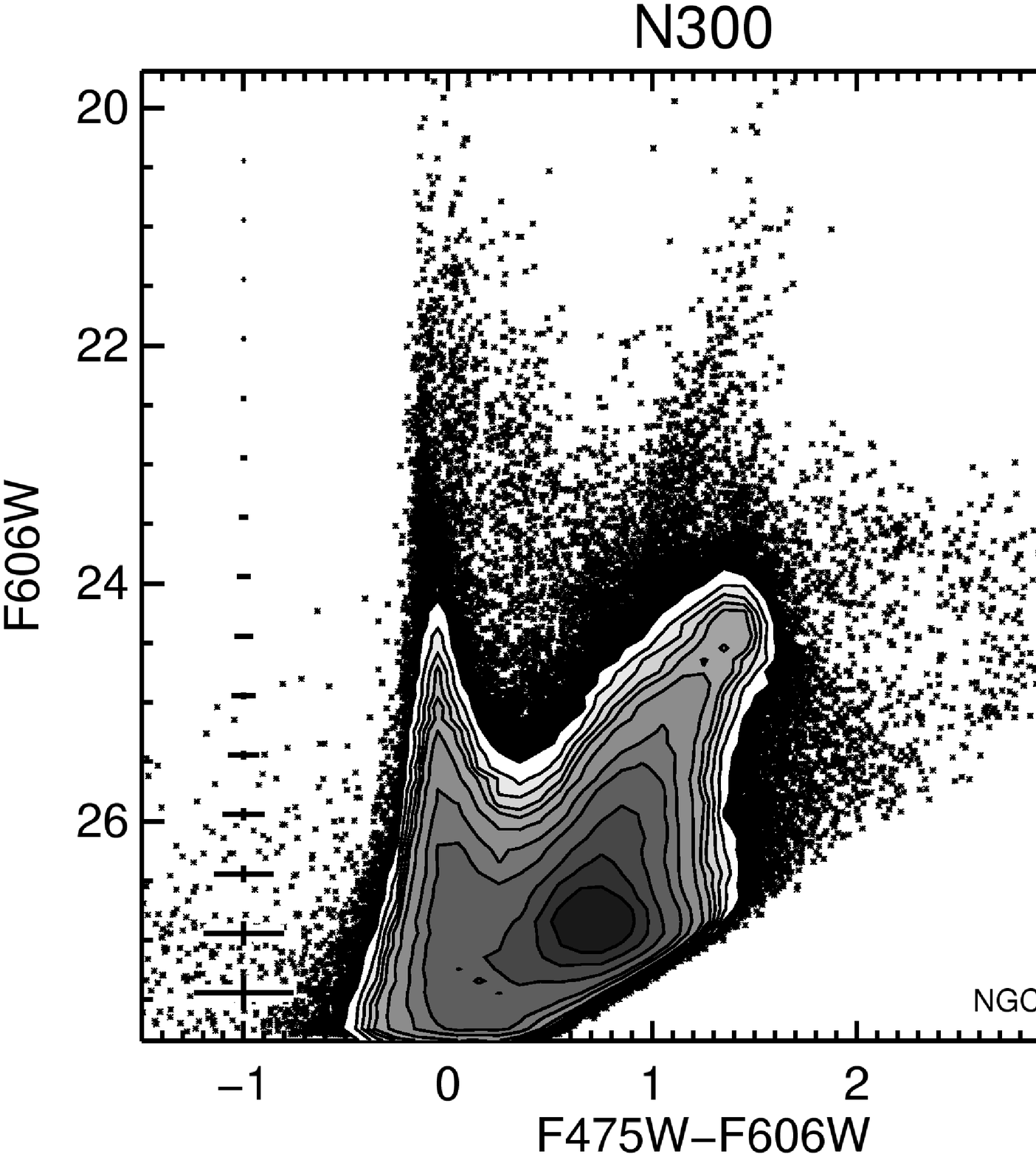}
\includegraphics[width=1.625in]{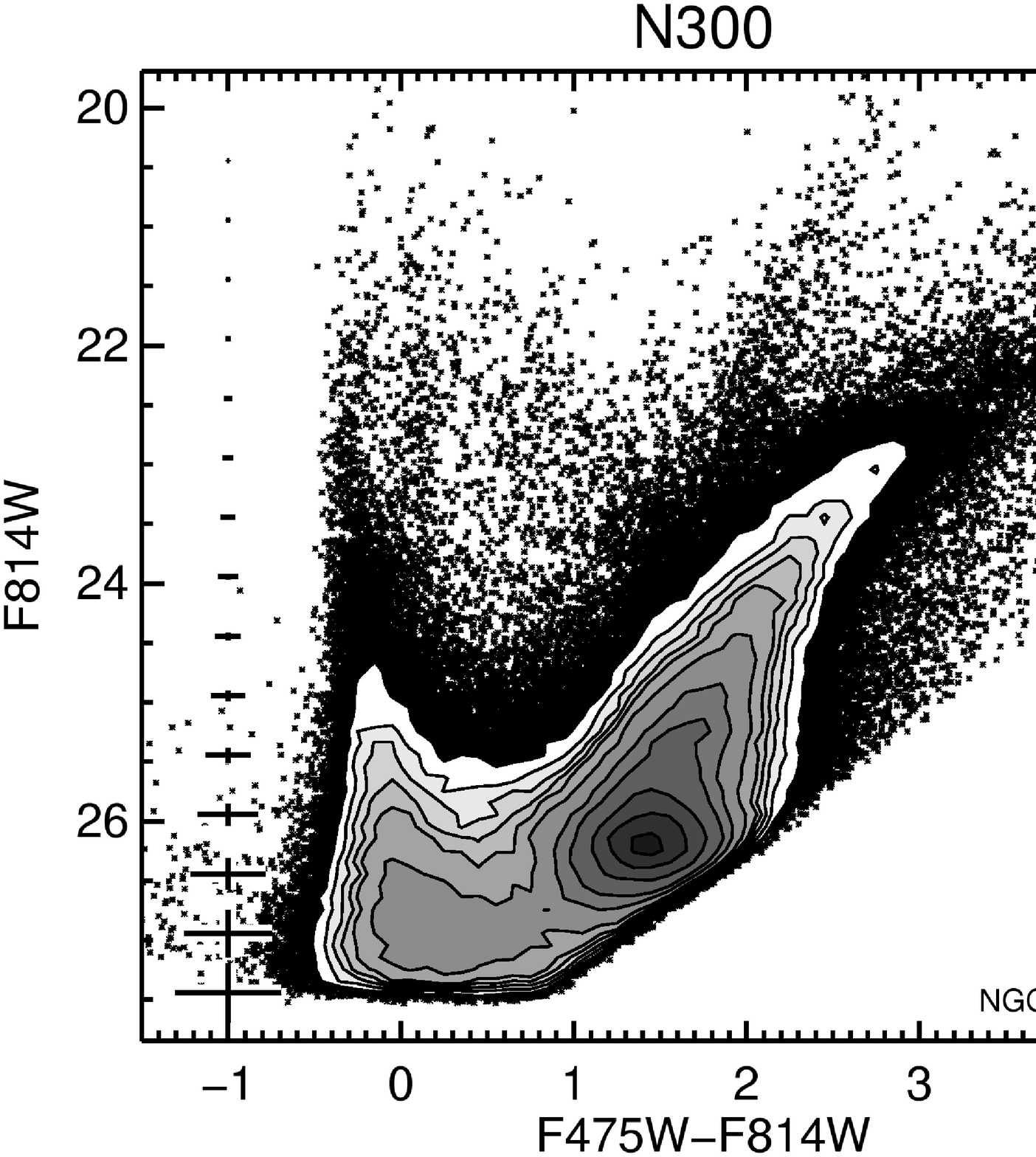}
\includegraphics[width=1.625in]{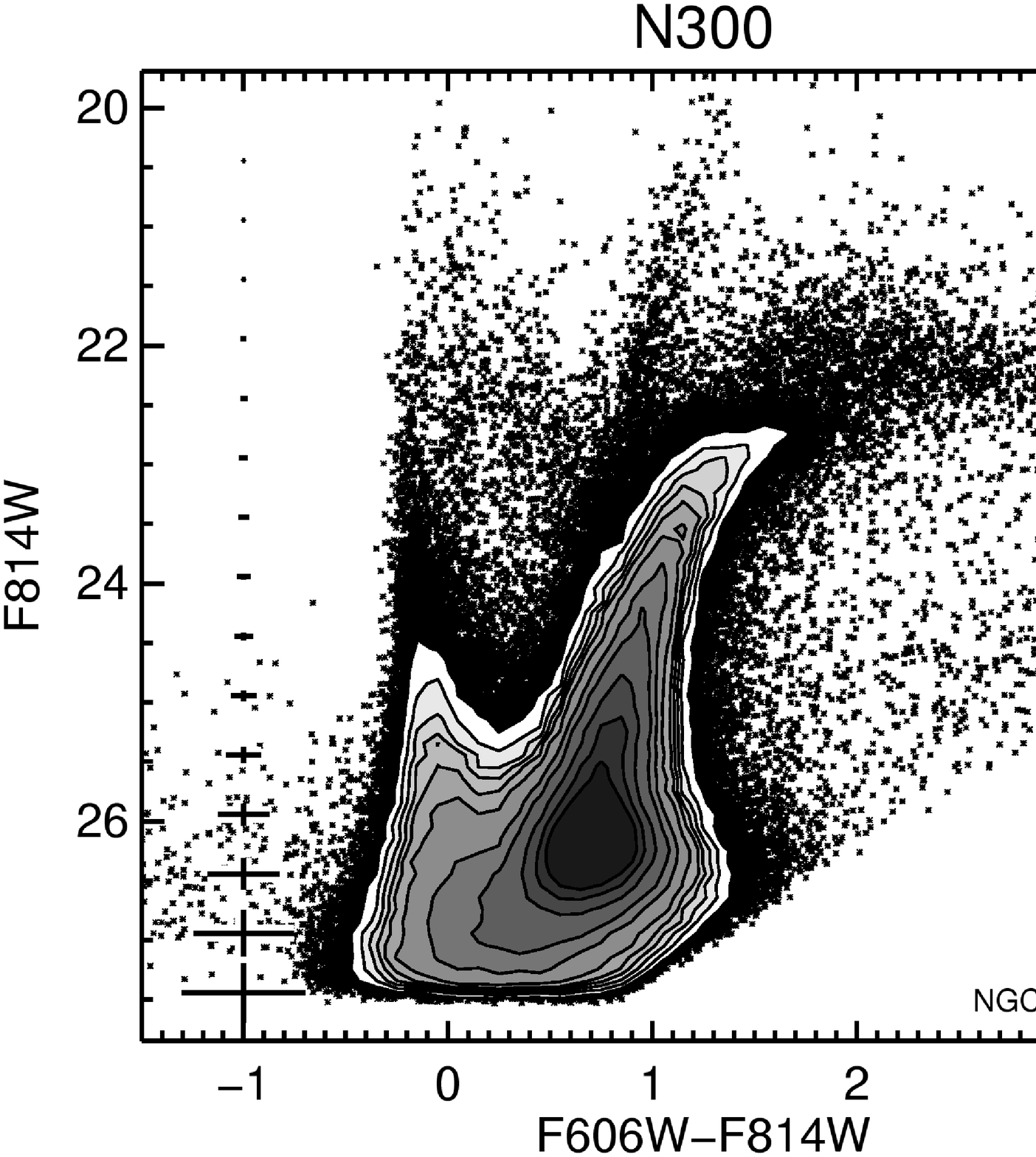}
\includegraphics[width=1.625in]{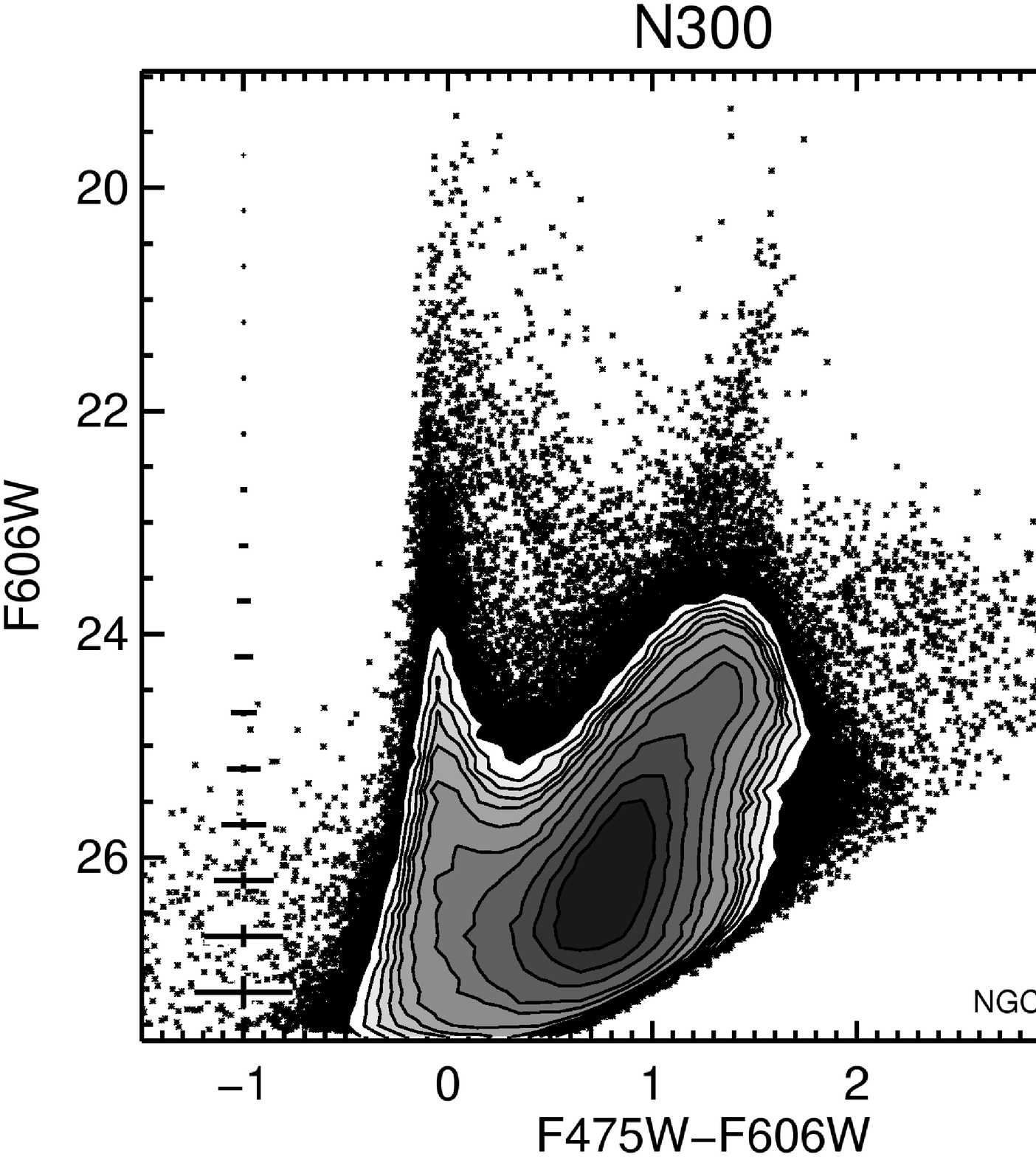}
}
\centerline{
\includegraphics[width=1.625in]{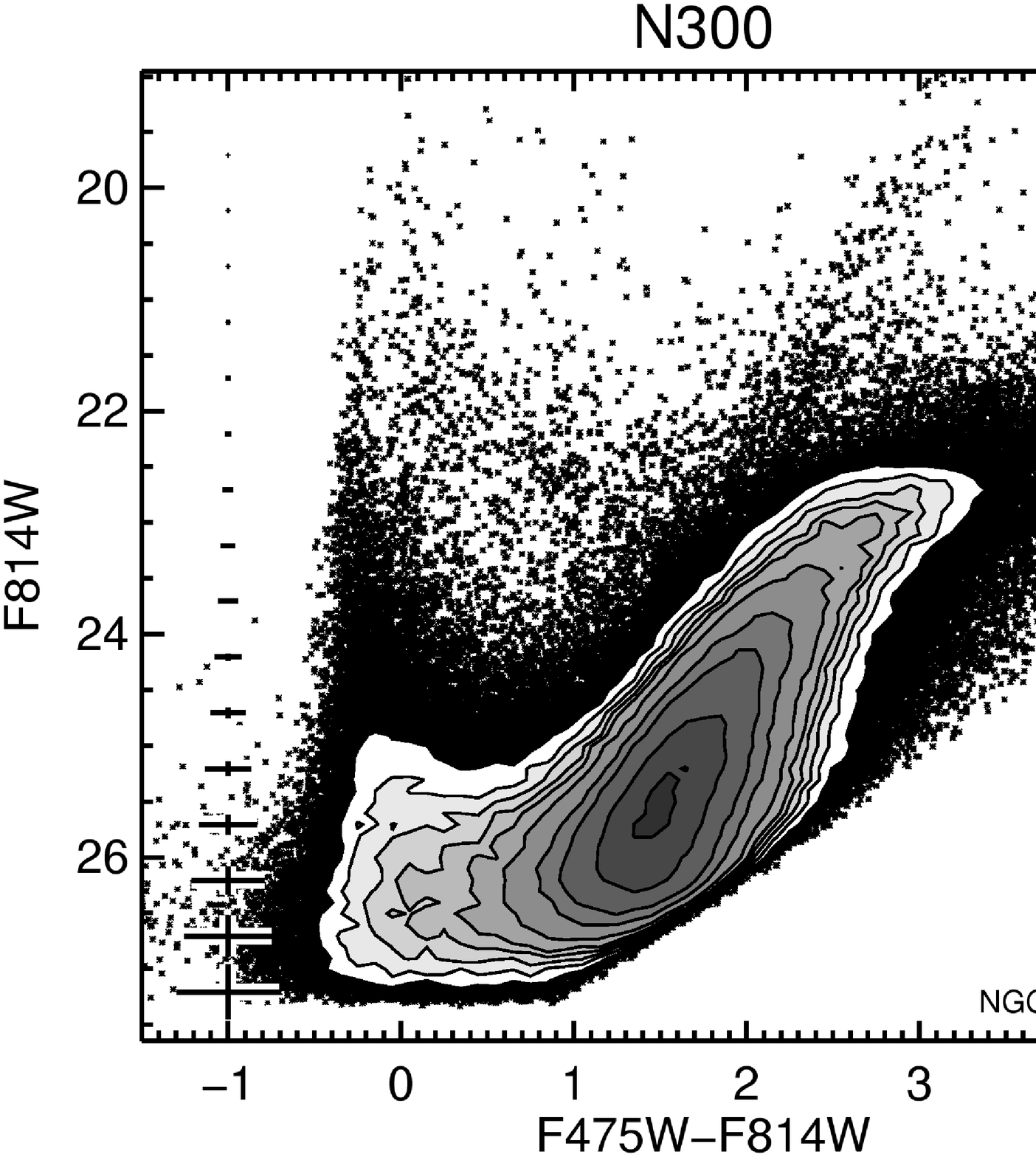}
\includegraphics[width=1.625in]{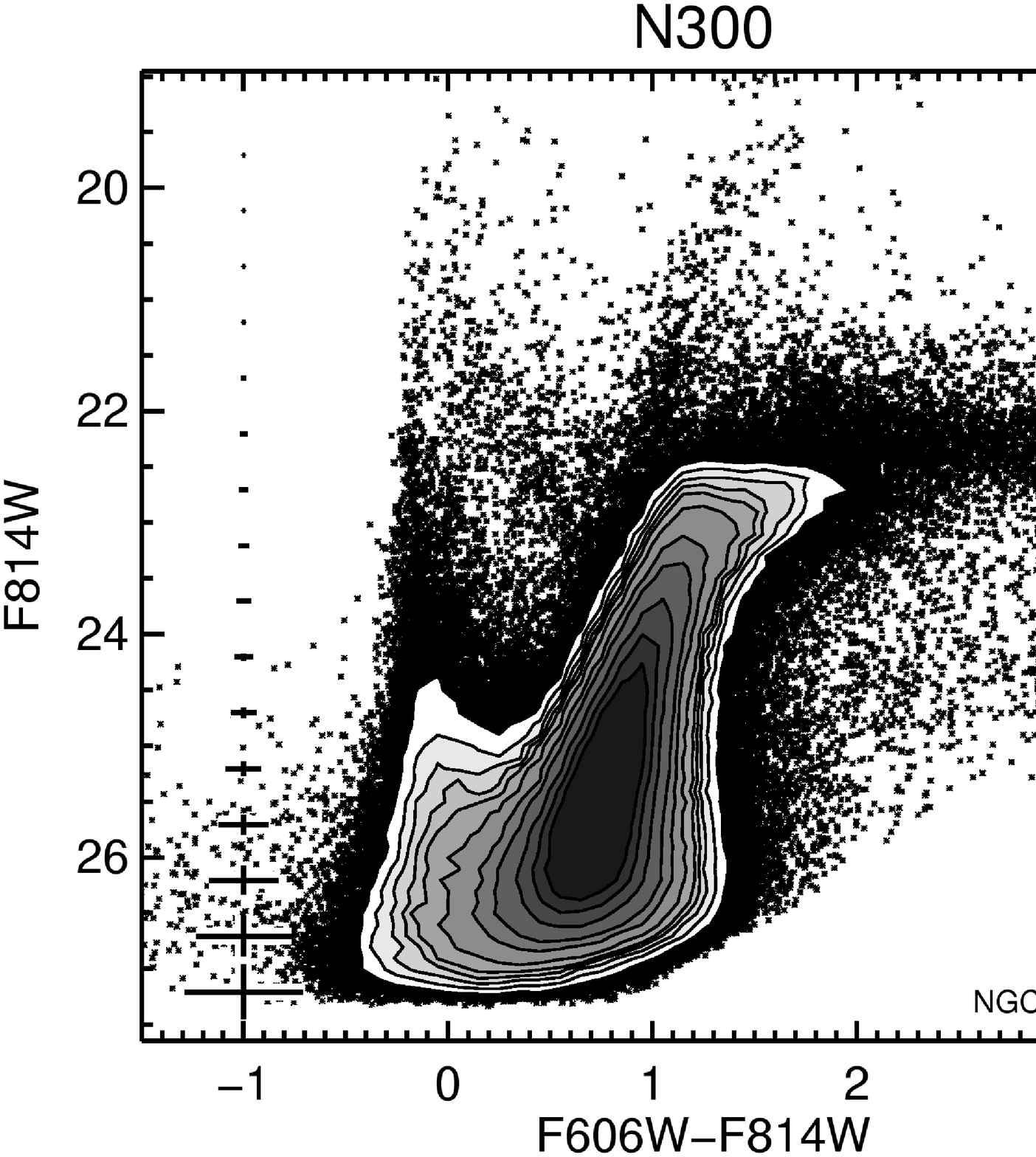}
\includegraphics[width=1.625in]{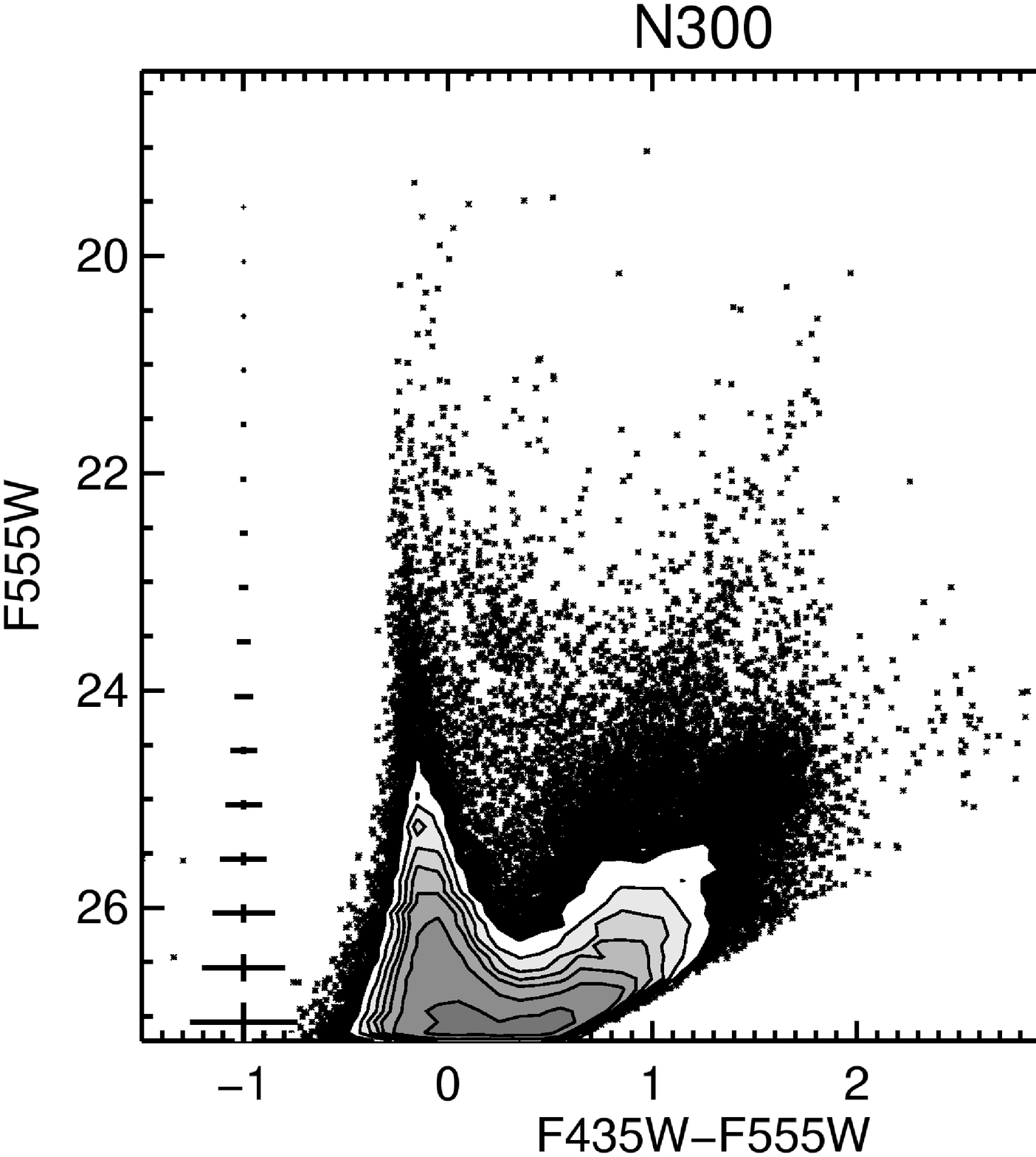}
\includegraphics[width=1.625in]{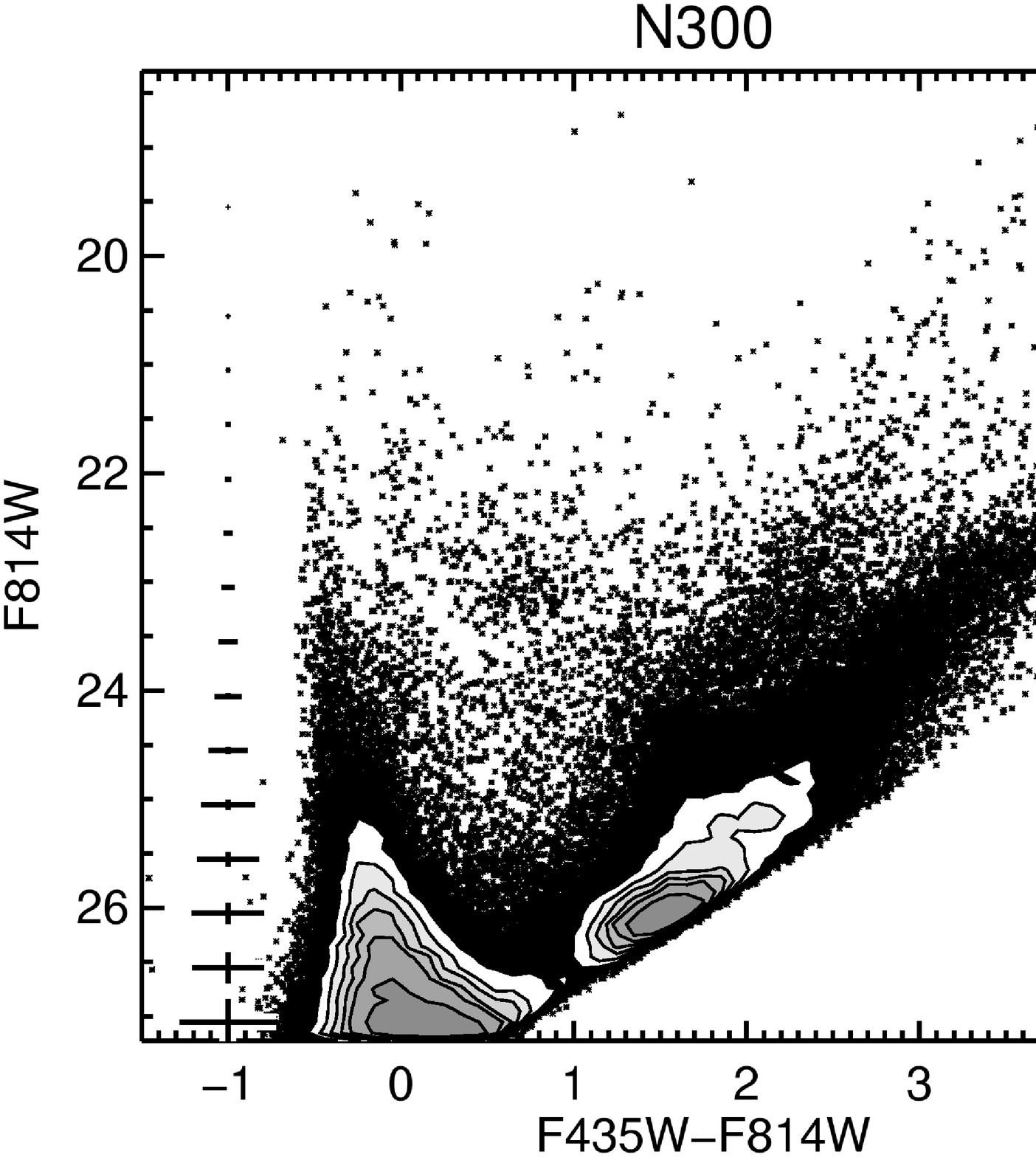}
}
\caption{
CMDs of galaxies in the ANGST data release,
as described in Figure~\ref{cmdfig1}.
Figures are ordered from the upper left to the bottom right.
(a) N55; (b) N55; (c) N55; (d) I5152; (e) GR8; (f) N300; (g) N300; (h) N300; (i) N300; (j) N300; (k) N300; (l) N300; (m) N300; (n) N300; (o) N300; (p) N300; 
    \label{cmdfig2}}
\end{figure}
\vfill
\clearpage
 
%-------------------
\begin{figure}[p]
\centerline{
\includegraphics[width=1.625in]{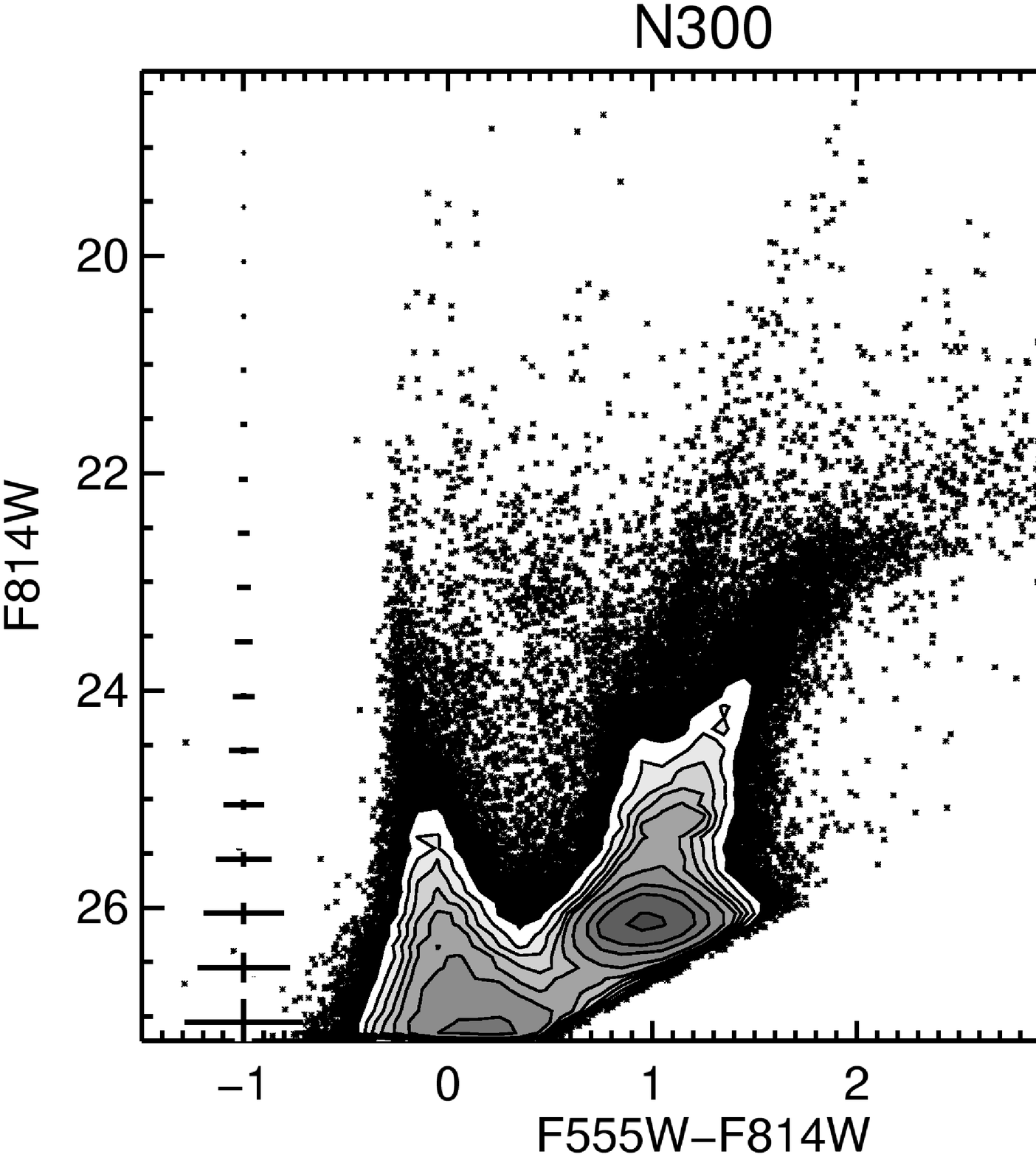}
\includegraphics[width=1.625in]{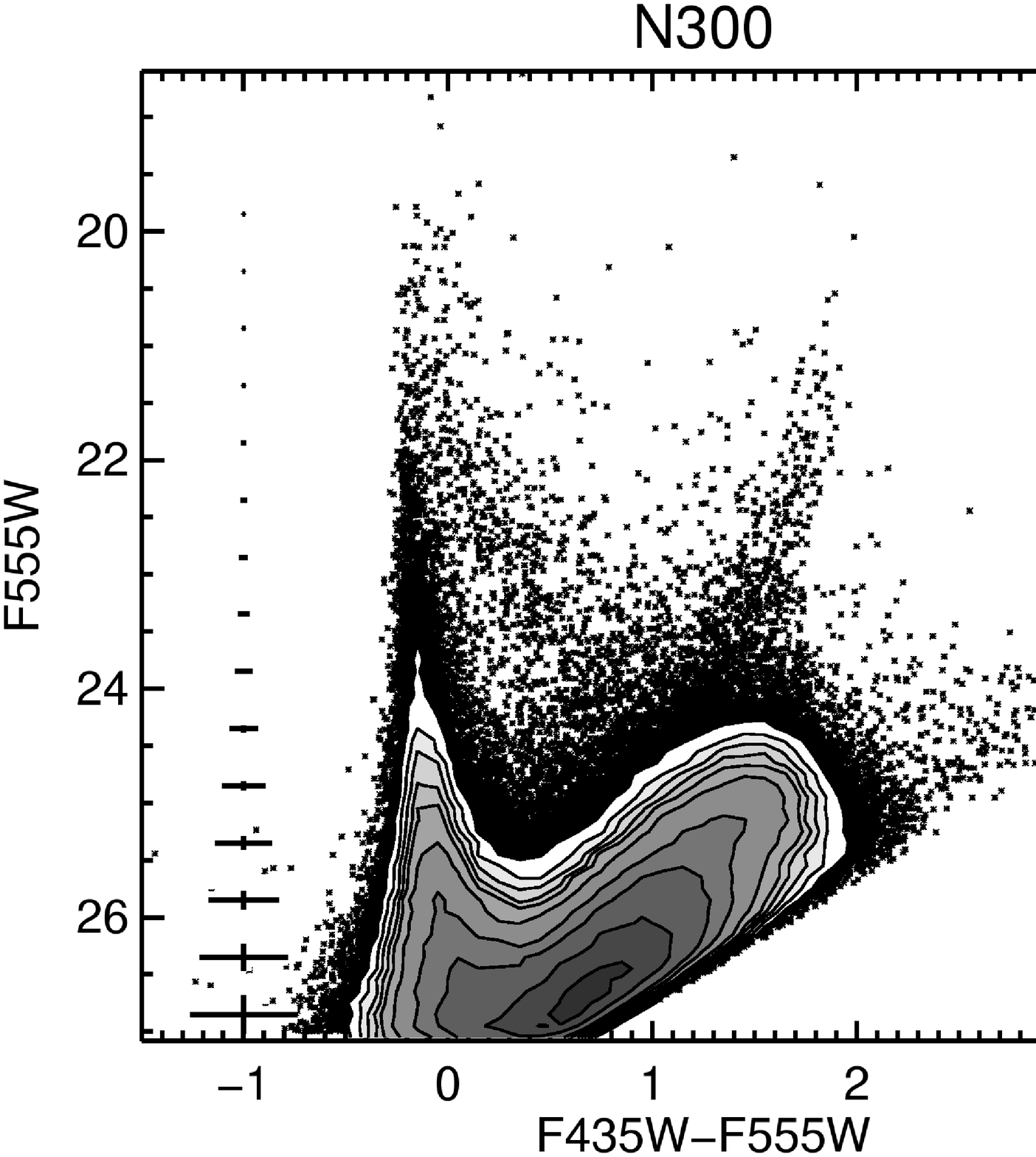}
\includegraphics[width=1.625in]{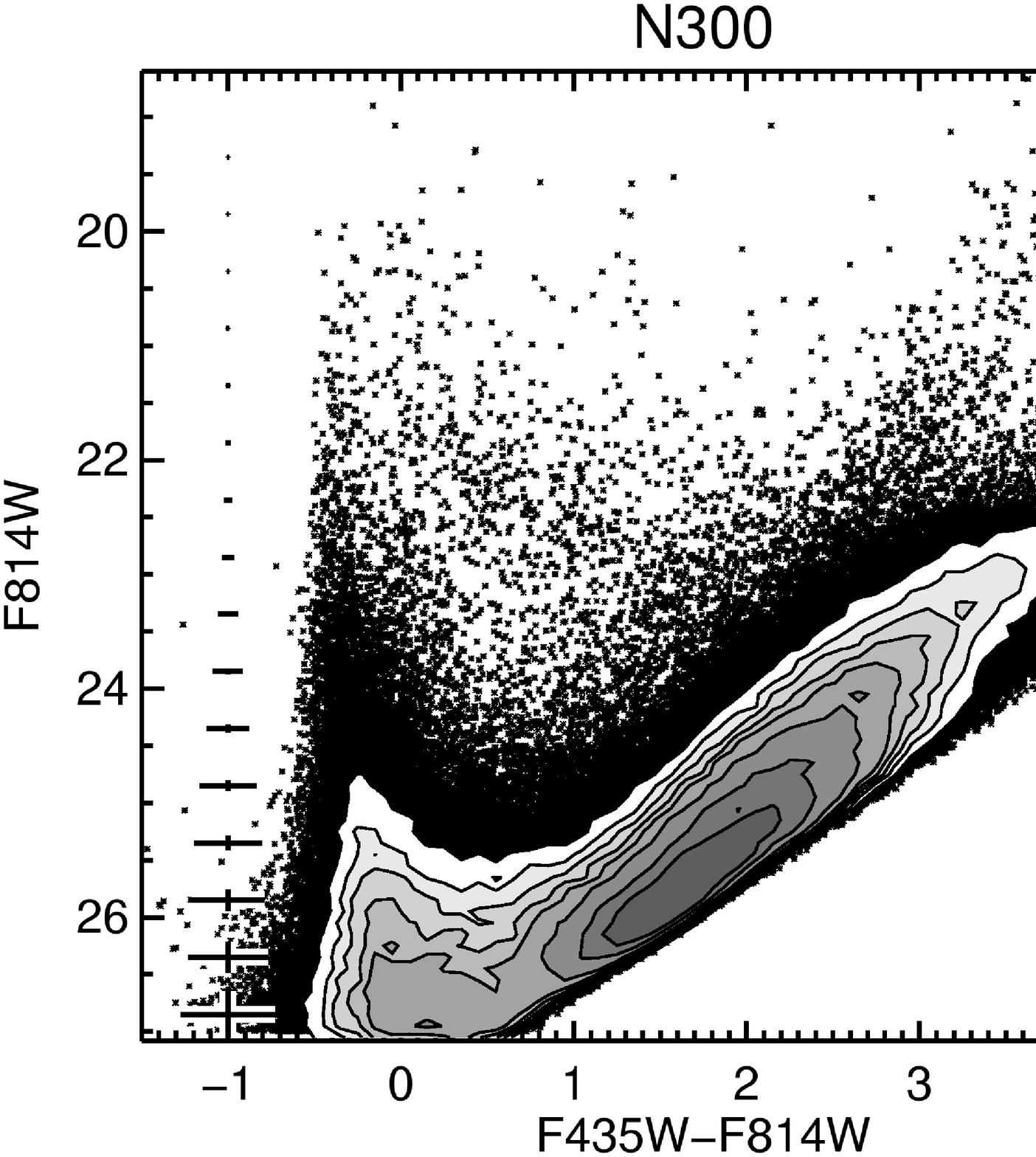}
\includegraphics[width=1.625in]{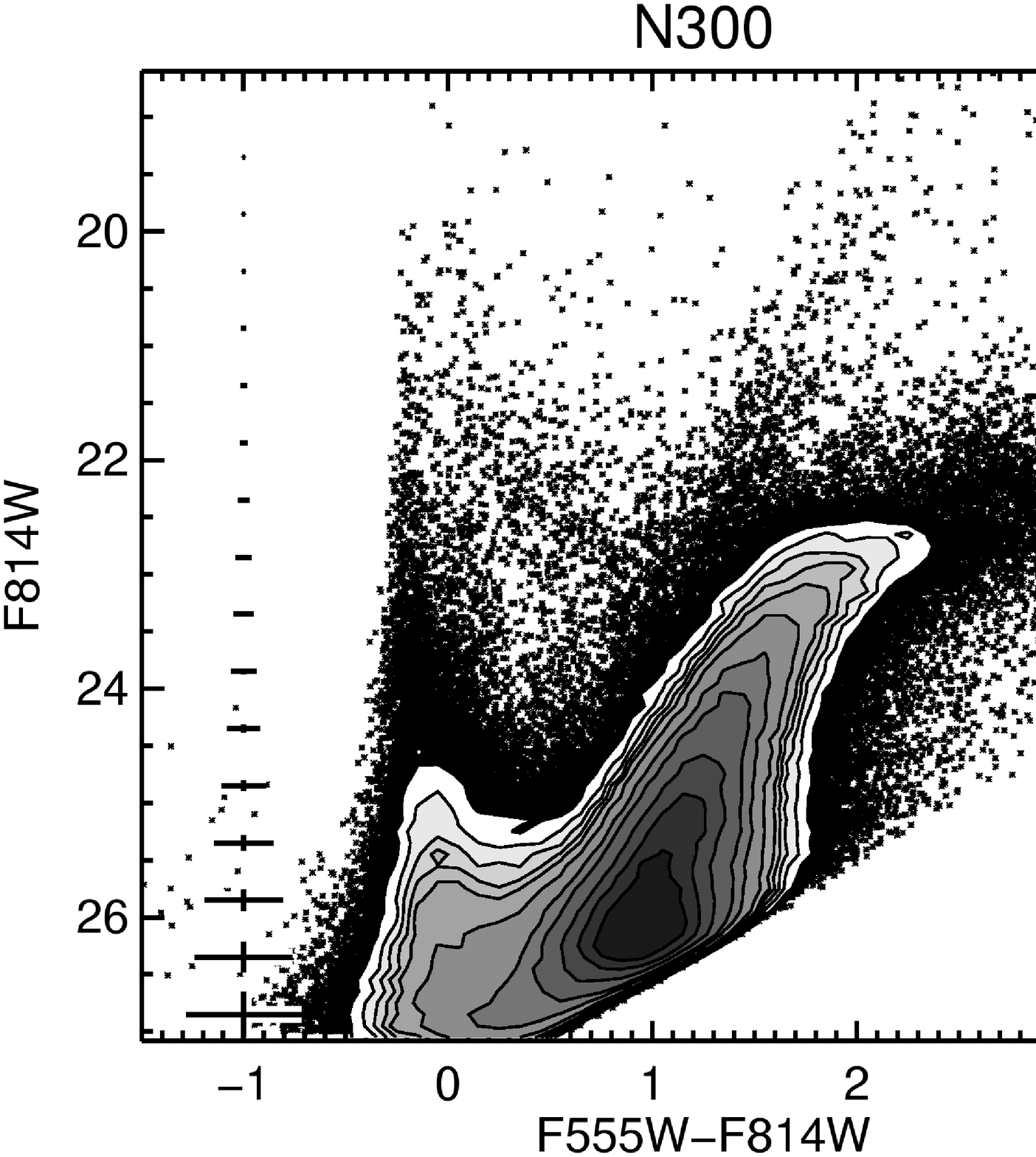}
}
\centerline{
\includegraphics[width=1.625in]{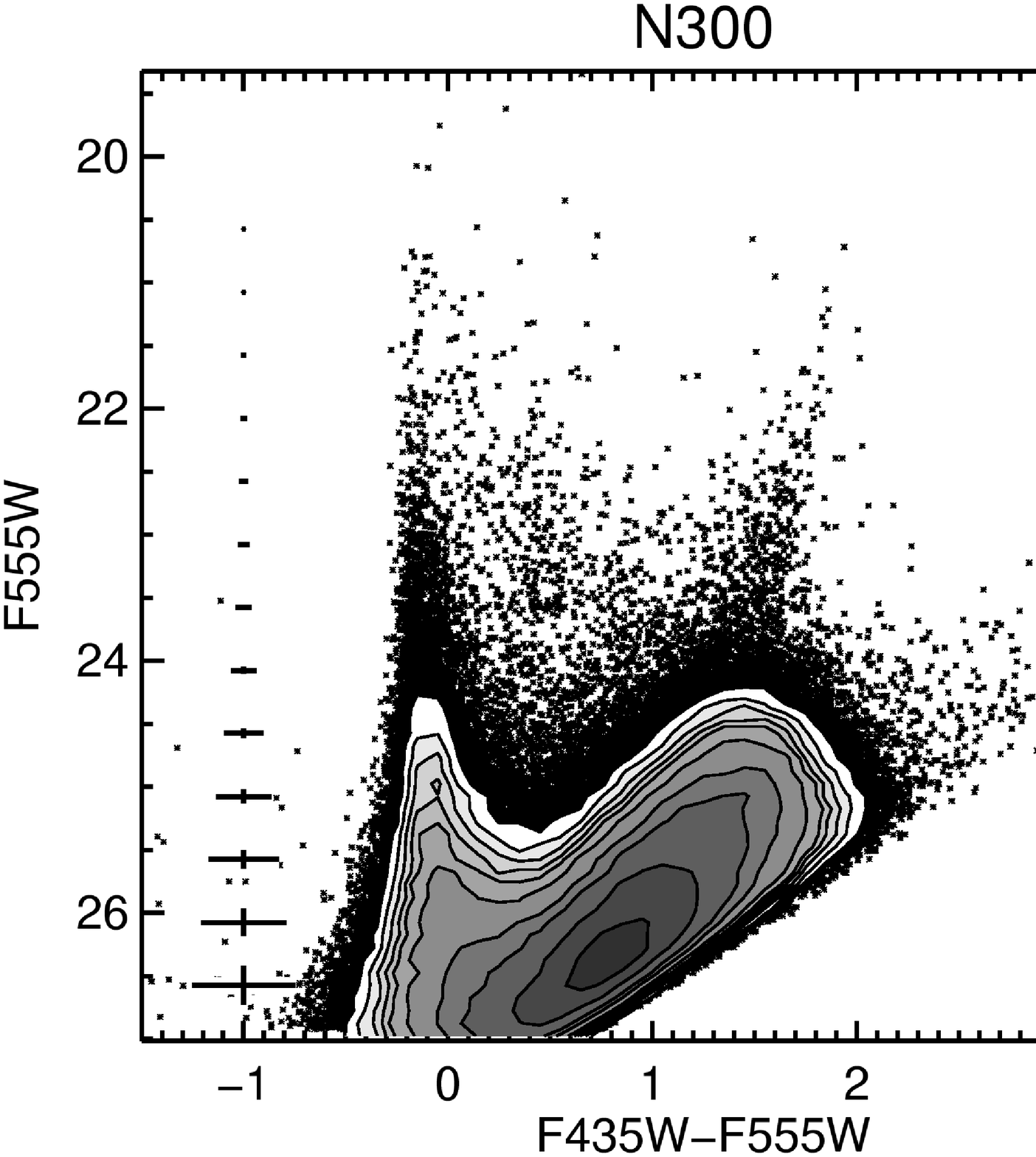}
\includegraphics[width=1.625in]{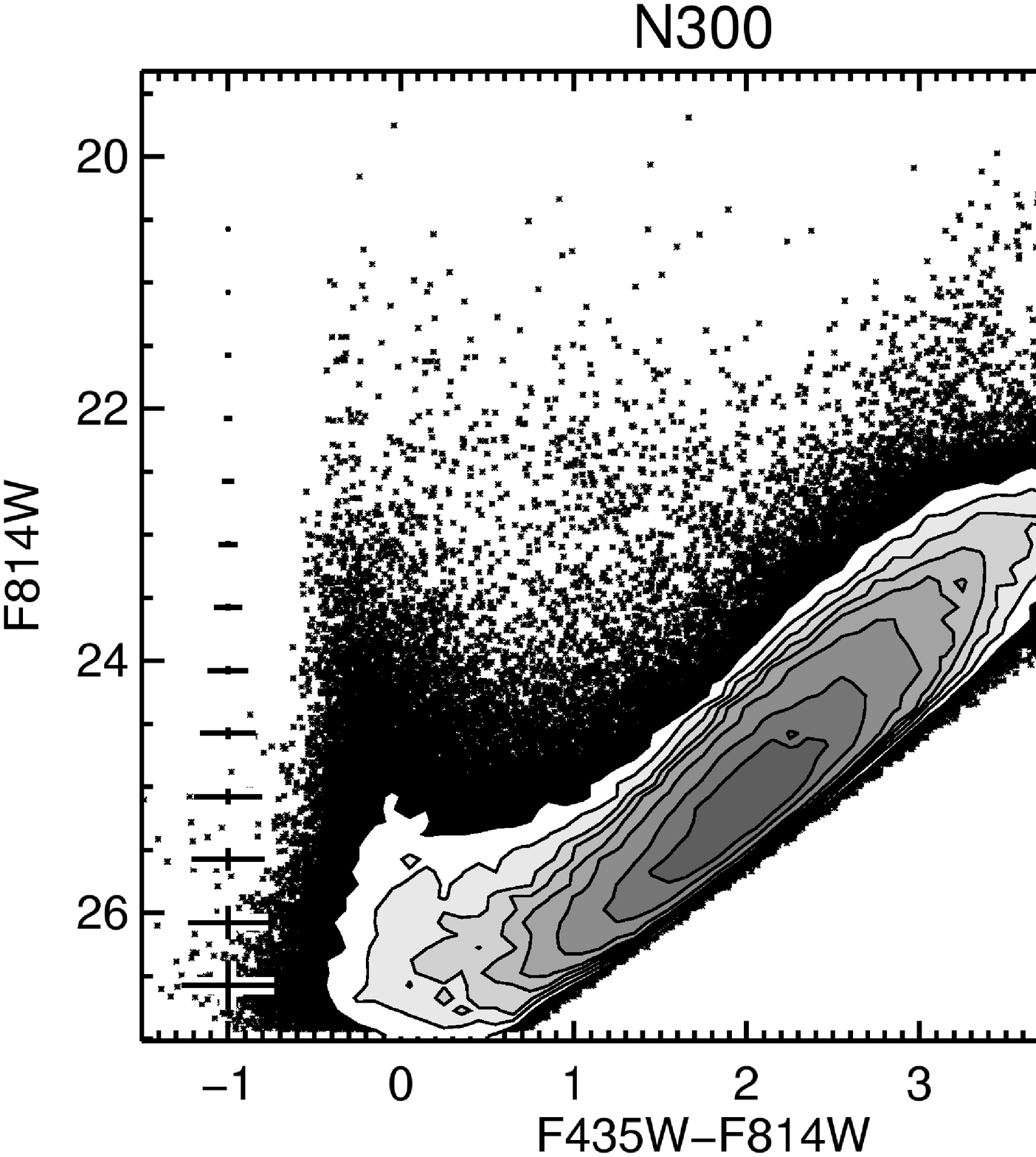}
\includegraphics[width=1.625in]{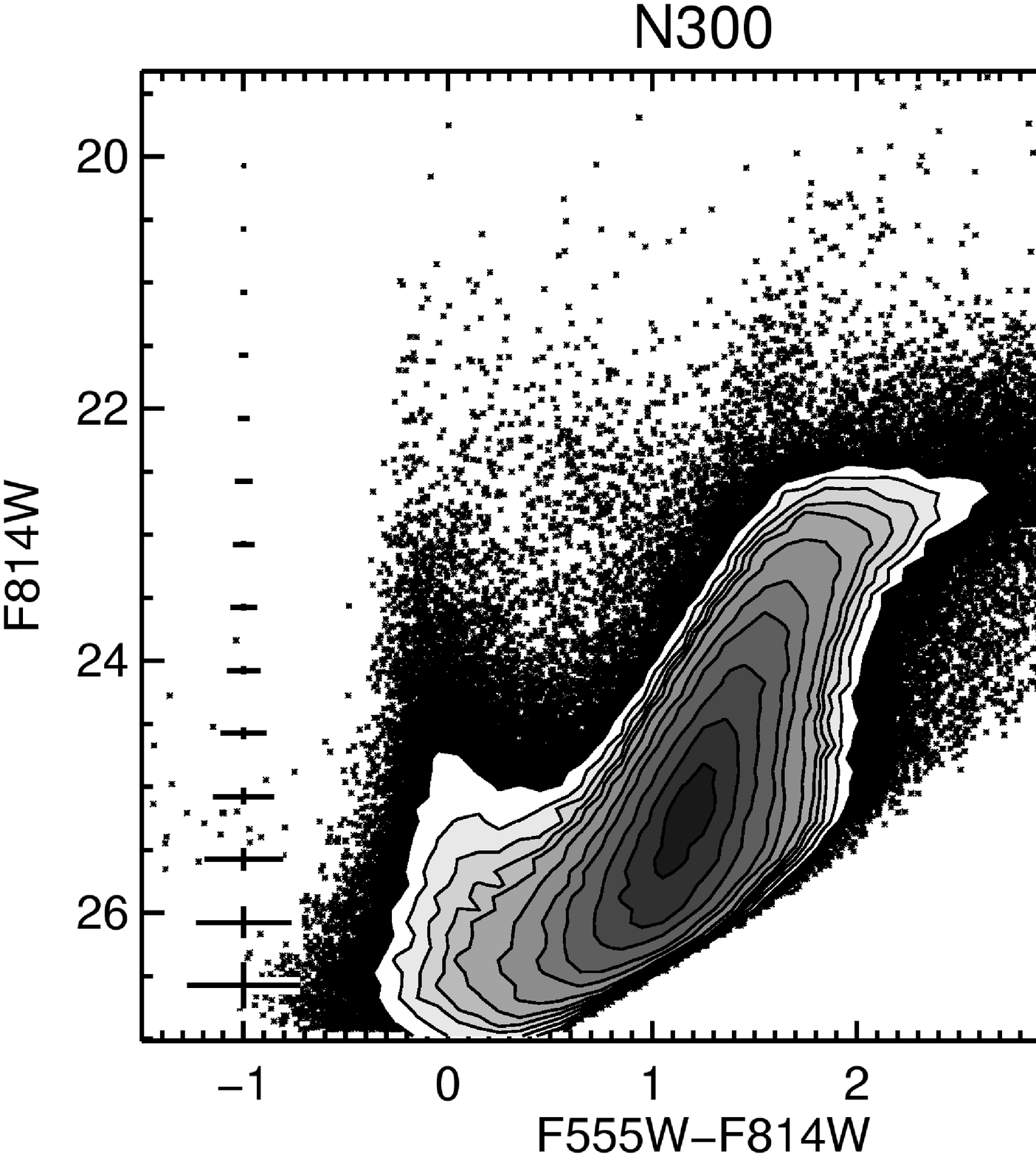}
\includegraphics[width=1.625in]{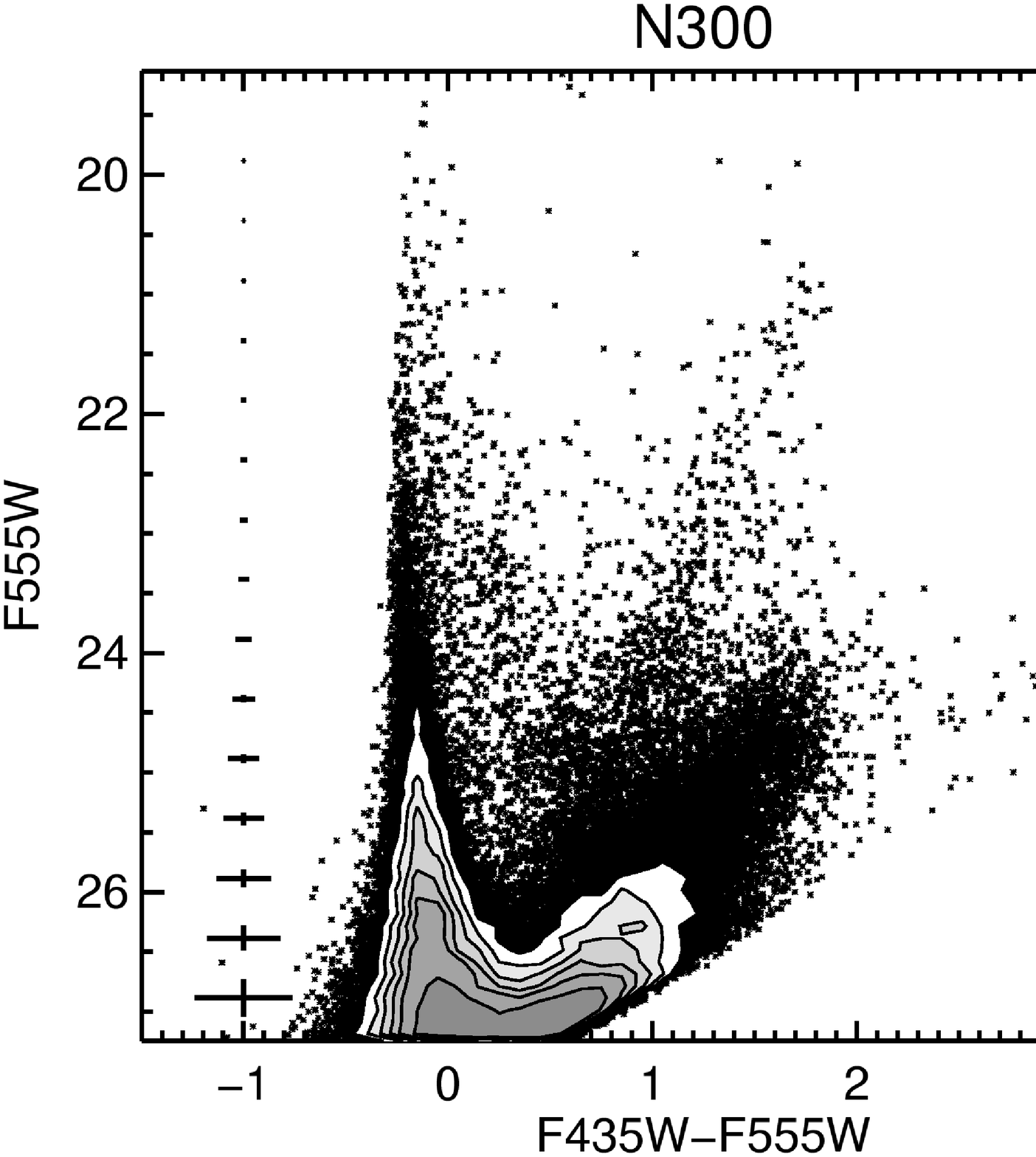}
}
\centerline{
\includegraphics[width=1.625in]{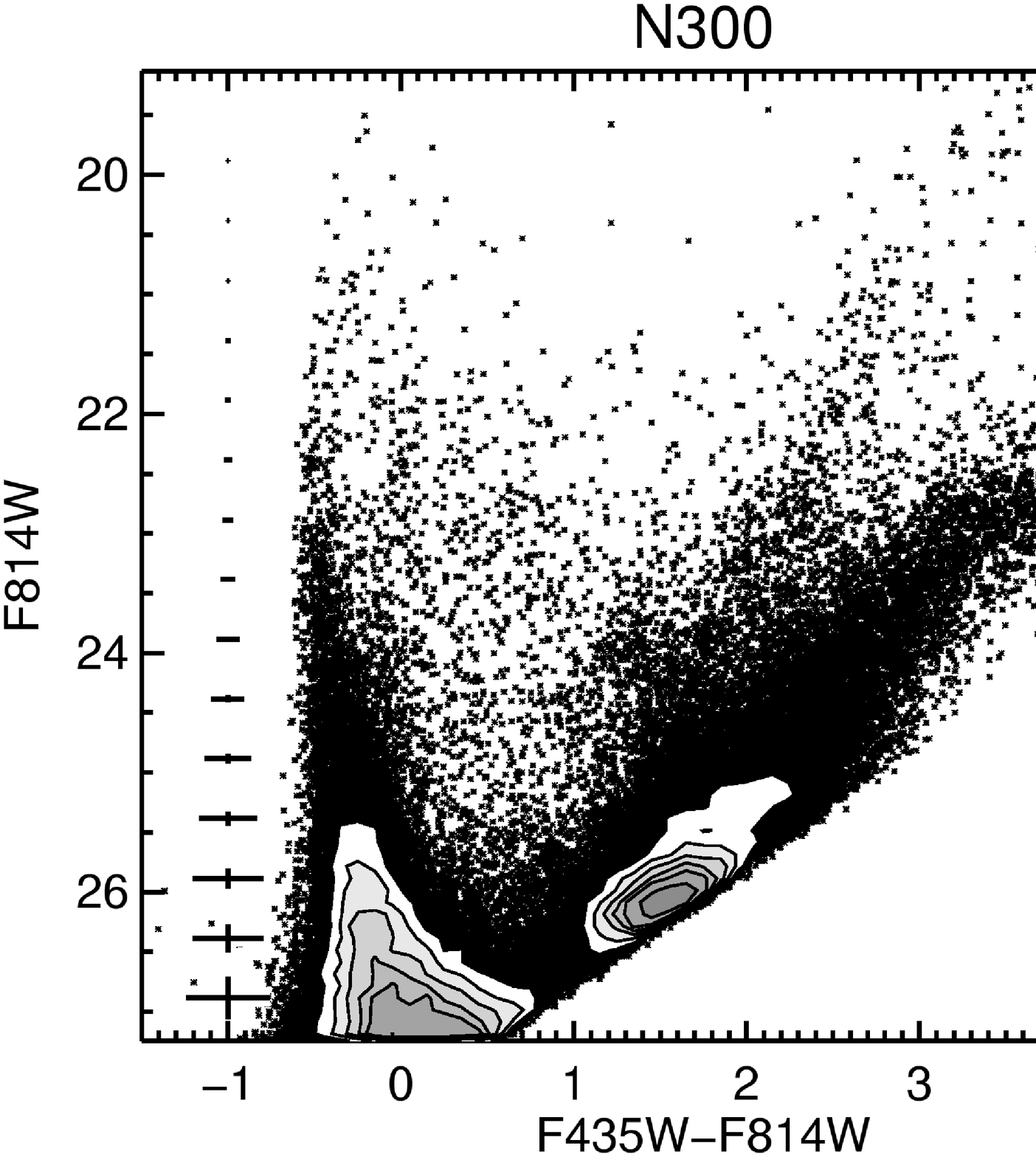}
\includegraphics[width=1.625in]{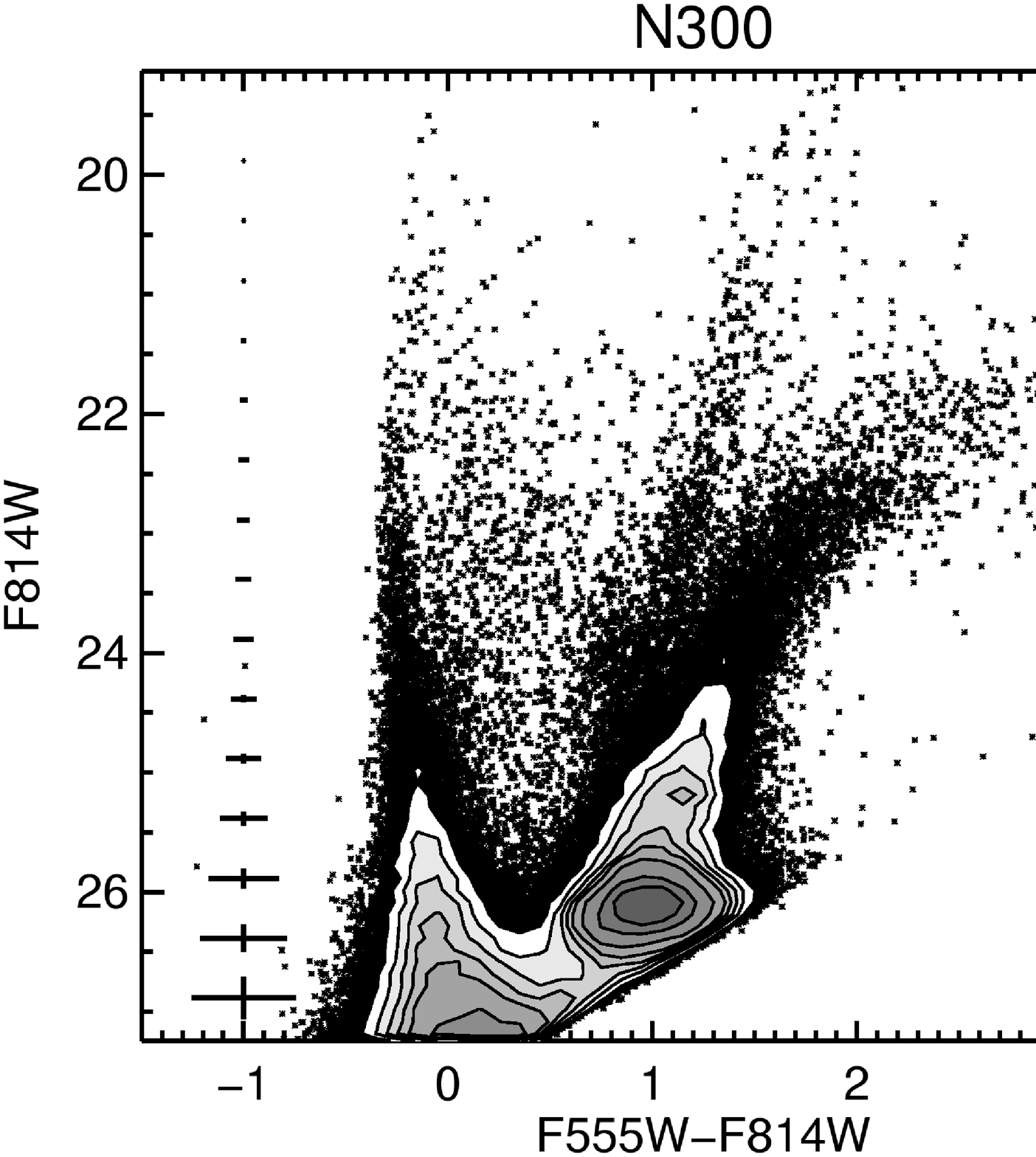}
\includegraphics[width=1.625in]{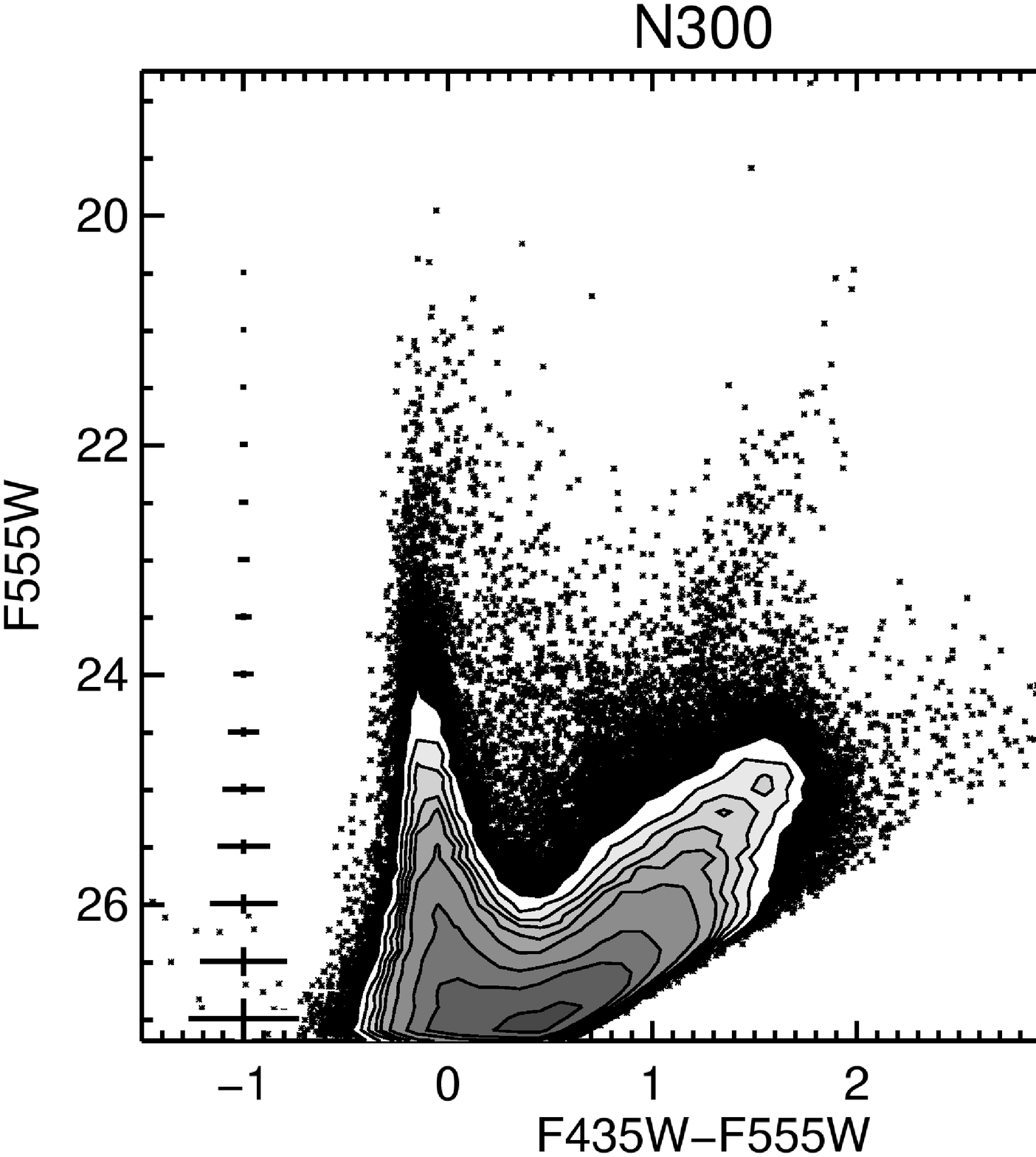}
\includegraphics[width=1.625in]{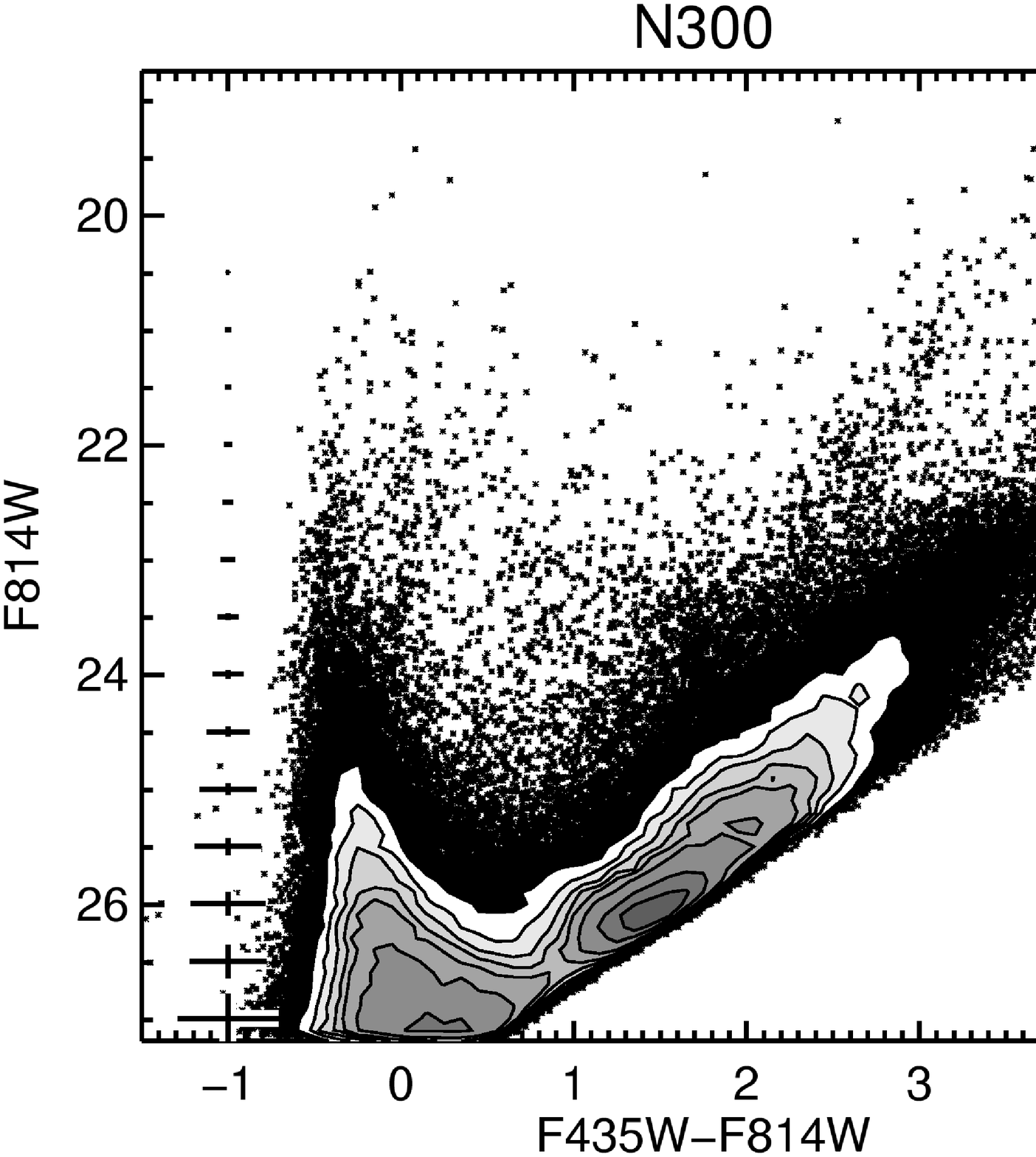}
}
\centerline{
\includegraphics[width=1.625in]{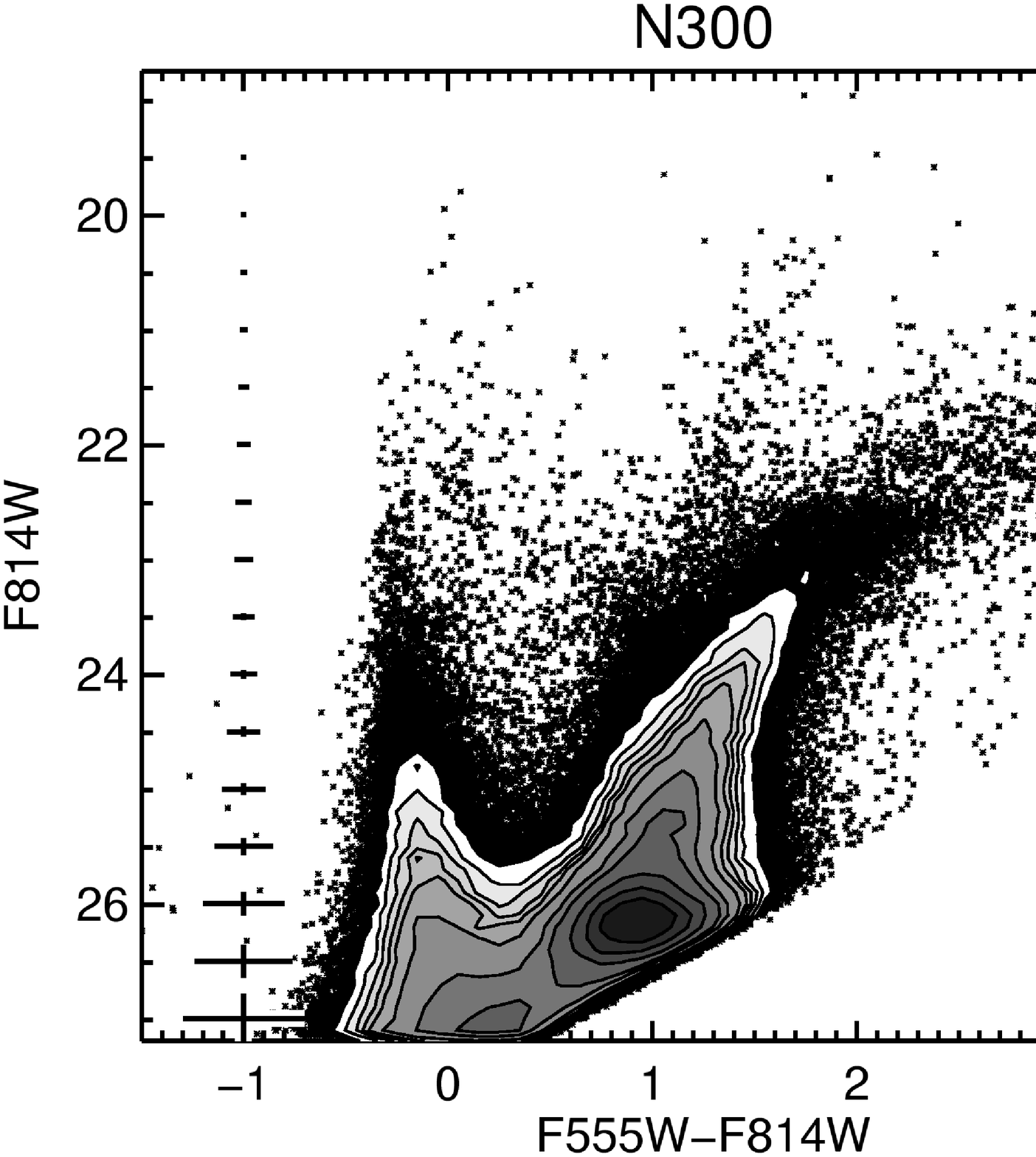}
\includegraphics[width=1.625in]{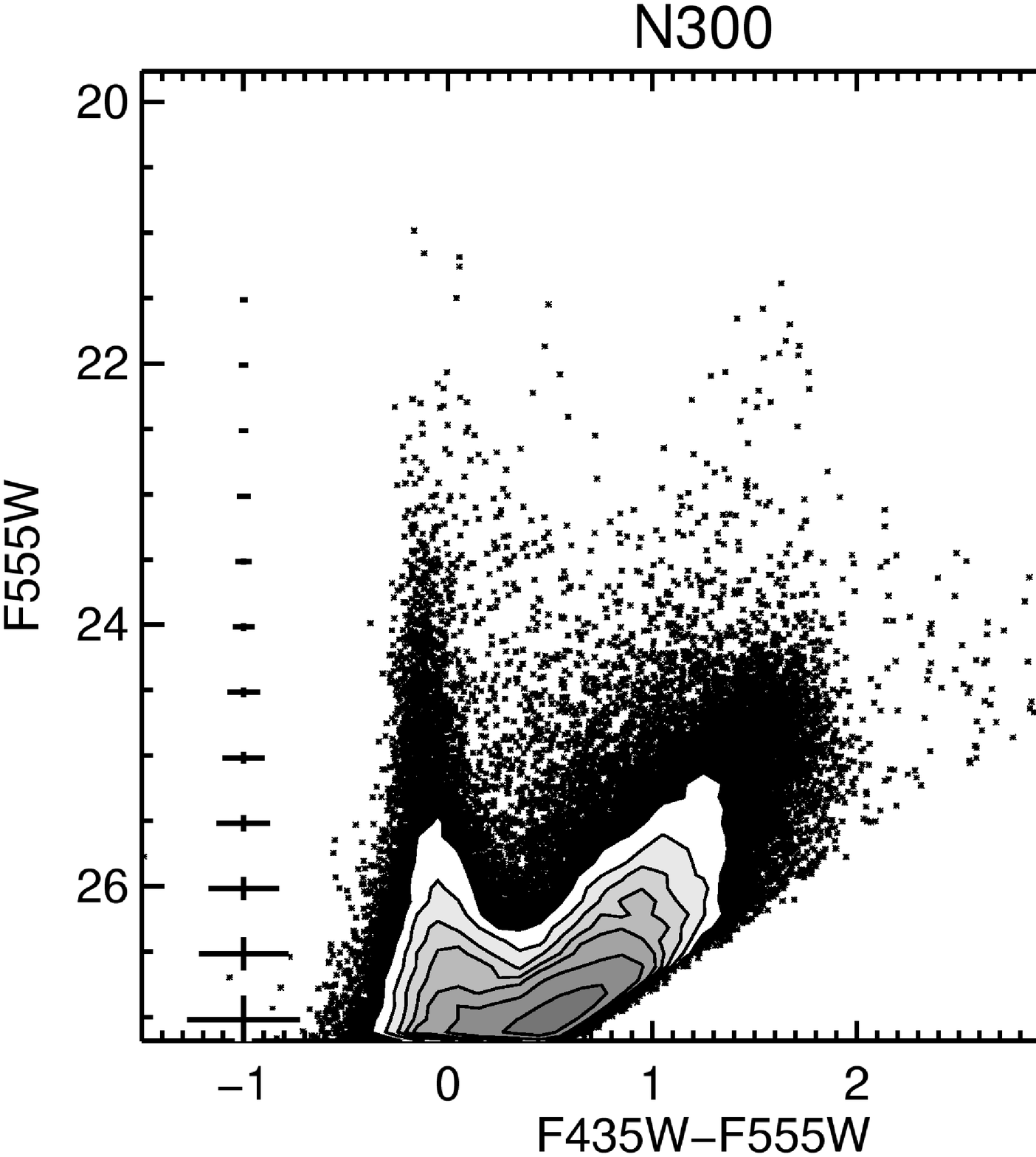}
\includegraphics[width=1.625in]{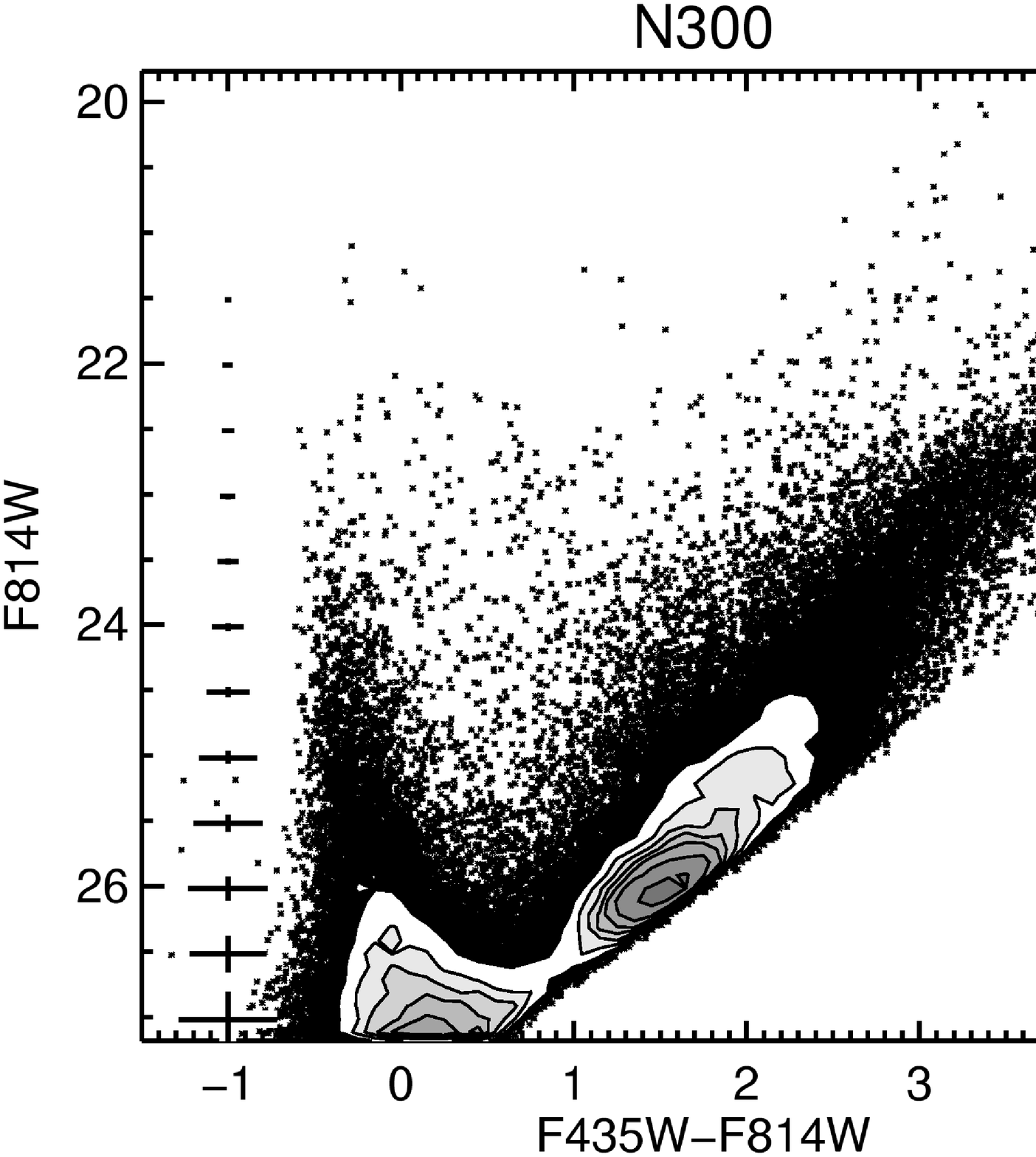}
\includegraphics[width=1.625in]{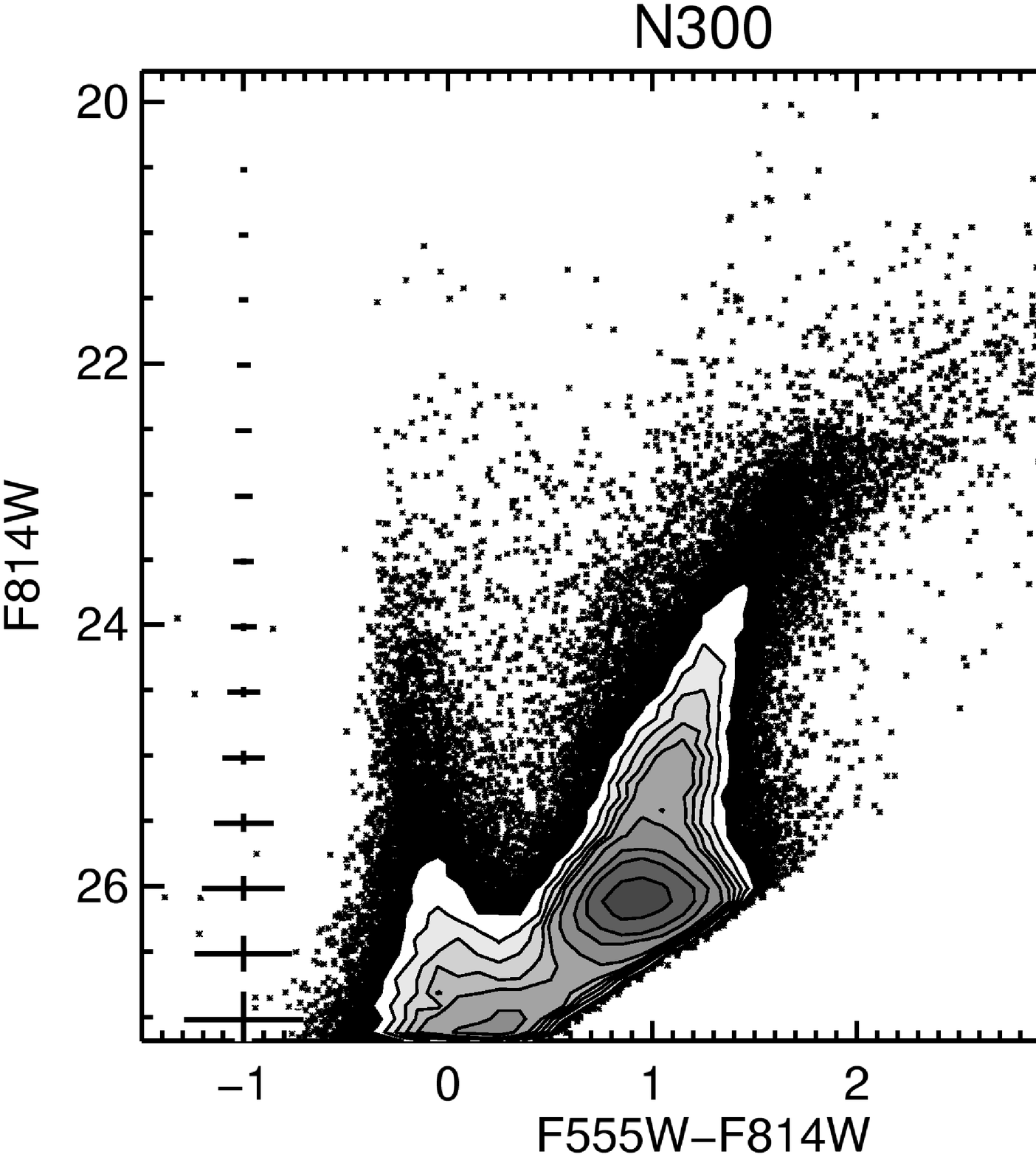}
}
\caption{
CMDs of galaxies in the ANGST data release,
as described in Figure~\ref{cmdfig1}.
Figures are ordered from the upper left to the bottom right.
(a) N300; (b) N300; (c) N300; (d) N300; (e) N300; (f) N300; (g) N300; (h) N300; (i) N300; (j) N300; (k) N300; (l) N300; (m) N300; (n) N300; (o) N300; (p) N300; 
    \label{cmdfig3}}
\end{figure}
\vfill
\clearpage
 
%-------------------
\begin{figure}[p]
\centerline{
\includegraphics[width=1.625in]{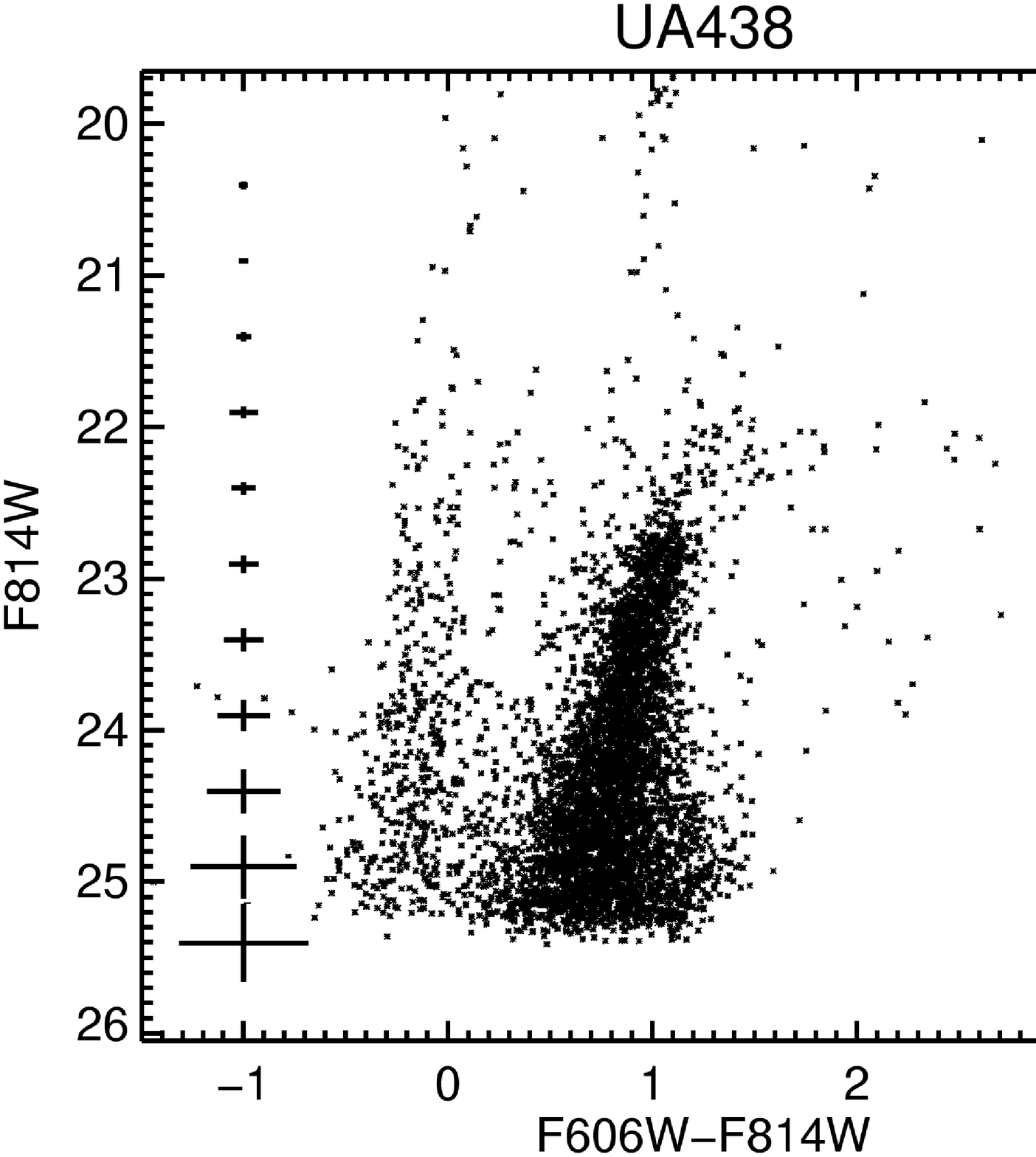}
\includegraphics[width=1.625in]{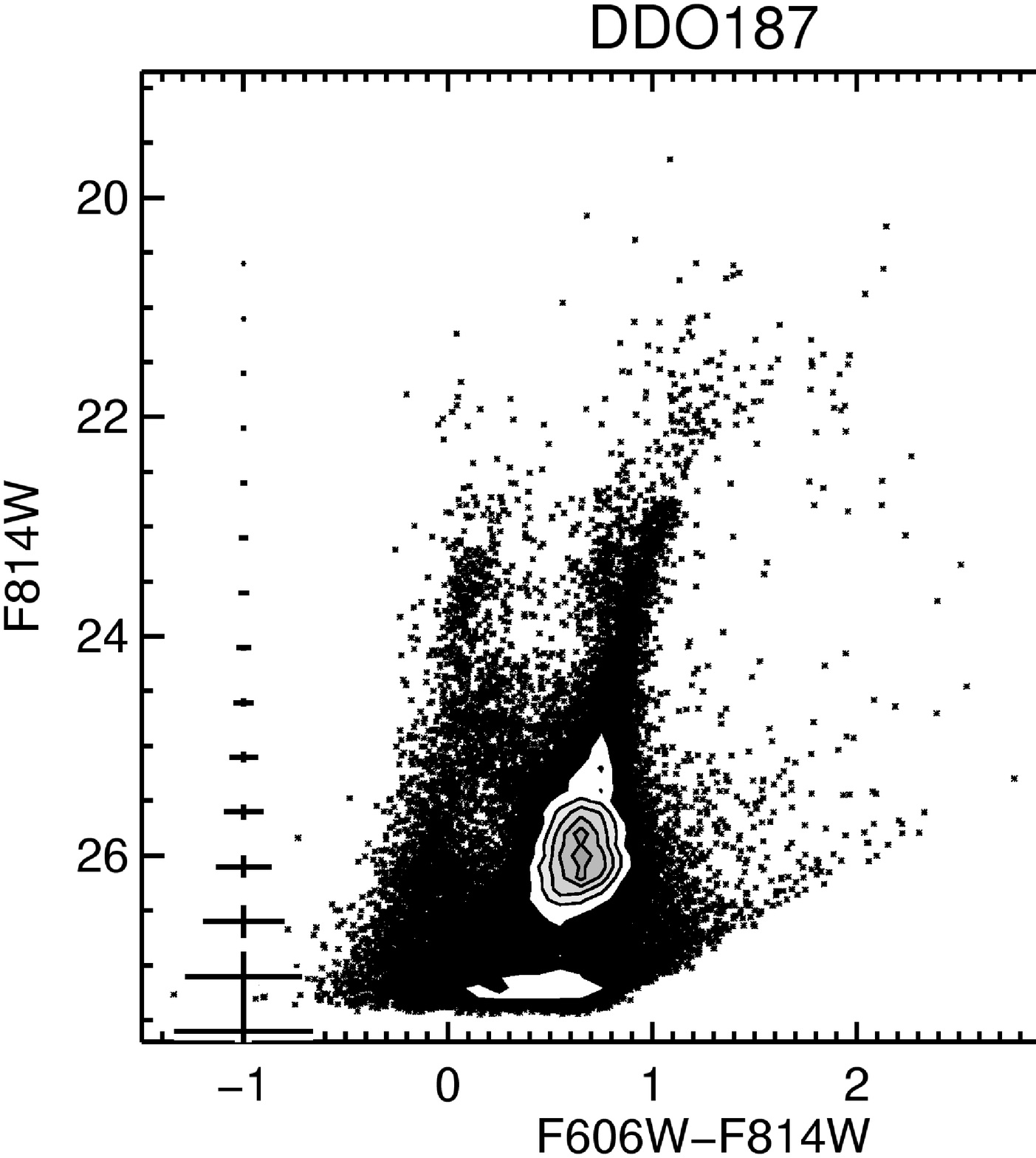}
\includegraphics[width=1.625in]{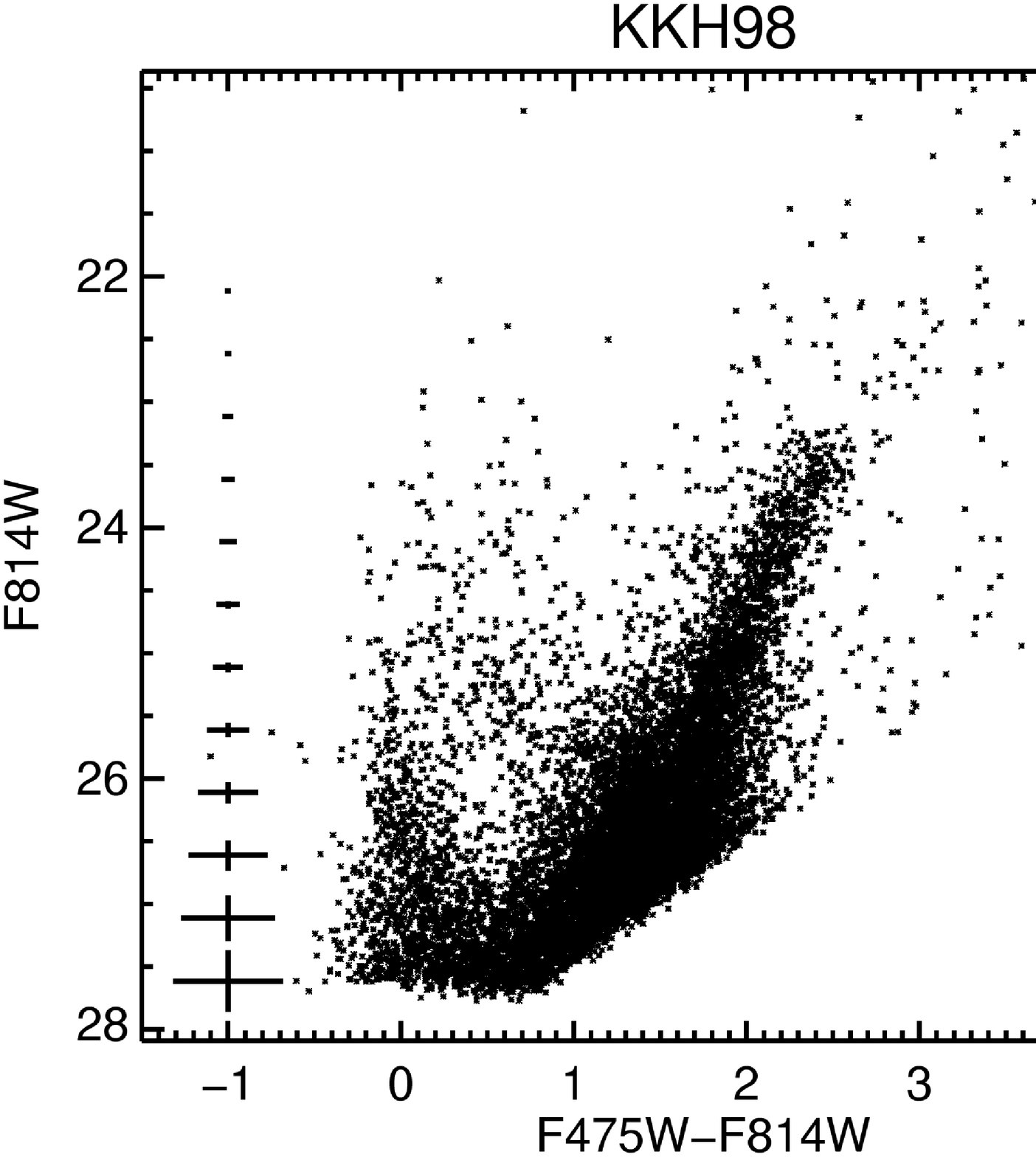}
\includegraphics[width=1.625in]{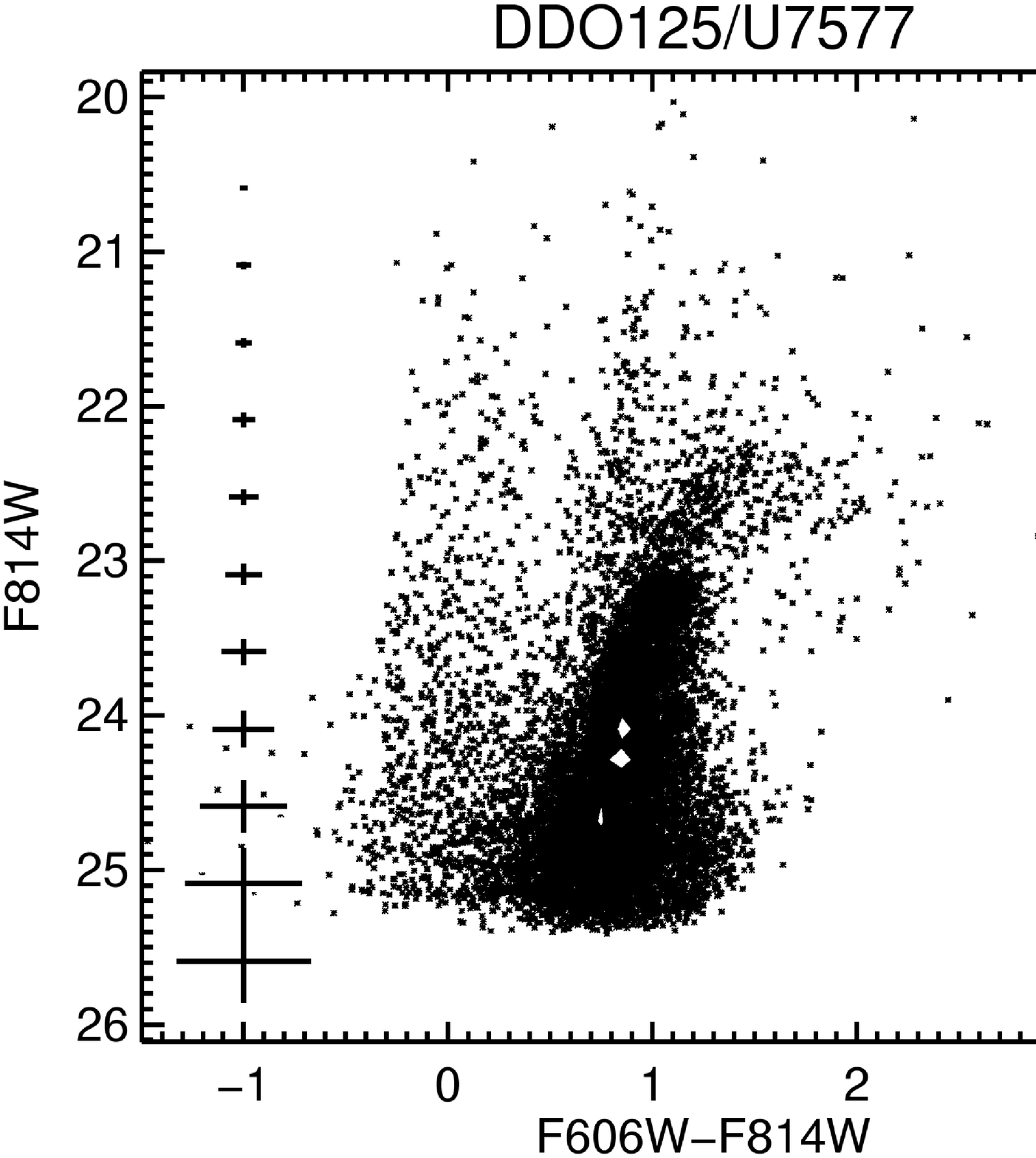}
}
\centerline{
\includegraphics[width=1.625in]{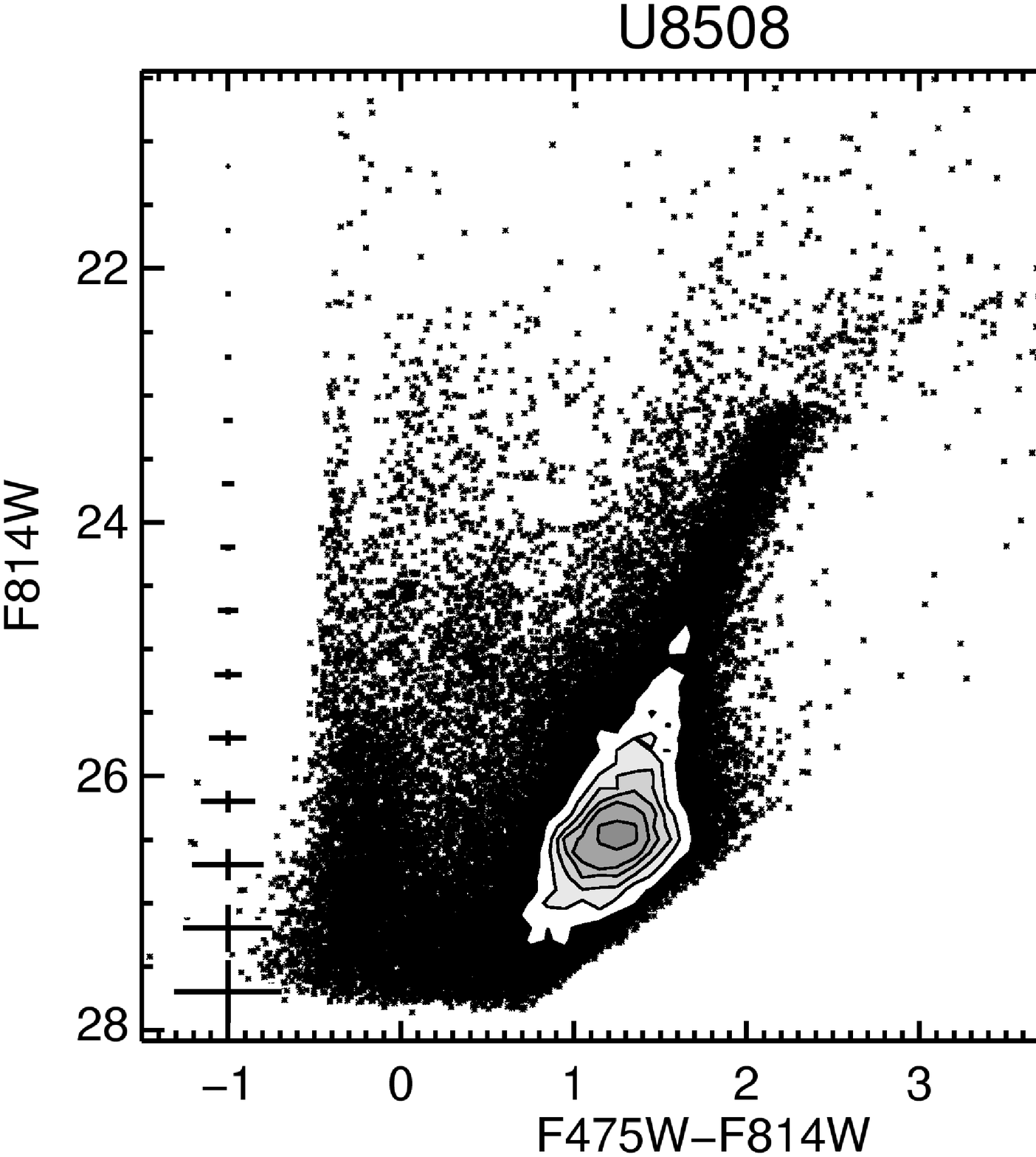}
\includegraphics[width=1.625in]{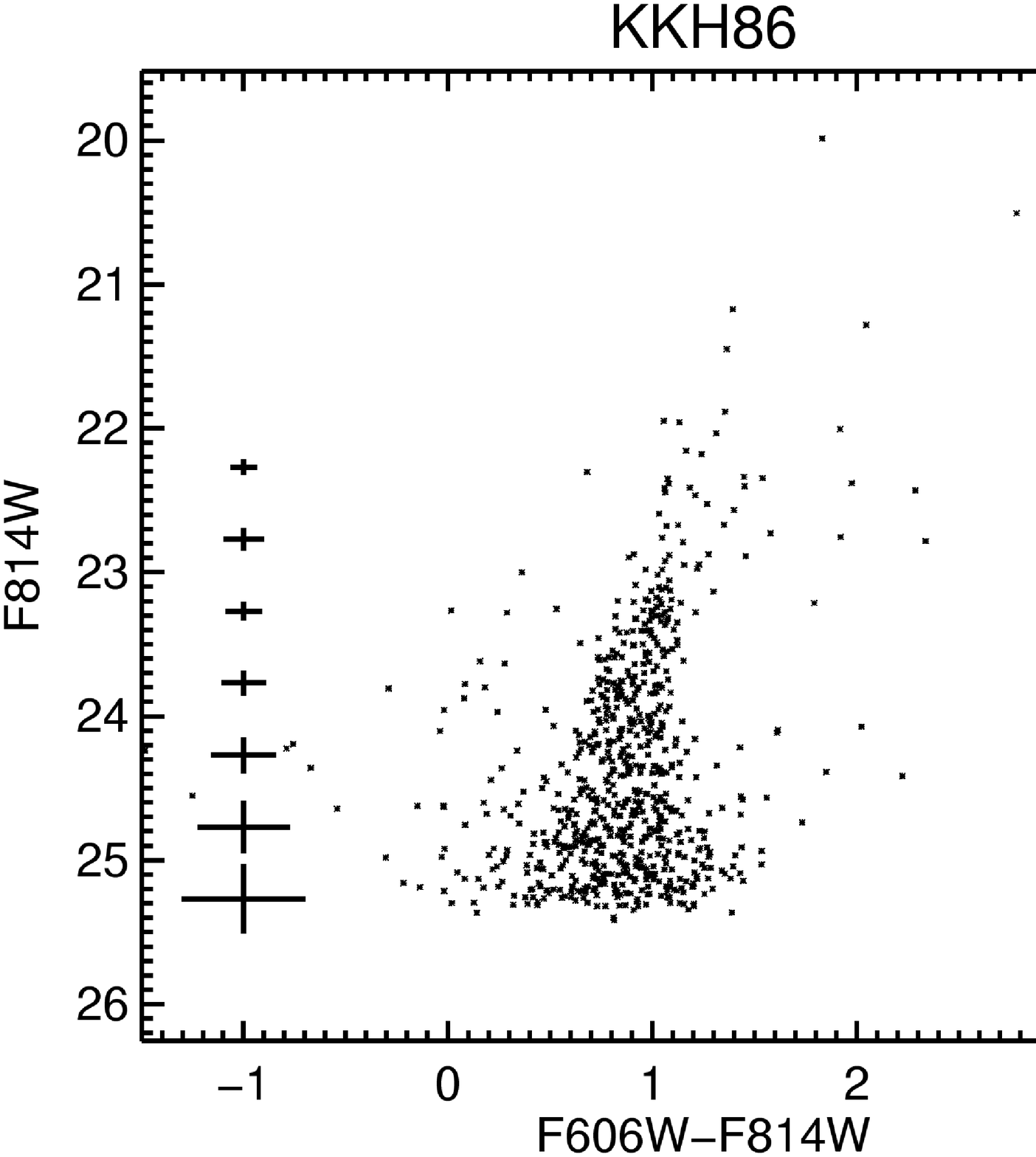}
\includegraphics[width=1.625in]{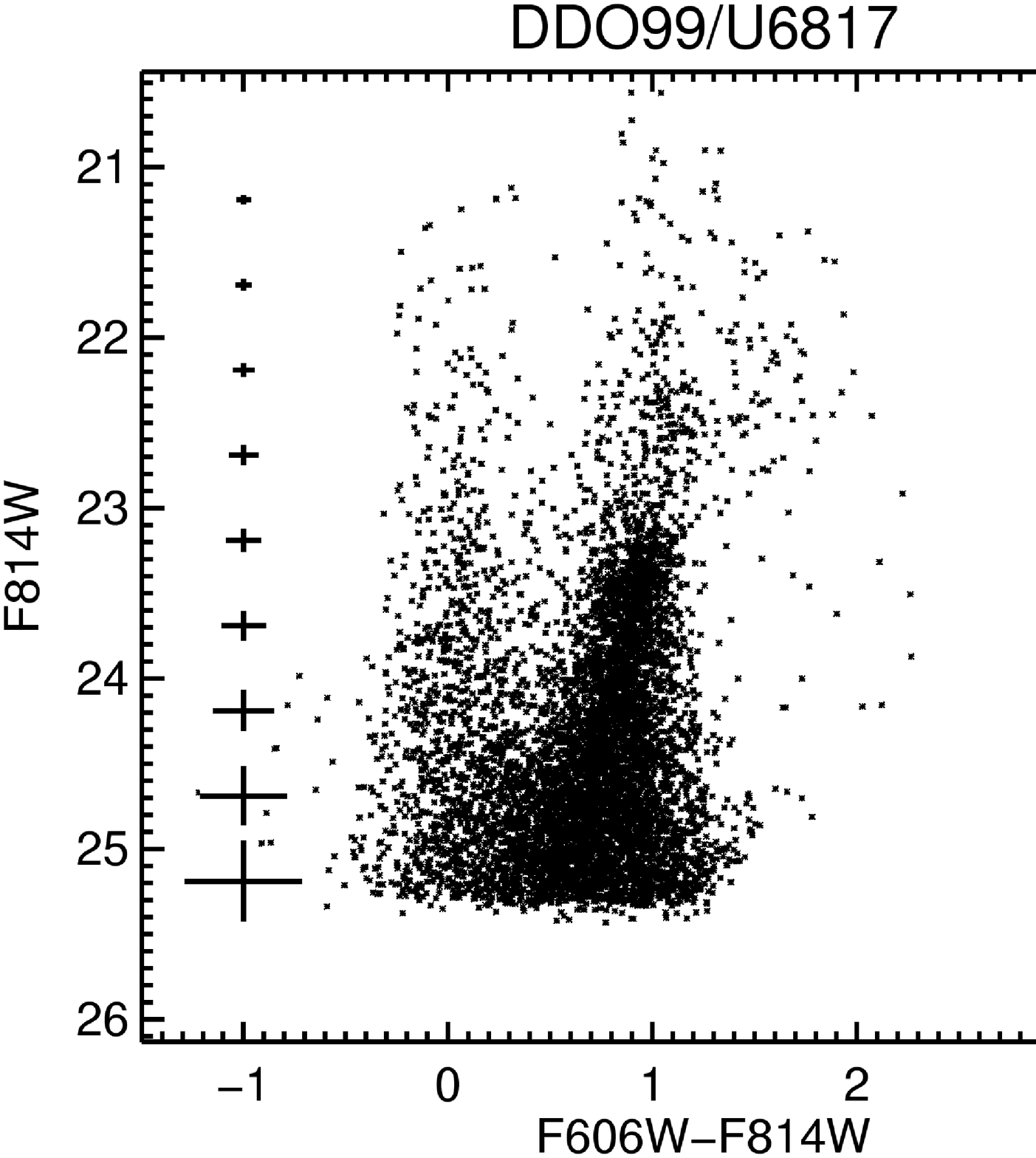}
\includegraphics[width=1.625in]{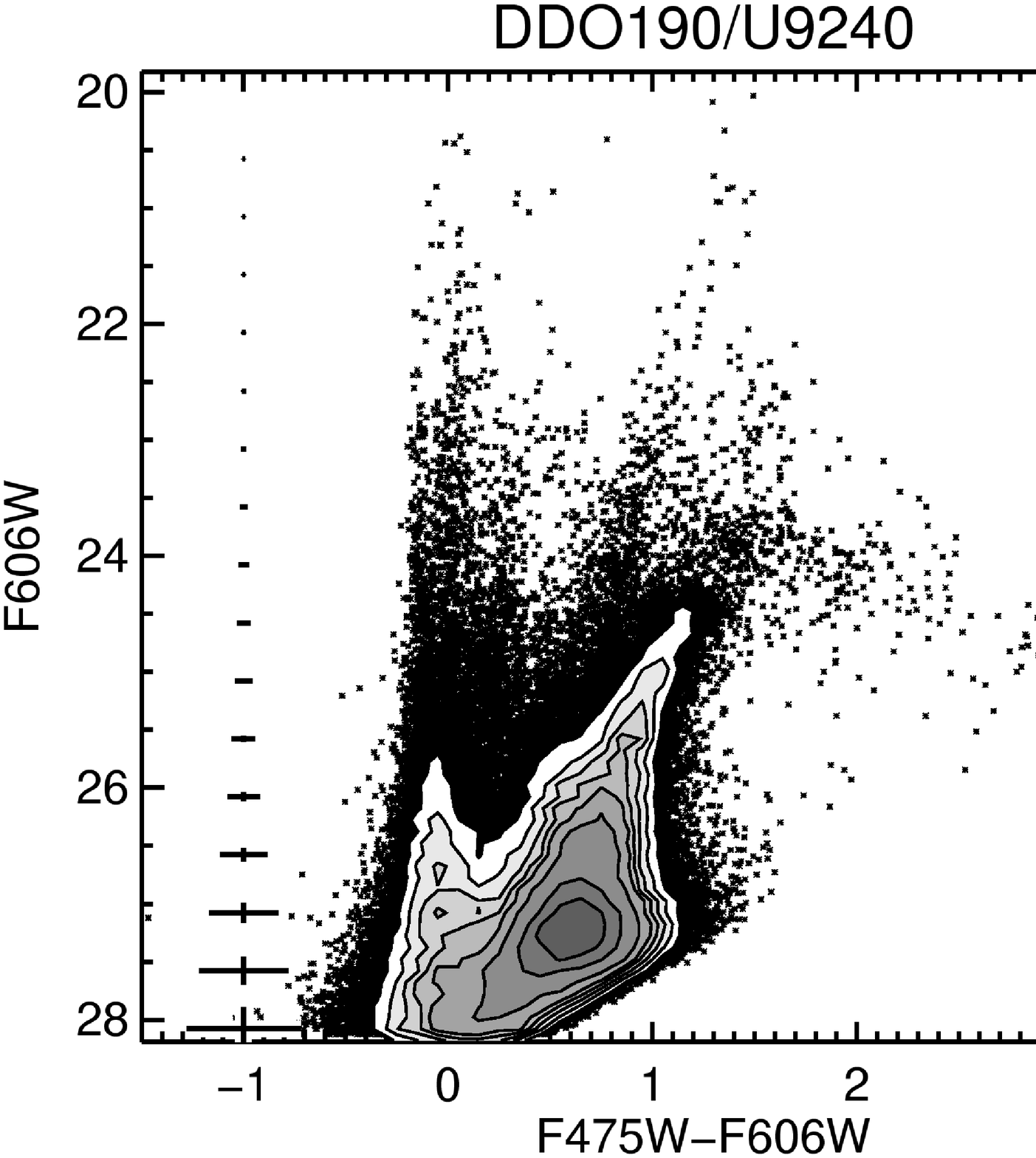}
}
\centerline{
\includegraphics[width=1.625in]{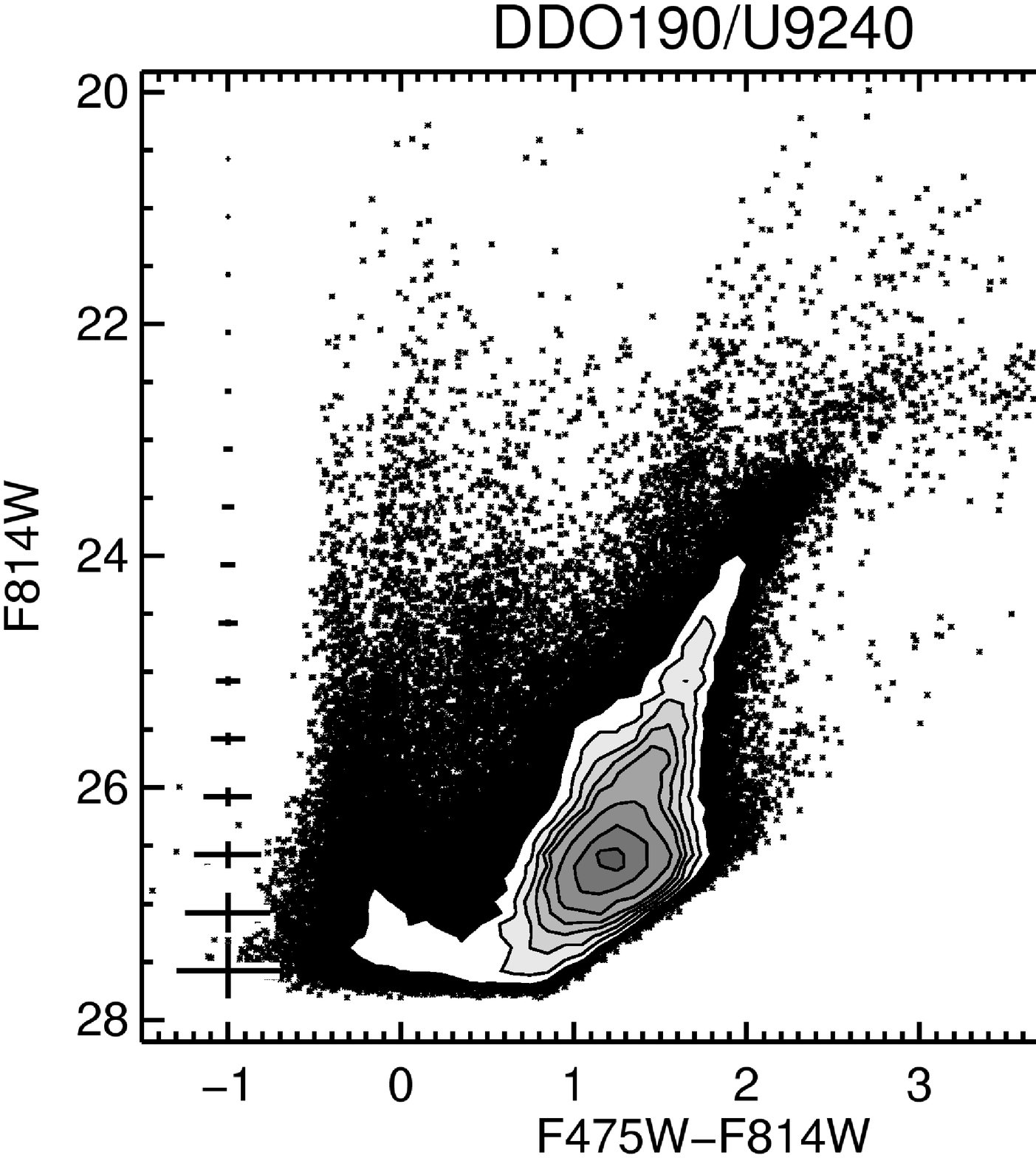}
\includegraphics[width=1.625in]{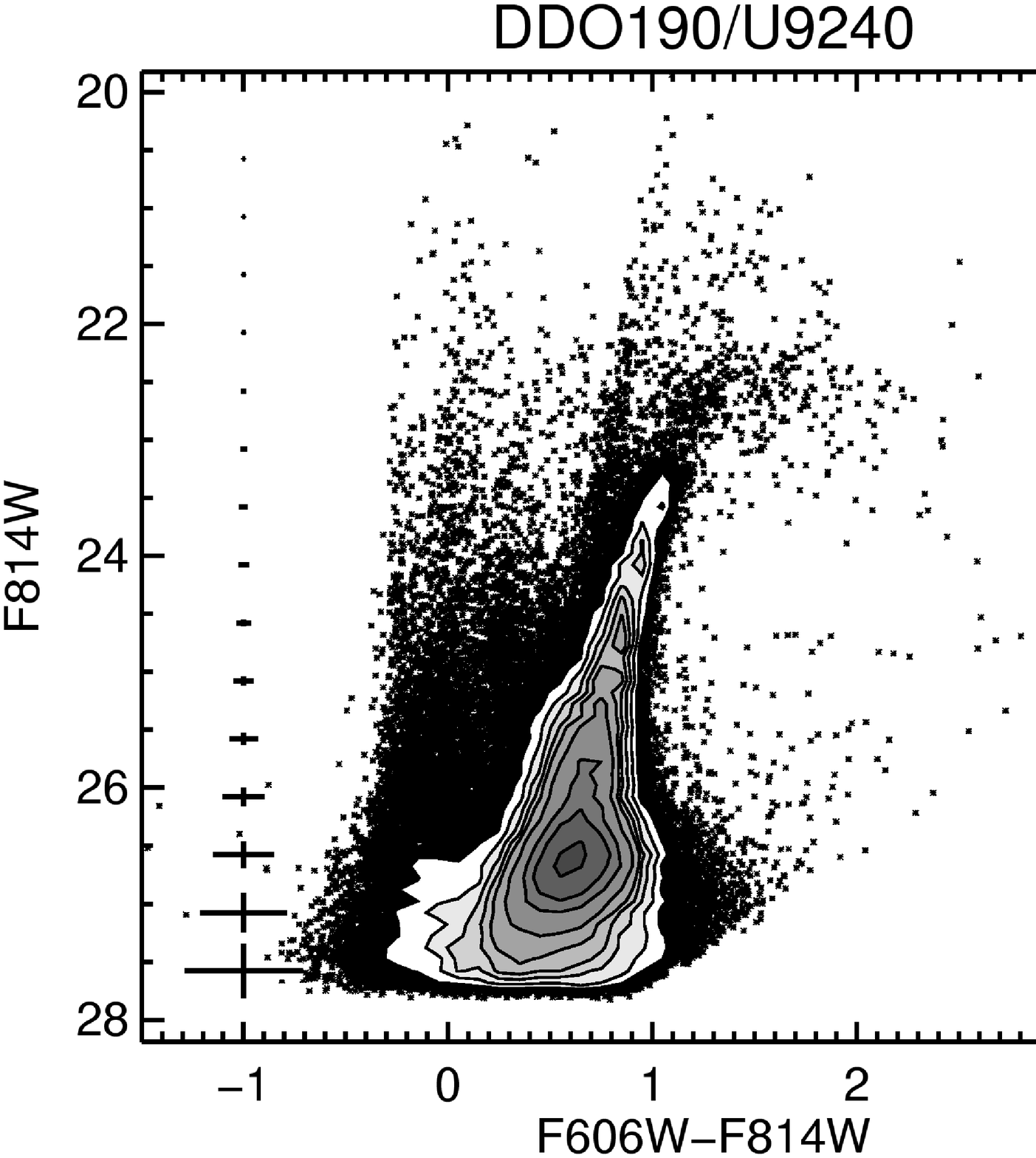}
\includegraphics[width=1.625in]{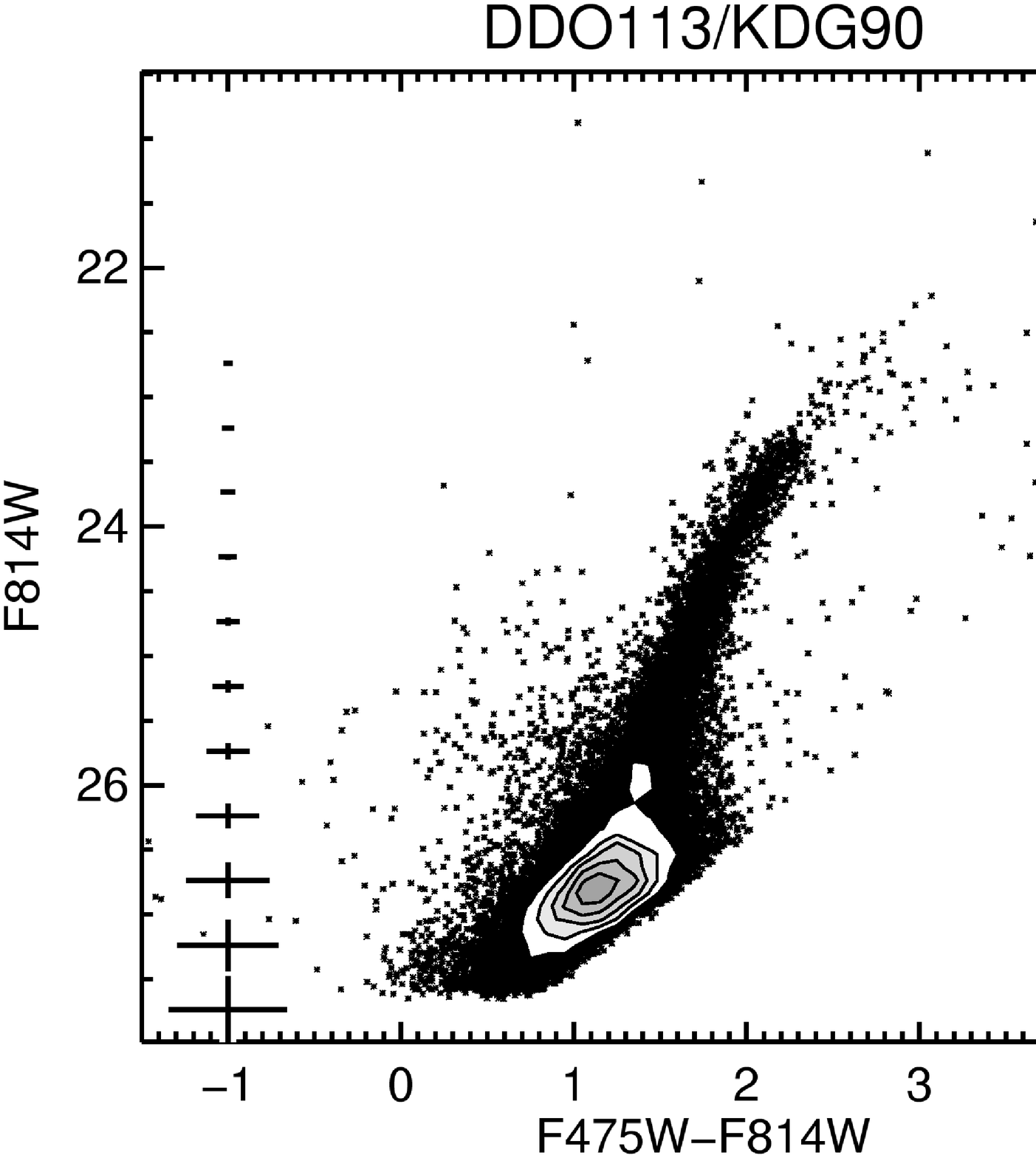}
\includegraphics[width=1.625in]{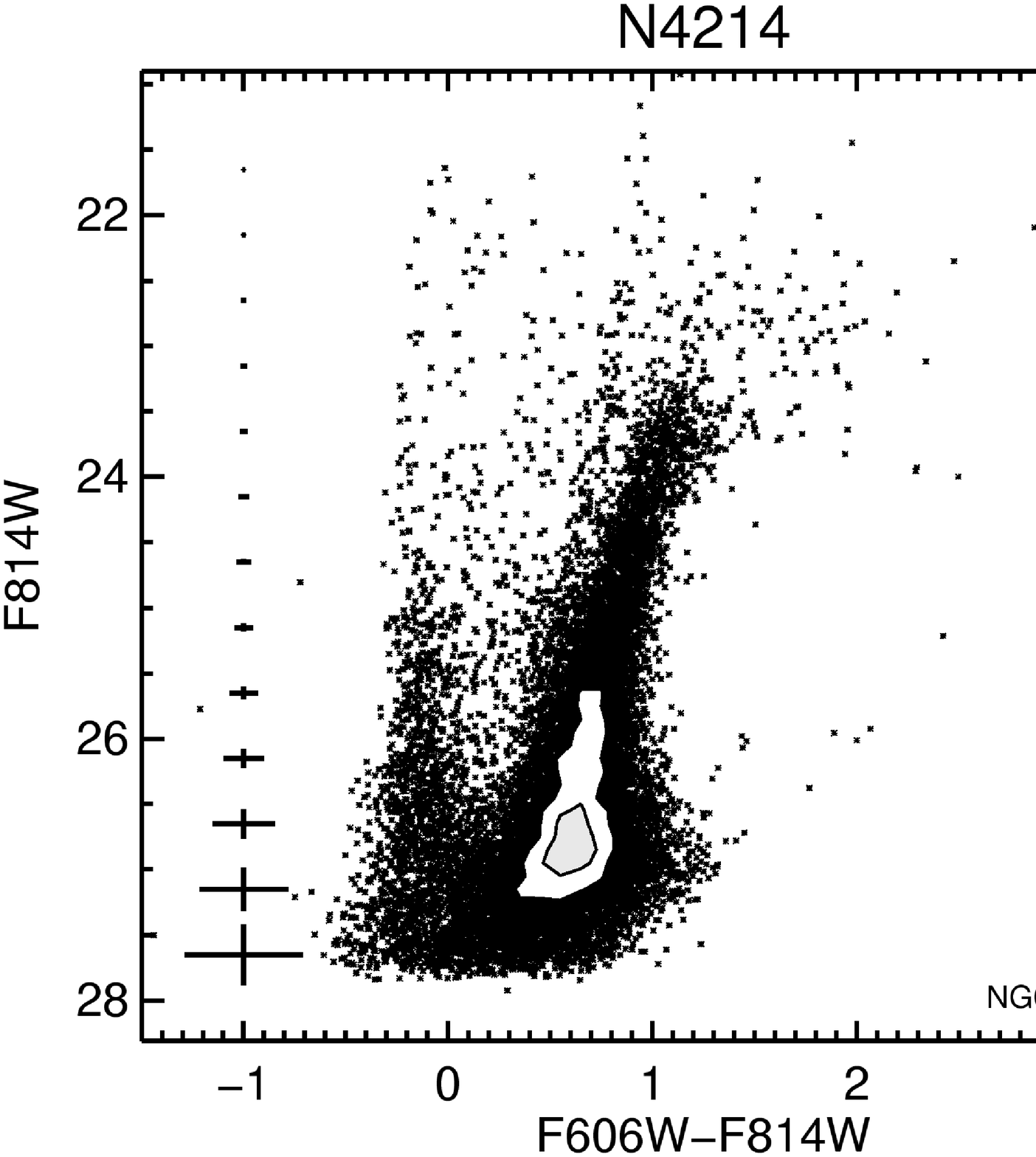}
}
\centerline{
\includegraphics[width=1.625in]{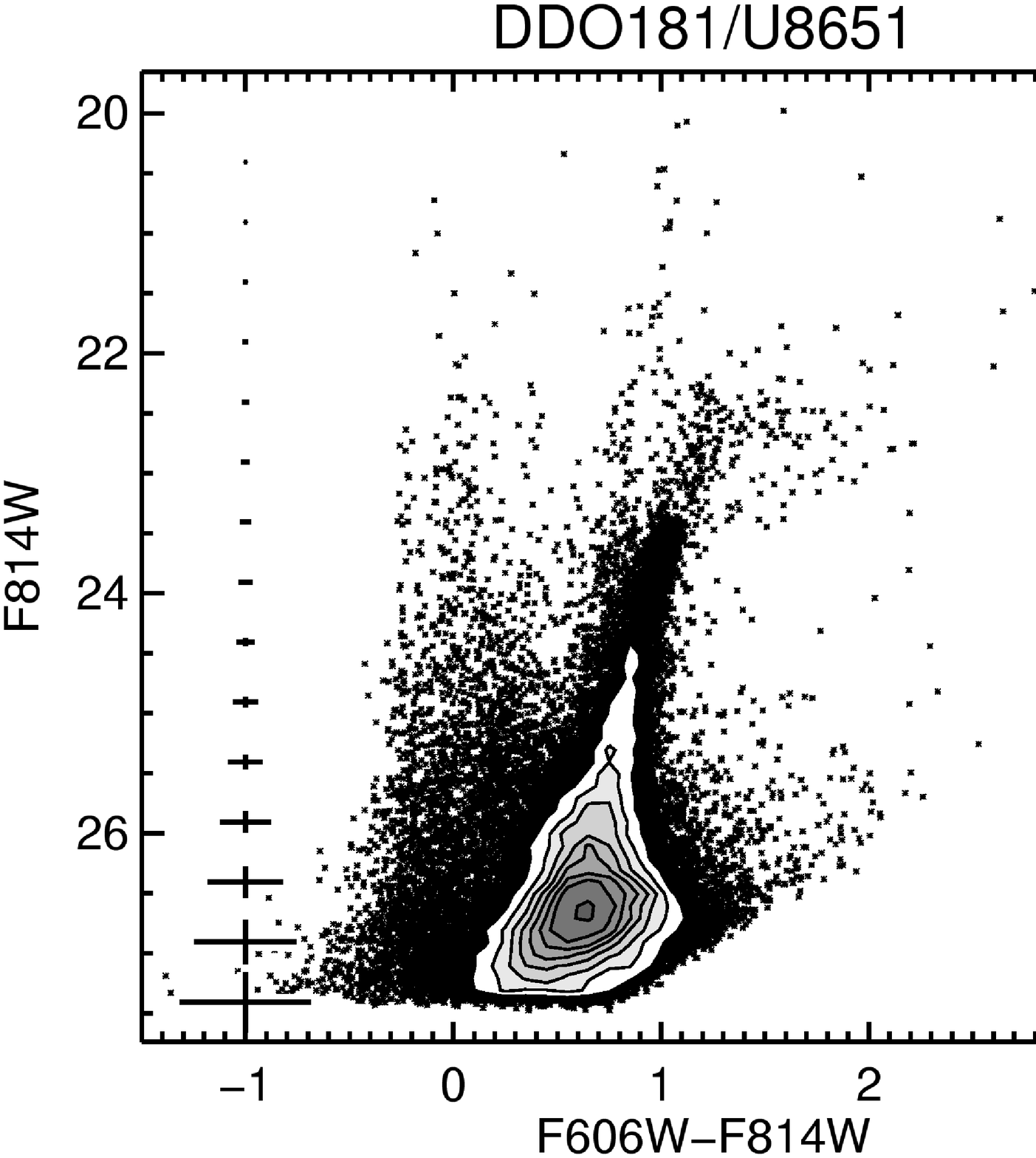}
\includegraphics[width=1.625in]{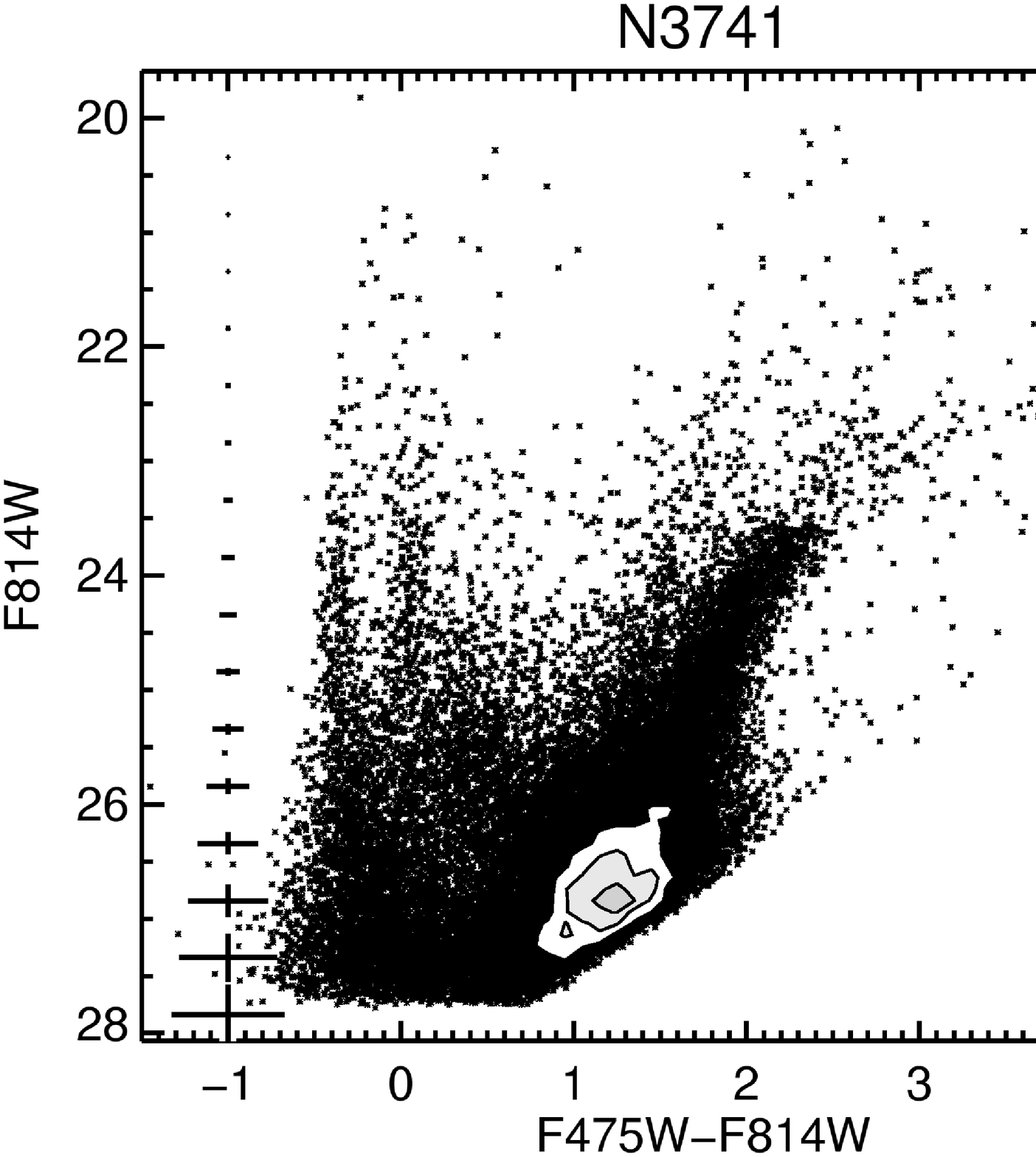}
\includegraphics[width=1.625in]{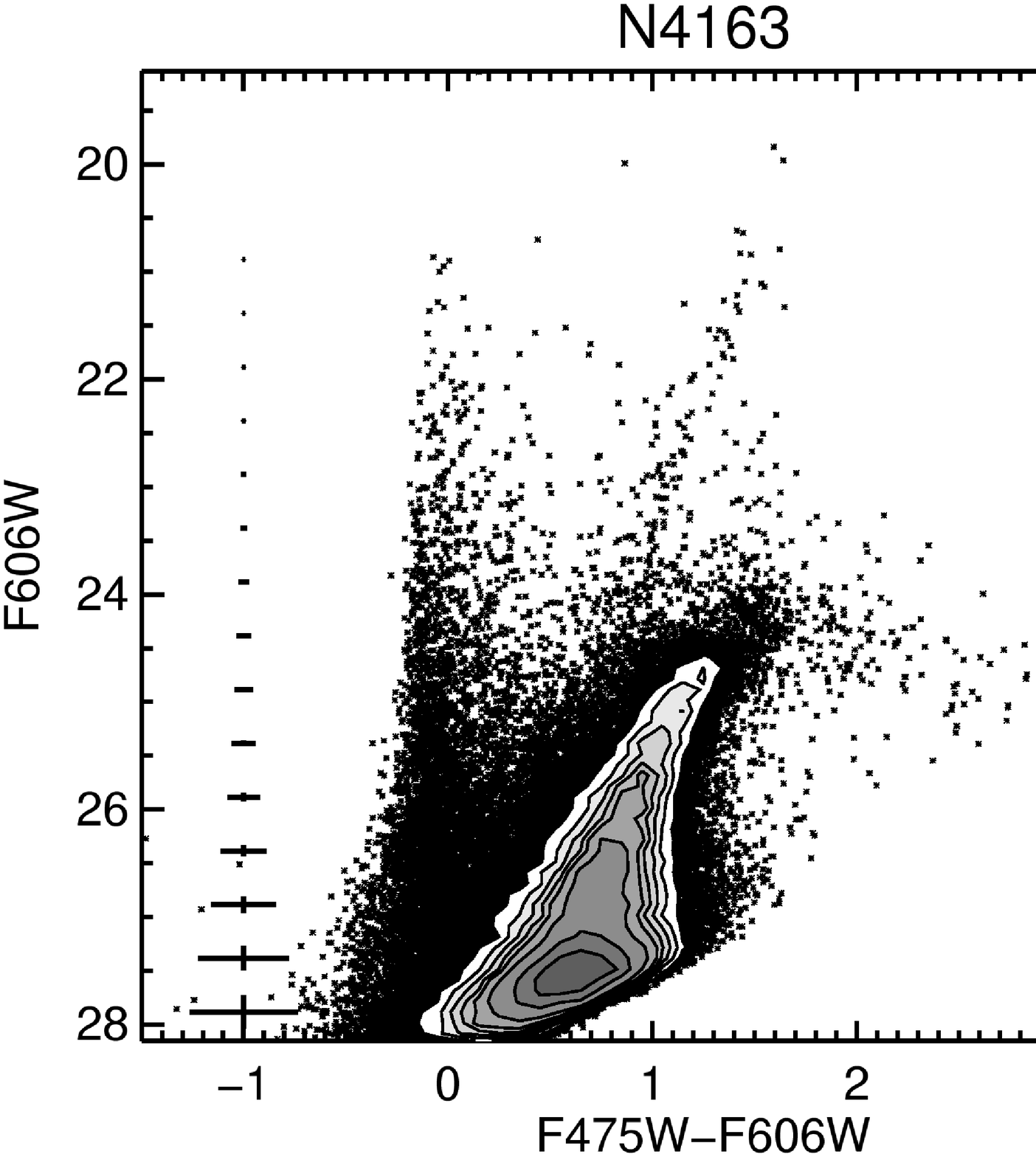}
\includegraphics[width=1.625in]{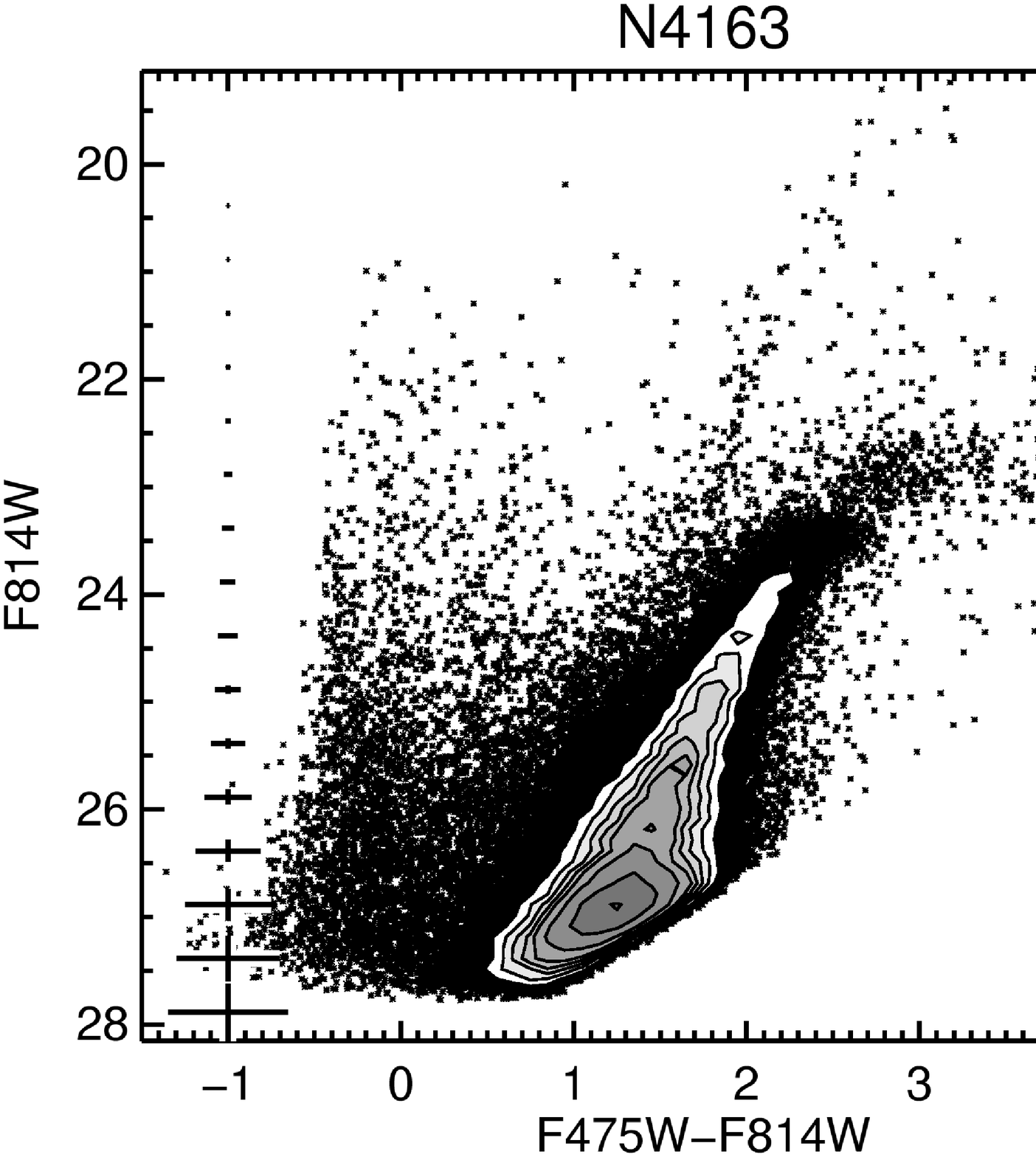}
}
\caption{
CMDs of galaxies in the ANGST data release,
as described in Figure~\ref{cmdfig1}.
Figures are ordered from the upper left to the bottom right.
(a) UA438; (b) DDO187; (c) KKH98; (d) DDO125; (e) U8508; (f) KKH86; (g) DDO99; (h) DDO190; (i) DDO190; (j) DDO190; (k) DDO113; (l) N4214; (m) DDO181; (n) N3741; (o) N4163; (p) N4163; 
    \label{cmdfig4}}
\end{figure}
\vfill
\clearpage
 
%-------------------
\begin{figure}[p]
\centerline{
\includegraphics[width=1.625in]{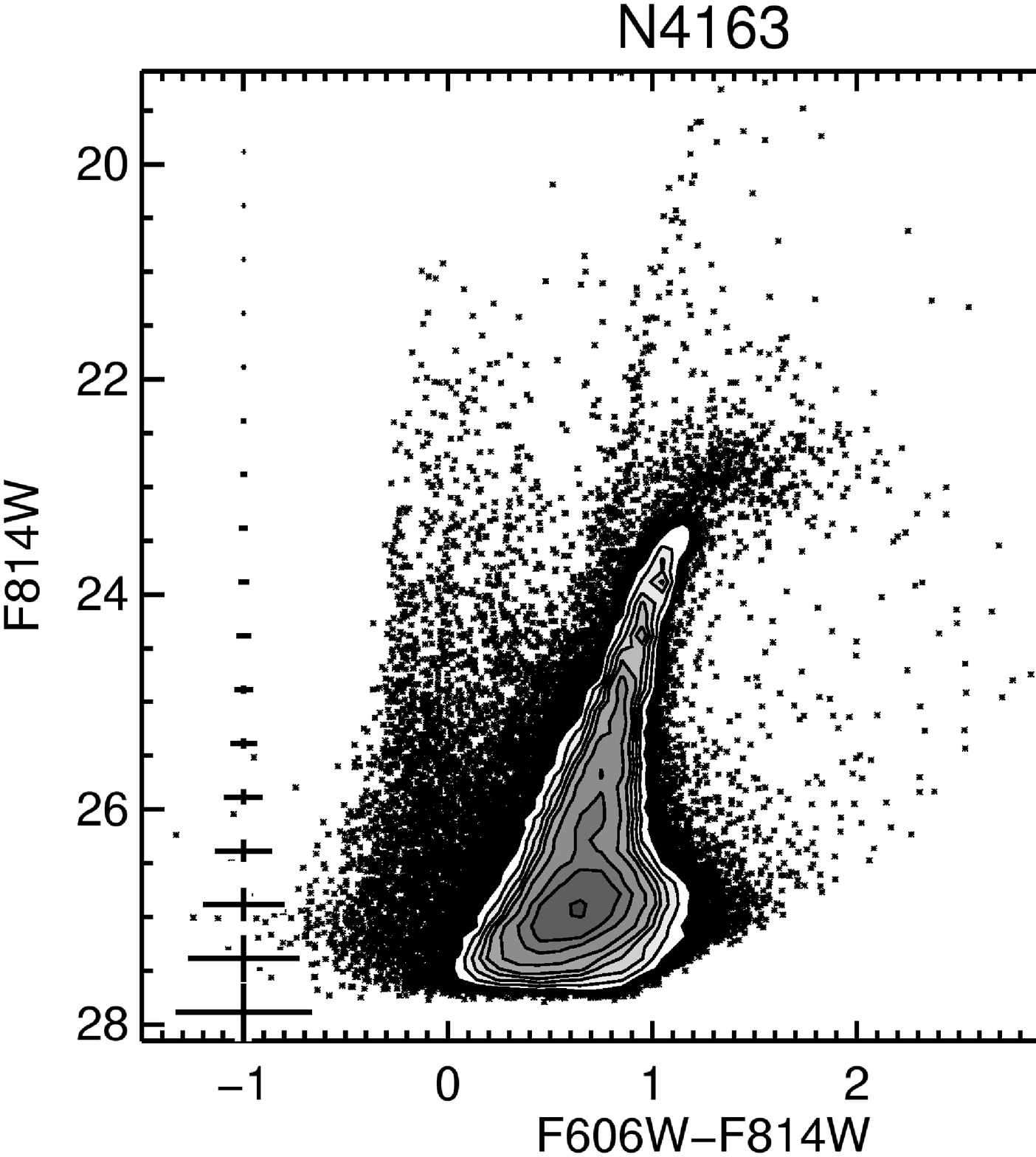}
\includegraphics[width=1.625in]{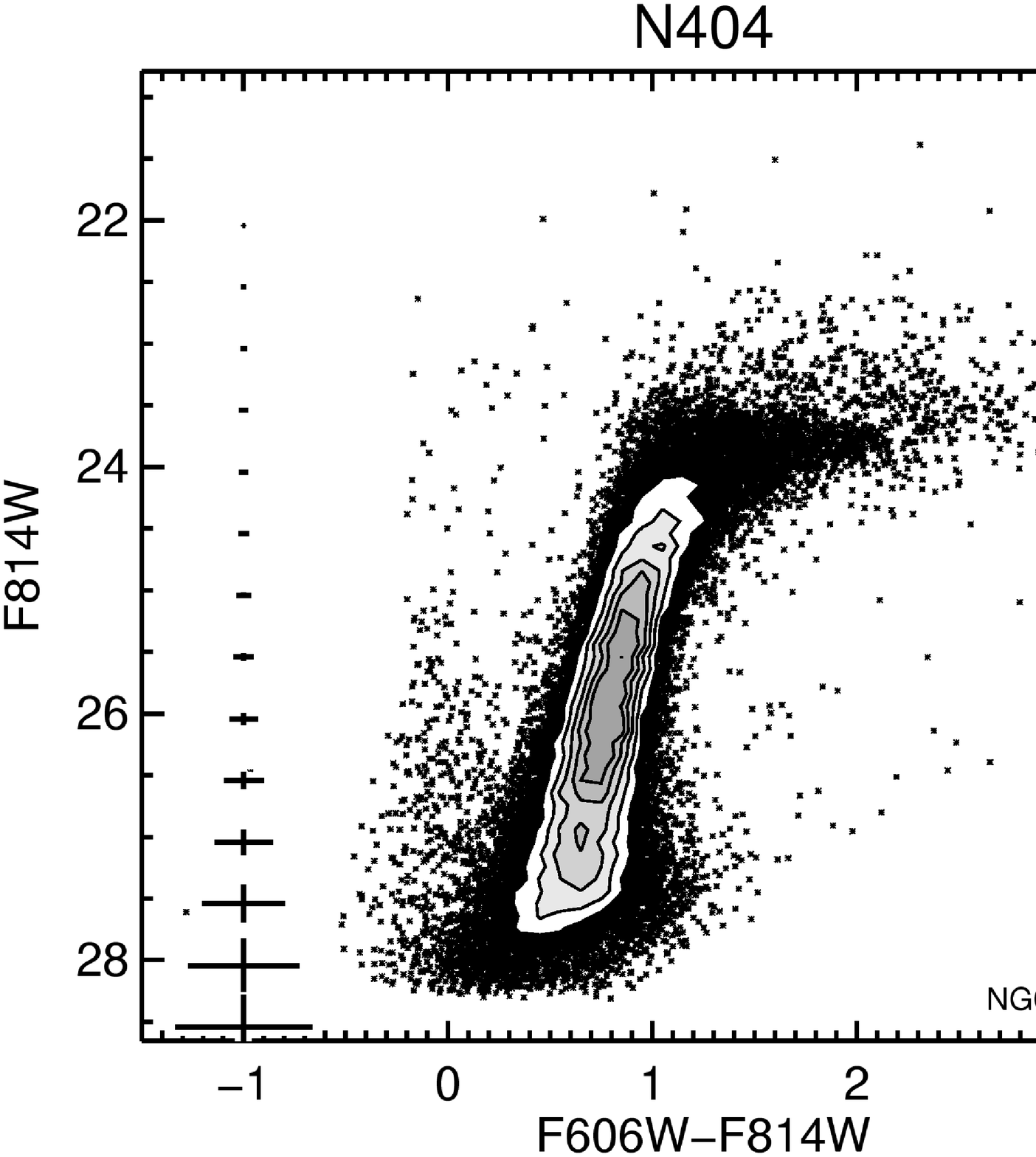}
\includegraphics[width=1.625in]{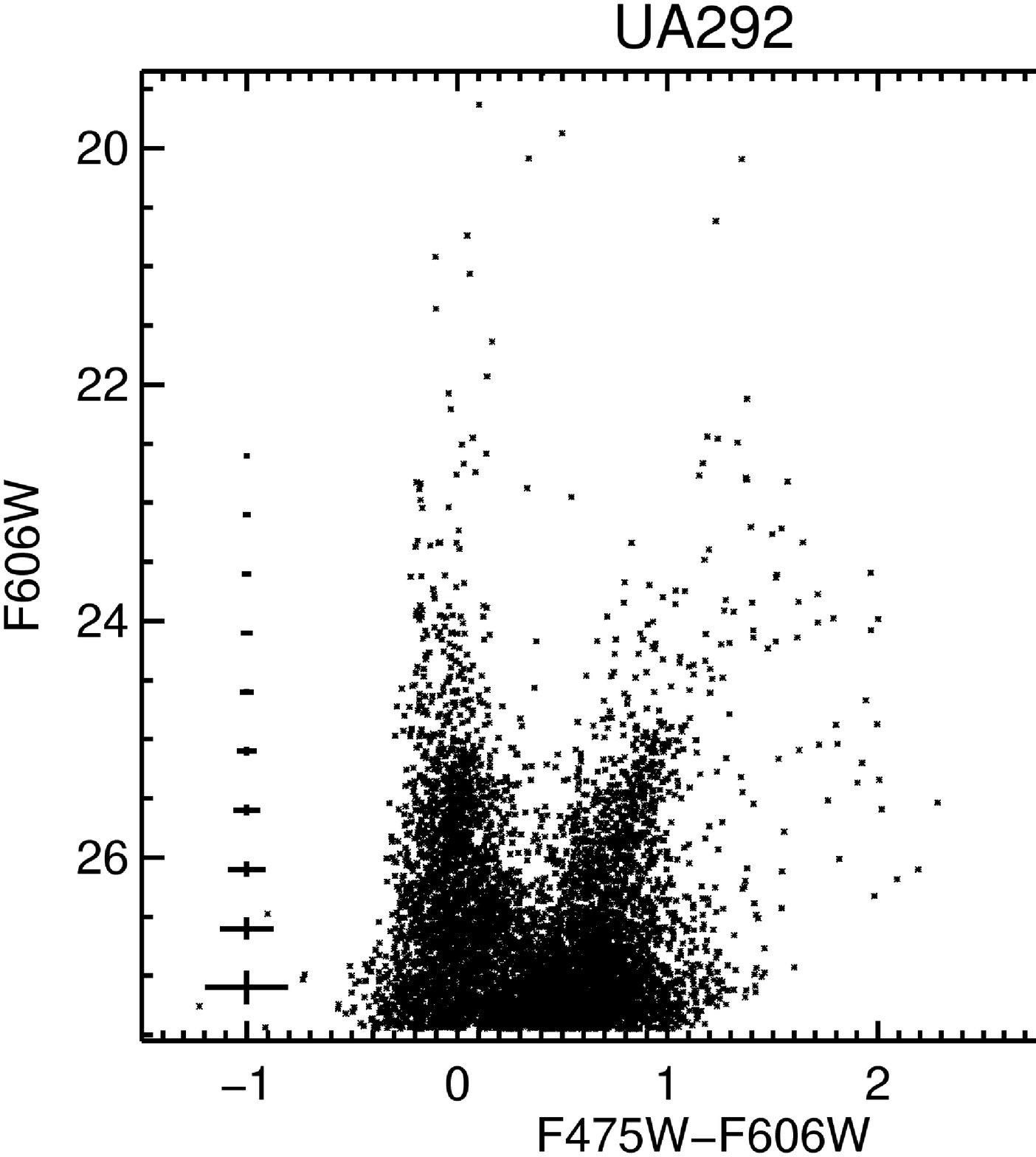}
\includegraphics[width=1.625in]{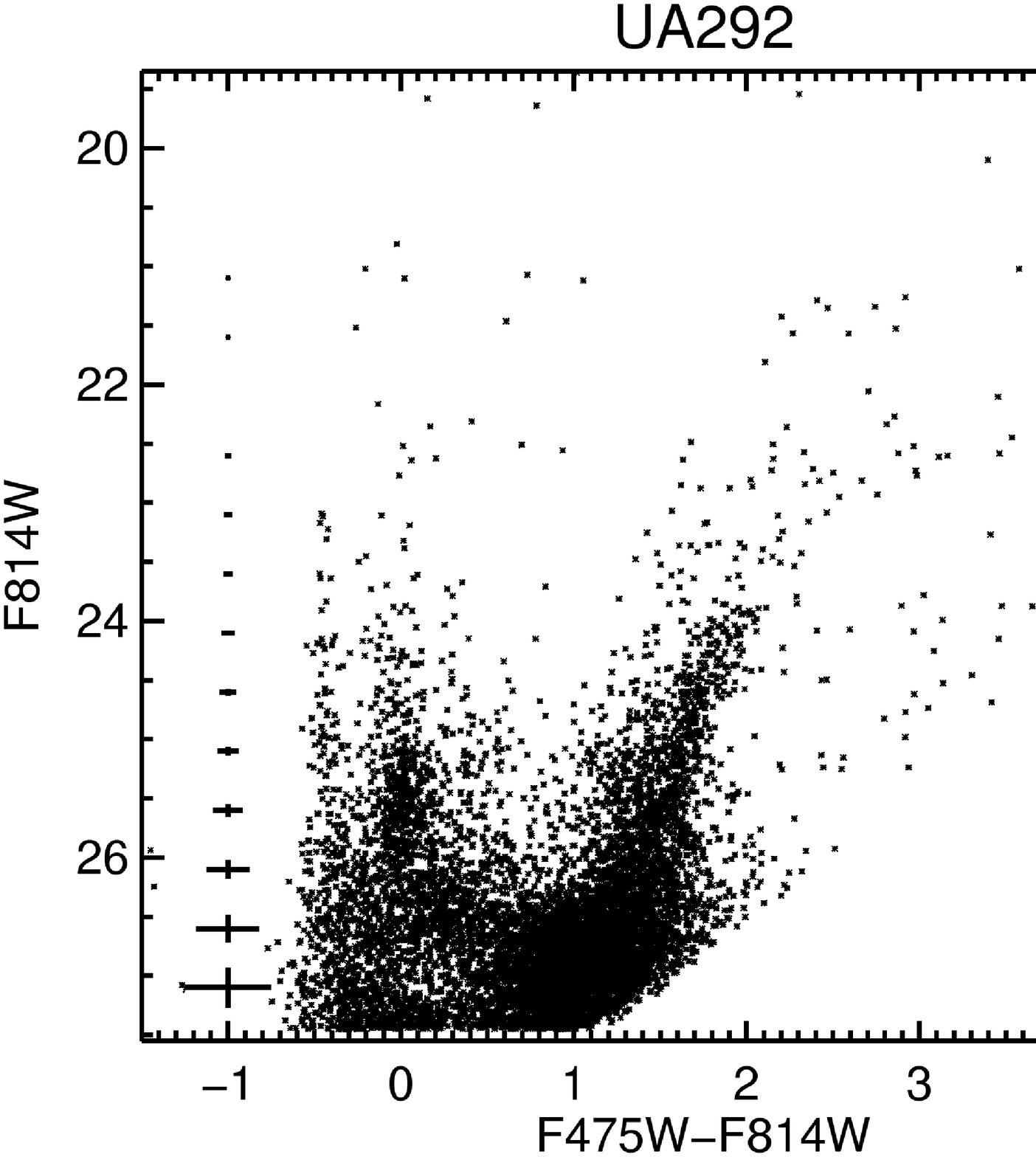}
}
\centerline{
\includegraphics[width=1.625in]{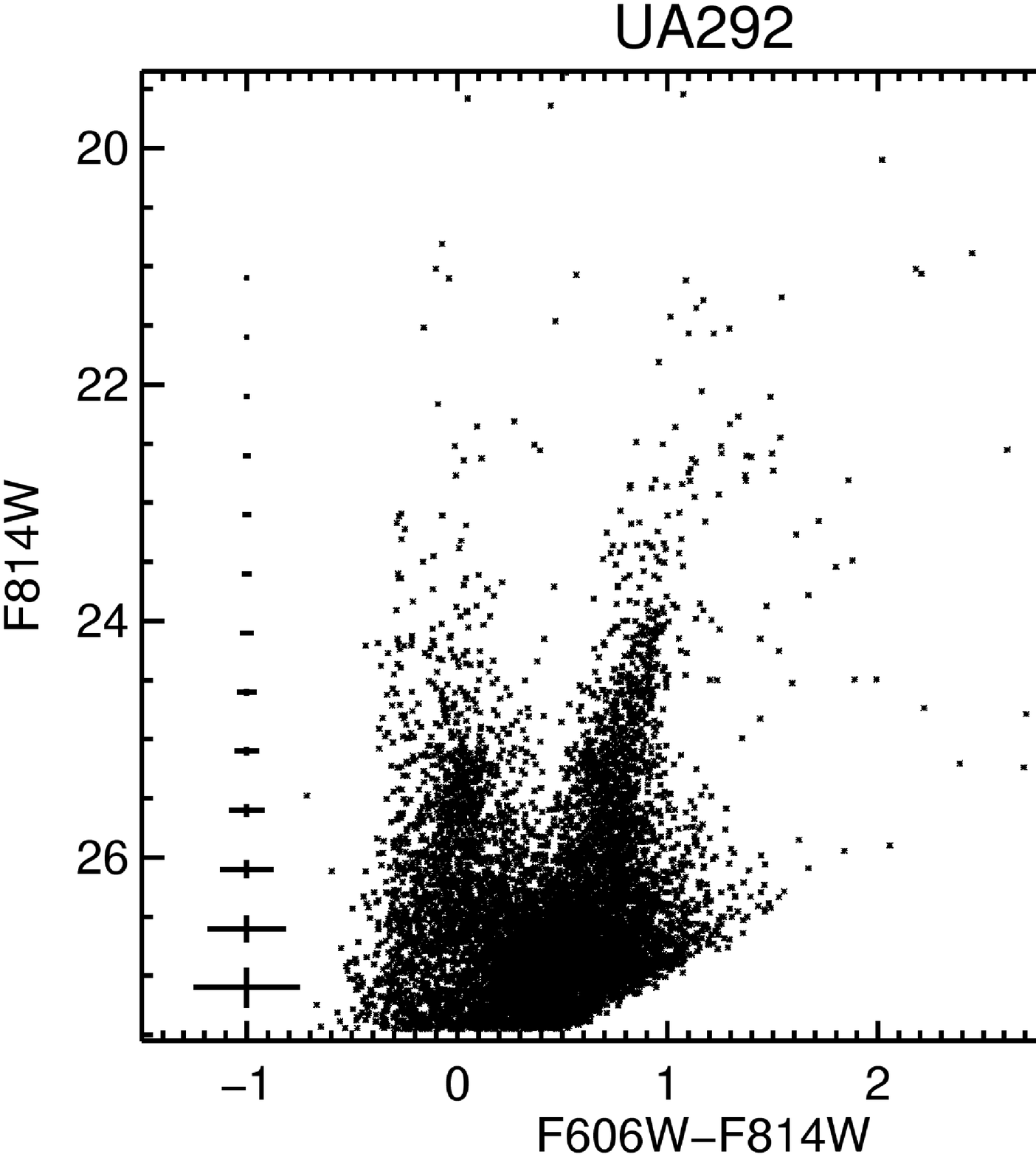}
\includegraphics[width=1.625in]{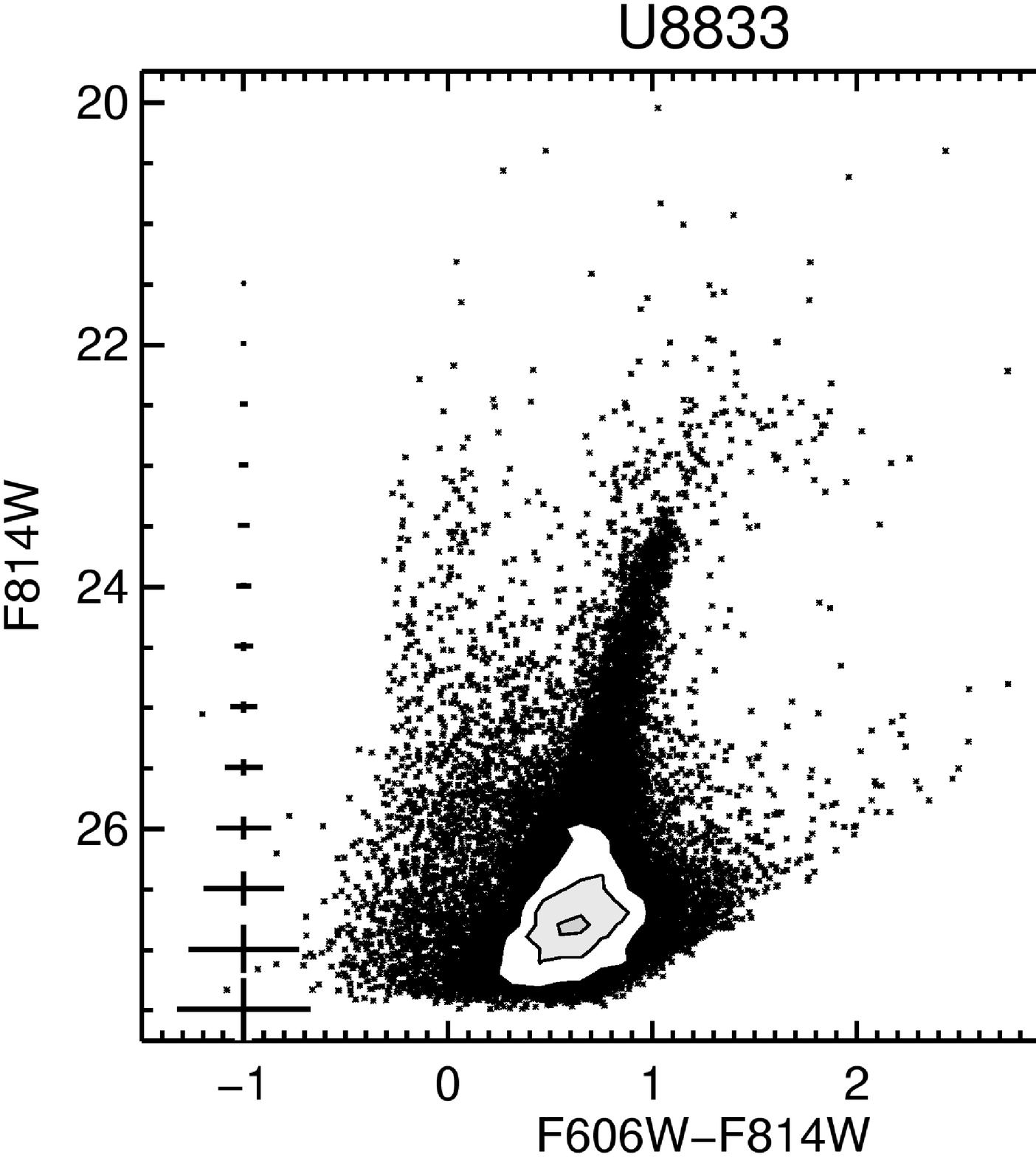}
\includegraphics[width=1.625in]{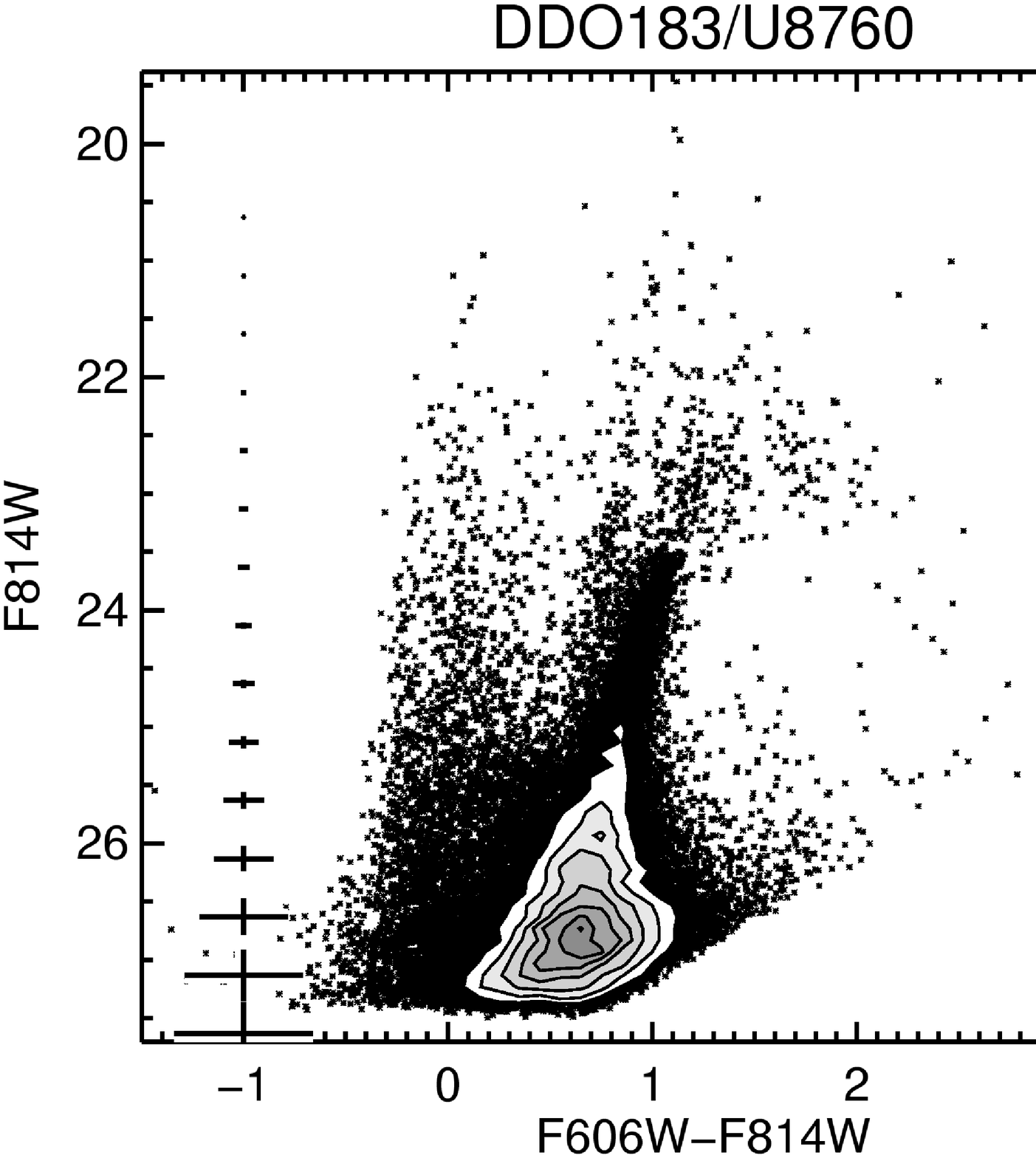}
\includegraphics[width=1.625in]{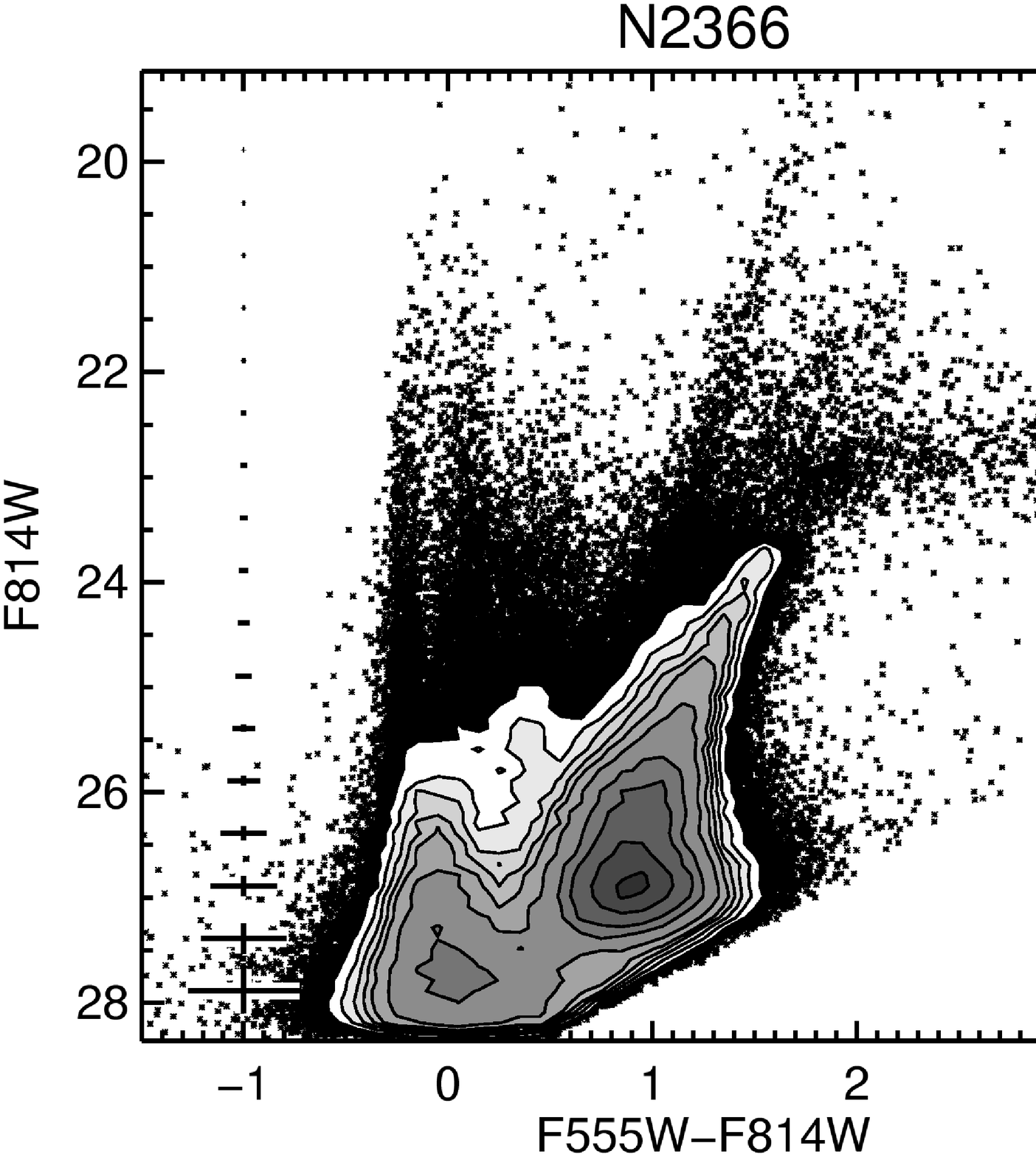}
}
\centerline{
\includegraphics[width=1.625in]{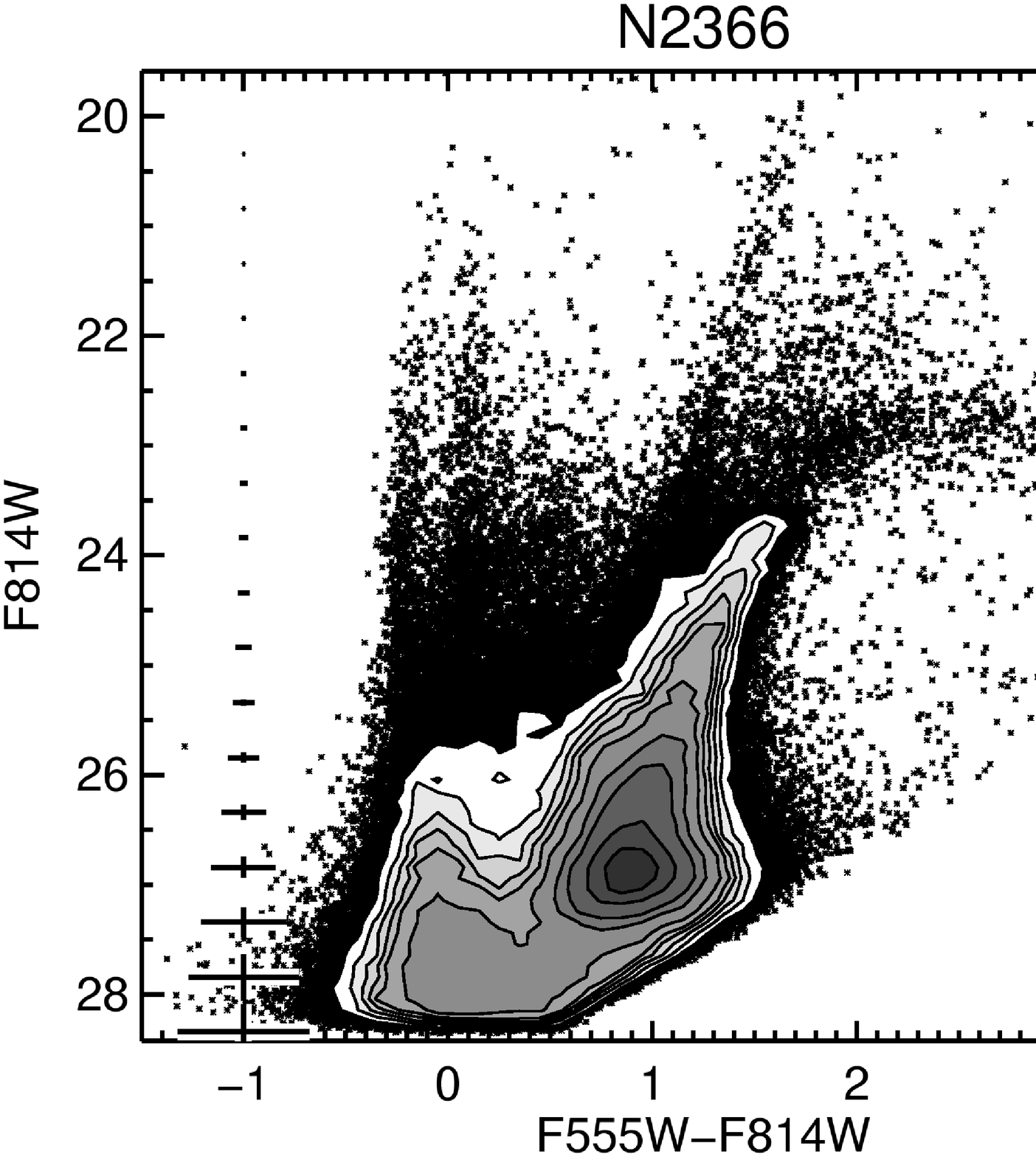}
\includegraphics[width=1.625in]{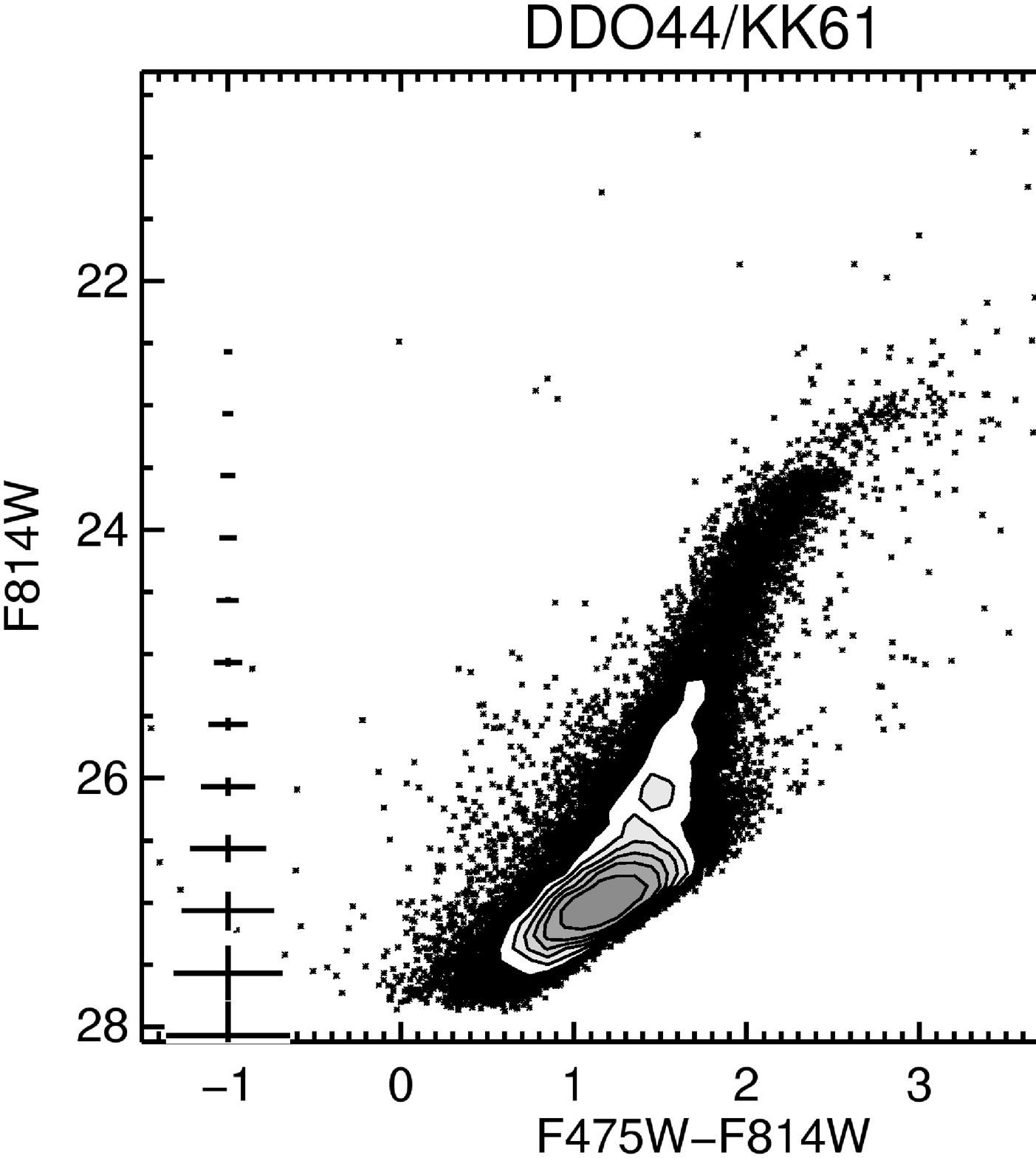}
\includegraphics[width=1.625in]{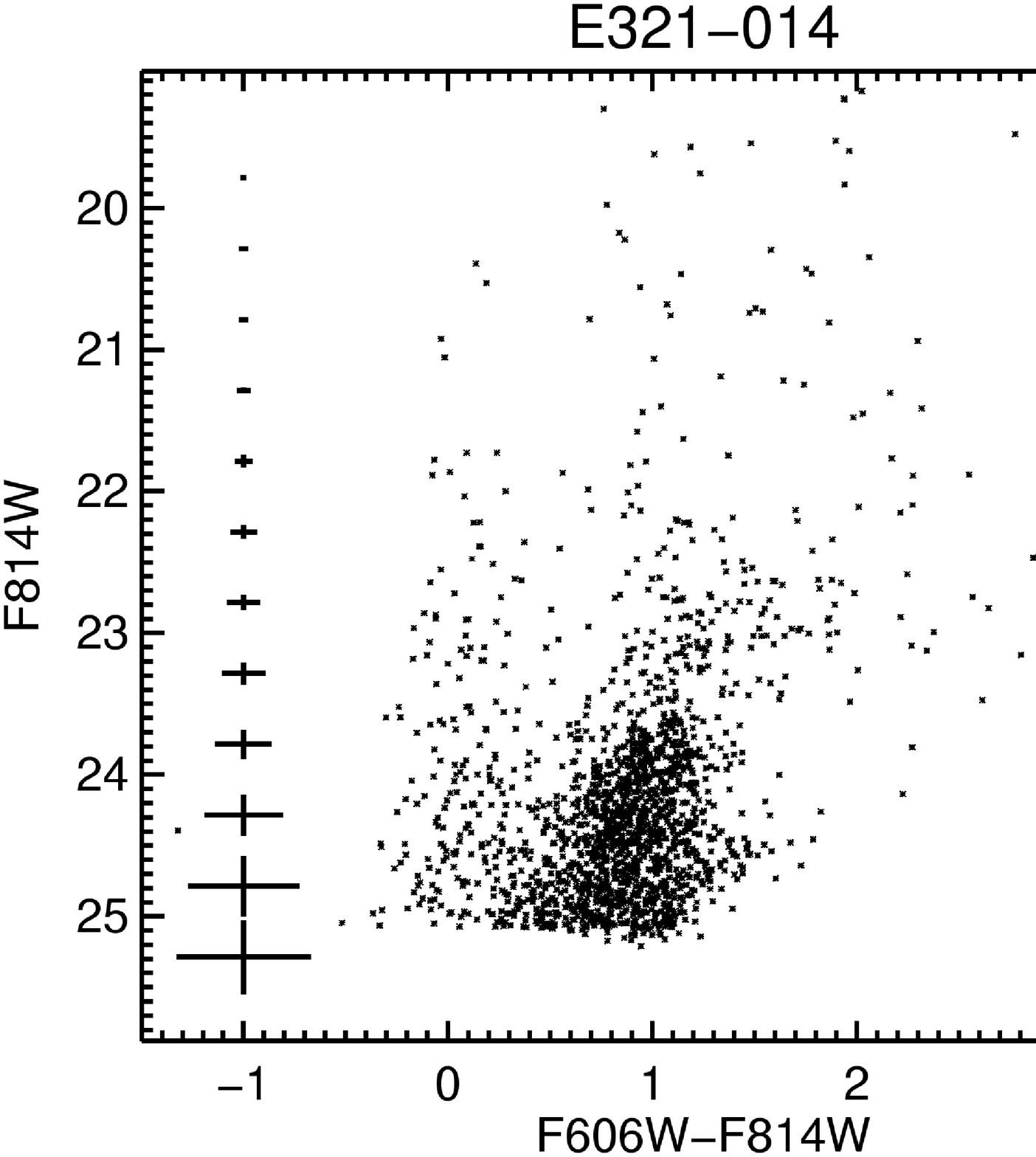}
\includegraphics[width=1.625in]{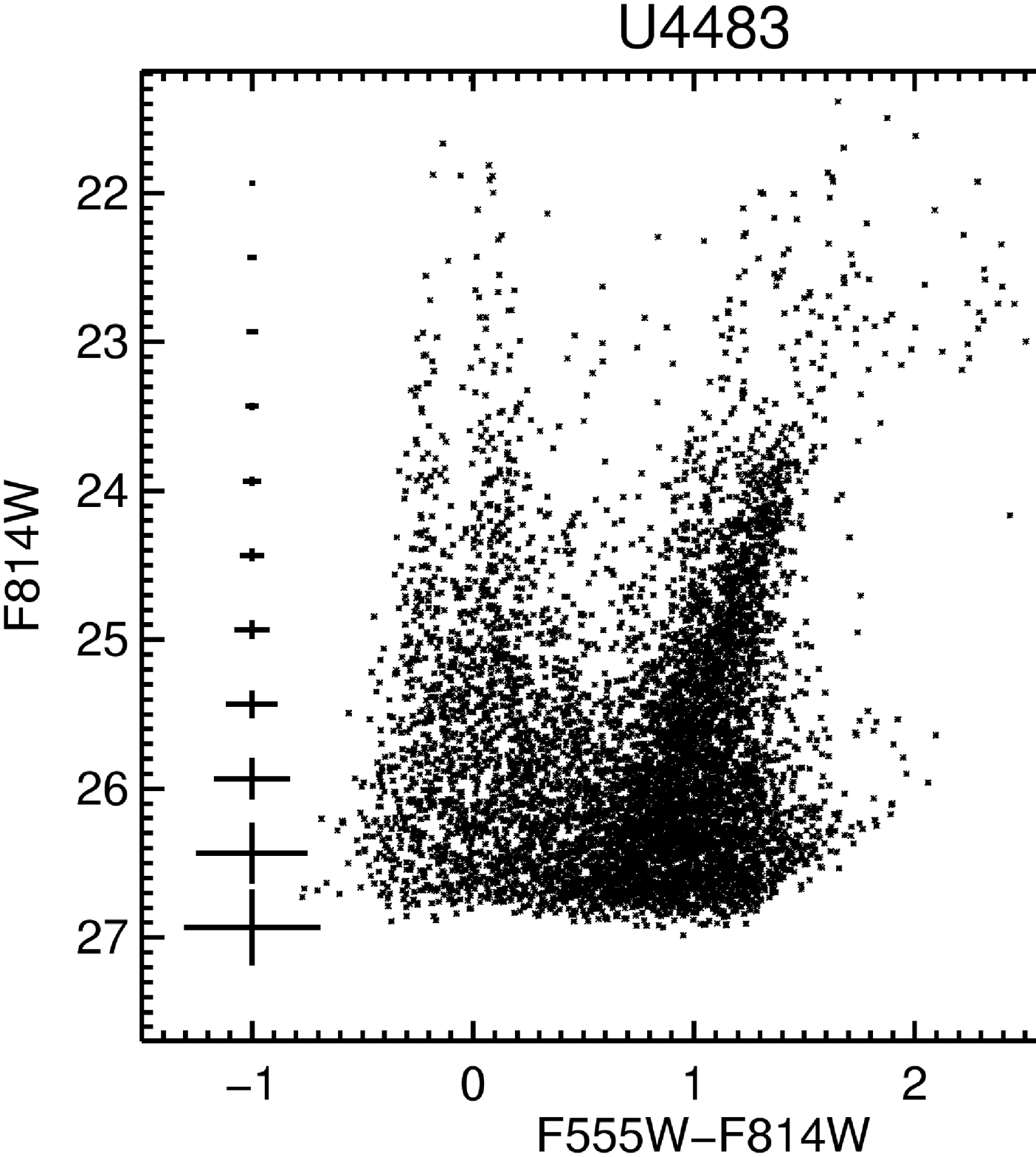}
}
\centerline{
\includegraphics[width=1.625in]{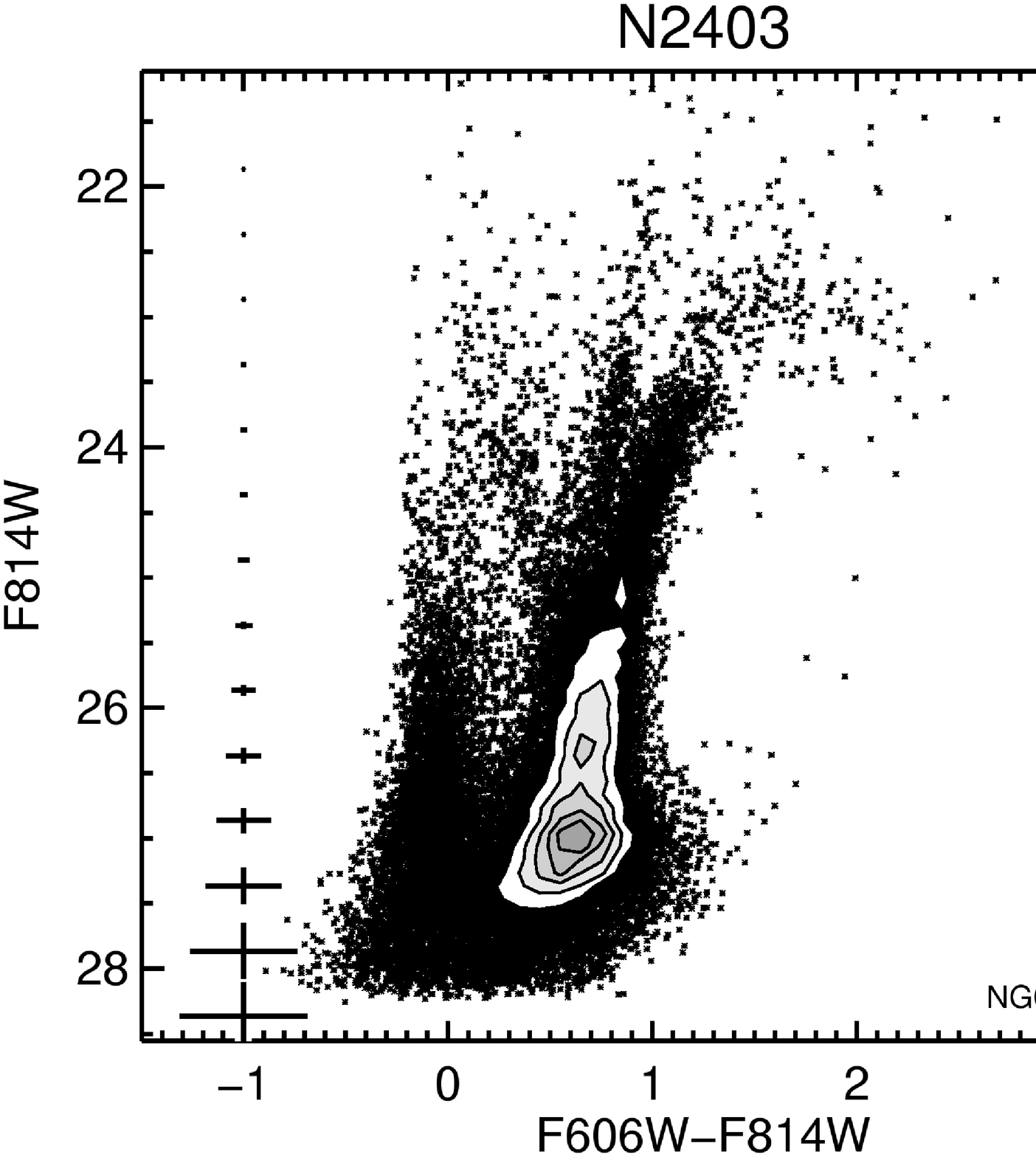}
\includegraphics[width=1.625in]{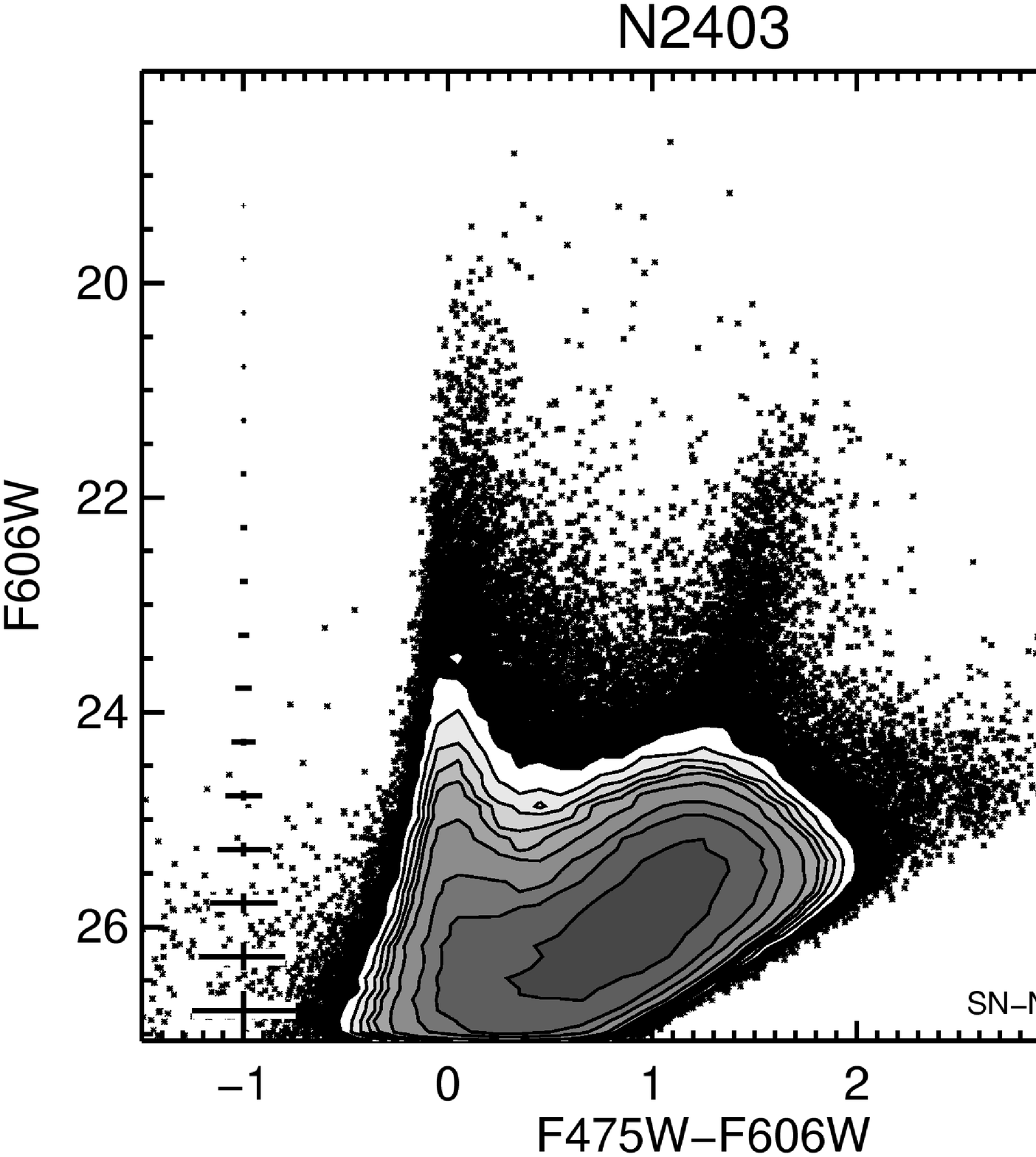}
\includegraphics[width=1.625in]{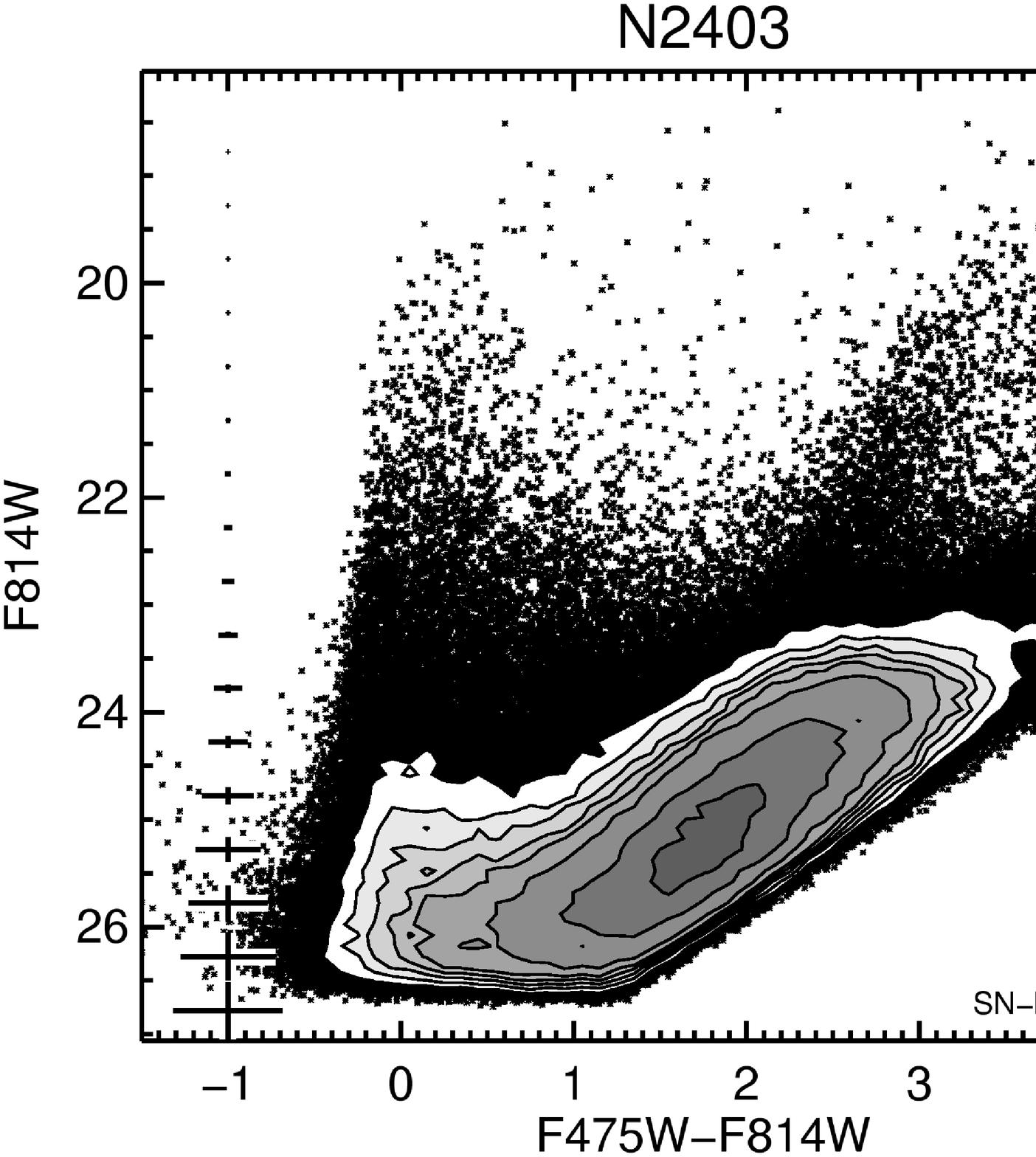}
\includegraphics[width=1.625in]{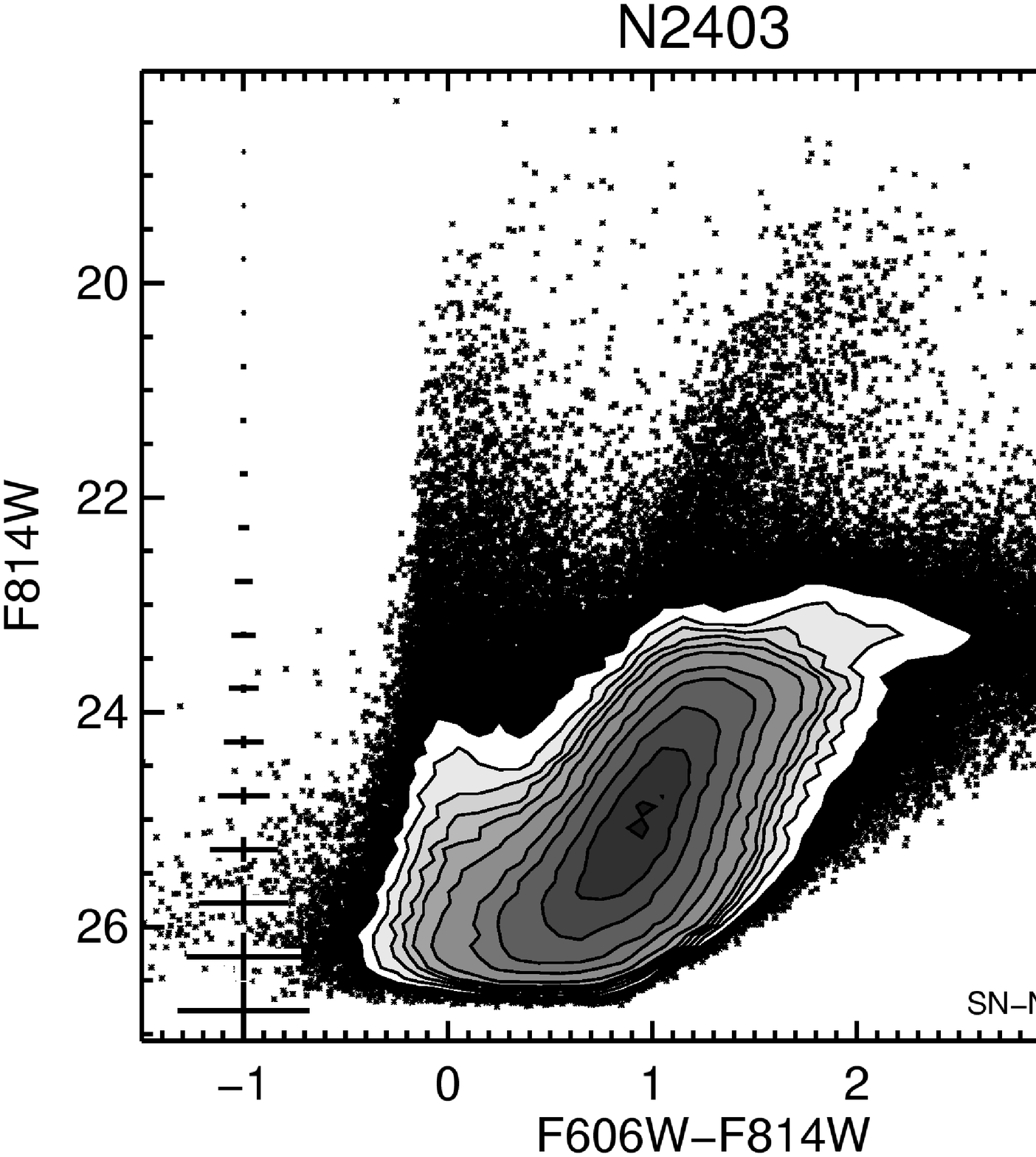}
}
\caption{
CMDs of galaxies in the ANGST data release,
as described in Figure~\ref{cmdfig1}.
Figures are ordered from the upper left to the bottom right.
(a) N4163; (b) N404; (c) UA292; (d) UA292; (e) UA292; (f) U8833; (g) DDO183; (h) N2366; (i) N2366; (j) DDO44; (k) E321-014; (l) U4483; (m) N2403; (n) N2403; (o) N2403; (p) N2403; 
    \label{cmdfig5}}
\end{figure}
\vfill
\clearpage
 
%-------------------
\begin{figure}[p]
\centerline{
\includegraphics[width=1.625in]{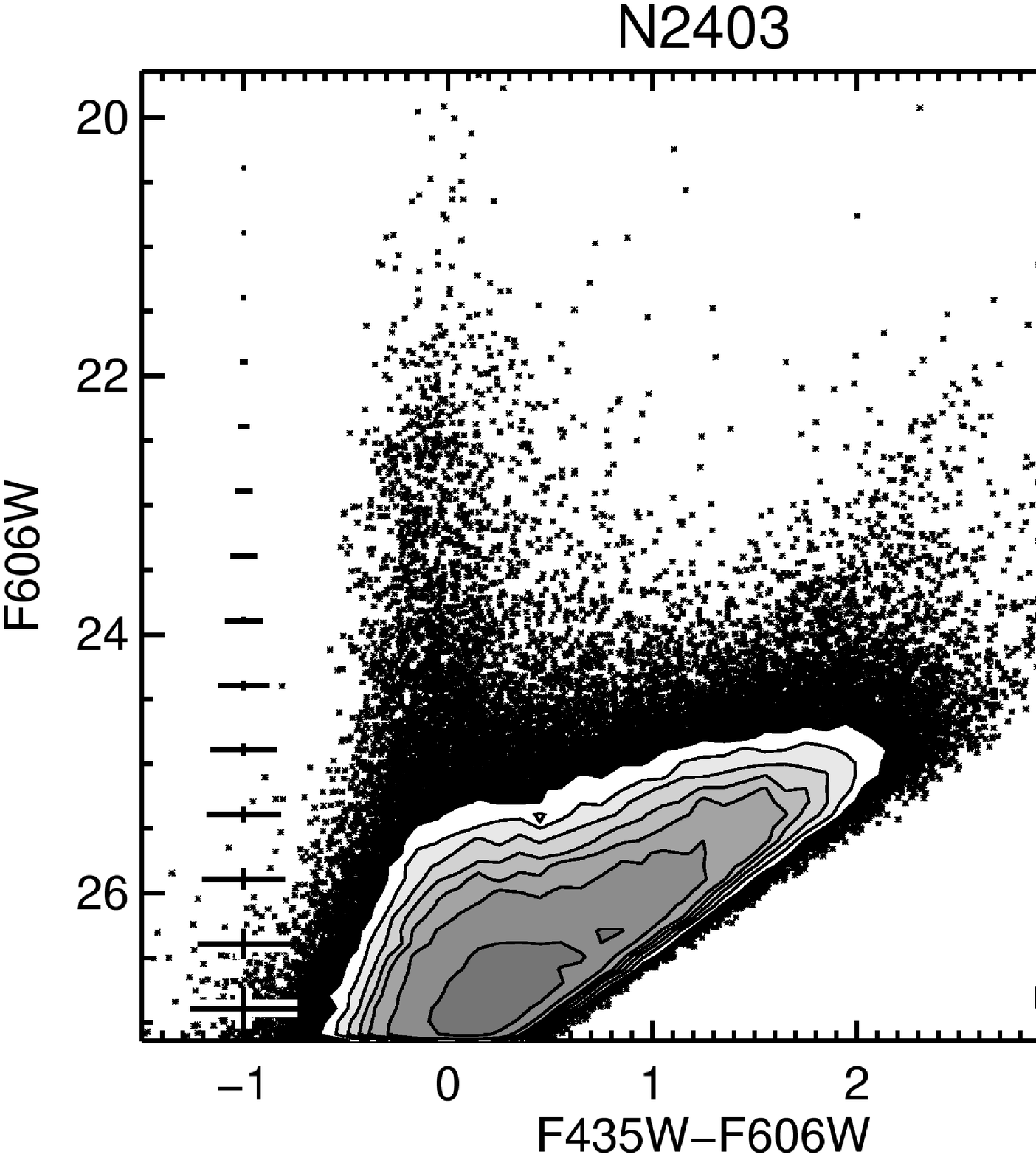}
\includegraphics[width=1.625in]{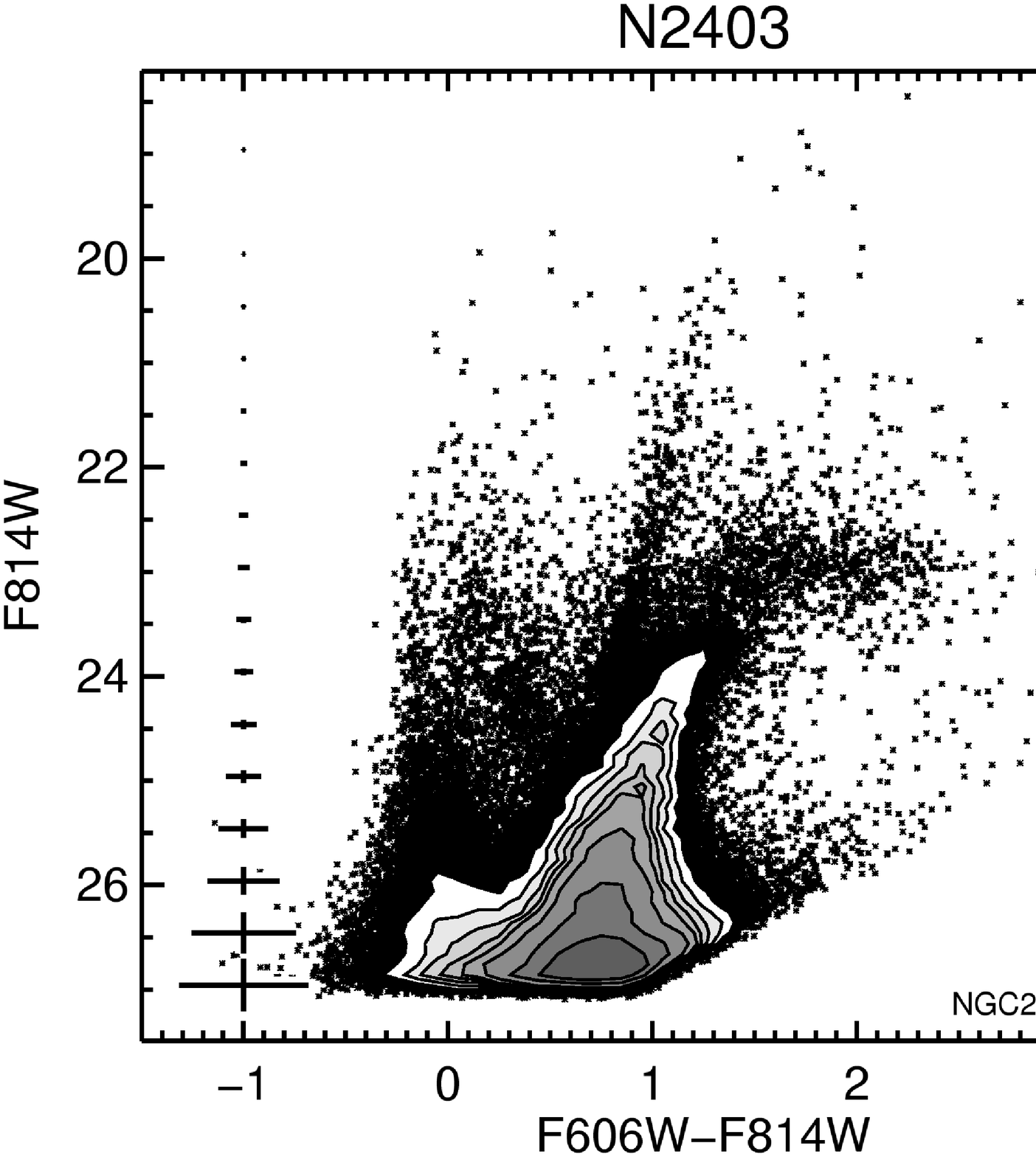}
\includegraphics[width=1.625in]{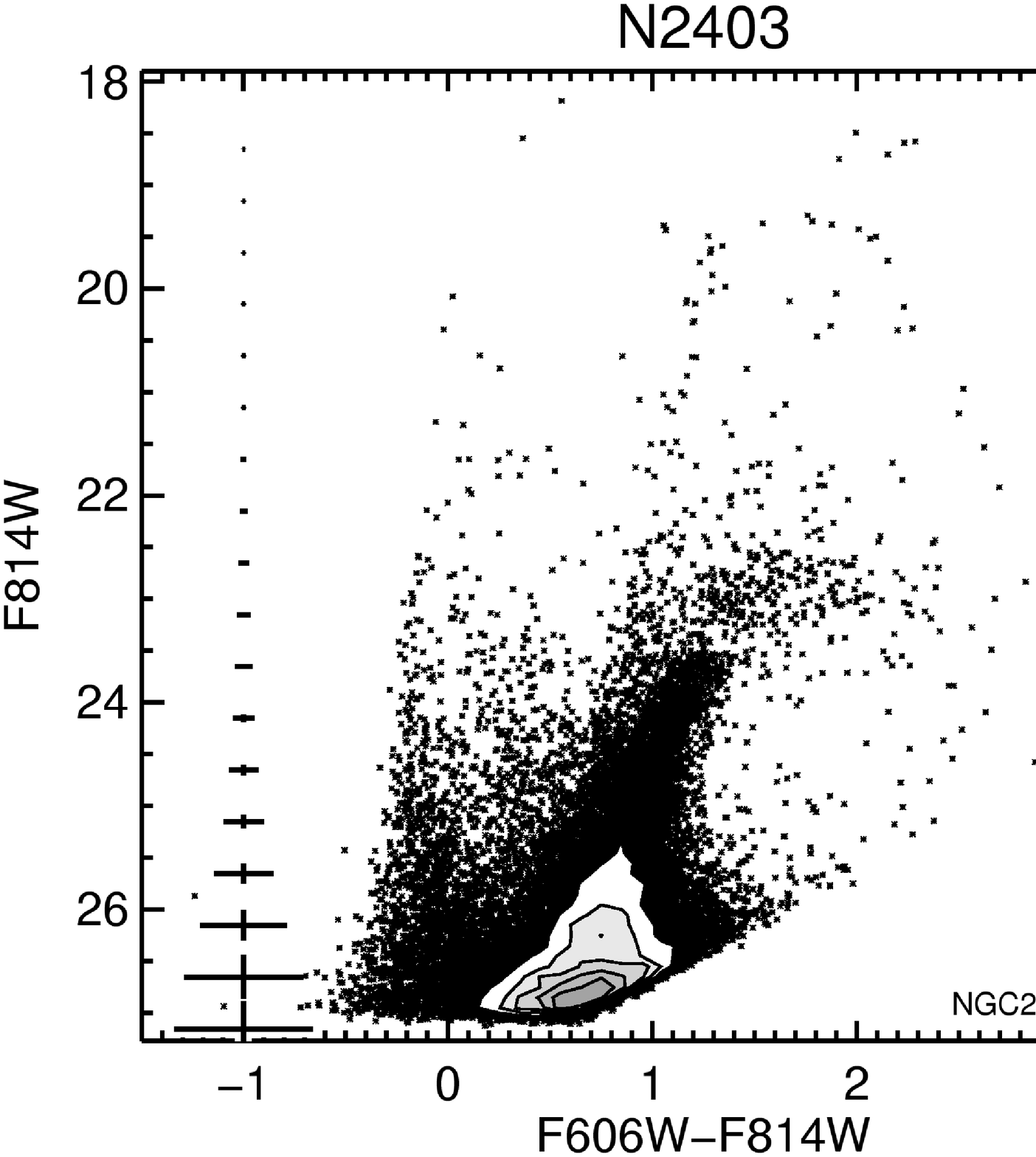}
\includegraphics[width=1.625in]{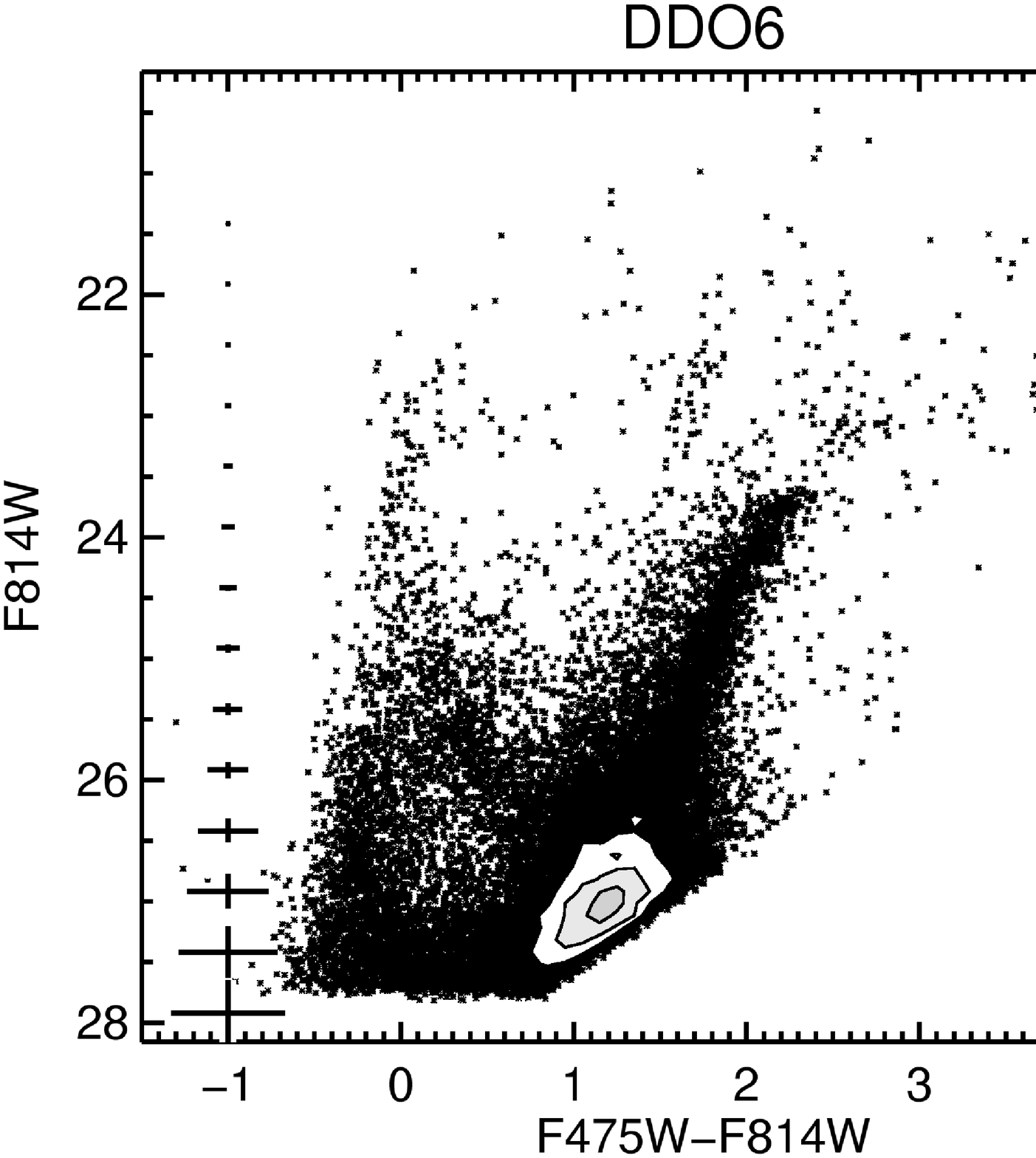}
}
\centerline{
\includegraphics[width=1.625in]{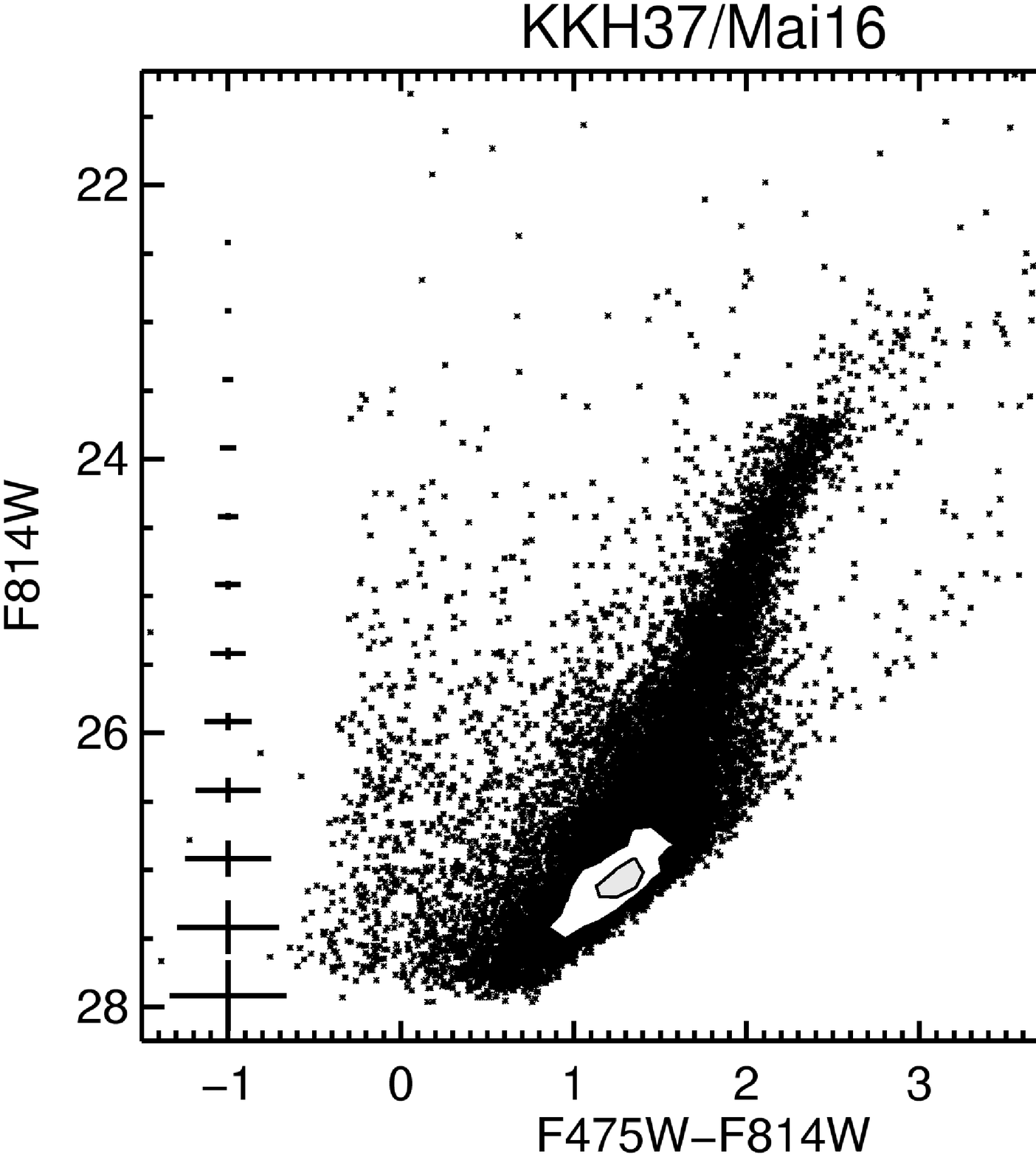}
\includegraphics[width=1.625in]{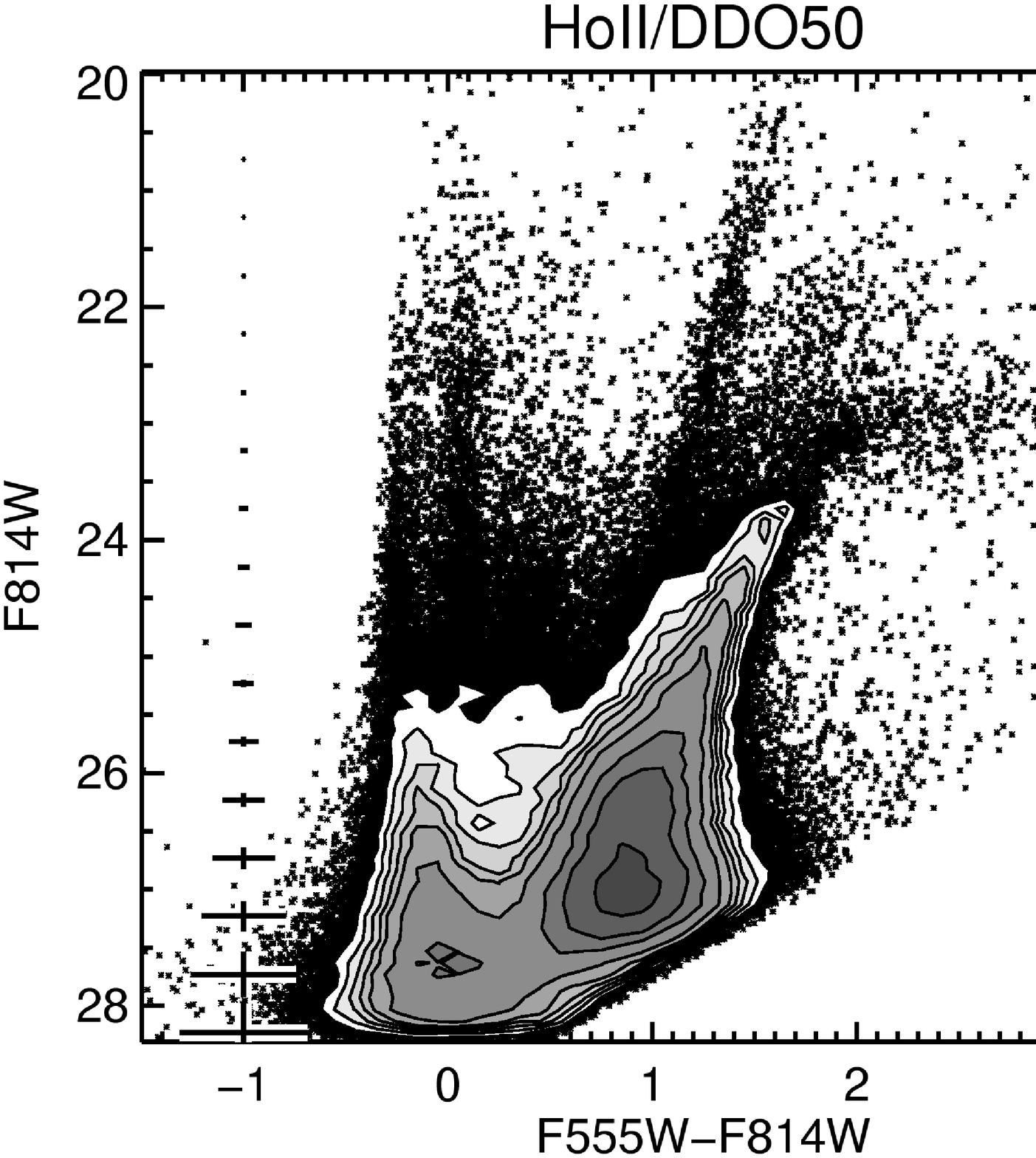}
\includegraphics[width=1.625in]{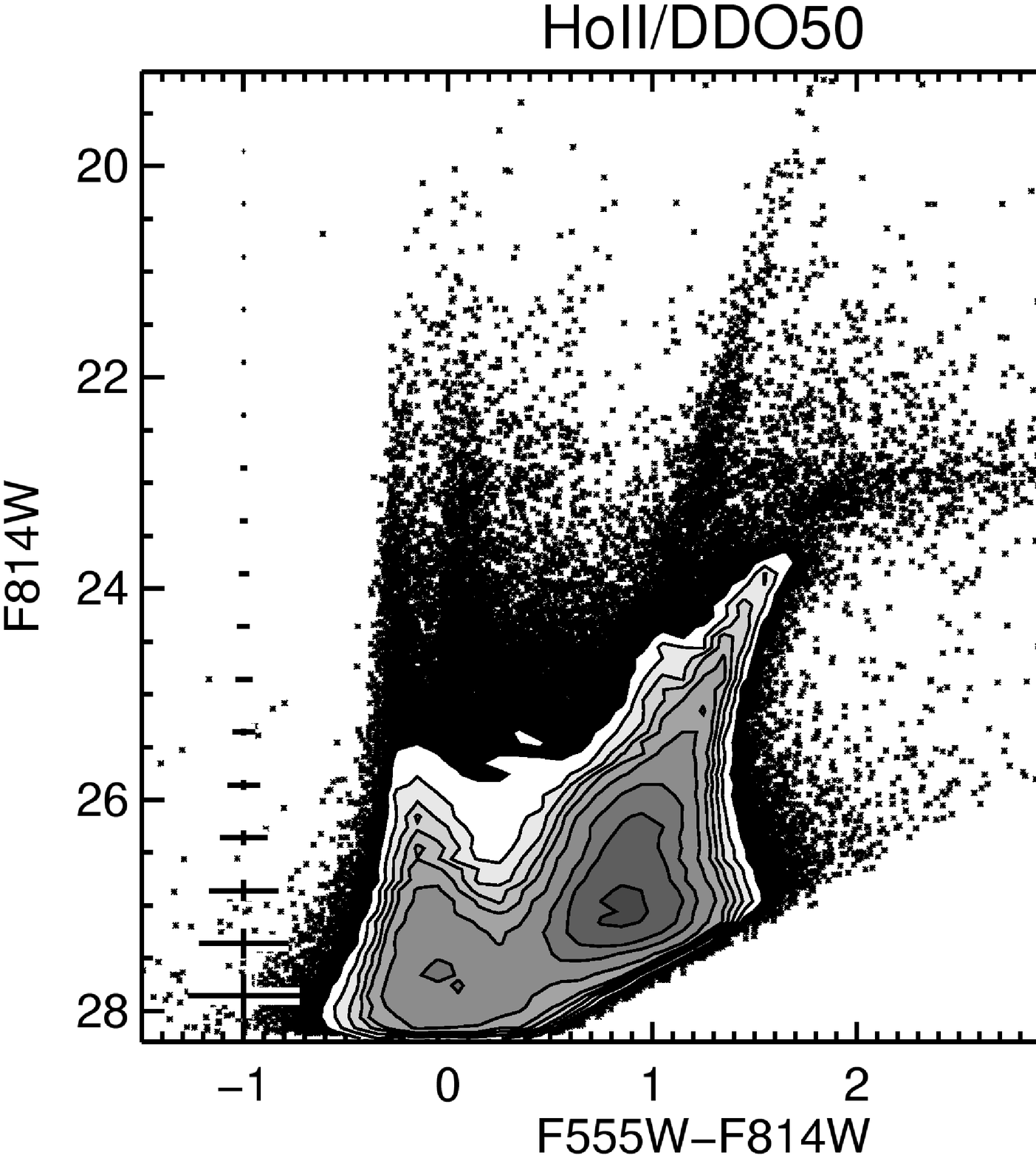}
\includegraphics[width=1.625in]{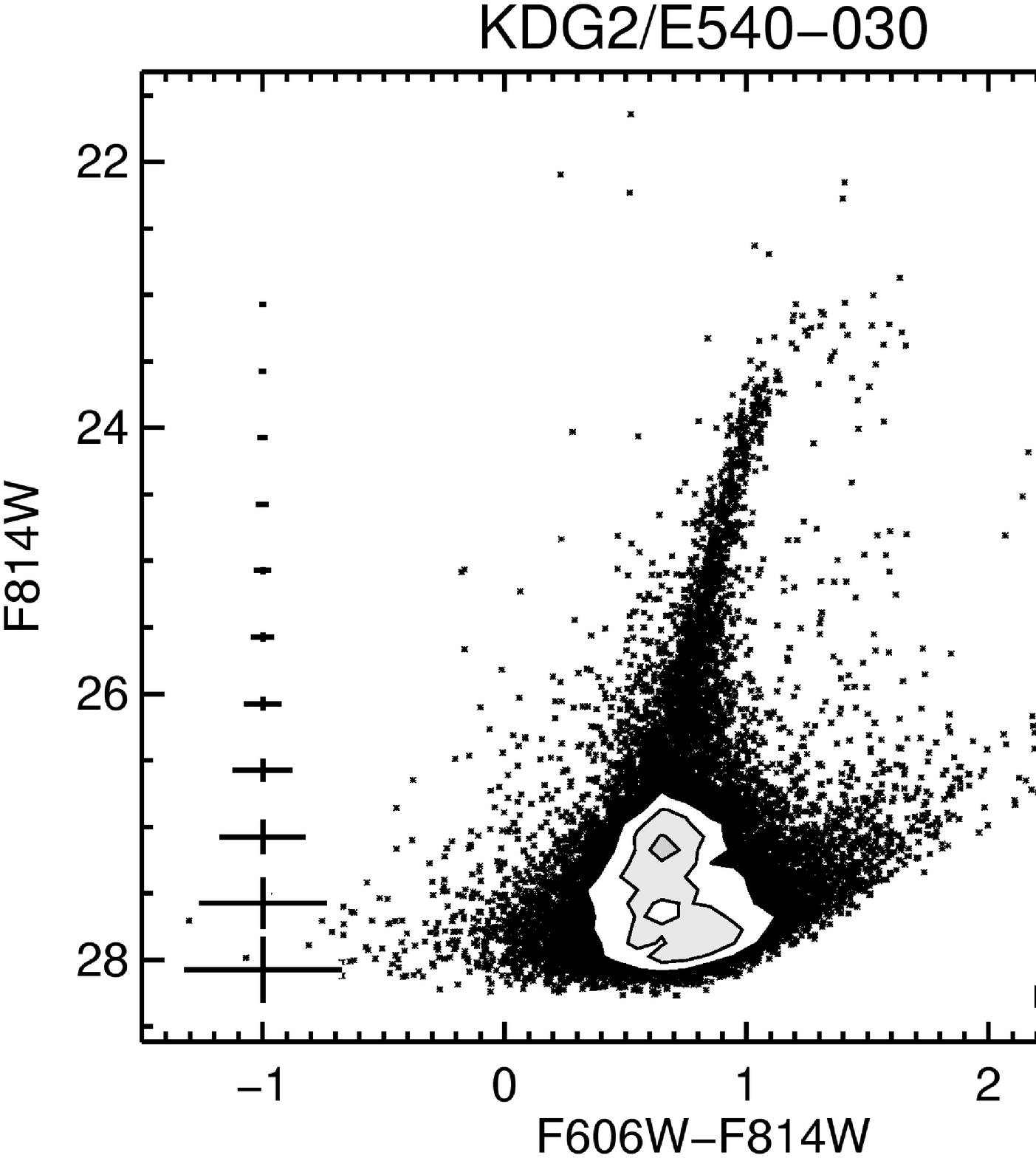}
}
\centerline{
\includegraphics[width=1.625in]{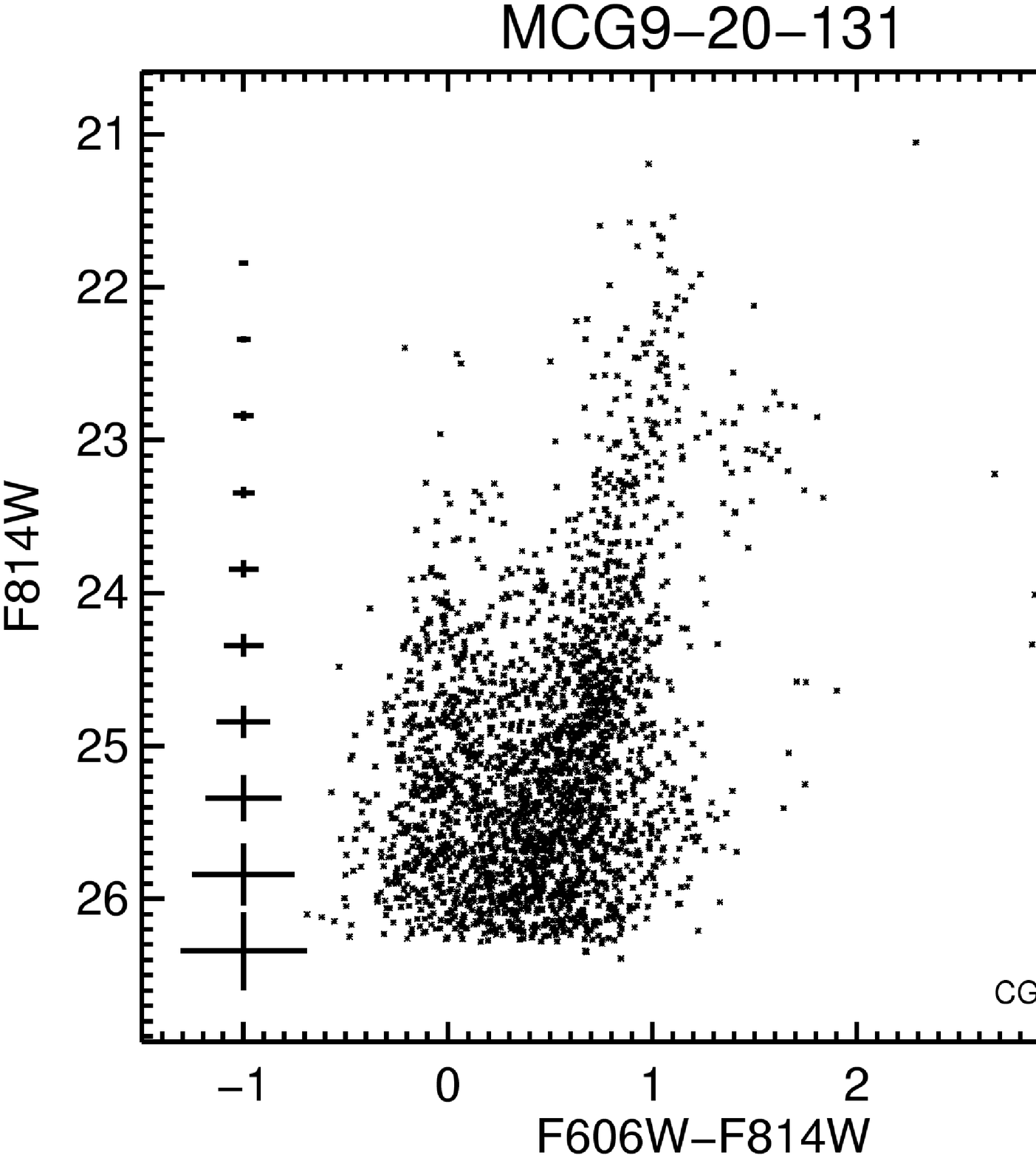}
\includegraphics[width=1.625in]{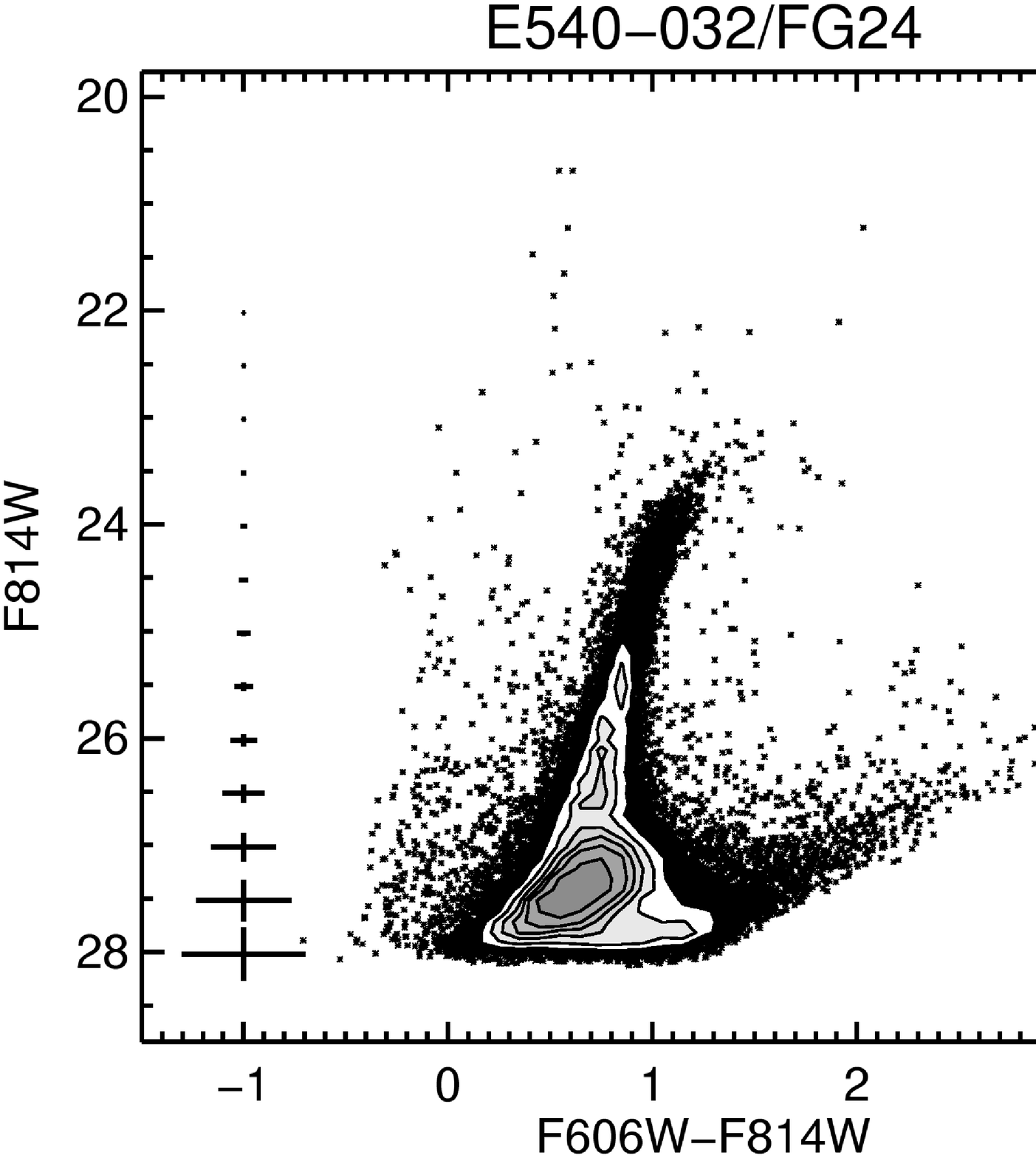}
\includegraphics[width=1.625in]{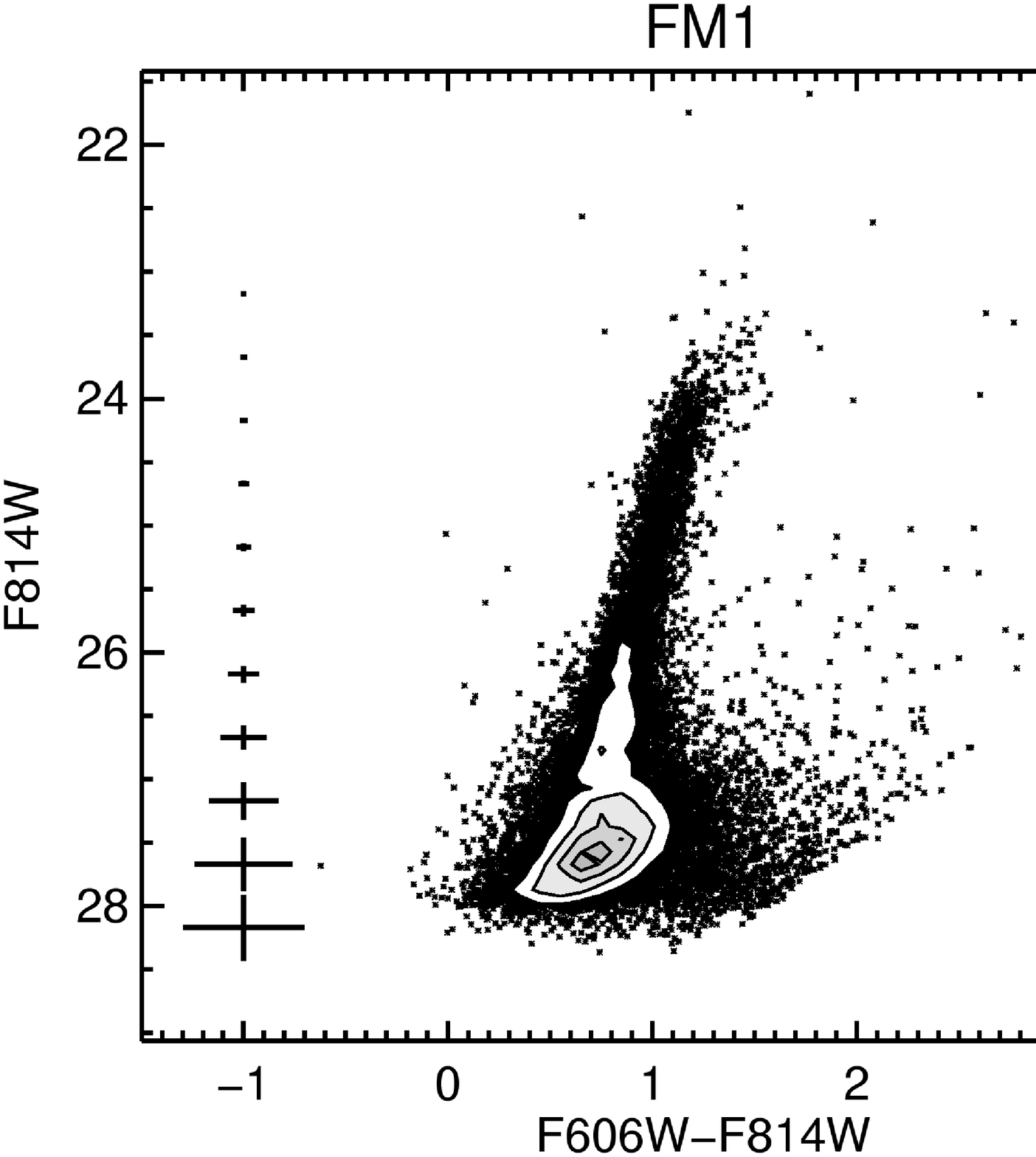}
\includegraphics[width=1.625in]{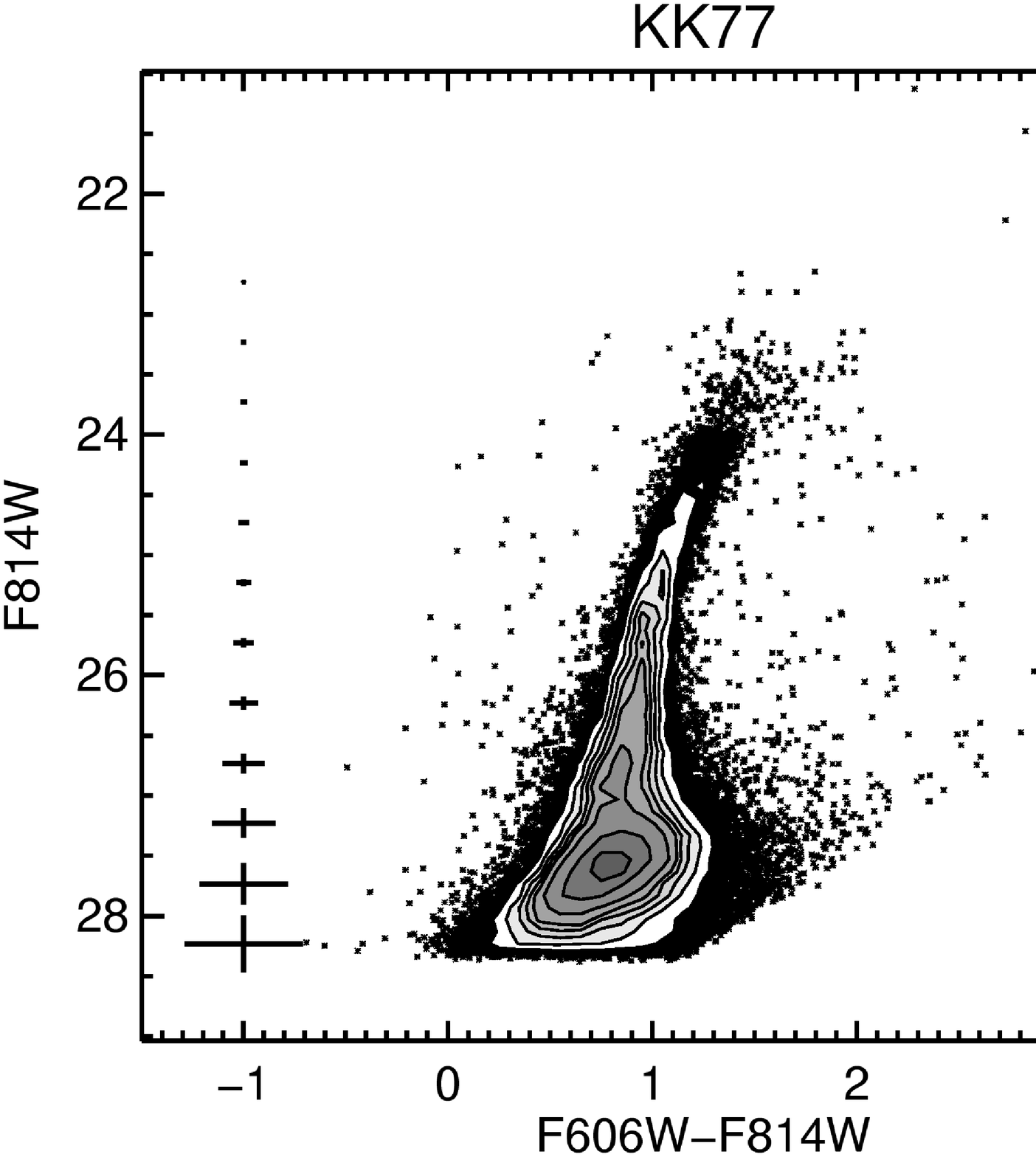}
}
\centerline{
\includegraphics[width=1.625in]{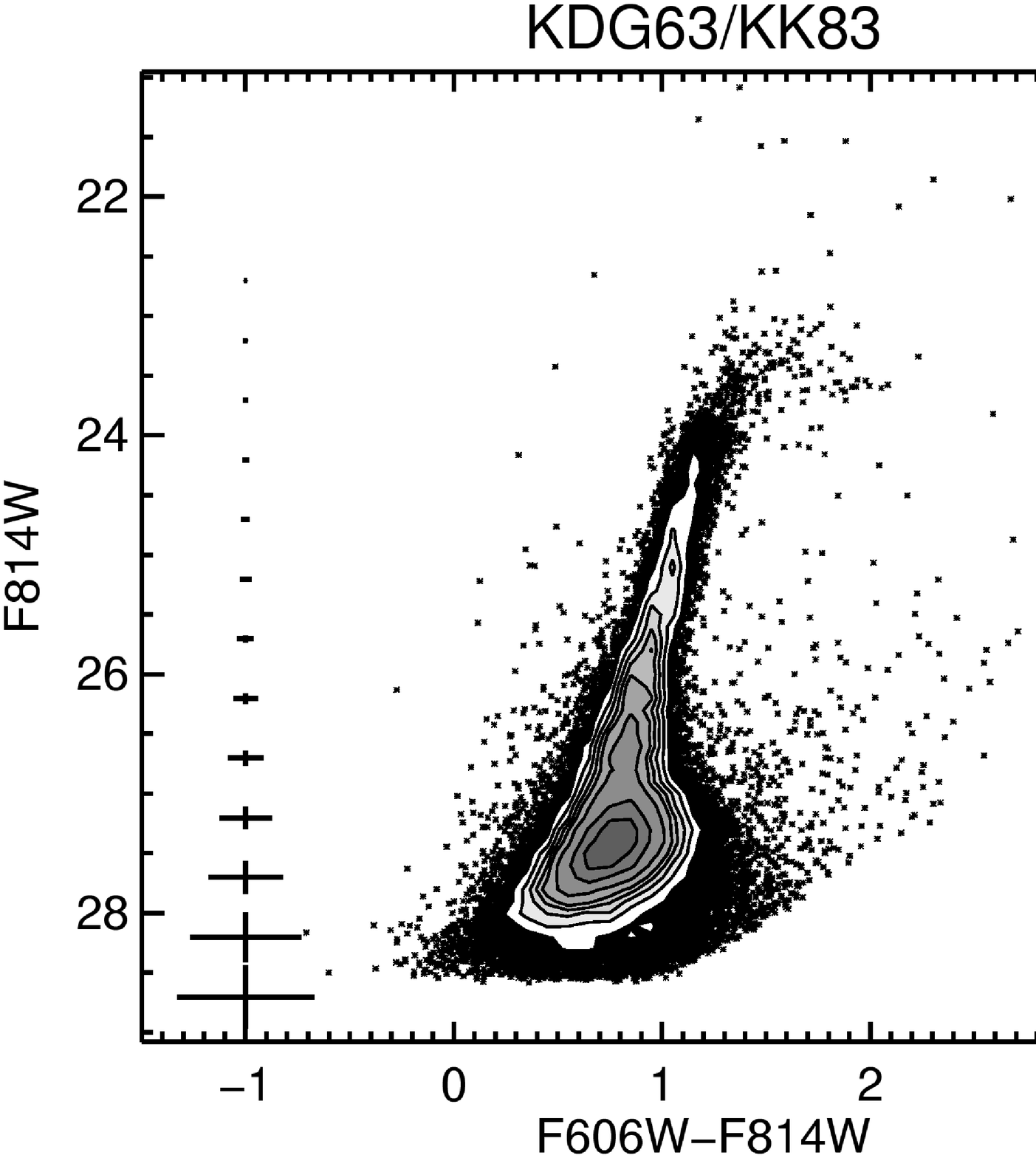}
\includegraphics[width=1.625in]{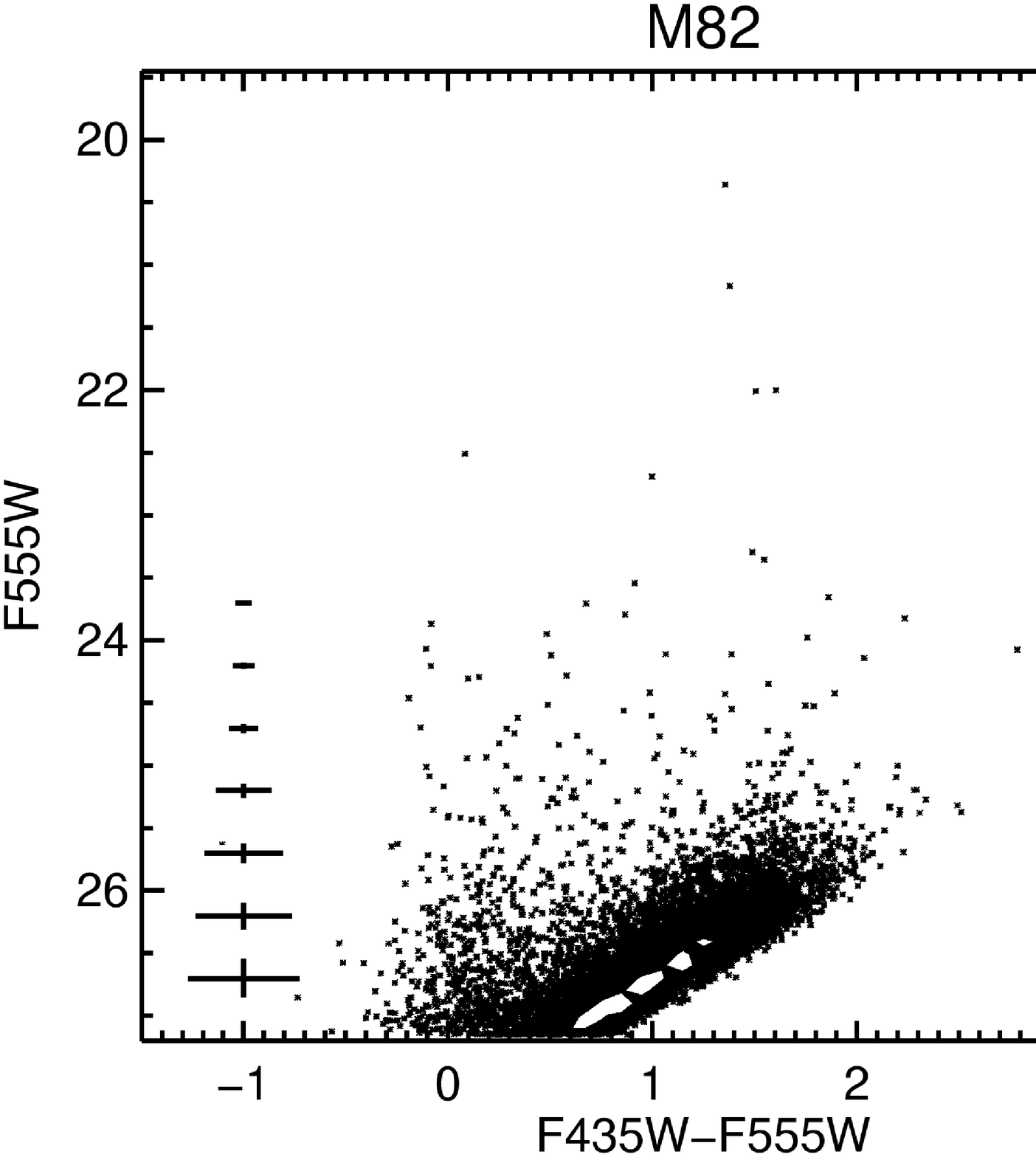}
\includegraphics[width=1.625in]{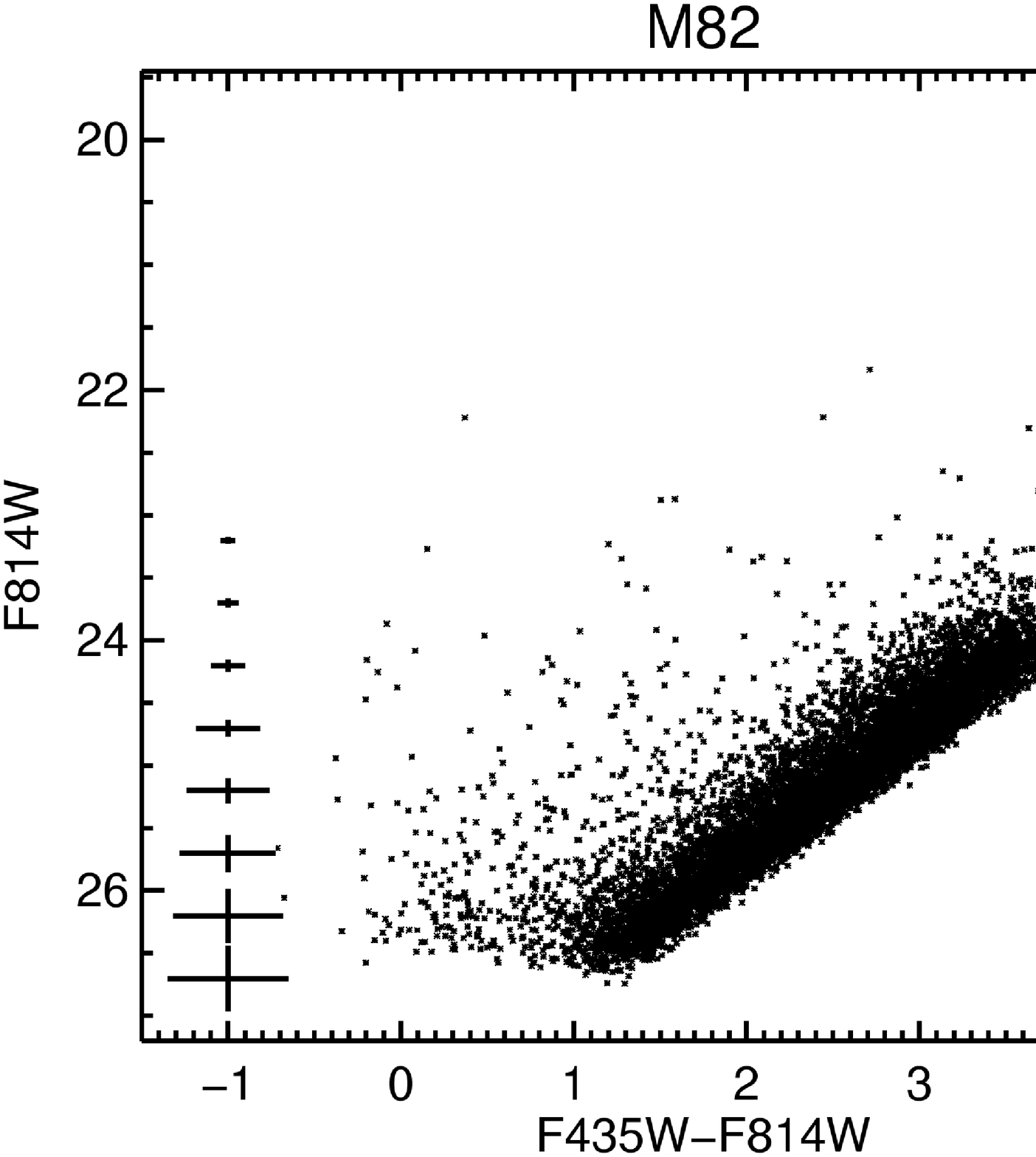}
\includegraphics[width=1.625in]{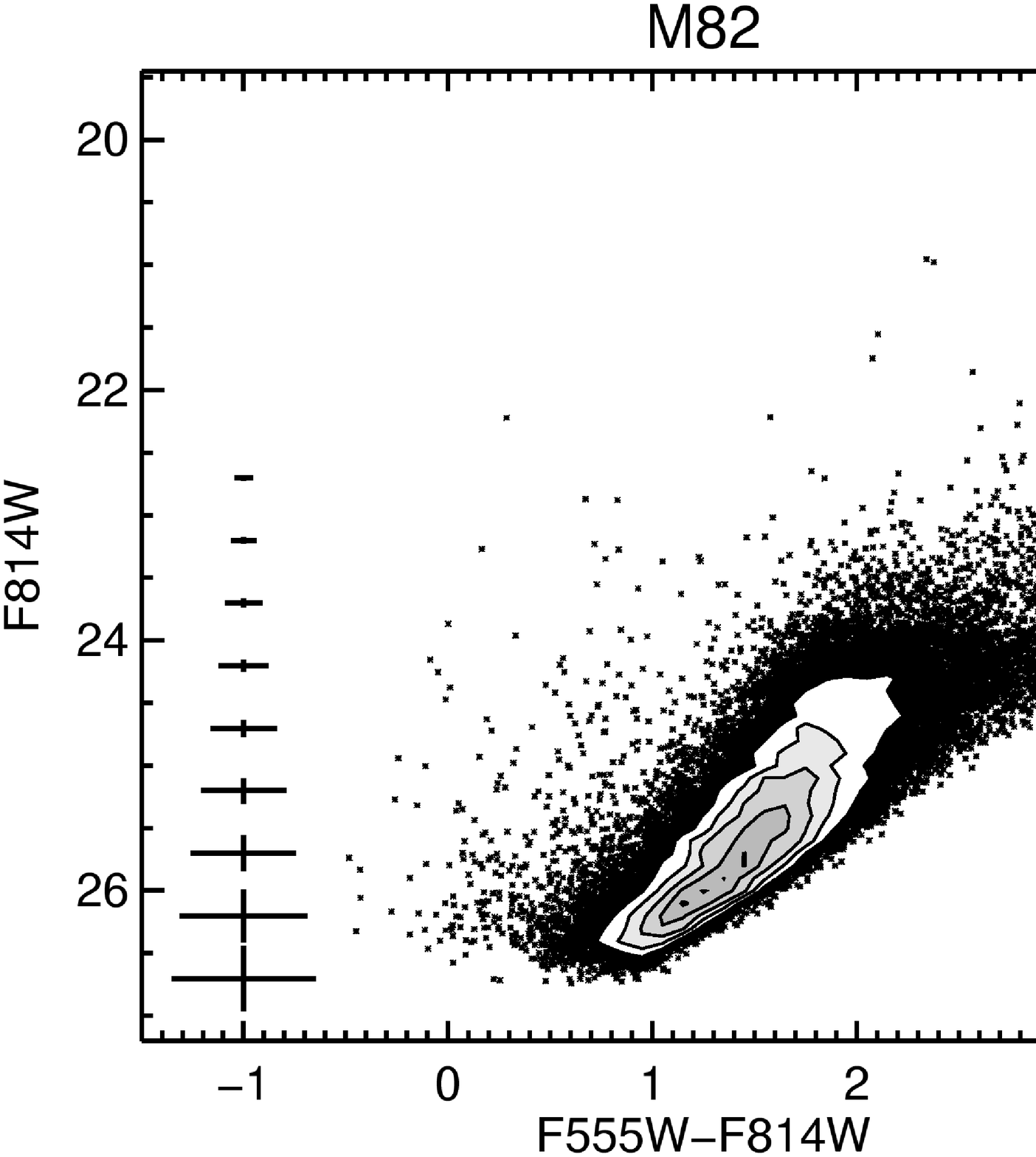}
}
\caption{
CMDs of galaxies in the ANGST data release,
as described in Figure~\ref{cmdfig1}.
Figures are ordered from the upper left to the bottom right.
(a) N2403; (b) N2403; (c) N2403; (d) DDO6; (e) KKH37; (f) HoII; (g) HoII; (h) KDG2; (i) MCG9-20-131; (j) E540-032; (k) FM1; (l) KK77; (m) KDG63; (n) M82; (o) M82; (p) M82; 
    \label{cmdfig6}}
\end{figure}
\vfill
\clearpage
 
%-------------------
\begin{figure}[p]
\centerline{
\includegraphics[width=1.625in]{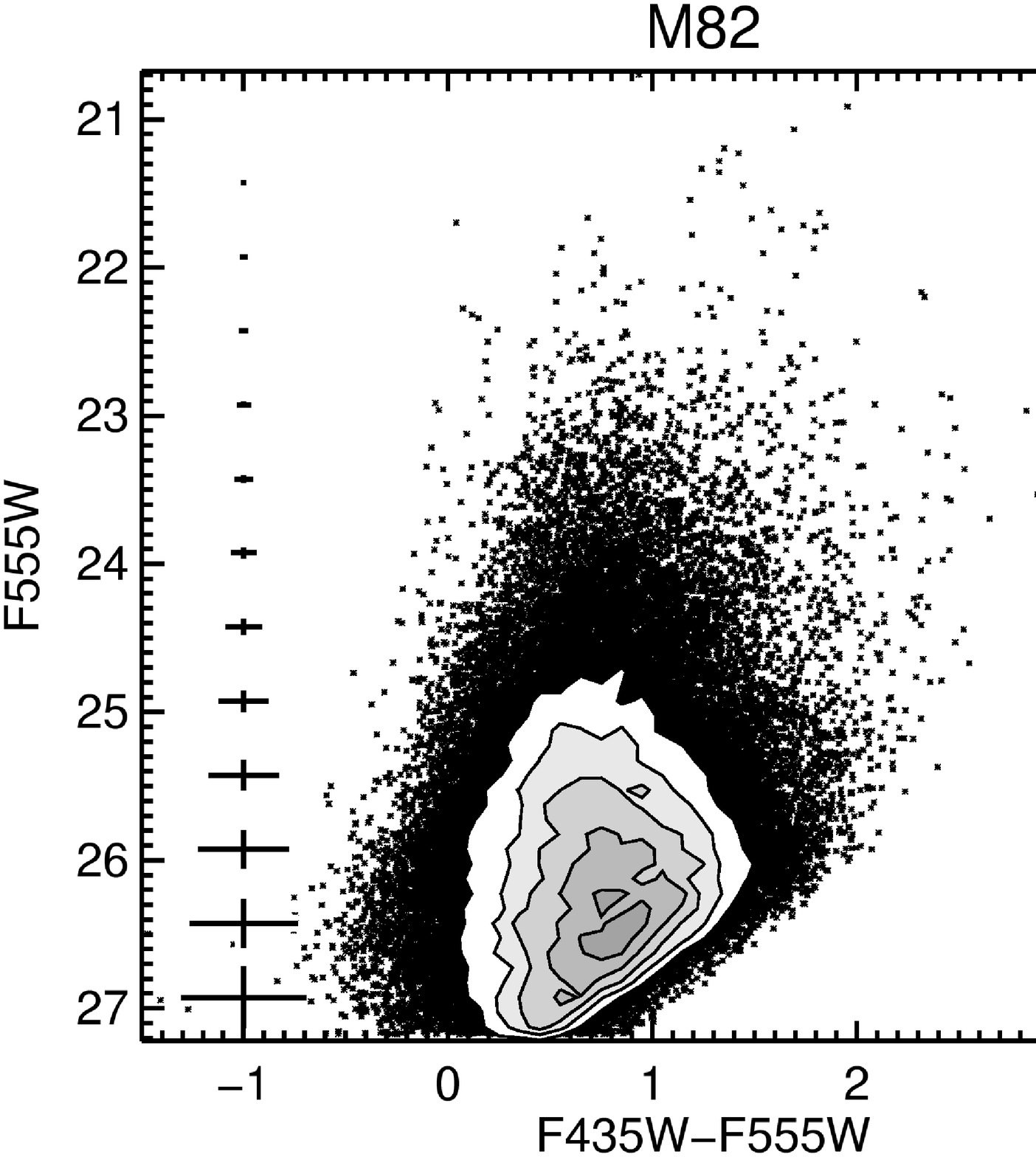}
\includegraphics[width=1.625in]{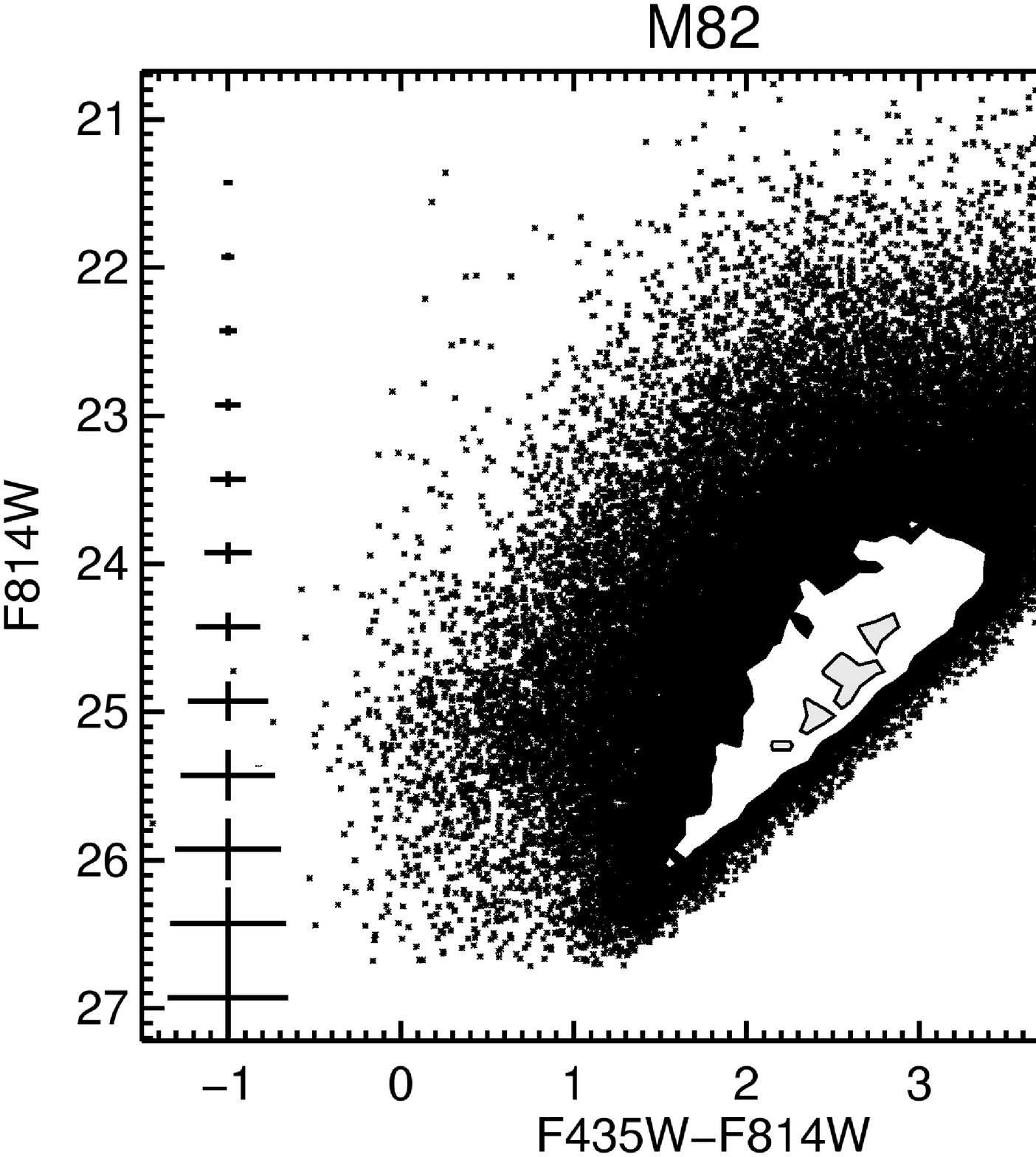}
\includegraphics[width=1.625in]{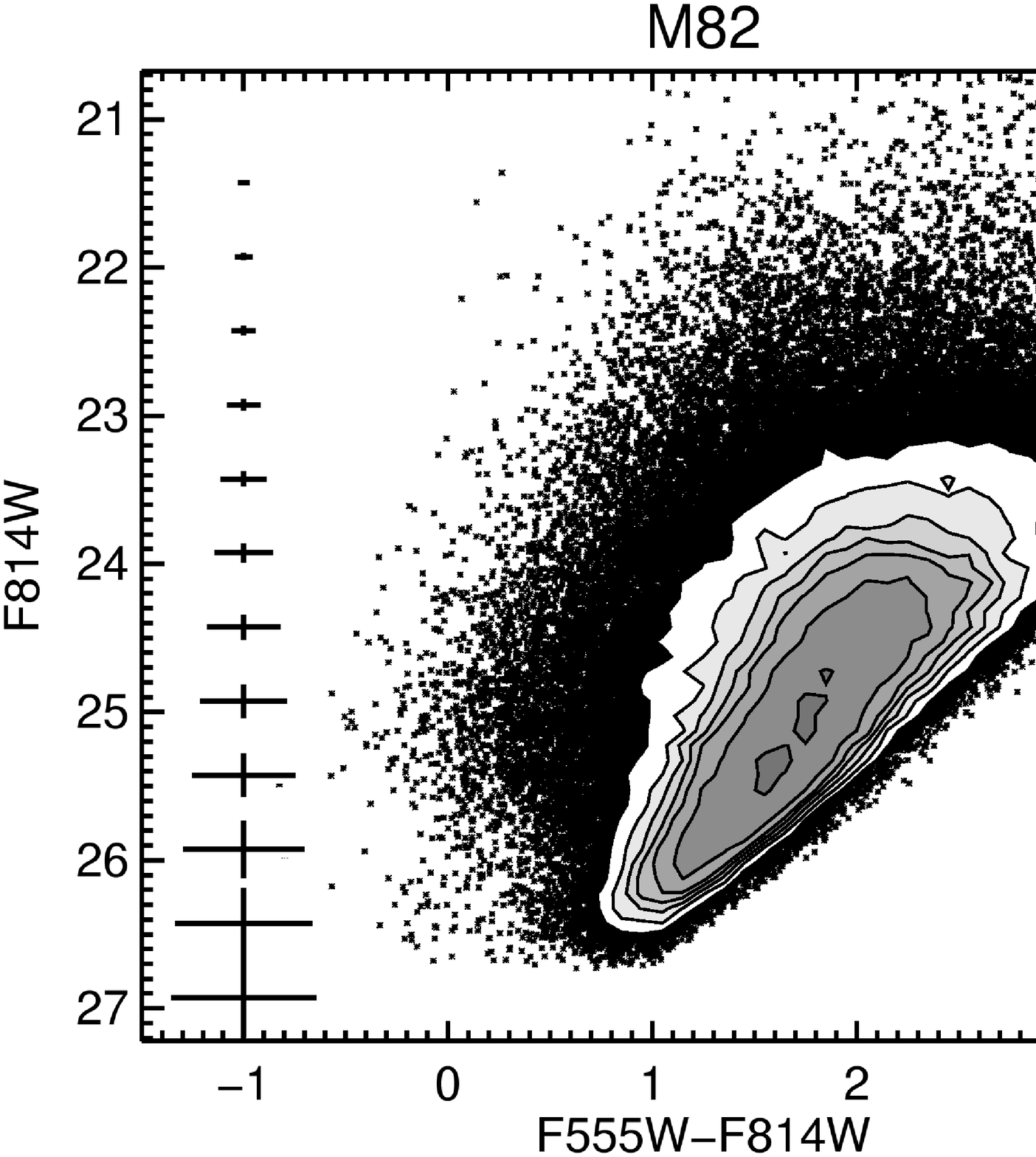}
\includegraphics[width=1.625in]{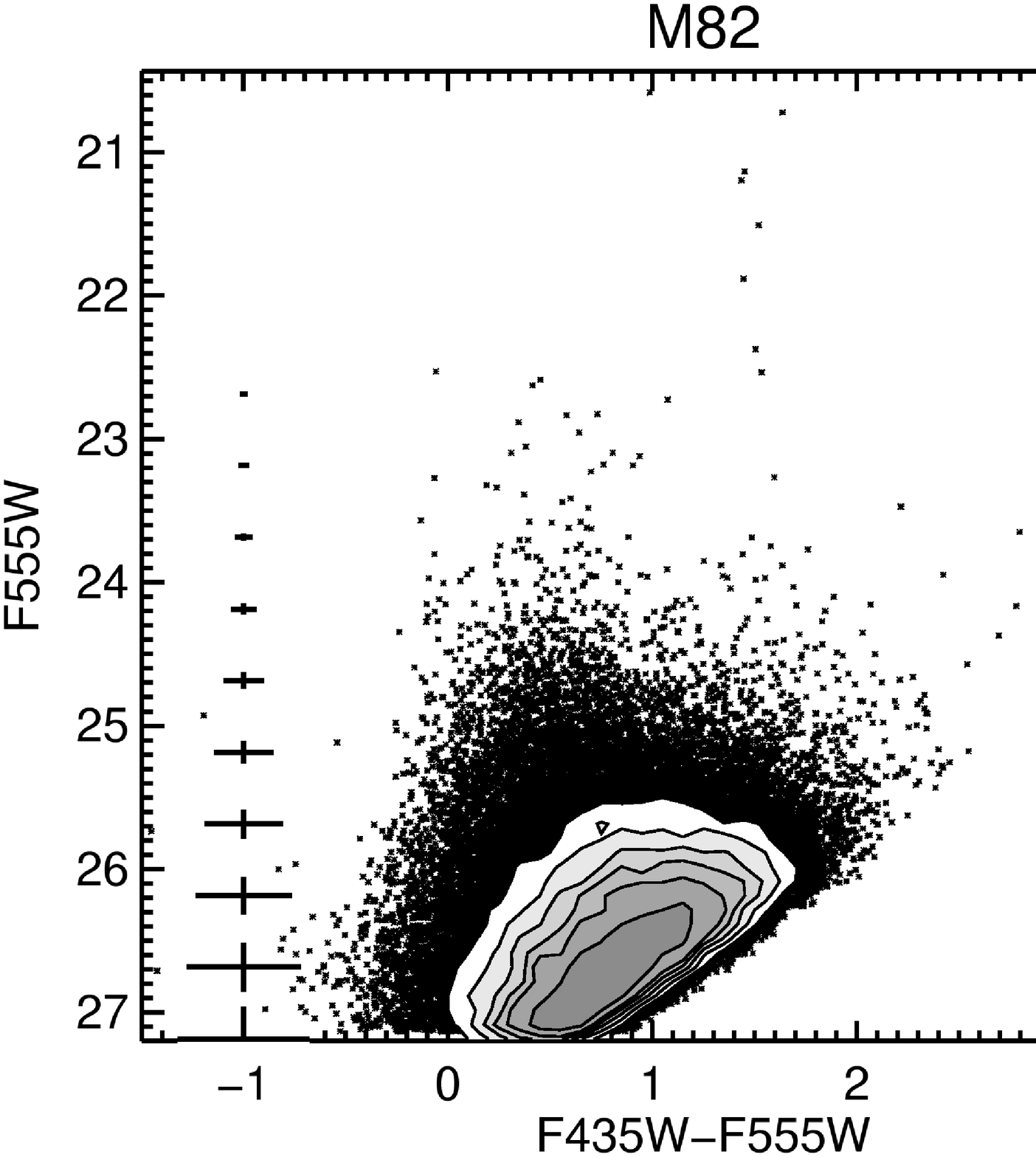}
}
\centerline{
\includegraphics[width=1.625in]{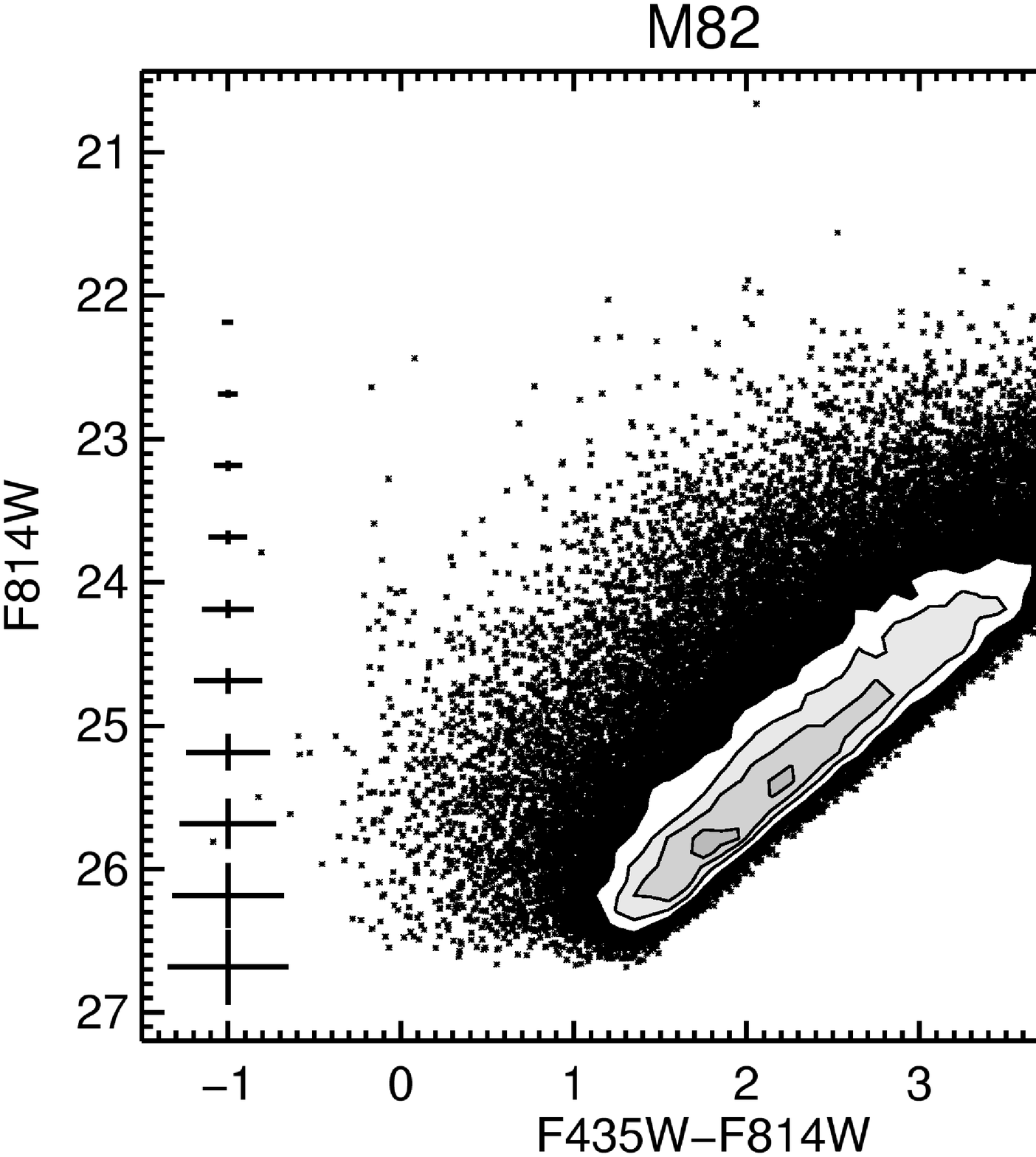}
\includegraphics[width=1.625in]{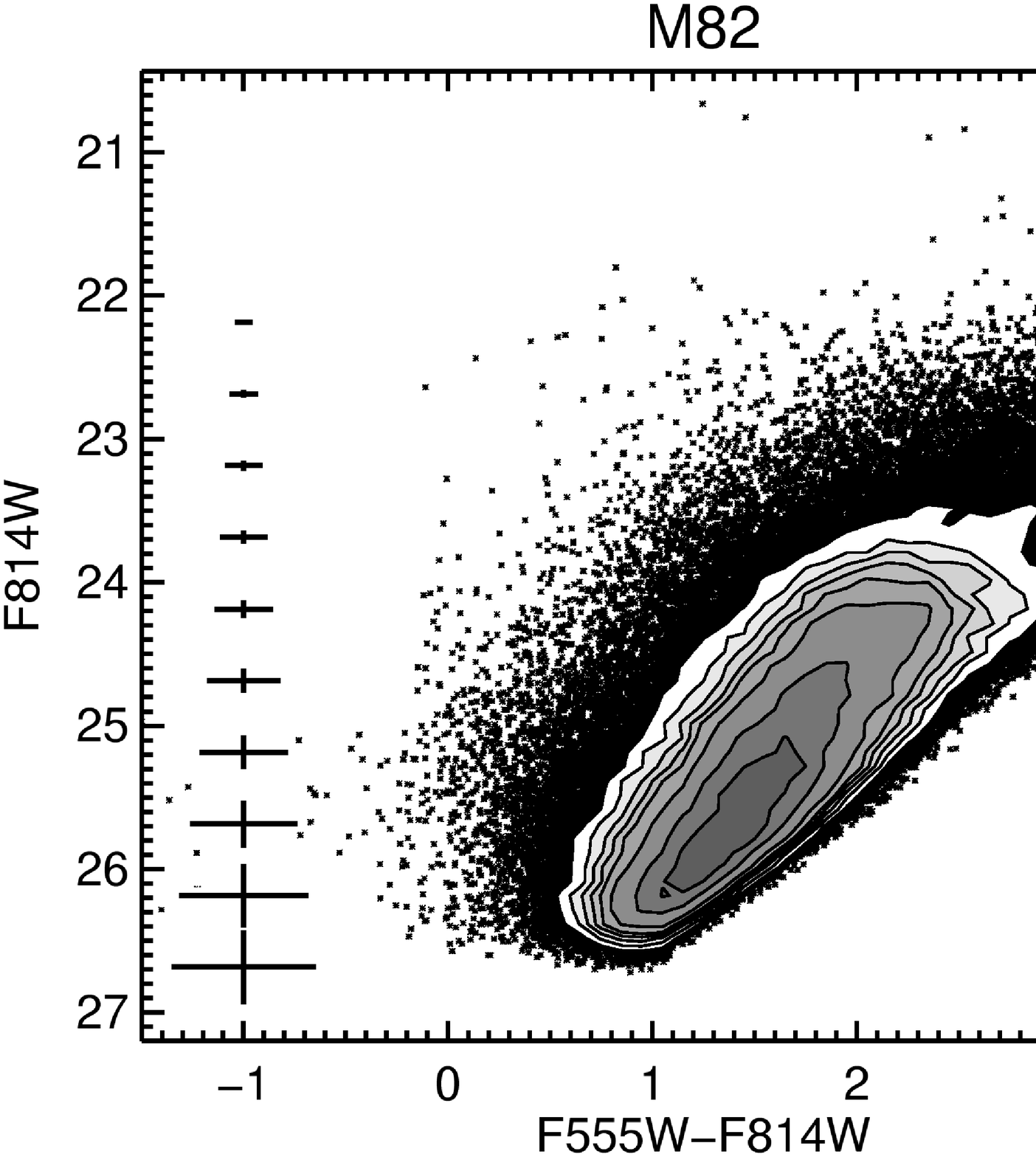}
\includegraphics[width=1.625in]{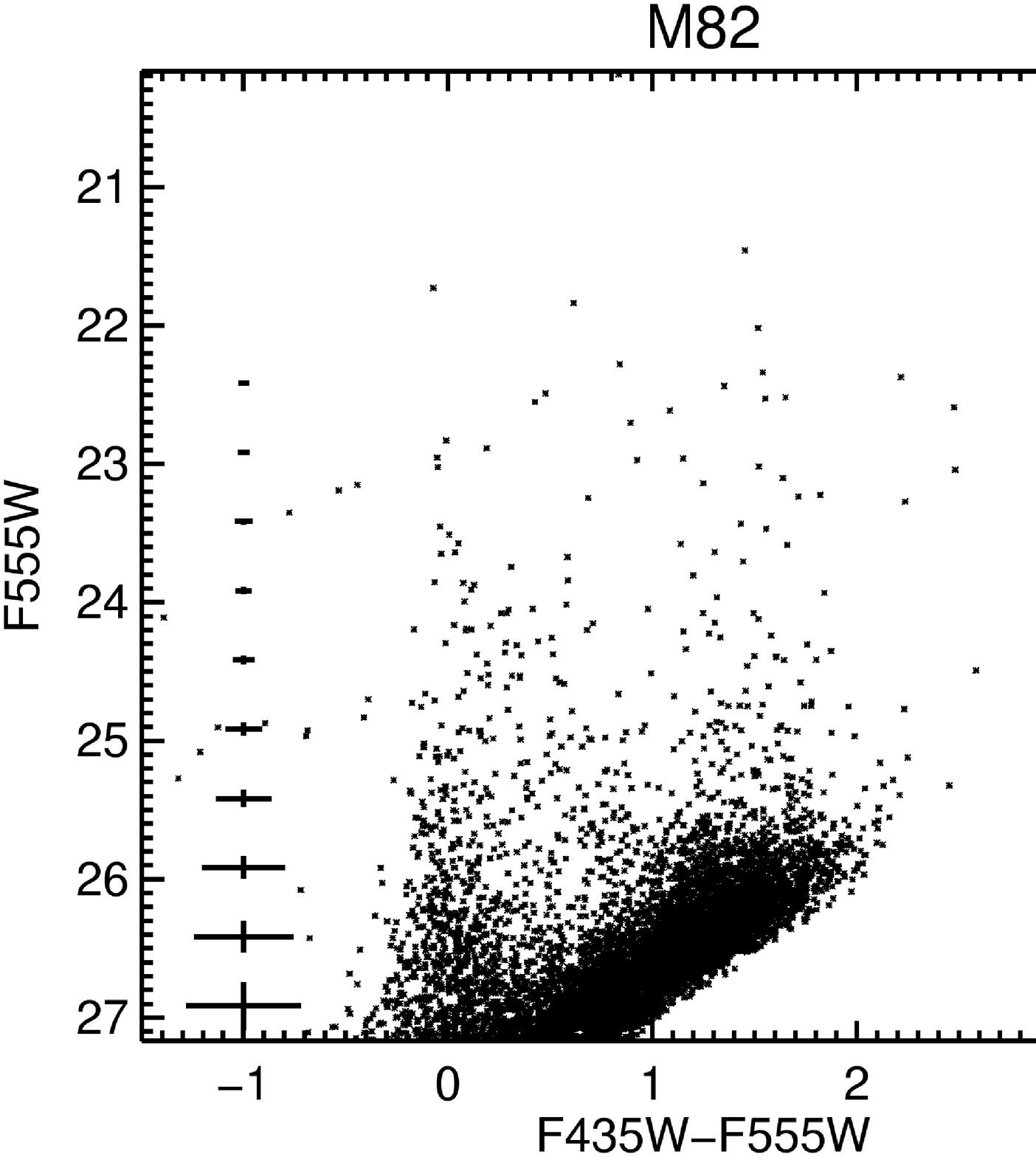}
\includegraphics[width=1.625in]{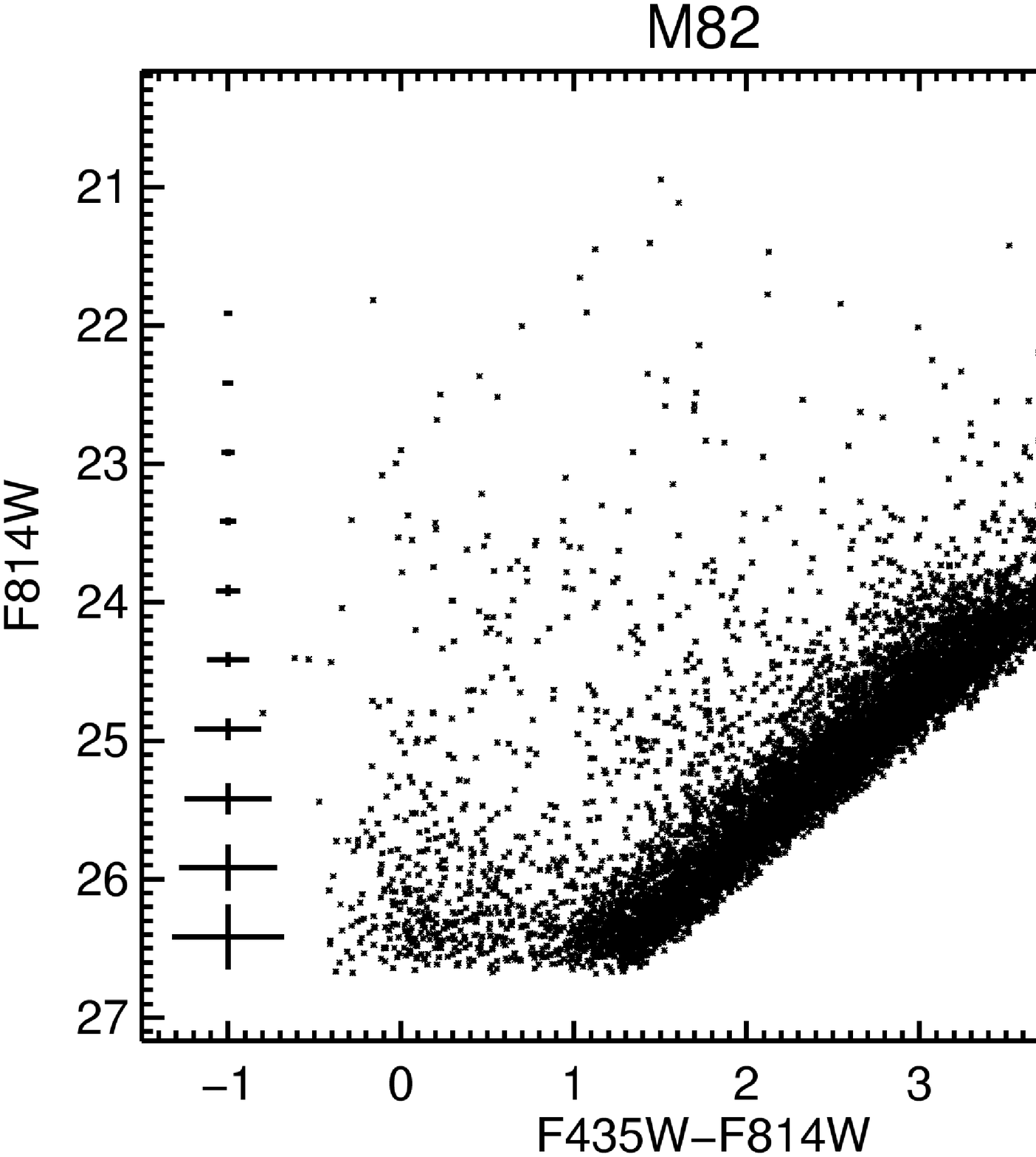}
}
\centerline{
\includegraphics[width=1.625in]{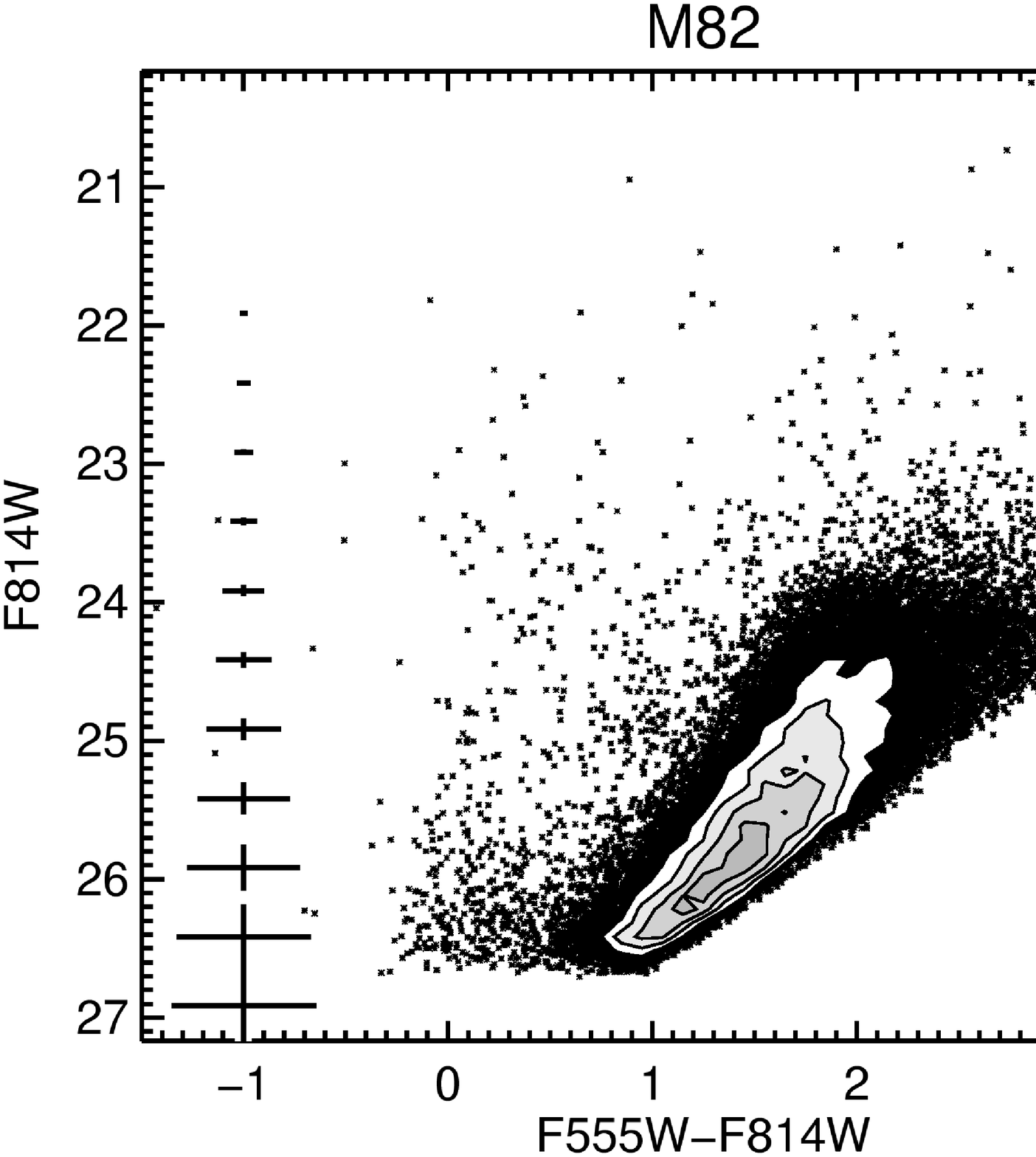}
\includegraphics[width=1.625in]{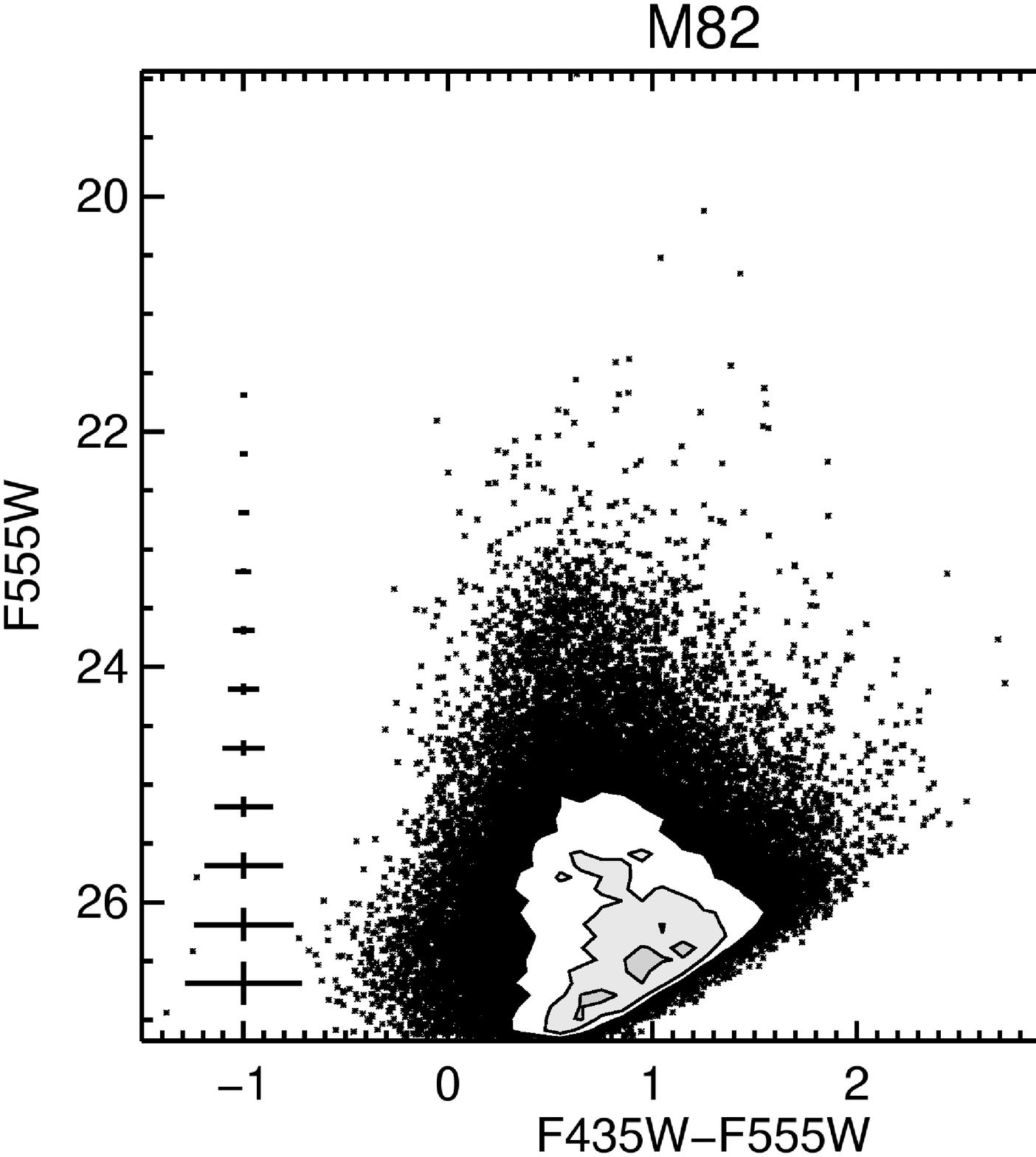}
\includegraphics[width=1.625in]{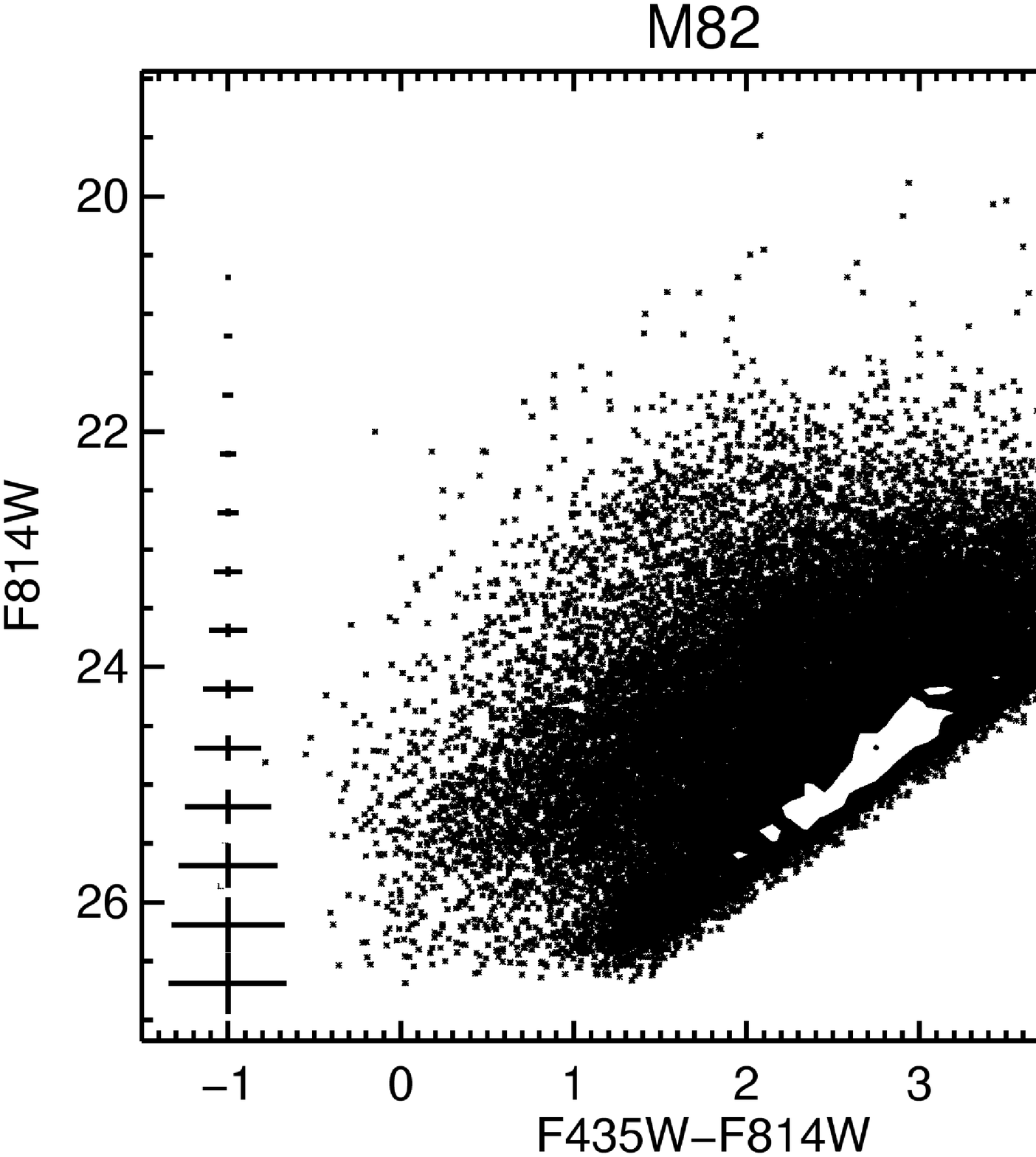}
\includegraphics[width=1.625in]{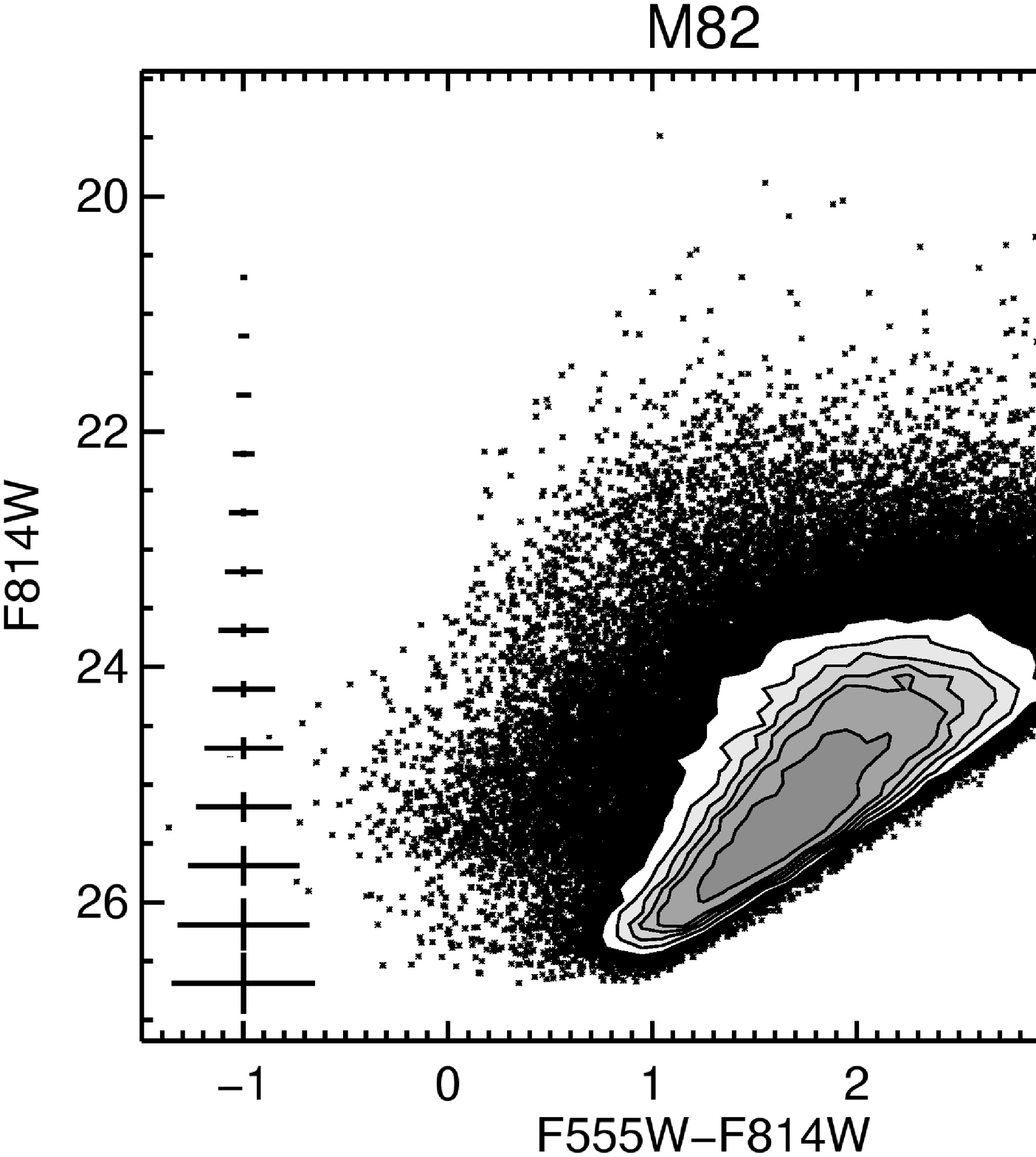}
}
\centerline{
\includegraphics[width=1.625in]{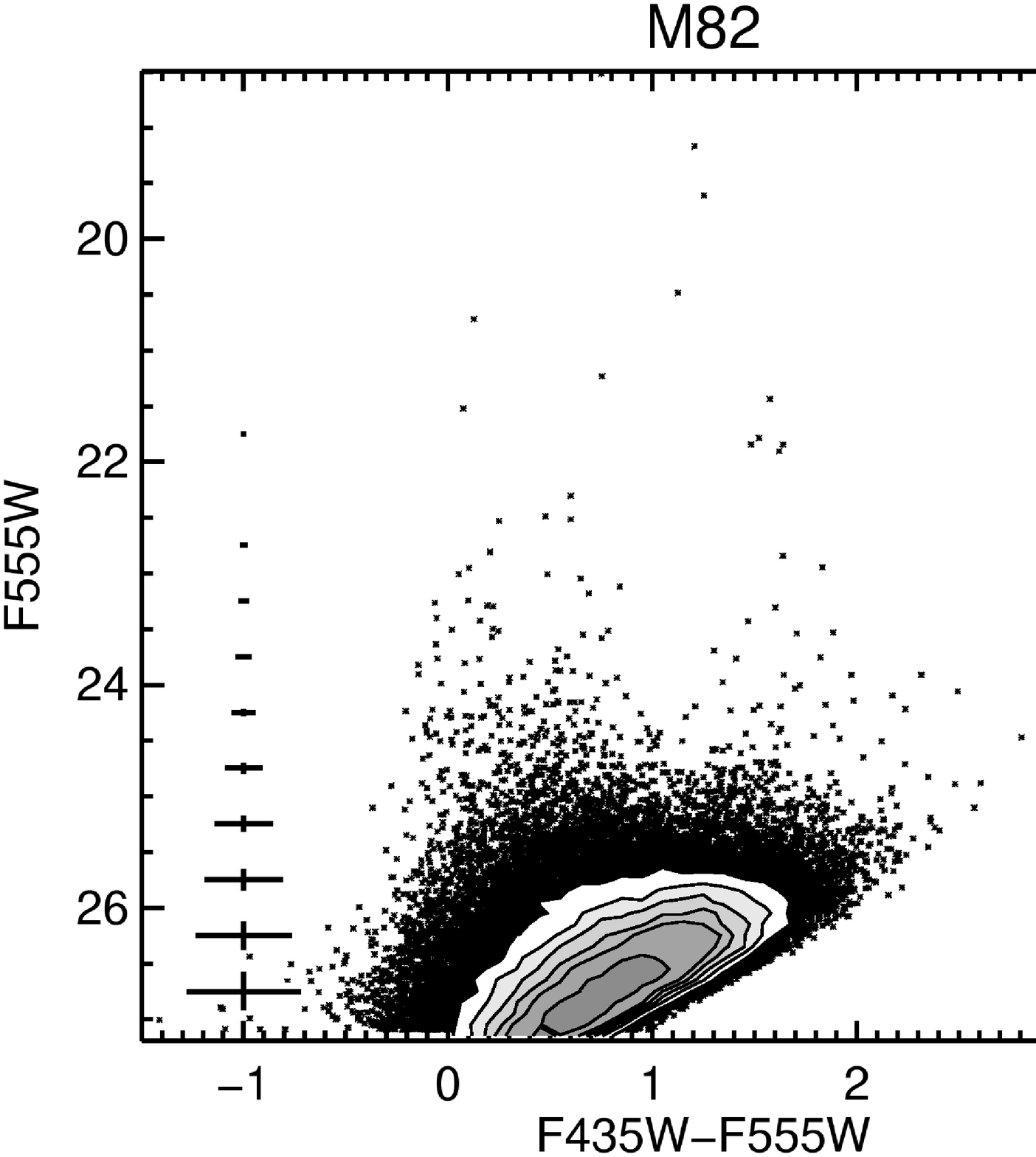}
\includegraphics[width=1.625in]{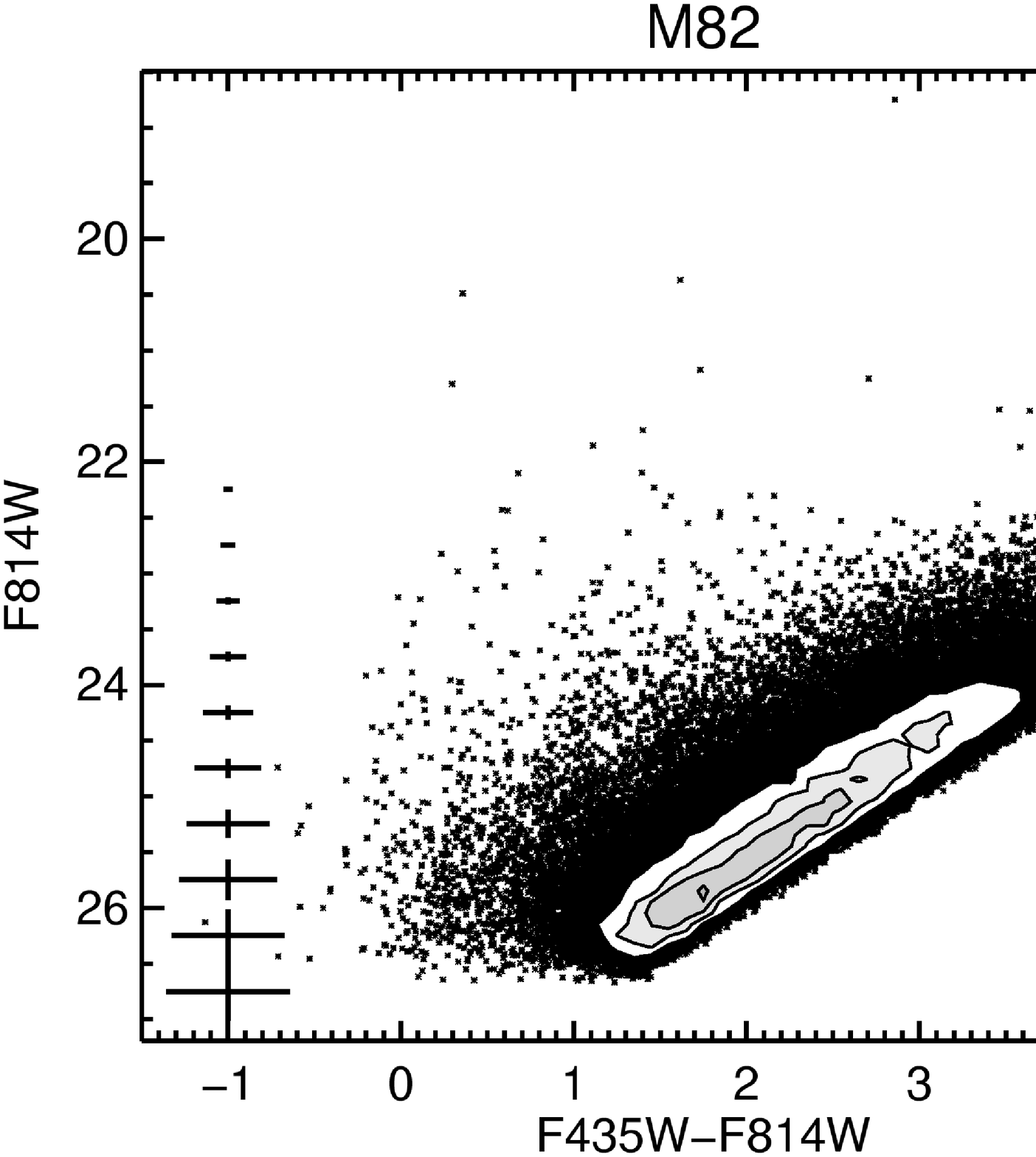}
\includegraphics[width=1.625in]{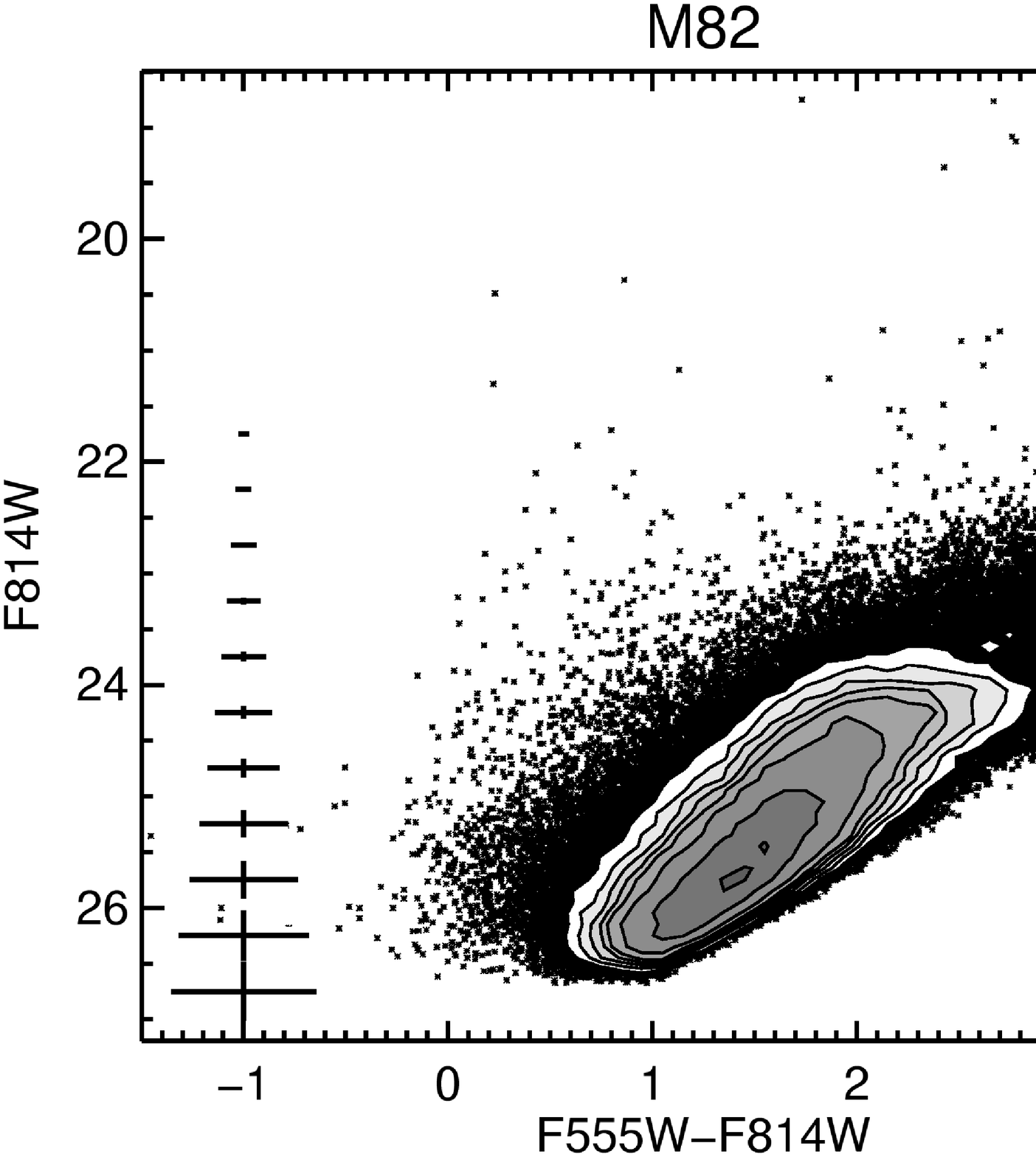}
\includegraphics[width=1.625in]{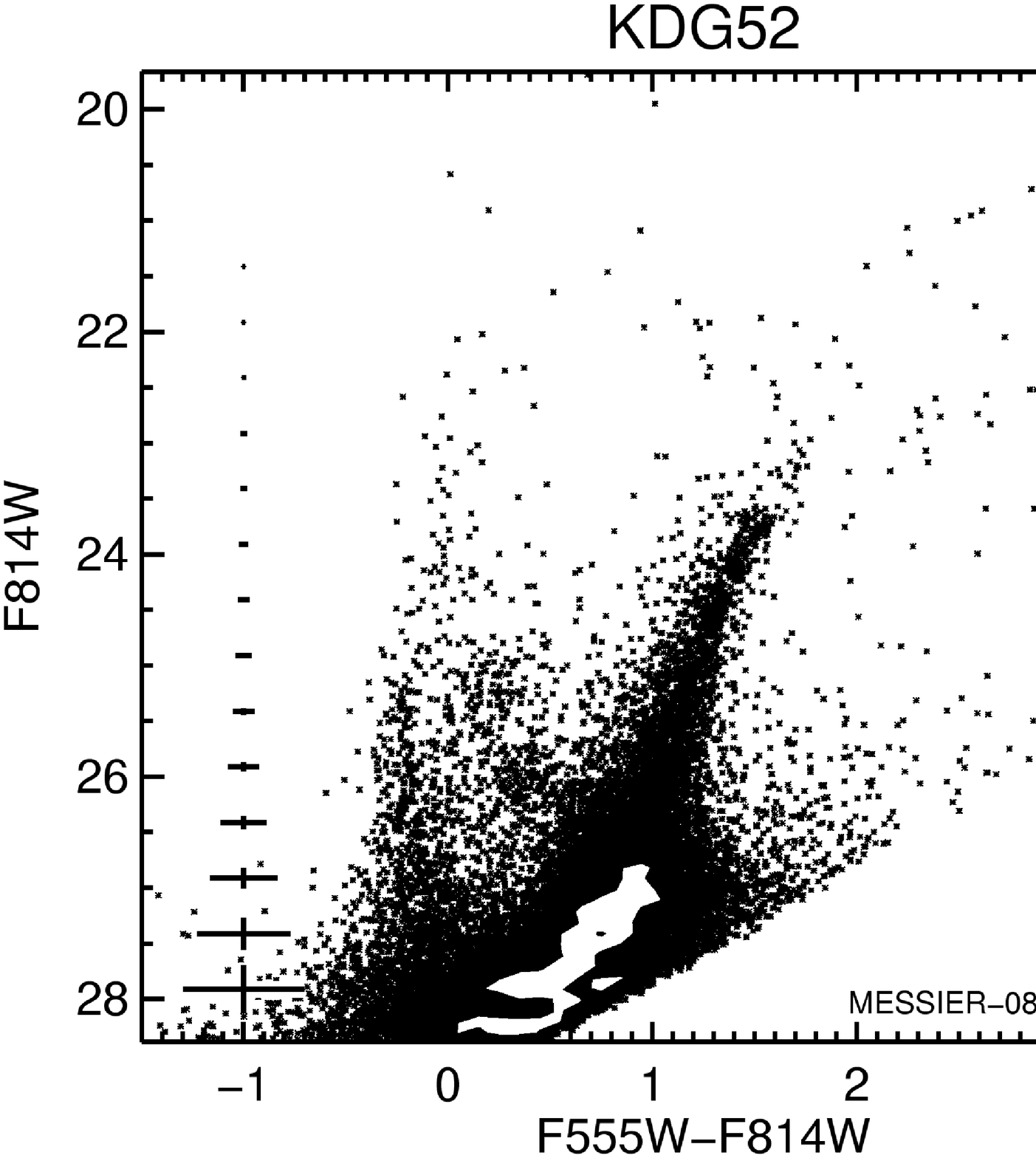}
}
\caption{
CMDs of galaxies in the ANGST data release,
as described in Figure~\ref{cmdfig1}.
Figures are ordered from the upper left to the bottom right.
(a) M82; (b) M82; (c) M82; (d) M82; (e) M82; (f) M82; (g) M82; (h) M82; (i) M82; (j) M82; (k) M82; (l) M82; (m) M82; (n) M82; (o) M82; (p) KDG52; 
    \label{cmdfig7}}
\end{figure}
\vfill
\clearpage
 
%-------------------
\begin{figure}[p]
\centerline{
\includegraphics[width=1.625in]{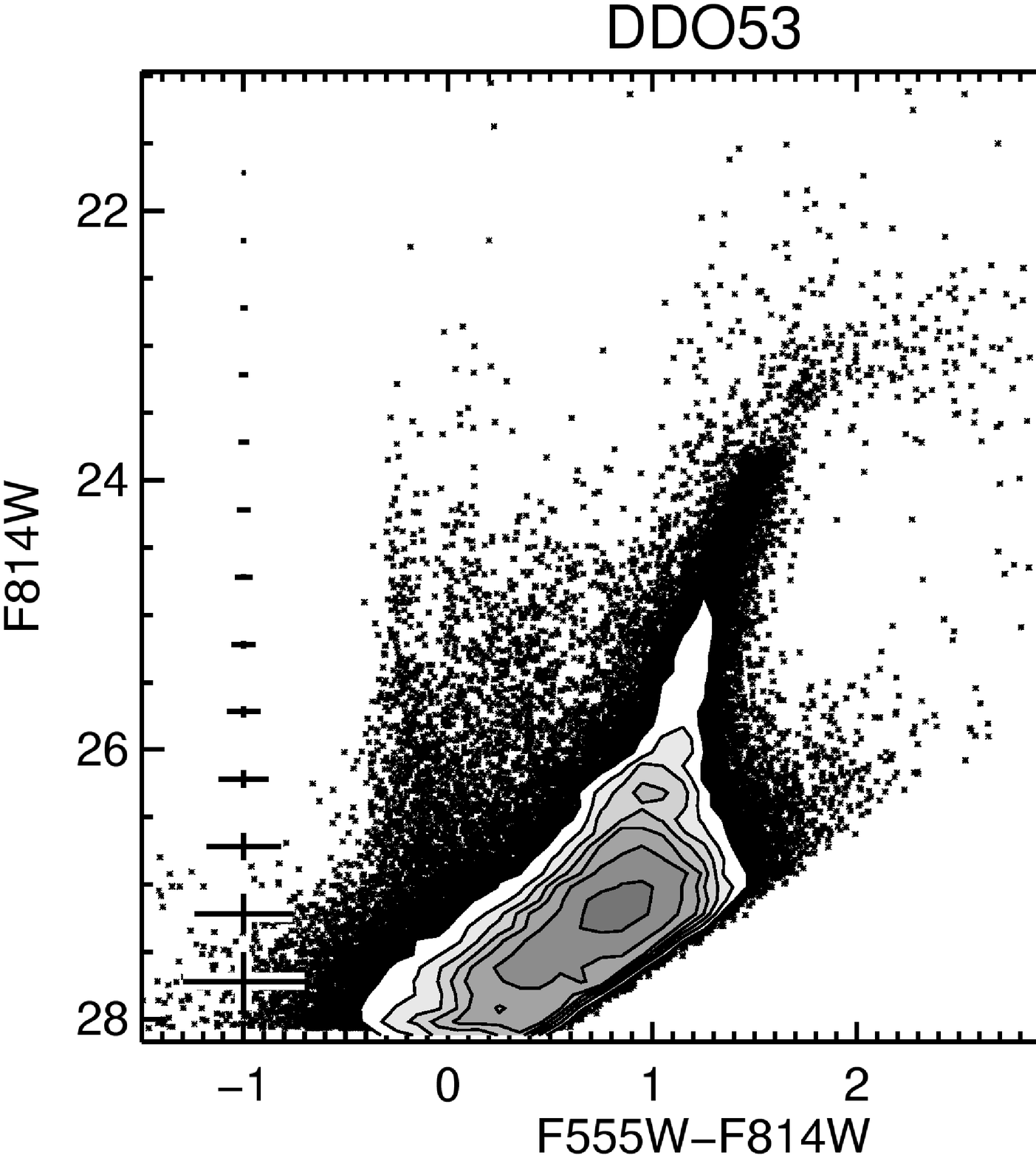}
\includegraphics[width=1.625in]{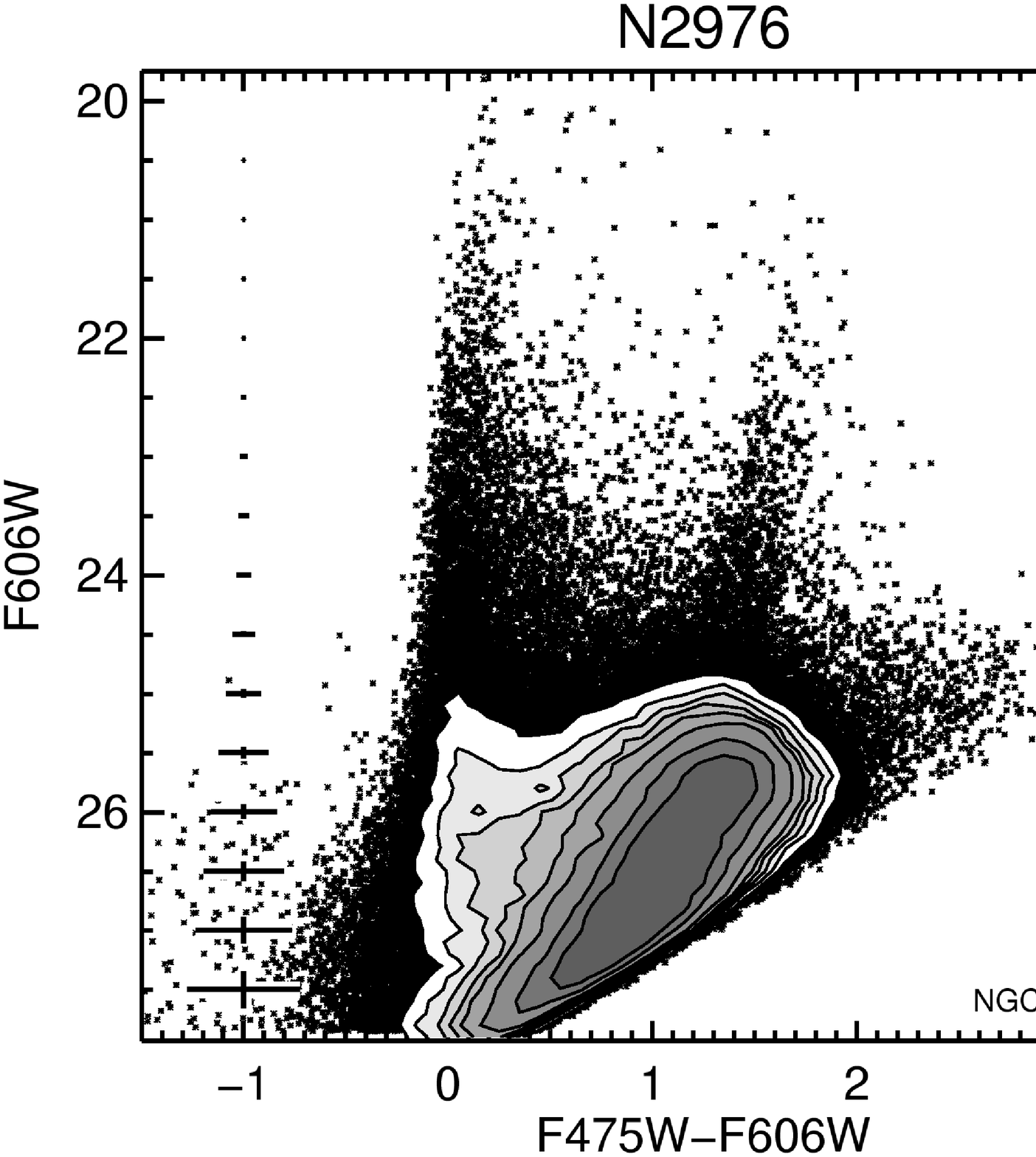}
\includegraphics[width=1.625in]{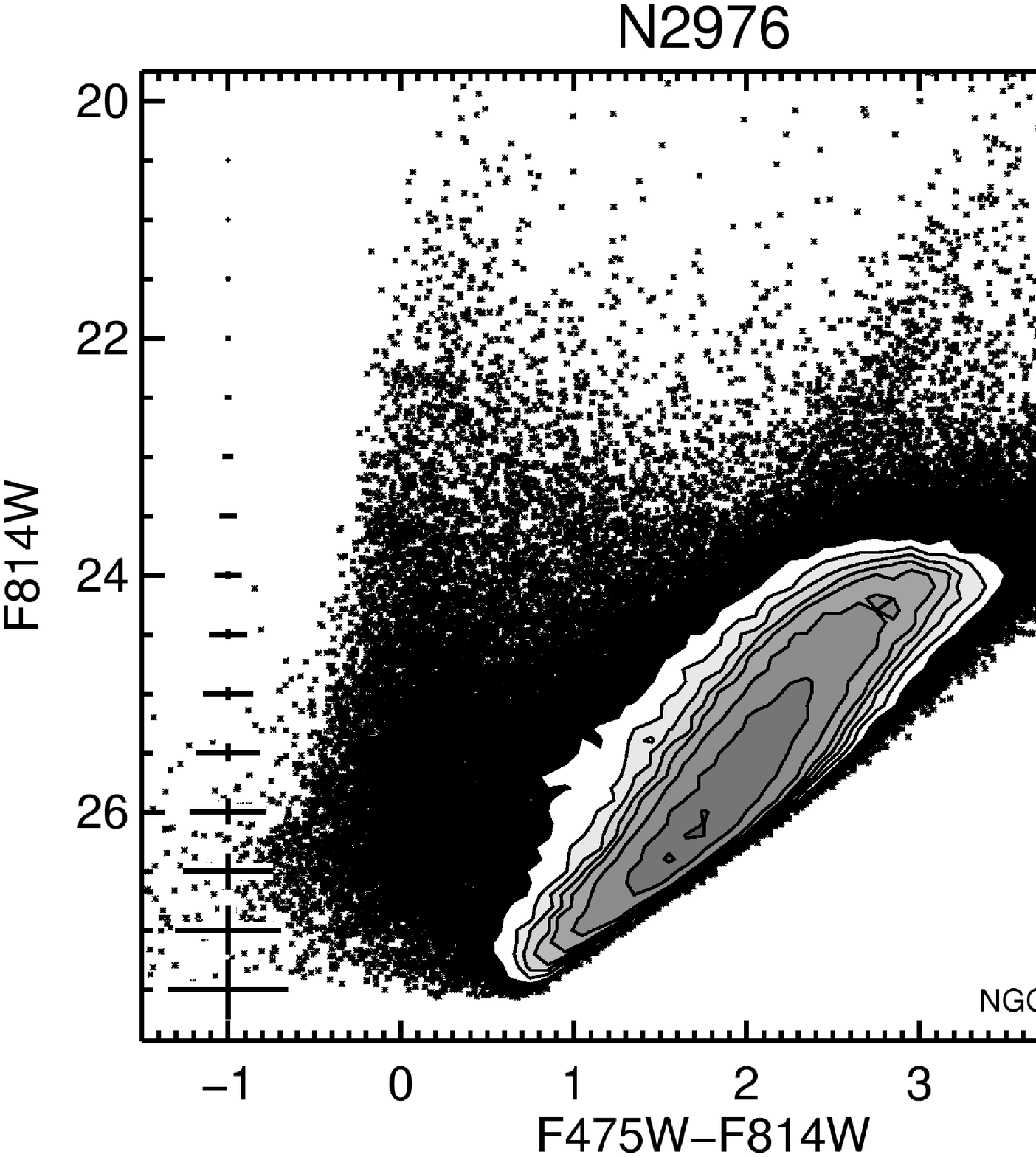}
\includegraphics[width=1.625in]{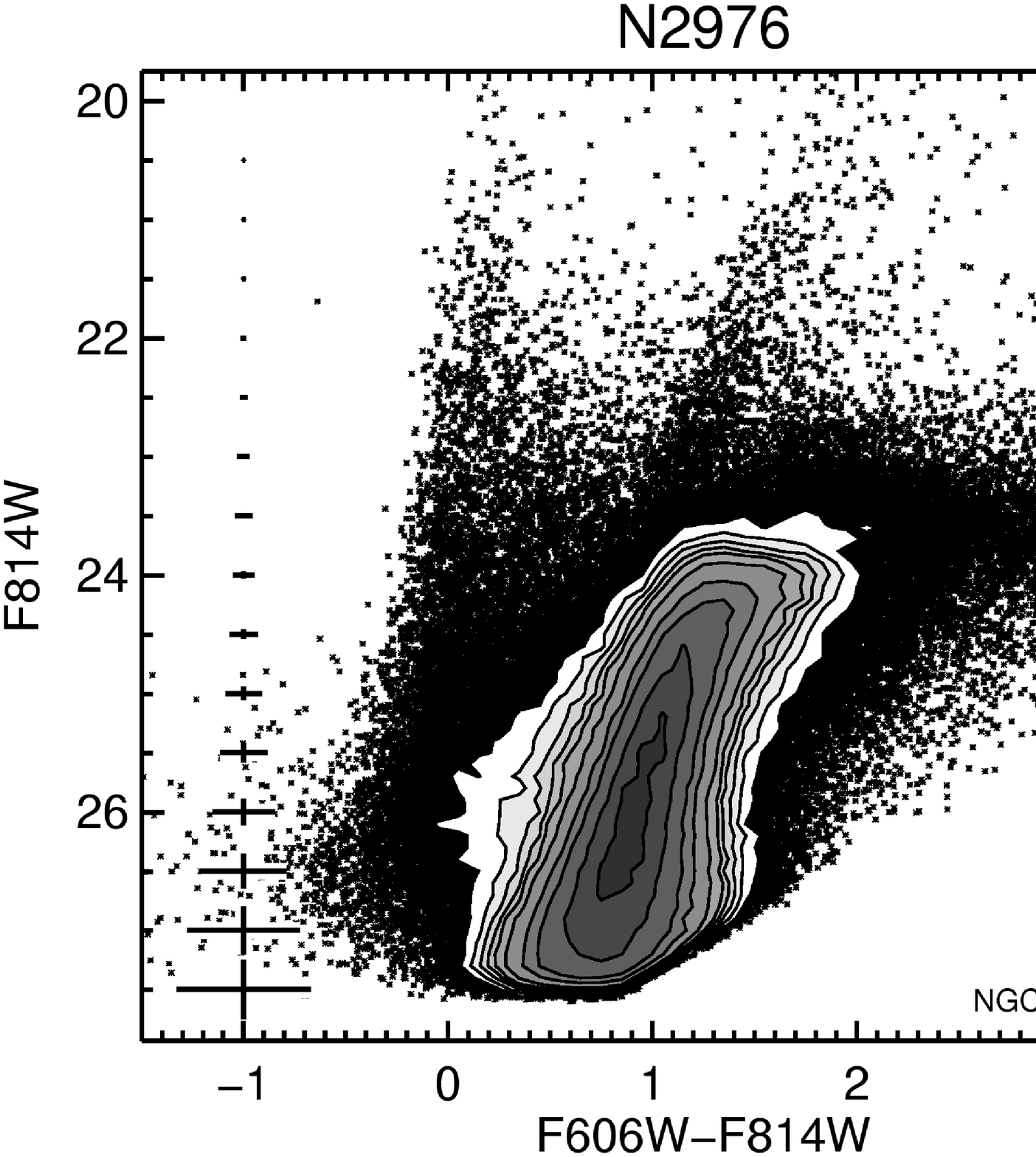}
}
\centerline{
\includegraphics[width=1.625in]{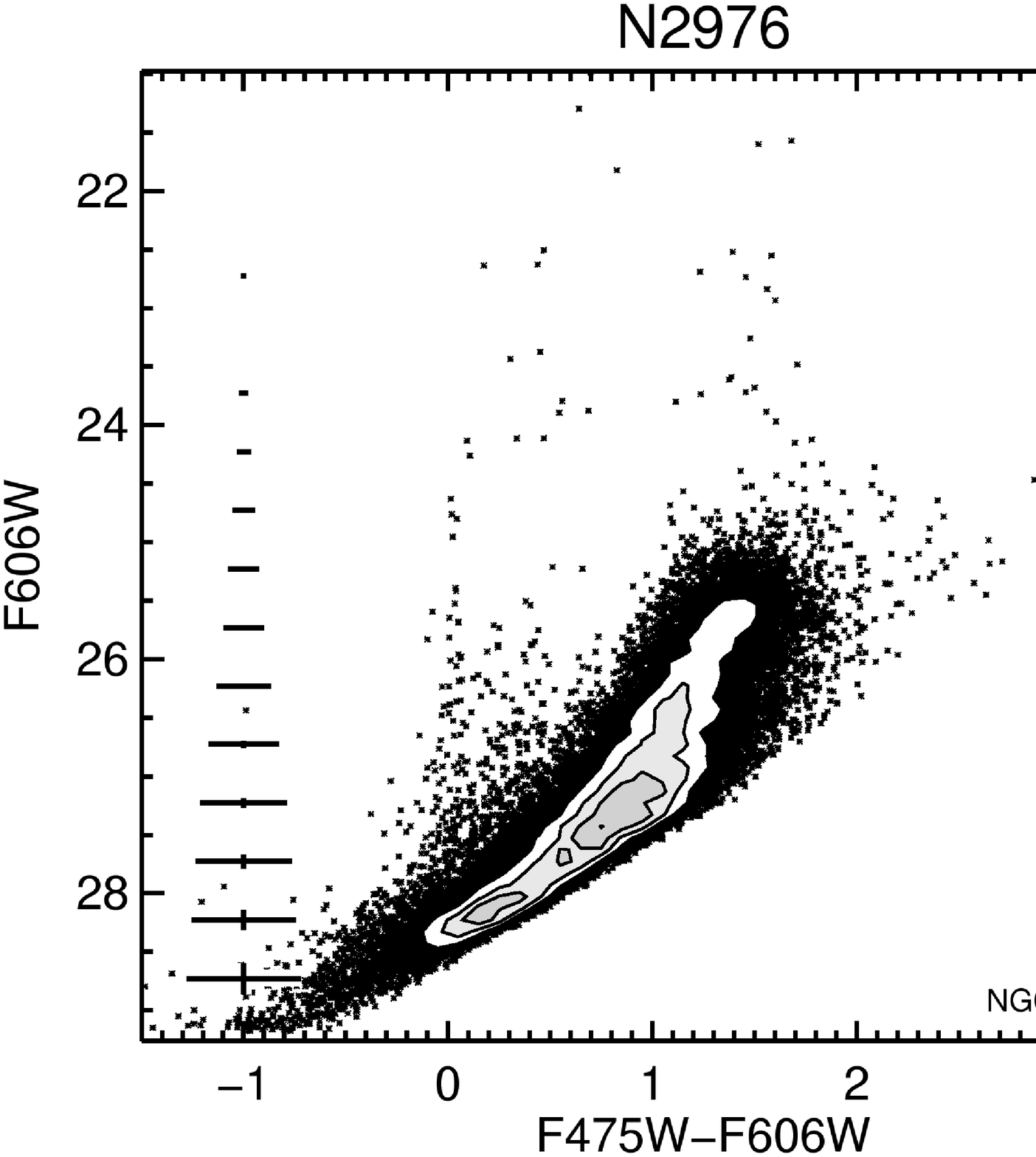}
\includegraphics[width=1.625in]{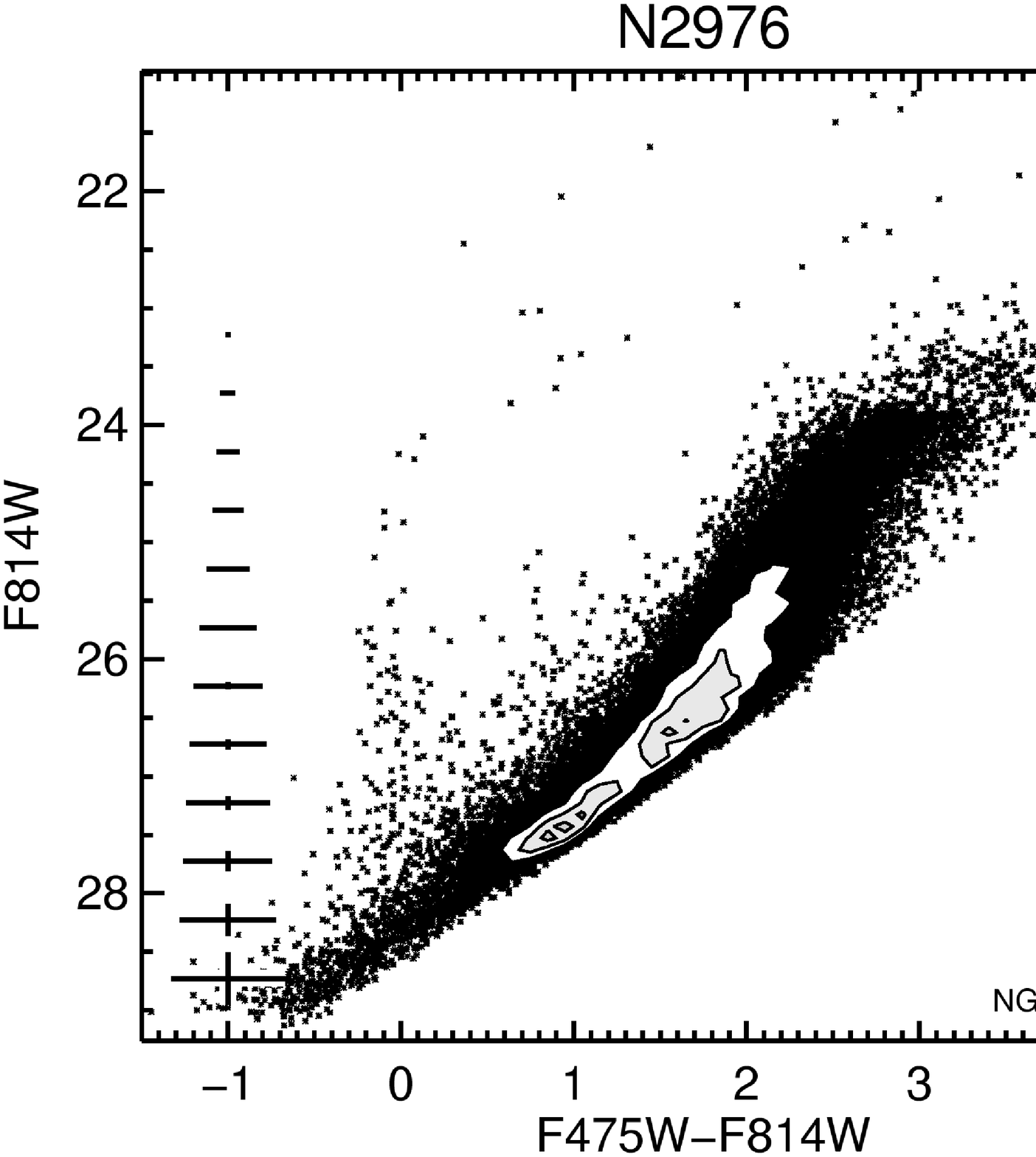}
\includegraphics[width=1.625in]{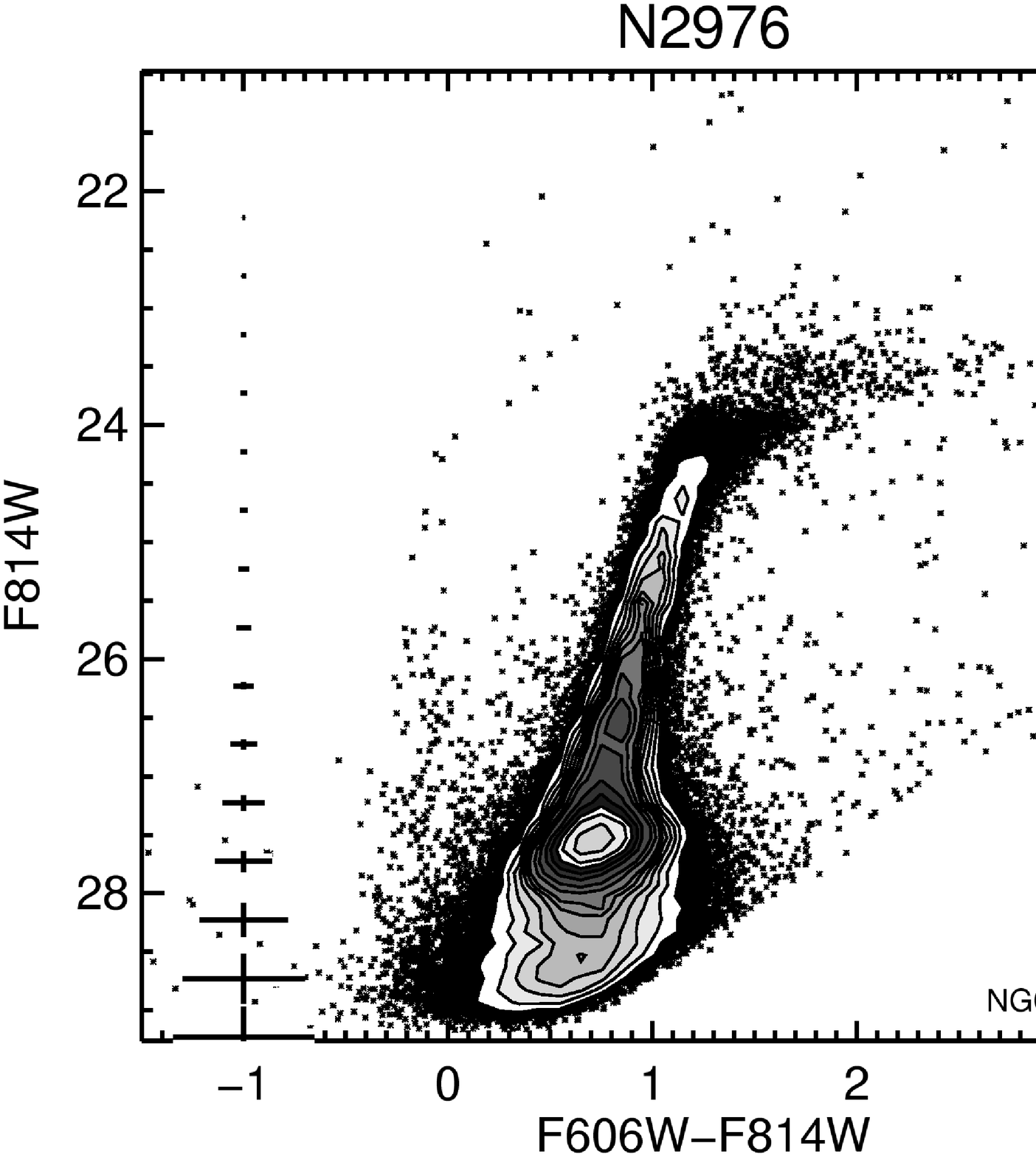}
\includegraphics[width=1.625in]{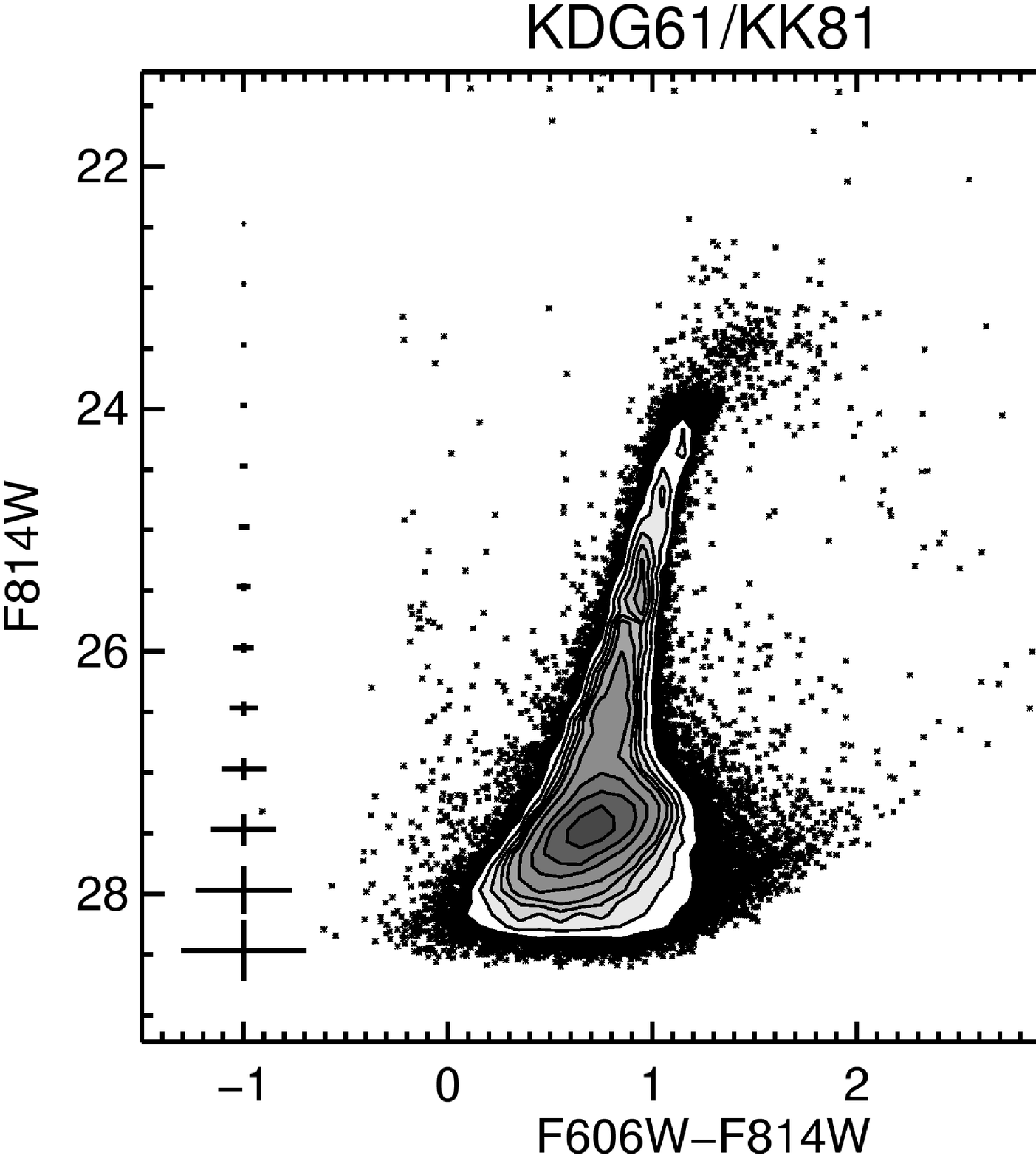}
}
\centerline{
\includegraphics[width=1.625in]{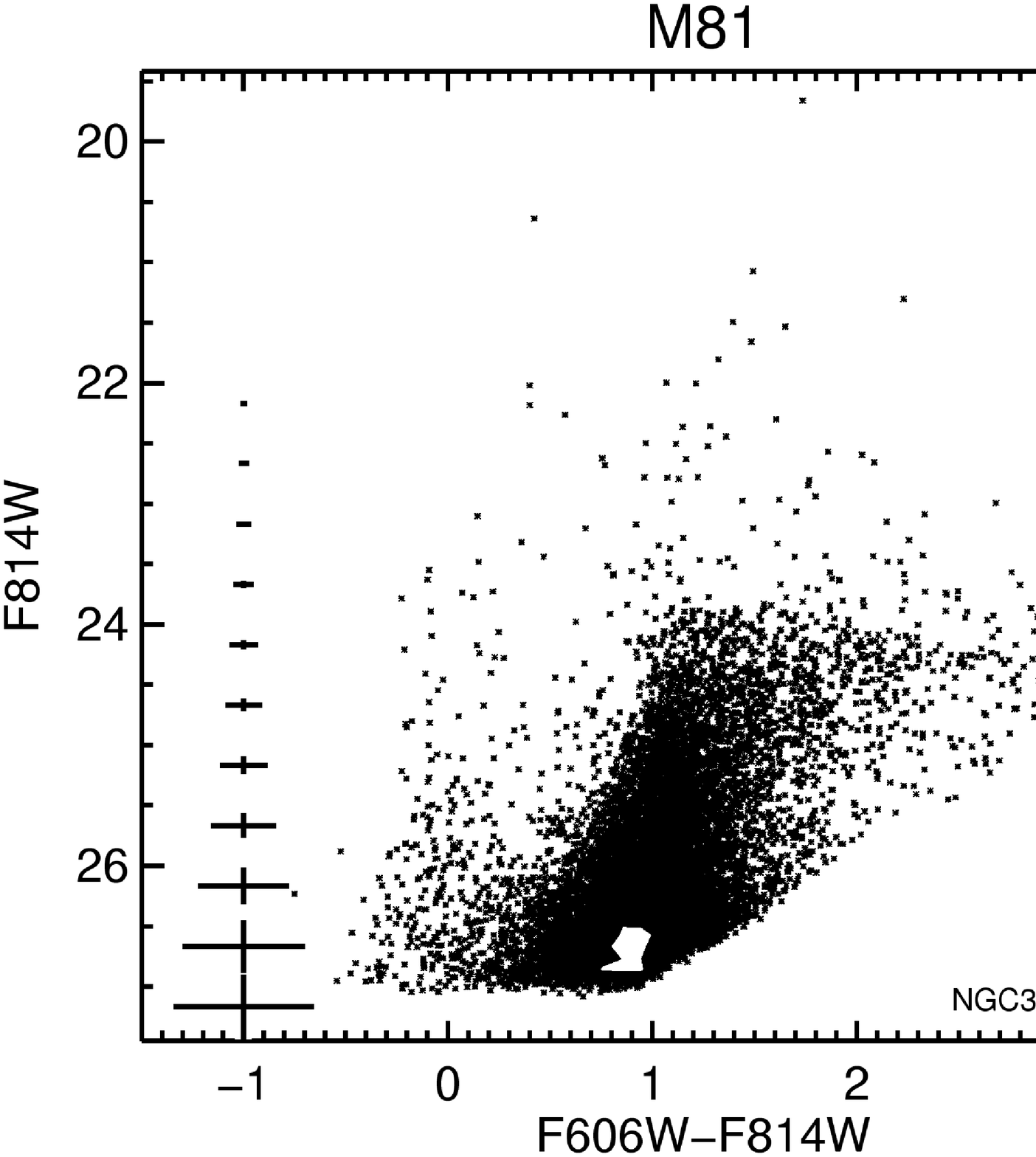}
\includegraphics[width=1.625in]{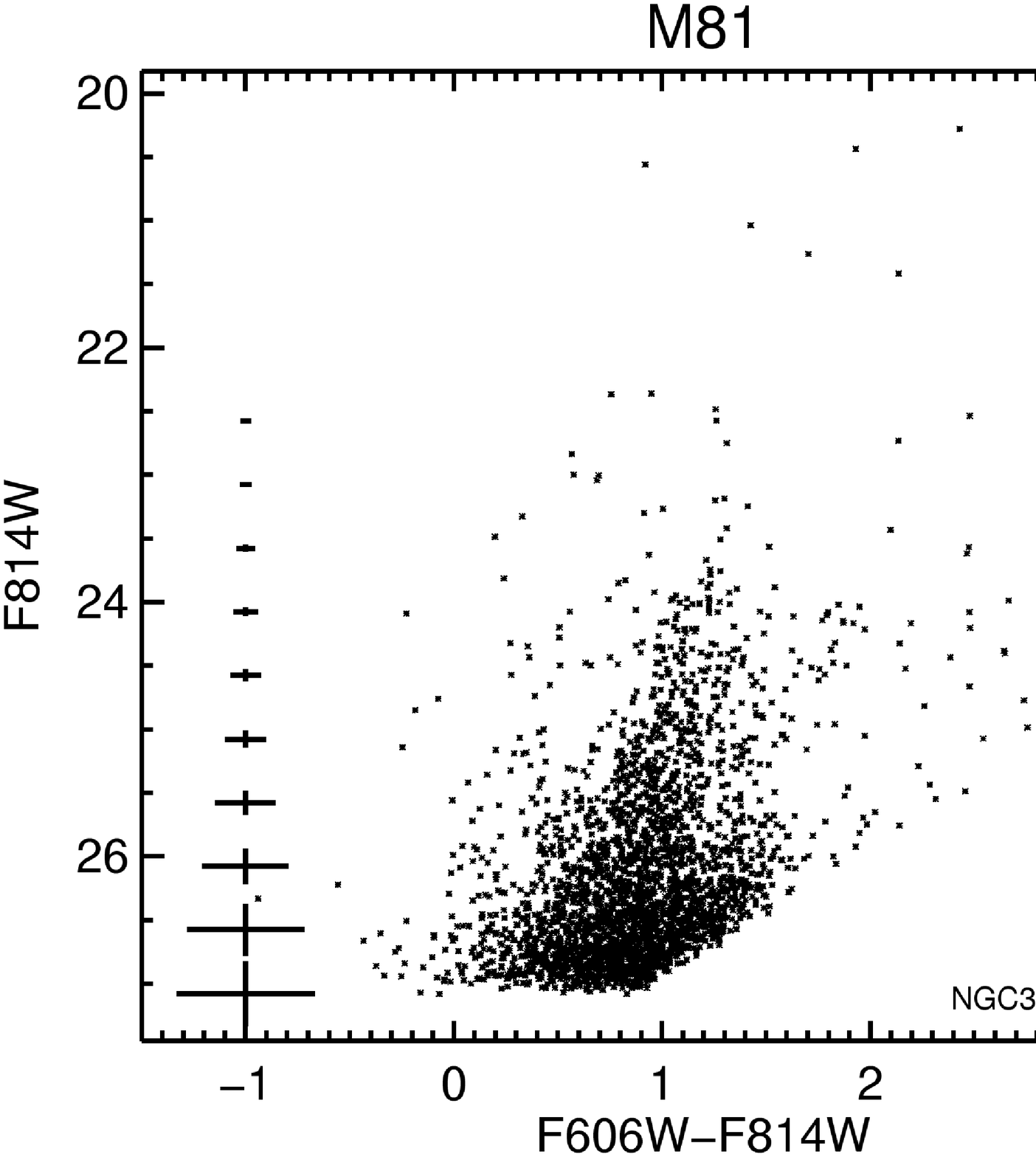}
\includegraphics[width=1.625in]{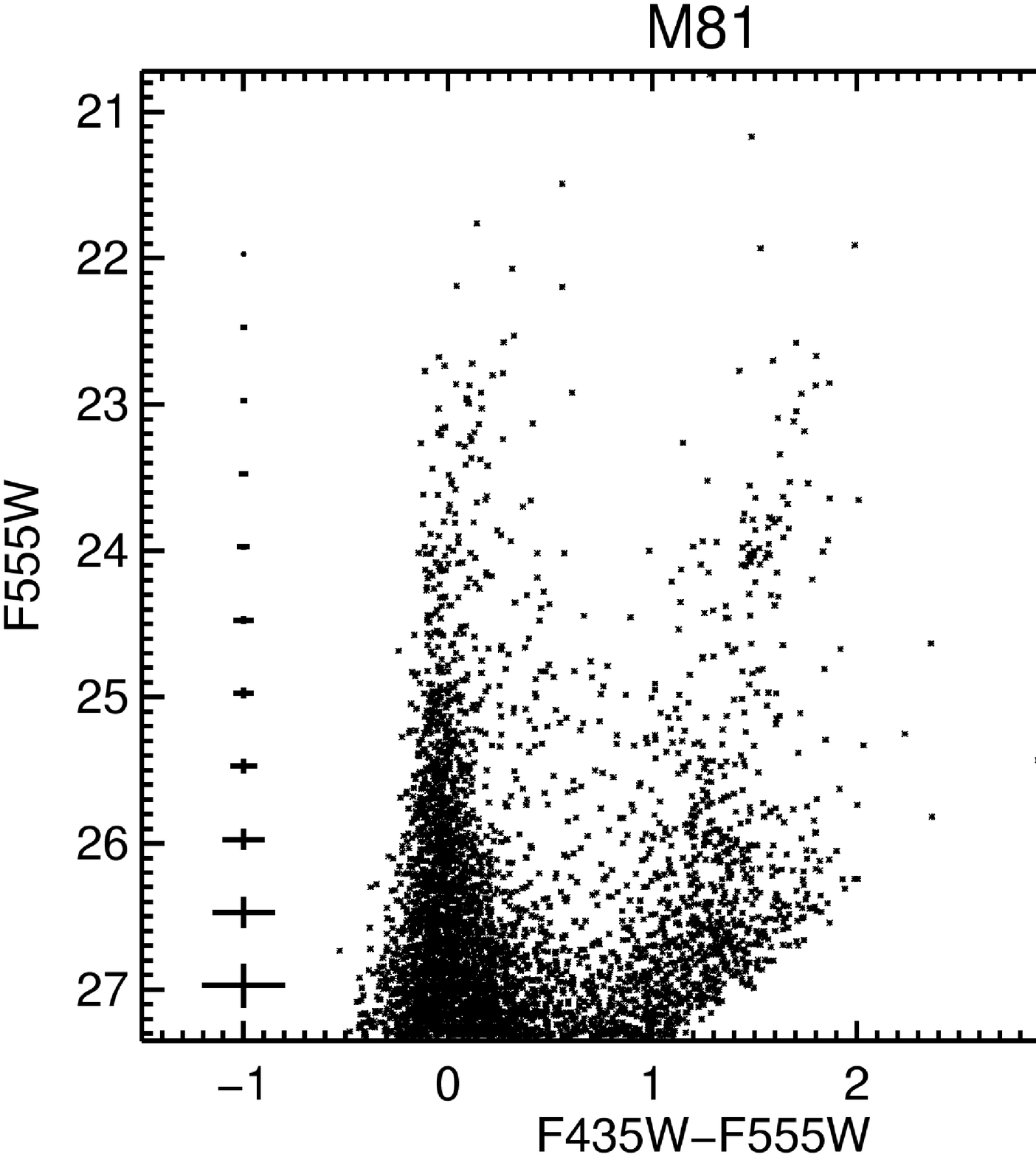}
\includegraphics[width=1.625in]{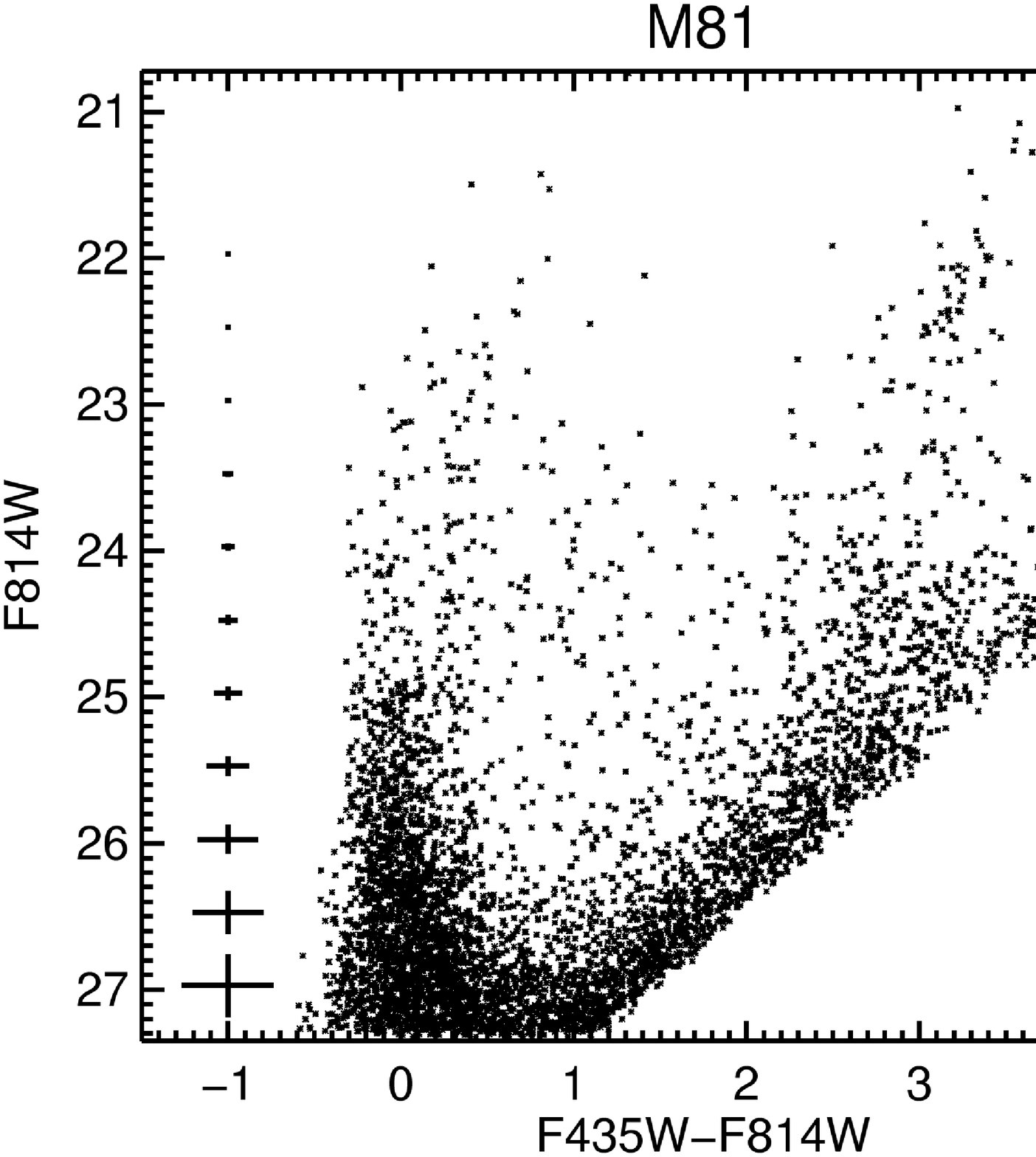}
}
\centerline{
\includegraphics[width=1.625in]{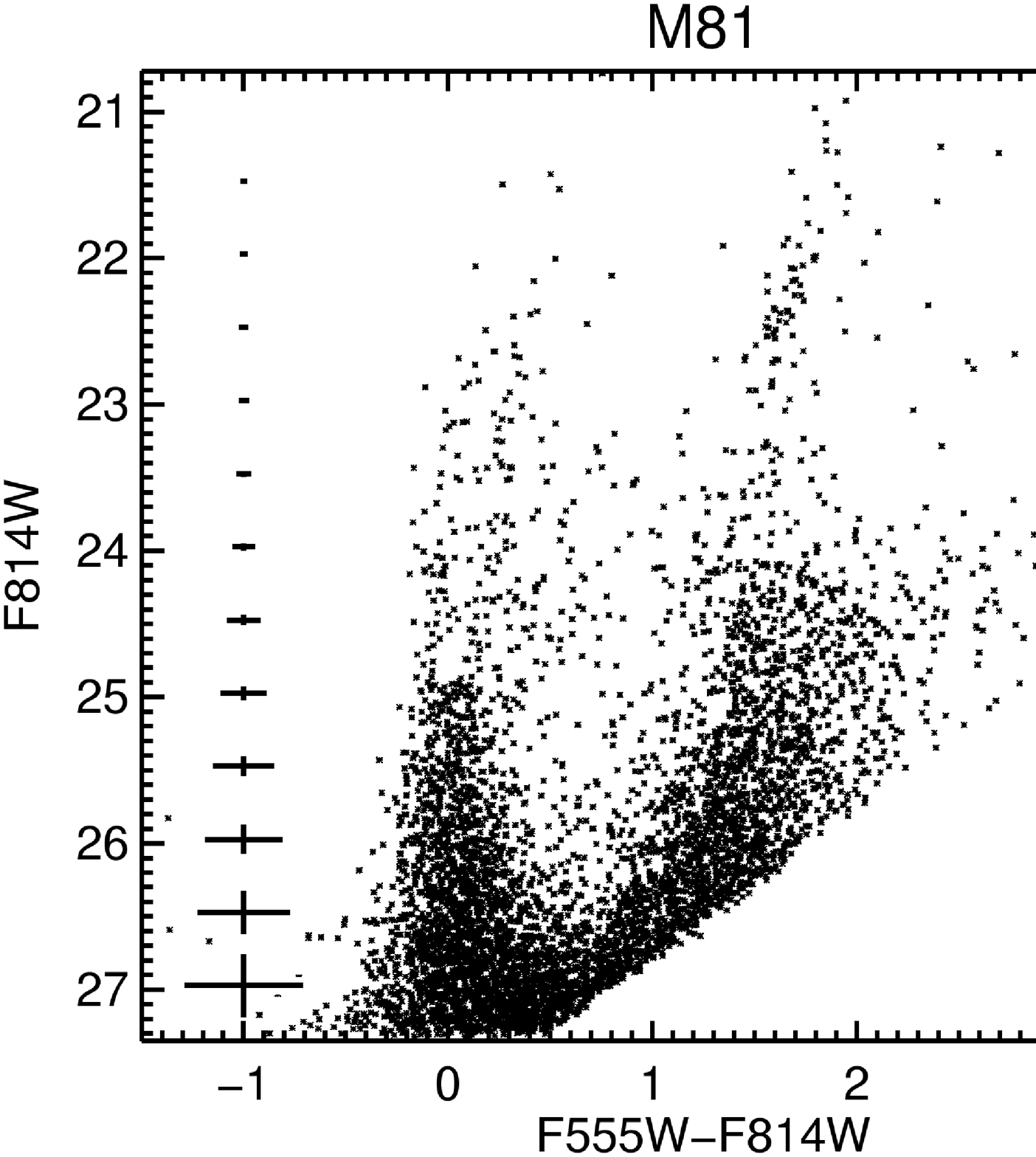}
\includegraphics[width=1.625in]{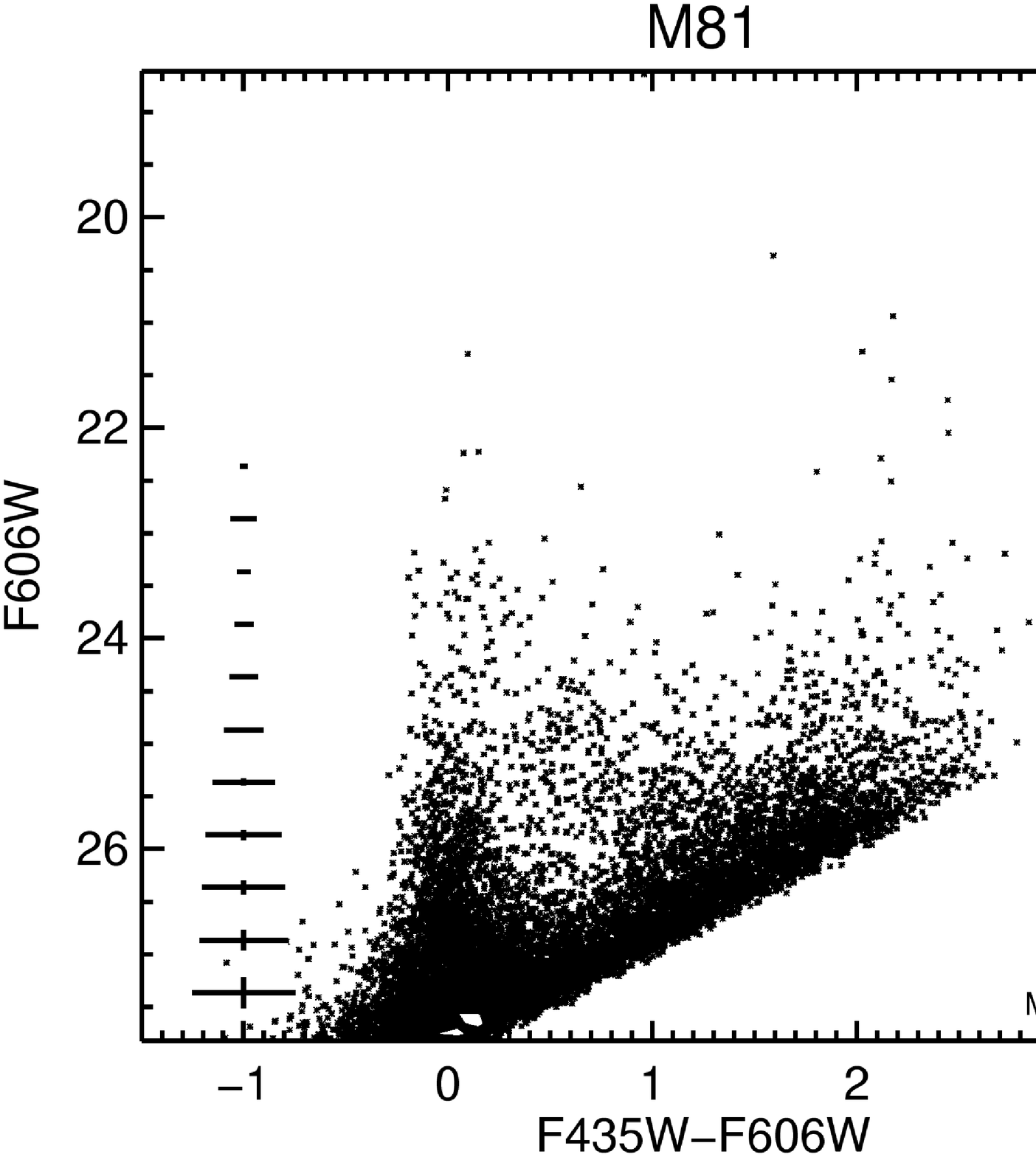}
\includegraphics[width=1.625in]{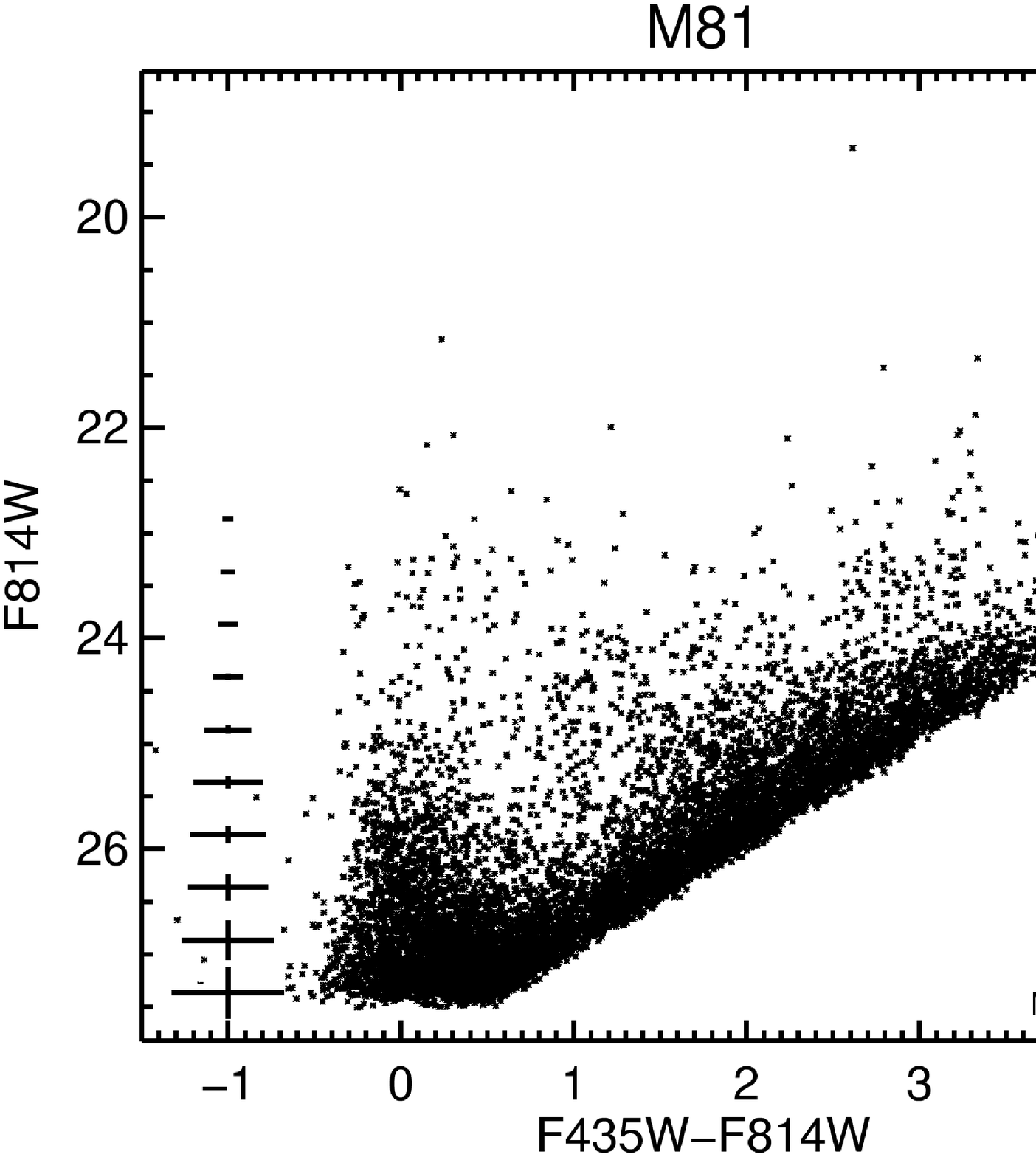}
\includegraphics[width=1.625in]{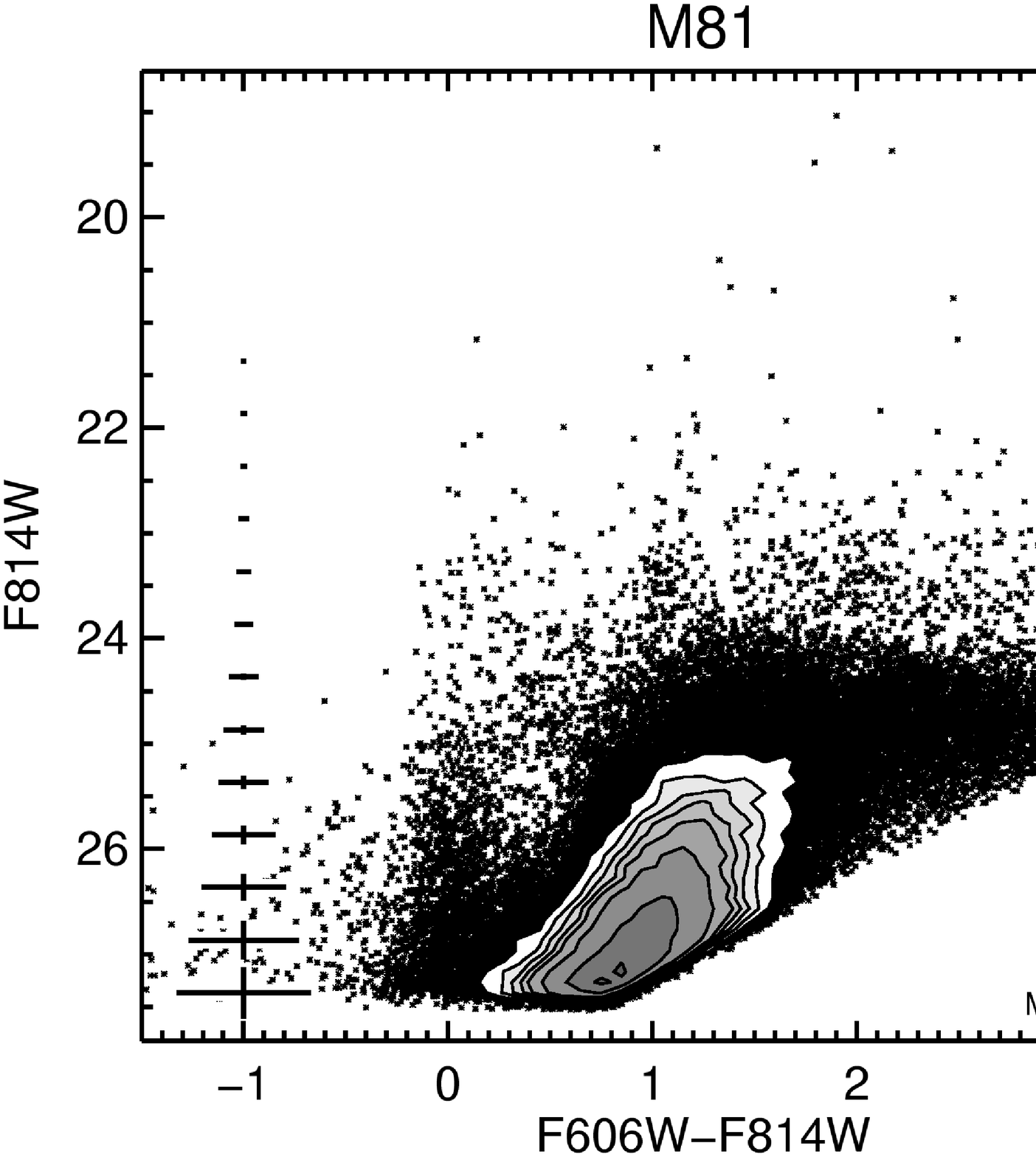}
}
\caption{
CMDs of galaxies in the ANGST data release,
as described in Figure~\ref{cmdfig1}.
Figures are ordered from the upper left to the bottom right.
(a) DDO53; (b) N2976; (c) N2976; (d) N2976; (e) N2976; (f) N2976; (g) N2976; (h) KDG61; (i) M81; (j) M81; (k) M81; (l) M81; (m) M81; (n) M81; (o) M81; (p) M81; 
    \label{cmdfig8}}
\end{figure}
\vfill
\clearpage
 
%-------------------
\begin{figure}[p]
\centerline{
\includegraphics[width=1.625in]{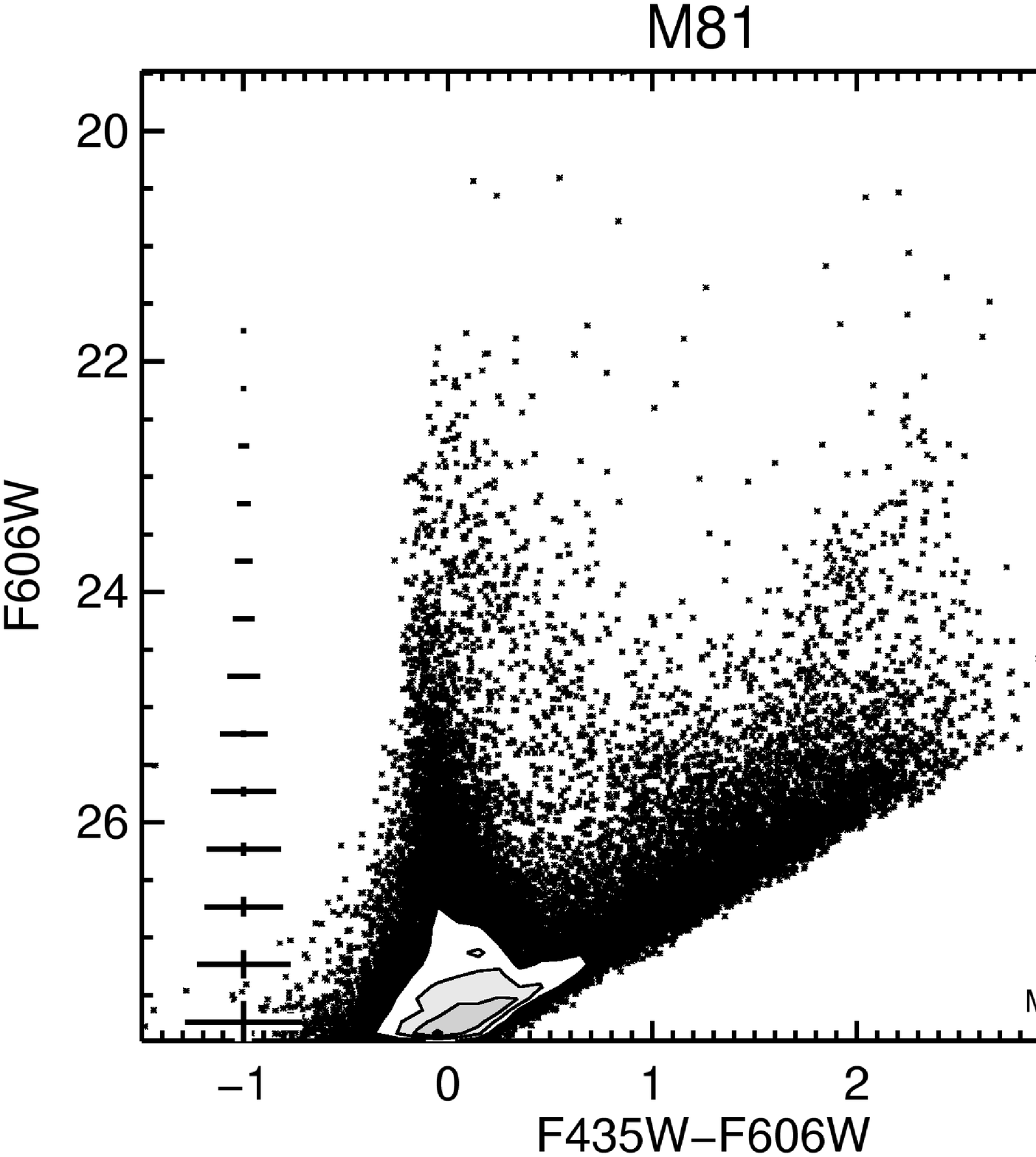}
\includegraphics[width=1.625in]{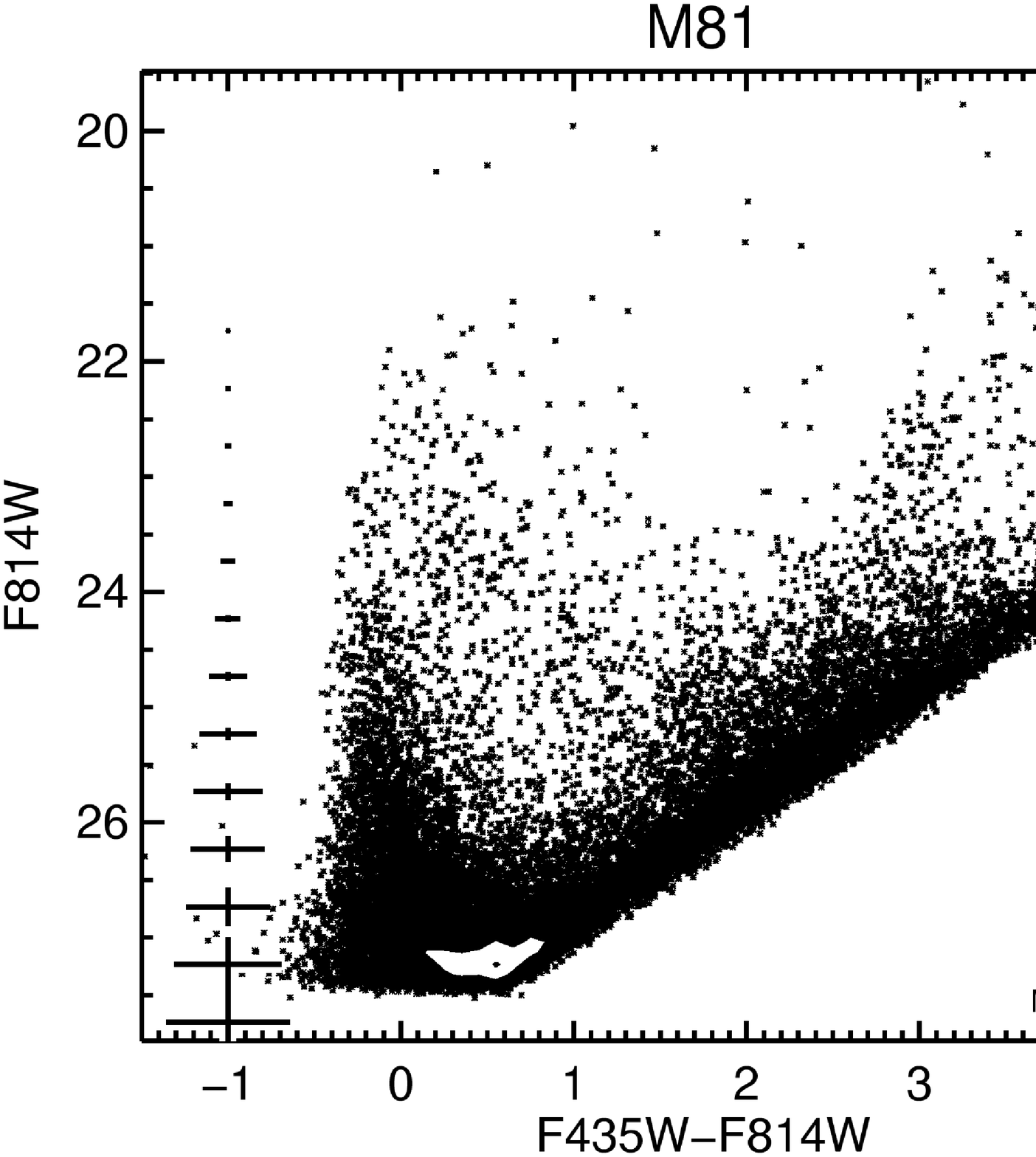}
\includegraphics[width=1.625in]{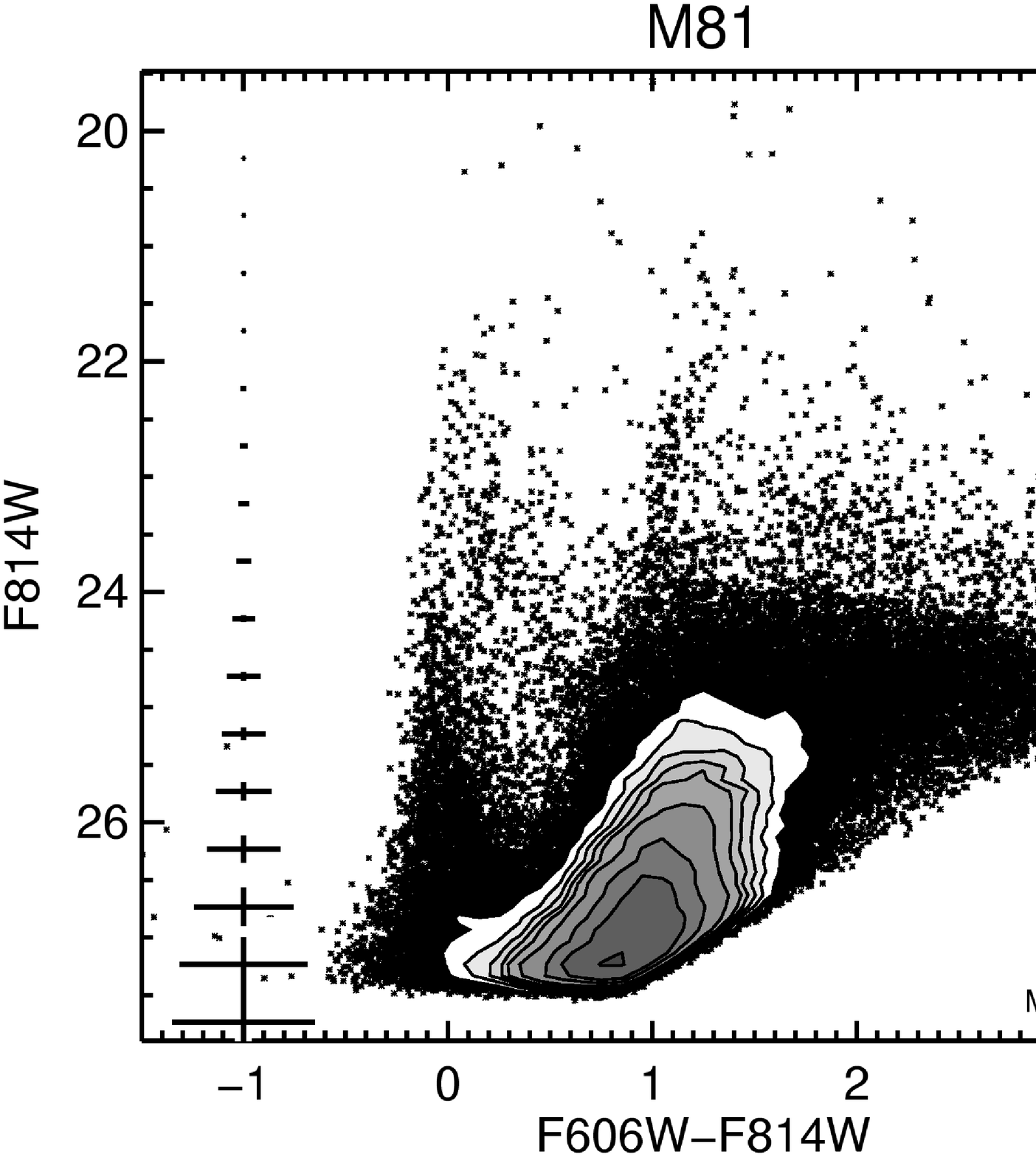}
\includegraphics[width=1.625in]{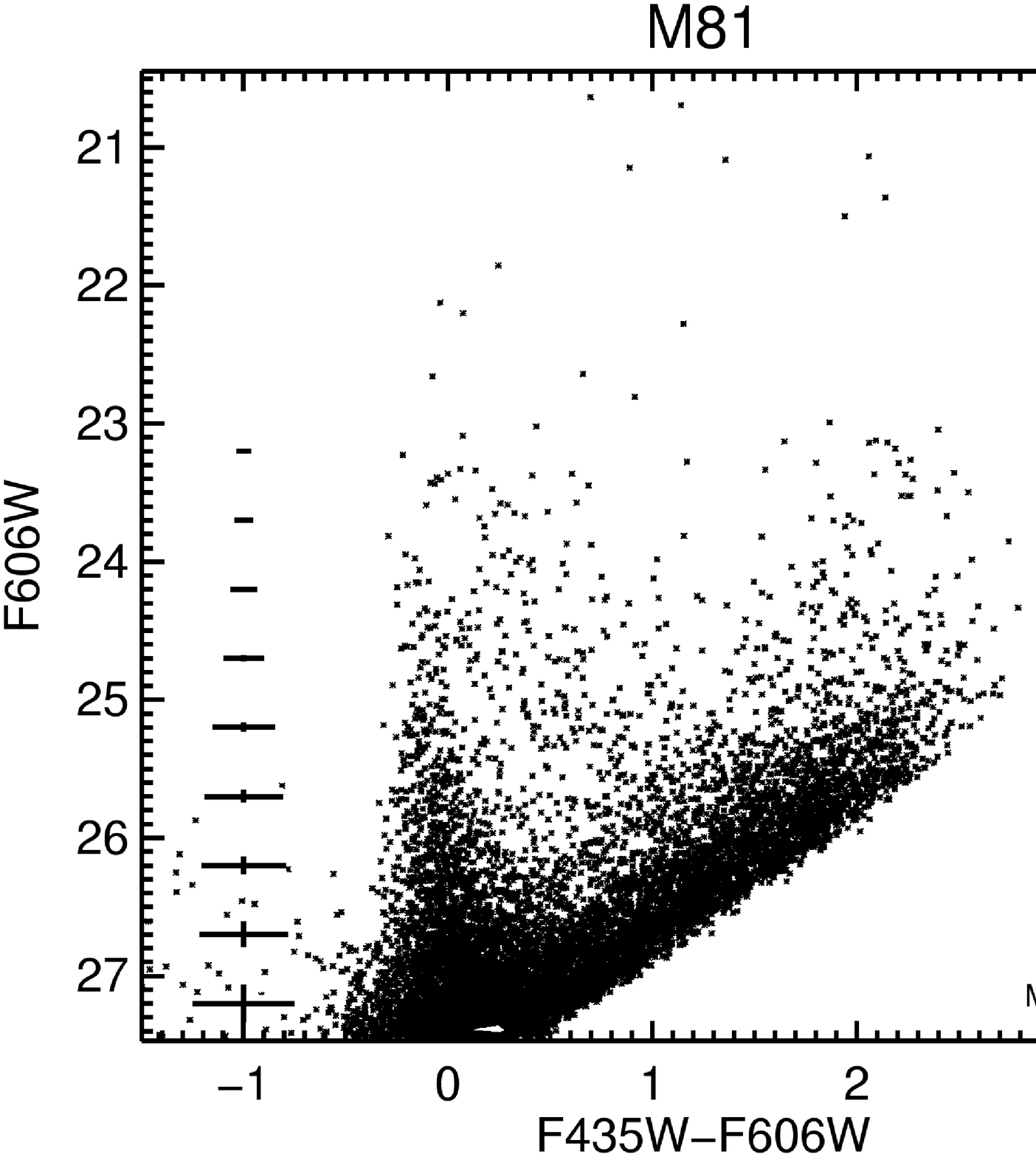}
}
\centerline{
\includegraphics[width=1.625in]{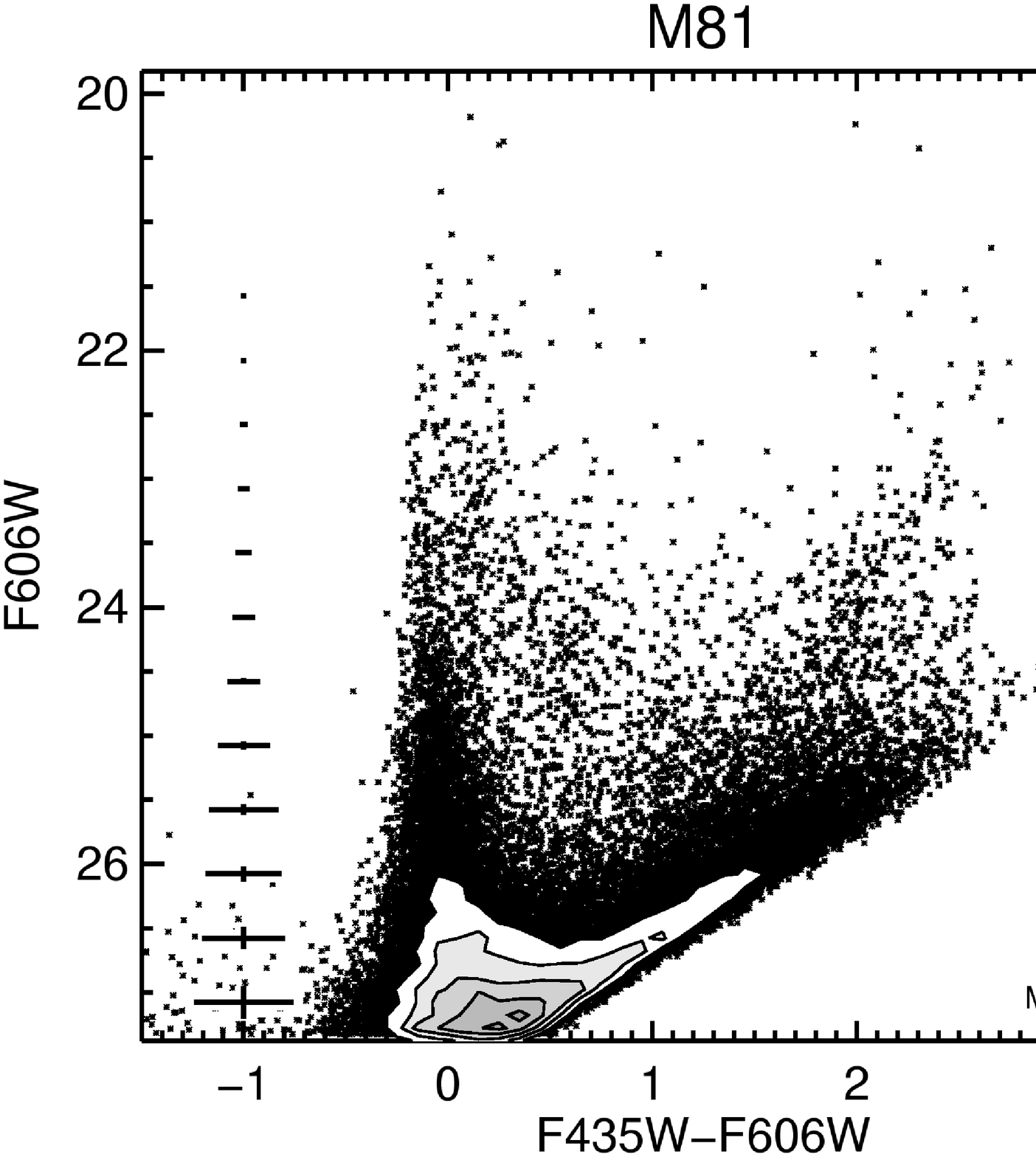}
\includegraphics[width=1.625in]{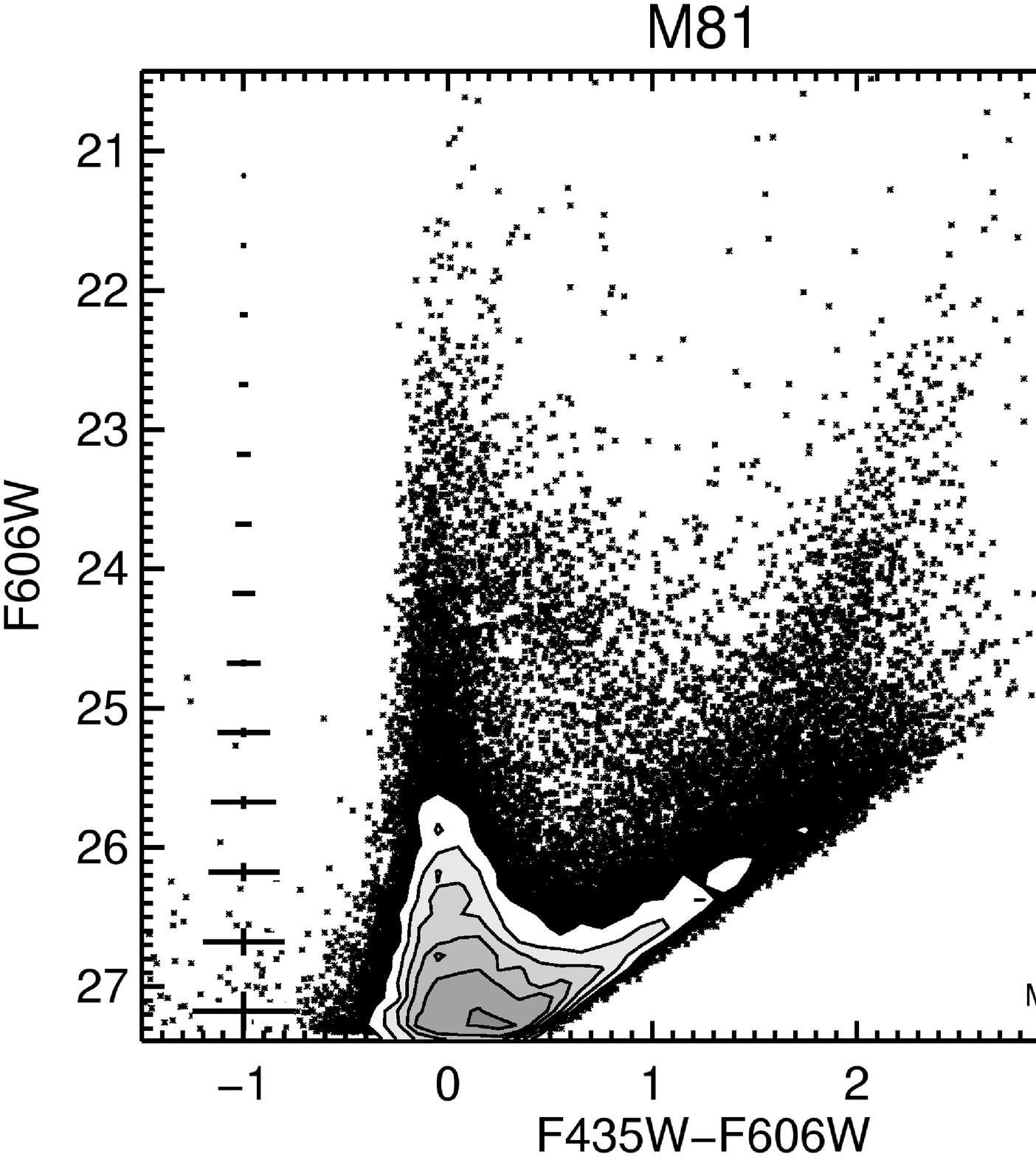}
\includegraphics[width=1.625in]{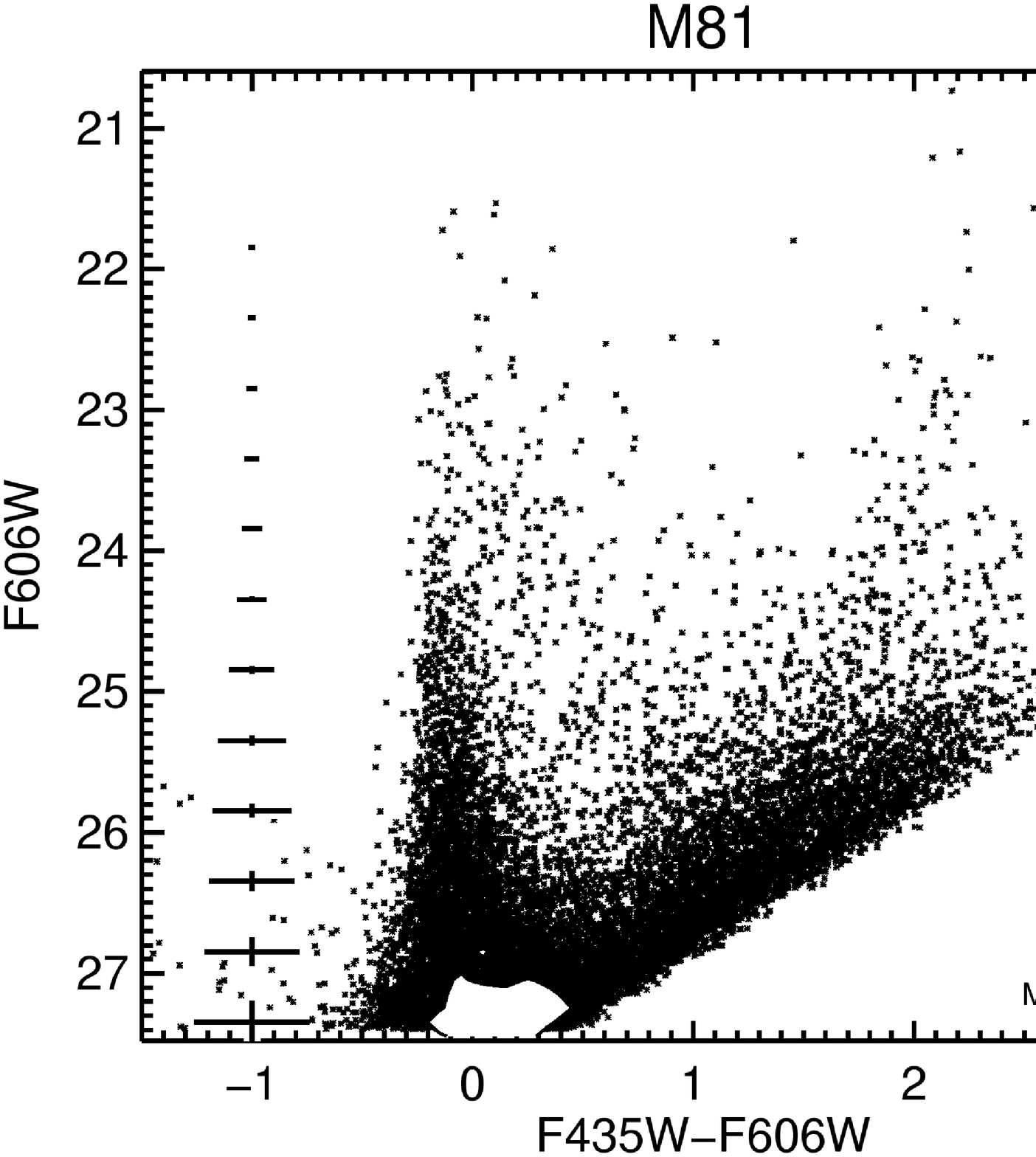}
\includegraphics[width=1.625in]{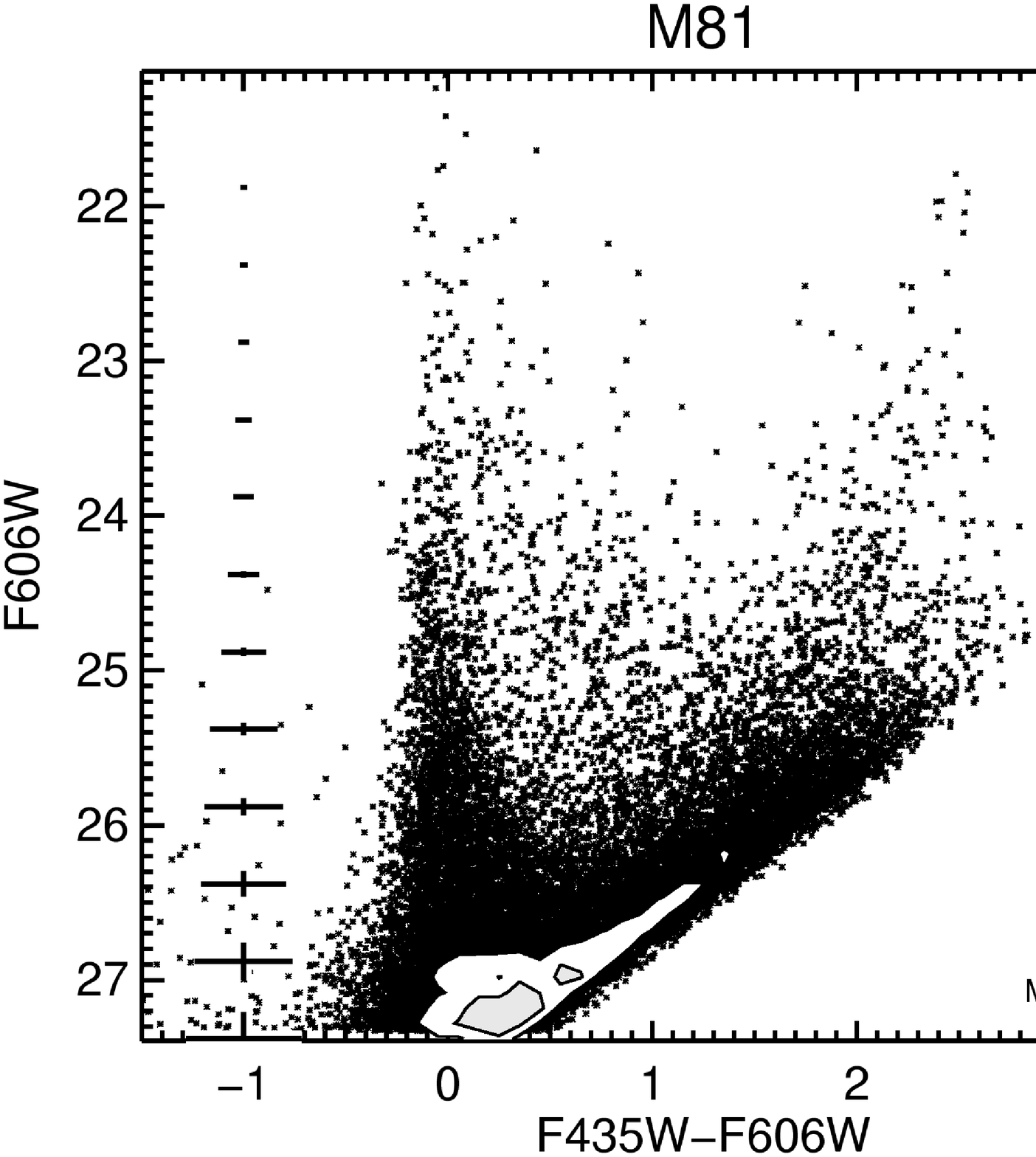}
}
\centerline{
\includegraphics[width=1.625in]{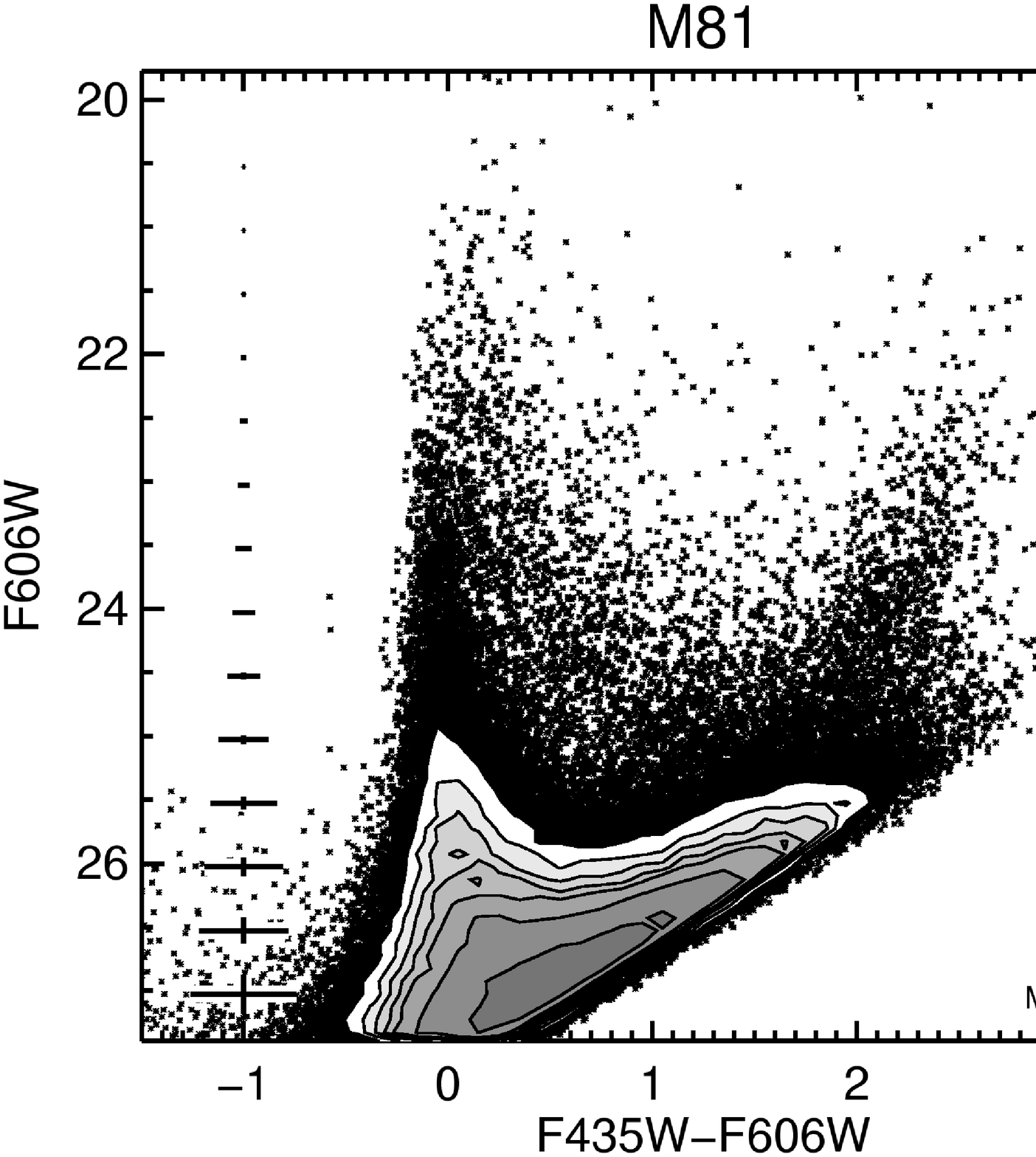}
\includegraphics[width=1.625in]{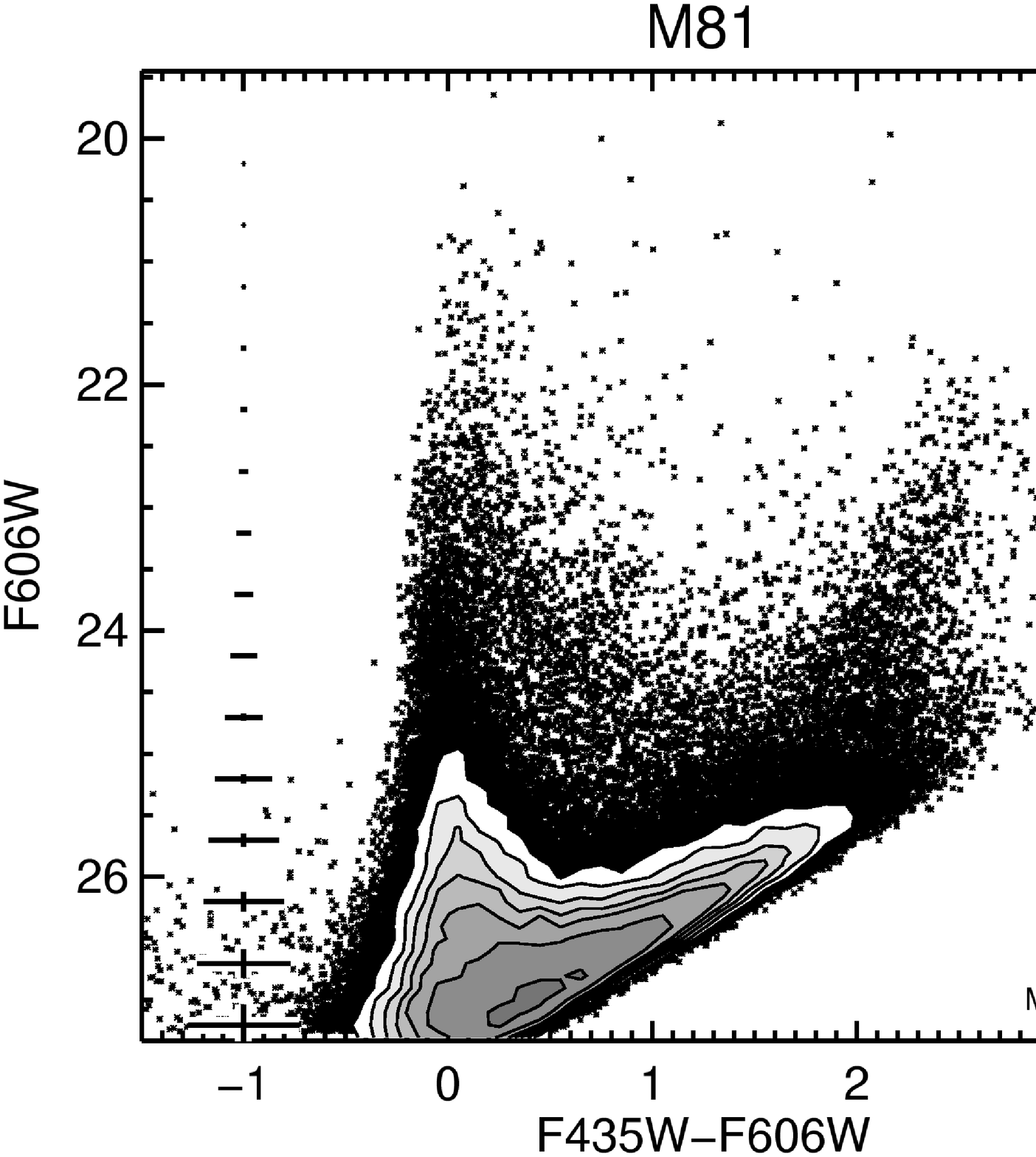}
\includegraphics[width=1.625in]{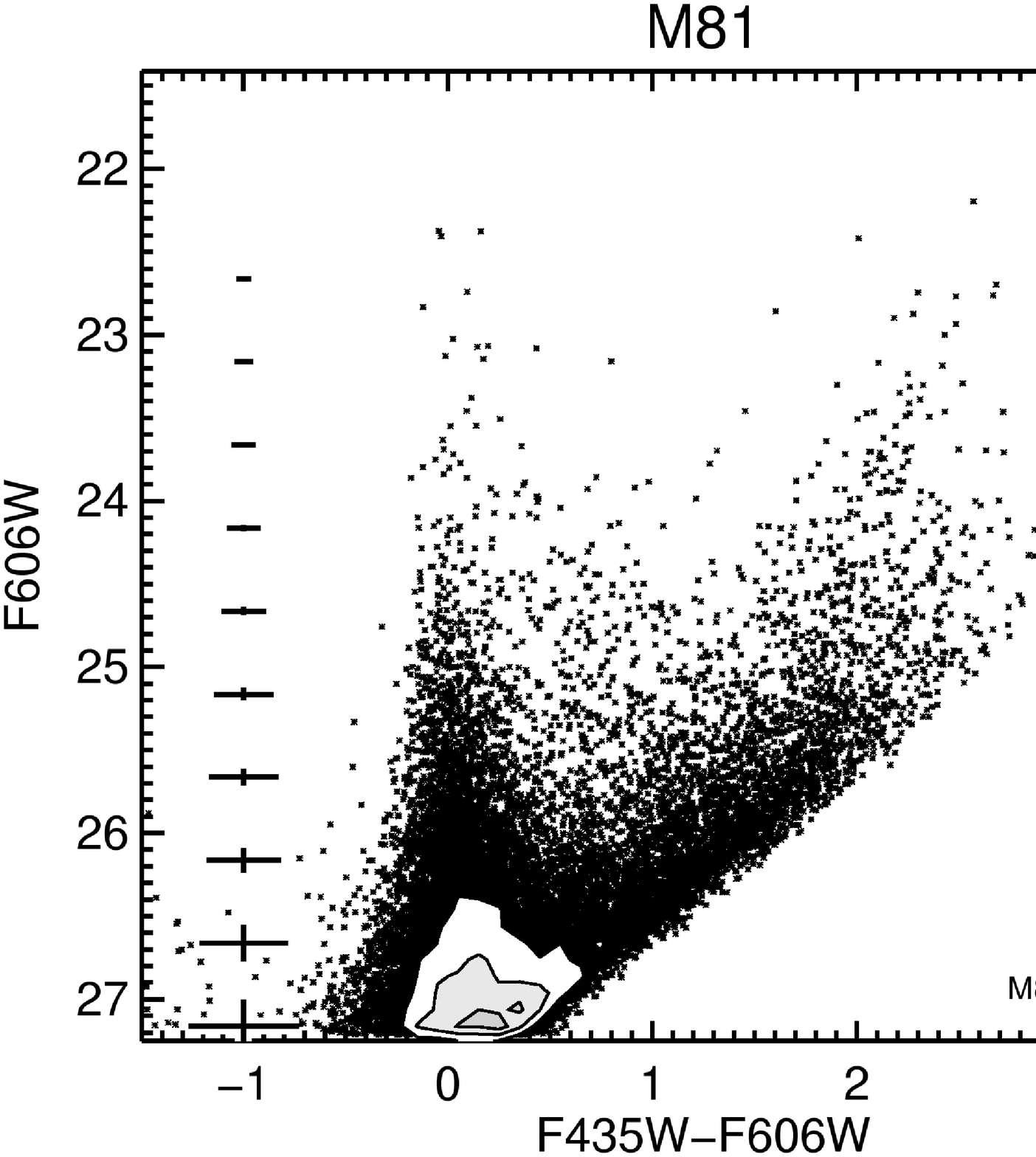}
\includegraphics[width=1.625in]{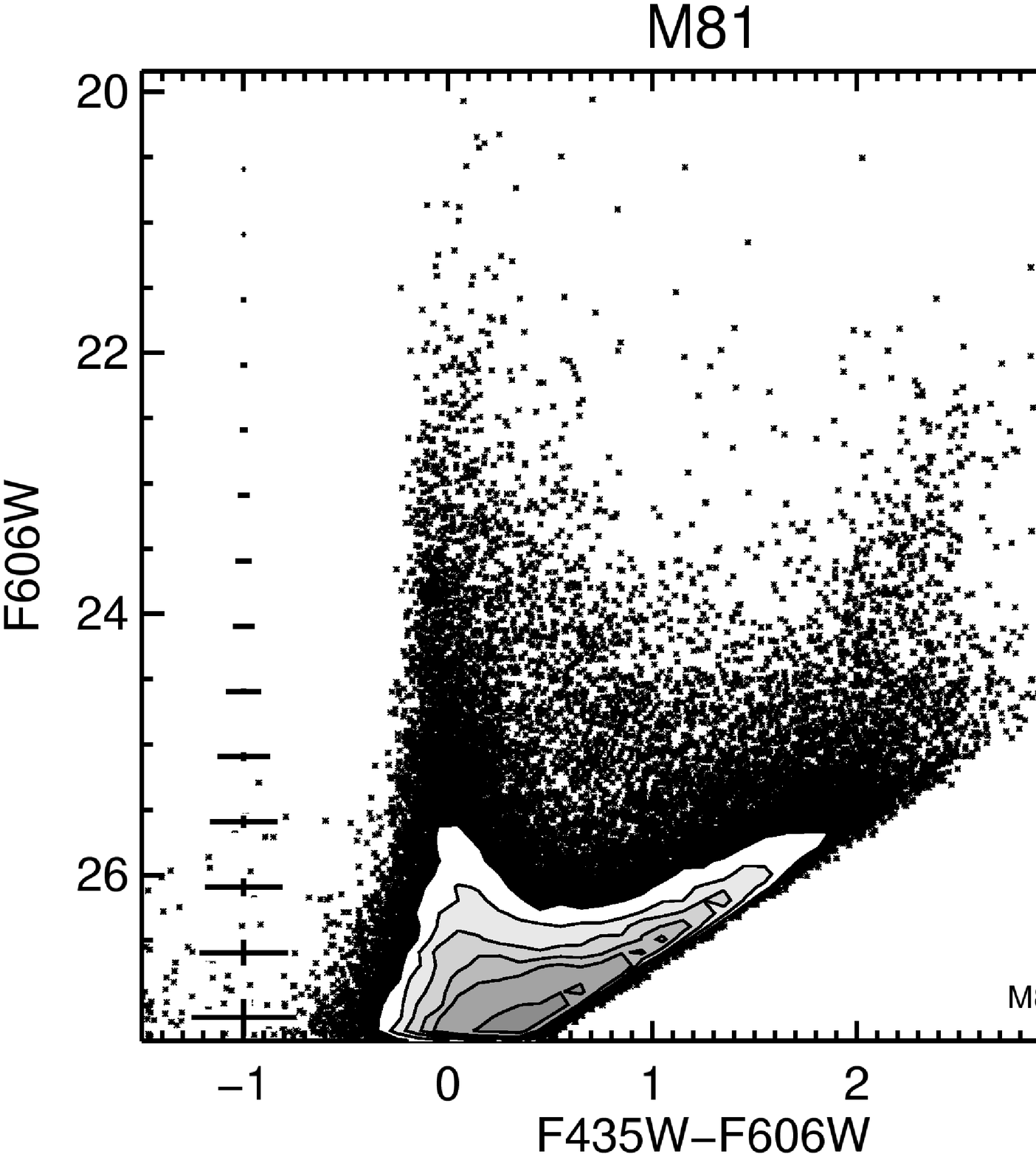}
}
\centerline{
\includegraphics[width=1.625in]{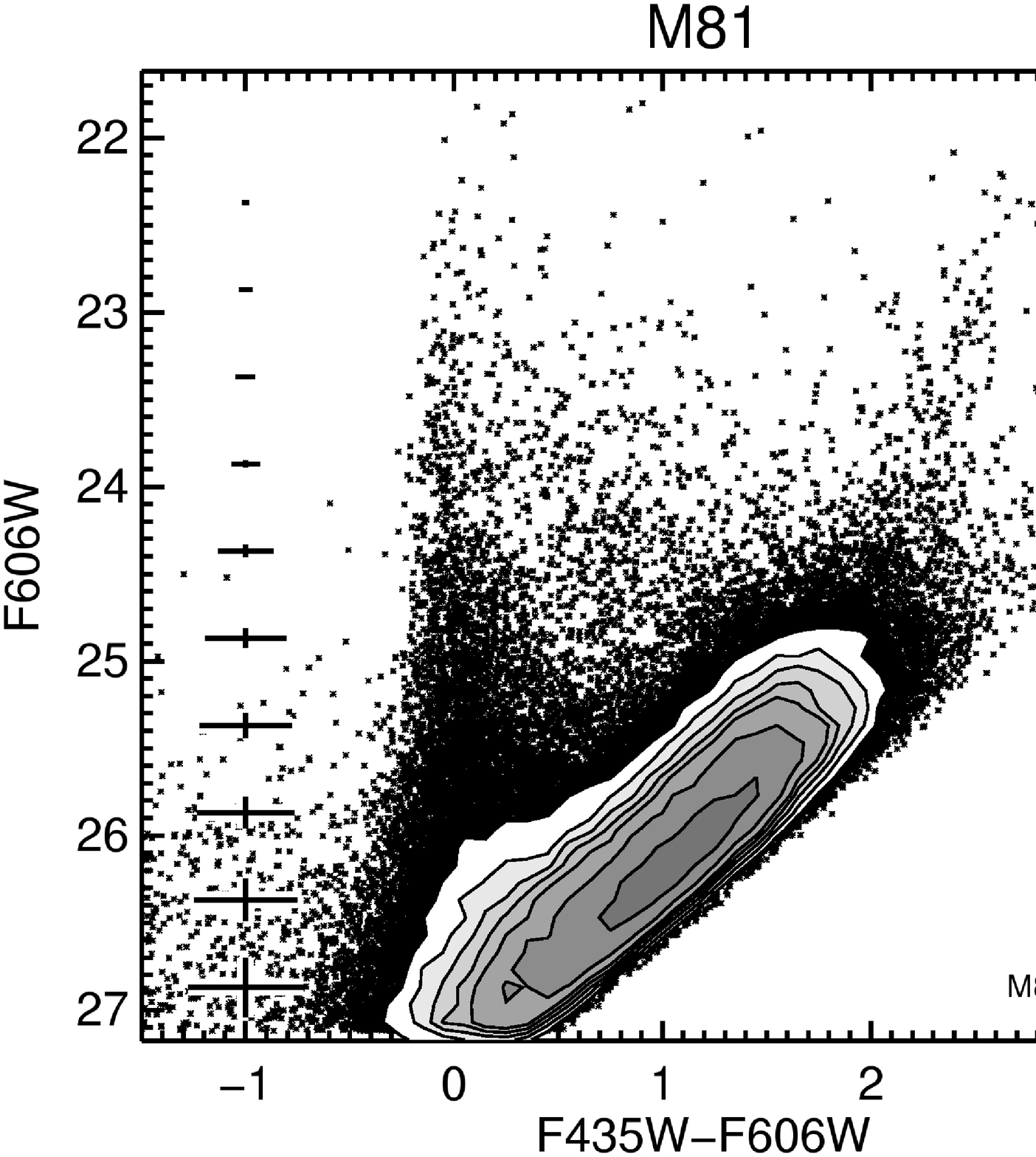}
\includegraphics[width=1.625in]{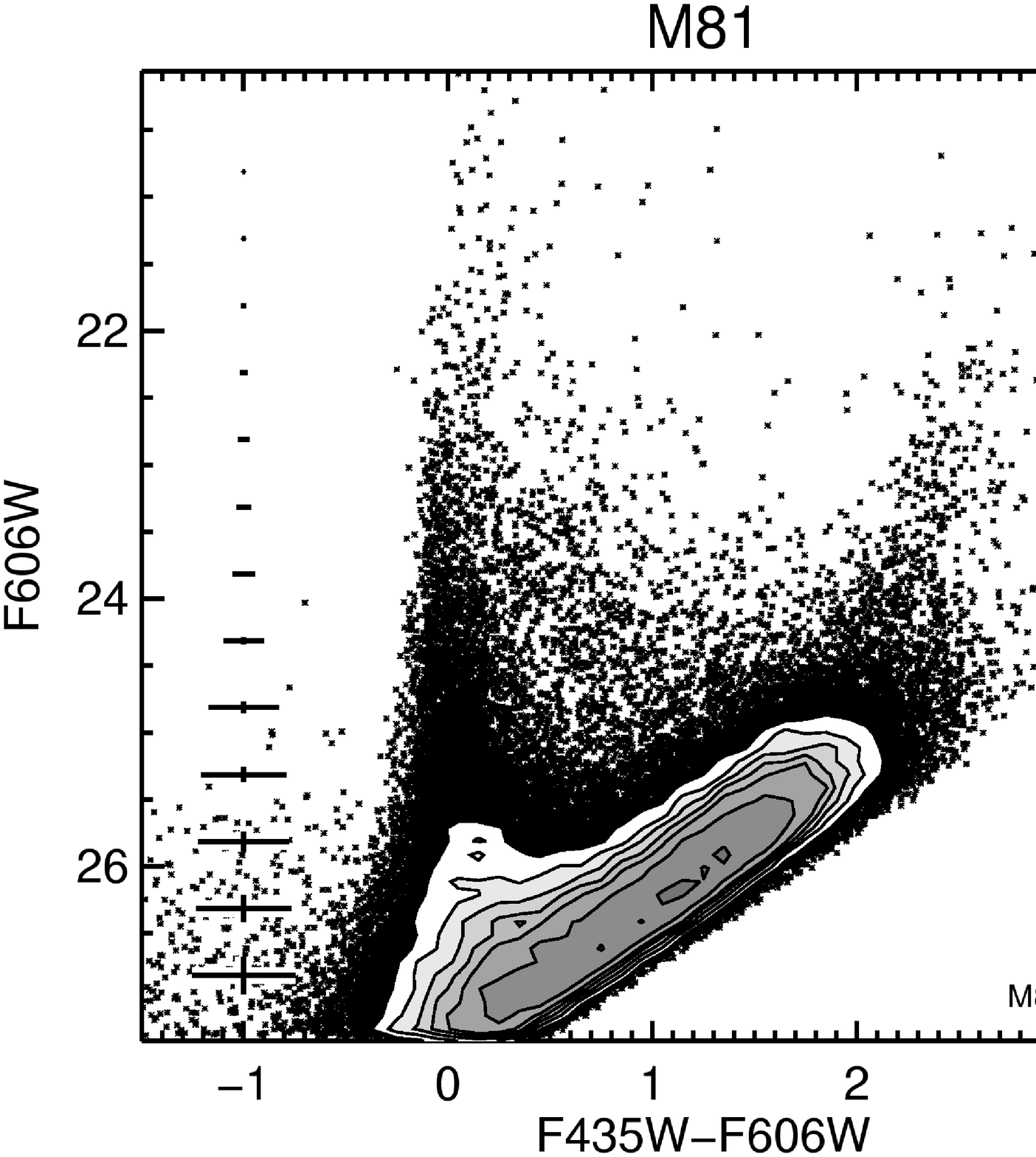}
\includegraphics[width=1.625in]{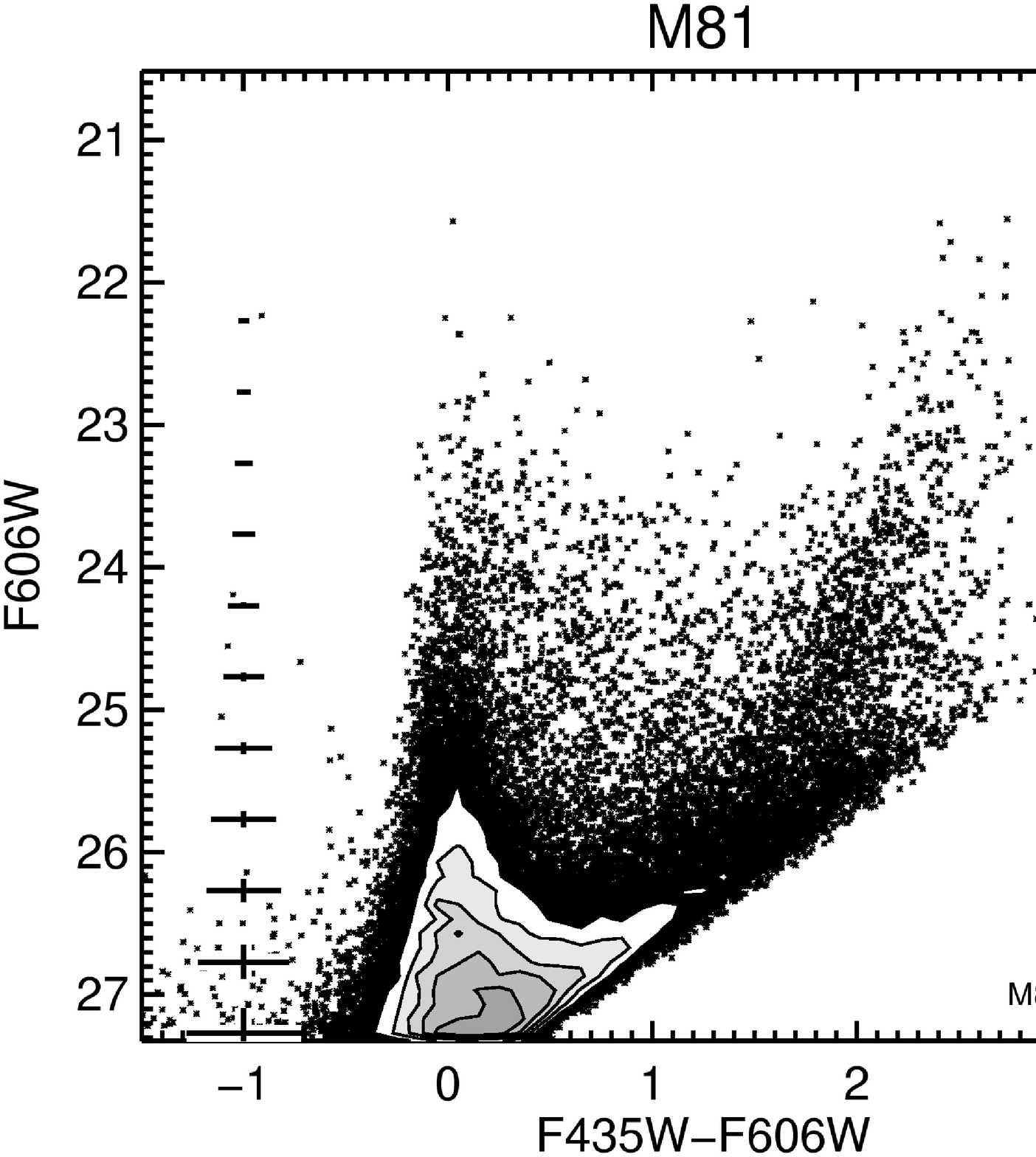}
\includegraphics[width=1.625in]{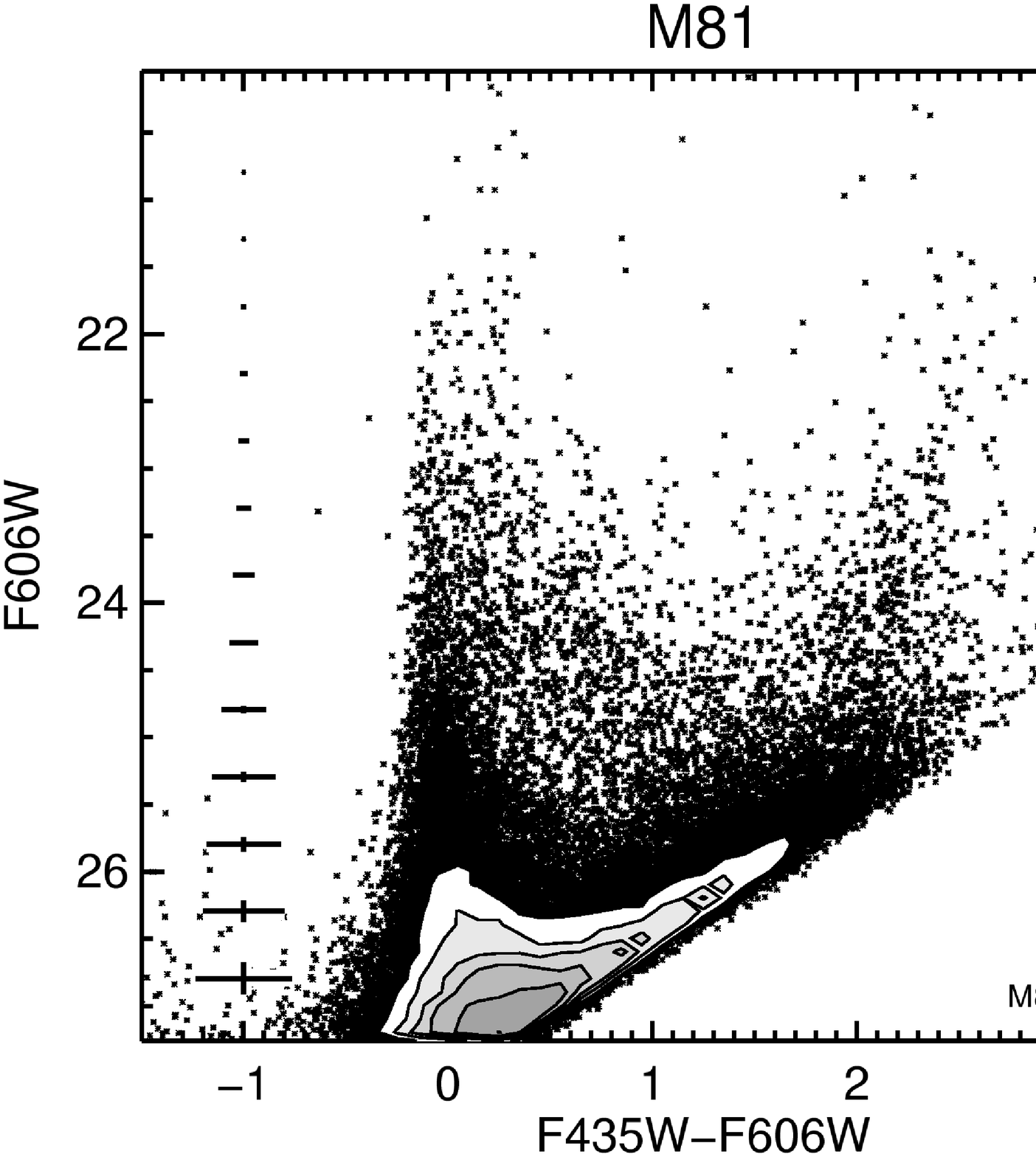}
}
\caption{
CMDs of galaxies in the ANGST data release,
as described in Figure~\ref{cmdfig1}.
Figures are ordered from the upper left to the bottom right.
(a) M81; (b) M81; (c) M81; (d) M81; (e) M81; (f) M81; (g) M81; (h) M81; (i) M81; (j) M81; (k) M81; (l) M81; (m) M81; (n) M81; (o) M81; (p) M81; 
    \label{cmdfig9}}
\end{figure}
\vfill
\clearpage
 
%-------------------
\begin{figure}[p]
\centerline{
\includegraphics[width=1.625in]{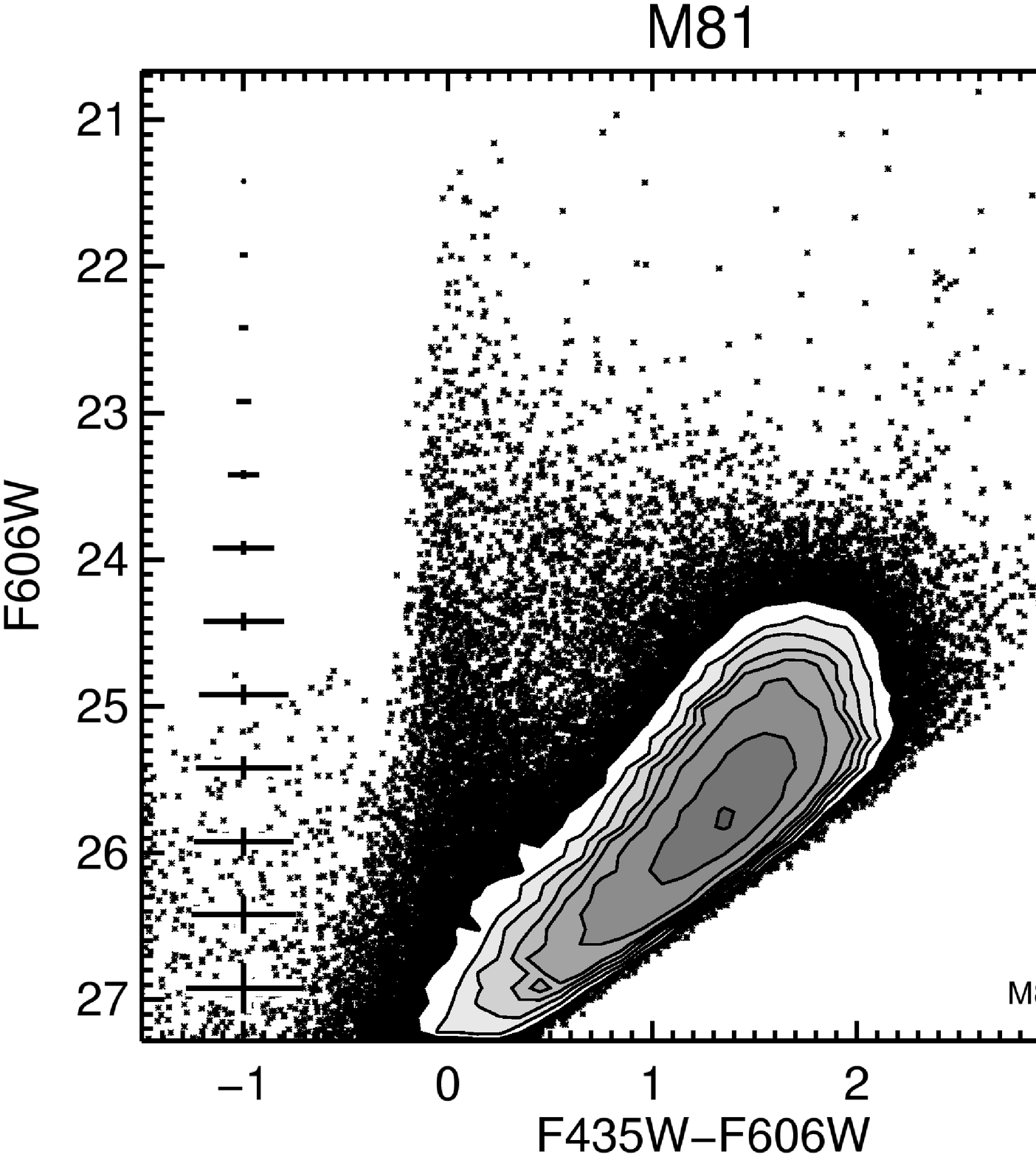}
\includegraphics[width=1.625in]{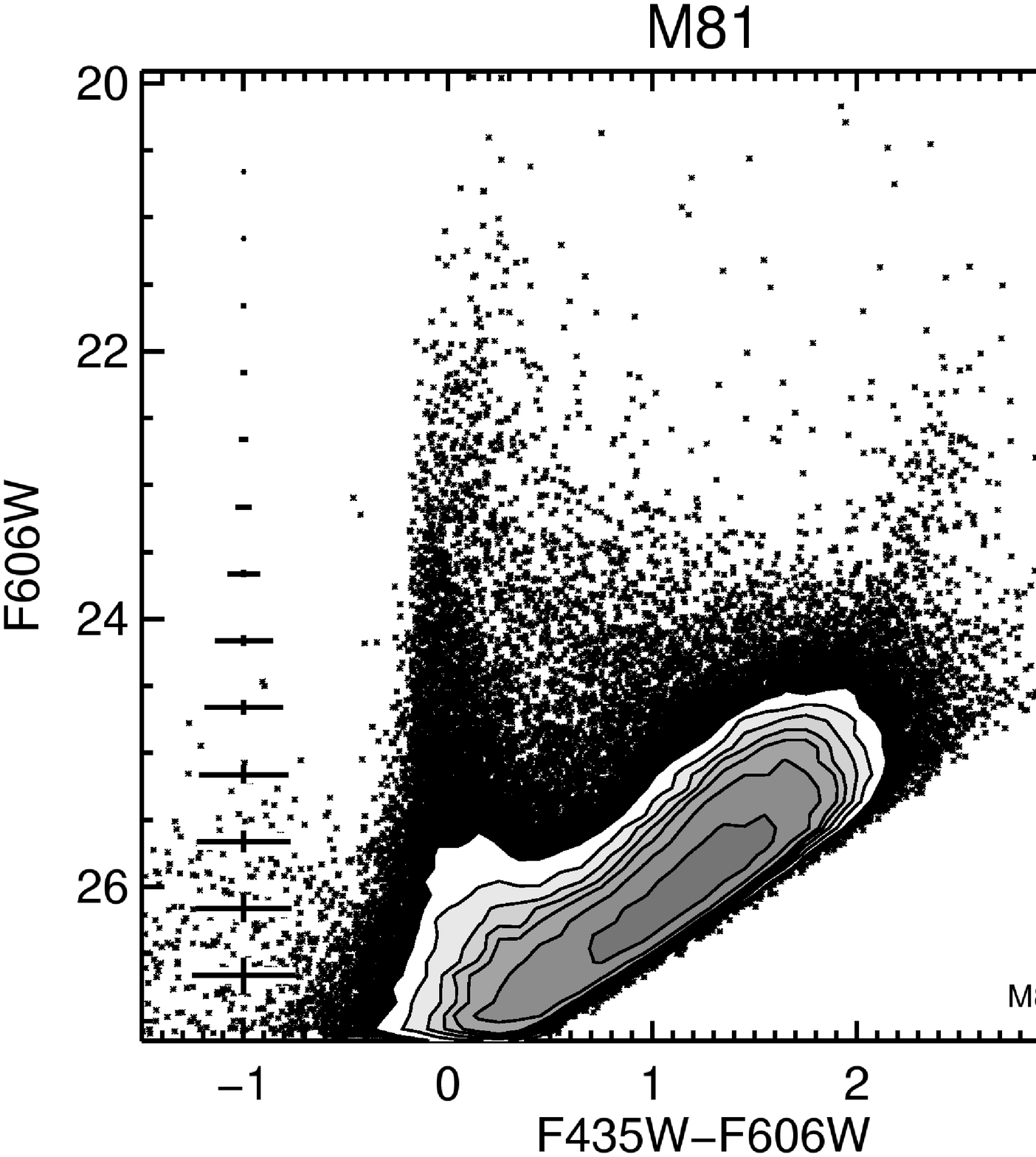}
\includegraphics[width=1.625in]{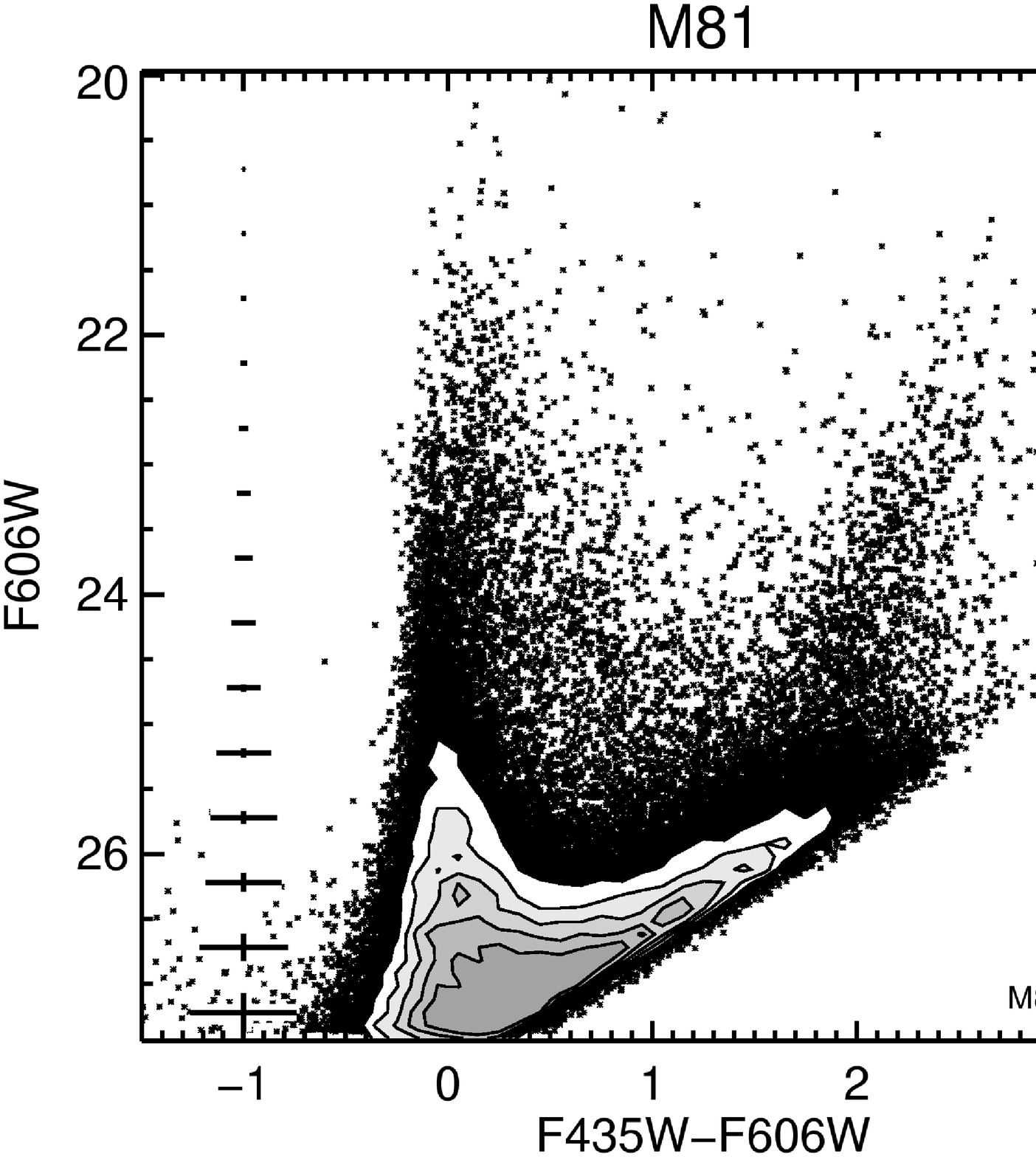}
\includegraphics[width=1.625in]{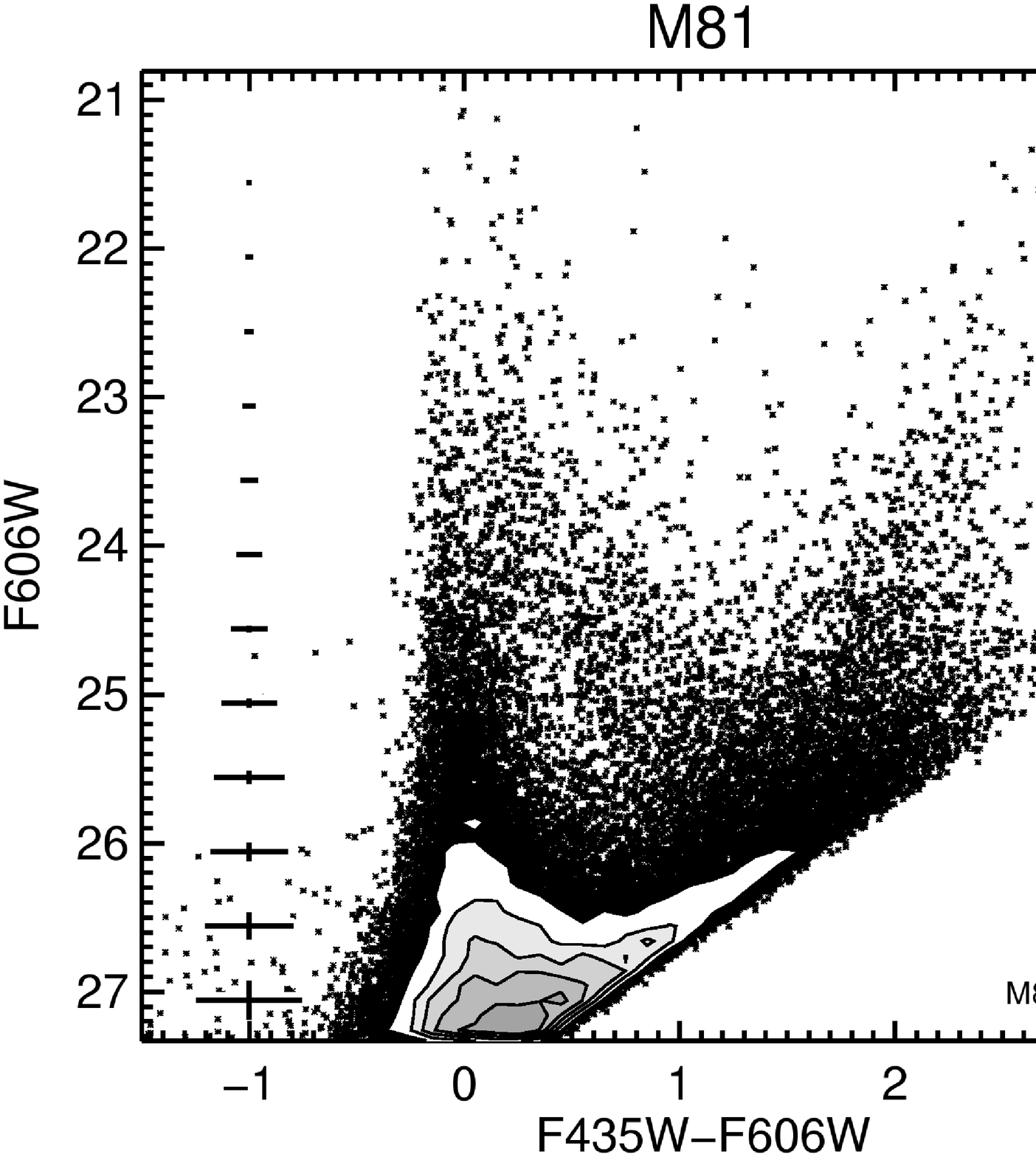}
}
\centerline{
\includegraphics[width=1.625in]{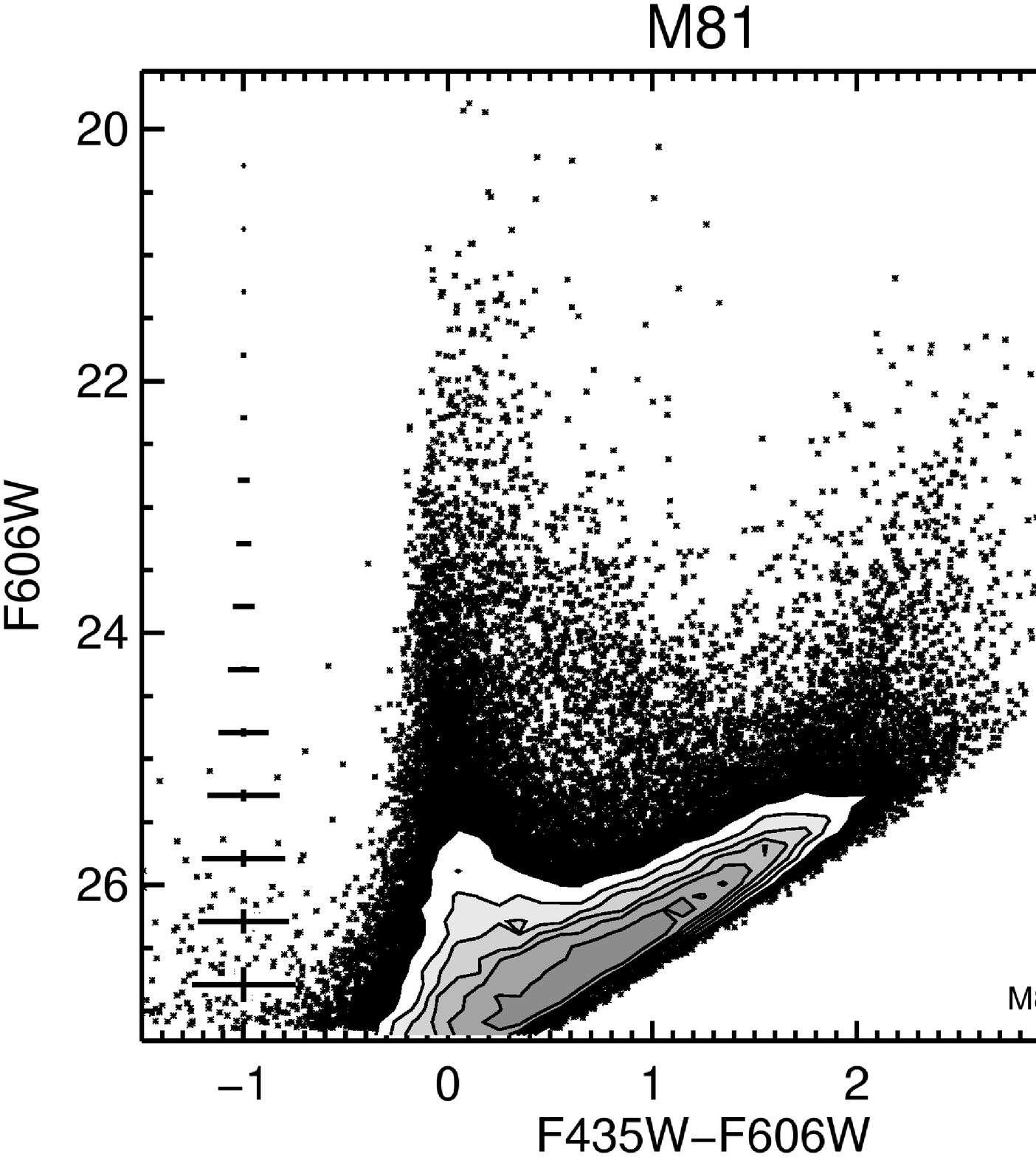}
\includegraphics[width=1.625in]{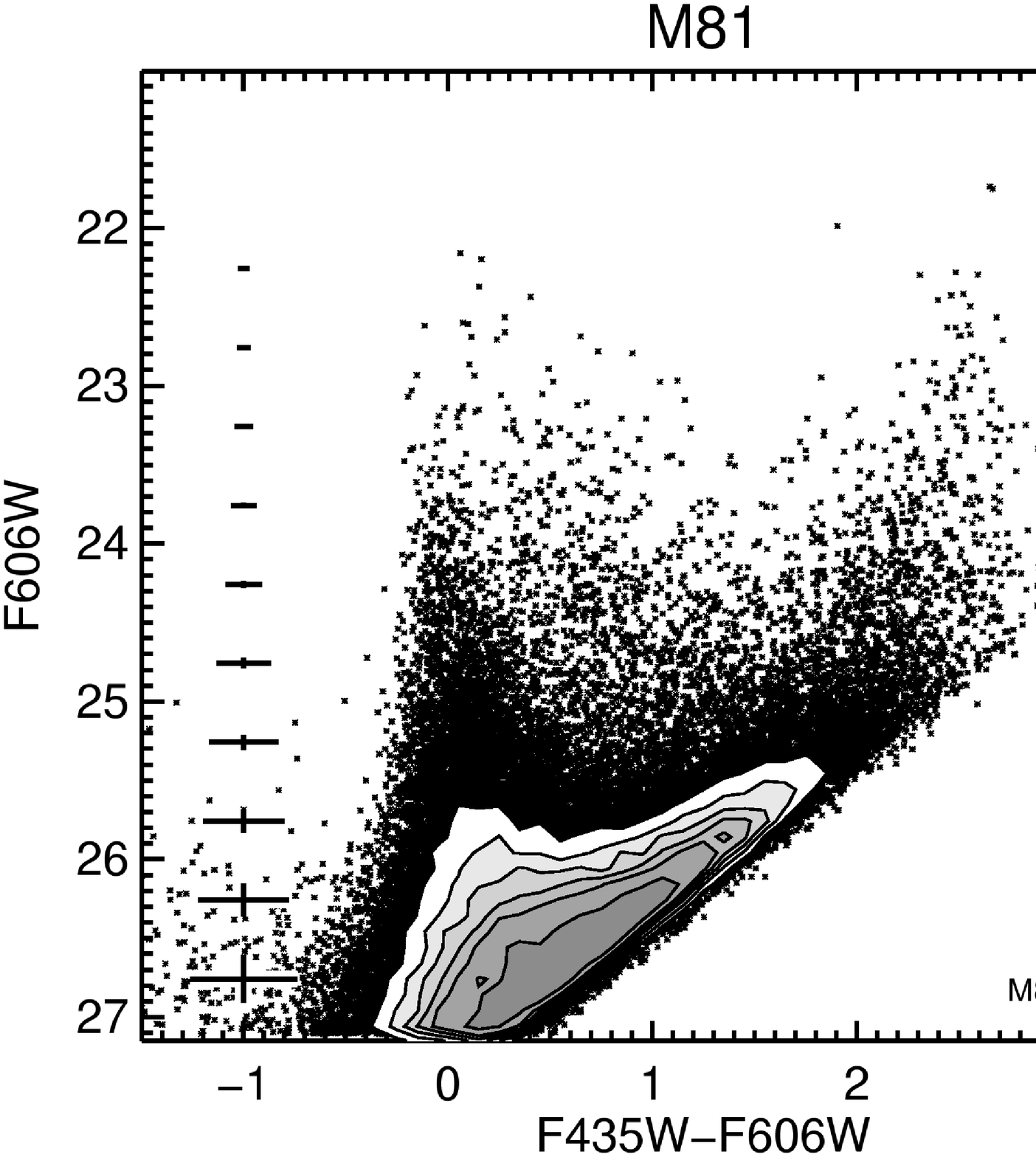}
\includegraphics[width=1.625in]{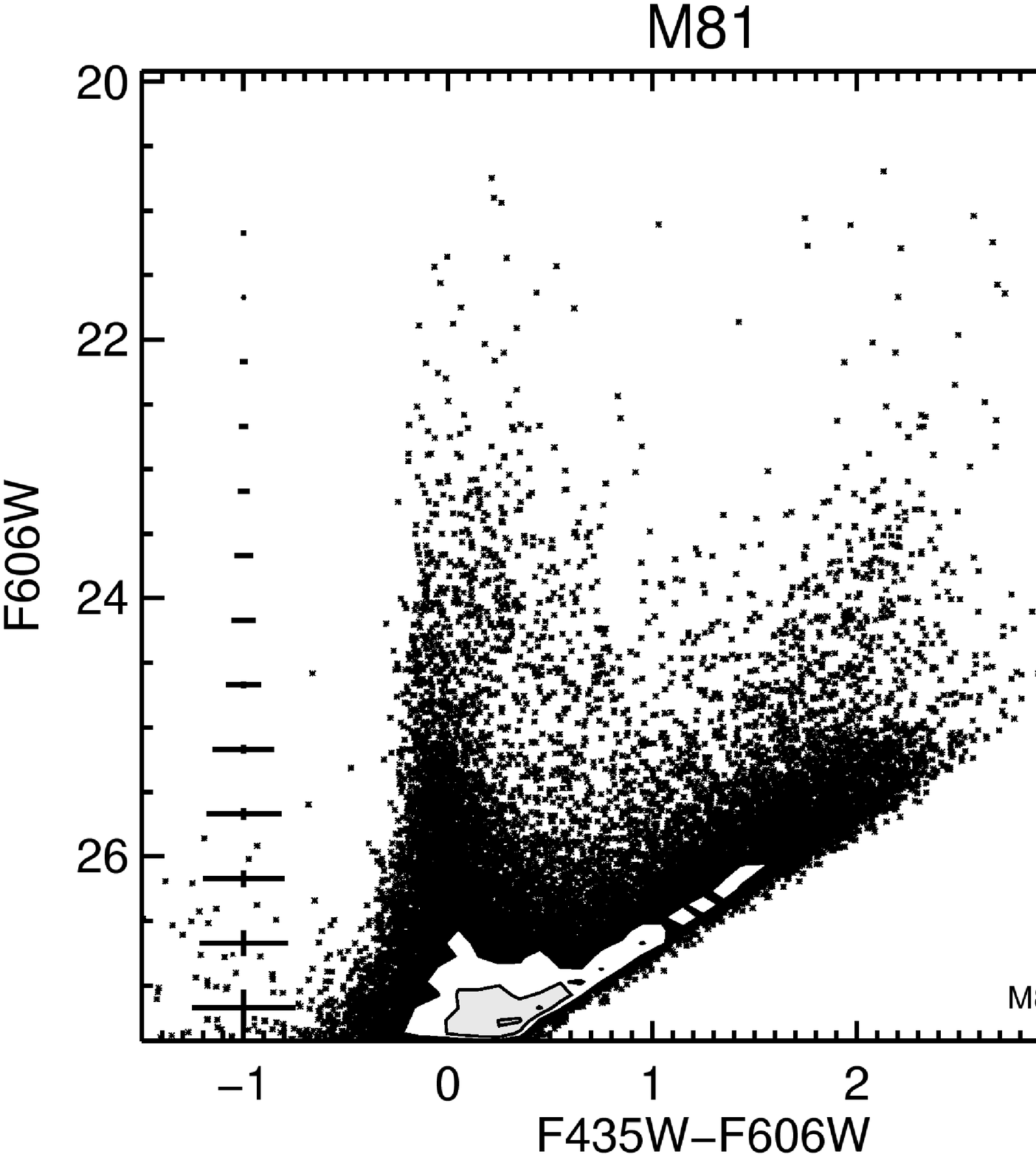}
\includegraphics[width=1.625in]{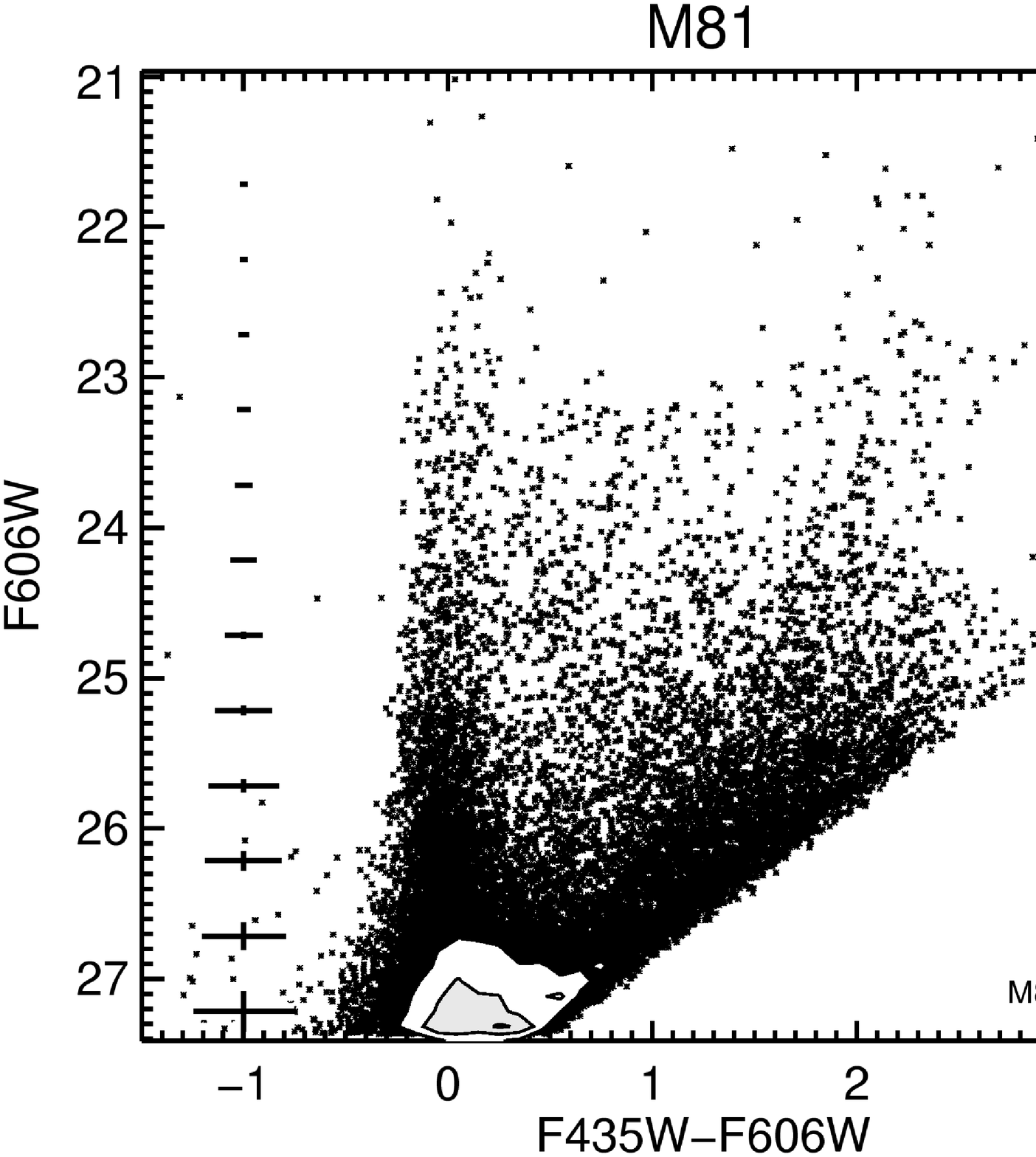}
}
\centerline{
\includegraphics[width=1.625in]{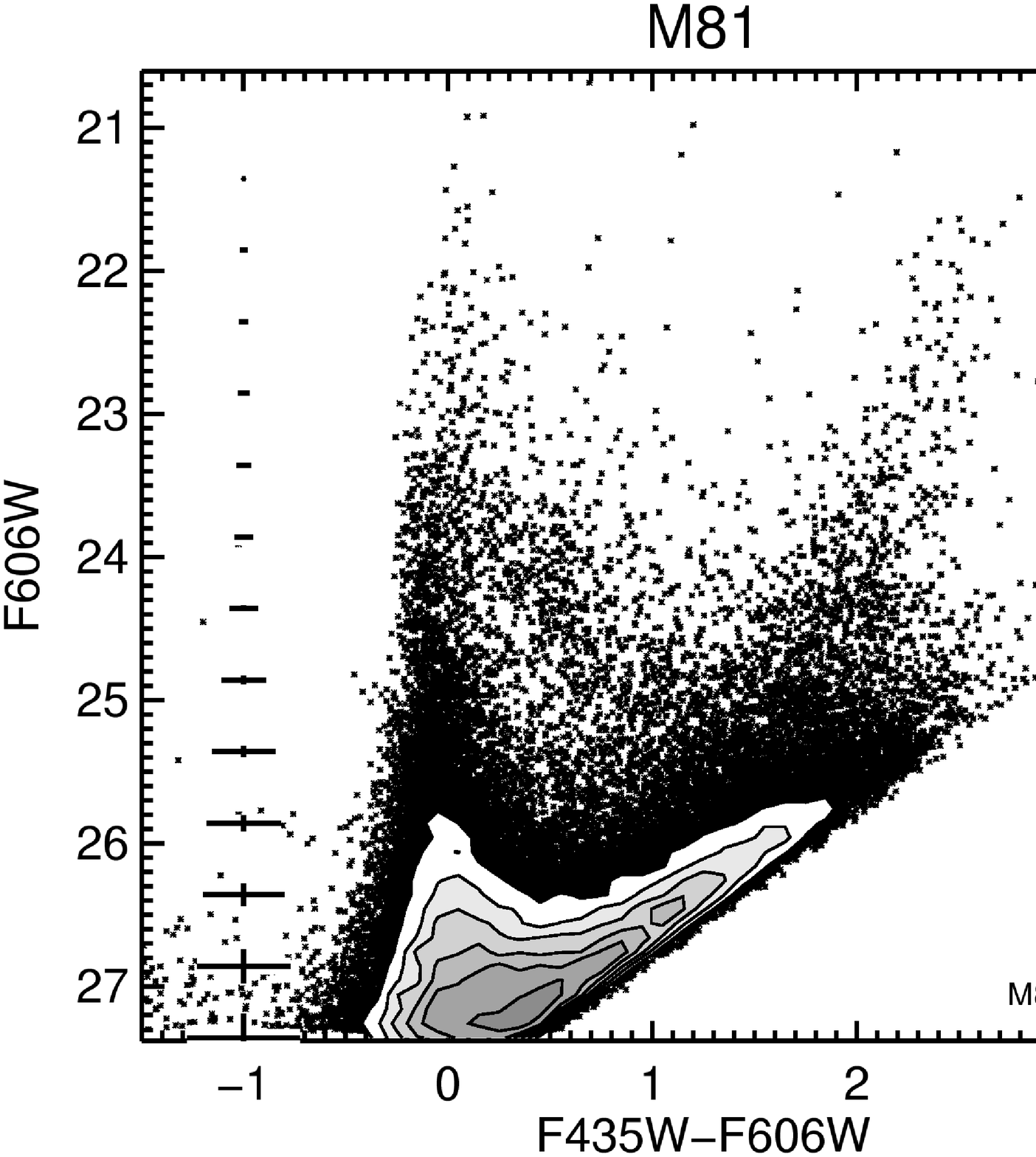}
\includegraphics[width=1.625in]{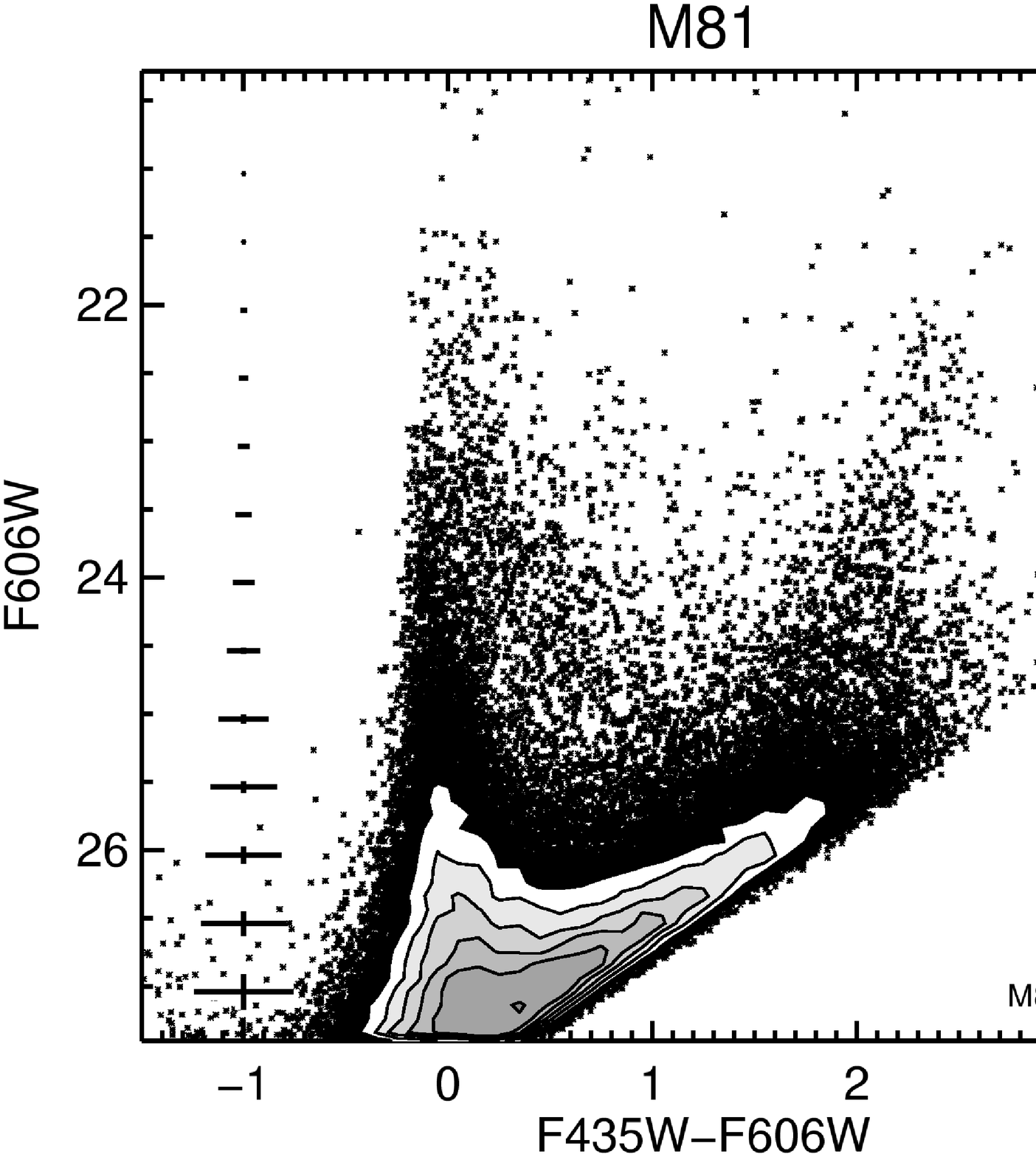}
\includegraphics[width=1.625in]{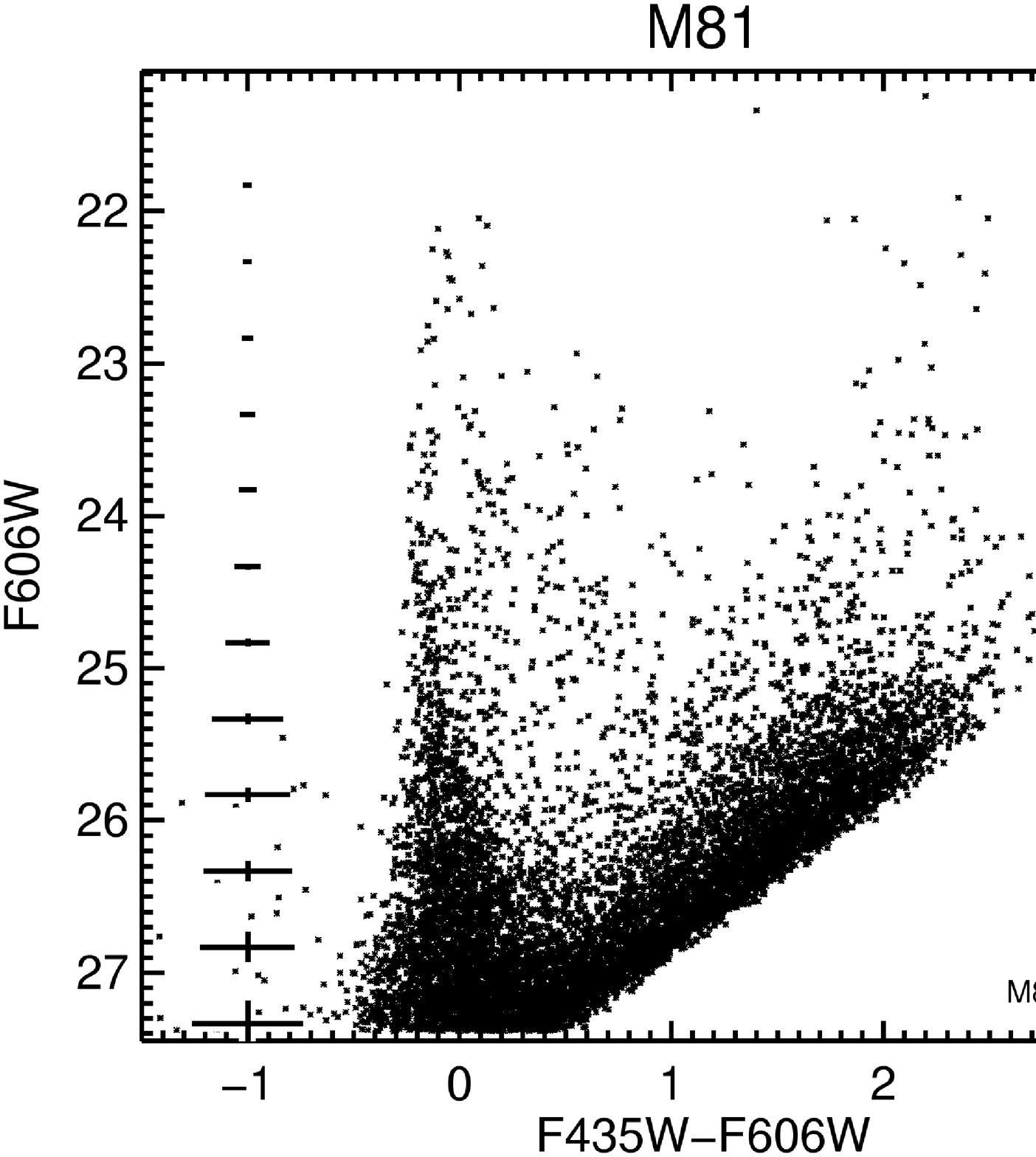}
\includegraphics[width=1.625in]{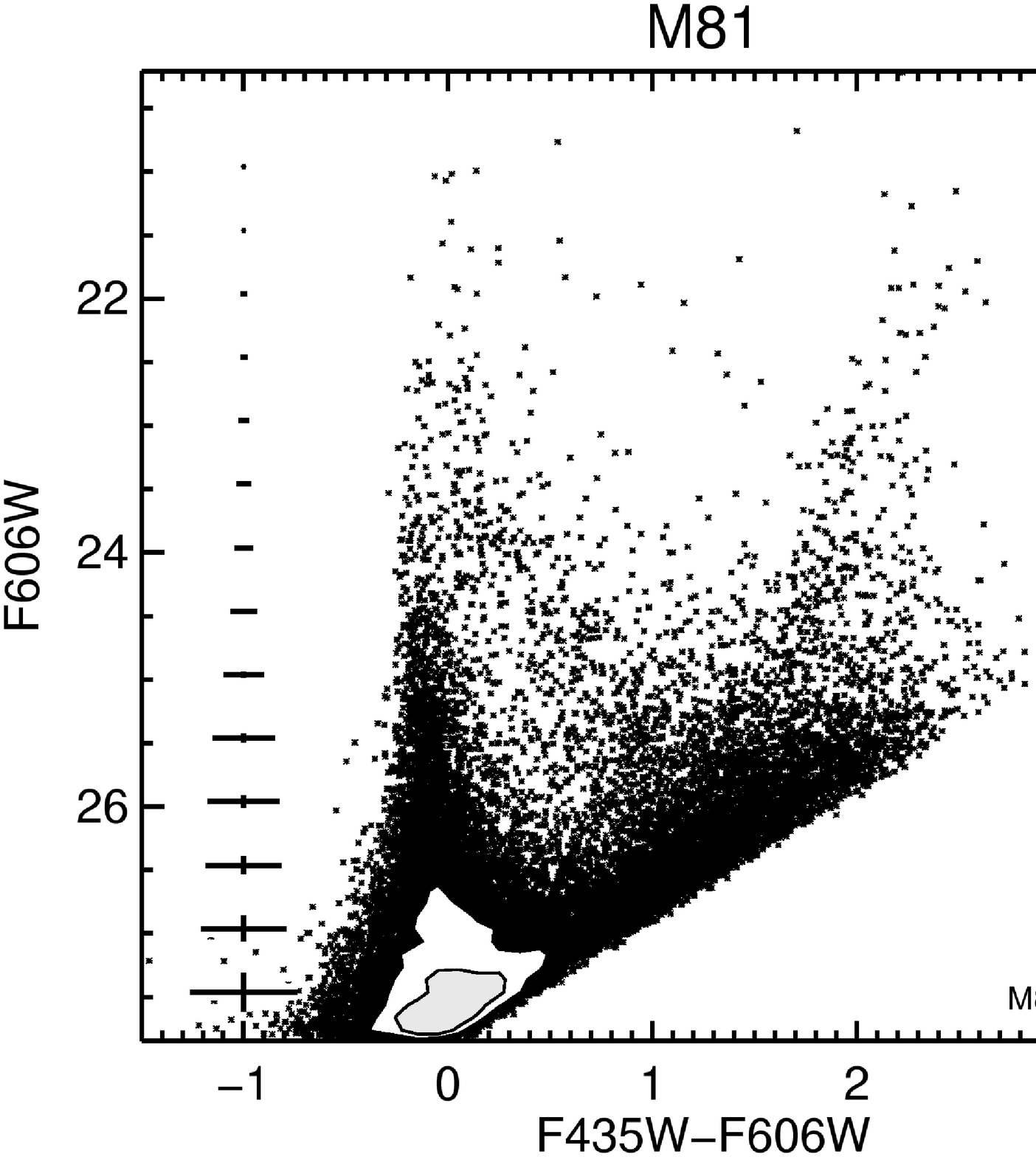}
}
\centerline{
\includegraphics[width=1.625in]{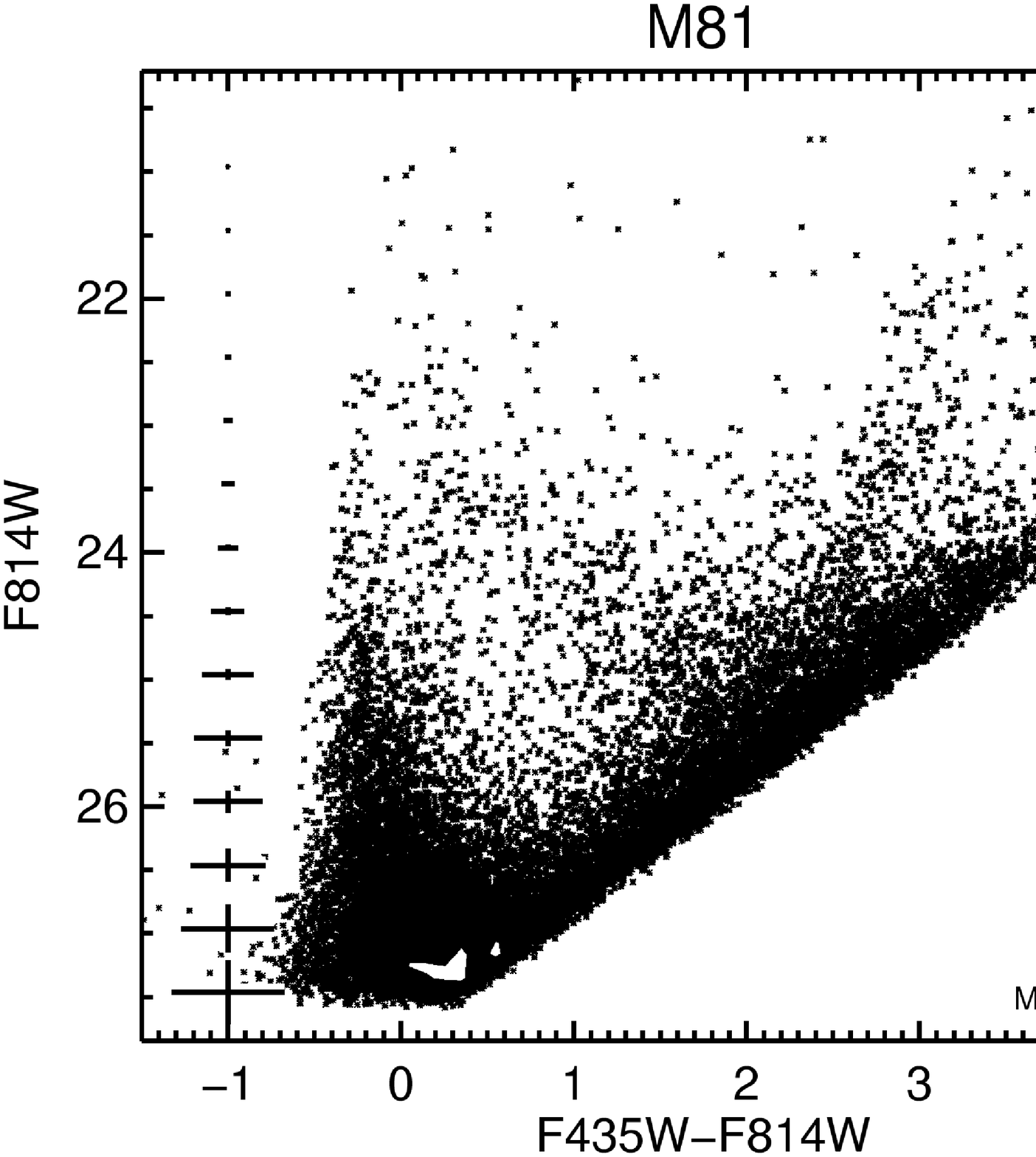}
\includegraphics[width=1.625in]{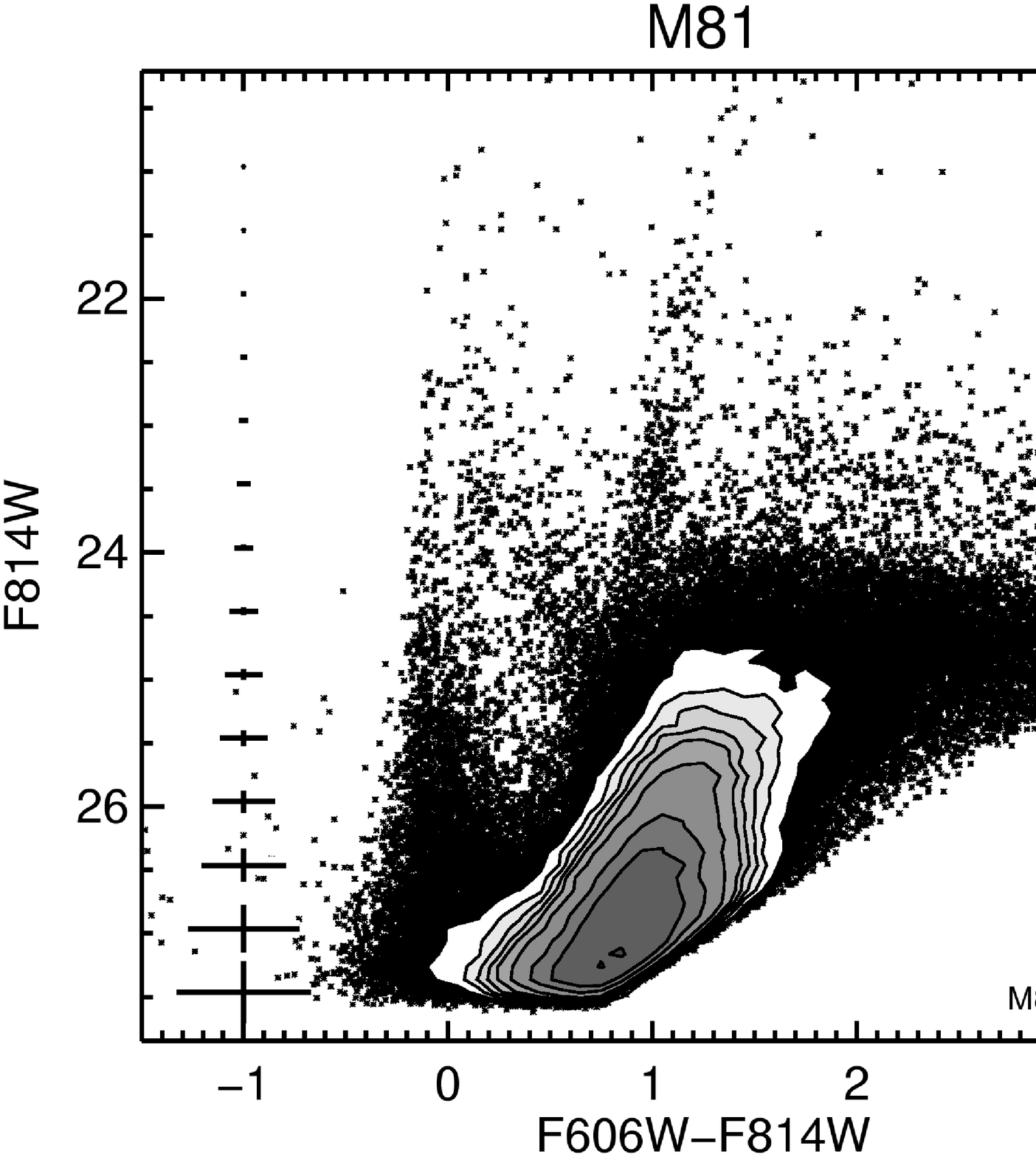}
\includegraphics[width=1.625in]{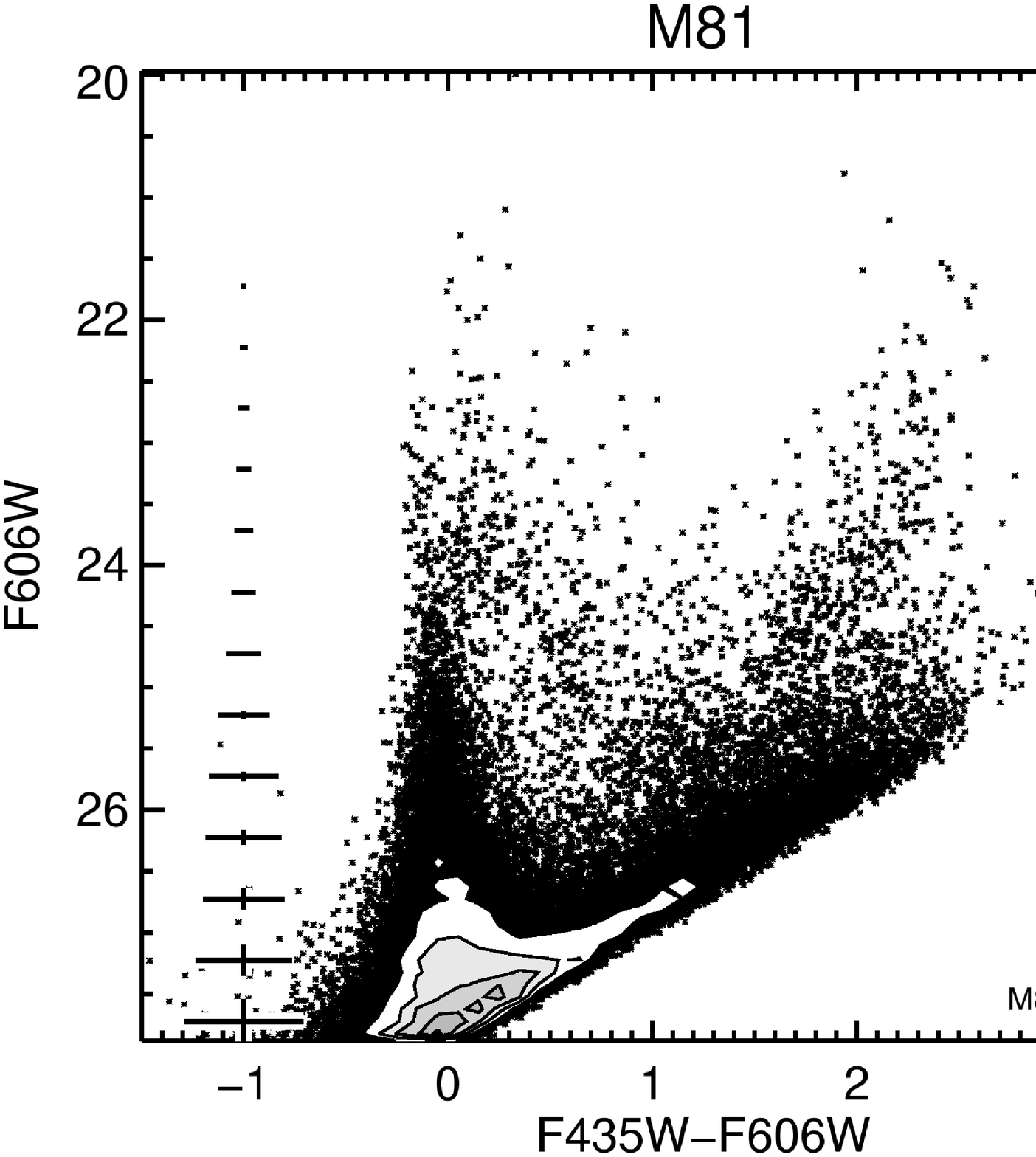}
\includegraphics[width=1.625in]{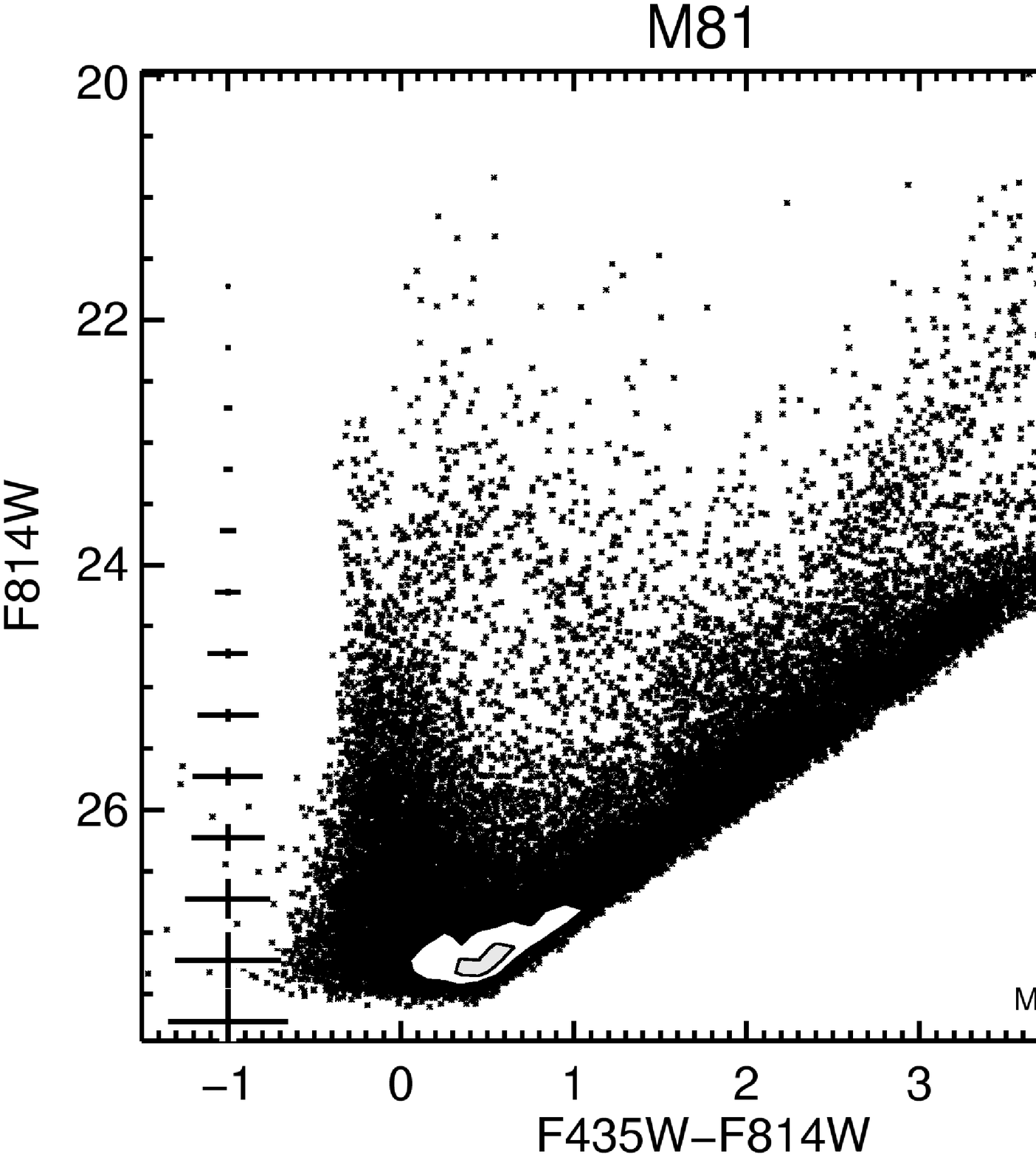}
}
\caption{
CMDs of galaxies in the ANGST data release,
as described in Figure~\ref{cmdfig1}.
Figures are ordered from the upper left to the bottom right.
(a) M81; (b) M81; (c) M81; (d) M81; (e) M81; (f) M81; (g) M81; (h) M81; (i) M81; (j) M81; (k) M81; (l) M81; (m) M81; (n) M81; (o) M81; (p) M81; 
    \label{cmdfig10}}
\end{figure}
\vfill
\clearpage
 
%-------------------
\begin{figure}[p]
\centerline{
\includegraphics[width=1.625in]{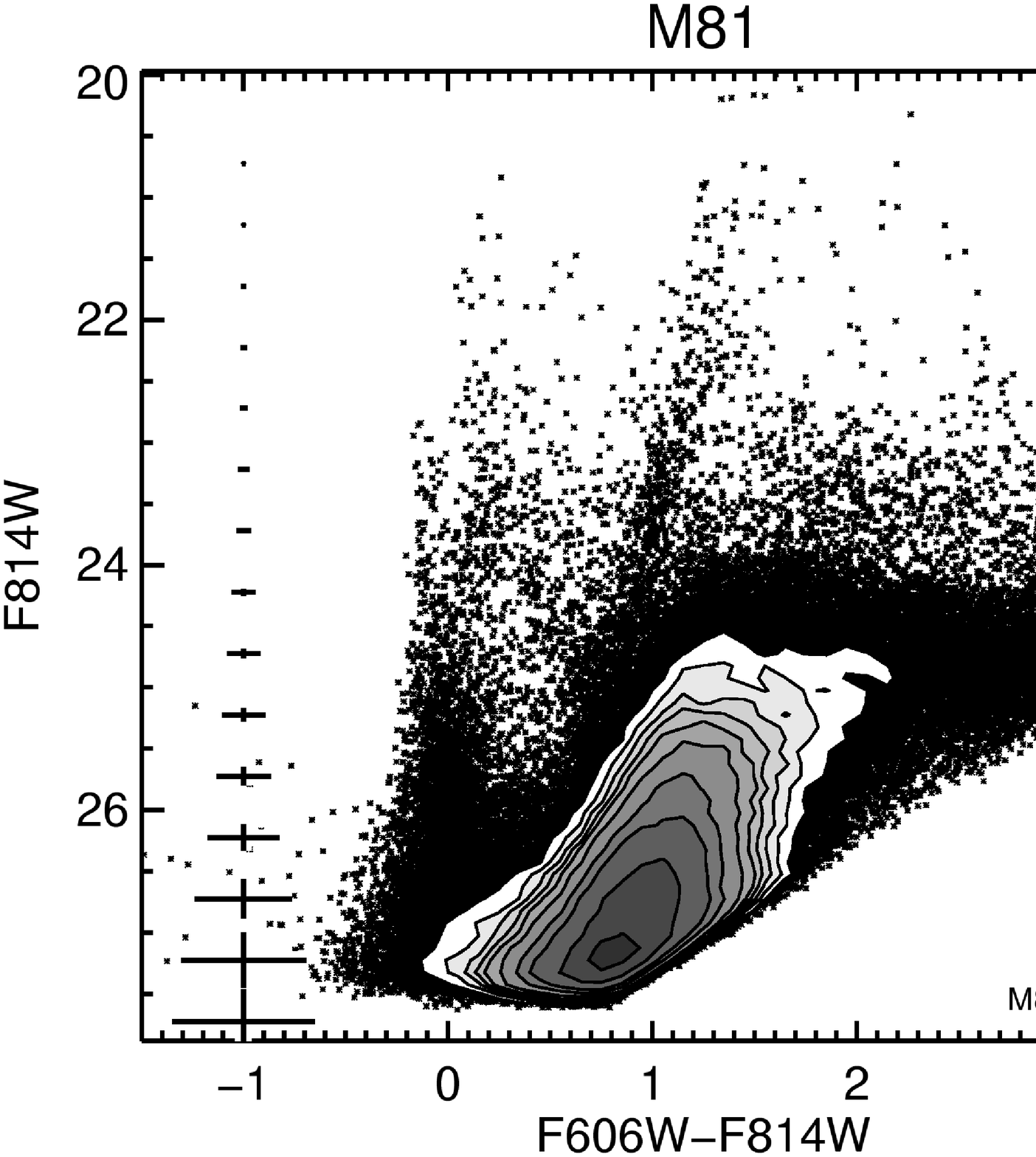}
\includegraphics[width=1.625in]{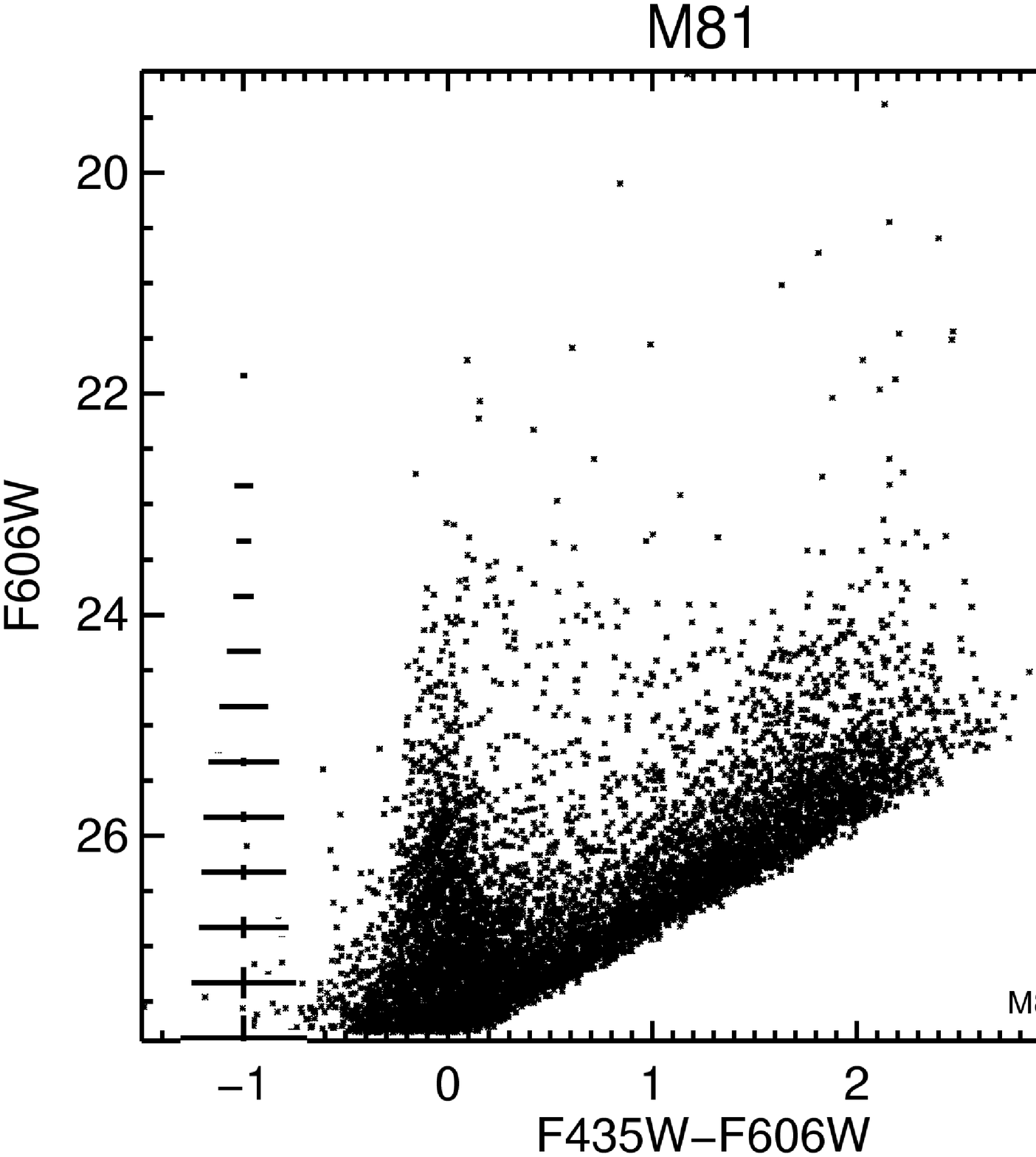}
\includegraphics[width=1.625in]{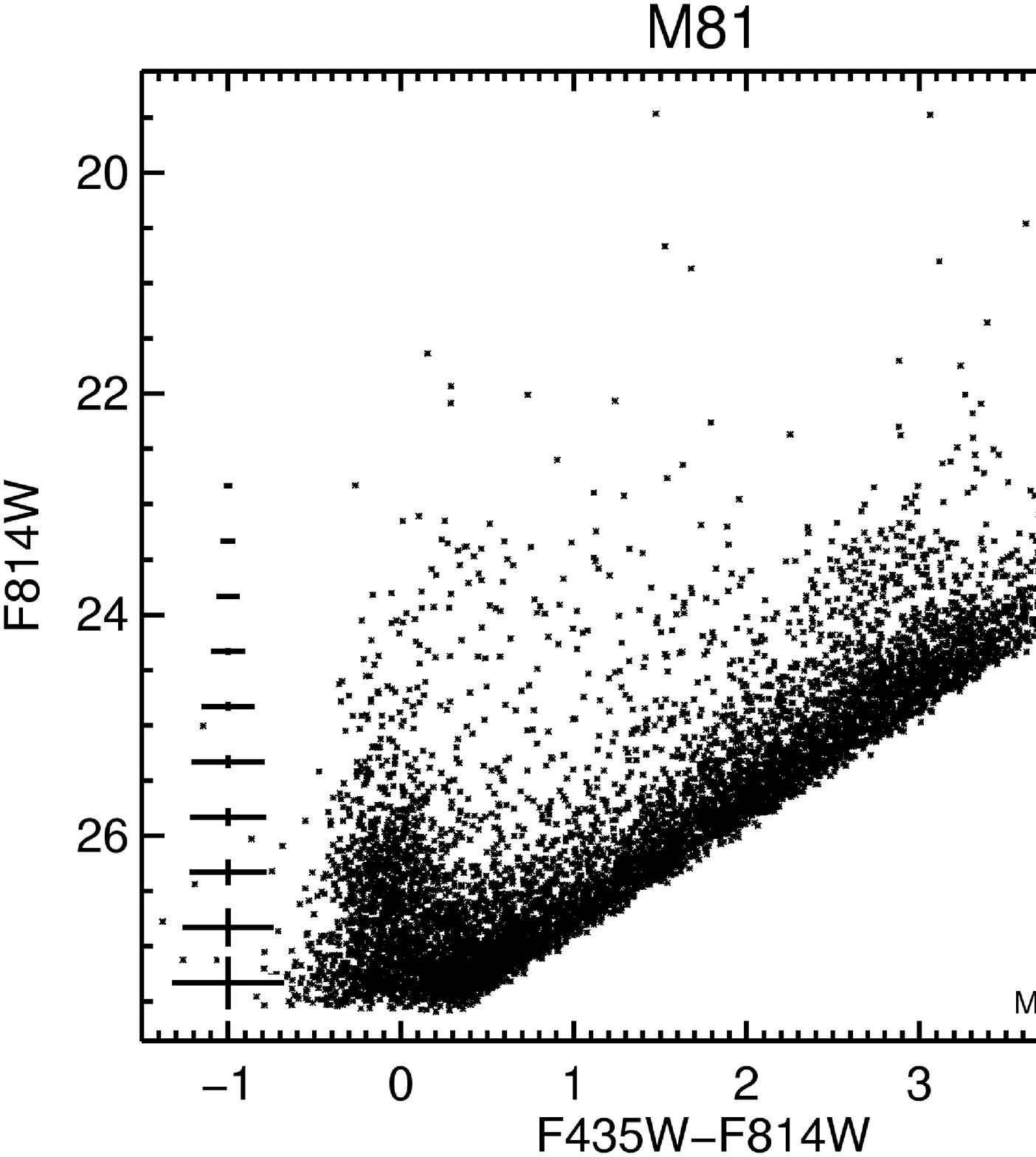}
\includegraphics[width=1.625in]{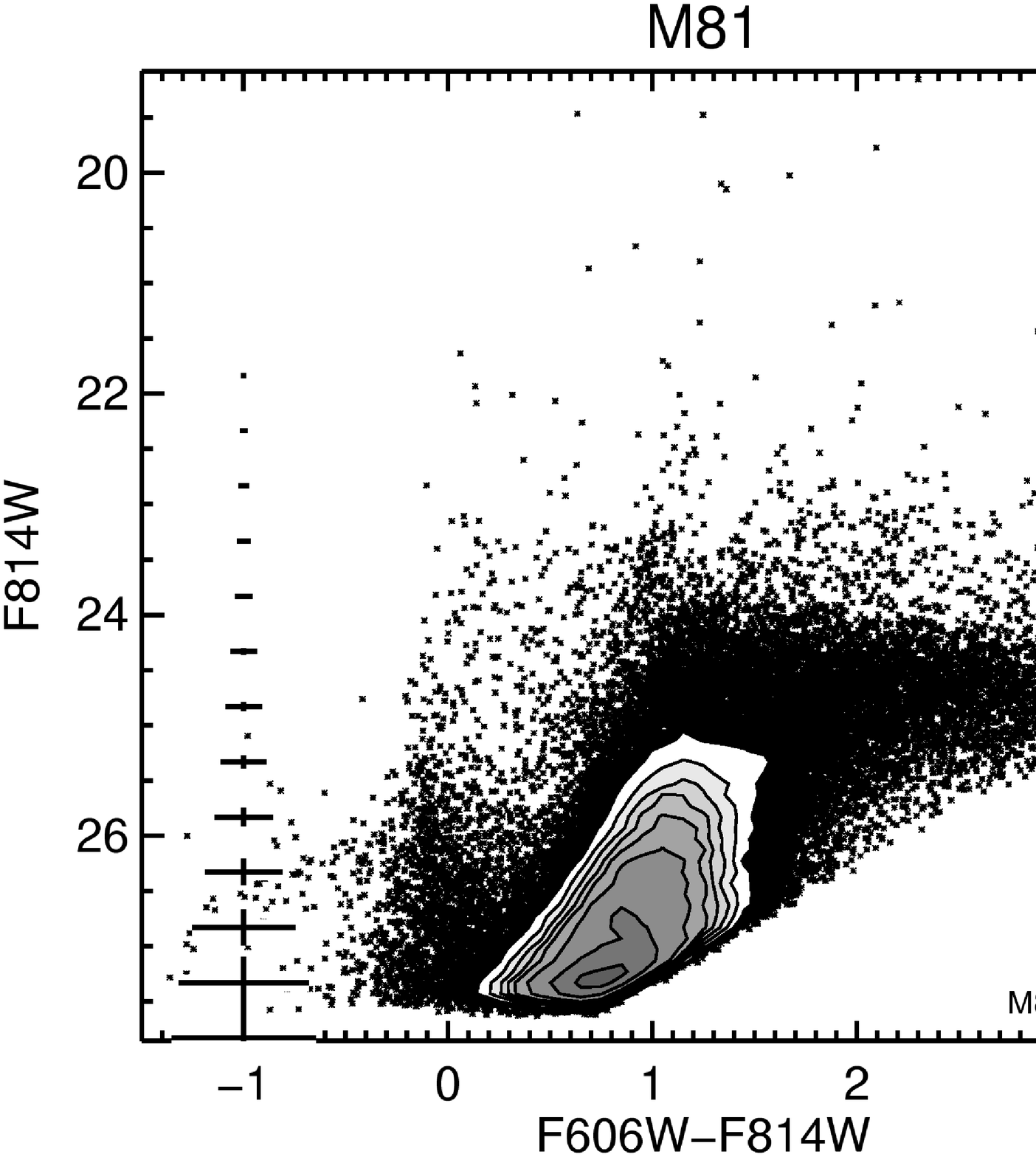}
}
\centerline{
\includegraphics[width=1.625in]{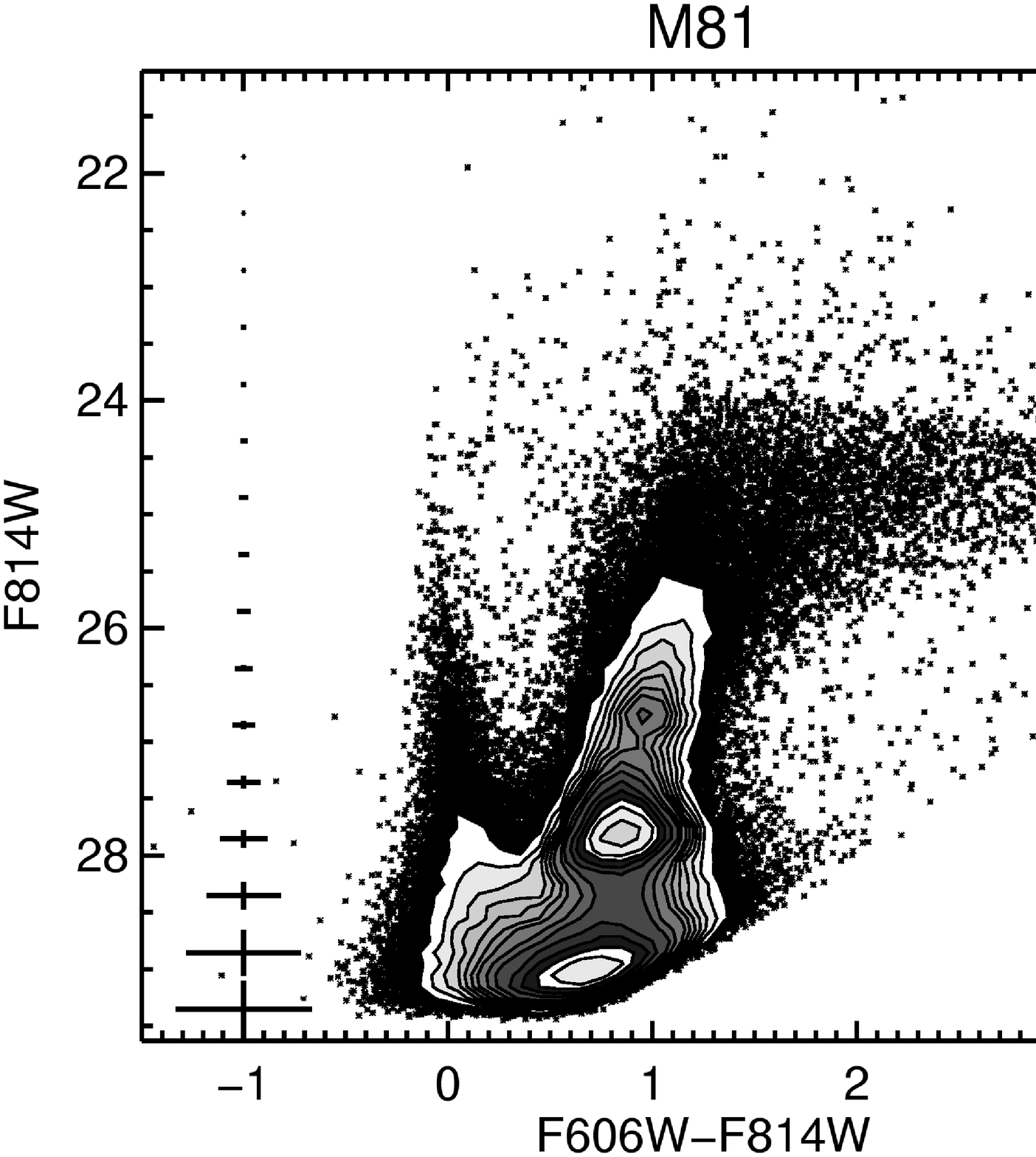}
\includegraphics[width=1.625in]{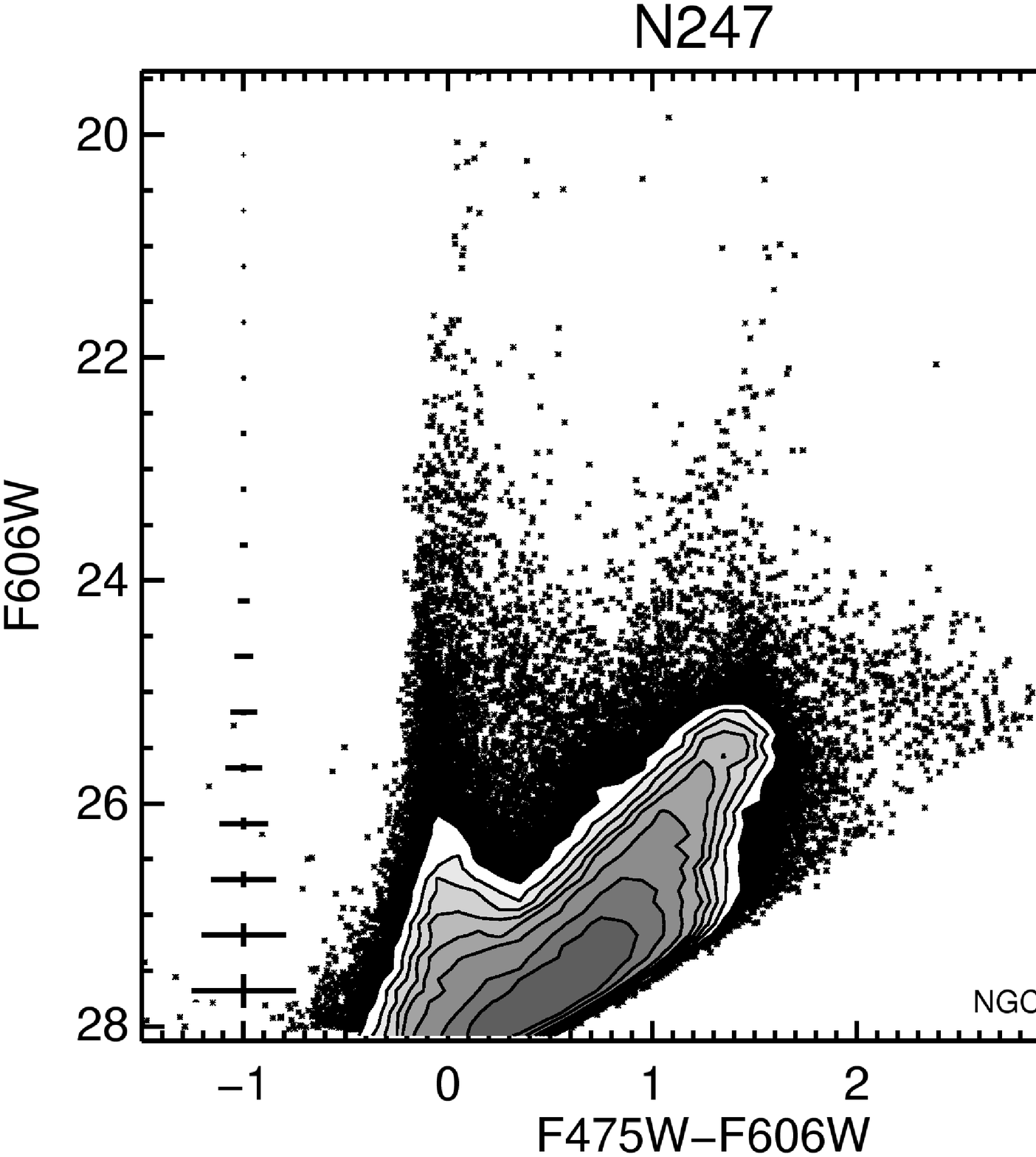}
\includegraphics[width=1.625in]{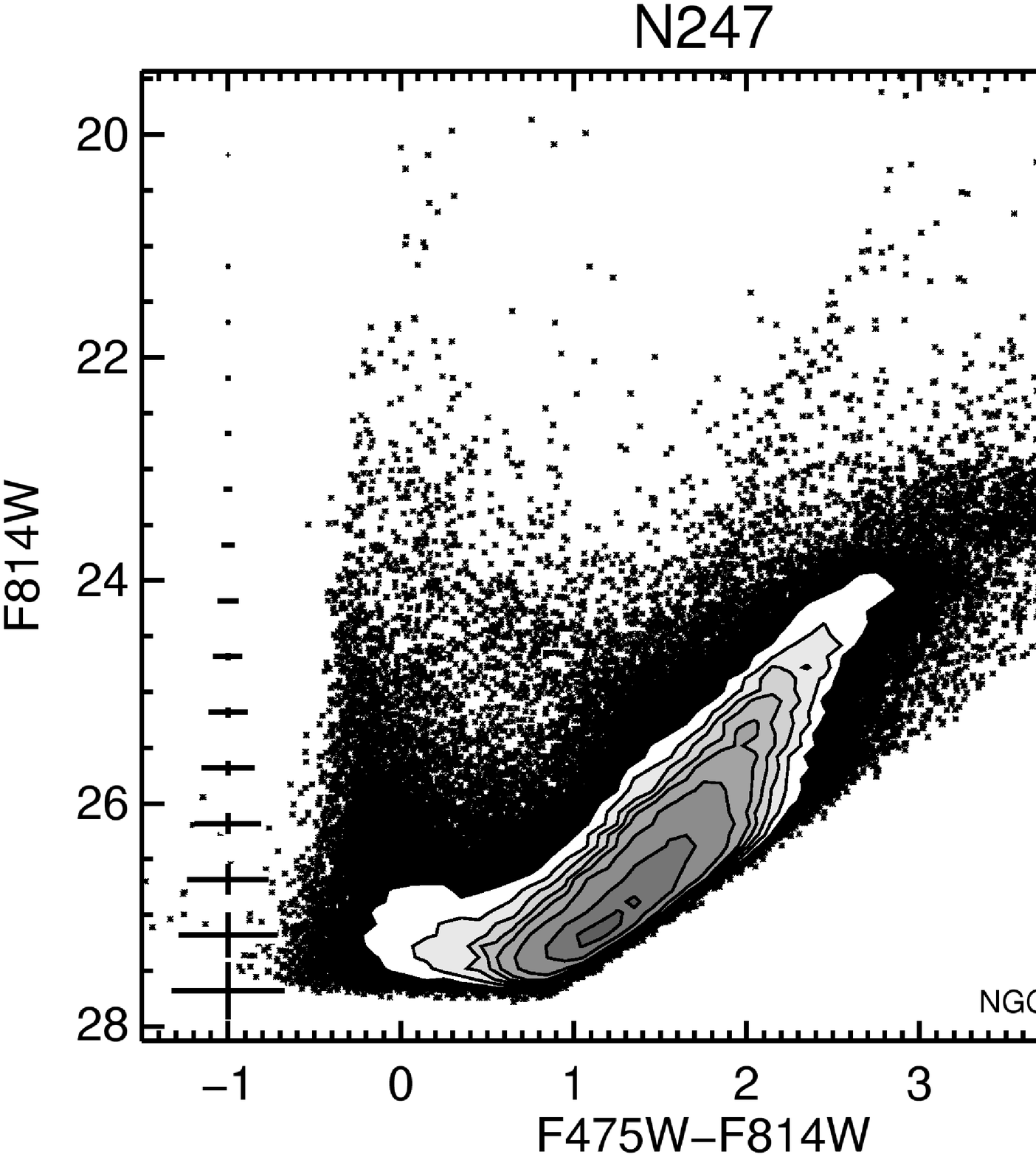}
\includegraphics[width=1.625in]{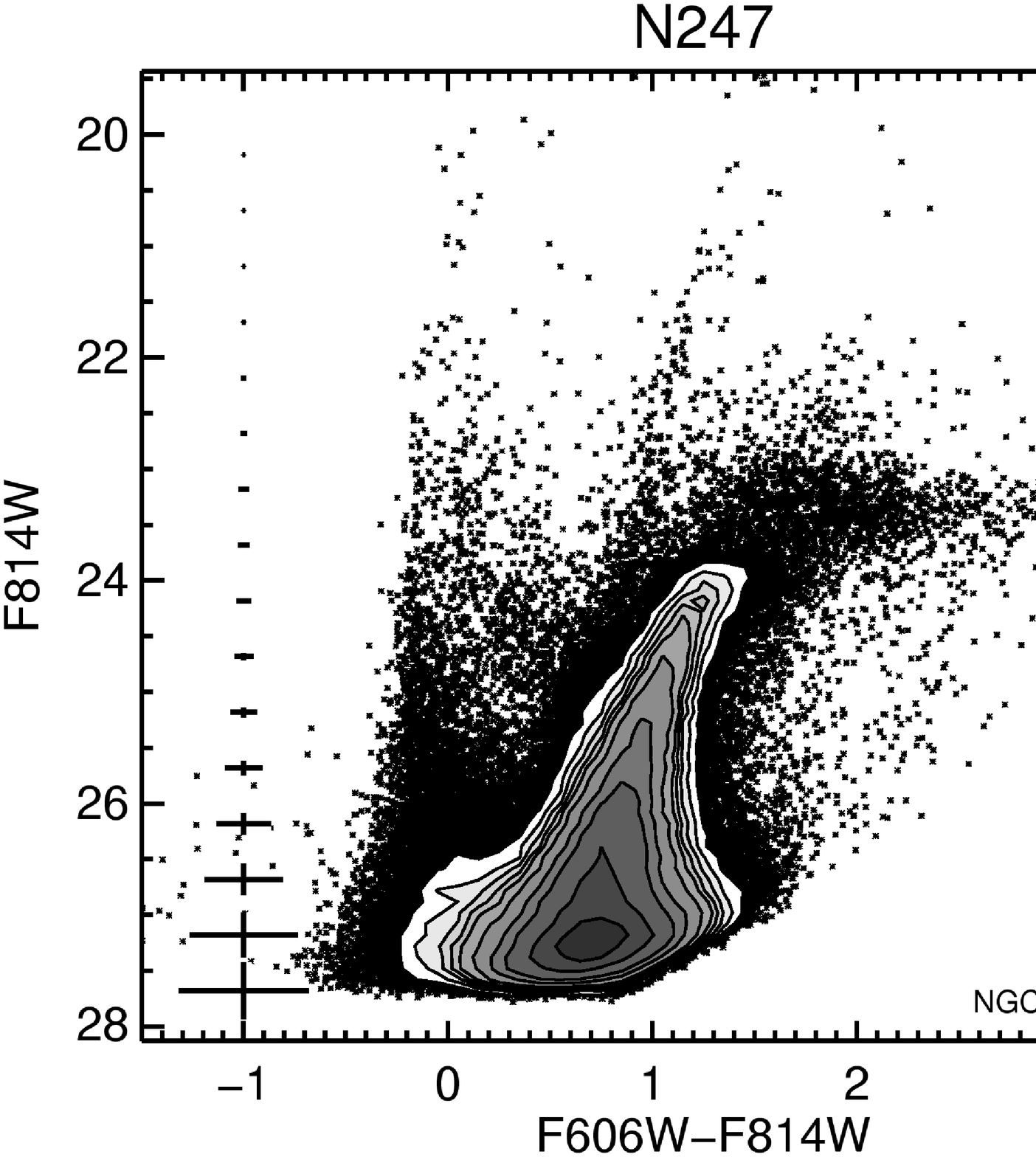}
}
\centerline{
\includegraphics[width=1.625in]{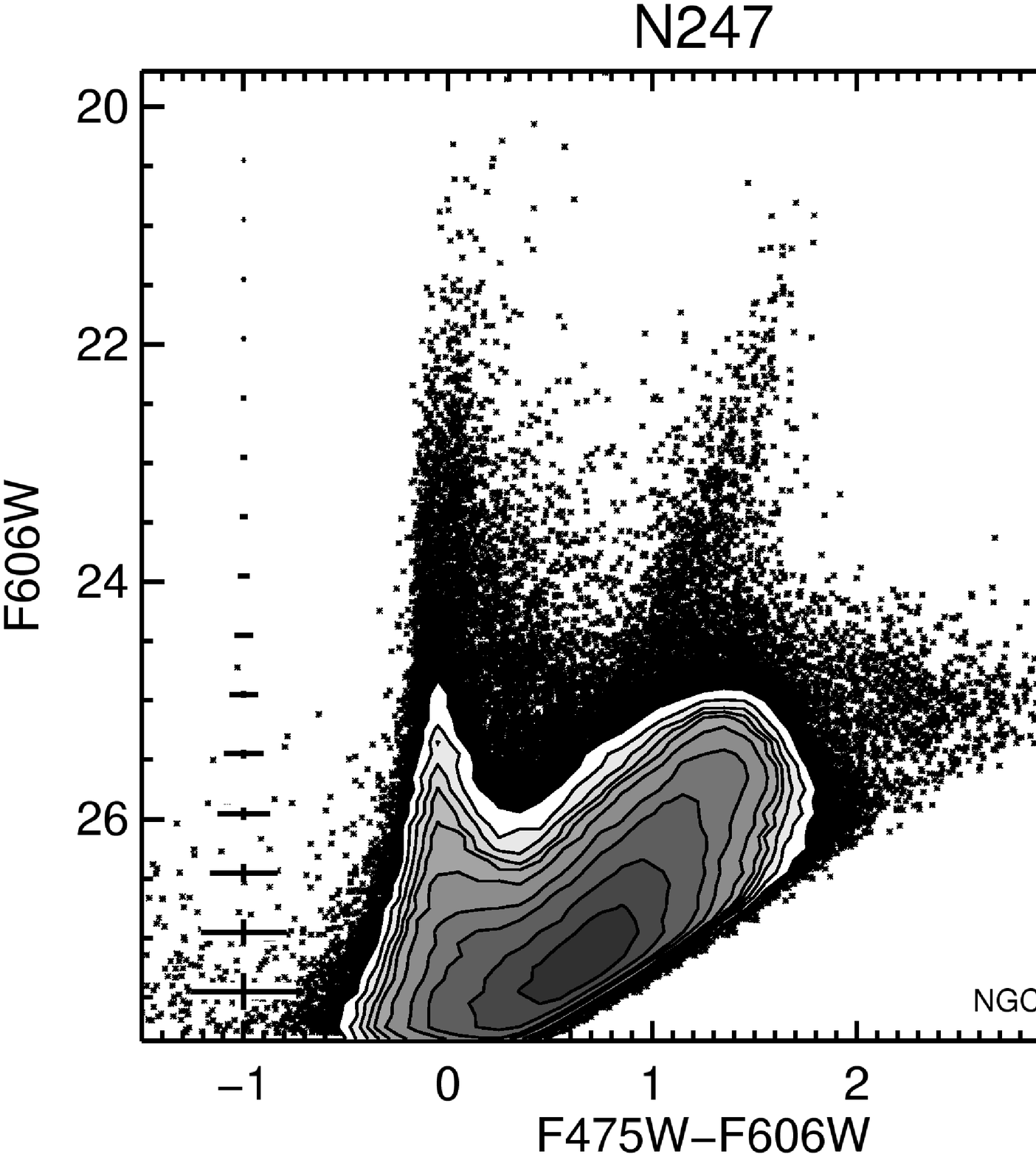}
\includegraphics[width=1.625in]{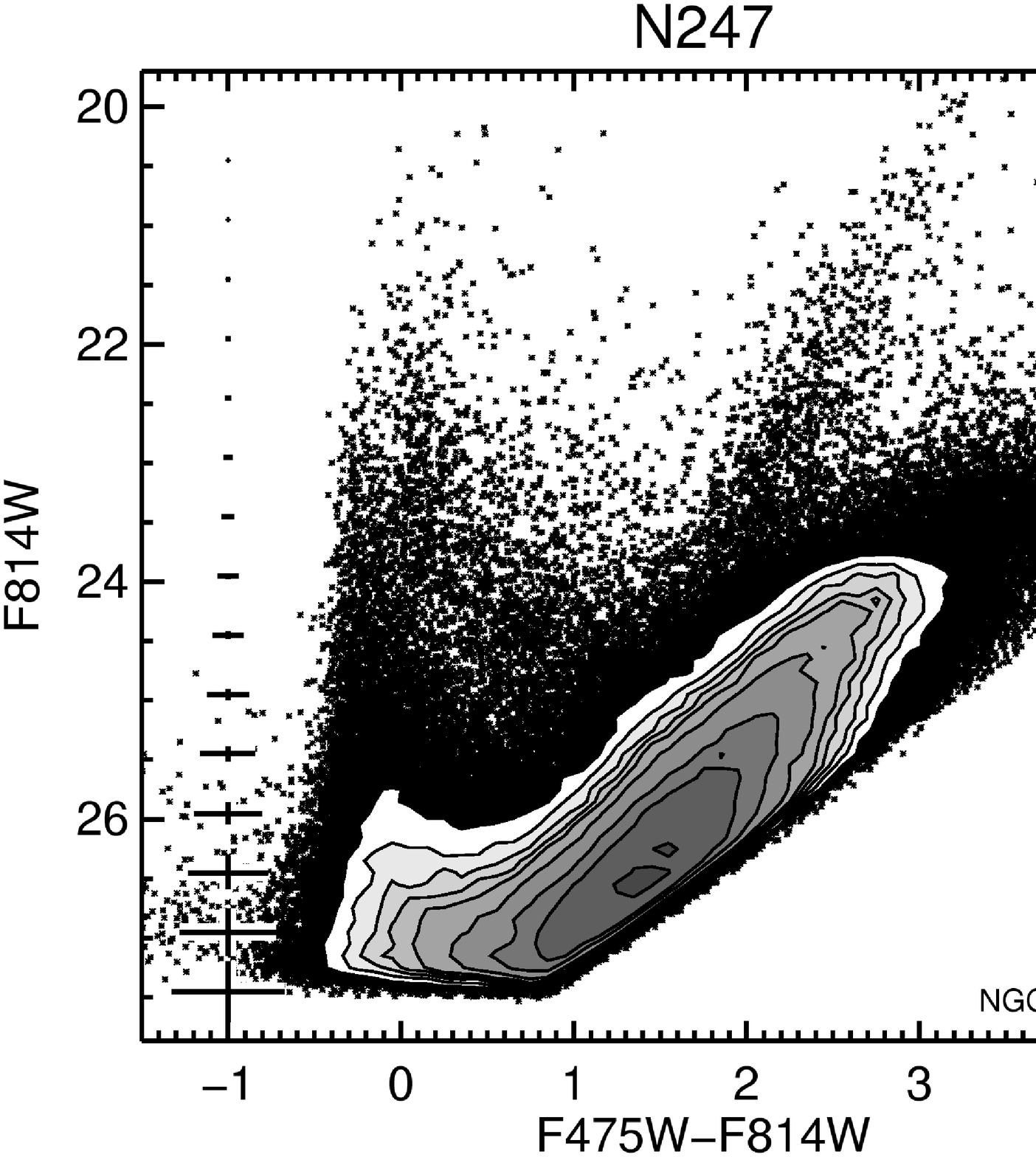}
\includegraphics[width=1.625in]{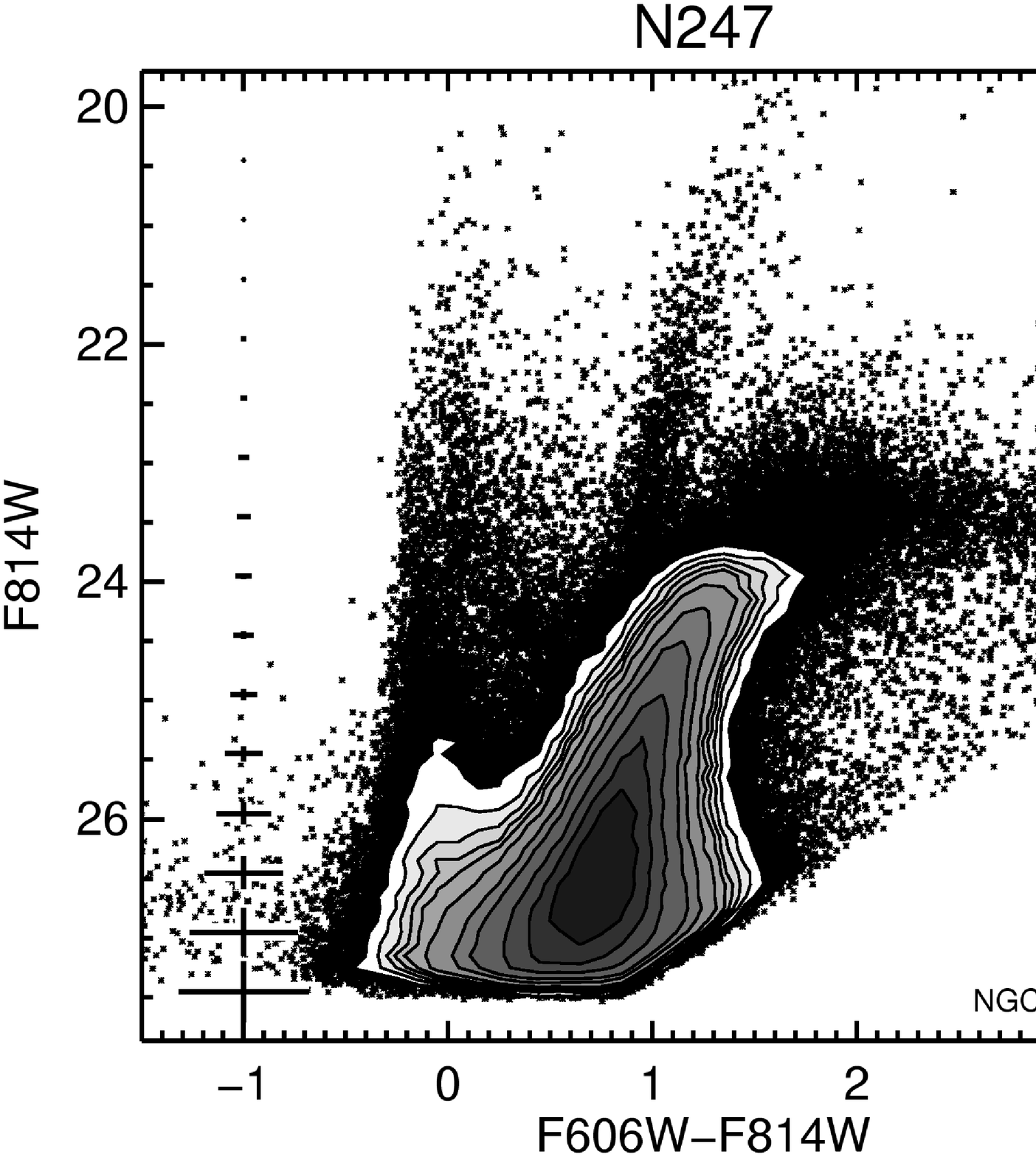}
\includegraphics[width=1.625in]{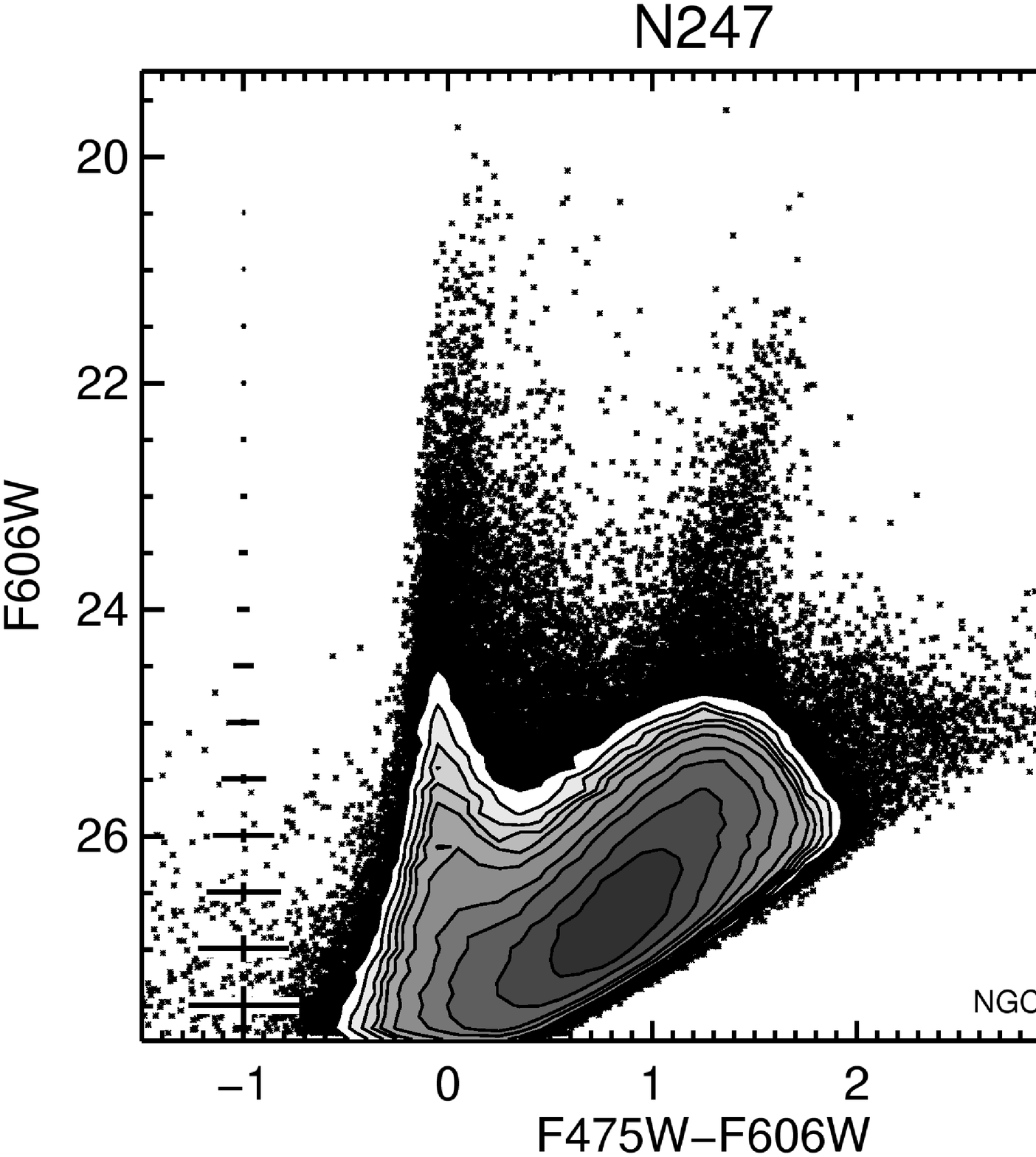}
}
\centerline{
\includegraphics[width=1.625in]{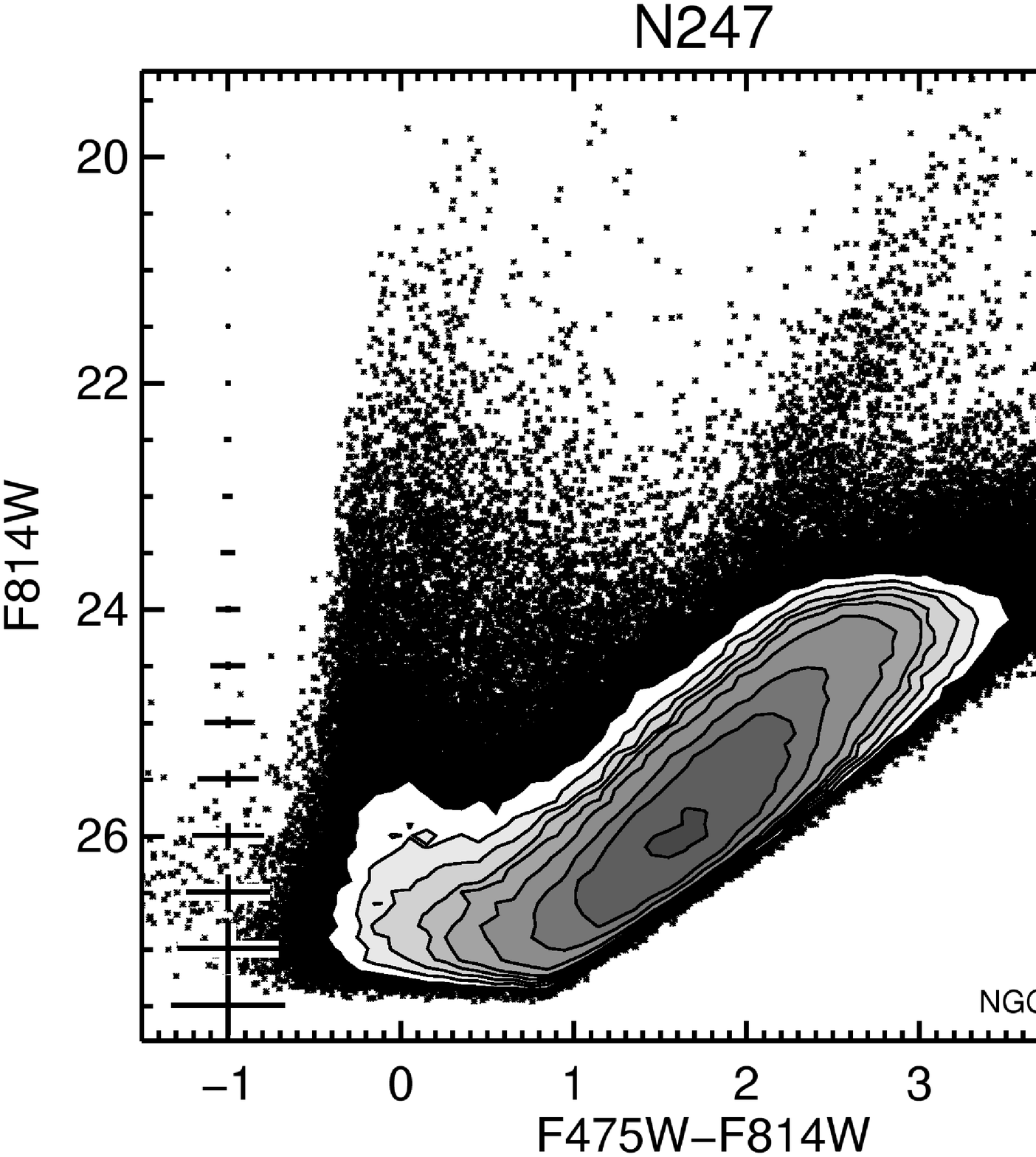}
\includegraphics[width=1.625in]{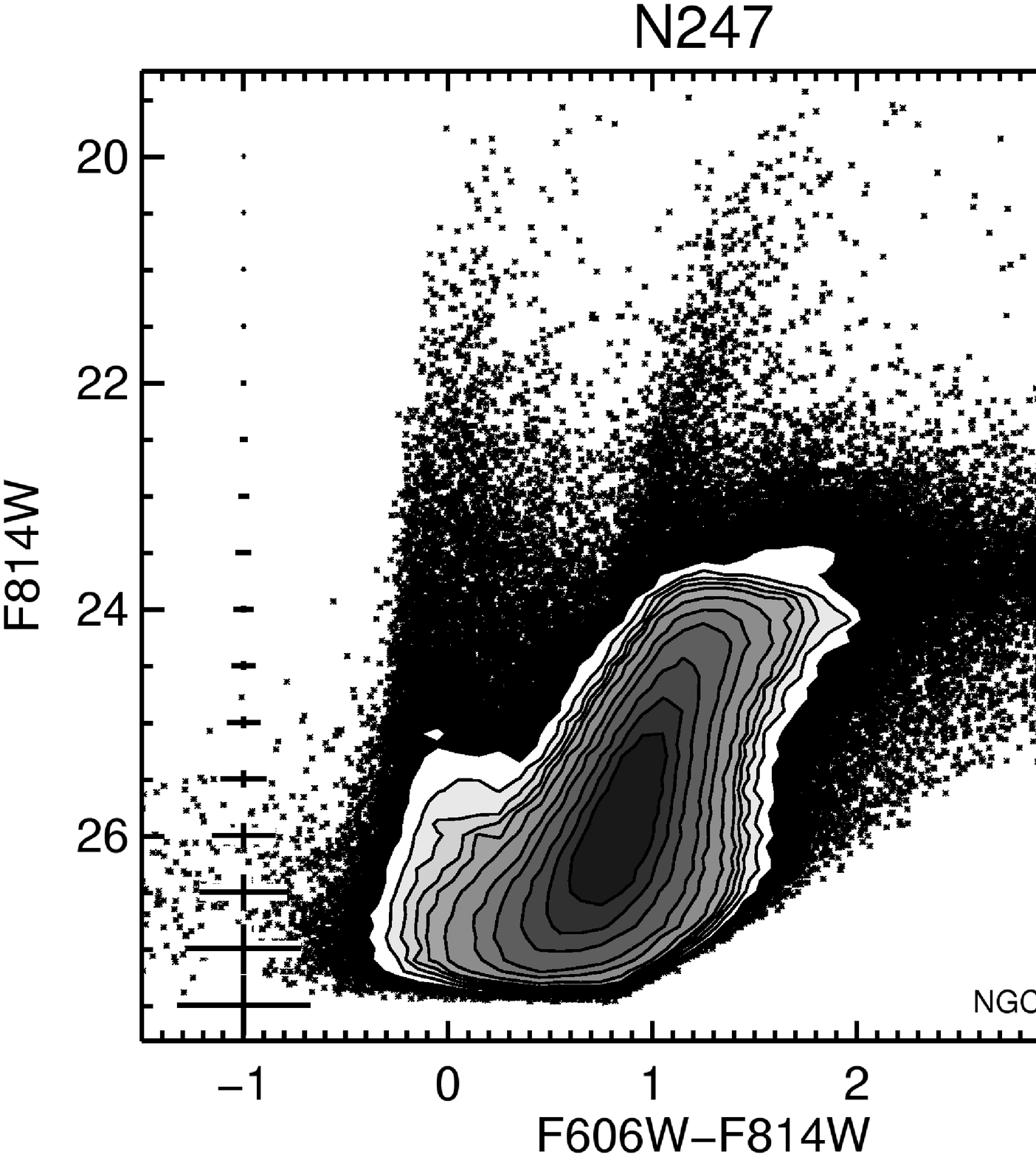}
\includegraphics[width=1.625in]{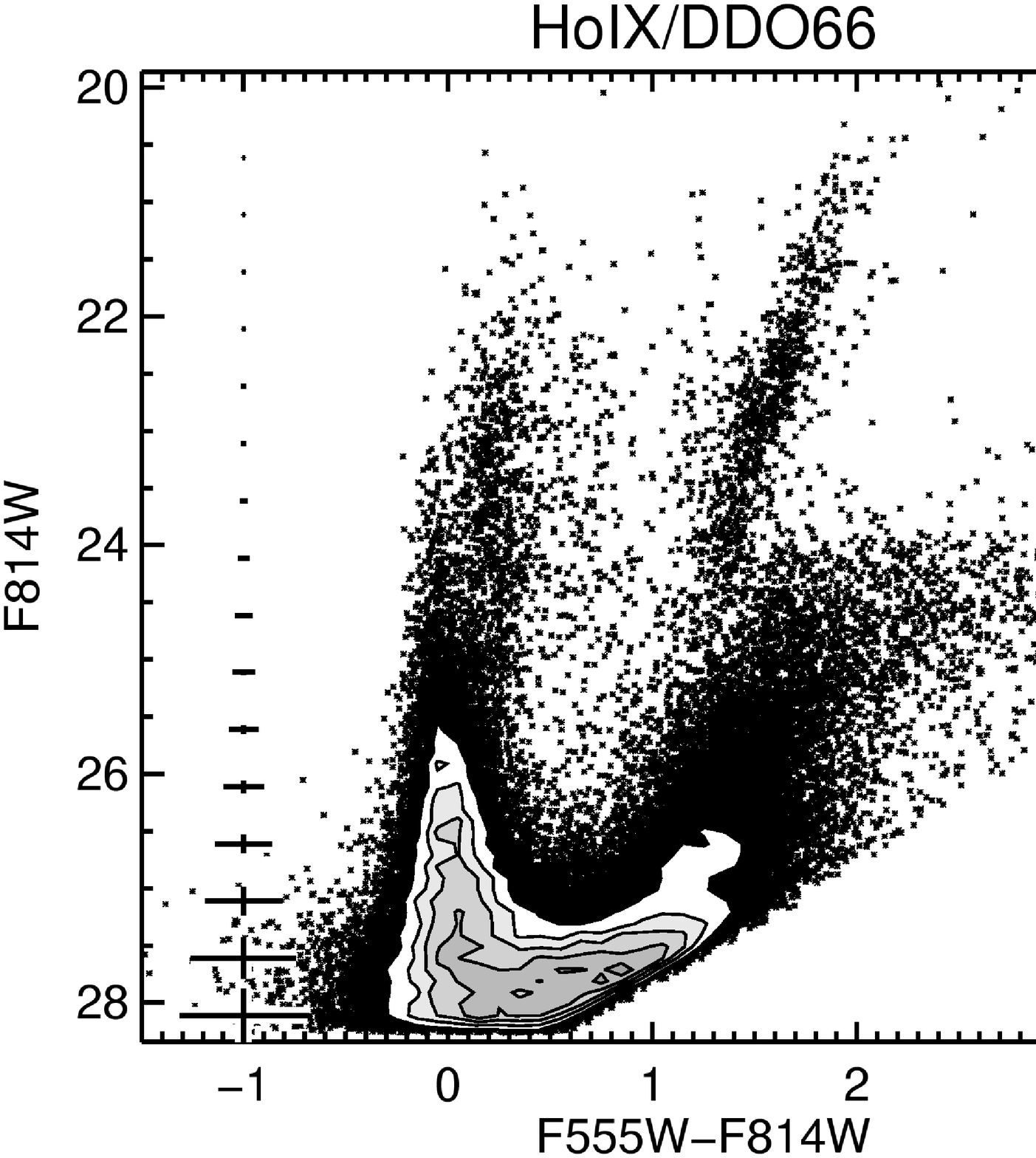}
\includegraphics[width=1.625in]{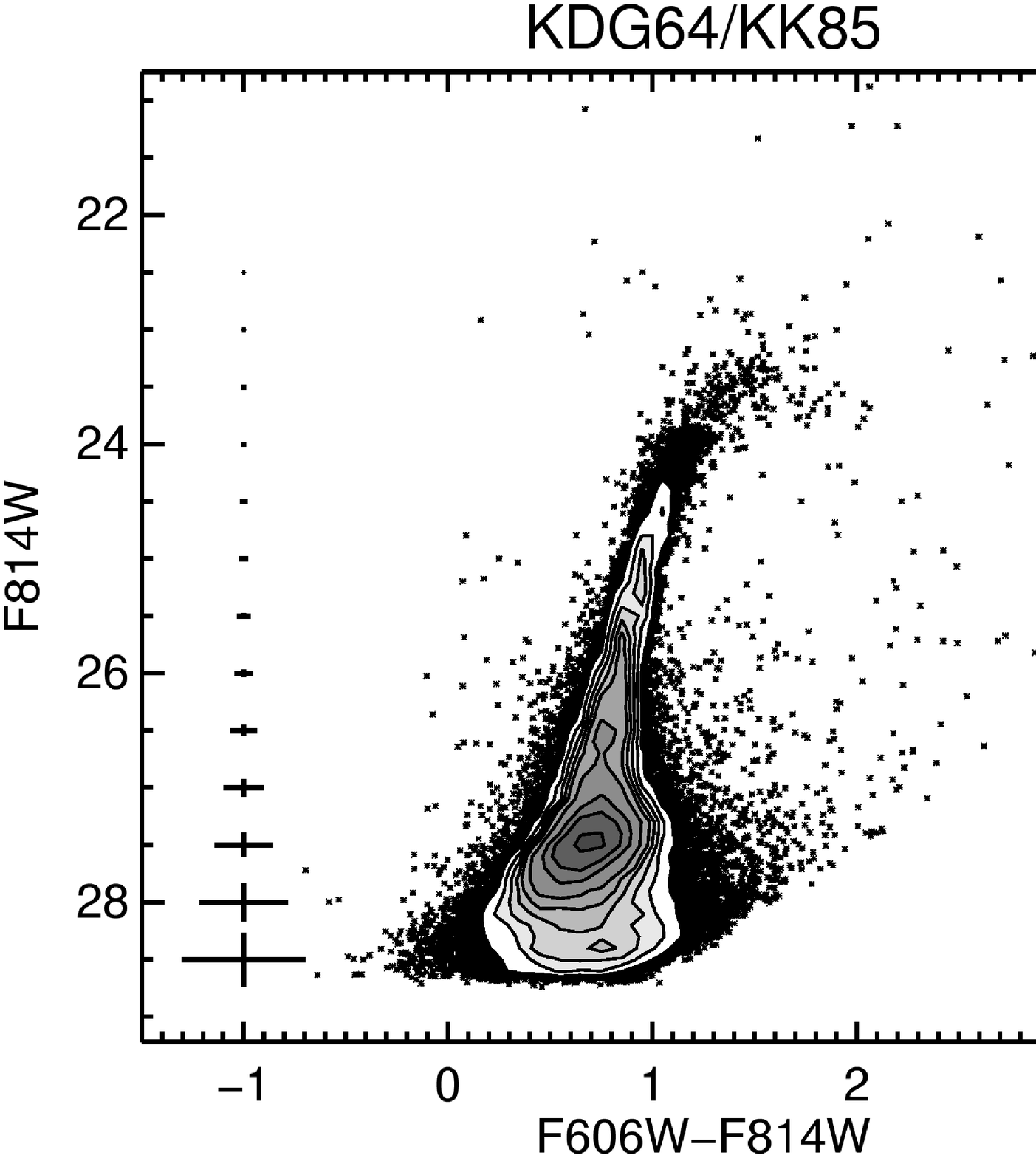}
}
\caption{
CMDs of galaxies in the ANGST data release,
as described in Figure~\ref{cmdfig1}.
Figures are ordered from the upper left to the bottom right.
(a) M81; (b) M81; (c) M81; (d) M81; (e) M81; (f) N247; (g) N247; (h) N247; (i) N247; (j) N247; (k) N247; (l) N247; (m) N247; (n) N247; (o) HoIX; (p) KDG64; 
    \label{cmdfig11}}
\end{figure}
\vfill
\clearpage
 
%-------------------
\begin{figure}[p]
\centerline{
\includegraphics[width=1.625in]{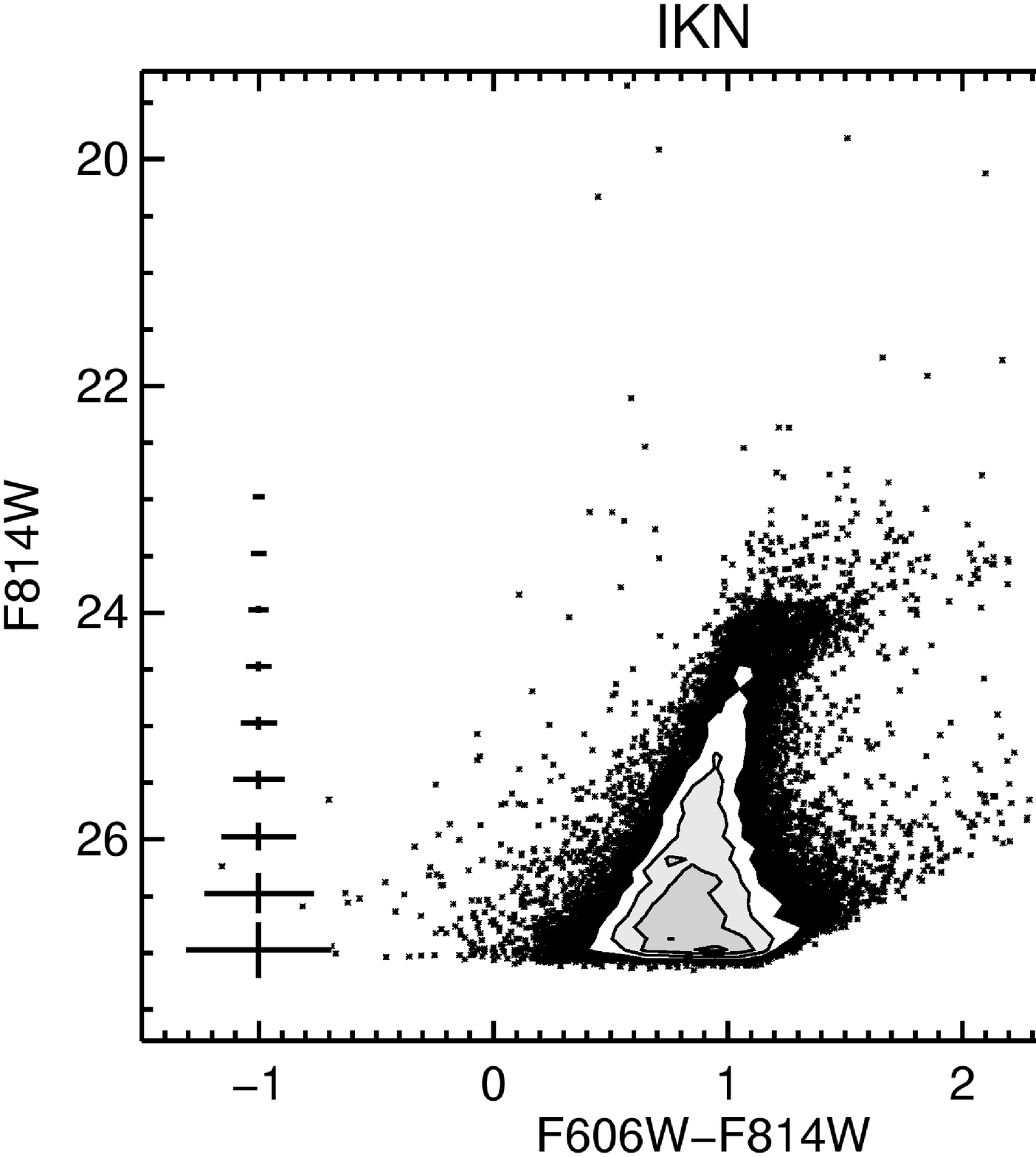}
\includegraphics[width=1.625in]{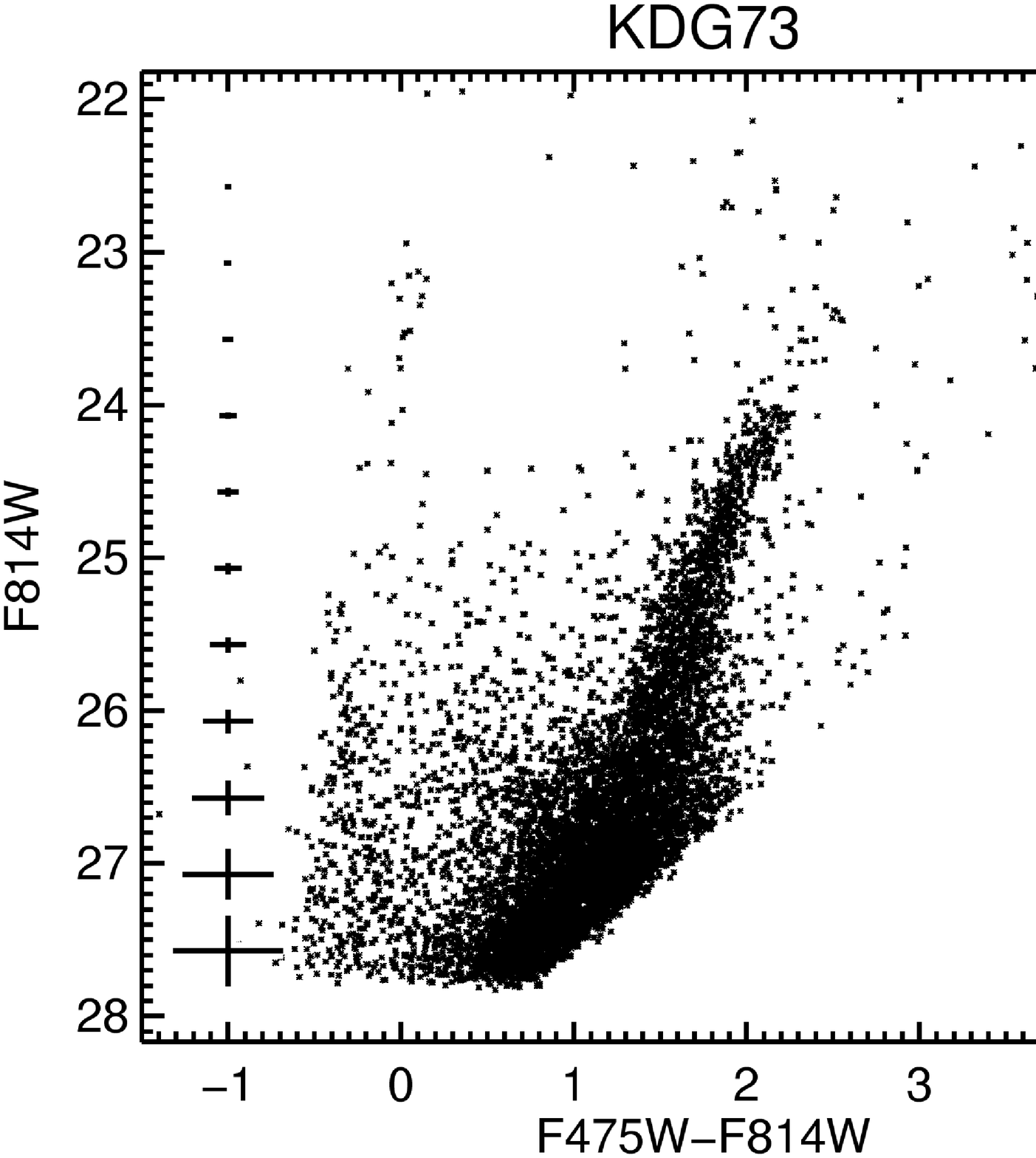}
\includegraphics[width=1.625in]{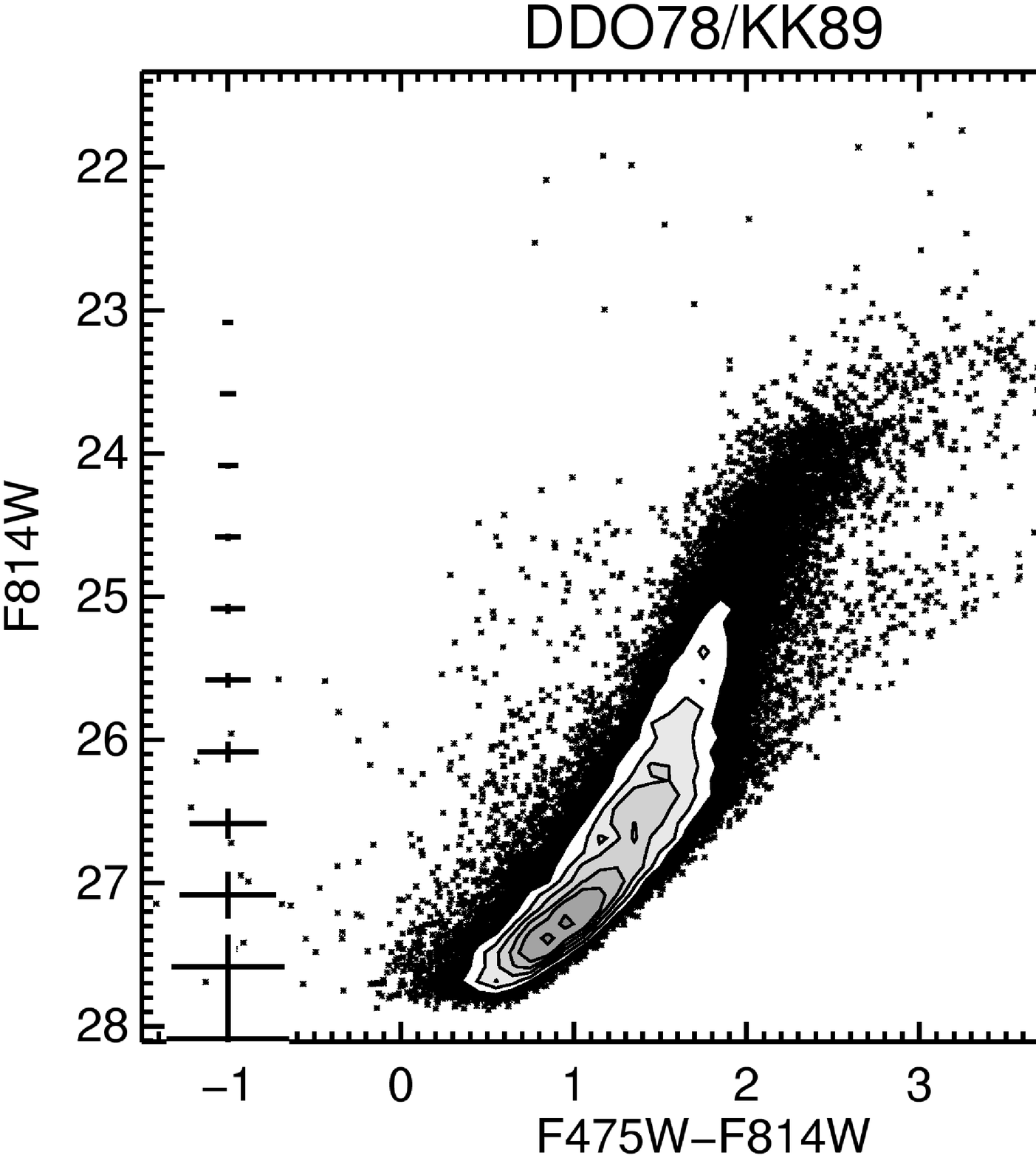}
\includegraphics[width=1.625in]{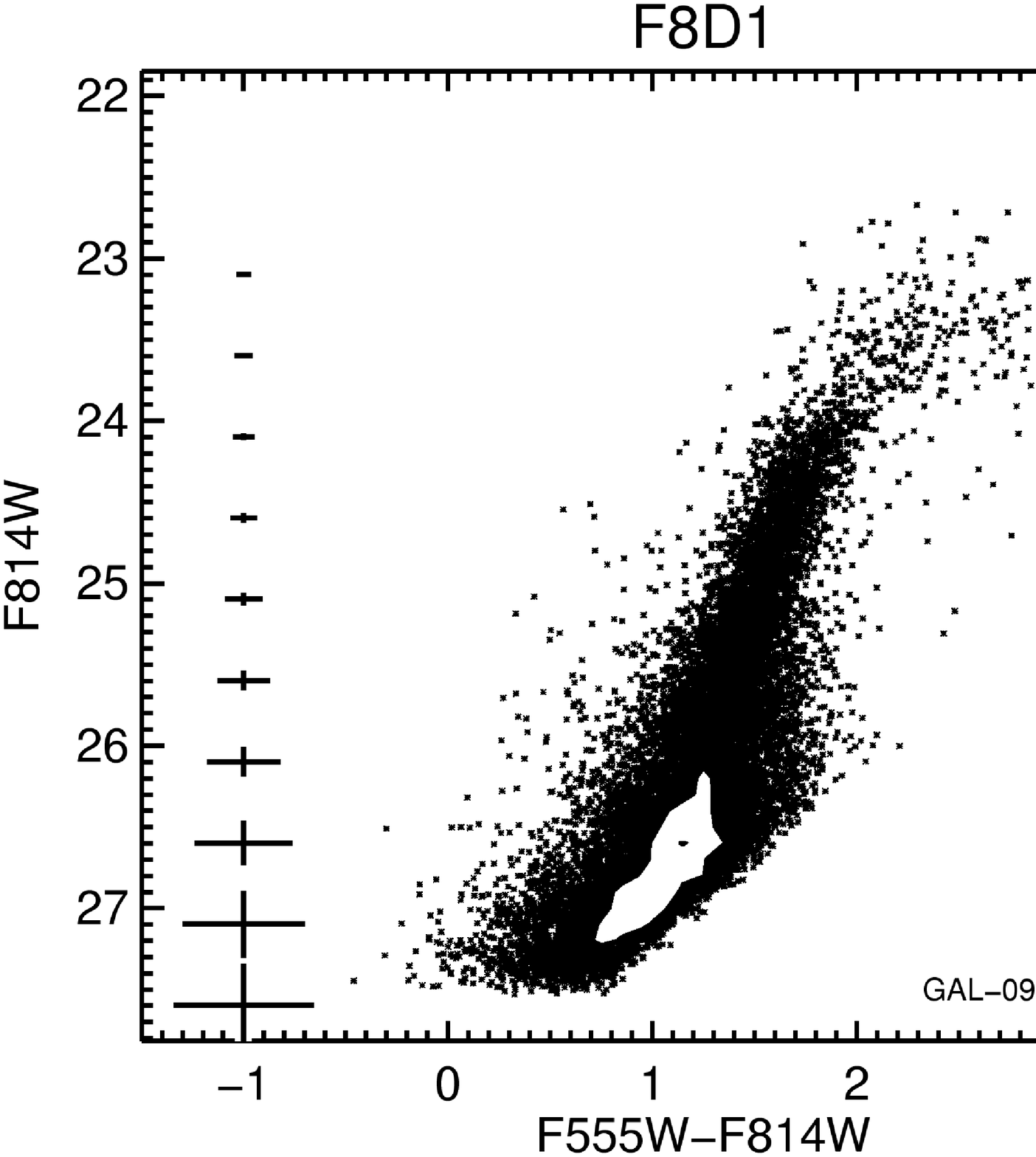}
}
\centerline{
\includegraphics[width=1.625in]{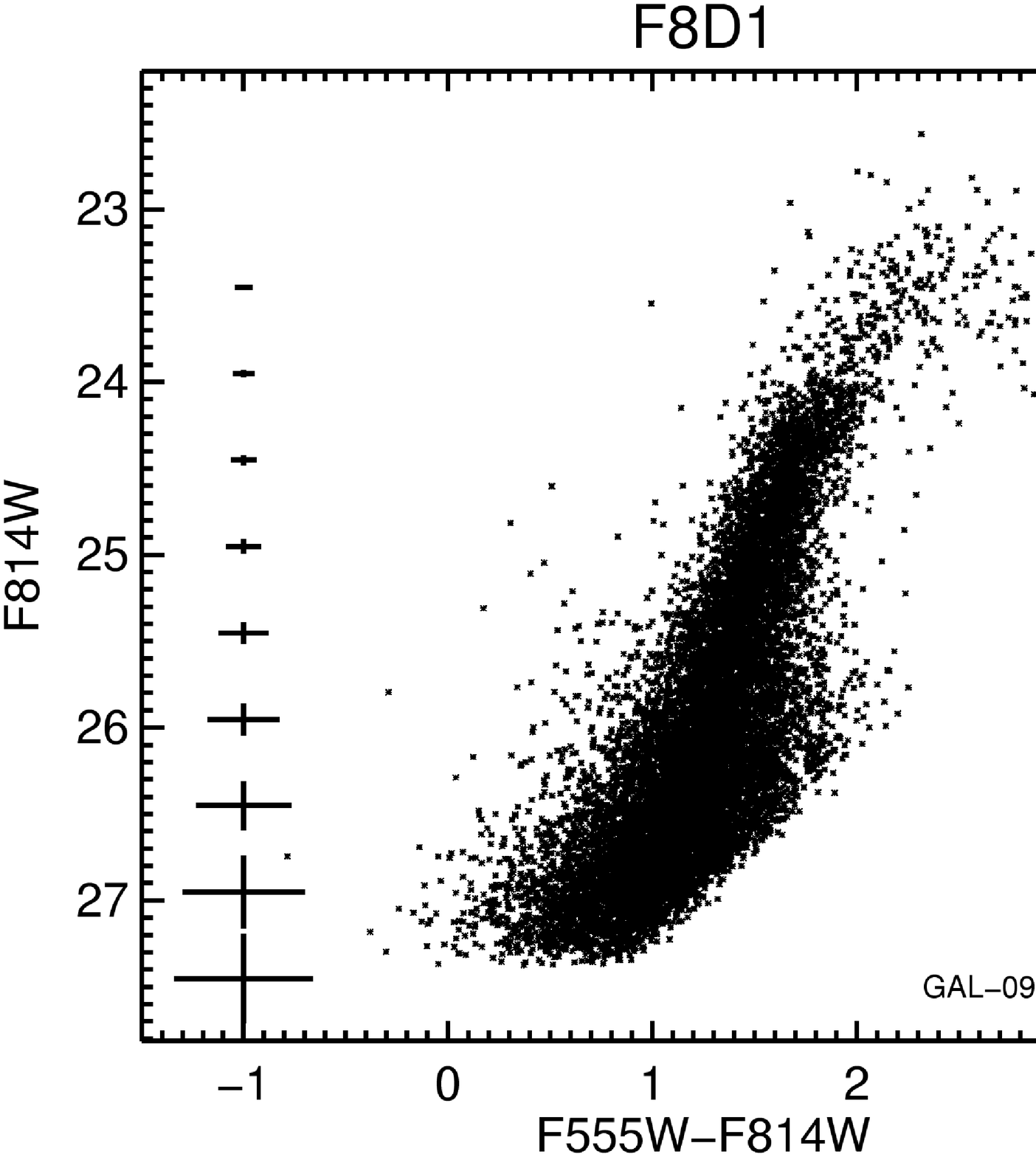}
\includegraphics[width=1.625in]{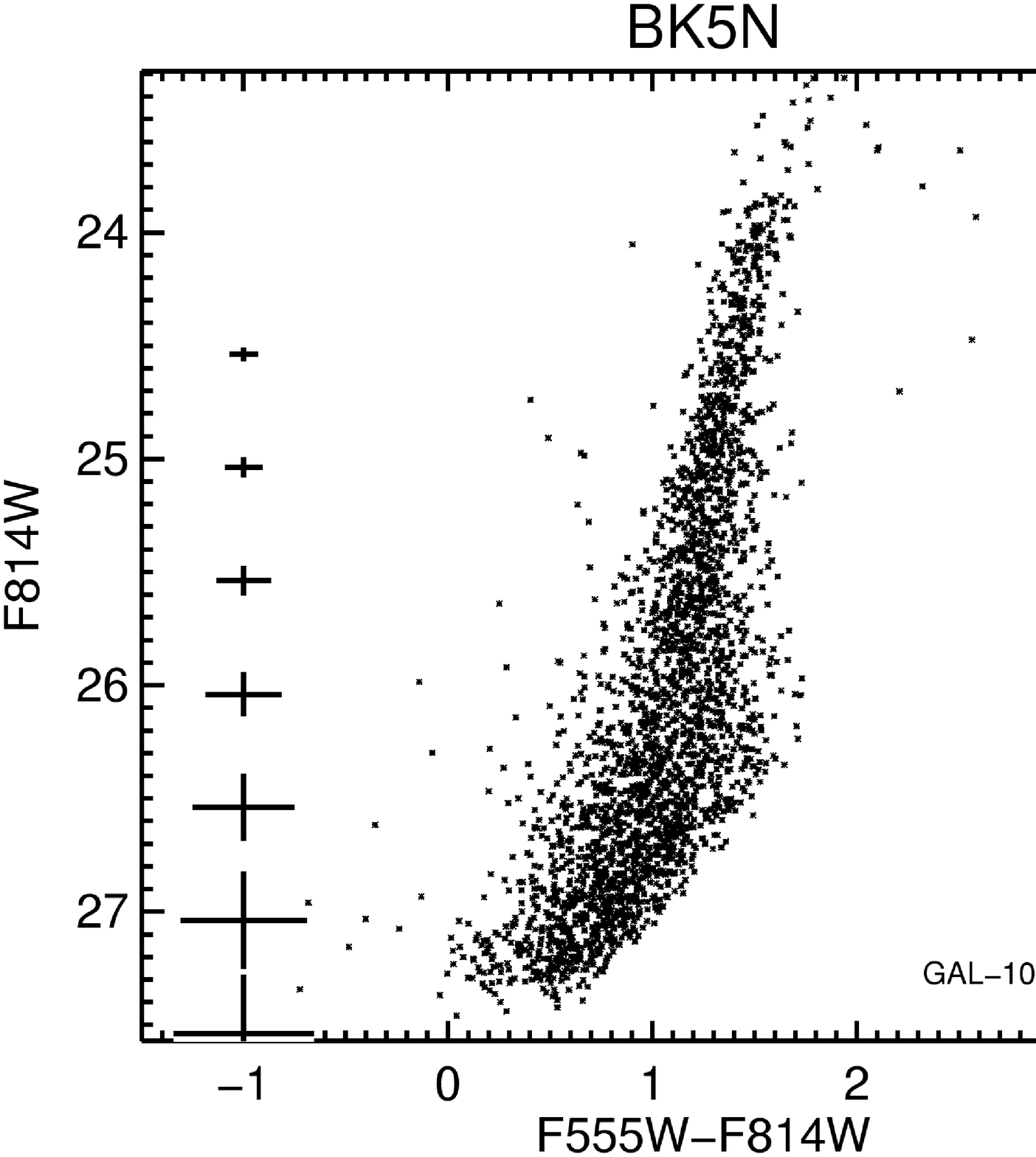}
\includegraphics[width=1.625in]{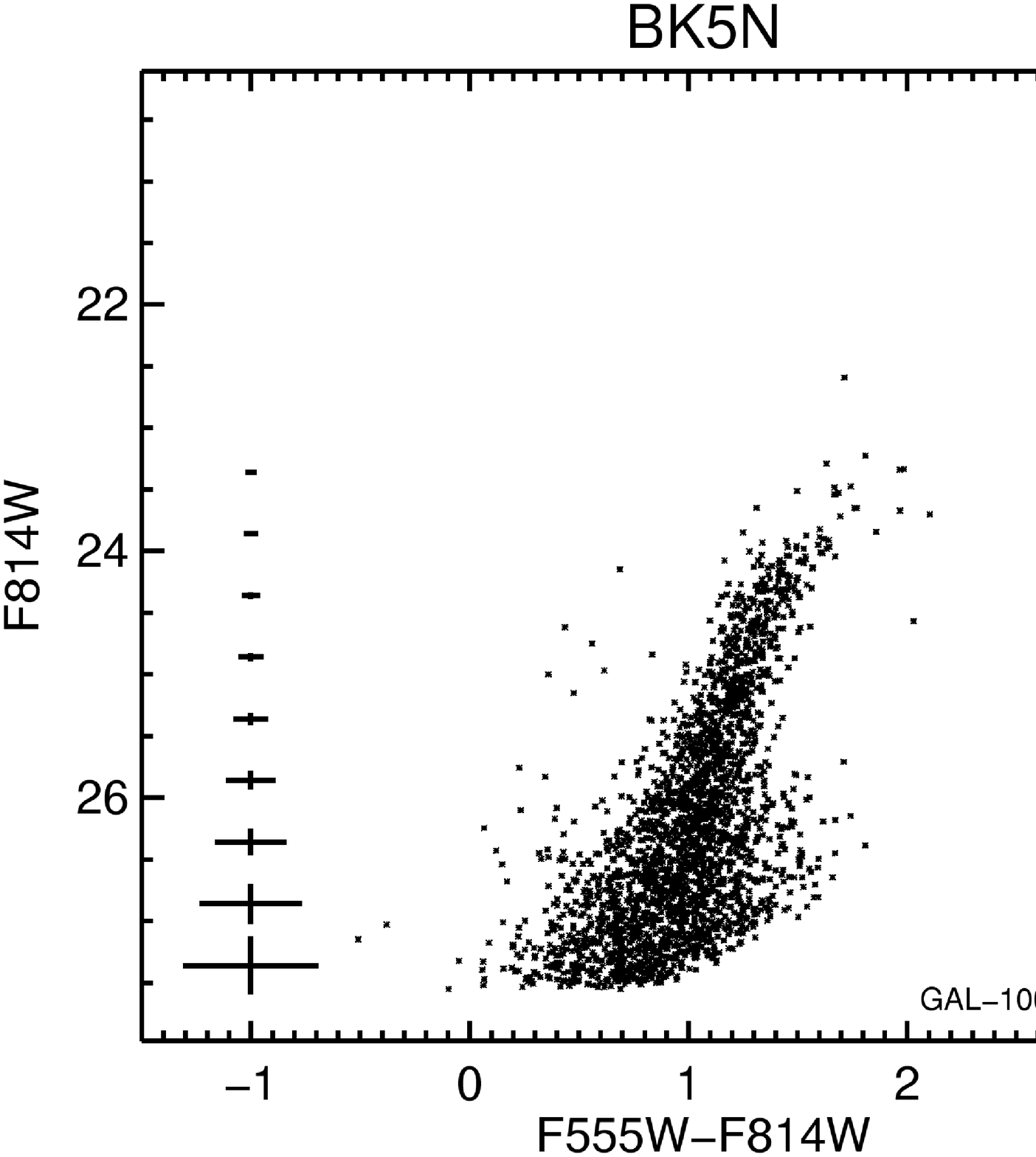}
\includegraphics[width=1.625in]{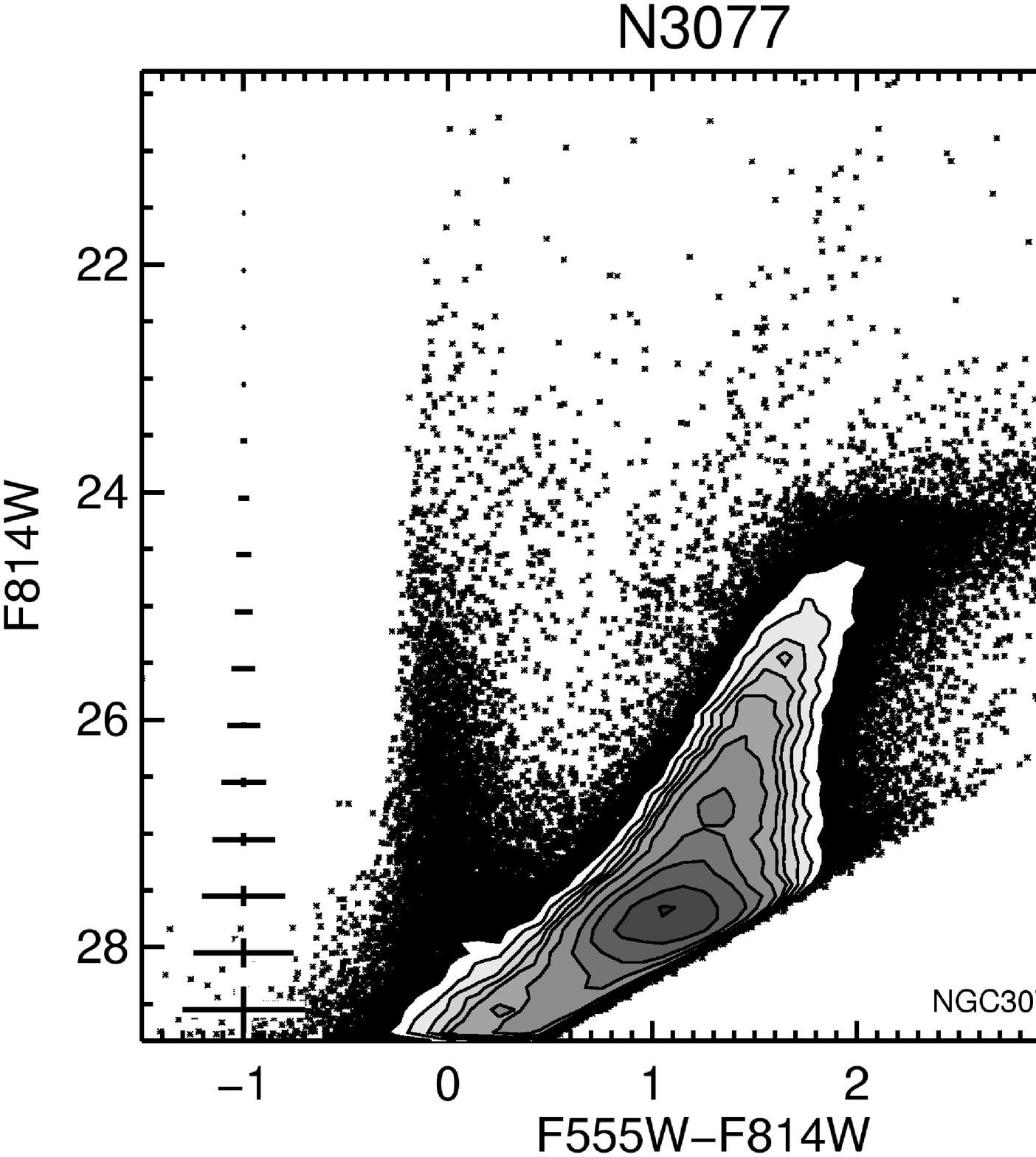}
}
\centerline{
\includegraphics[width=1.625in]{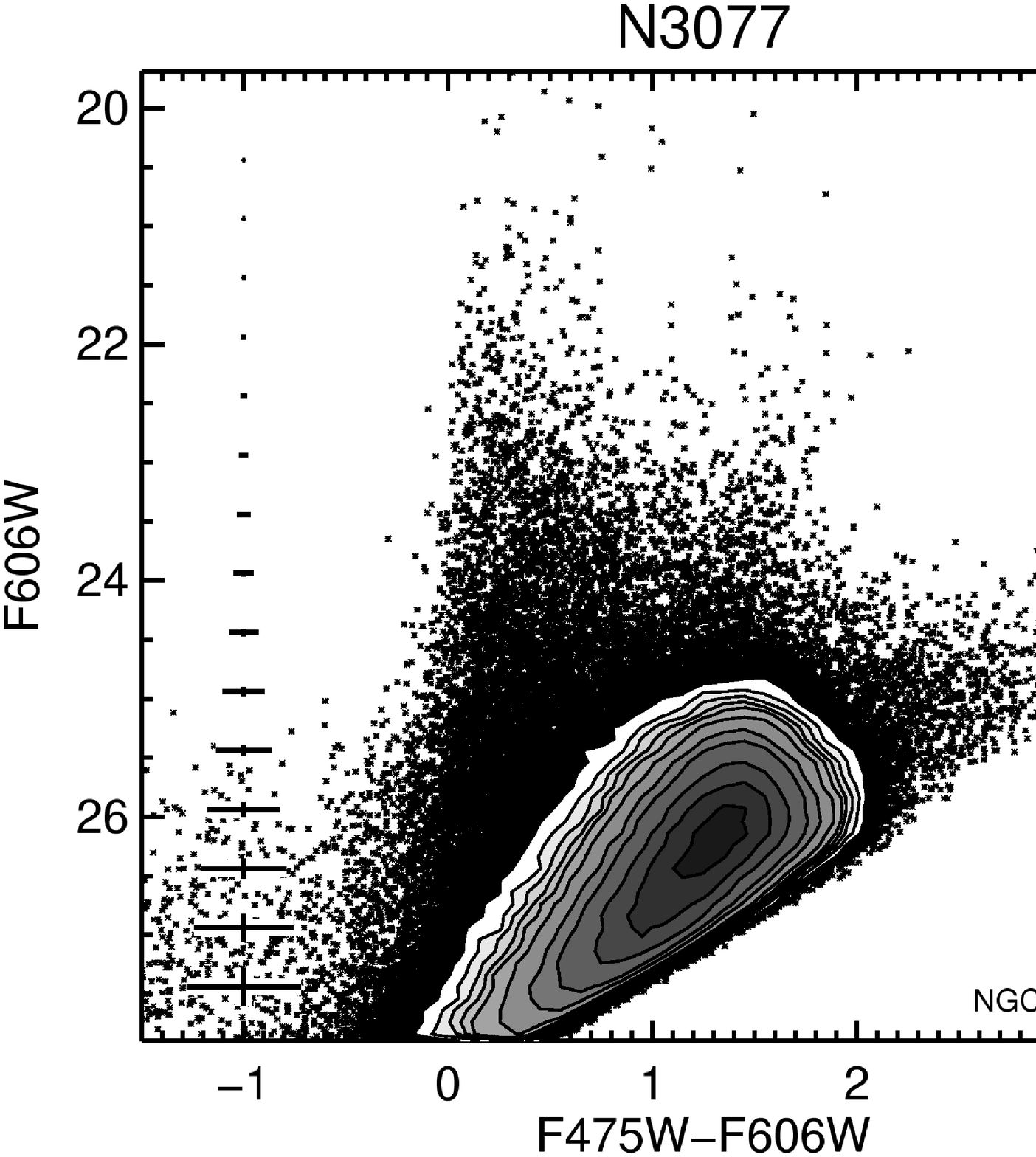}
\includegraphics[width=1.625in]{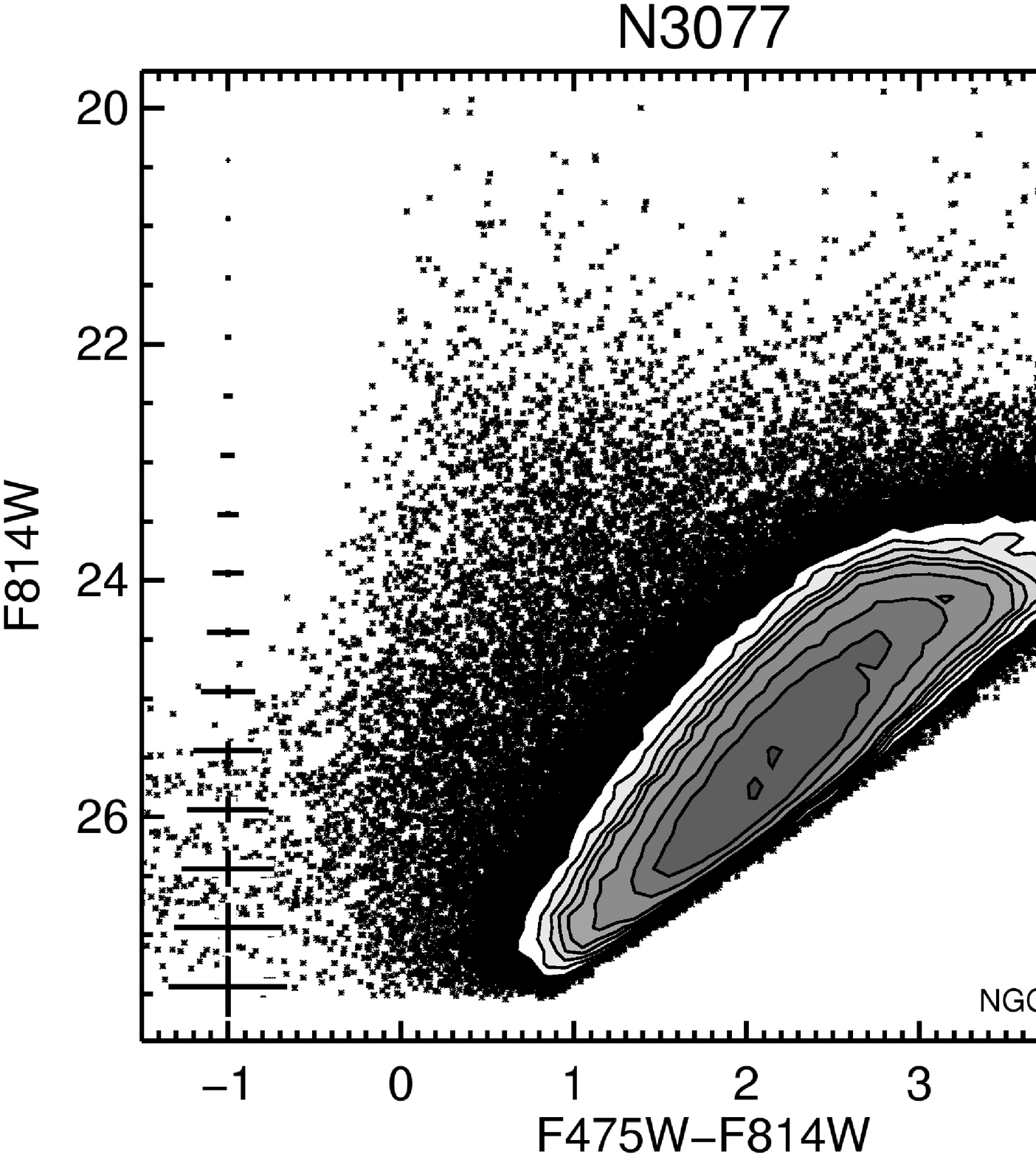}
\includegraphics[width=1.625in]{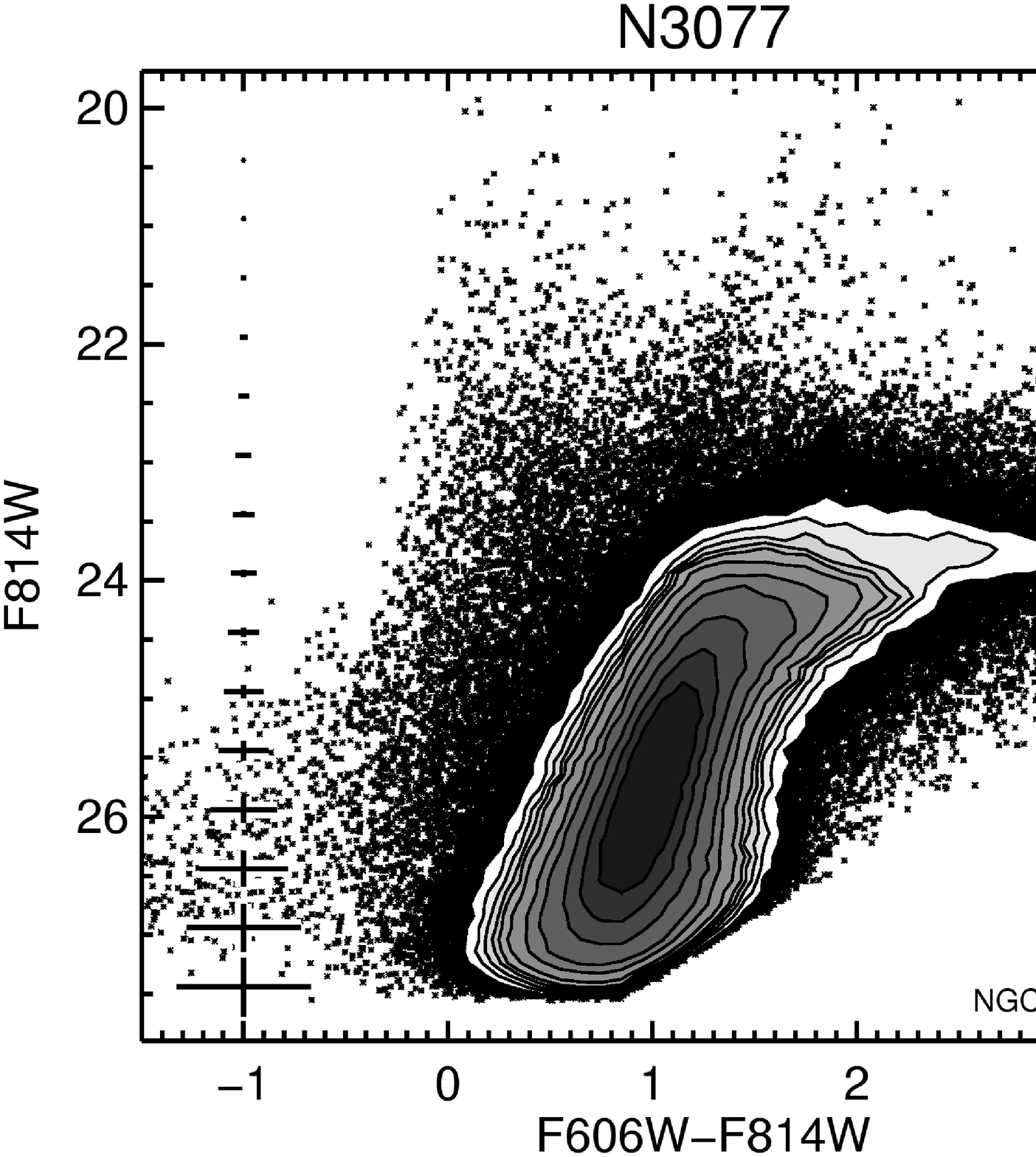}
\includegraphics[width=1.625in]{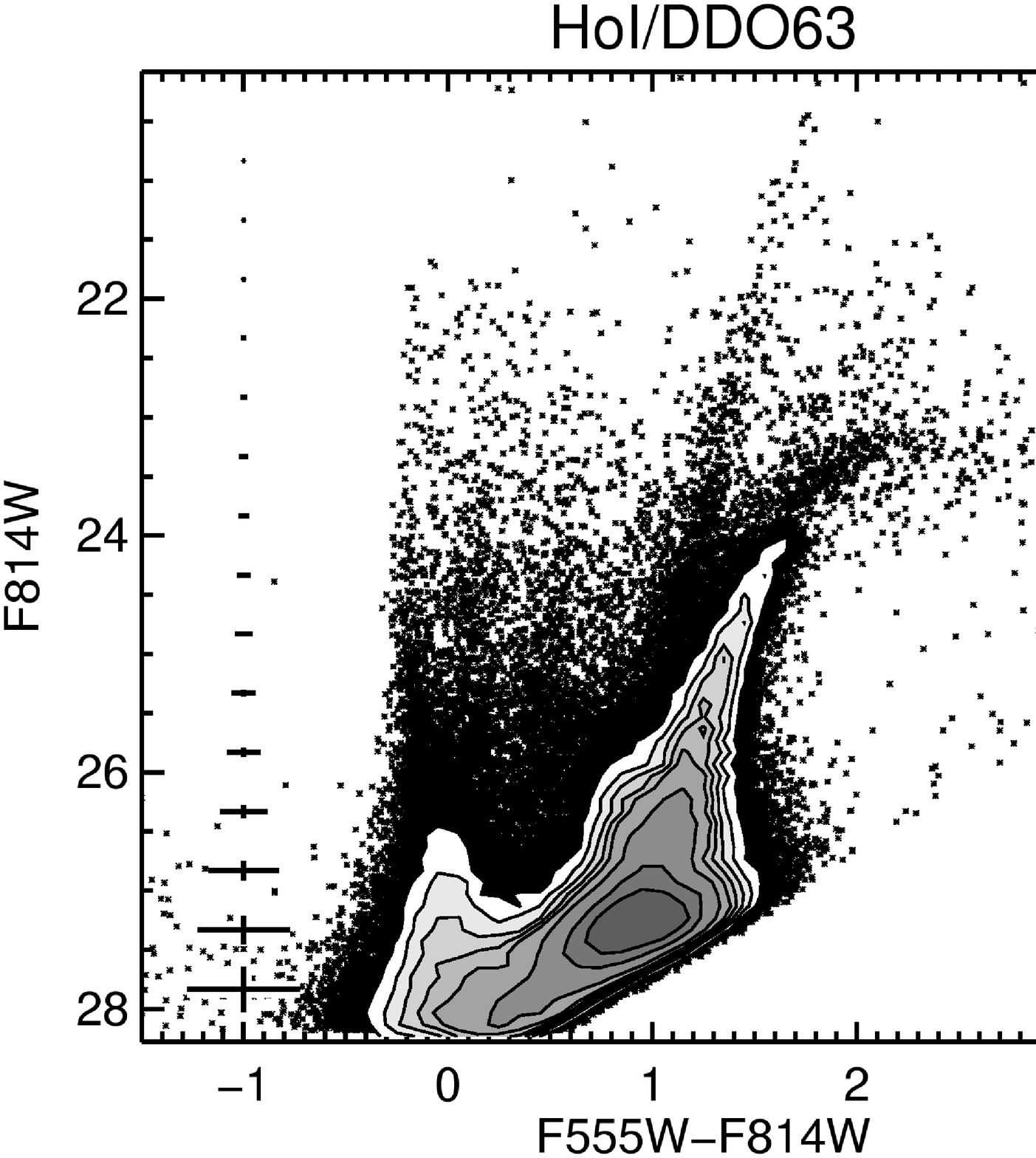}
}
\centerline{
\includegraphics[width=1.625in]{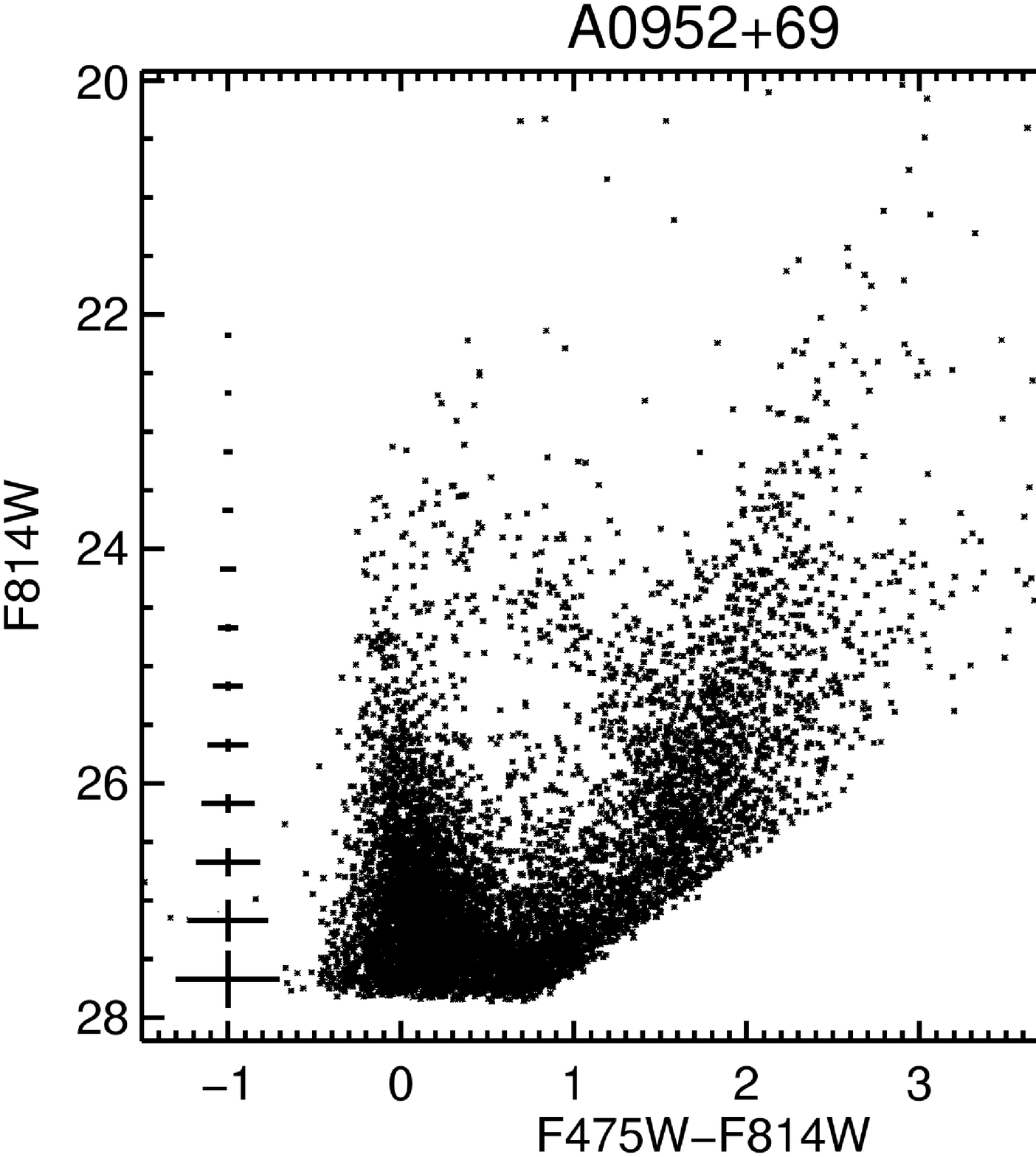}
\includegraphics[width=1.625in]{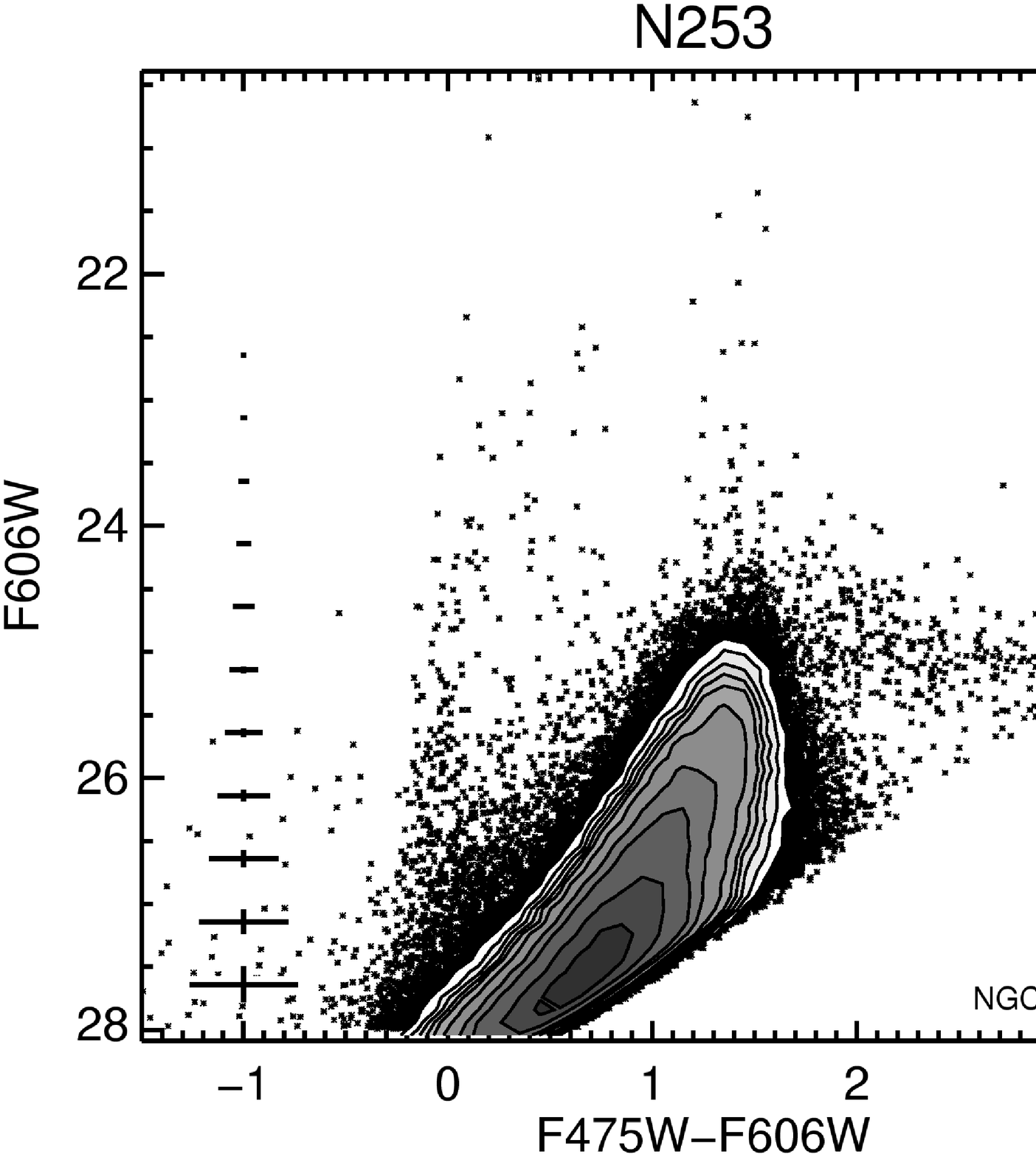}
\includegraphics[width=1.625in]{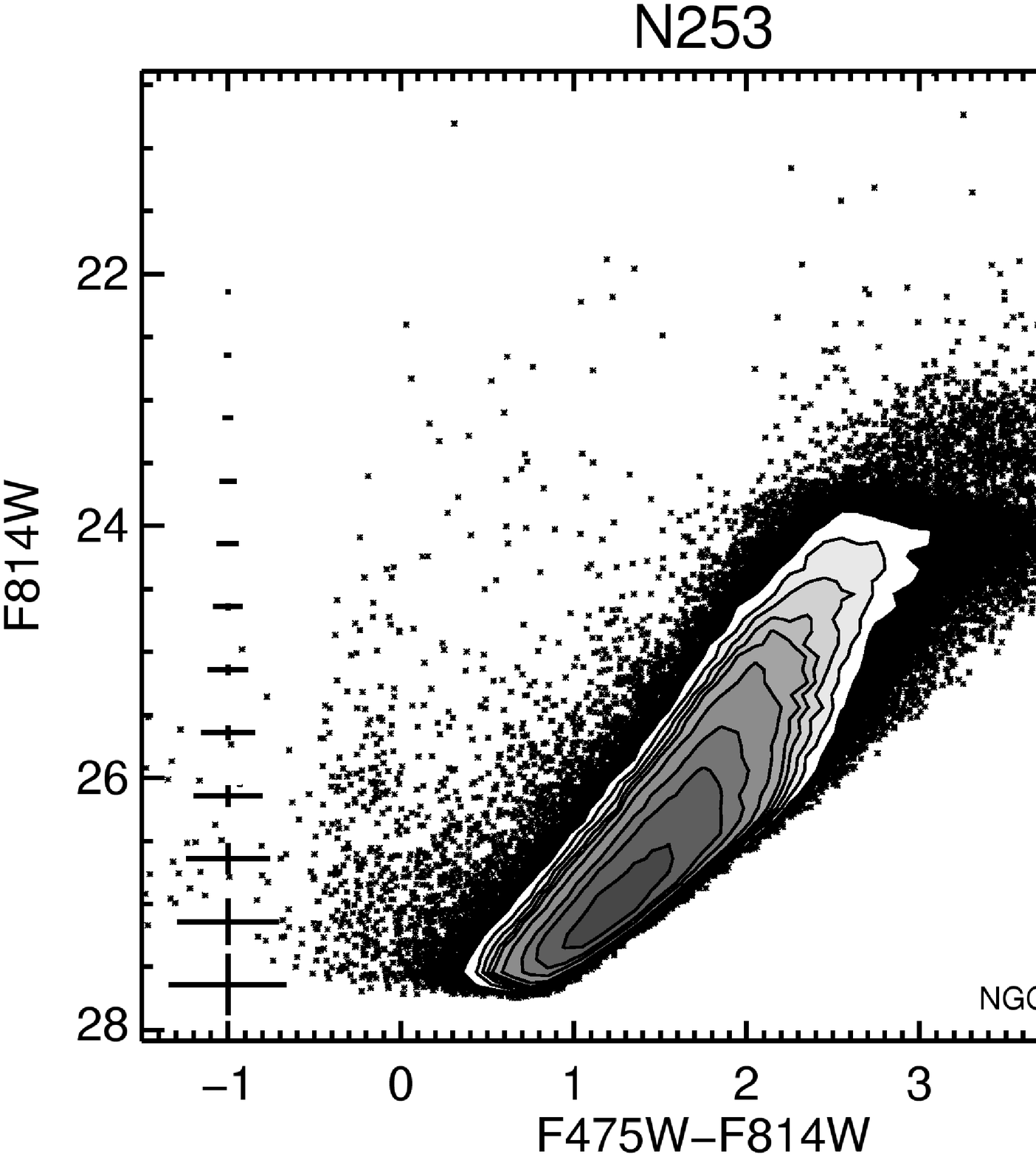}
\includegraphics[width=1.625in]{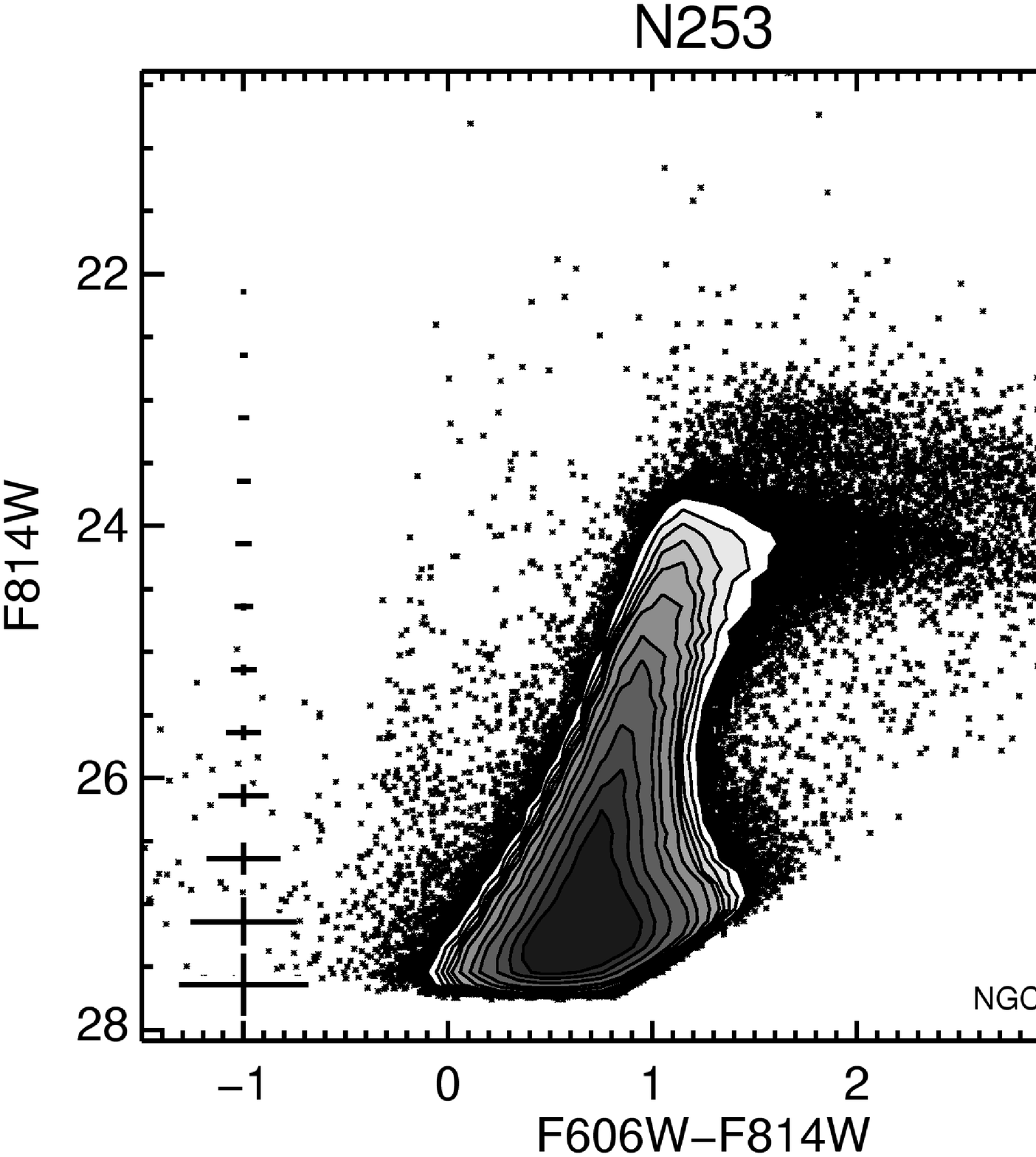}
}
\caption{
CMDs of galaxies in the ANGST data release,
as described in Figure~\ref{cmdfig1}.
Figures are ordered from the upper left to the bottom right.
(a) IKN; (b) KDG73; (c) DDO78; (d) F8D1; (e) F8D1; (f) BK5N; (g) BK5N; (h) N3077; (i) N3077; (j) N3077; (k) N3077; (l) HoI; (m) A0952+69; (n) N253; (o) N253; (p) N253; 
    \label{cmdfig12}}
\end{figure}
\vfill
\clearpage
 
%-------------------
\begin{figure}[p]
\centerline{
\includegraphics[width=1.625in]{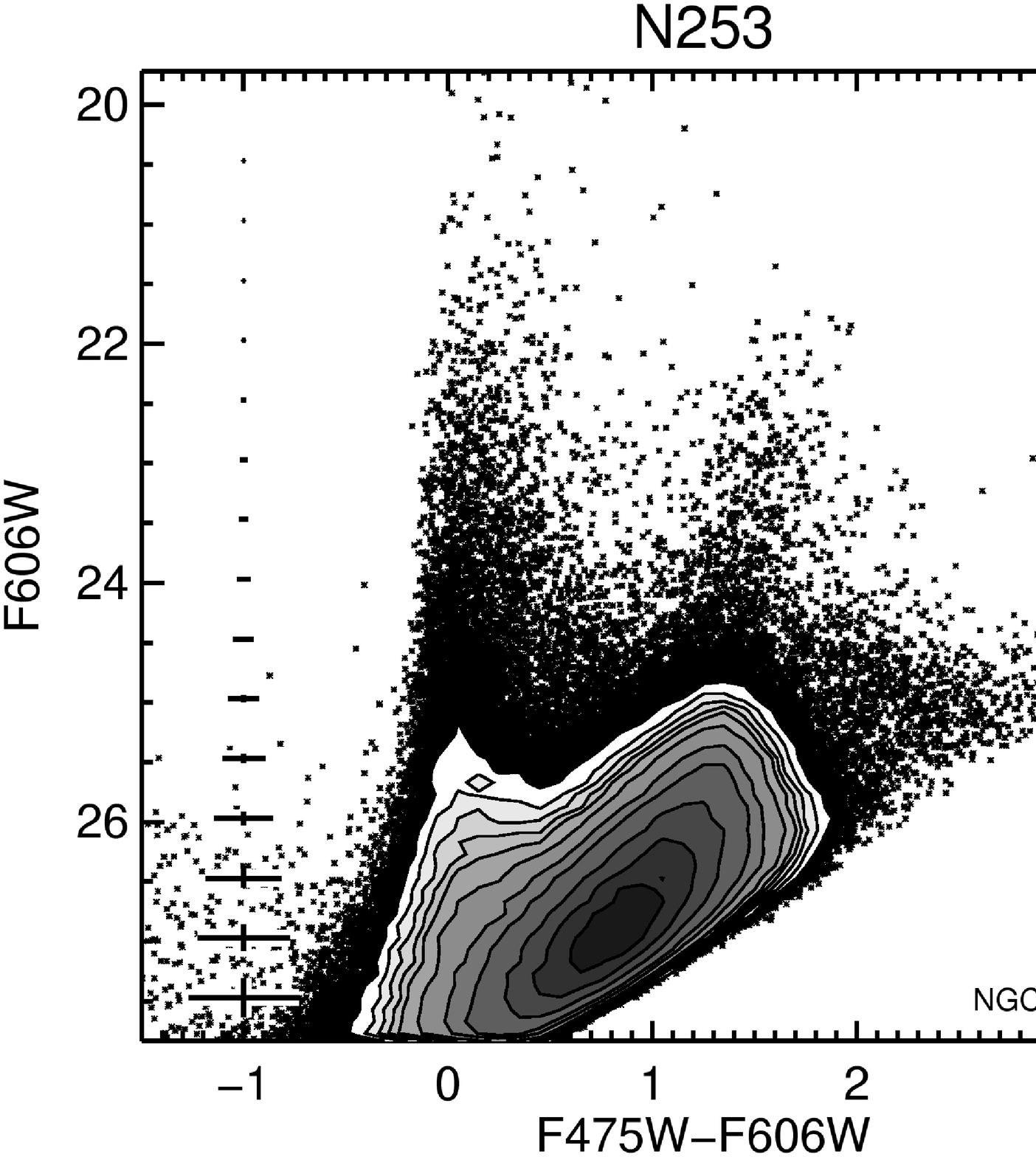}
\includegraphics[width=1.625in]{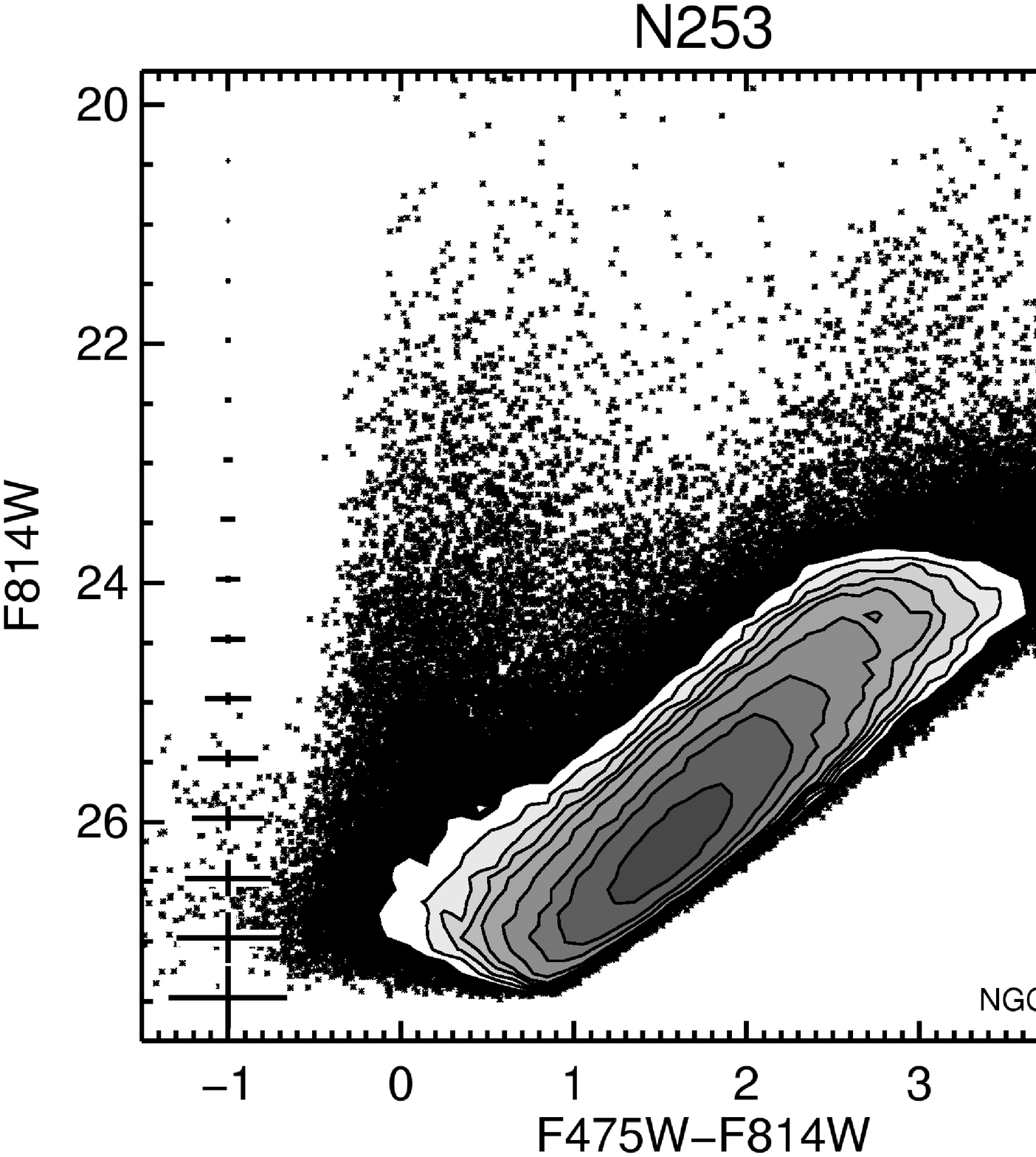}
\includegraphics[width=1.625in]{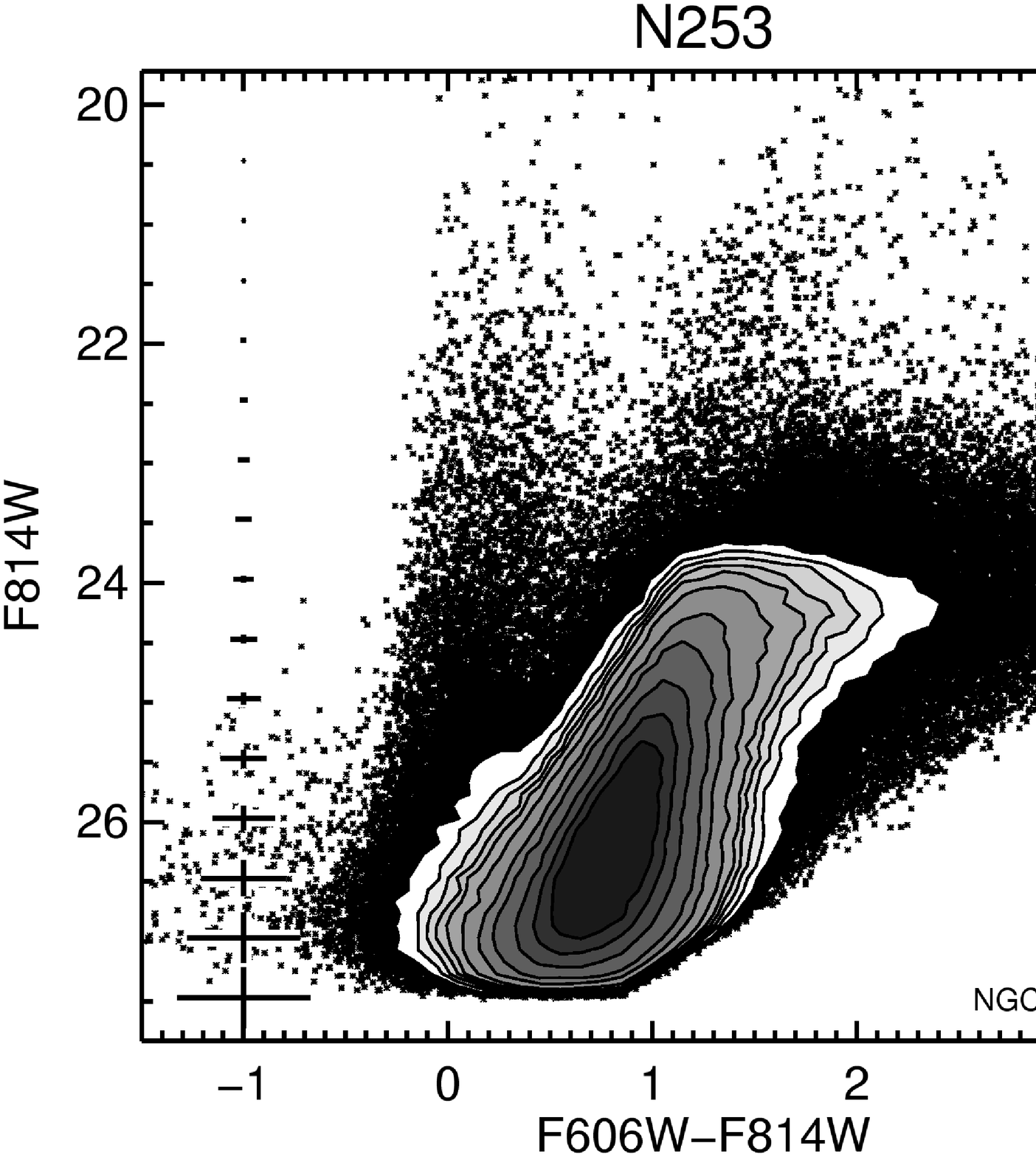}
\includegraphics[width=1.625in]{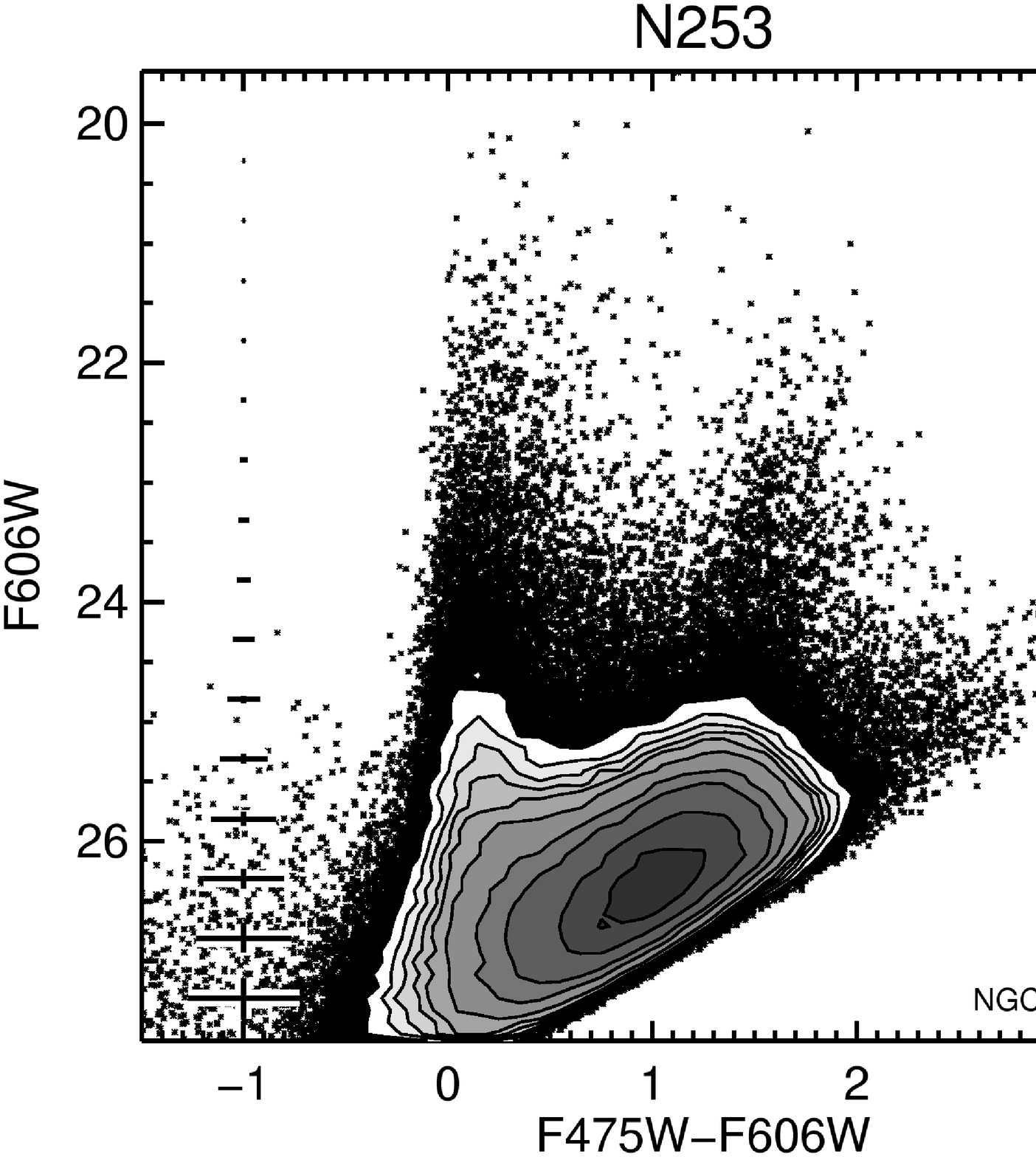}
}
\centerline{
\includegraphics[width=1.625in]{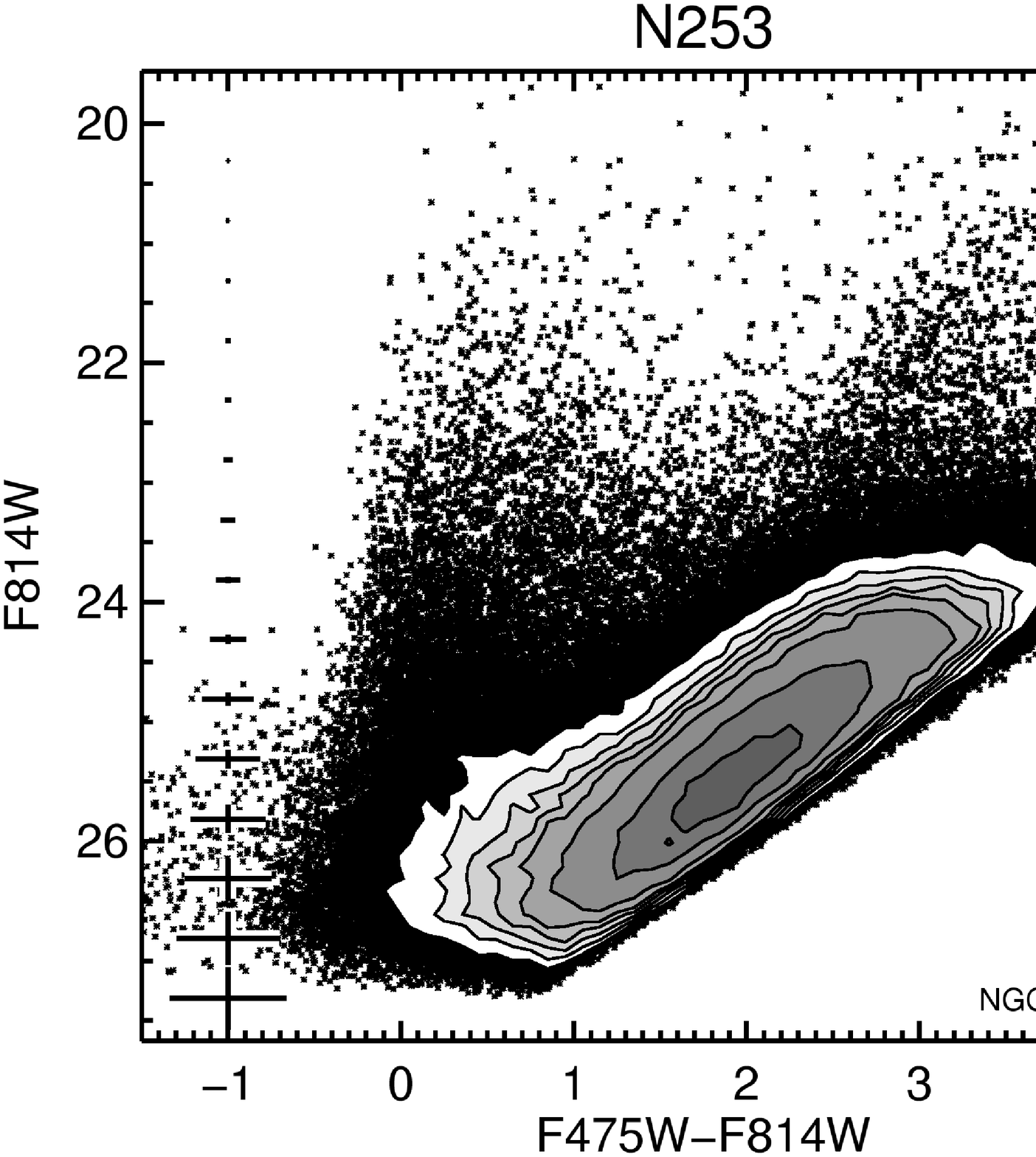}
\includegraphics[width=1.625in]{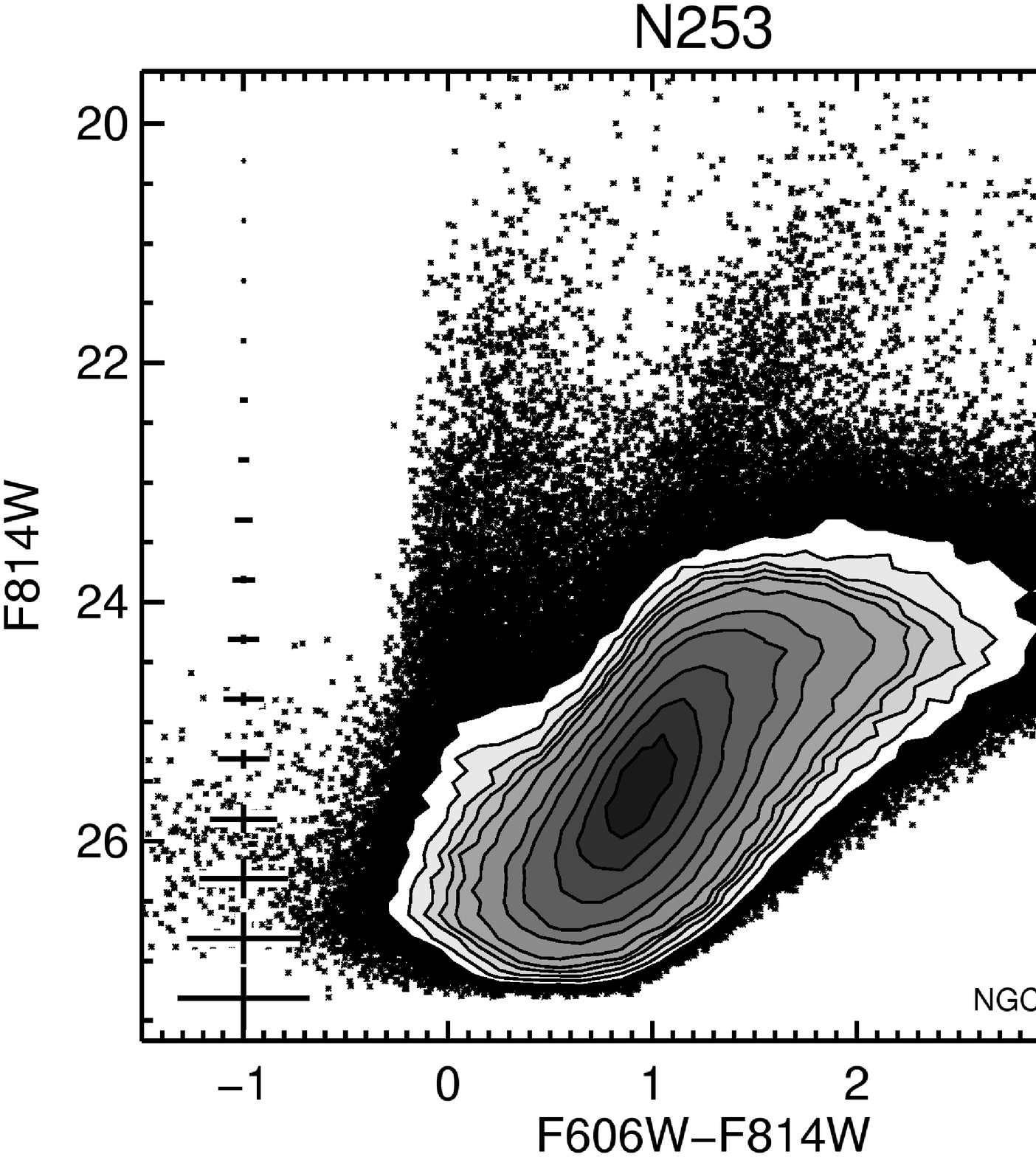}
\includegraphics[width=1.625in]{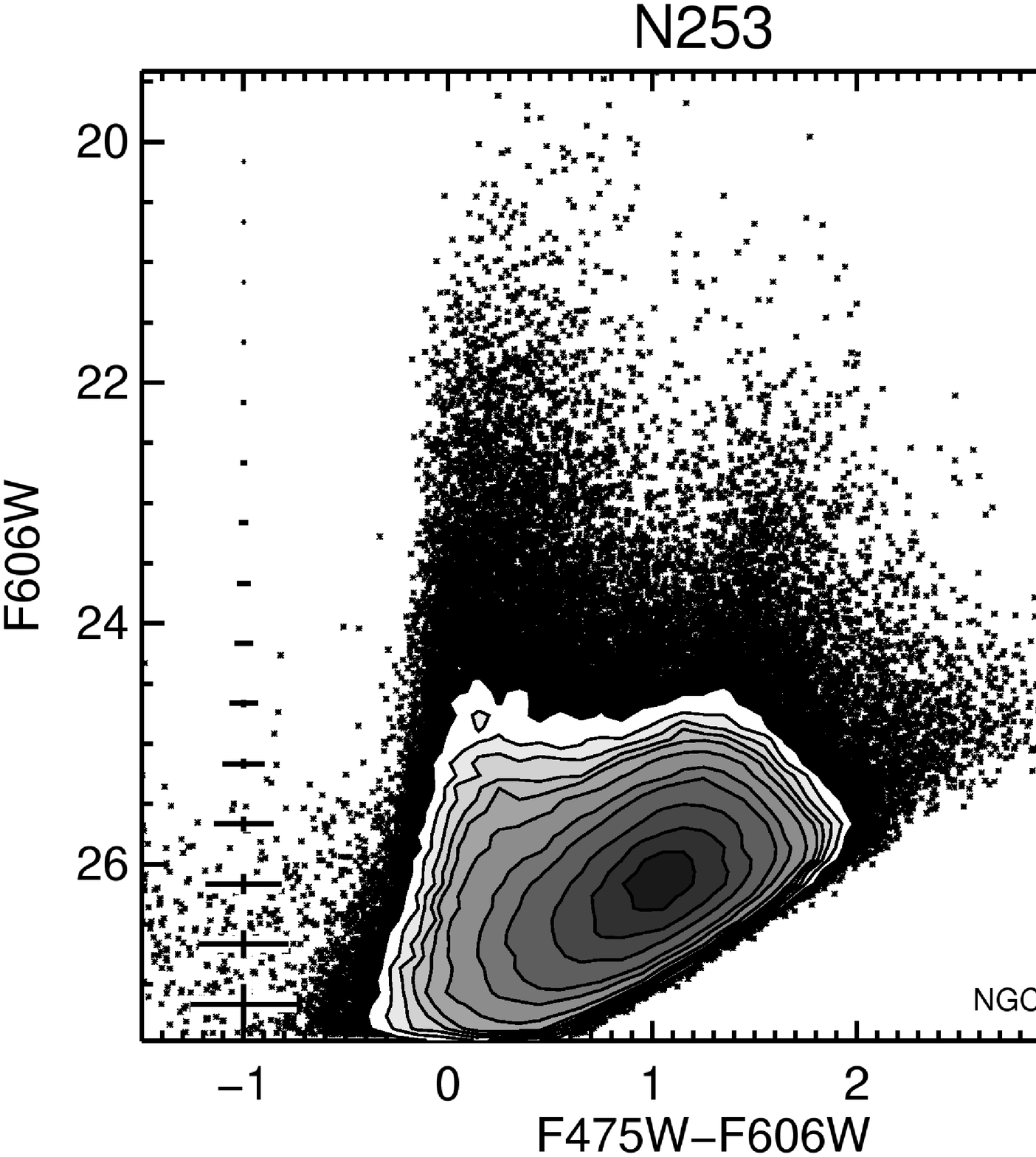}
\includegraphics[width=1.625in]{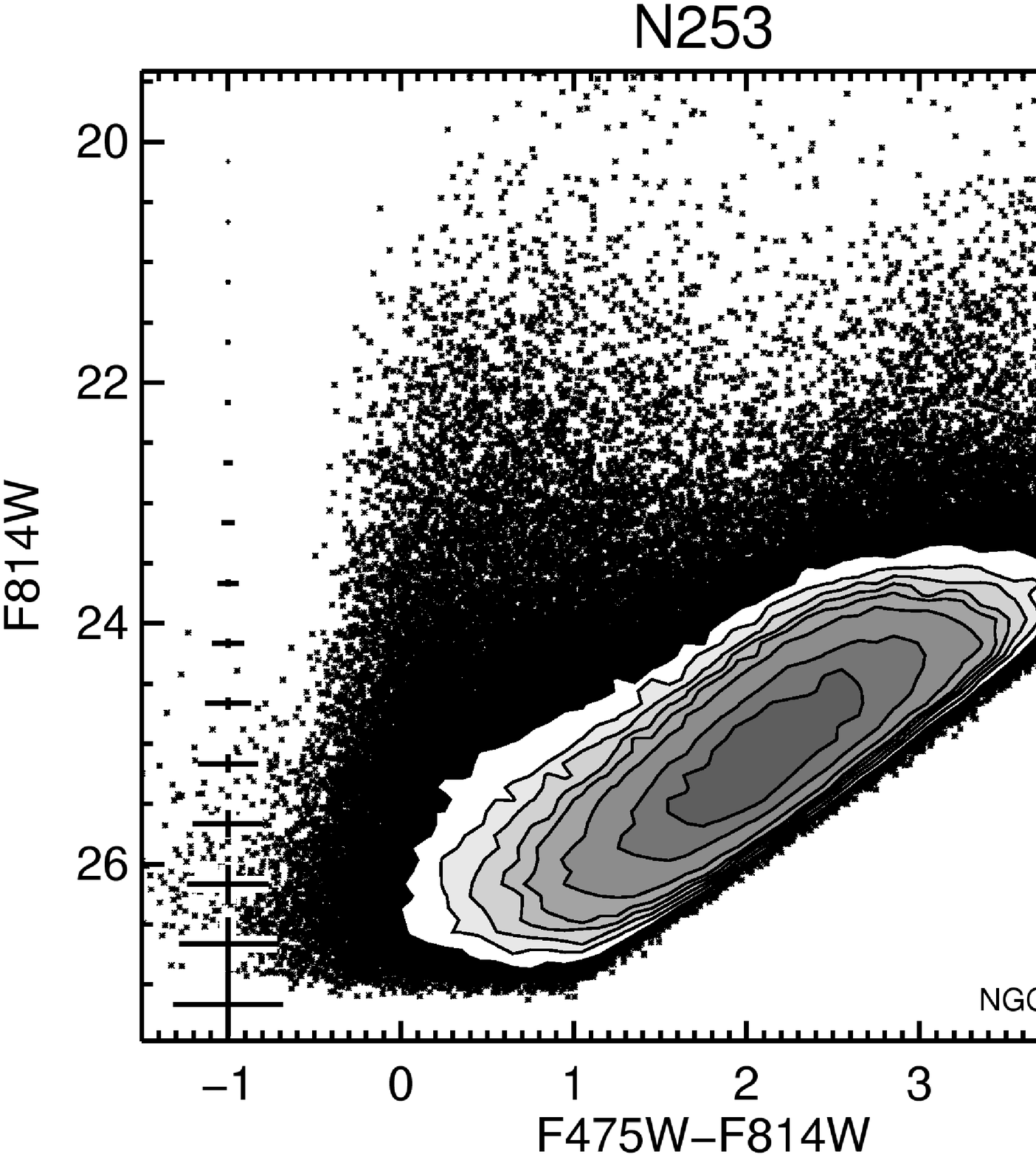}
}
\centerline{
\includegraphics[width=1.625in]{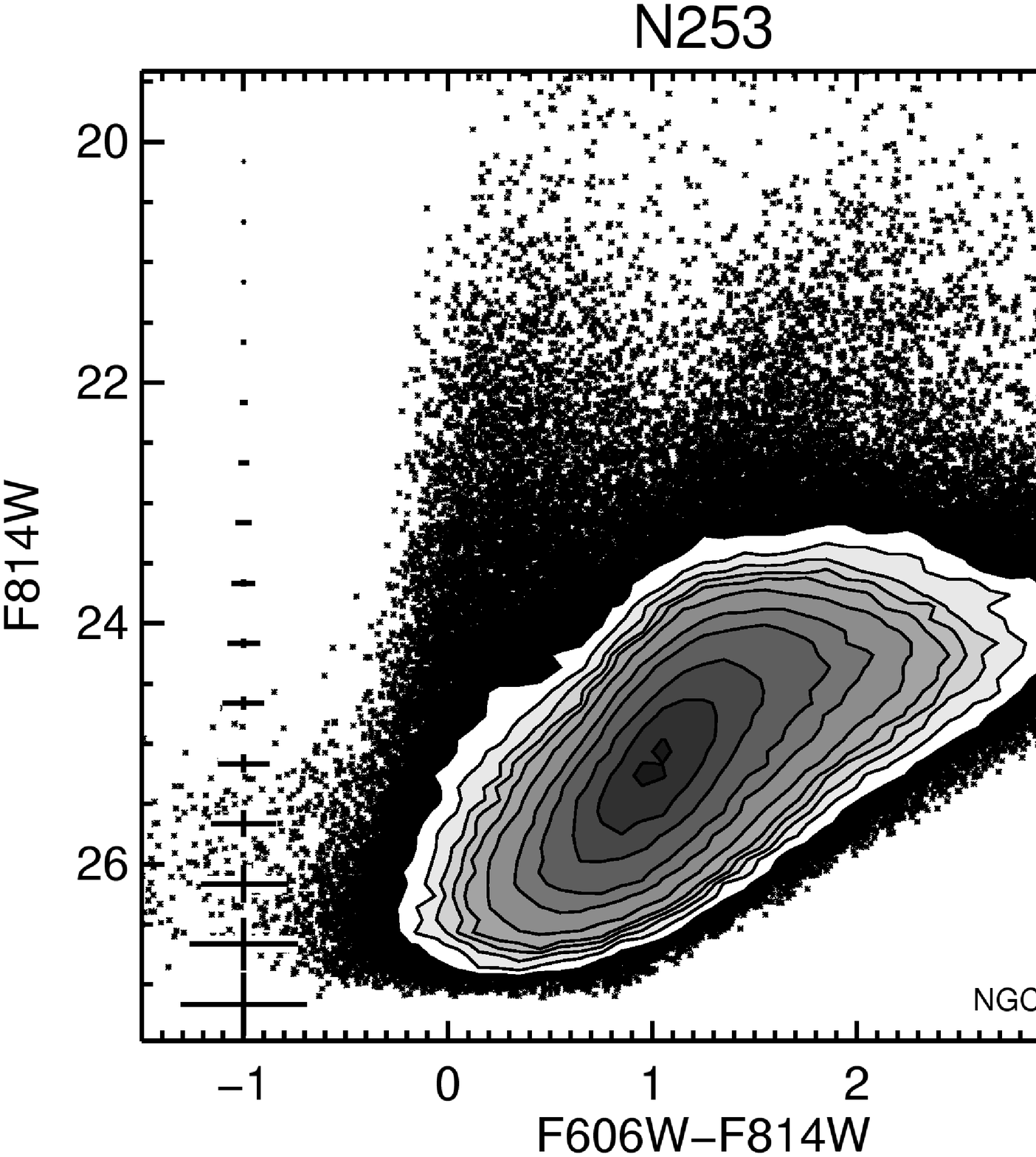}
\includegraphics[width=1.625in]{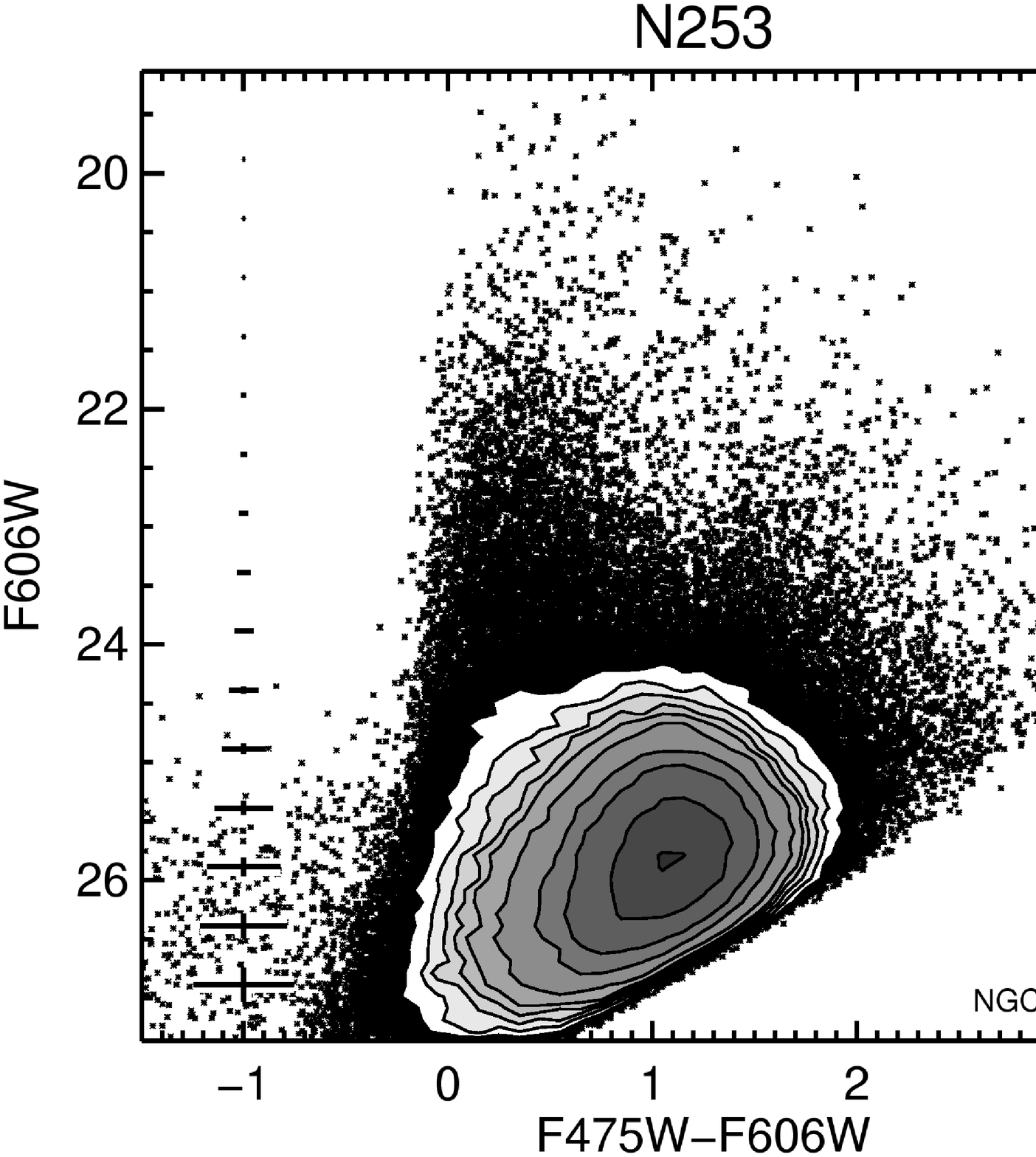}
\includegraphics[width=1.625in]{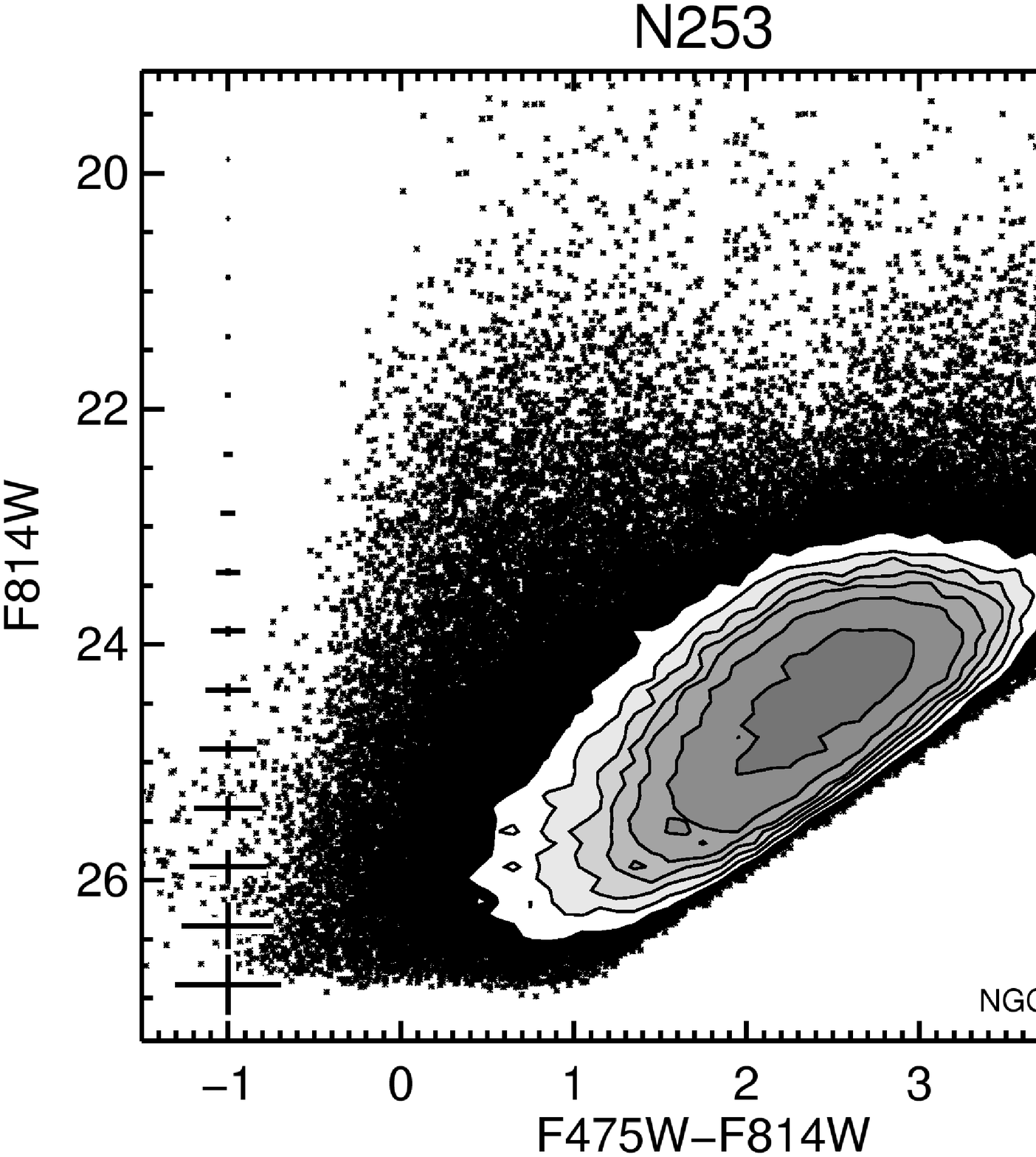}
\includegraphics[width=1.625in]{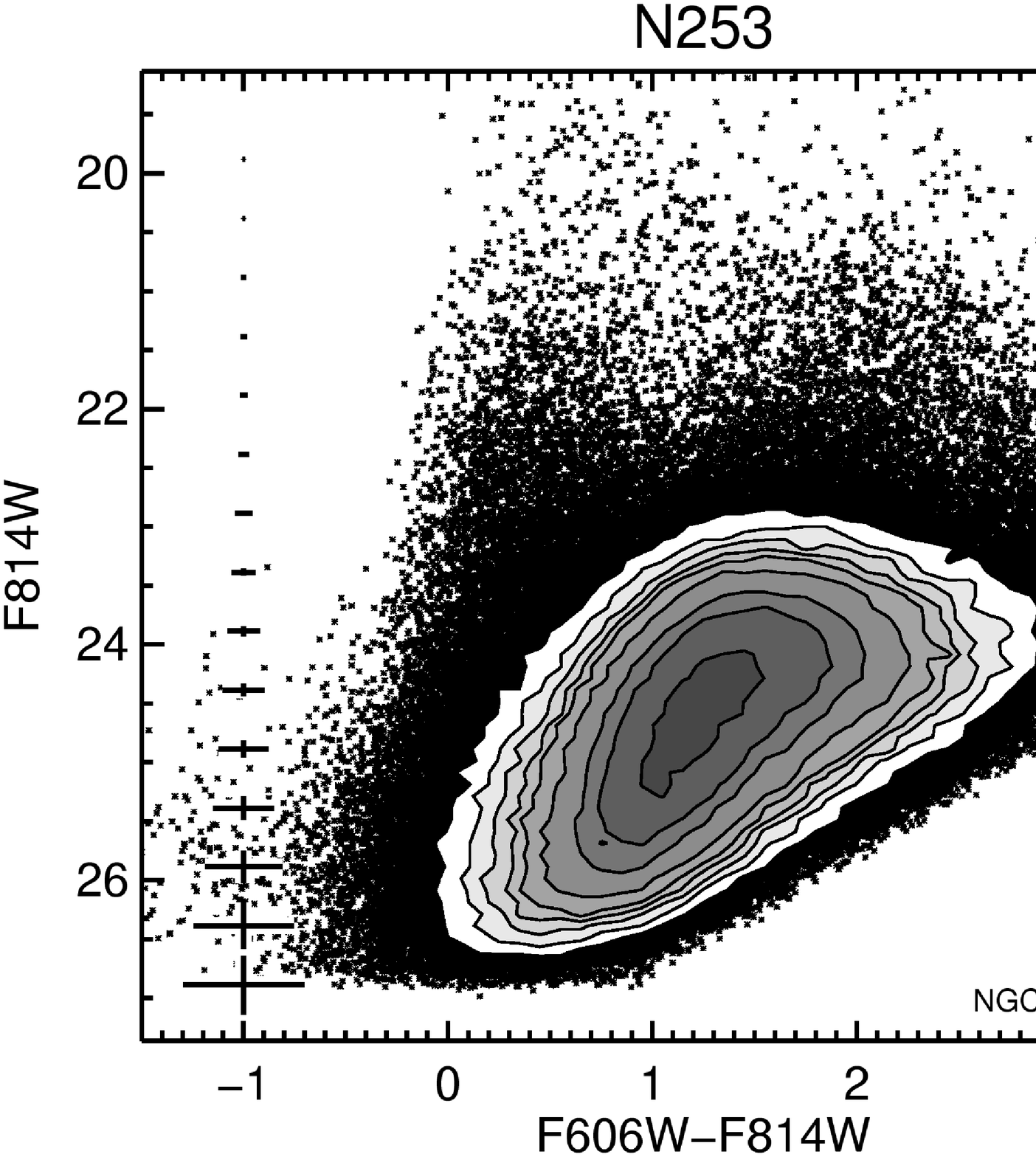}
}
\centerline{
\includegraphics[width=1.625in]{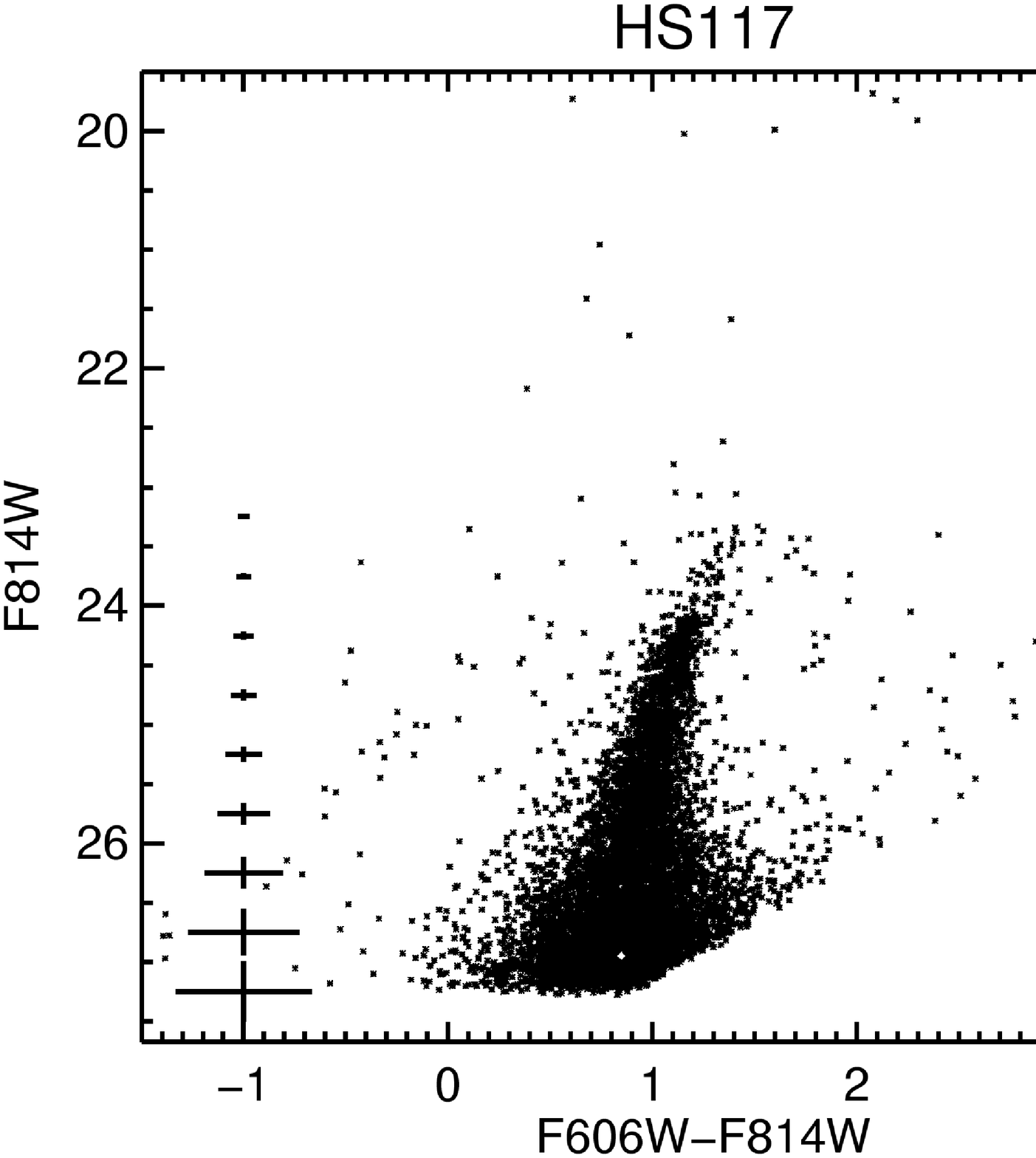}
\includegraphics[width=1.625in]{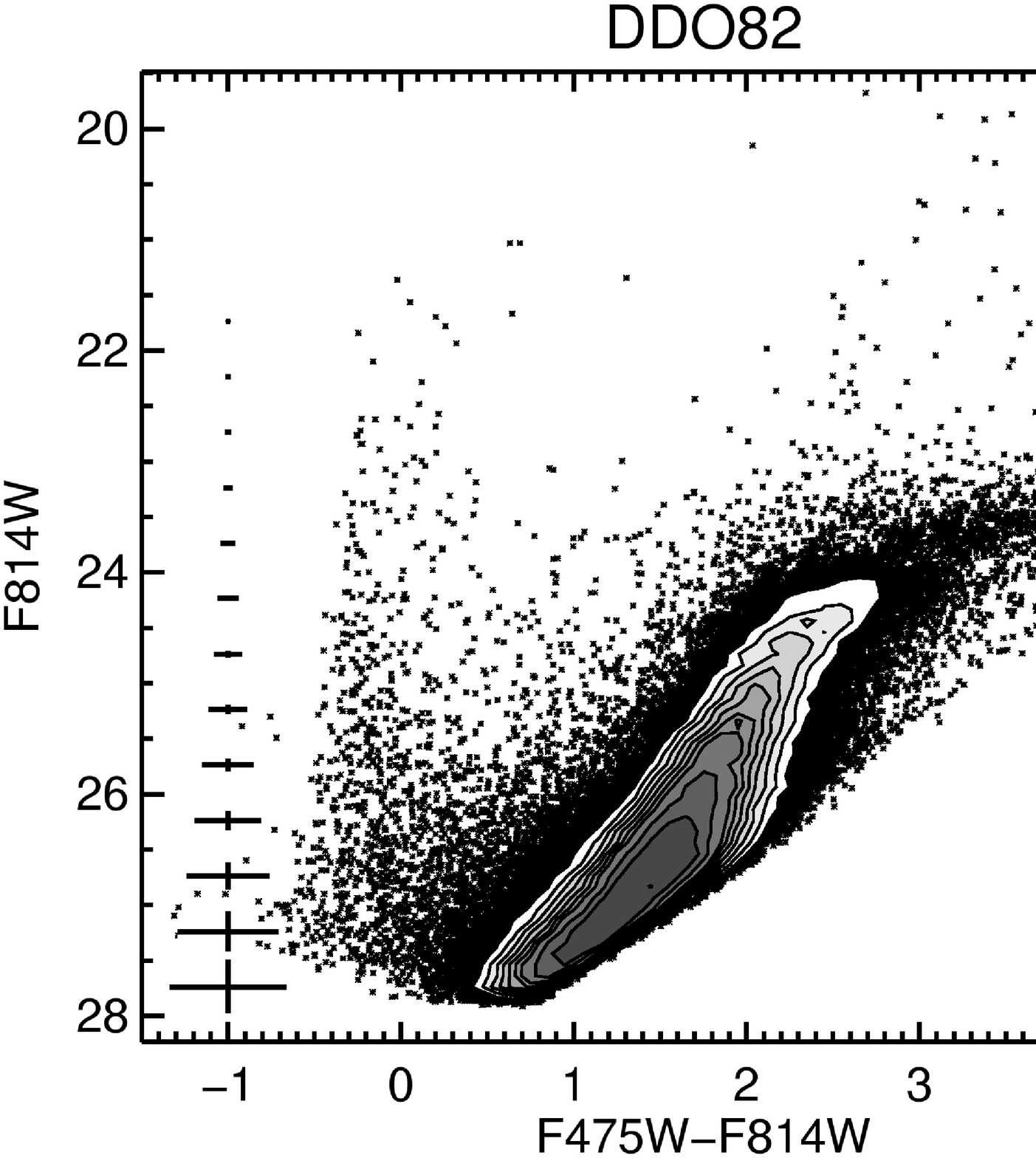}
\includegraphics[width=1.625in]{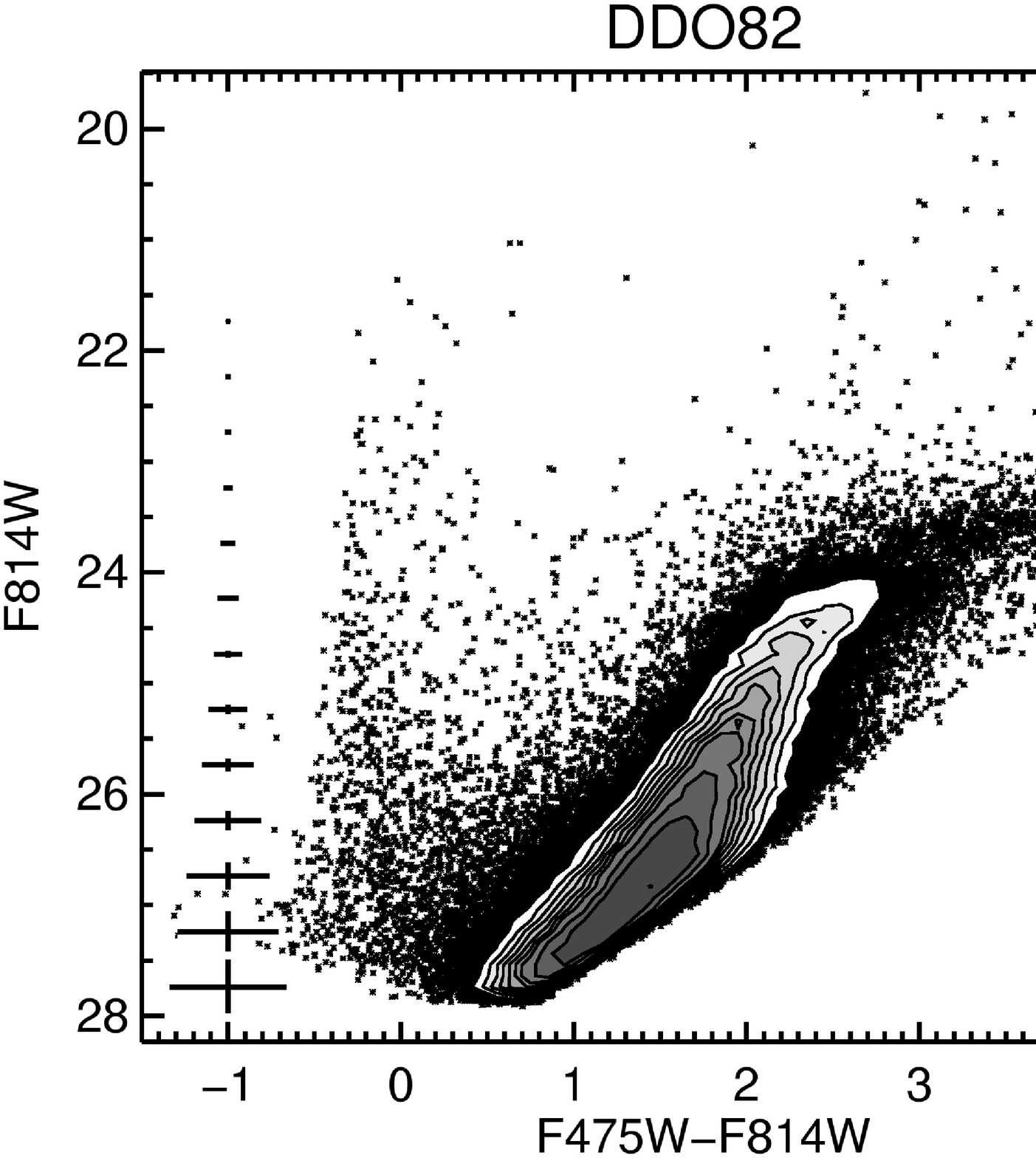}
\includegraphics[width=1.625in]{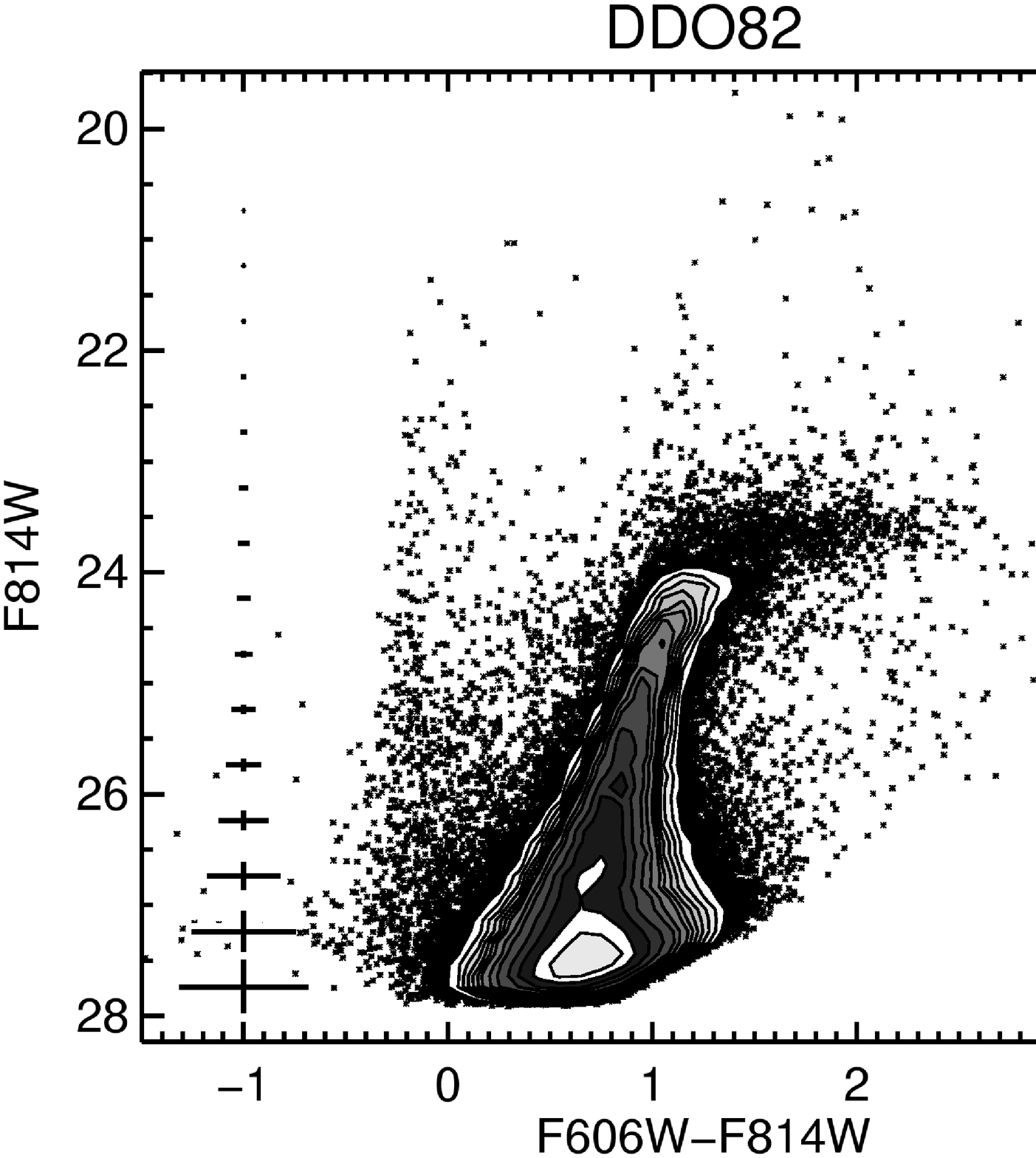}
}
\caption{
CMDs of galaxies in the ANGST data release,
as described in Figure~\ref{cmdfig1}.
Figures are ordered from the upper left to the bottom right.
(a) N253; (b) N253; (c) N253; (d) N253; (e) N253; (f) N253; (g) N253; (h) N253; (i) N253; (j) N253; (k) N253; (l) N253; (m) HS117; (n) DDO82; (o) DDO82; (p) DDO82; 
    \label{cmdfig13}}
\end{figure}
\vfill
\clearpage
 
%-------------------
\begin{figure}[p]
\centerline{
\includegraphics[width=1.625in]{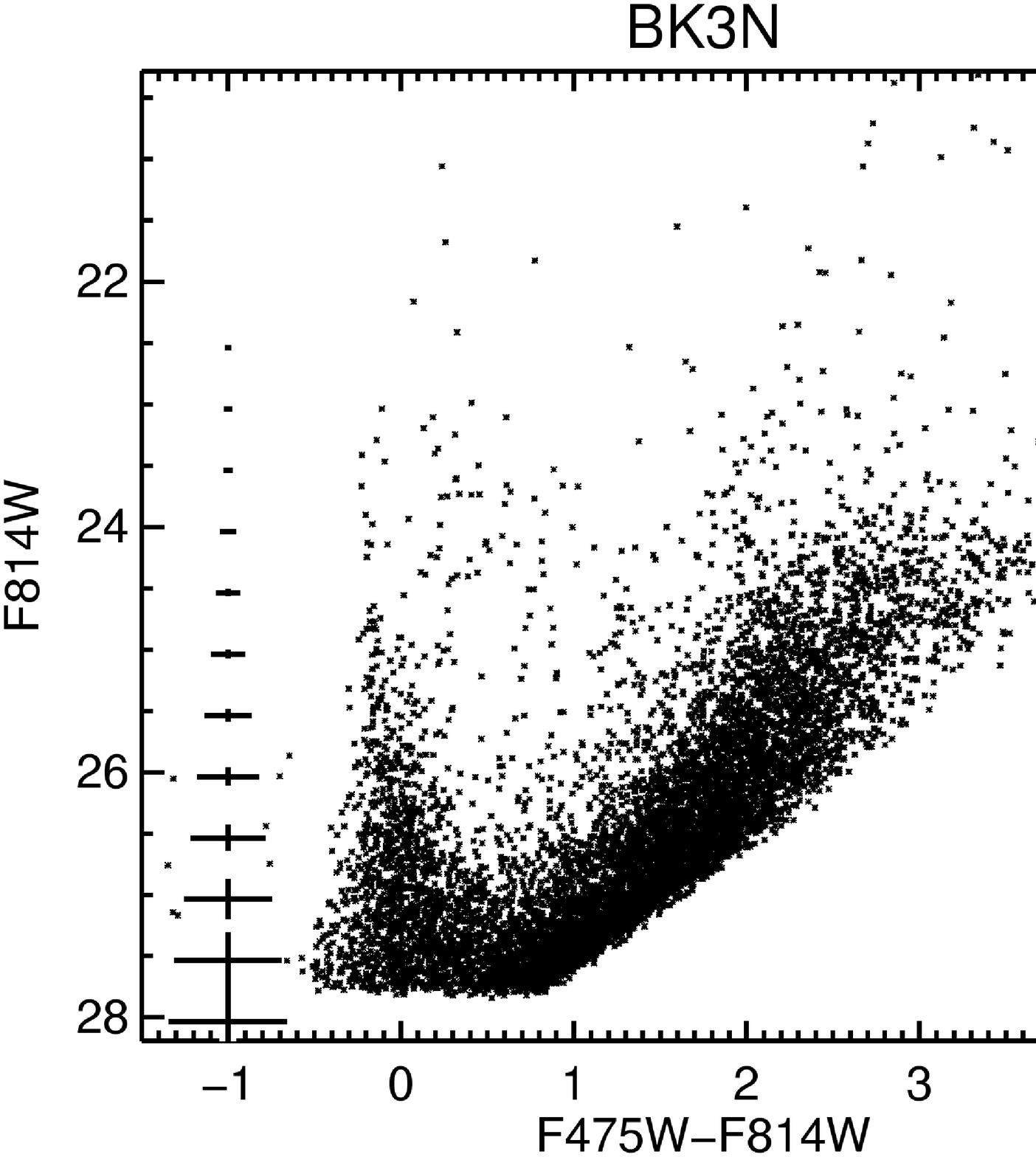}
\includegraphics[width=1.625in]{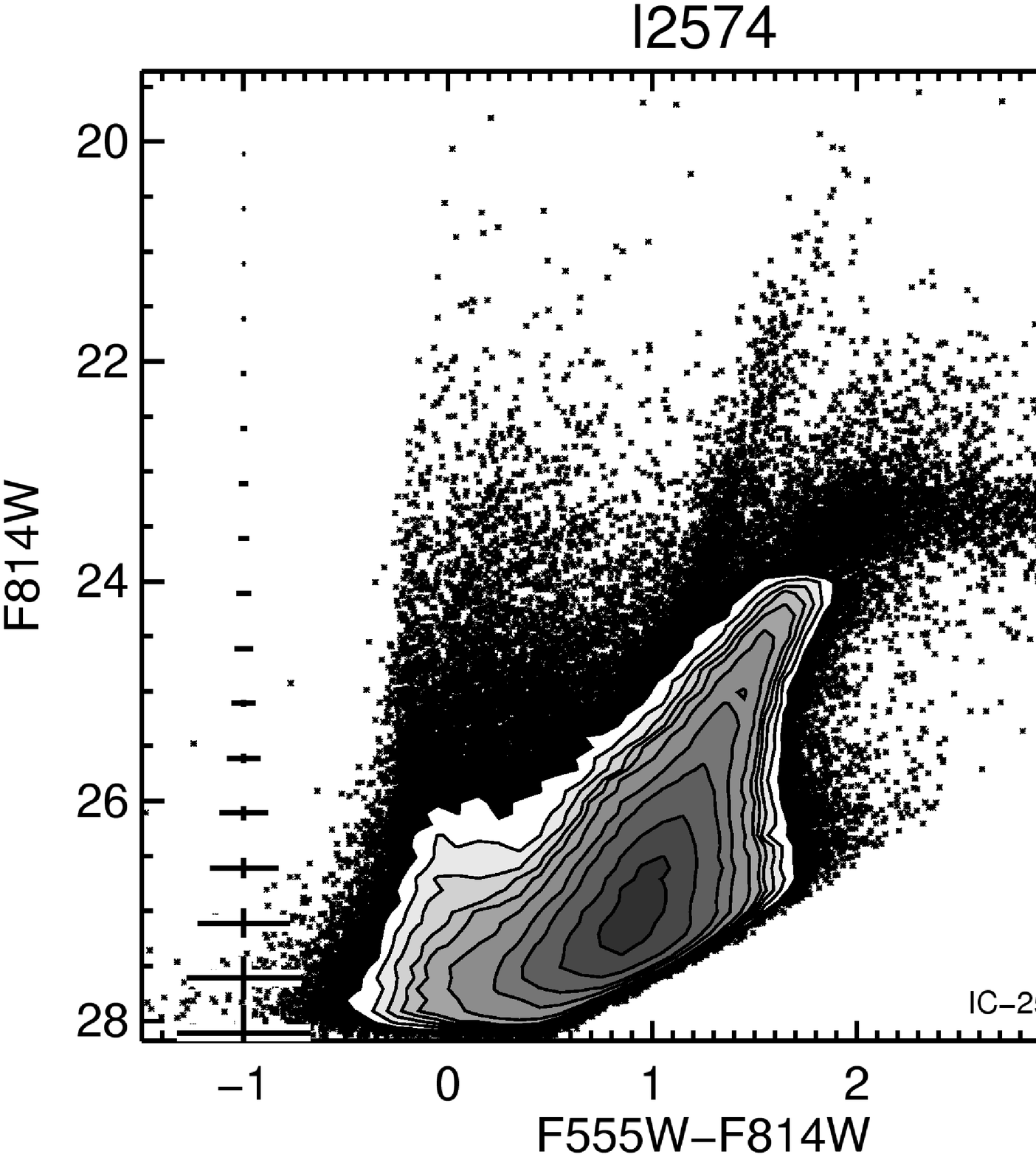}
\includegraphics[width=1.625in]{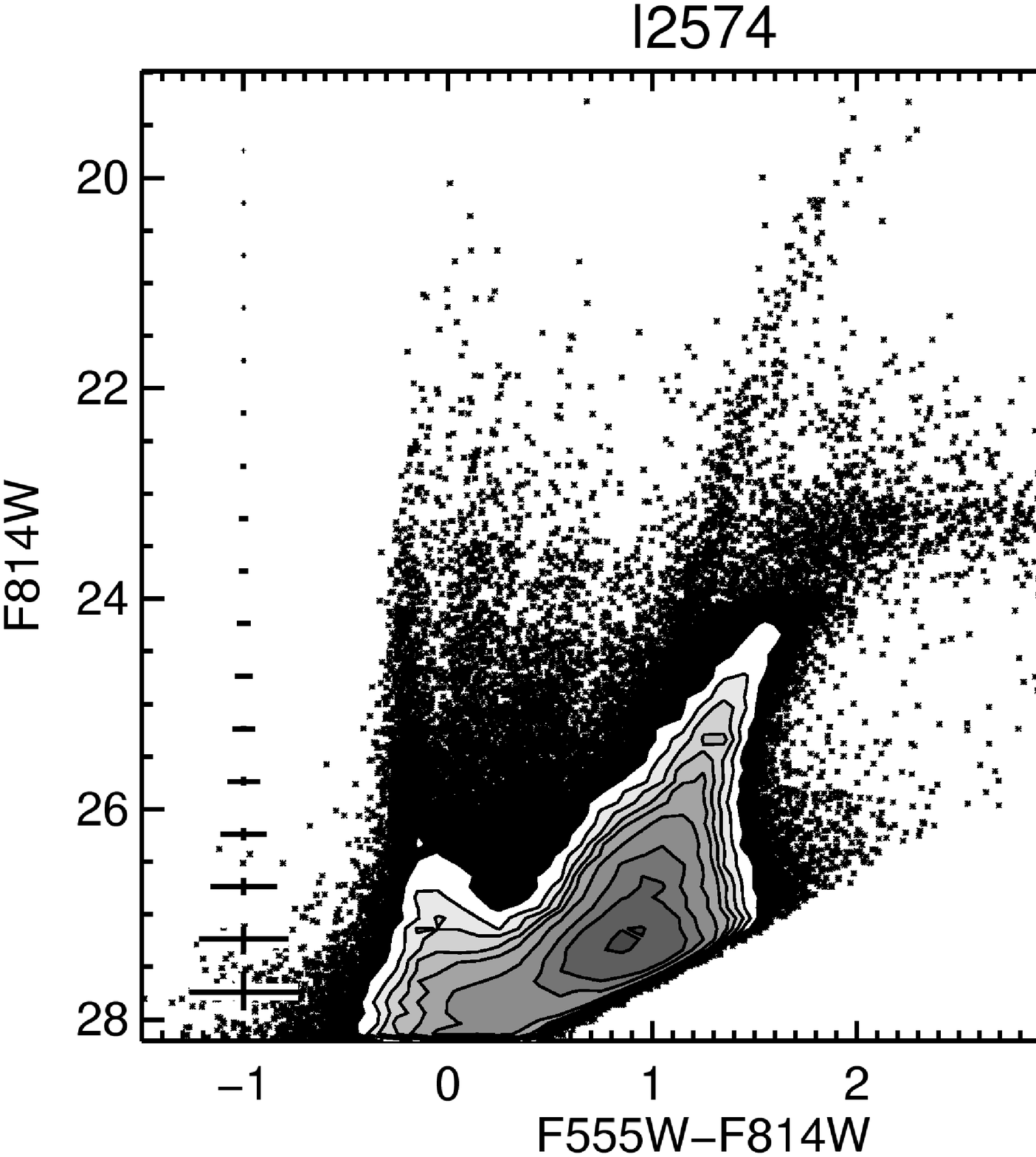}
\includegraphics[width=1.625in]{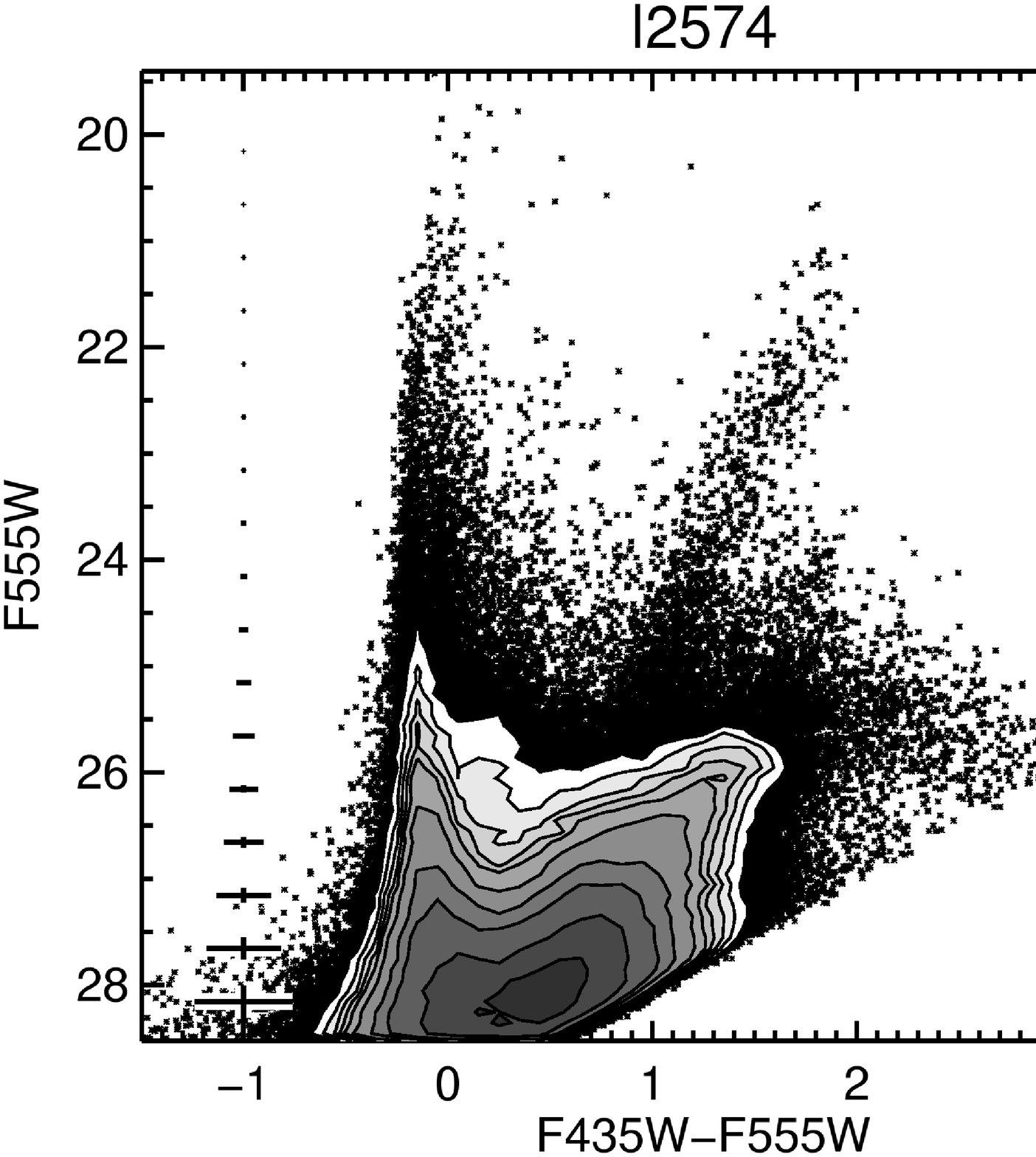}
}
\centerline{
\includegraphics[width=1.625in]{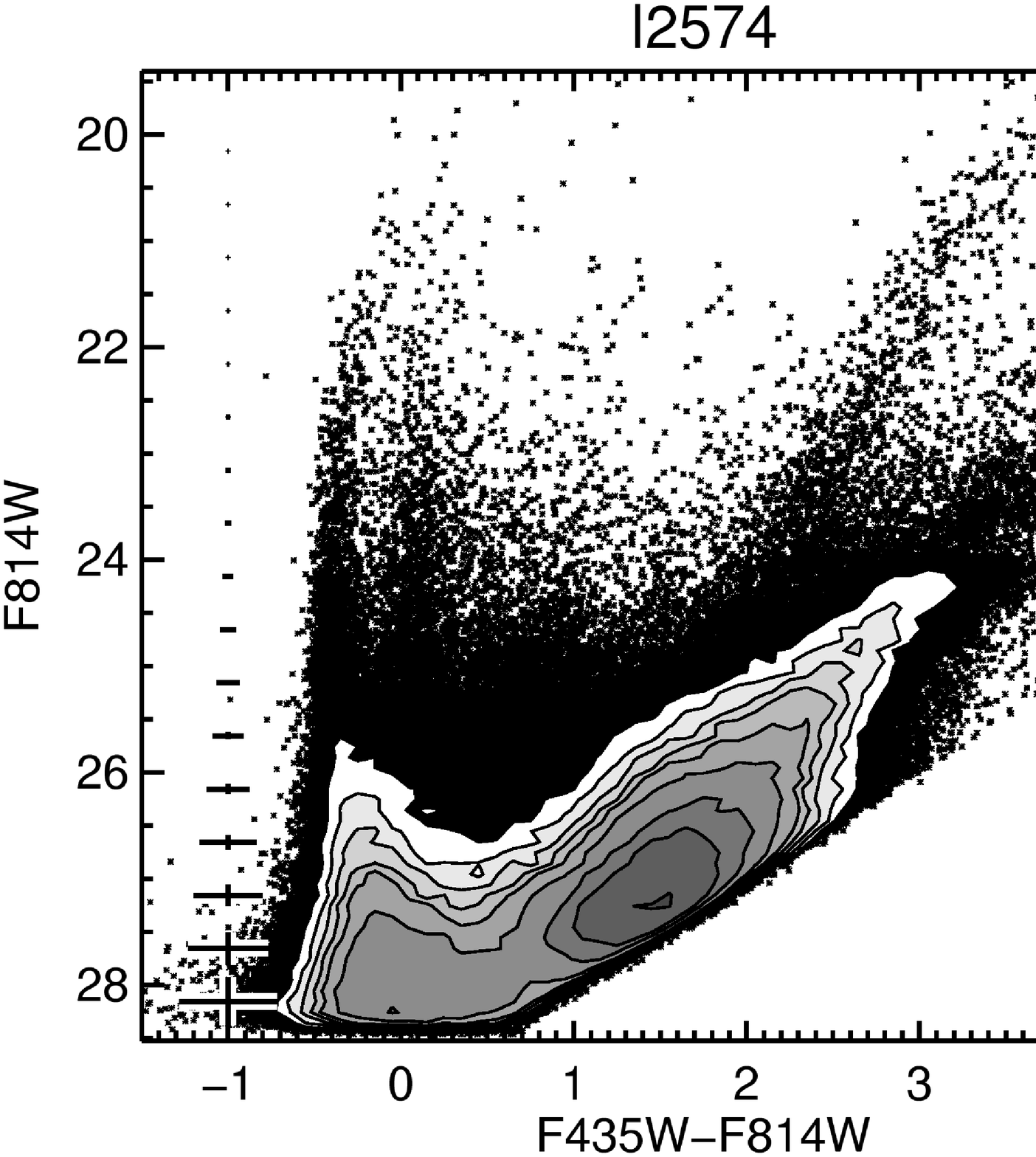}
\includegraphics[width=1.625in]{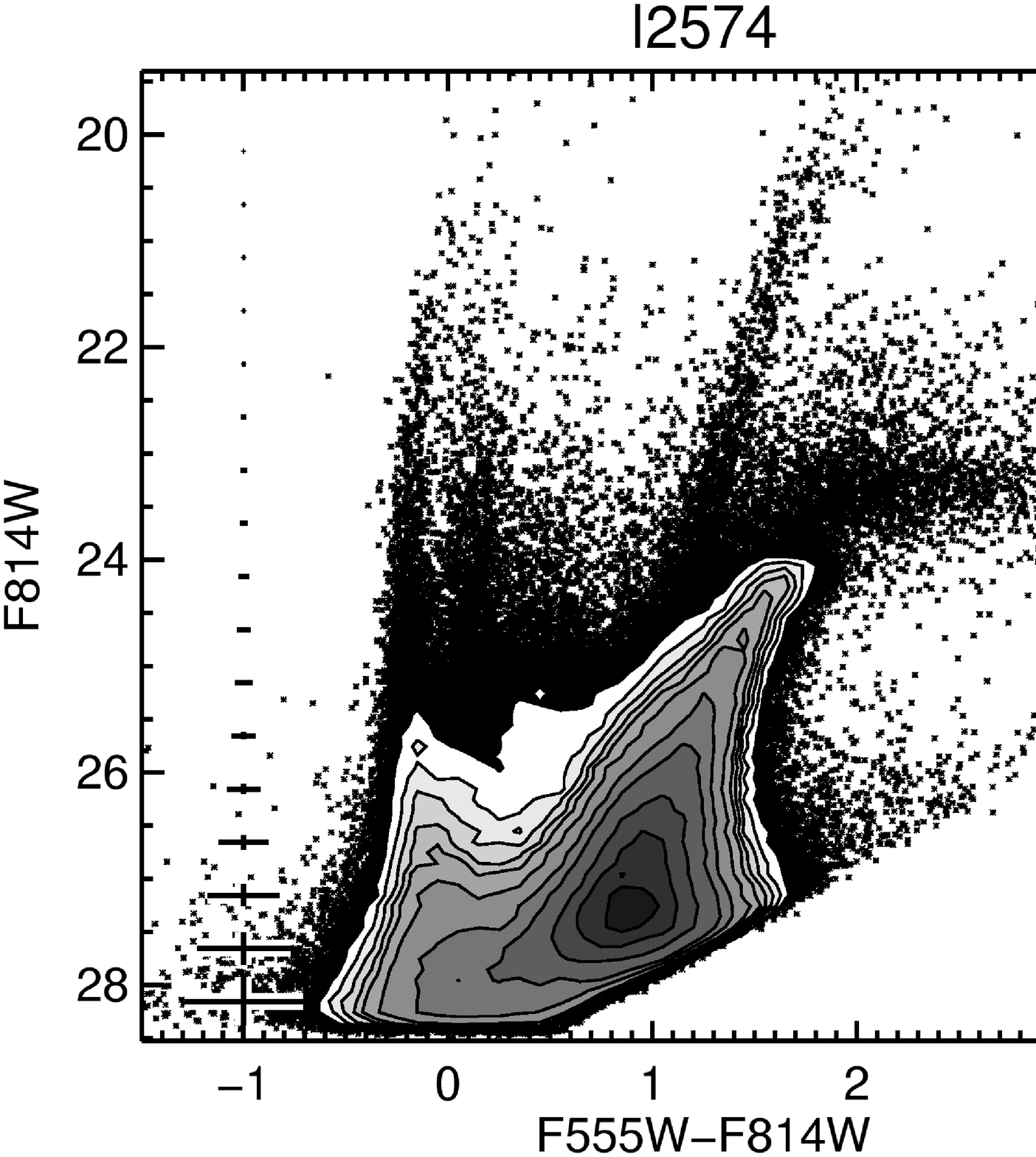}
\includegraphics[width=1.625in]{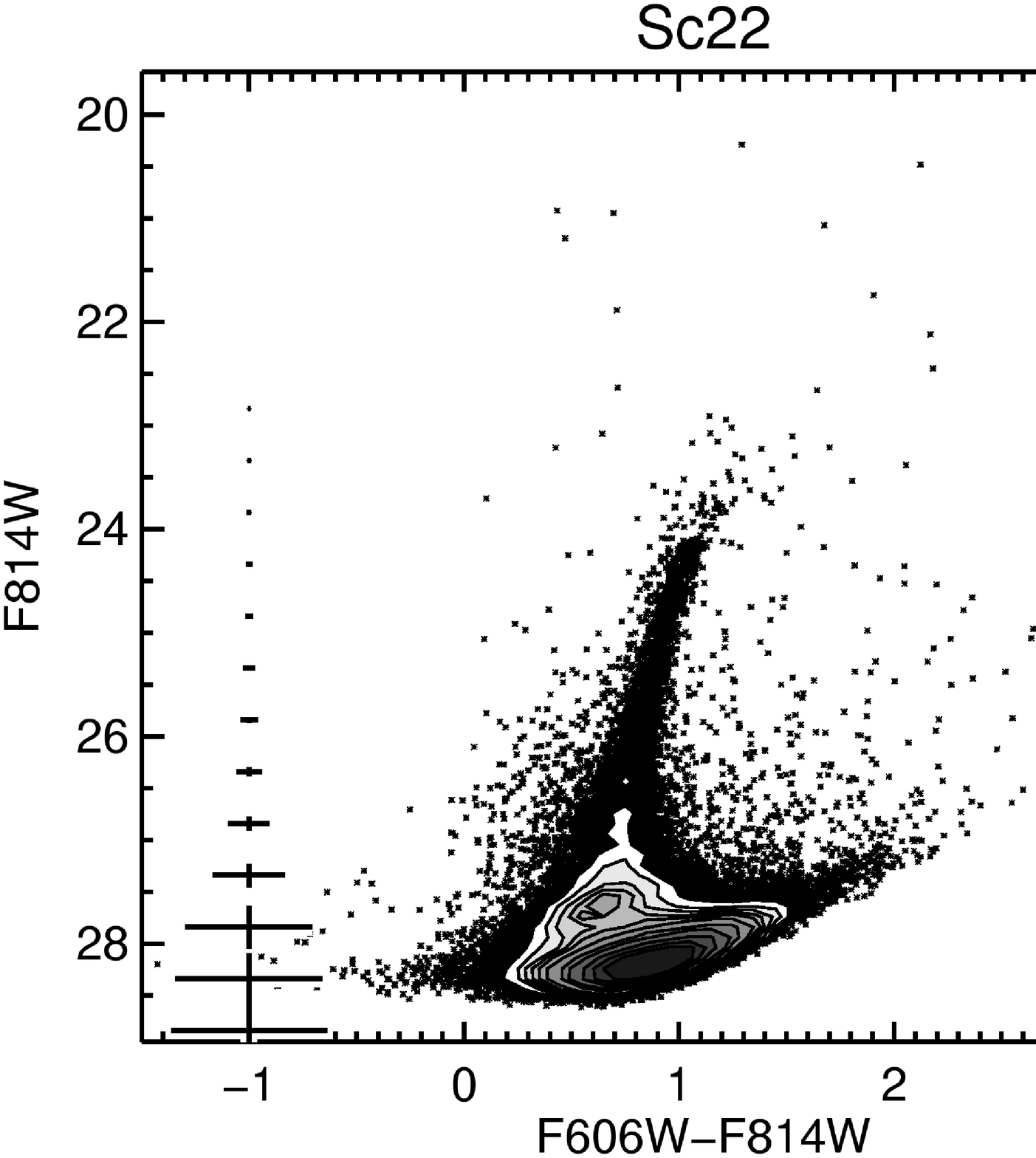}
}
\caption{
CMDs of galaxies in the ANGST data release,
as described in Figure~\ref{cmdfig1}.
Figures are ordered from the upper left to the bottom right.
(a) BK3N; (b) I2574; (c) I2574; (d) I2574; (e) I2574; (f) I2574; (g) Sc22; 
    \label{cmdfig14}}
\end{figure}
\vfill
%\clearpage

%------------------------
\begin{figure}[p]
\centerline{
\includegraphics[width=3.25in]{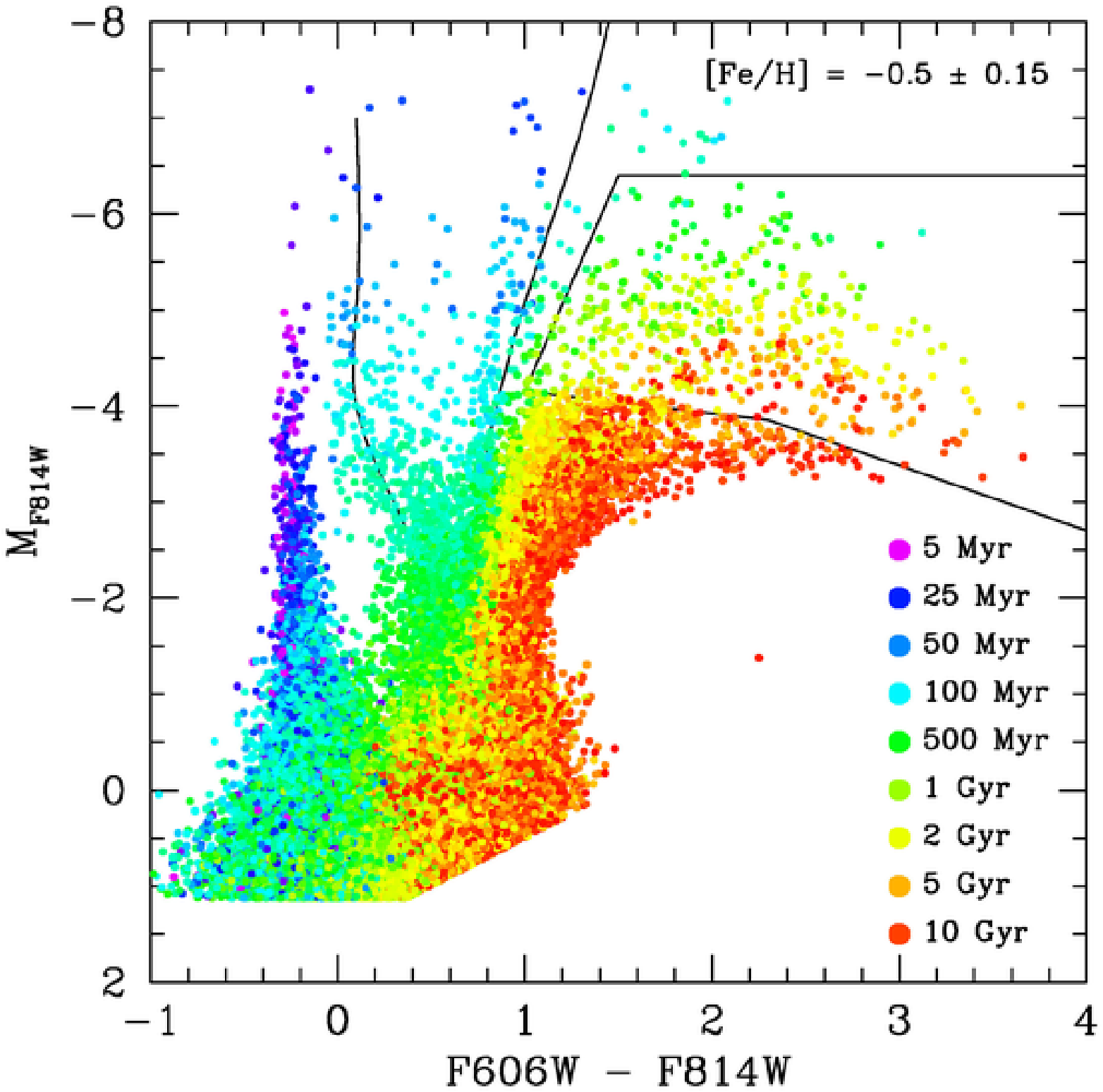}
\includegraphics[width=3.25in]{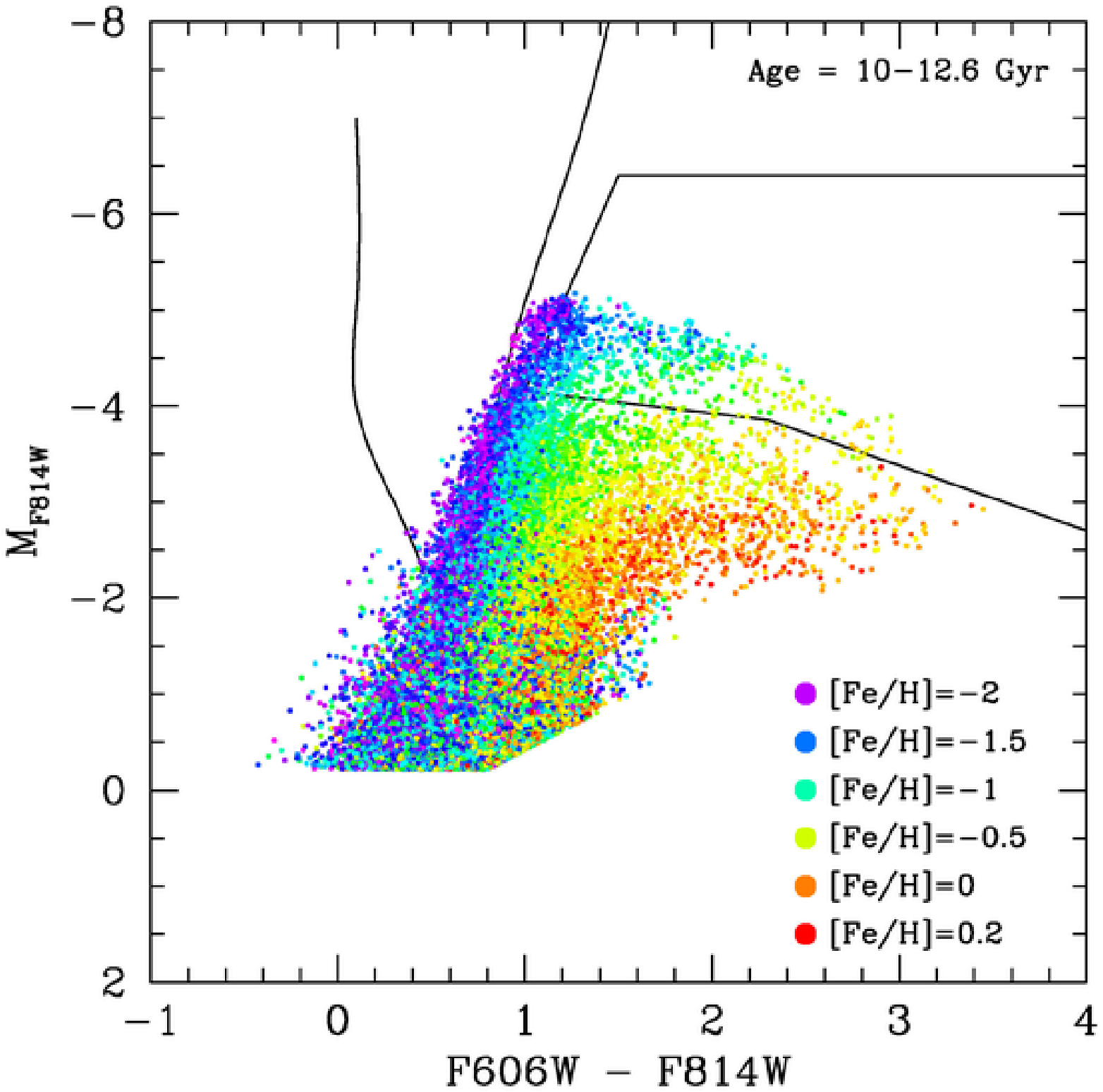}
}
\caption{Simulated CMDs for a constant star formation rate color-coded by age
(left) and for a uniform old age color-coded by metallicity (right).
The panels adopt the photometric errors and biases for the
NGC0300-WIDE1 (left) and NGC0253-WIDE1 (right) targets.  The three
(largely vertical) solid lines indicate several prominant sequences
identified with young stellar populations: the main sequence (leftmost
line), the blue core helium burning sequence (middle line), and the
red core helium burning sequence (rightmost line).  The enclosed polygon in the
upper right indicates the region typically occupied by AGB stars.  The
simulated CMDs assume the most recent \citet{girardi08} isochrone set.
\label{simcmdfig}}
\end{figure}
\vfill
\clearpage

%------------------------
\begin{figure}[p]
\centerline{
\includegraphics[width=6.5in]{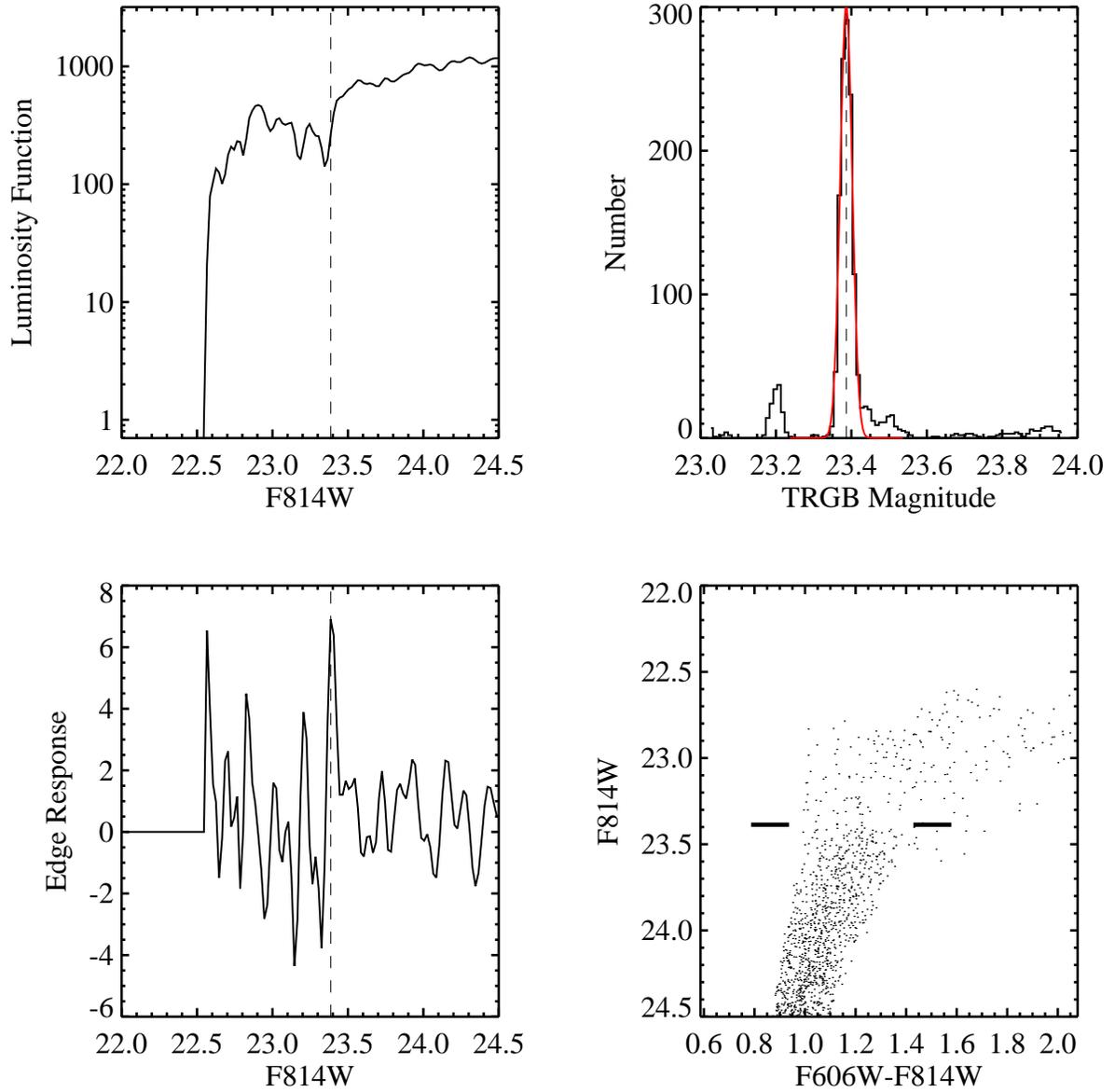}
}
\caption{Results of TRGB fitting for NGC~2403, showing the $F814W$
luminosity function (upper left), edge-detection response (lower
left), distribution TRGB magnitudes from of Monte Carlo trials (upper
right), and the CMD of stars used in the TRGB determination (lower
right).  The adopted TRGB magnitude is shown as the vertical line in
the first three panels, and as the two horizontal tic marks in the lower
right panel.
	 \label{TRGBfig}}
\end{figure}
\vfill
\clearpage

%------------------------
\begin{figure}[p]
\centerline{
\includegraphics[width=6.5in]{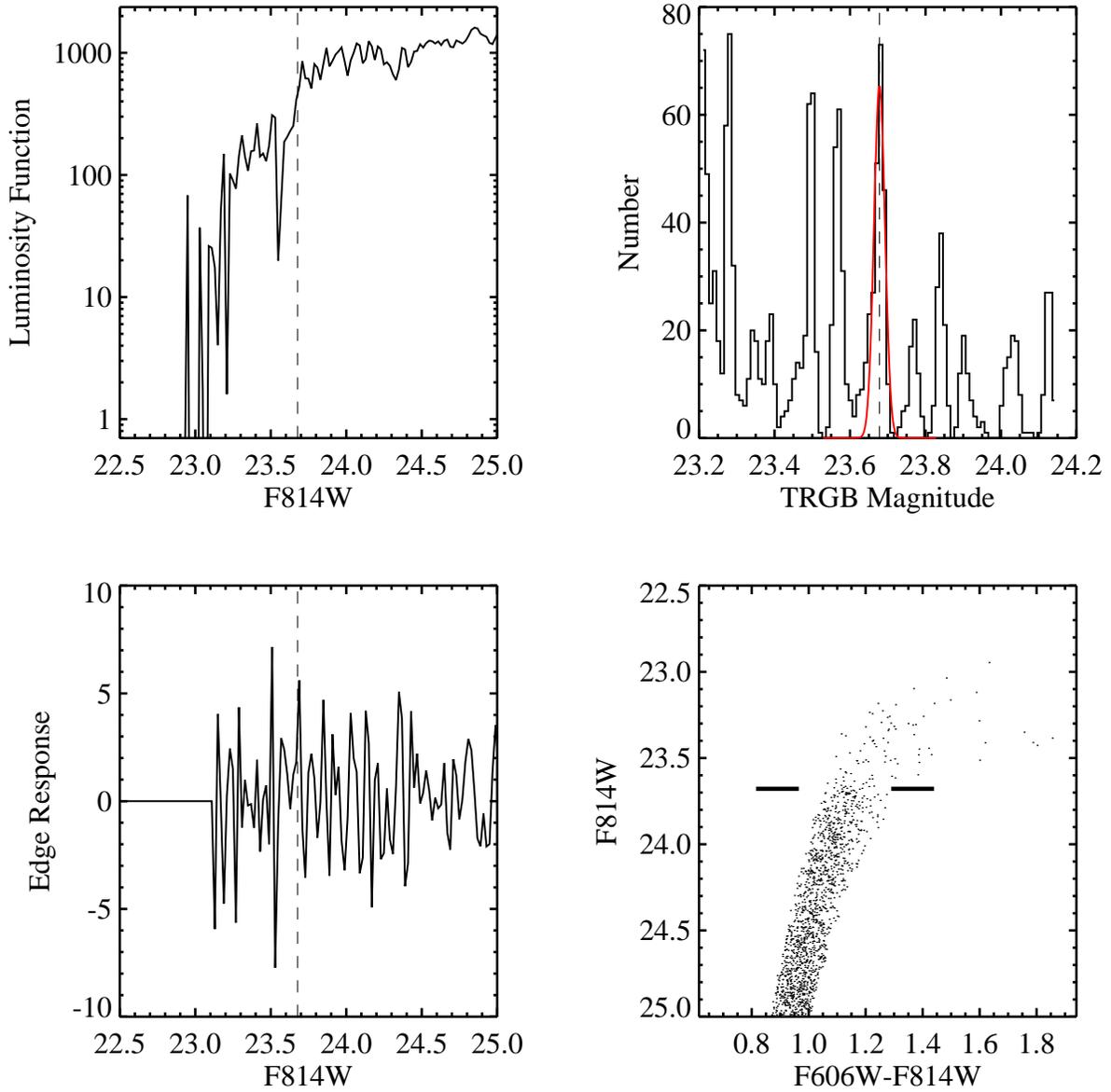}
}
\caption{Results of TRGB fitting for KDG63.  Panels are the same as in
Figure~\ref{TRGBfig}.  The histogram of Monte Carlo values is more
complicated than in Figure~\ref{TRGBfig}, due to the smaller number of
stars.
	 \label{TRGBbadfig}}
\end{figure}
\vfill
\clearpage

%------------------------
\begin{figure}[p]
\centerline{
\includegraphics[width=3.25in]{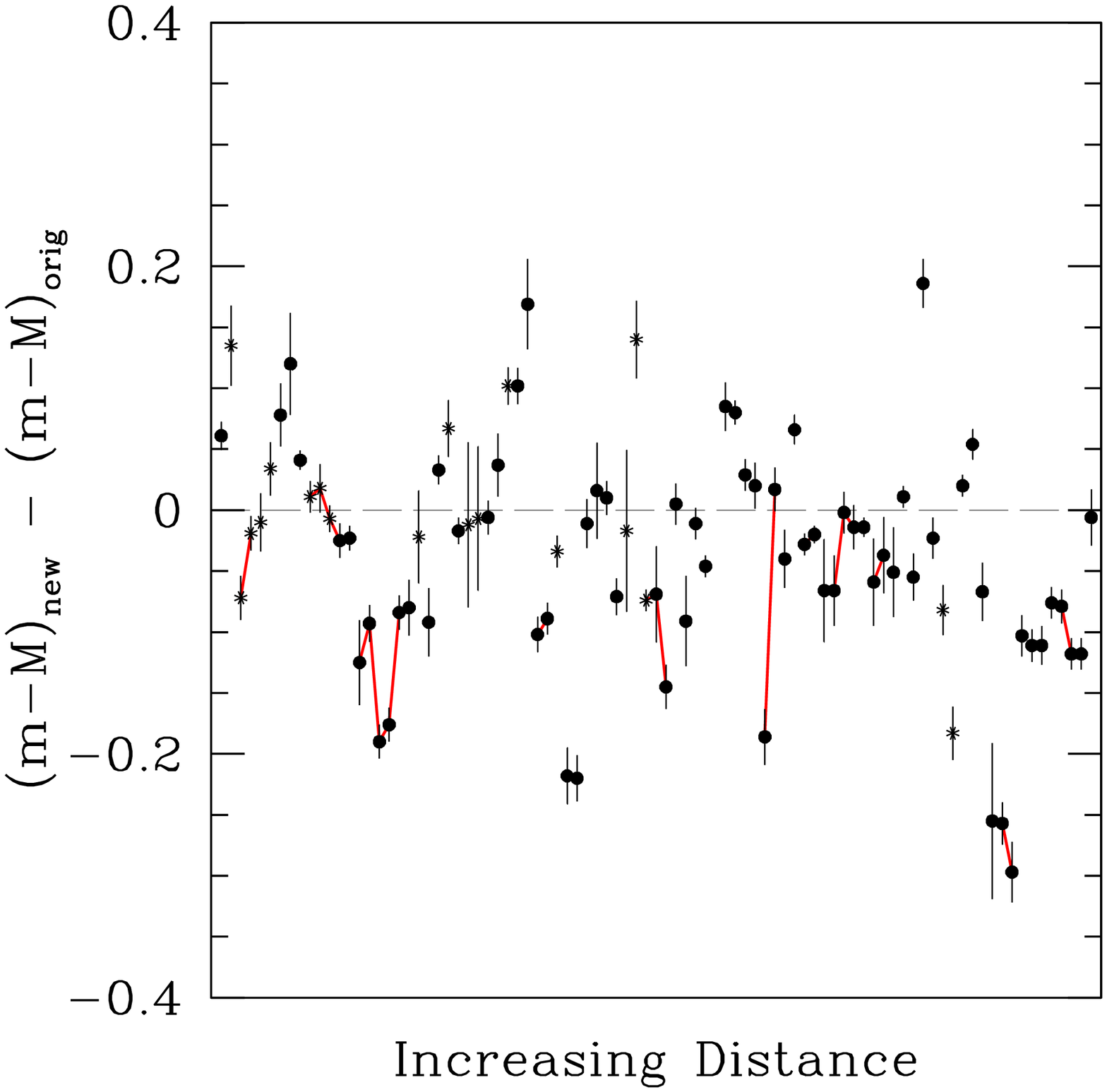}
\includegraphics[width=3.25in]{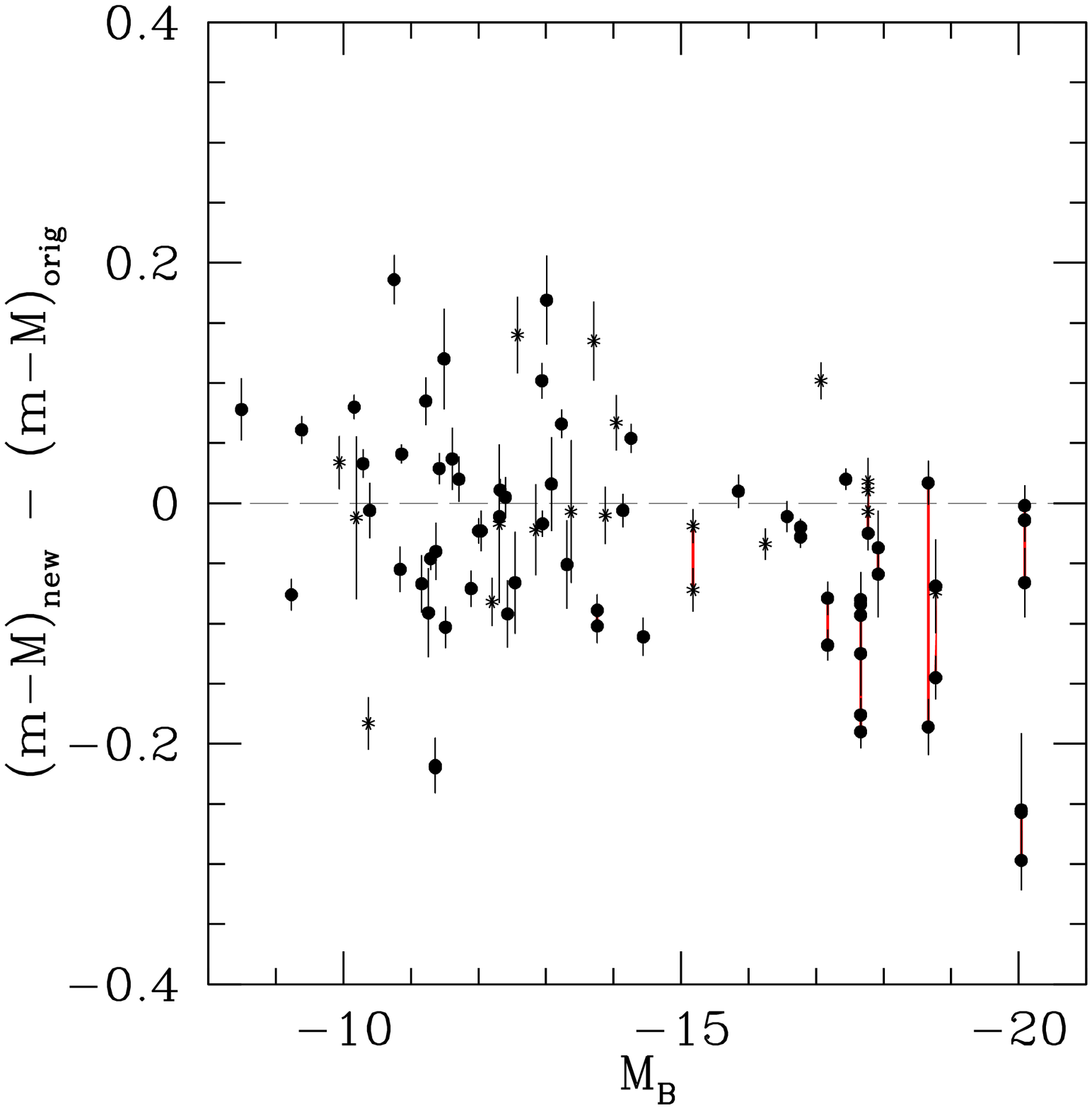}
}
\caption{Differences between the new TRGB distance moduli and those
  used in Table~\ref{sampletable} at the time of survey selection, as
  a function of either each target's position in
  Table~\ref{trgbtable}, which was sorted by the initial distance
  estimate (left) or absolute magnitude (right).  Solid circles are
  used for ACS data and asterix for WFPC2 data.  Error bars indicate
  the Monte Carlo uncertainties reported in Table~5, but do not
  include systematic uncertainties due to dust extinction or the
  adopted TRGB absolute magnitudes.  The median change in distance
  modulus is only $-0.02$ mag, and the dispersion about the mean is
  $0.05$ mag.  Multiple observations of the same galaxy are connected
  with a solid line (for NGC~3109, NGC~55, NGC~300, NGC~4163, UA292,
  NGC~2403, M82, NGC~2976, M81, NGC~247, NGC~253, DDO82, and IC~2574),
  and show differences of typically less than $0.1$ mag (i.e.\ 10\% in
  distance).  This variation is likely to be dominated by differences
  in internal extinction at different locations within the galaxy, with
  the outermost distance measurement being least likely to be
  affected by dust but more likely to be affected by Poisson 
  uncertainties due to reduced numbers of stars.  
  \label{TRGBdifffig}}
\end{figure}
\vfill
\clearpage

\fi

\end{document}